\RequirePackage{lineno}
\documentclass[prd,twocolumn,showpacs,amsmath,amssymb]{revtex4}
\usepackage{graphicx}
\usepackage{soul}
\usepackage{dcolumn}
\usepackage{bm}
\usepackage{overpic}
\usepackage{verbatim}
\usepackage[colorlinks,linkcolor=blue,anchorcolor=red,citecolor=blue,urlcolor=black]{hyperref}
\usepackage{float}
\usepackage{xcolor}
\usepackage{soul}
\floatstyle{plaintop}
\restylefloat{table}
\usepackage[justification=raggedright]{caption}
\usepackage{babel}
\usepackage[export]{adjustbox}

\usepackage{rotating}

\newcommand{\bfg}{\begin{figure}}
\newcommand{\efg}{\end{figure}}
\newcommand{\bitm}{\begin{itemize}}
\newcommand{\eitm}{\end{itemize}}
\newcommand{\bnum}{\begin{enumerate}}
\newcommand{\enum}{\end{enumerate}}
\newcommand{\btbl}{\begin{table}}
\newcommand{\etbl}{\end{table}}
\newcommand{\btbu}{\begin{tabular}}
\newcommand{\etbu}{\end{tabular}}

\def\beq{\begin{equation}}
\def\eeq{\end{equation}}

\def\BF{\mathcal{B}}
\def\semilep{\Ds\to Xe^{+}\nu_e}
\def\semilepBF{\BF\left(\semilep\right)}
\def\taucont{\Ds\to\tau^+\nu_
\tau\to e^+\nu_e\nu_\tau\overline{\nu}_\tau}

\def\D0{D^{0}}

\def\Ds{D_{s}^{+}}

\def\DsSTDs{D_{s}^{*+}D_{s}^{-}}
\def\DsDs{D_{s}^{+}D_{s}^{-}}

\def\ee{e^+e^-}

\def\Ecm{E_{\text{cm}}}

\def\Mrec{M_{\text{Rec}}}
\def\btag{b_{\text{tag}}}

\def\invfb{\text{ fb}^{-1}}

\def\GeV{\text{ GeV}}

\def\EcmA{\Ecm=4.178\GeV}
\def\EcmB{\Ecm=4.189-4.219\GeV}
\def\EcmC{\Ecm=4.225-4.230\GeV}

\begin{document}

\normalsize
\parskip=5pt plus 1pt minus 1pt

\title{\boldmath Measurement of the absolute branching fraction of inclusive semielectronic $\Ds$ decays}

\author{
M.~Ablikim$^{1}$, M.~N.~Achasov$^{10,b}$, P.~Adlarson$^{67}$, S. ~Ahmed$^{15}$, M.~Albrecht$^{4}$, R.~Aliberti$^{28}$, A.~Amoroso$^{66A,66C}$, M.~R.~An$^{32}$, Q.~An$^{63,49}$, X.~H.~Bai$^{57}$
% ...
\\
\vspace{0.2cm}
(BESIII Collaboration)\\
\vspace{0.2cm} {\it
$^{1}$ Institute of High Energy Physics, Beijing 100049, People's Republic of China\\
% ...
$^{4}$ Bochum Ruhr-University, D-44780 Bochum, Germany\\
% ...
$^{10}$ G.I. Budker Institute of Nuclear Physics SB RAS (BINP), Novosibirsk 630090, Russia\\
% ...
$^{15}$ Helmholtz Institute Mainz, Staudinger Weg 18, D-55099 Mainz, Germany\\
% ...
$^{28}$ Johannes Gutenberg University of Mainz, Johann-Joachim-Becher-Weg 45, D-55099 Mainz, Germany\\
% ...
$^{32}$ Liaoning Normal University, Dalian 116029, People's Republic of China\\
% ...
$^{49}$ State Key Laboratory of Particle Detection and Electronics, Beijing 100049, Hefei 230026, People's Republic of China\\
% ...
$^{57}$ University of Jinan, Jinan 250022, People's Republic of China\\
% ...
$^{63}$ University of Science and Technology of China, Hefei 230026, People's Republic of China\\
% ...
$^{66}$ University of Turin and INFN, (A)University of Turin, I-10125, Turin, Italy; (B)University of Eastern Piedmont, I-15121, Alessandria, Italy; (C)INFN, I-10125, Turin, Italy\\
$^{67}$ Uppsala University, Box 516, SE-75120 Uppsala, Sweden\\
% ...
$^{b}$ Also at the Novosibirsk State University, Novosibirsk, 630090, Russia\\
% ...
}
}

\date{\today}

\begin{abstract}
We measure the inclusive semielectronic decay branching fraction of
the $\Ds$ meson. A double-tag technique is applied to $\ee$
annihilation data collected by the BESIII experiment at the BEPCII
collider, operating in the center-of-mass energy range $4.178 - 4.230$
GeV. We select positrons from $\semilep$ with momenta greater than 200
MeV/$c$, and determine the laboratory momentum spectrum, accounting for
the effects of detector efficiency and resolution. The total positron
yield and semielectronic branching fraction are determined by
extrapolating this spectrum below the momentum cutoff.  We measure the
$\Ds$ semielectronic branching fraction to be
$\semilepBF=\left(6.30\pm0.13\;(\text{stat.})\pm
0.10\;(\text{syst.})\right)\%$, showing no evidence for unobserved
exclusive semielectronic modes. We combine this result with external
data taken from literature to determine the ratio of the $\Ds$ and
$D^0$ semielectronic widths, $\frac{\Gamma(D_{s}^{+}\rightarrow
  Xe^+\nu_e)}{\Gamma(D^0\rightarrow Xe^+\nu_e)}=0.790\pm
0.016\;(\text{stat.})\pm0.020\;(\text{syst.})$. Our results are
consistent with and more precise than previous measurements.
\end{abstract} 


\maketitle

\section{Introduction}

The $\Ds$ meson is the ground state of charmed-strange mesons, and
precise measurements of its semileptonic decays allow for crucial
tests of Standard Model predictions of flavor-changing
interactions. This article presents a new measurement of the
$\Ds$-meson inclusive semielectronic branching fraction and positron
momentum spectrum. (Here and throughout this article charge conjugate
modes are implied.)  A previous measurement by the CLEO-c
experiment reported the first results for these quantities,
including the measurement of
$\semilepBF=\left(6.52\pm0.39\;(\text{stat.})\pm
0.15\;(\text{syst.})\right)\%$~\cite{ref:CLEOMeas}.  Measuring this
branching fraction with improved precision contributes to a
comprehensive understanding of $D_s^+$ decays and is an important
component of the overall experimental and theoretical heavy-flavor
physics program.

\renewcommand{\arraystretch}{1.25}
\begin{table*}[hbt]
\centering{
\begin{tabular}{c|cc}
Mode                                         &  \multicolumn{2}{c}{Averaged Branching Fraction}      \\ \hline
  $\Ds\to\phi e^+\nu_e$                        & $(2.37\pm 0.11)\%$& \cite{ref:BESIII2020phiBF,ref:BESIII2018phiBF,ref:f0PDGBF,ref:BABAR2008phiBF} \\ \hline
$\Ds\to\eta e^+\nu_e$                          & $(2.32\pm 0.08)\%$& \cite{ref:BESIII2019etaBF,ref:BESIII2016etaBF,ref:f0PDGBF} \\ \hline
$\Ds\to\eta'e^+\nu_e$                          & $(0.80\pm 0.07)\%$ &\cite{ref:BESIII2019etaBF,ref:BESIII2016etaBF,ref:f0PDGBF} \\ \hline
$\Ds\to K^0 e^+\nu_e$                          & $(0.34\pm 0.04)\%$&\cite{ref:BESIII2019KBF,ref:f0PDGBF} \\ \hline
$\Ds\to K^{*}(892)^0 e^+\nu_e$                 & $(0.21\pm 0.03)\%$& \cite{ref:BESIII2019KBF,ref:f0PDGBF} \\\hline 
$\Ds\to f_{0}(980)e^+\nu_e, \;f_{0}(980)\to \pi\pi$                 & $(0.30\pm 0.05)\%$&\cite{ref:newf0PDGBF}\\ \hline\hline
Sum of Semielectronic Modes                  & $(6.34\pm 0.17)\%$& \\ \hline\hline
$\semilepBF$                                 & $(6.5 \pm 0.4)\%$& \cite{ref:CLEOMeas}                    \\ \hline\hline
$\Ds\to\tau^+\nu_\tau\to e^+\overline{\nu}_\tau\nu_e\nu_\tau$ & $(0.96\pm 0.04)\%$& \cite{ref:BESIII2016tauBF,ref:Belle2013tauBF,ref:BABAR2010tauBF,
ref:CLEO2009taurhoBF,ref:CLEO2009taupiBF,ref:CLEO2009taueBF}     \\\hline
\end{tabular}
}

\raggedright
\caption{Branching fractions for observed $\Ds$ semielectronic decays and for $\taucont$. The listed uncertainties are the total uncertainties. The $\Ds\to f_{0}(980)e^+\nu_e, \;f_{0}(980)\to \pi\pi$ branching fraction is calculated based on the measurements of  $\Ds\to\pi^+\pi^-e^+\nu_e$ from~\cite{ref:newf0PDGBF} and corrected for the corresponding $\pi^0\pi^0$ branching fraction by an isospin correction factor of $\frac{3}{2}$.}
\label{table:BFS}
\end{table*}

\renewcommand{\arraystretch}{1.}

Table~\ref{table:BFS} lists the six exclusive $D_s^+$ semielectronic
modes that have been observed to date and their branching
fractions, as well as the branching fraction of $\taucont$
and the previously measured $\semilep$ branching fraction
\cite{ref:CLEOMeas,ref:BESIII2020phiBF,ref:BESIII2018phiBF,ref:f0PDGBF,ref:newf0PDGBF,ref:BABAR2008phiBF,ref:BESIII2019etaBF,ref:BESIII2016etaBF,ref:BESIII2019KBF,ref:BESIII2016tauBF,ref:Belle2013tauBF,ref:BABAR2010tauBF,ref:CLEO2009taurhoBF,ref:CLEO2009taupiBF,ref:CLEO2009taueBF}. By
comparing the inclusive semielectronic branching fraction with the sum
of all measured exclusive semielectronic branching fractions, we can
estimate the branching fraction for further unobserved
$D_s^+$ semielectronic decays.  Measurements of the semielectronic
branching fractions for different charmed mesons can be combined to
probe for non-spectator eﬀects in heavy-meson decays
\cite{ref:Bigi,ref:Voloshin}.  The CLEO-c measurement of the ratio of
the $D_s^+$ and $D^0$ semielectronic widths,
$\frac{\Gamma(D_s^+\rightarrow X e^+ \nu_e)}{\Gamma(D^0\rightarrow X
  e^+ \nu_e)} = 0.815 \pm 0.052$, is in agreement with predictions
employing an eﬀective quark model~\cite{ref:Rosner} and shows that
non-spectator eﬀects are present in semielectronic charmed-meson
decays.  It has also been demonstrated with CLEO-c data that the
inclusive semielectronic momentum spectrum can be used to make
sensitive tests for specific non-spectator processes, such as Weak
Annihilation (WA)~\cite{ref:Manohar,ref:Gambino}. Strong understanding of these processes are required for application of heavy-quark-expansion in extracting CKM elements from inclusive semileptonic $B$ meson decays~\cite{ref:Vos}.  Thus, the improved
precision of both the inclusive branching fraction and the momentum
spectrum of $D_s^+$ semielectronic decays reported in this article
have potential to contribute to reducing theoretical uncertainties in
determining CKM parameters with heavy-meson decays.

The remainder of the article is organized in seven sections. The BESIII detector, the analyzed data, and the Monte Carlo (MC) simulation samples are described in Sec.~\ref{sec:detector}. An overview of the measurement technique is presented in Sec.~\ref{sec:technique}. Event-selection requirements based on full reconstruction of hadronic $D_s^-$ decays are discussed in Sec.~\ref{sec:stag}. Semielectronic decay selection requirements and further analysis of candidate signal events are presented in Sec.~\ref{sec:analysis}. The systematic uncertainties of our measurement are evaluated in Sec.~\ref{sec:sys}. We conclude with a summary of our results in Sec.~\ref{sec:conc} and acknowledgements in Sec.~\ref{sec:acknowledgements}.

\section{Detector and data samples}\label{sec:detector}

The BESIII detector records the results of symmetric $e^+e^-$ collisions provided by the BEPCII collider~\cite{ref:BEPCII}. BEPCII produces collisions at center-of-mass energies ($\Ecm$) between $2$ and $4.9$ GeV, and BESIII has collected the world's largest data samples near a number of pair-production threshold energies for charmed hadrons. The BESIII detector is composed of the following sub-systems for particle detection and identification: a helium-based multilayer drift chamber (MDC), a plastic scintillator time-of-flight system (TOF), a CsI(Tl) electromagnetic calorimeter (EMC), a 1.0 T superconducting solenoid, and a set of resistive plate chambers for muon identification (MUC). The acceptance of the BESIII detector for charged particles is $93\%$ of the full solid angle. Ionization energy deposits in the MDC are used to reconstruct charged particle tracks and determine particle momenta from the curvature in the magnetic field. The MDC provides a momentum resolution of $0.5\%$ for particles with a momentum of $1$ GeV/$c$. The end cap TOF system was upgraded in 2015 with multi-gap resistive plate chamber technology, providing a time resolution of 60 ps~\cite{ref:TOF}. Measurements of charged particle specific ionization in the MDC ($dE/dx$), the flight time in the TOF, and the energy deposit in the EMC associated with a track are combined to identify particles. A detailed description of the BESIII detector is given in~\cite{ref:BESIII}.

Inclusive MC samples with the equivalent of five times the luminosity of
data and exclusive
MC samples with the equivalent of thirty-five times the luminosity of
data are used to optimize selection criteria, investigate
distributions of signal and background processes, and determine the
efficiencies of our selection criteria. Hadronic simulation samples
are produced using the event generators KKMC~\cite{ref:KKMC} and
EvtGen
\cite{ref:EvtGen,ref:BesEvtGen}. BabayagaNLO~\cite{ref:BabayagaNLO} is
used to produce radiative Bhabha $(e^+e^-\to\gamma e^+e^-)$
samples. Final-state radiation (FSR) of particles is simulated with
PHOTOS~\cite{ref:PHOTOS}. Interaction of the simulated particles with
the detector material is handled by GEANT4 \cite{ref:GEANT4}, which
uses a detailed XML-based description of the detector
geometry~\cite{ref:Bes3Geometry}.

In 2016, BESIII collected $3.19 \text{ fb}^{-1}$ of data at $\EcmA$, which provides approximately $6.4 \times 10^6$ $\Ds$ mesons primarily through the process $e^+e^-\to\DsSTDs$, with a small contribution from $e^+e^-\to\DsDs$. We analyze the entirety of this sample, as well as $2.08\invfb$ of data collected at center-of-mass energies in the range $\EcmB$ in 2017, and $1.05\invfb$ of data collected in the range $\EcmC$ in 2013. These three samples are analyzed separately due to differing detector and running conditions. A summary of the data sets with their $\Ecm$~\cite{ref:4230EBeam}, integrated luminosities~\cite{ref:4230Lum}, and estimated number of $\Ds$ mesons produced is shown in Table~\ref{table:samples}.   

\begin{table}[hbt]
\centering{
\begin{tabular}{c|c|c}
$\Ecm$ (MeV)& $\int \mathcal L \;dt$ ($\text{pb}^{-1}$)& $N_{D_s} (\times 10^6)$\\ \hline
$ 4178$ & $3189.0 \pm 0.9 \pm 31.9$ & $6.4$ \\  \hline
$4189$   & $526.7 \pm 0.1 \pm 2.2$  &$1.0$\\
$4199$   & $526.0 \pm 0.1 \pm 2.1$  &$1.0$\\ 
$4209$  & $517.1 \pm 0.1 \pm 1.8$  &$0.9$\\ 
$4219 $   & $514.6 \pm 0.1 \pm 1.8$  &$0.8$\\ \hline
$4225-4230$ \cite{ref:4230EBeam}  &$1047.3 \pm 0.1 \pm 10.2$ \cite{ref:4230Lum}&$1.3$
\end{tabular}
}

\raggedright
\caption{$\Ecm$, integrated luminosities, and estimated number of $D_s$ mesons $(N_{D_s})$ for the analyzed data samples. In each case, the first listed uncertainty is statistical and the second is systematic.}
\label{table:samples}
\end{table}

\section{Measurement Technique}\label{sec:technique}

Our analysis procedure employs the double-tag technique pioneered by the MARKIII collaboration~\cite{ref:MARKIII}. We fully reconstruct hadronic $D_s^-$ mesons (the tag or tag-side), and determine the number of signal events by analyzing the remaining charged tracks unused in the reconstruction of the tags (the recoil or recoil-side). We refer to events where a tag meson is found as single-tag events, and events where a semielectronic decay is identified on the recoil side in addition to the tag meson as double-tag events. Our single-tag selection is described in more detail in Sec.~\ref{sec:stag}. 

Since identification and reconstruction of positron tracks is only possible above a certain momentum threshold with the BESIII detector, double-tag candidate tracks are only accepted with $p > 200 \text{ MeV}/c$. The differential inclusive semielectronic branching fraction as a function of the electron momentum $p_e$ ($ \frac{d\mathcal B_\text{SL}}{dp_e}$)  is given as

\begin{equation}
\frac{d\mathcal B_\text{SL}}{dp_e}=\frac{\Delta\left(\frac{ n_\text{DT}}{\epsilon_\text{DT}}\right)/\Delta p_{e}}{n_\text{ST}/\epsilon_\text{ST}}=\frac{\left(\frac{\Delta n_\text{DT}}{\epsilon_\text{sig}}\right)/\Delta p_e}{n_\text{ST}\frac{\epsilon'_\text{ST}}{\epsilon_\text{ST}}}=\frac{\Delta N_\text{DT}/\Delta p_e}{n_\text{ST}\btag}.
\label{eqn:DiffBRF}
\end{equation}
In this equation $n_\text{ST}$ is the number of the observed single-tag events, $\Delta n_\text{DT}$ is the number of the observed (semielectronic) double-tag events in a particular $p_e$ momentum bin, and $\epsilon_\text{ST}$ and $\epsilon_\text{DT}$ are the single-tag and double-tag reconstruction efficiencies, respectively. We define $\epsilon_\text{DT}=\epsilon'_\text{ST}\times \epsilon_\text{sig}$, where $\epsilon_\text{sig}$ is the momentum-dependent efficiency of reconstructing only the signal side, and $\epsilon'_\text{ST}$ is the tag-side reconstruction efficiency given that a signal event is present on the recoil-side. We define $\btag\equiv \frac{\epsilon'_\text{ST}}{\epsilon_\text{ST}}$ as the tag bias, which accounts for the difference in the single-tag detection efficiency given that a semielectronic decay is present on the recoil side, and $\Delta N_\text{DT}$ as the true number of double-tag signal events in our single-tag sample for a particular $p_e$ momentum bin.

We begin our signal selection by sorting recoil-side tracks with $p>
200 \text{ MeV}/c$ into momentum bins.  For a given momentum bin, we
further sort these tracks by their charge: tracks with charge opposite
to the tag (right-sign or RS) and tracks with the same charge as the
tag (wrong-sign or WS). The WS sample is used to estimate
charge-symmetric backgrounds in the RS distributions, which include
the true signal. For both charge categories in each momentum bin, we
sort tracks into three mutually exclusive particle identification
hypotheses: electron/positron ($e$ ID), pion ($\pi$ ID), and kaon ($K$
ID). The similar detector response to muons and pions allows us to
treat muons as identical to pions with negligible uncertainty. For
each of these six categories in a given momentum bin, we determine the
number of tracks that originate from a true $D_s^+$ meson by fitting
to the invariant-mass distribution of the tag $D_s^-$ candidate.

We relate the true number of double-tag signal events in our single-tag sample and the observed number of signal-candidate tracks from double-tag events through detector response matrices for charged particle tracking ($A_\text{trk}$) and PID ($A_\text{PID}$):

\begin{equation}\label{eqn:unfolding}
n^b_\text{DT}(p_{i})=\sum_a A_\text{PID}(b|a,p_{i})\sum_{j}A_\text{trk}(a|p_{i},p_{j})N_\text{DT}^a(p_{j}).
\end{equation}
Here, $A_\text{trk}(a|p_{i},p_{j})$ gives the probability of a
particle of type $a$ ($a\in \{e,\pi,K\}$) tracked in momentum bin
$p_{i}$ to have true momentum in bin
$p_{j}$. $A_\text{PID}(b|a,p_{i})$ gives the probability of a particle
of type $a$ passing the PID requirements of particle type $b$ ($b\in
\{e,\pi,K\}$) in track momentum bin $p_{i}$.


We determine the observed double-tag yields $n^b_\text{DT}(p_{i})$ by fitting the invariant-mass distribution of tag-side mesons. For both RS and WS tracks in each momentum bin, we account for $e$ ID efficiencies and particles faking $e$ ID with the PID unfolding method introduced in Eq.~\eqref{eqn:unfolding}:

\begin{equation}
\small
\label{eqn:PIDUnf}
\begin{bmatrix}
n^e_\text{trk}\\
n^\pi_\text{trk}\\
n^K_\text{trk}\\
\end{bmatrix}=A^{-1}_\text{PID}
\begin{bmatrix}
n^e_\text{DT}\\
n^\pi_\text{DT}\\
n^K_\text{DT}\\
\end{bmatrix},
 \;  A_\text{PID}=\begin{bmatrix}
\epsilon_{e} & \text{P}_{\pi\to e} & P_{K\to e} \\
P_{e\to \pi} & \epsilon_{\pi} & P_{K\to \pi} \\ 
P_{e\to K} & P_{\pi\to K} & \epsilon_{K}
\end{bmatrix}.
\end{equation}
Here, $\epsilon_a$ is the efficiency of our PID selection requirements for the particles of species $a$ and $P_{a\to b}$ is the probability of a particle of species $a$ passing the selection requirements for a particle of species $b$. We apply this PID unfolding to the fitted RS and WS yields for each momentum bin and take the $e$ solution to determine the number of tracked positrons originating from a $D_s^+$ meson. We then take the difference of the number of tracked RS positrons and WS positrons to subtract the contributions of positrons from Dalitz decays of light mesons in the final state of $\Ds$ decay such as $\pi^0\to\gamma e^+e^-$ and any other charge-symmetric backgrounds.

This leaves us with the momentum spectrum of tracked positrons from $\semilep$ events and a smaller contribution from $D_s^+\to\tau^+\nu_\tau\to e^+\nu_e\nu_\tau\overline\nu_\tau$. To account for tracking inefficiency and mismeasurement, we execute the second unfolding from Eq.~\eqref{eqn:unfolding}:

\begin{equation} 
\label{eqn:TrkUnf}
N_\text{DT}(p_j)=A^{-1}_\text{trk}n^e_\text{trk}(p_i).
\end{equation}
This provides the true sum of $\semilep$ and $ \taucont$ positron momentum spectra. We then subtract the contribution of $\taucont$ based on the branching fraction in Table~\ref{table:BFS} to obtain the $\semilep$ momentum spectrum for $p_{e}>200\text{ MeV}/c$. 

To account for positrons with $p_{e}\leq 200 \text{ MeV}/c$, we fit the expected $\semilep$ momentum spectrum with an assumed spectrum, described in more detail in Sec.~\ref{sec:analysis}, and add the fitted yields in the region $p_{e}\leq 200 \text{ MeV}/c$ to the measured yields with $p_{e}>200\text{ MeV}/c$. This gives us the signal-efficiency-corrected number of signal events 
\begin{equation}\label{eq:extrap}
N_\text{DT}=N_{p_e>200 \text{ MeV/} c}+N_{p_e\leq 200 \text{ MeV/} c},
\end{equation} which allows us to calculate the branching fraction similarly to Eq.~\eqref{eqn:DiffBRF}  through

\begin{equation}
\semilepBF=\frac{N_\text{DT}}{n_\text{ST}\btag},
\label{eqn:DTagBRF}
\end{equation}
with the MC simulation prediction of $\btag$ and the determined number of single-tag events $n_\text{ST}$.

\section{Single-Tag Event Selection}\label{sec:stag}

\begin{figure}[hbt]
\centering{
\begin{tabular}{c}
\includegraphics[width=3.25in]{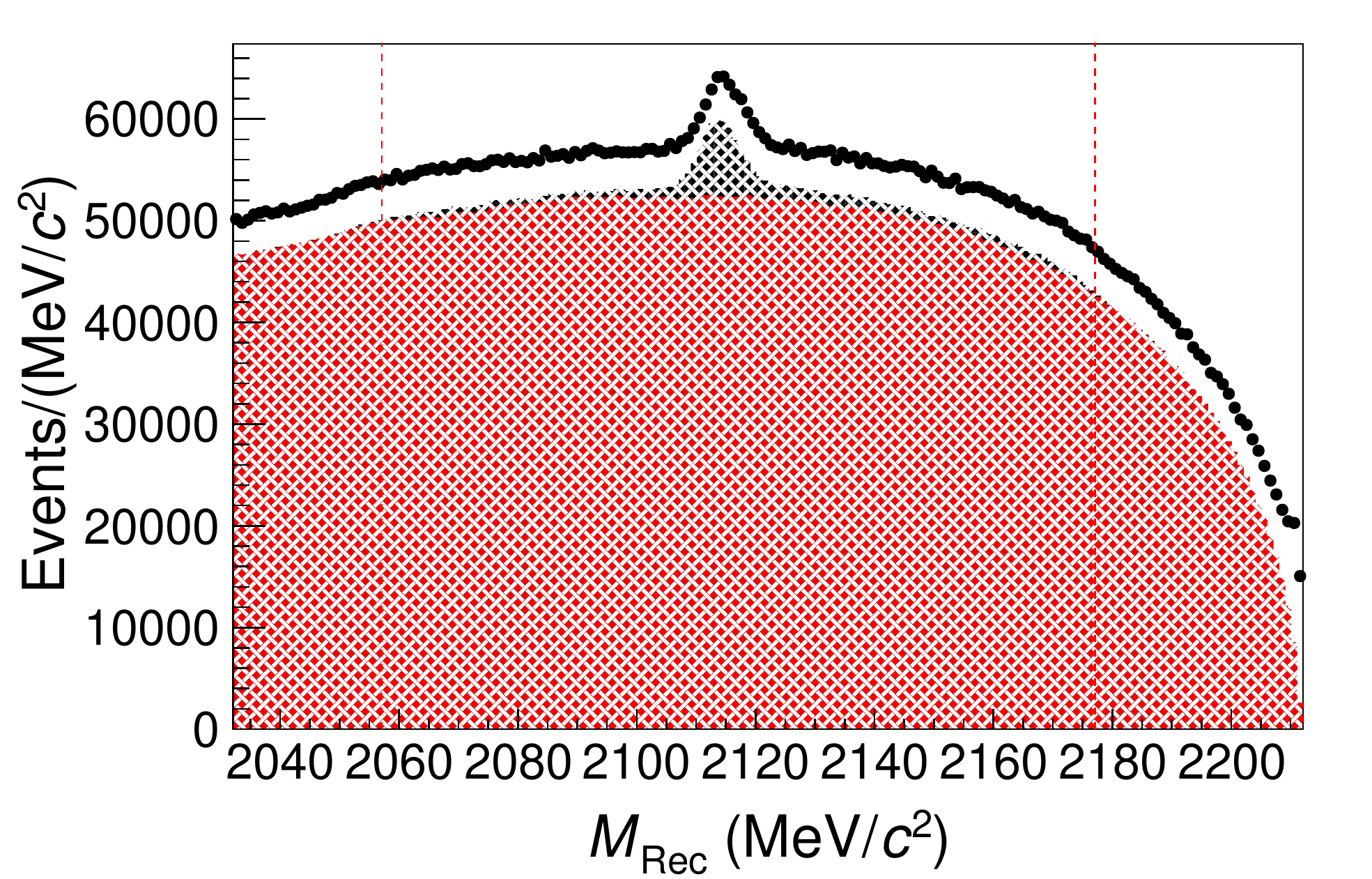}\\\includegraphics[width=3.25in]{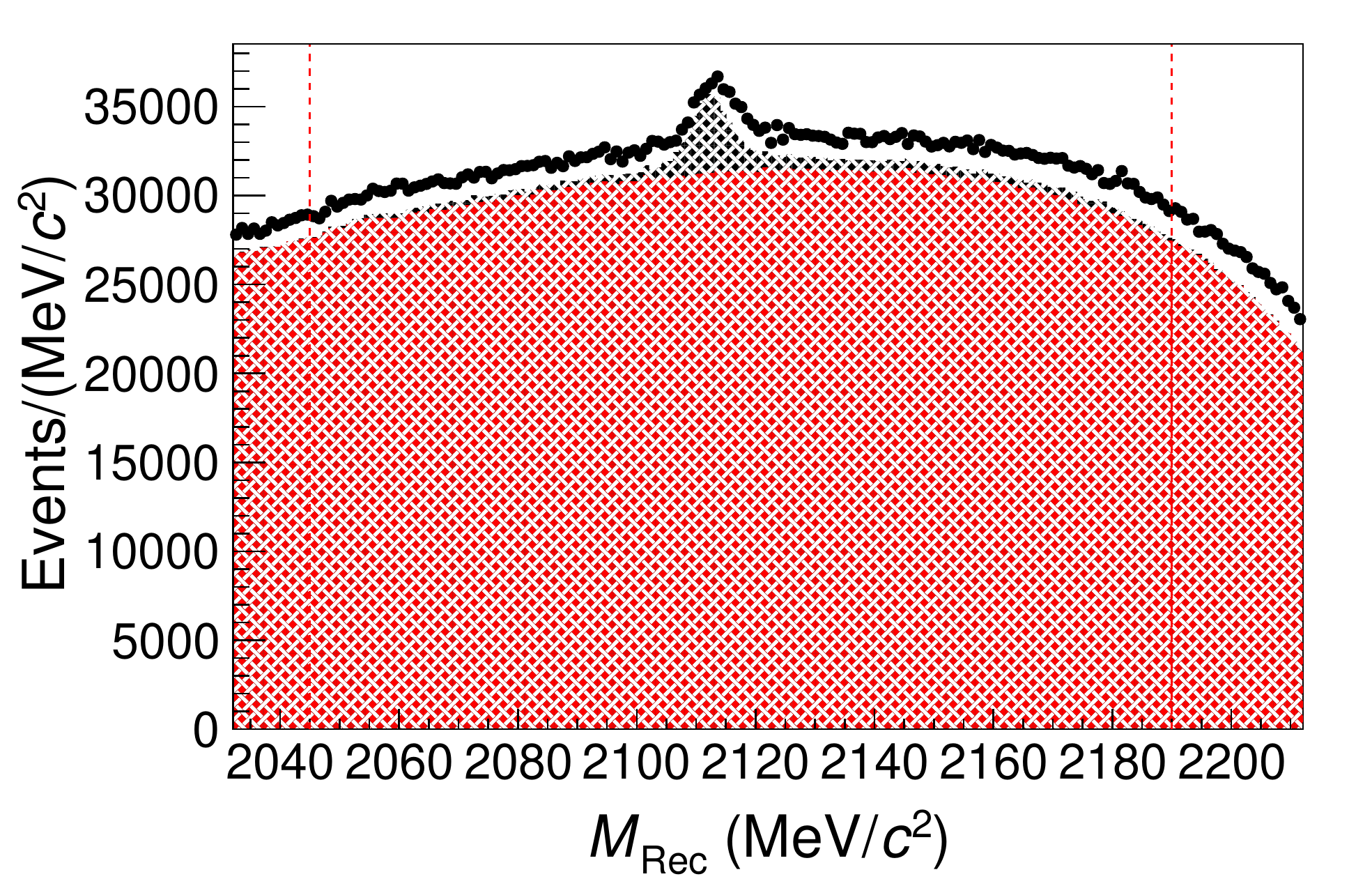}\\\includegraphics[width=3.25in]{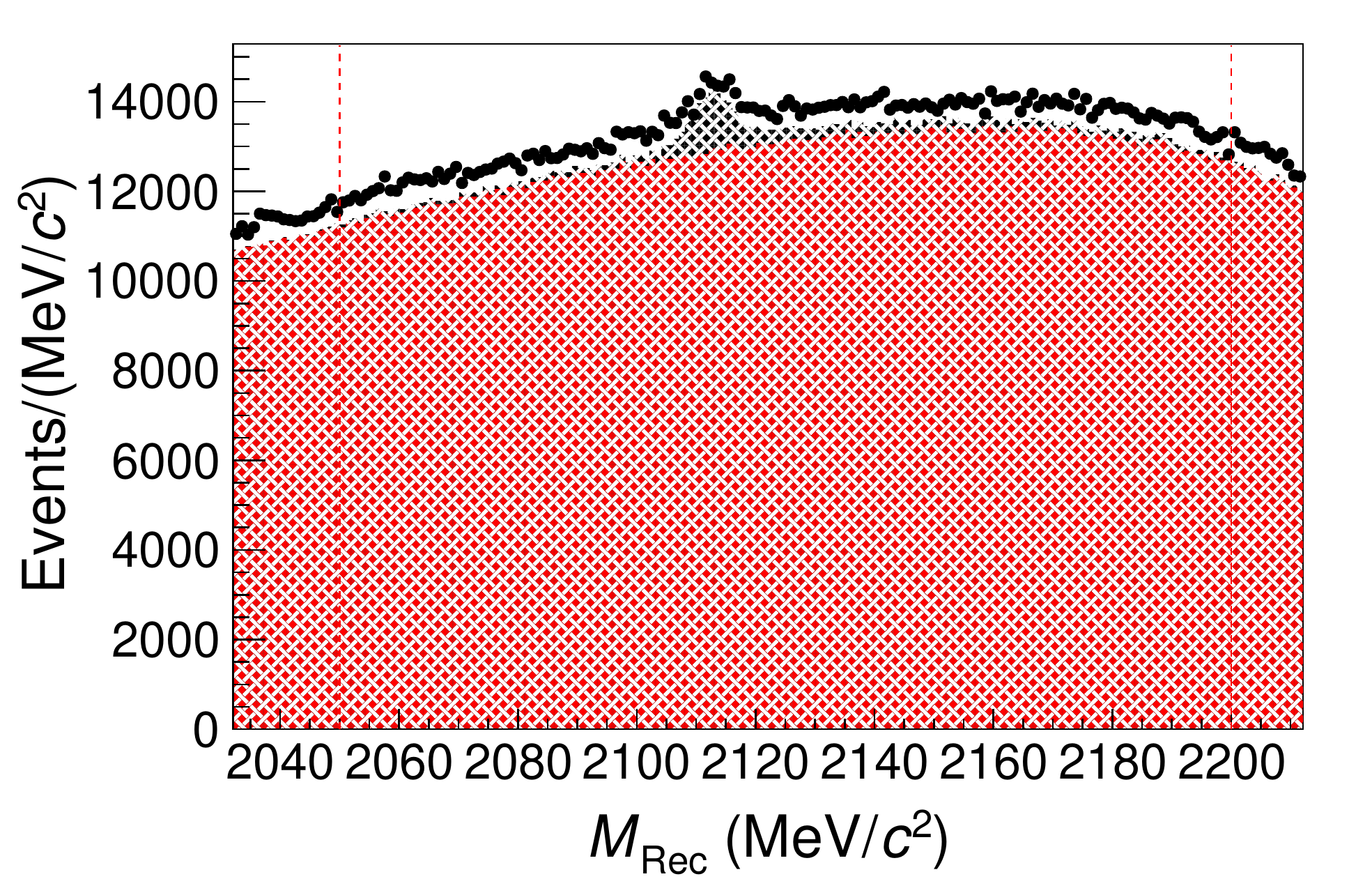}\\
\end{tabular}
}

\caption{Recoil-mass distribution of $D_s^-\to K^+K^-\pi^-$ candidates from the  $\EcmA$ (top), $\EcmB$ (center), and $\EcmC$ (bottom) data sets. The black points are data, and the black and red histograms are the signal and  background distribution predicted by the MC simulation, scaled by integrated luminosity.  The dashed red lines indicate the selection conditions.}
\label{fig:mRec}
\end{figure}

We select single-tag candidates using only the $D_s^-\to K^+K^-\pi^-$
hadronic decay mode because it is unique in having sufficient
statistics and well-known backgrounds. All tag candidate daughters are
required to satisfy the following track-quality requirements: the
track must originate from a region within 1 (10) cm of the $e^+e^-$
interaction point perpendicular (parallel) to the $z$ axis, which is
the symmetry axis of the MDC, and must be tracked with an angle
$\theta$ with respect to the $z$ axis that satisfies $|\cos
\theta|<0.93$. We apply $\pi$/$K$ ID based on $dE/dx$ and TOF
measurements to all tag-candidate daughters to maximize the purity of
the tag sample.  Multiple candidates in a single event are allowed,
both to increase the single-tag selection efficiency and to minimize
the tag bias, $\btag$. We calculate the recoil mass against the tag
$D_s^-$ candidate,

\begin{equation}
\small
 \Mrec c^2=\left[\left(\Ecm-\sqrt{\left|\vec{p}_{D_s}c\right|^2+m_{D_s}^2c^4}\right)^2-\left|\vec{p}_{D_s}c\right|^2\right]^{\frac 12},
 \end{equation}
where $\vec{p}_{D_s}$ is the momentum of the reconstructed tag $D_s^-$ candidate, and $m_{D_s}$ is the known $D_s$ mass~\cite{ref:PDG}.  We require the recoil mass to be consistent with a $\DsSTDs$ event hypothesis by imposing requirements for each data set: $2057 <\Mrec<2177 \text{ MeV}/c^2$ for $\EcmA$, $2145 <\Mrec< 2190  \text{ MeV}/c^2$ for $\EcmB$, and $2150 <\Mrec< 2200 \text{ MeV}/c^2$ for $\EcmC$. For about 20\% of cases, our event selection identifies more than one candidate per event after these recoil-mass requirements. Background due to extra candidates is subtracted as part of the fitting procedure described later in this section. The $\Mrec$ distributions for each data set can be seen in Fig.~\ref{fig:mRec}. While differences between the MC simulation and data can be seen in the figure, our measurement is not sensitive to the MC simulation of this distribution.

\begin{figure}[h!]
\centering{
\begin{tabular}{ccc}
\includegraphics[width=3.25in]{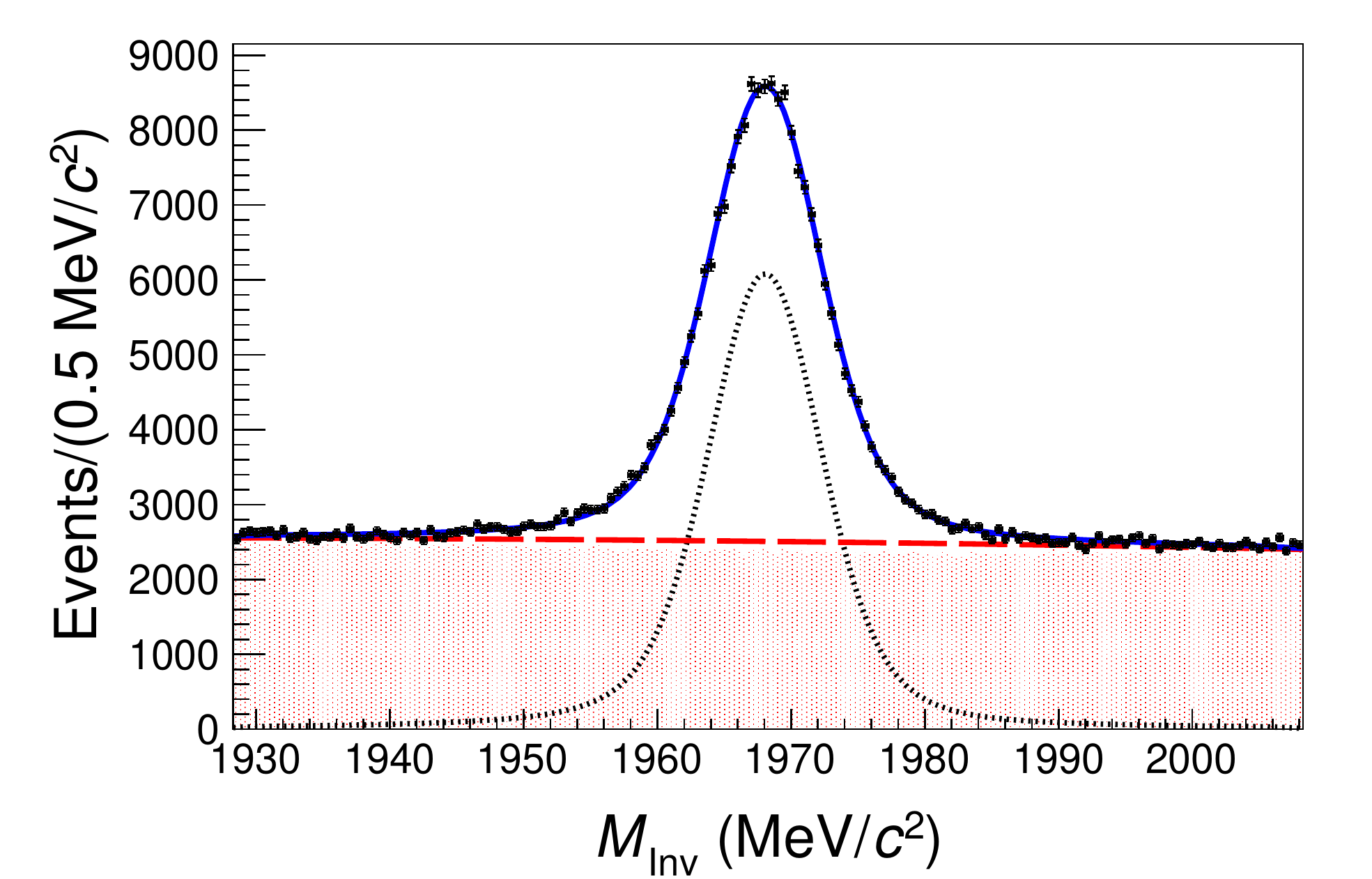}\\
\includegraphics[width=3.25in]{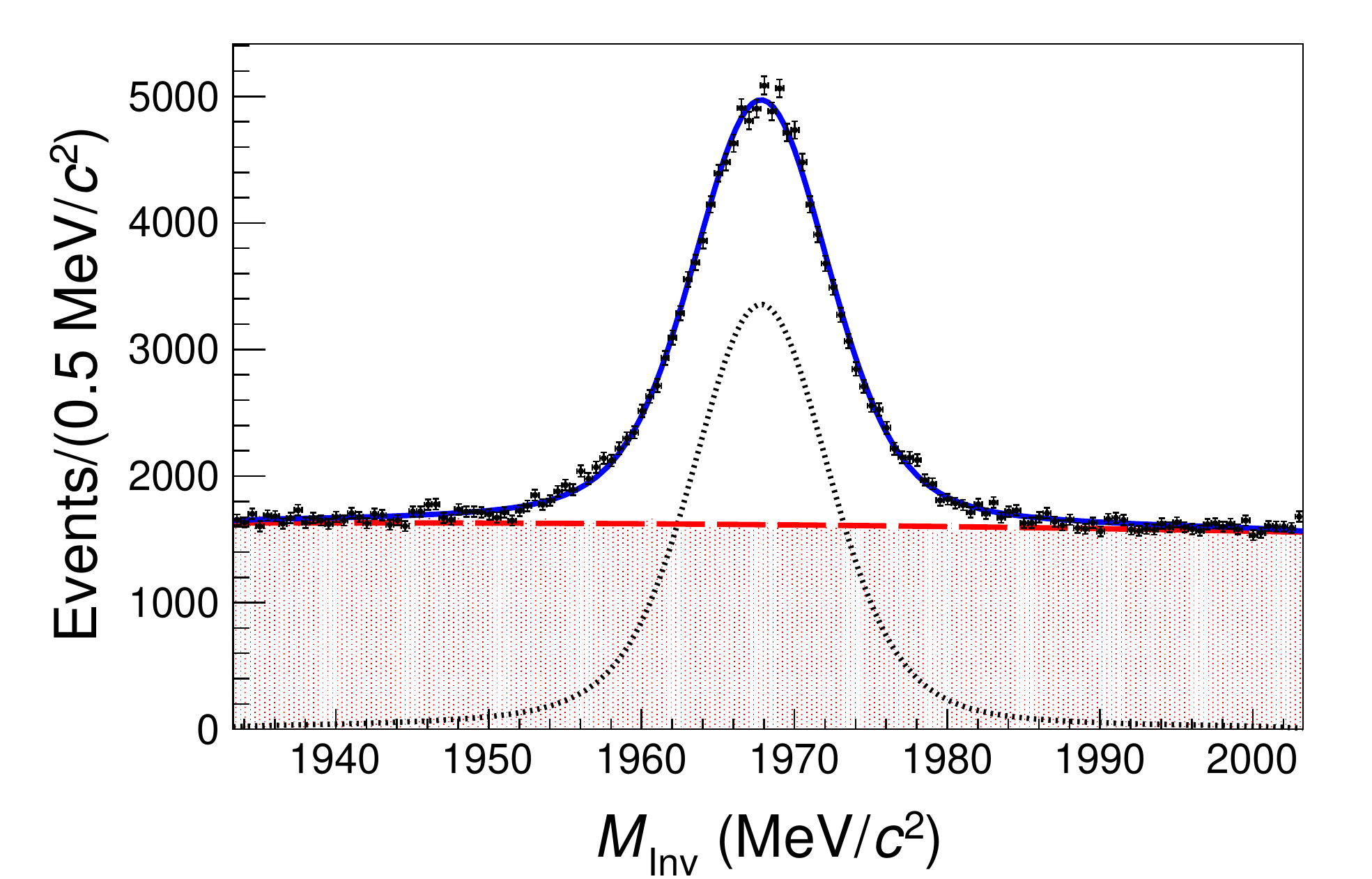}\\
\includegraphics[width=3.25in]{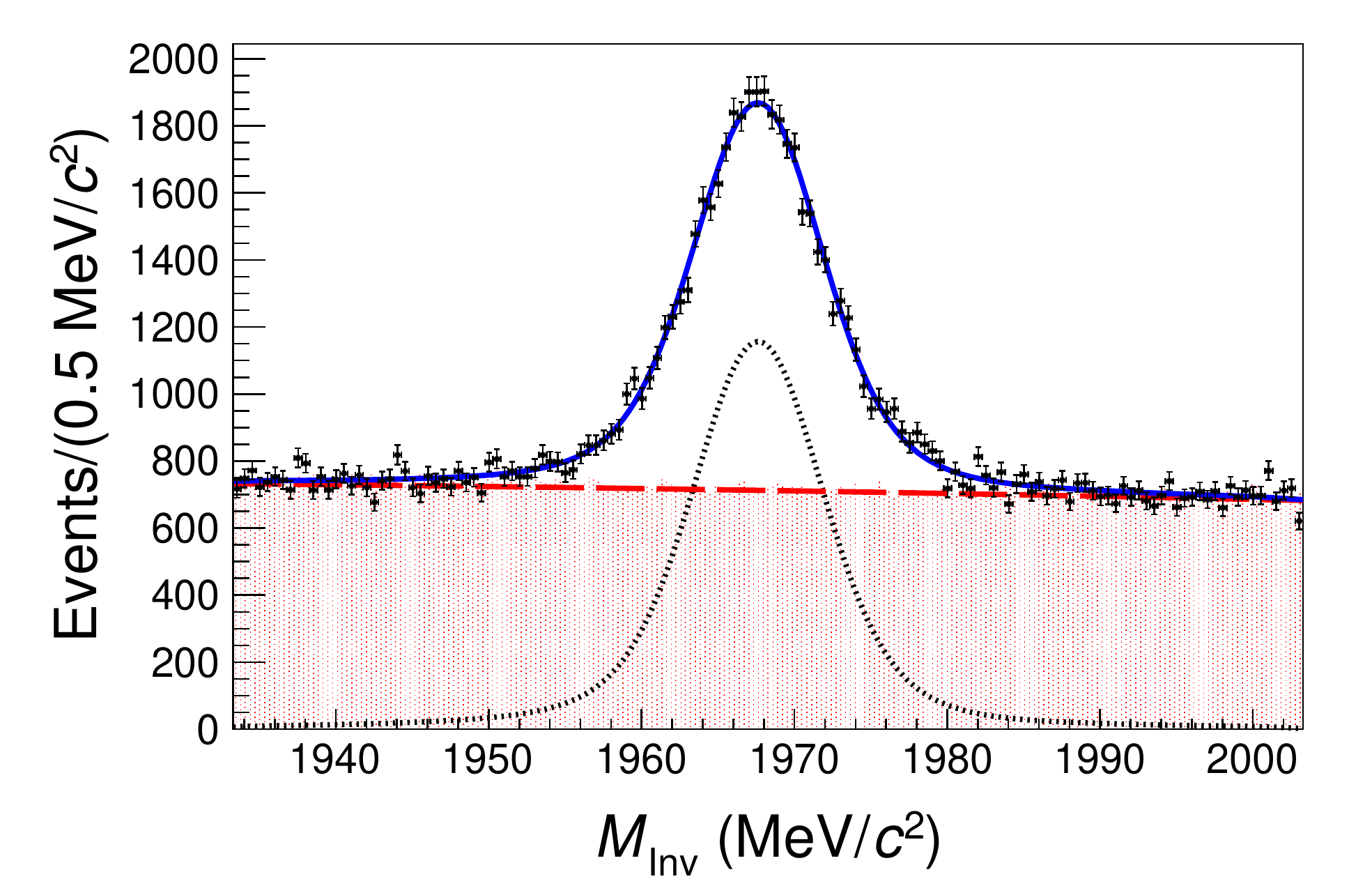}
\end{tabular}
}
\caption{Results of fits to invariant-mass distributions of $D_s^-\to K^+K^-\pi^-$ single-tag candidates from the $\EcmA$ data set (top), $\EcmB$ data set (center),  and $\EcmC$ data (bottom). In each plot, the solid blue line is the result of the total fit, the dashed red line is the fitted distribution of non-$D_s^-\to K^+K^-\pi^-$ backgrounds, the dotted black line is the fitted signal distribution, the filled red histogram is the contribution from backgrounds predicted by MC simulation, and the black points are data. The choice of binning is arbitrary and used solely for display. MC simulation distributions are scaled by luminosity.}
\label{fig:mInv}
\end{figure}

We determine the number of true single-tag candidates by performing an
unbinned fit to the distribution of the invariant mass of the tag
$D_s^-$ candidates, $M_\text{Inv}$. The signal shape is based on MC
simulation and convolved with a Gaussian function whose width and mean
are left free in the fit to account for a possible difference in
resolution and calibration between data and the MC simulation. The
distribution of backgrounds in this variable is modeled using a
second-order Chebyshev polynomial. The fit range is chosen to be
within $\pm40 \text{ MeV}/c^2$ of the known $\Ds$ mass for the $\EcmA$
distribution corresponding to approximately $5\sigma$ of the simulated
$M_\text{Inv}$ distribution at $\EcmA$. In the $\EcmB$ and $\EcmC$
data sets, the fit range is reduced to be within $\pm35 \text{ MeV}/c^2$
of the known $D_s$ mass due to non-polynomial background structures
appearing at the edges of the fit range. The single-tag fits to each
data set are shown in Fig.~\ref{fig:mInv} and the fitted yields are
listed in Table~\ref{table:STYields}. The $\EcmA$ data set provides
approximately $56\%$ of our single-tag sample, and the $\EcmB$ and
$\EcmC$ data sets provide $33\%$ and $11\%$, respectively. Within each
data set, $99\%$ of our $D_s^{-}$ candidates come from $D_s^*D_s$
events and $1\%$ come from $D_sD_s$ events. $D_s^*D_s^*$ events can be produced at $\EcmC$, but contribute negligibly to our candidate sample.

The single-tag reconstruction efficiency depends on the topology of the recoil-side decay, which we account for through $\btag$. We determine the tag bias from MC samples for double-tag events from each of the observed semielectronic modes listed in Table~\ref{table:BFS} and for $\taucont$. The $\semilep$ tag bias is determined by averaging the tag biases for the six observed modes weighted by the branching fractions in Table~\ref{table:BFS}. The single-tag efficiencies with no specification on the signal-side and the determined tag biases for $\semilep$ and $\taucont$ for each data set are shown in Table~\ref{table:bias}.
\begin{table}[H]
\centering{
\begin{tabular}{c|c}
Data Set & Fitted Single-Tag Yields\\ \hline
$\EcmA$ & $147581\pm 779$\\ \hline
$\EcmB$ & $85845\pm 705$\\ \hline
$\EcmC$ & $29234\pm 435$\\ \hline\hline
Sum &  $262660 \pm 1137$ \\ \hline
\end{tabular}
}

\raggedright
\caption{Fitted single-tag yields from each data set. Only statistical uncertainties are shown.}
\label{table:STYields}
\end{table}

\begin{table*}[hbtp]
\centering{
\begin{tabular}{c|c|c|c}
Data Set&$\epsilon_\text{ST}$ &$\btag$ &$b^\tau_{\rm{tag}}$ \\ \hline
$\EcmA$ & $\left(43.10 \pm 0.01\right)\%$ & $1.007 \pm 0.001$ & $1.037 \pm 0.003$ \\ \hline
$\EcmB$ & $\left(42.46 \pm 0.02\right)\%$ & $1.004 \pm 0.002$ & $1.034 \pm 0.004$ \\ \hline
$\EcmC$ & $\left(40.48 \pm 0.03\right)\%$ & $1.005 \pm 0.003$ & $1.044 \pm 0.007$ 
\end{tabular}
}
\caption{Single-tag efficiencies for all events ($\epsilon_\text{ST}$), and tag bias for $\semilep$ events ($\btag$) and for $\taucont$ events ($b^\tau_{\rm{tag}}$) from each data set determined with MC  samples. Listed uncertainties are only statistical.}
\label{table:bias}
\end{table*}

\section{Double-Tag Selection And Analysis}\label{sec:analysis}
  After a single-tag candidate is found, we begin searching for positron candidates among the recoil-side tracks. We sort recoil-side tracks that satisfy track requirements defined in Sec.~\ref{sec:stag} into eighteen 50-MeV$/c$ momentum bins between $200 \text{ MeV}/c$ and $1100 \text{ MeV}/c$. Mutually exclusive PID hypotheses are assigned to each track based on information from the MDC, TOF, and EMC.
  
\begin{figure}[hbt]
\centering{
\begin{tabular}{c}
\includegraphics[width=3.25in]{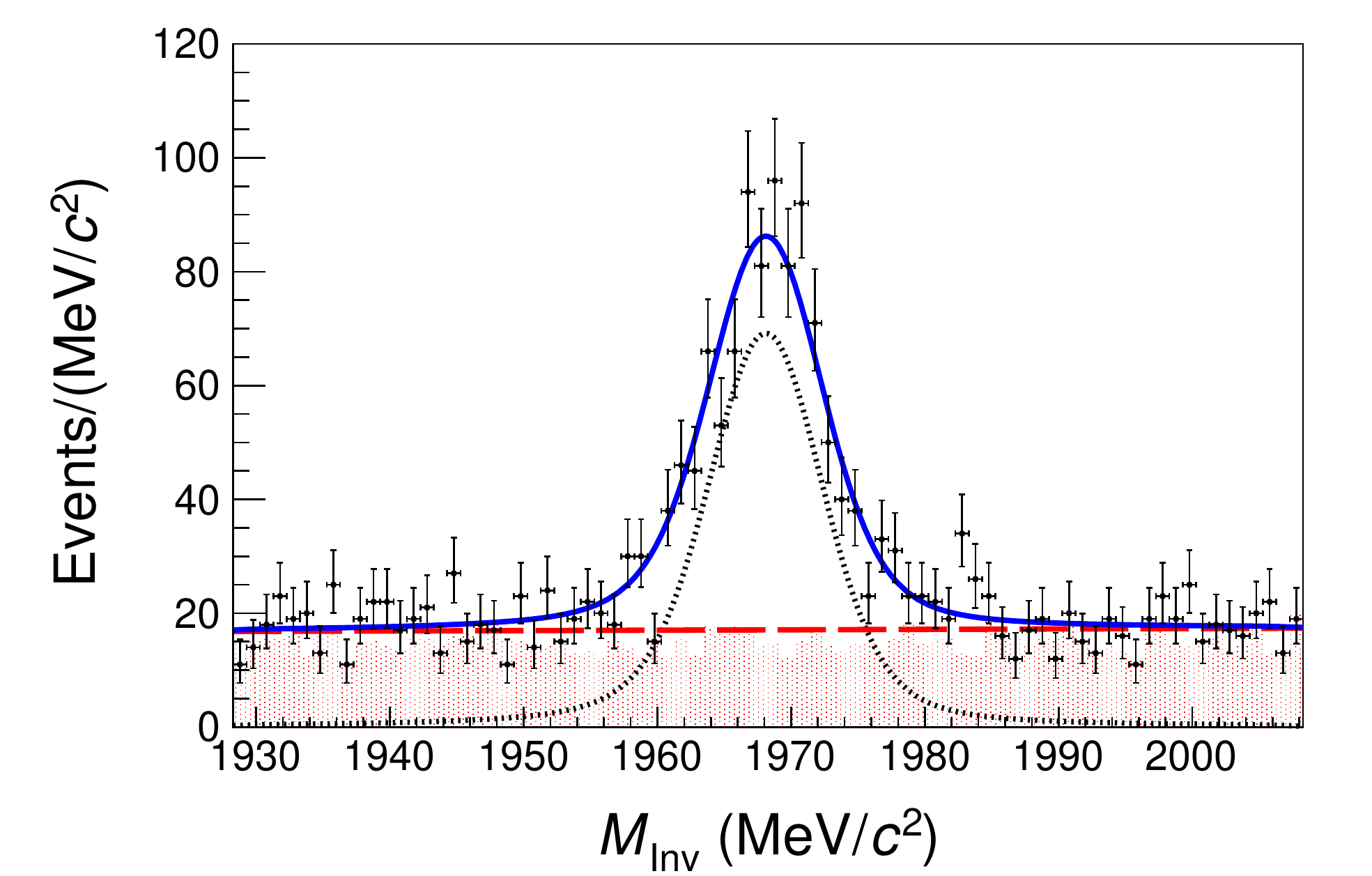}\\
\includegraphics[width=3.25in]{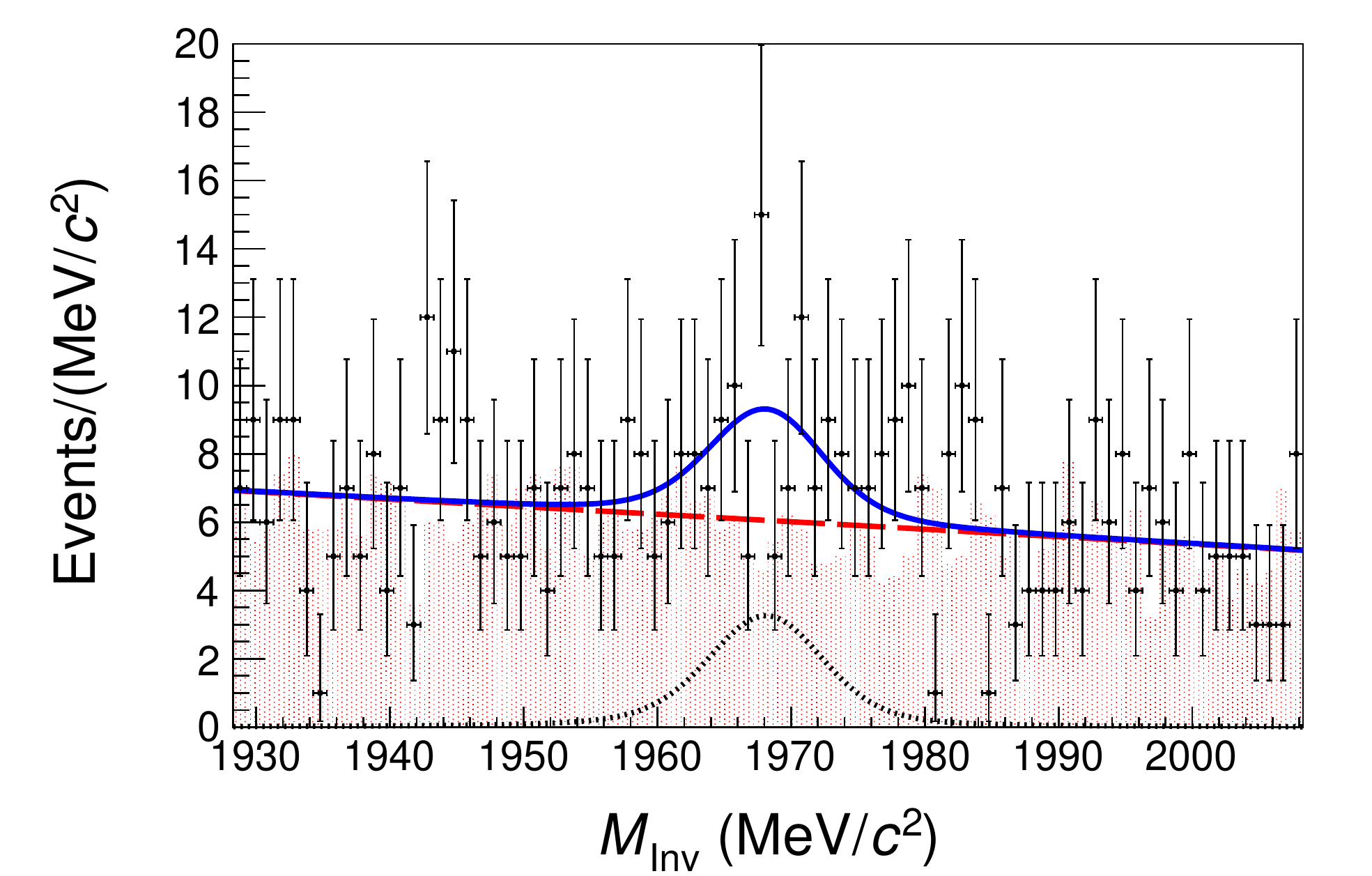}
\end{tabular}
}

\caption{Example tag invariant-mass fits for double-tag RS (top) and
  WS (bottom) positron candidates with momentum in the range 450-500
  MeV$/c$ from $\EcmA$ data. In both plots, the solid blue line is the
  result of the total fit, the dashed red line is the fitted
  distribution of non-$D_s^-\to K^+K^-\pi^-$ backgrounds, the dotted
  black line is the fitted signal distribution, the filled red
  histogram is the contribution from backgrounds predicted by MC
  simulation, and the black points are data. The choice of binning is
  arbitrary and used solely for display.  MC simulation distributions
  are scaled by the number of single-tag events.}
\label{fig:DatamInvFits}
\end{figure}

\begin{figure*}[hbt]
\centering{
\large
\begin{tabular}{cc}
RS $e$ yields& WS $e$ yields\\
\includegraphics[width=3.00in]{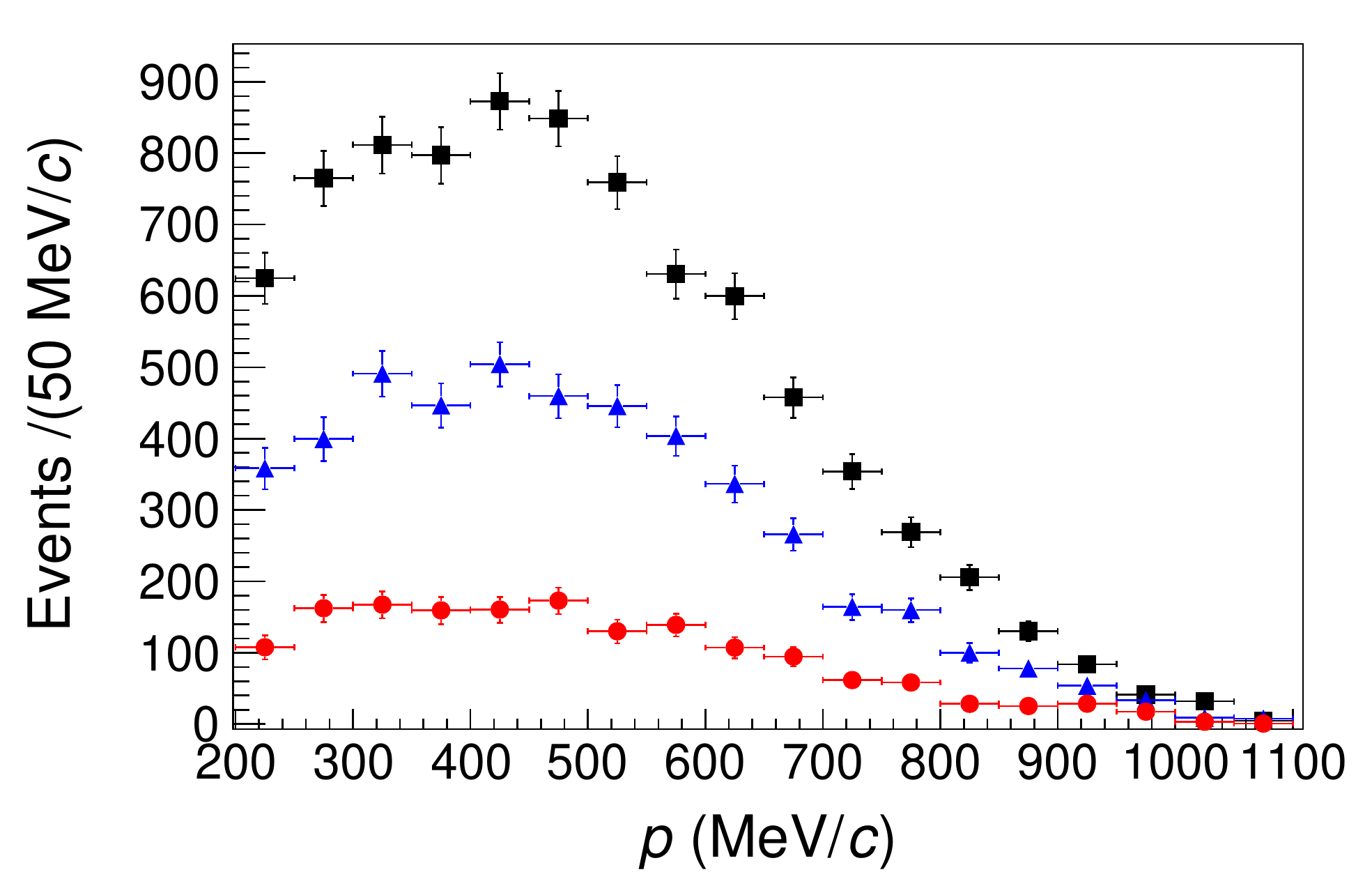}&\includegraphics[width=3.00in]{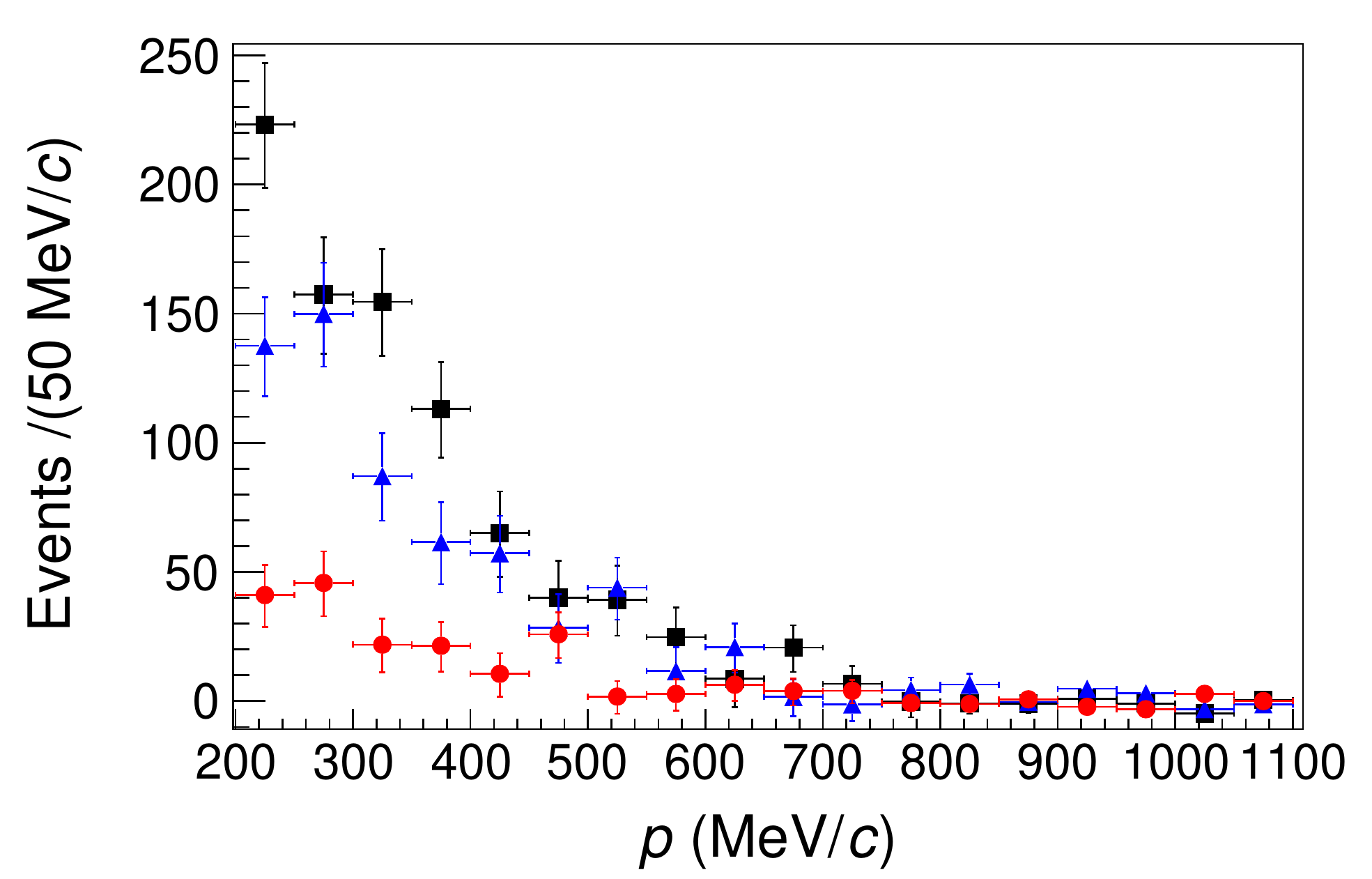}\\
 RS $\pi$ yields& WS $\pi$ yields\\
\includegraphics[width=3.00in]{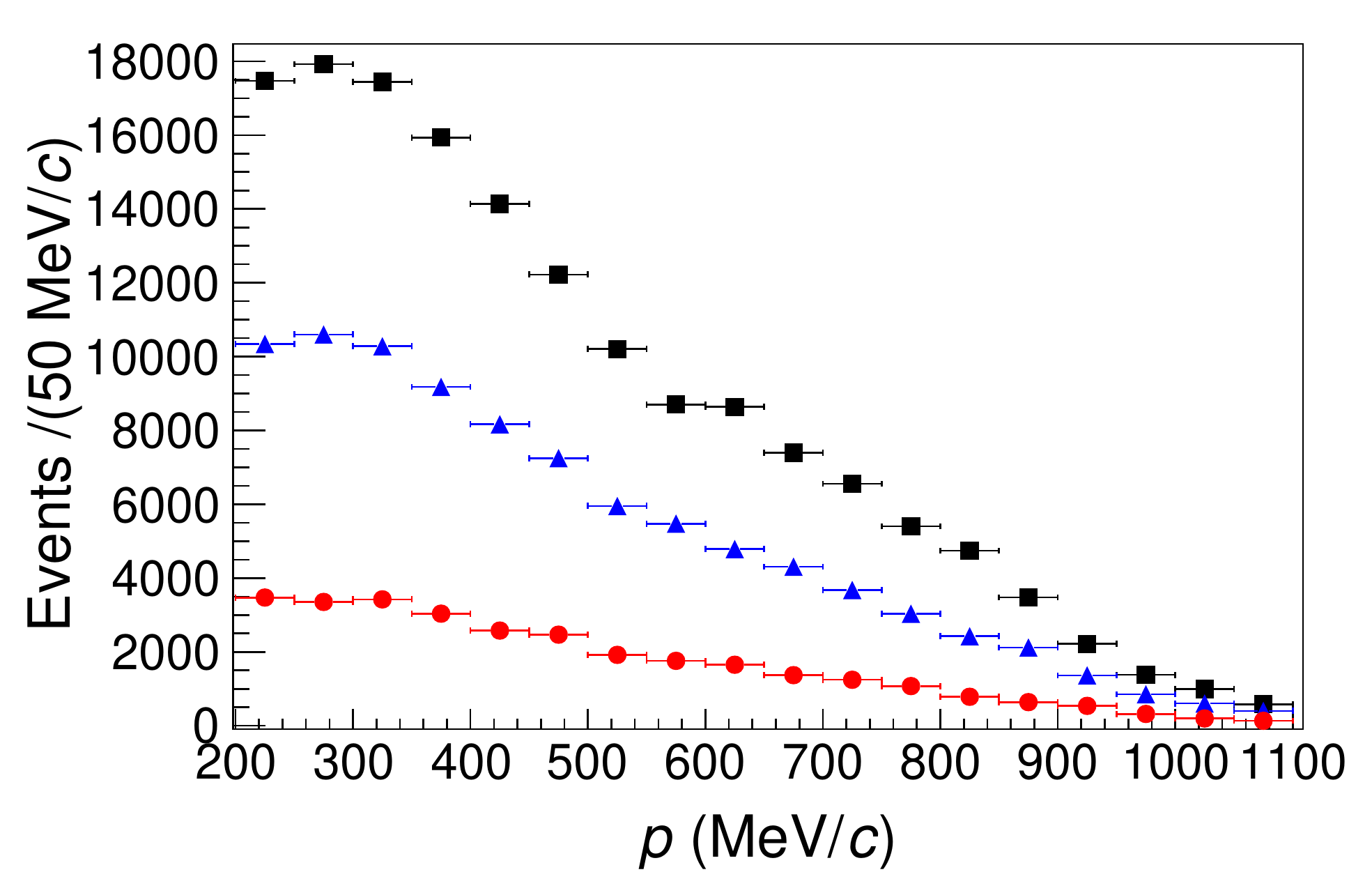}&\includegraphics[width=3.00in]{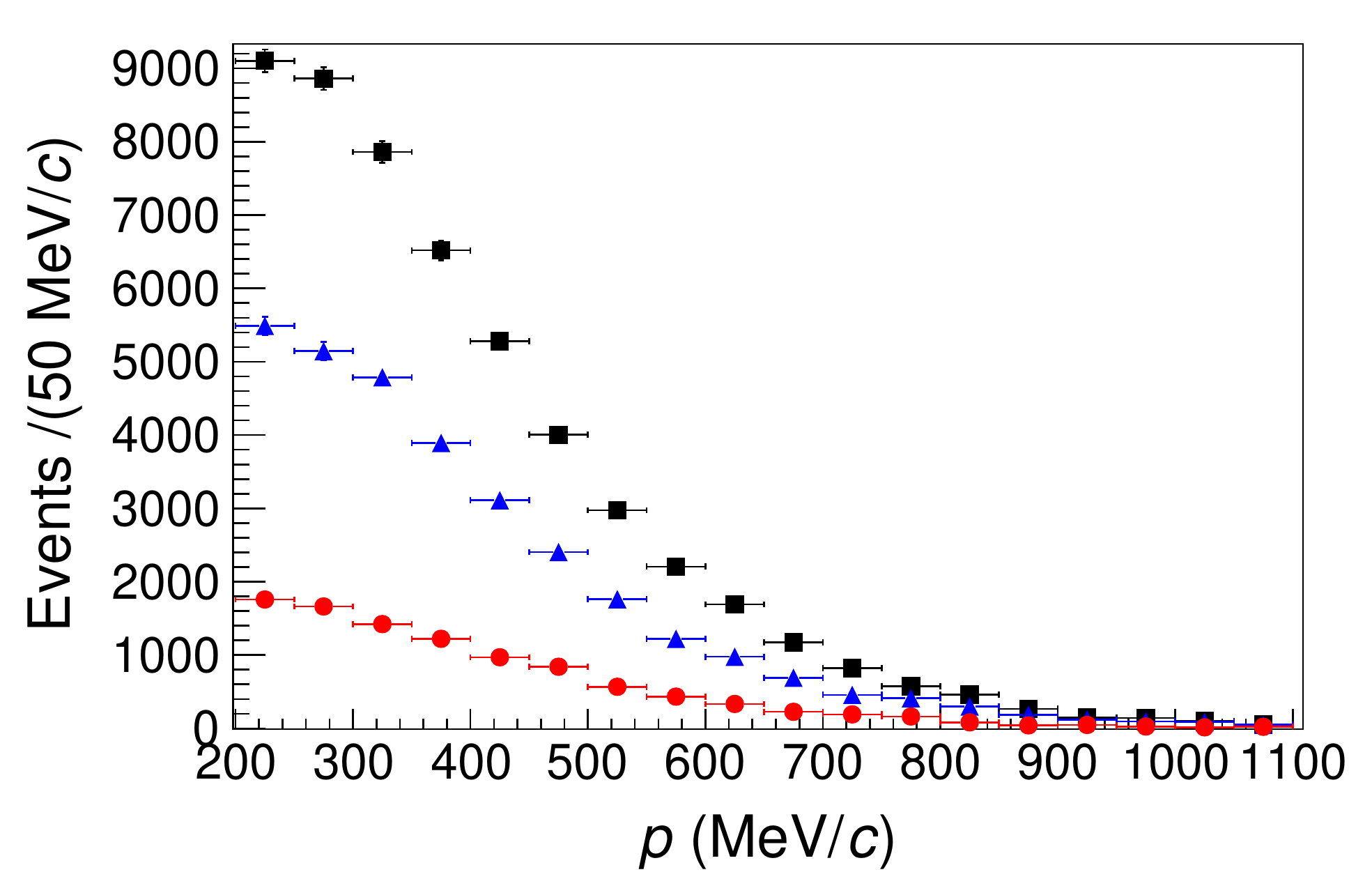}\\
 RS $K$ yields& WS $K$ yields\\
\includegraphics[width=3.00in]{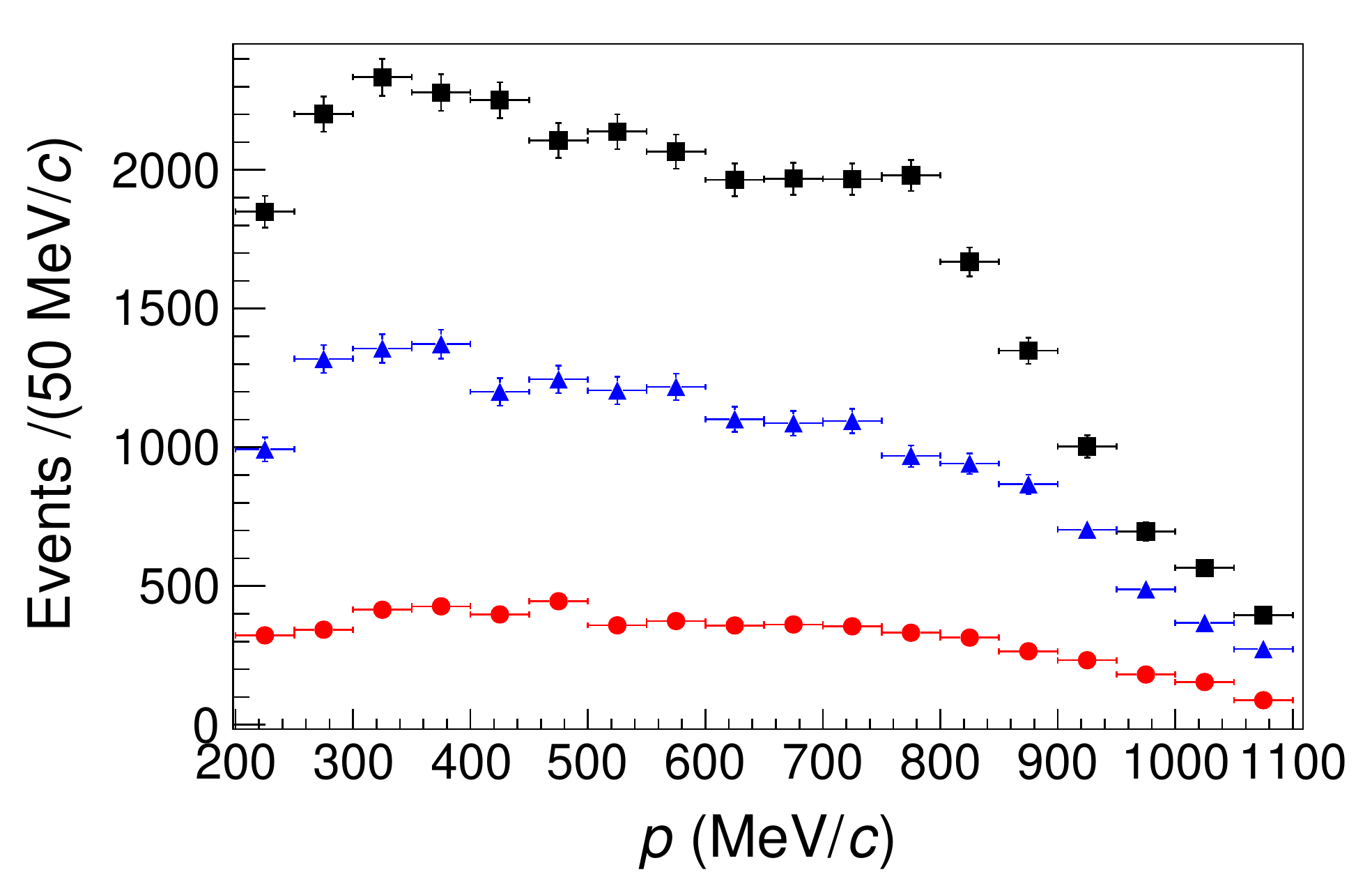}&\includegraphics[width=3.00in]{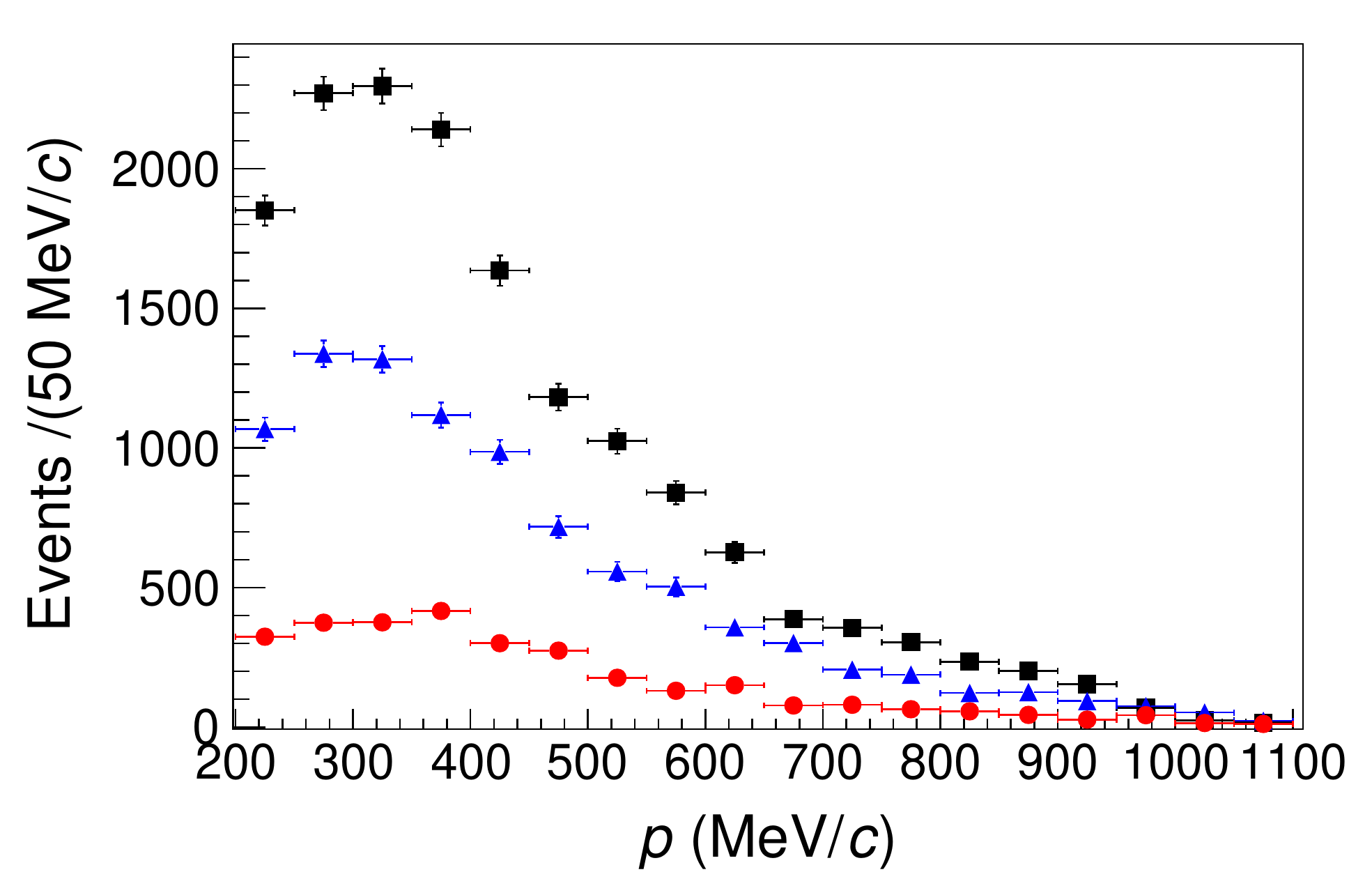}
\end{tabular}
}
\caption{Measured yields for each track category as a function of momentum. The results from $\EcmA$ are shown as black squares, those from $\EcmB$ as blue triangles, and those from $\EcmC$ as red circles.}
\label{fig:yields}
\end{figure*}

Based on our PID assigment, we fill tag-side invariant-mass
distributions in each momentum bin for each of the six categories of
recoil-side tracks defined in Sec.~\ref{sec:technique}. We determine
the number of tracks in each category originating from true $\Ds$
events by performing an unbinned fit to the tag-side invariant mass
distribution with the same signal shape used for the single-tag fit,
with the Gaussian parameters fixed to those determined in the
single-tag fit.  We employ a first-order Chebyshev polynomial to model
the distribution of backgrounds. Two examples of the 172 fits we
perform are shown in Fig.~\ref{fig:DatamInvFits}. The full set of fits
is made available as supplemental
material~\cite{ref:supplement}. Measured yields as a function of
momentum bin for each category are shown in Fig.~\ref{fig:yields}.

\begin{figure*}[hbt]
\centering{
\large
\begin{tabular}{ccc}
$\epsilon_{e}$&$P_{\pi\to e}$&$P_{K\to e}$\\
\includegraphics[width=2in]{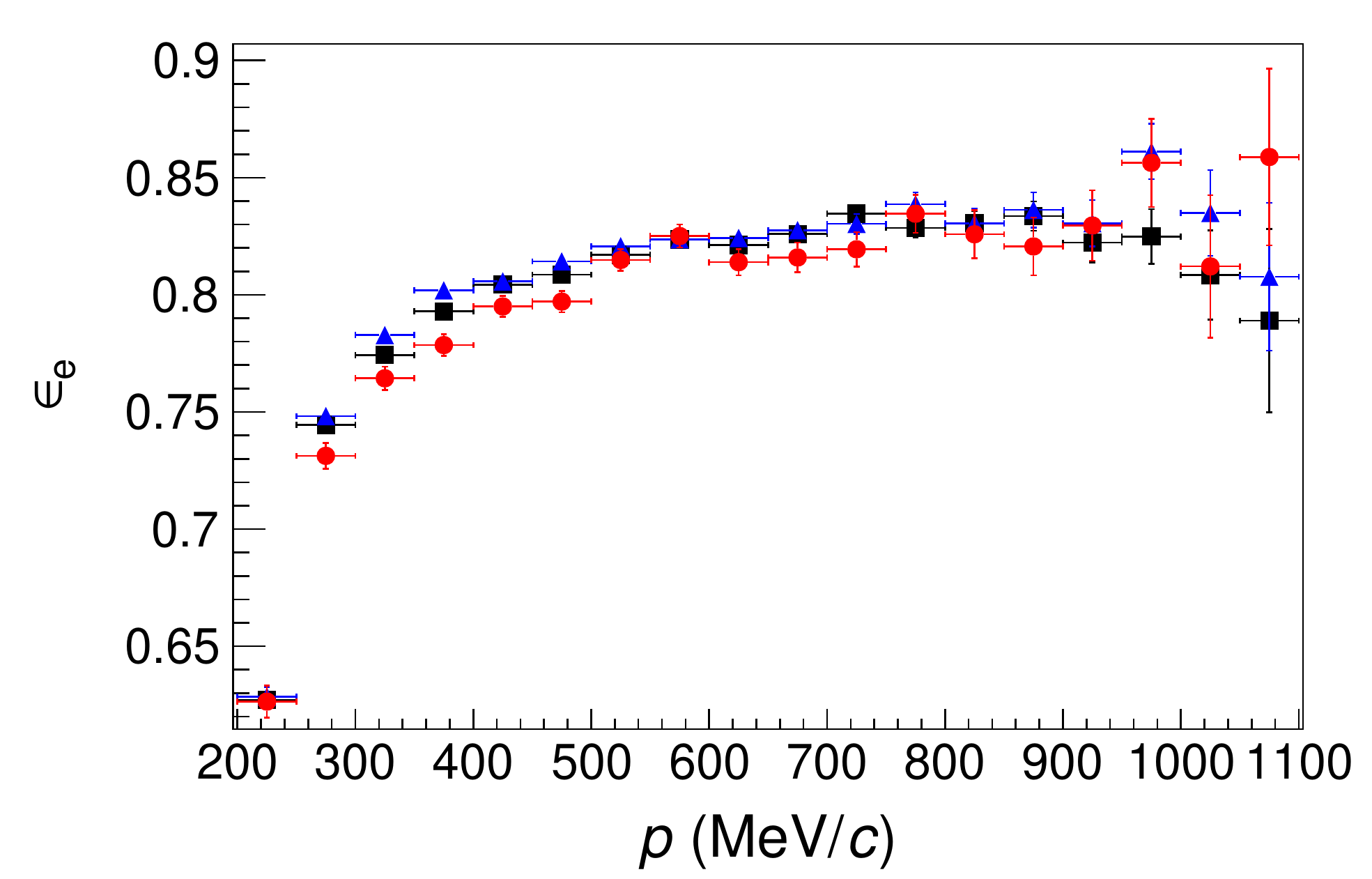}&\includegraphics[width=2in]{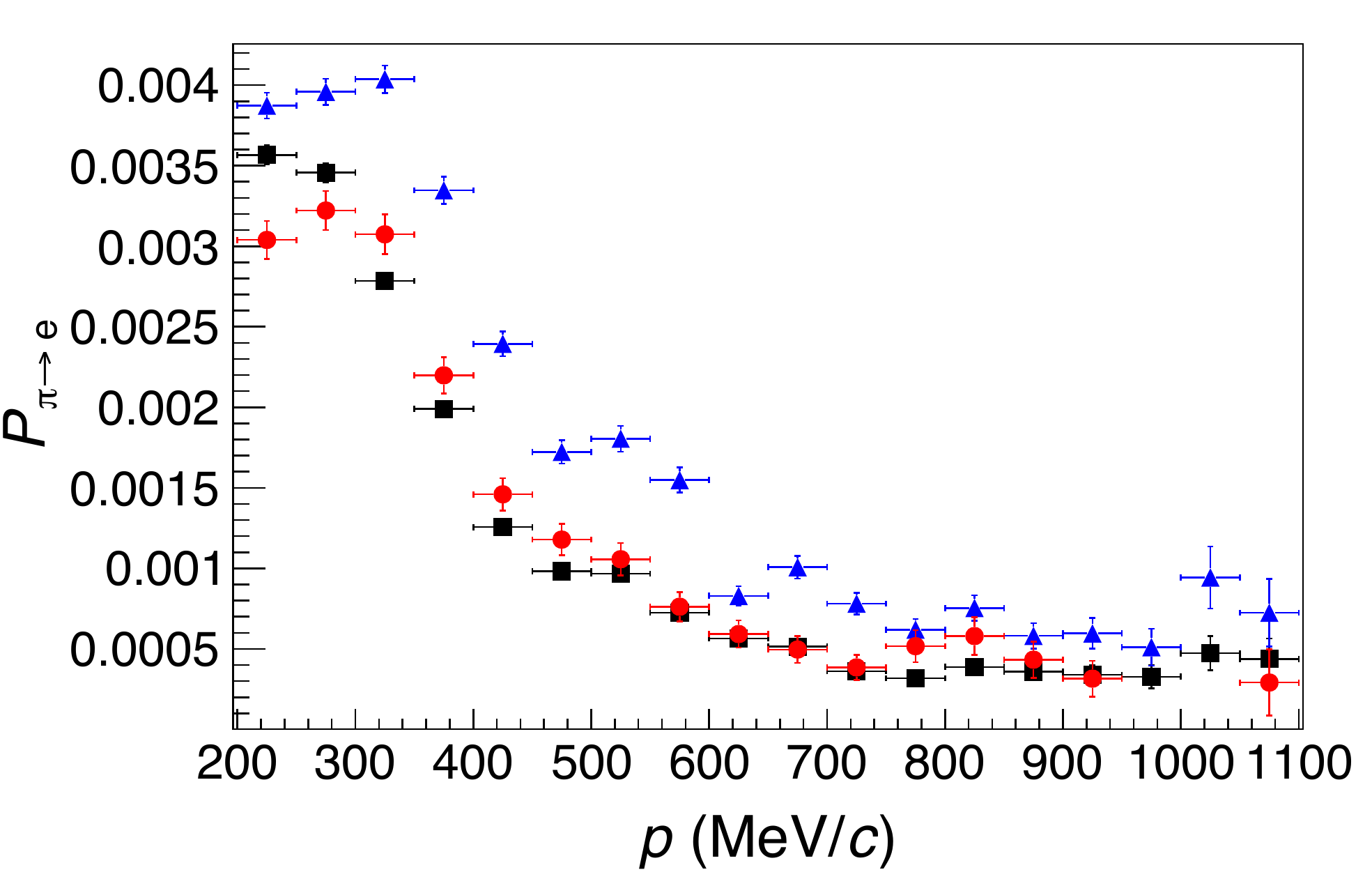}&\includegraphics[width=2in]{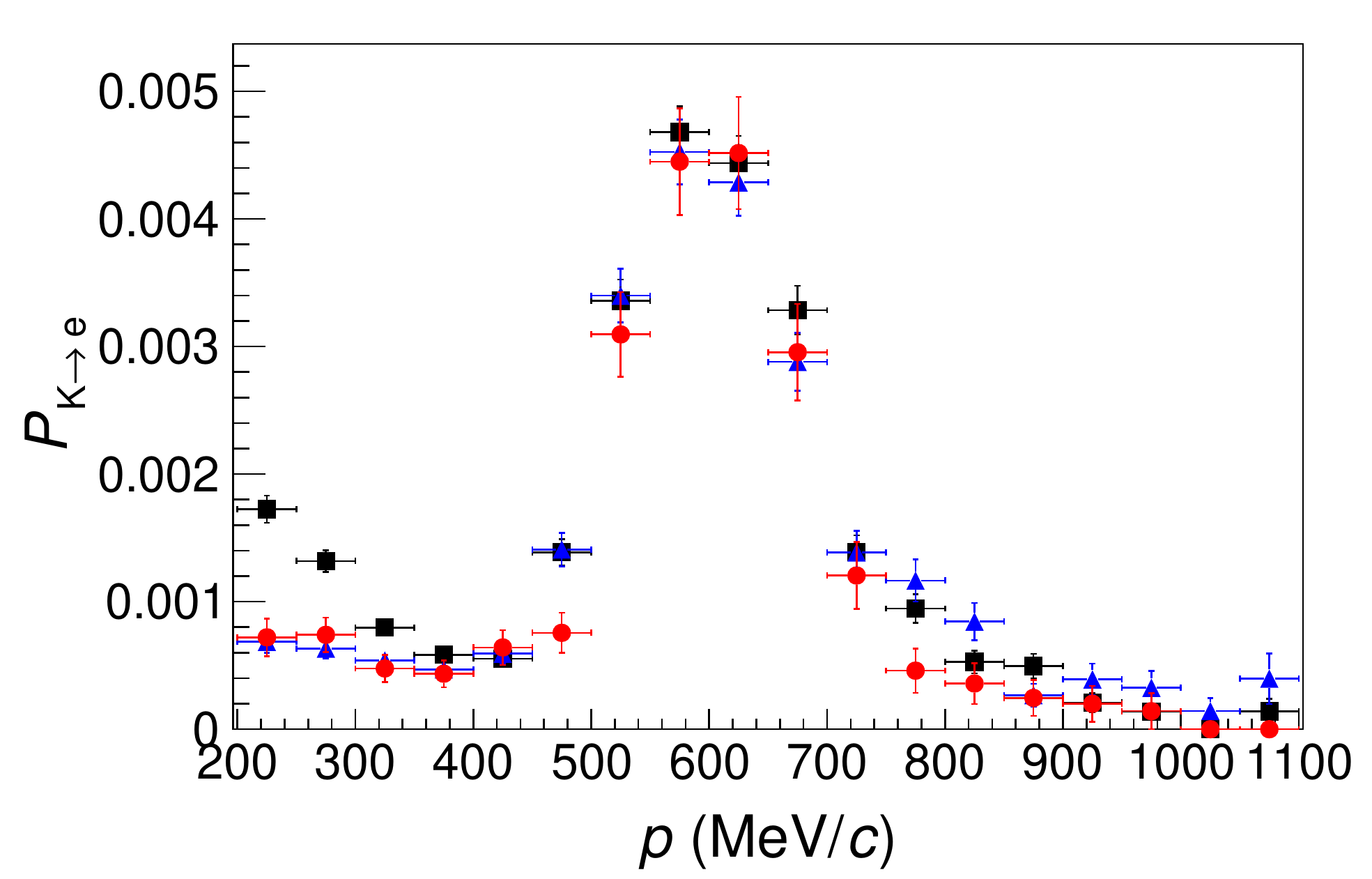}\\
$P_{e\to \pi}$&$\epsilon_{\pi}$&$P_{K\to \pi}$\\
\includegraphics[width=2in]{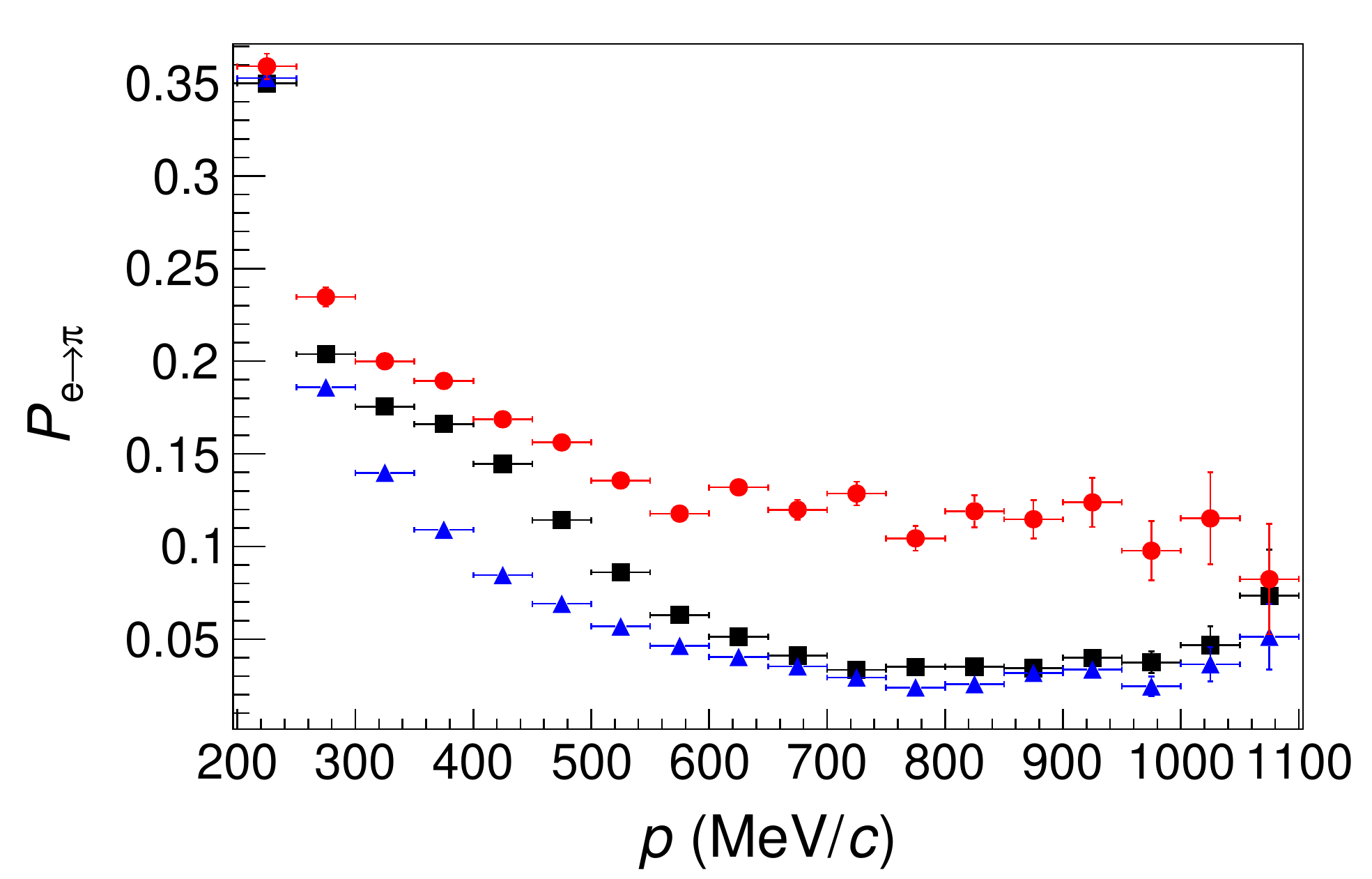}&\includegraphics[width=2in]{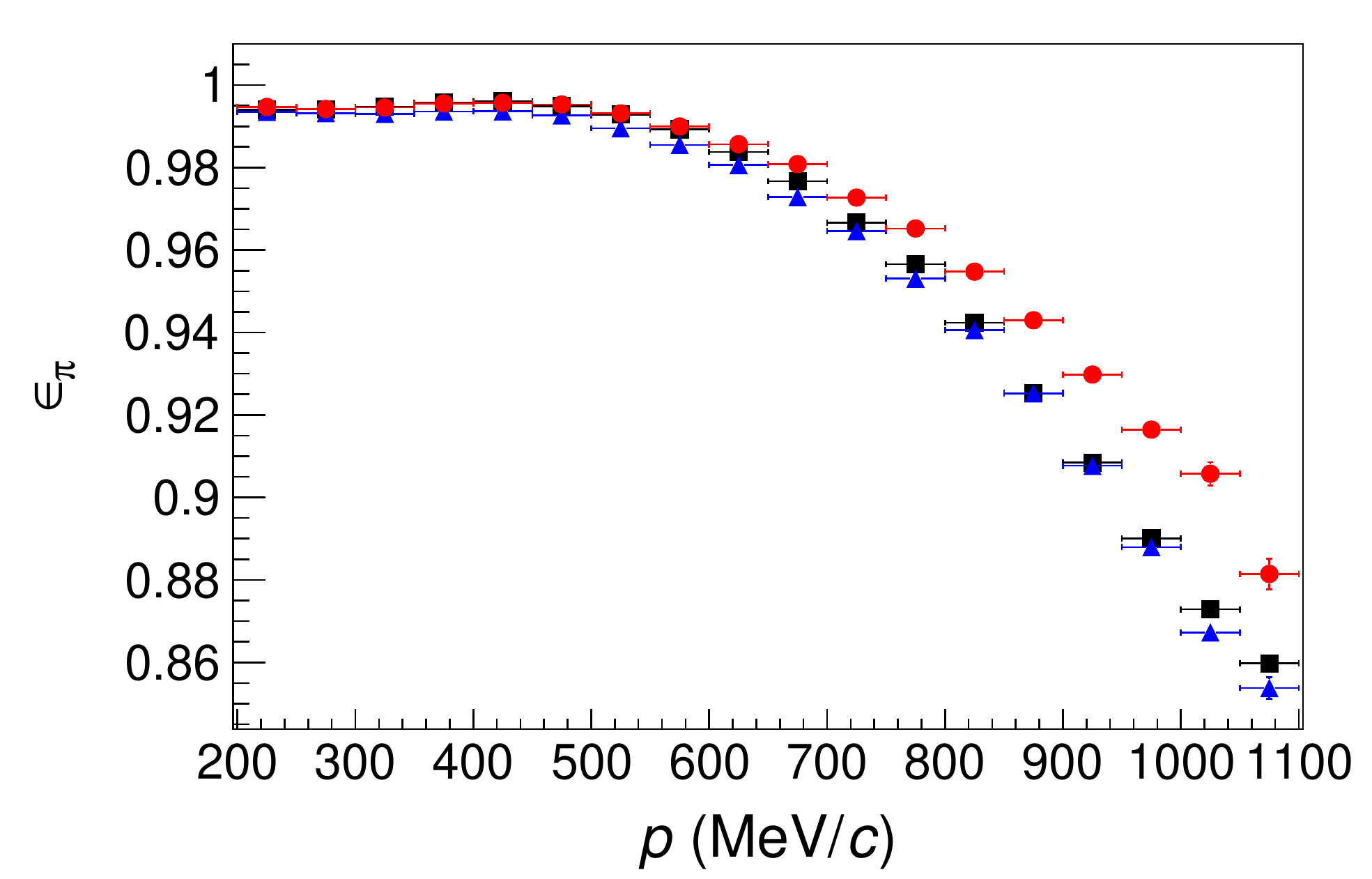}&\includegraphics[width=2in]{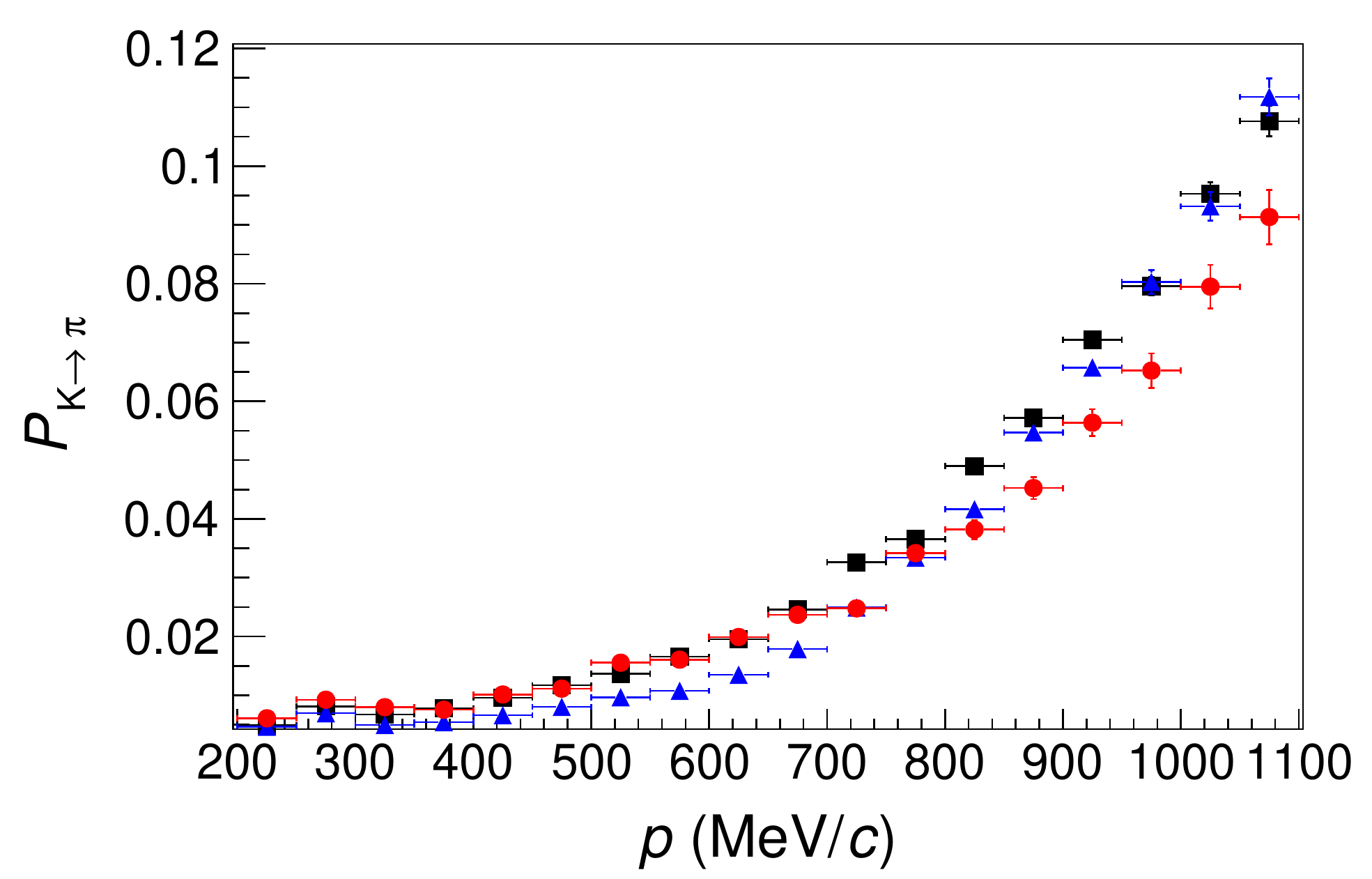}\\
$P_{e\to K}$&$P_{\pi\to K}$&$\epsilon_{K}$\\
\includegraphics[width=2in]{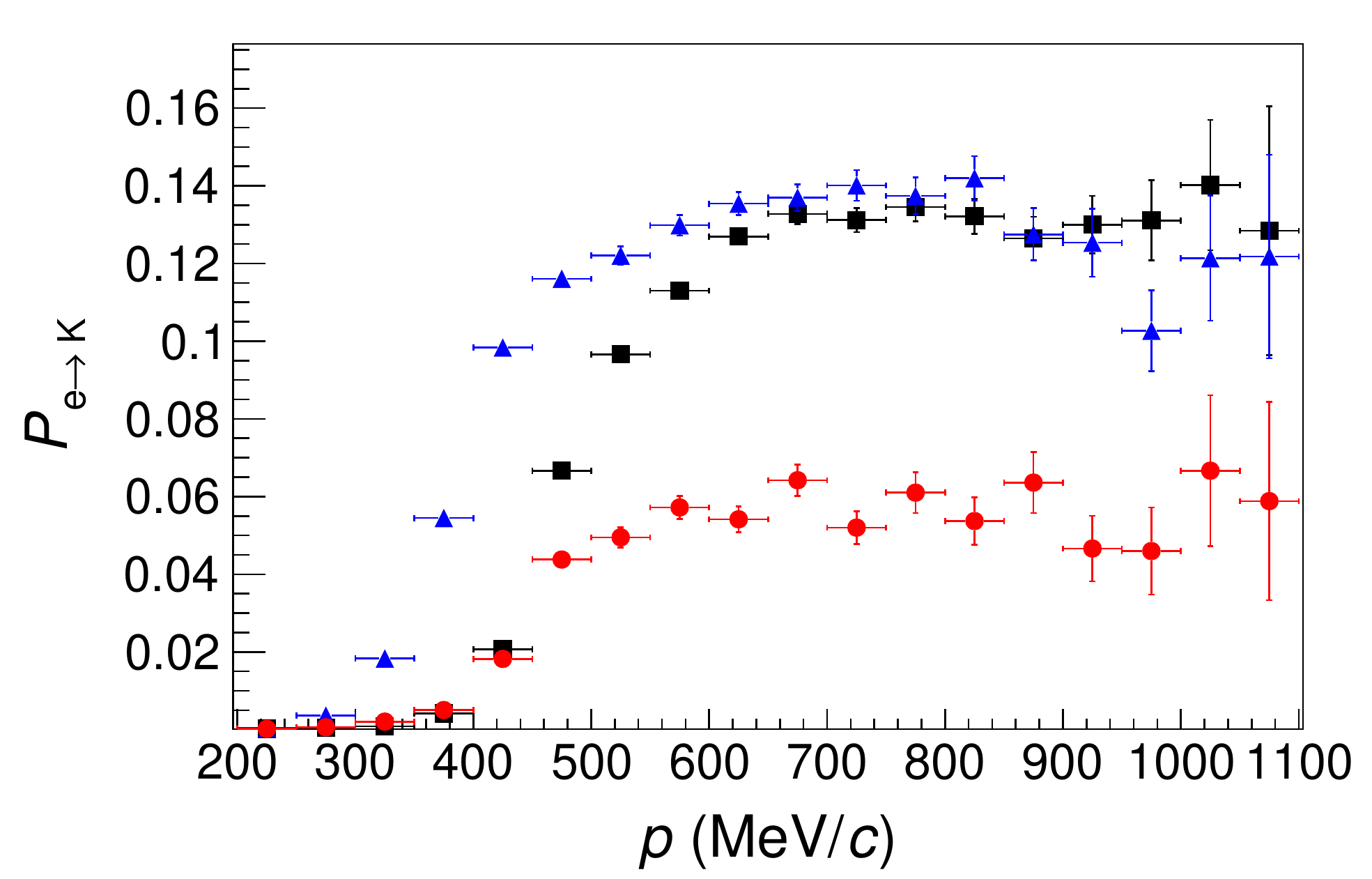}&\includegraphics[width=2in]{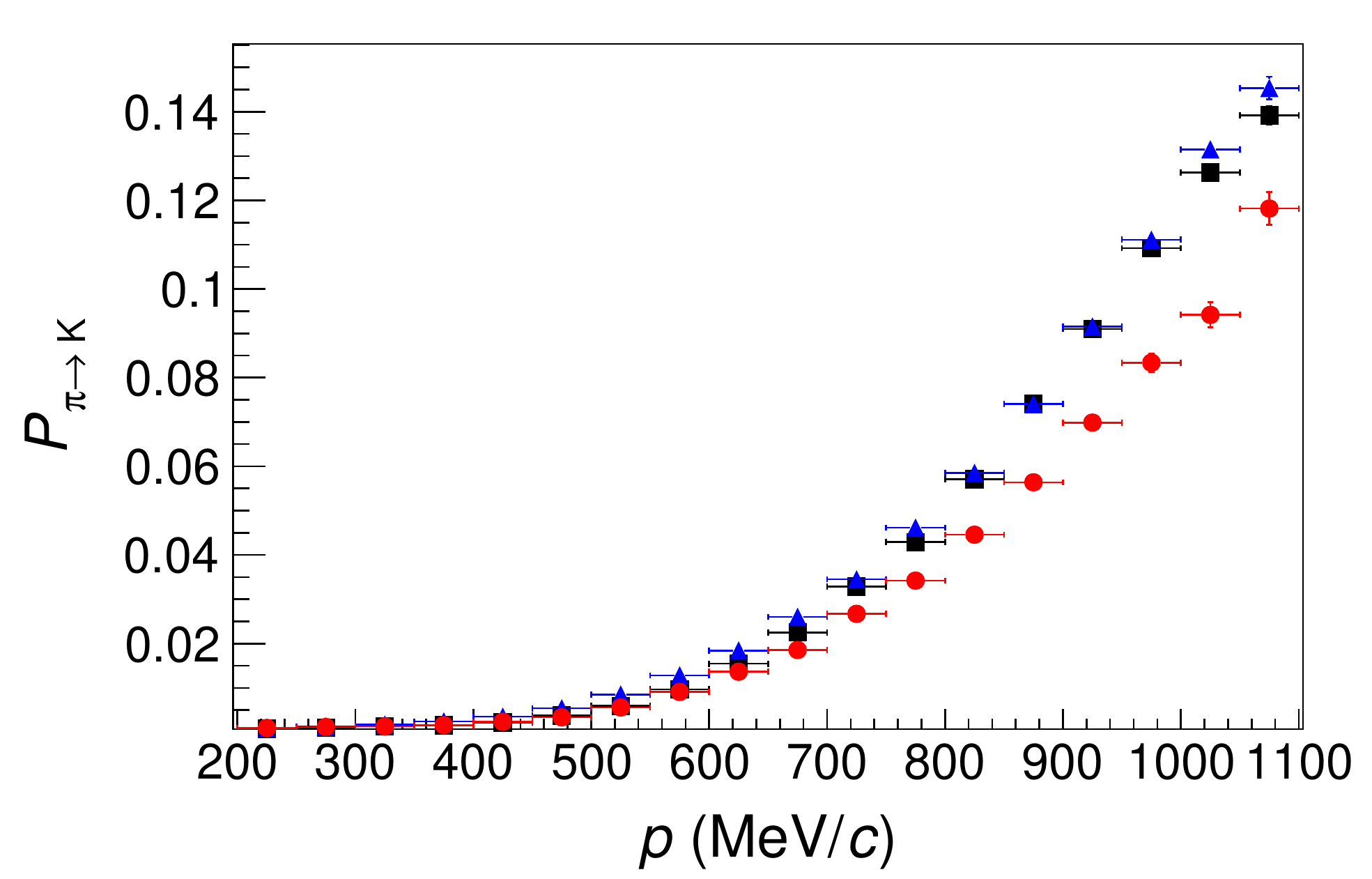}&\includegraphics[width=2in]{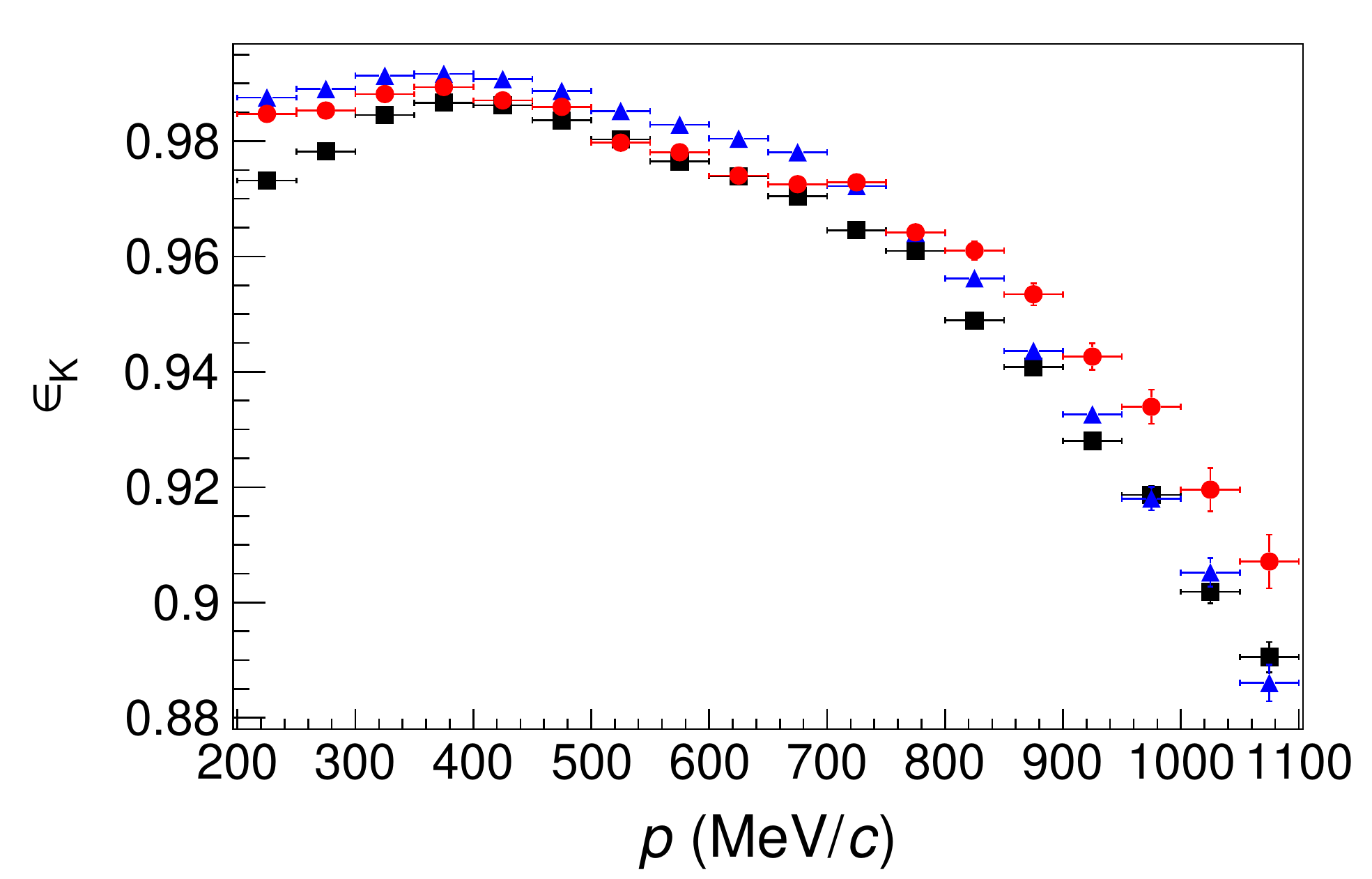}
\end{tabular}
}

\caption{PID rates as a function of momentum used to populate the $A_\text{PID}$ matrices. Entries from $\EcmA$ are shown as black squares, those from $\EcmB$ as blue triangles, and those from $\EcmC$ as red circles.}
\label{fig:PIDeffs}
\end{figure*}

\begin{figure}[!hbt]
\centering{
\begin{tabular}{c}
 \includegraphics[width=3.25in]{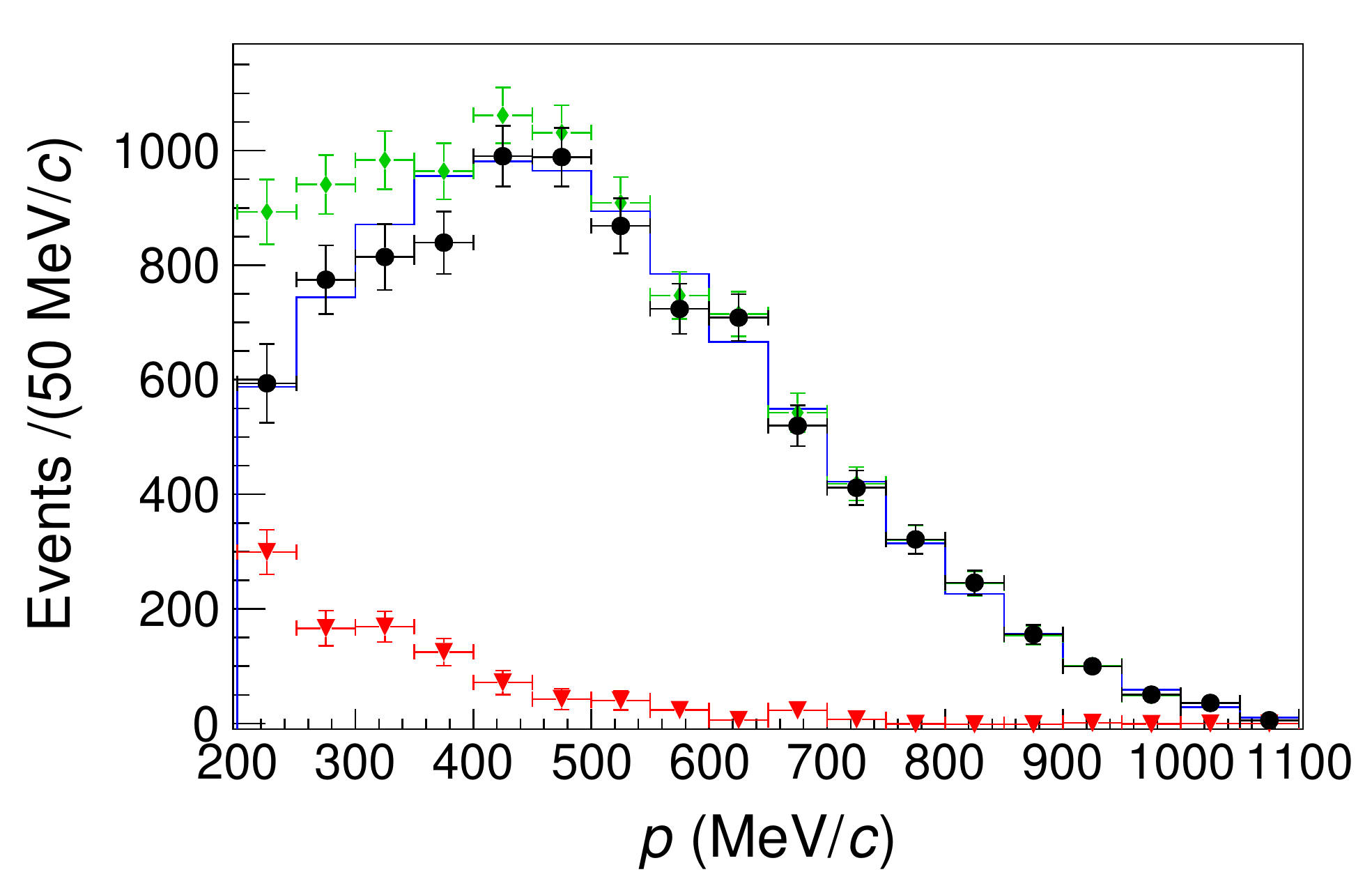}\\
 \includegraphics[width=3.25in]{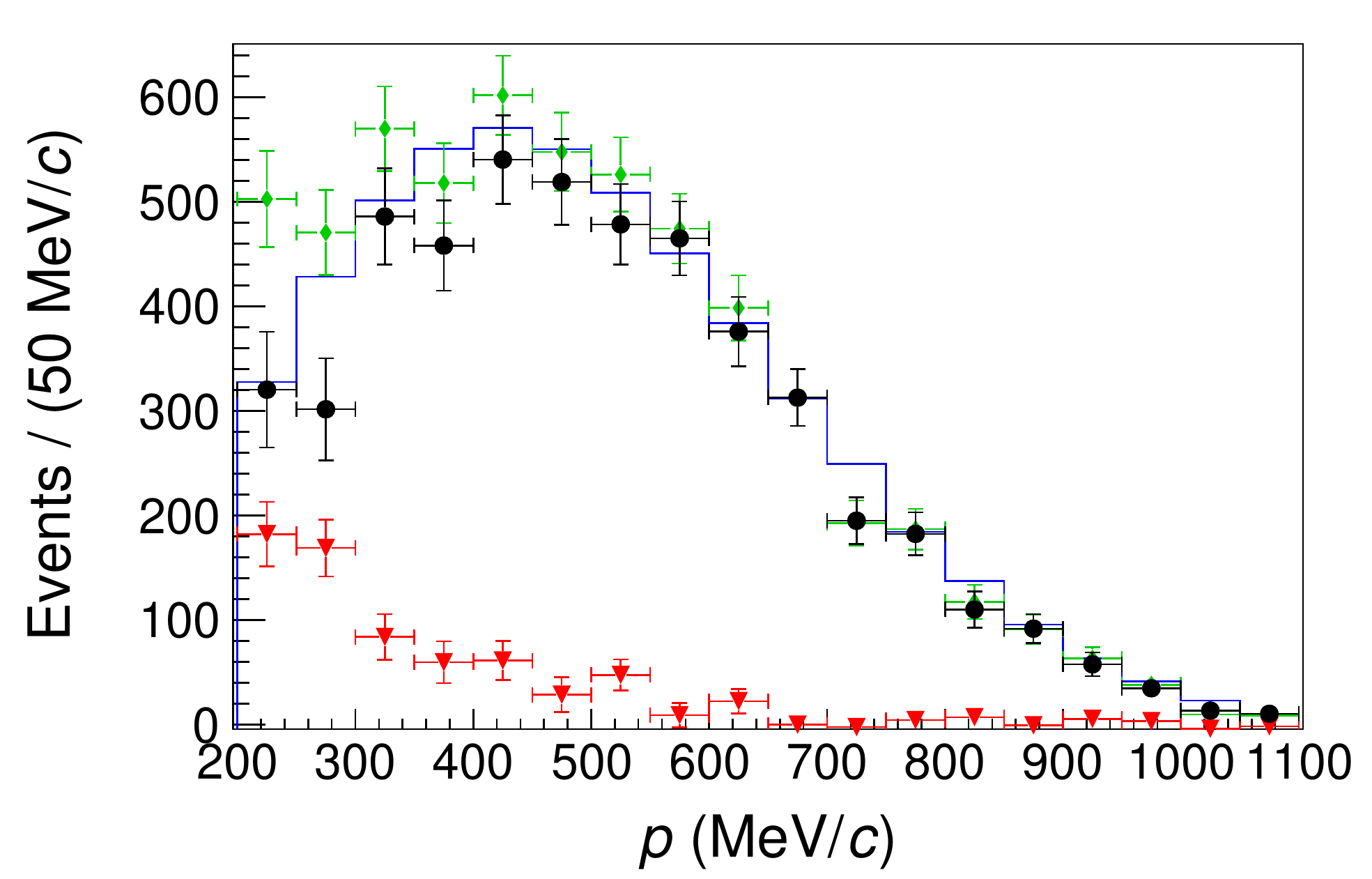}\\
\includegraphics[width=3.25in]{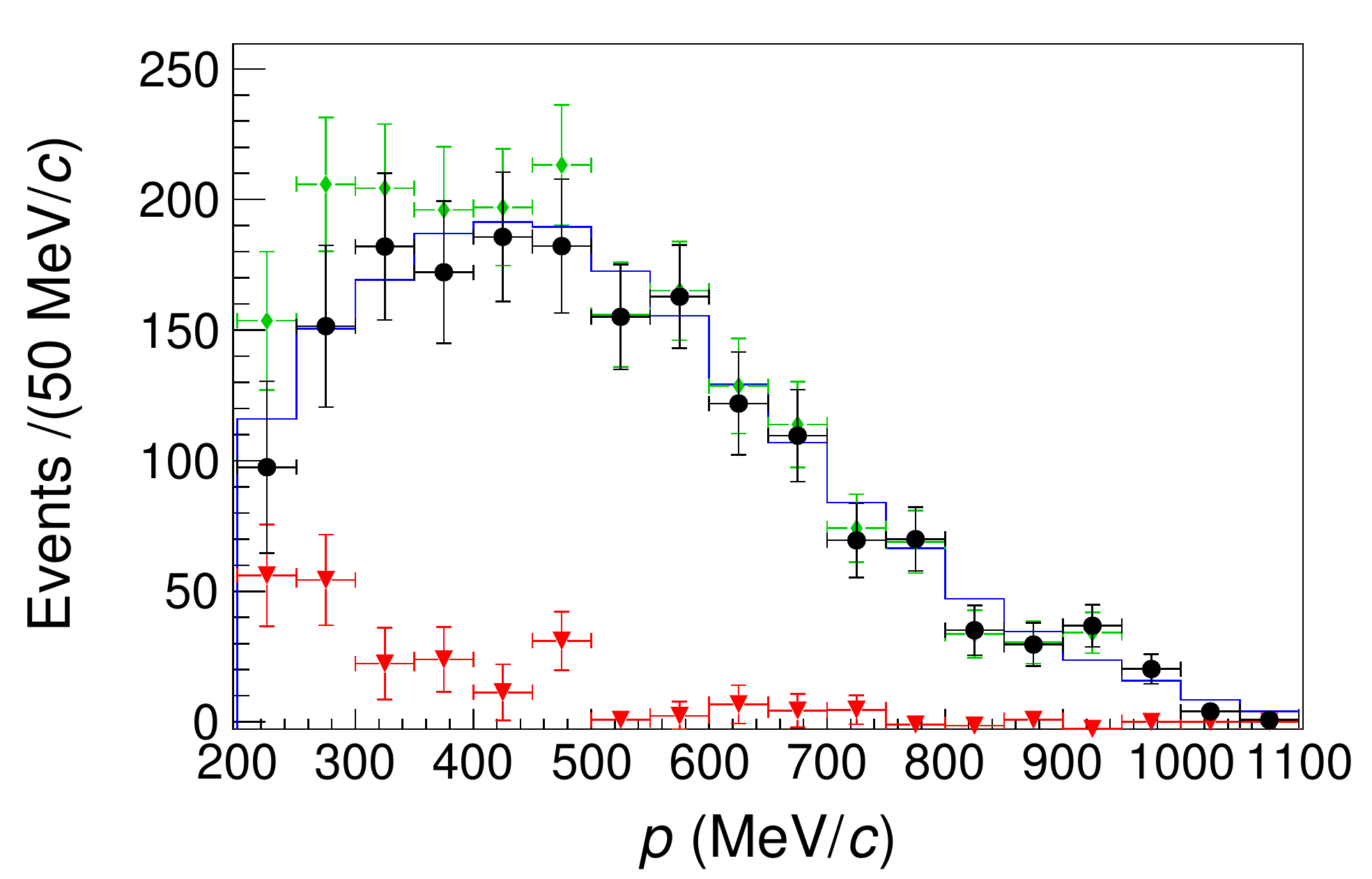}
\end{tabular}
}
\caption{PID unfolding results of positrons from
  $\semilep$ and $\taucont$ events for the data sets with $\EcmA$
  (top), $\EcmB$ (center), and $\EcmC$ (bottom). RS PID-corrected
  yields are shown as green diamonds, WS PID-corrected yields are
  shown as red triangles, their differences are shown as black
  circles, and the predictions from MC samples scaled by the number of
  single-tag events are shown as blue histograms.}
\label{fig:PIDUnfolding}
\end{figure}

With the yields determined from the fits, we perform the matrix unfolding procedure described in Sec.~\ref{sec:technique} to correct for the inefficiencies of our electron/positron identification and for the misidentification of pions and kaons as electrons in each momentum bin for both RS and WS tracks, as described in Eq.~\eqref{eqn:PIDUnf}. The elements of $A_\text{PID}$ for each data set are determined by applying our PID requirements to MC samples of particles originating from $\Ds$ decays. Differing detector conditions of the three data sets introduce deviations among the PID rates from the data sets. The rates that populate the $A_\text{PID}$ matrices are shown in Fig.~\ref{fig:PIDeffs} for each data set. The results of the unfolding procedure for both RS and WS tracks are shown in Fig.~\ref{fig:PIDUnfolding}, as well as the difference of the PID-unfolded right-sign and wrong-sign yields, which gives the PID-unfolded momentum spectra for tracked positrons originating from $\semilep$ and $\taucont$ events.  The assessment of systematic uncertainties related to the two rates to which we are most sensitive (the electron identification efficiency and the pion-faking-electron rate) are discussed in Sec.~\ref{sec:sys}.  

After taking the difference of the RS and WS PID-corrected yields, we
apply the tracking unfolding matrix to correct for tracking
reconstruction efficiencies and momentum bin mis-assignment
(Eq.~\eqref{eqn:TrkUnf}). The momentum bin mis-assignment is caused not
only by imperfect detector resolution, but also by FSR from electrons
and positrons, which increases the likelihood for the track momentum
to be less than the momentum produced in the $\Ds$ decay. The tracking
unfolding matrices are consistent among the MC samples. 



\begin{figure}[!h]
\centering{
\begin{tabular}{c}
\includegraphics[width=3.25in]{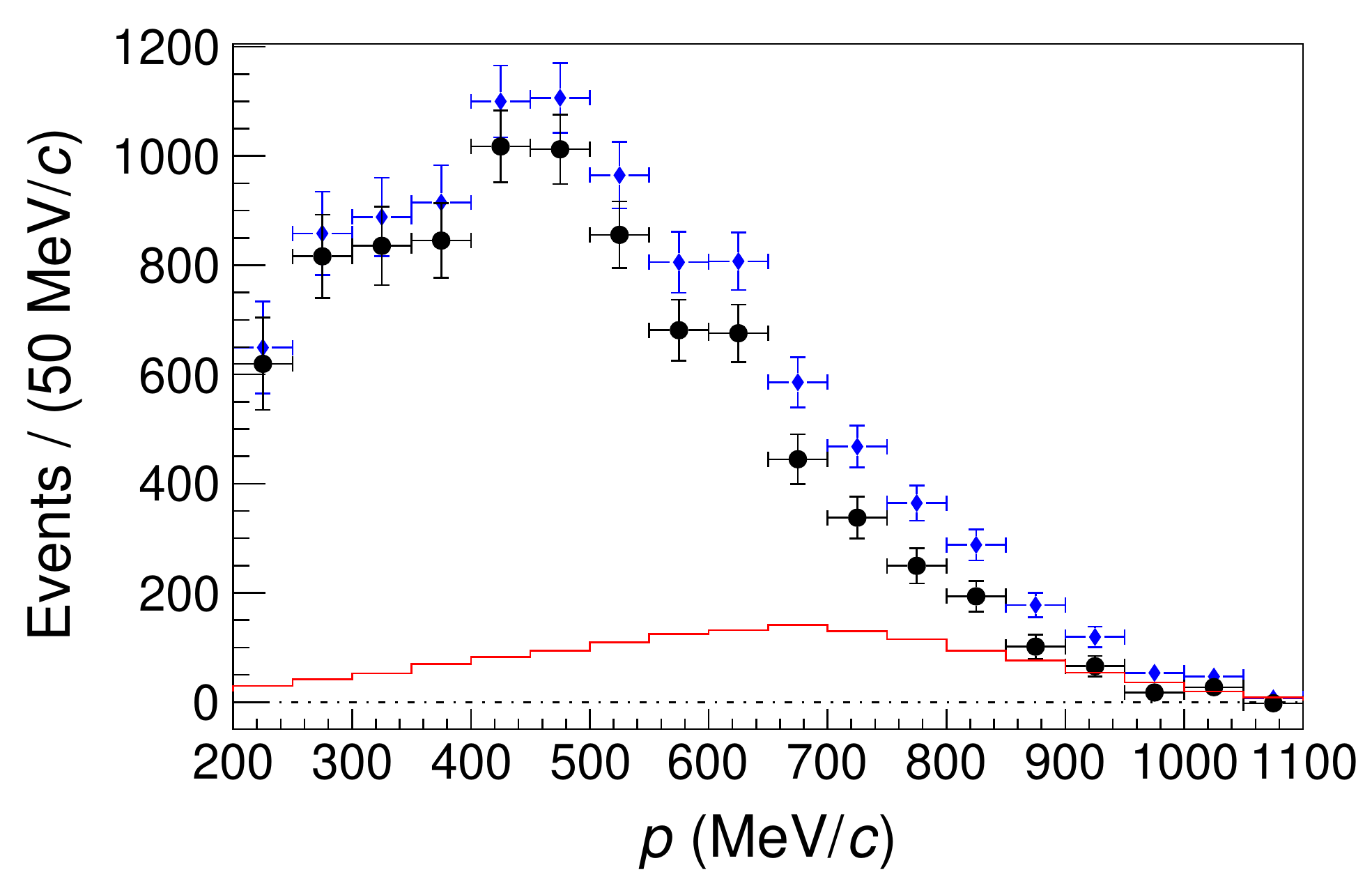}\\
\includegraphics[width=3.25in]{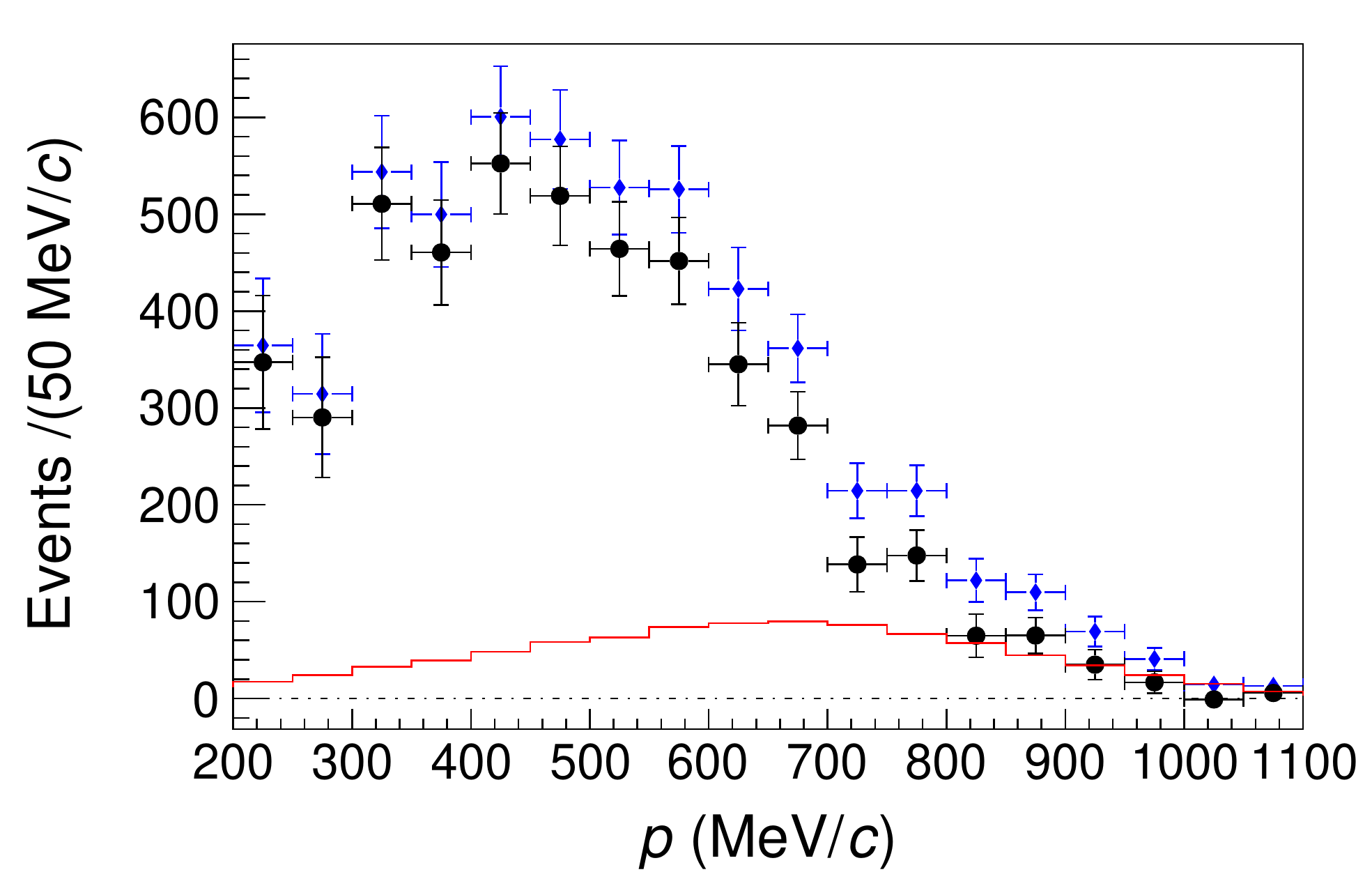}\\
\includegraphics[width=3.25in]{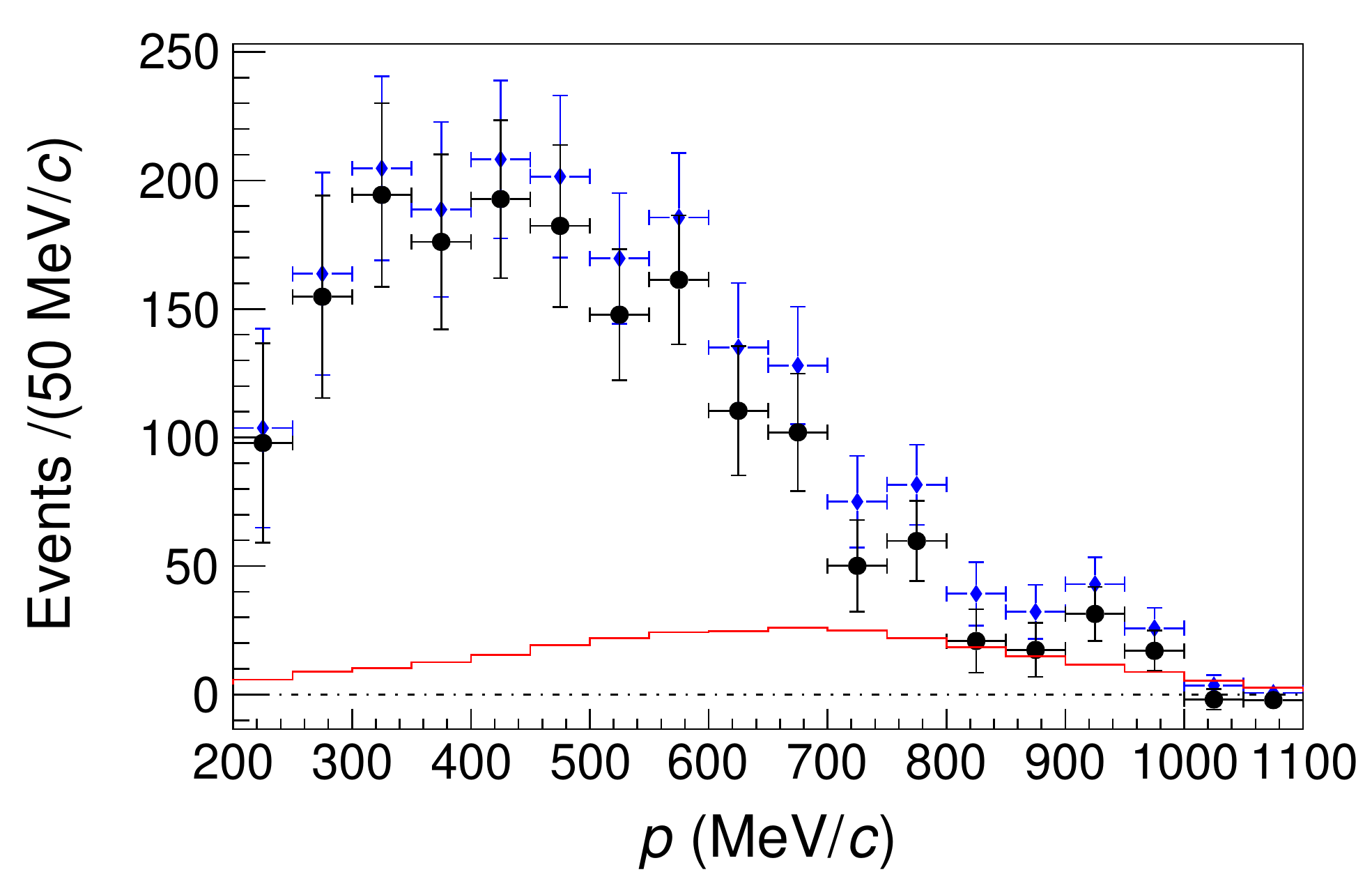}
\end{tabular}
}
\caption{Results of subtracting the contribution of $D_s^+\to\tau^+\nu_\tau\to e^+\nu_e\overline{\nu}_\tau\nu_\tau$ in $\EcmA$ data (top), $\EcmB$ data (center), and $\EcmC$ data (bottom). Tracking-corrected yields are shown as blue diamonds, and the predicted contributions from $\taucont$ are shown as red histograms. The differences of the two, the $\semilep$ yields, are shown as black circles. The dotted black line indicates zero events.}
\label{fig:TauSub}
\end{figure}

We determine the $\semilep$ positron momentum spectrum for each data sample by subtracting the contribution of positrons from $\taucont$ events from the spectra obtained from the tracking unfolding procedure. We take the $\taucont$ positron momentum spectra from MC samples. We fix the normalization according to 

\begin{equation}
N_{DT,\tau}=\mathcal B(\taucont)\times n_\text{ST}\times b_\text{tag,$\tau$},
\end{equation}
where the values of $\BF(\taucont)$ and $b_\text{tag,$\tau$}$ are given in Table~\ref{table:BFS}  and Table~\ref{table:bias}, respectively. The subtracted $\tau$ component is shown for each data set in Fig.~\ref{fig:TauSub}.

By determining the signal-efficiency-corrected number of $\semilep$ events in each data set with $p_{e}> 200 \text{ MeV}/c$, all effects of the detector response, except for small effects in $\btag$, have been accounted for. As such, we sum the determined yields from each data set in each momentum bin to produce a combined $\semilep$ momentum spectrum. A table with the total summed yields as a function of momentum along with their statistical uncertainties is provided in the supplemental material~\cite{ref:supplement}. We determine the number of $\semilep$ events with $p_{e}\leq 200 \text{ MeV}/c$ by fitting a shape based on MC simulation to the combined yields with $p_{e}>200 \text{ MeV}/c$. The shape is constructed by adding the momentum spectra predicted by MC simulation of the six observed exclusive modes ($\phi e^+\nu_e$, $\eta e^+\nu_e$, $\eta' e^+\nu_e$, $K^0 e^+\nu_e$, $K^*(892)^0 e^+\nu_e$, and $f_0(980) e^+\nu_e$) in proportion to the branching fractions listed in Table~\ref{table:BFS}. The momentum spectra for the $\eta' e^+\nu$, $K^0 e^+\nu_e$, $K^*(892)^0 e^+\nu_e$, and $f_0(980) e^+\nu_e$ modes are taken from MC samples generated with the ISGW2 model~\cite{ref:ISGW2} for the decay of the $\Ds$ meson.
We generate separate MC samples for the two largest modes, $\phi e^+\nu_e$ and $\eta e^+\nu_e$, using simple-pole parameterizations of the respective
decay form factors as functions of the lepton-neutrino system squared
four-momentum $q^2$. The form-factor parameters are taken from
measurements of the BABAR collaboration for $\phi e^+\nu_e$~\cite{ref:BABAR2008phiBF} and the BESIII
collaboration for $\eta e^+\nu_e$~\cite{ref:BESIII2019etaBF}.
%
%

\begin{figure}[hbt]
\centering{
\includegraphics[width=3.25in]{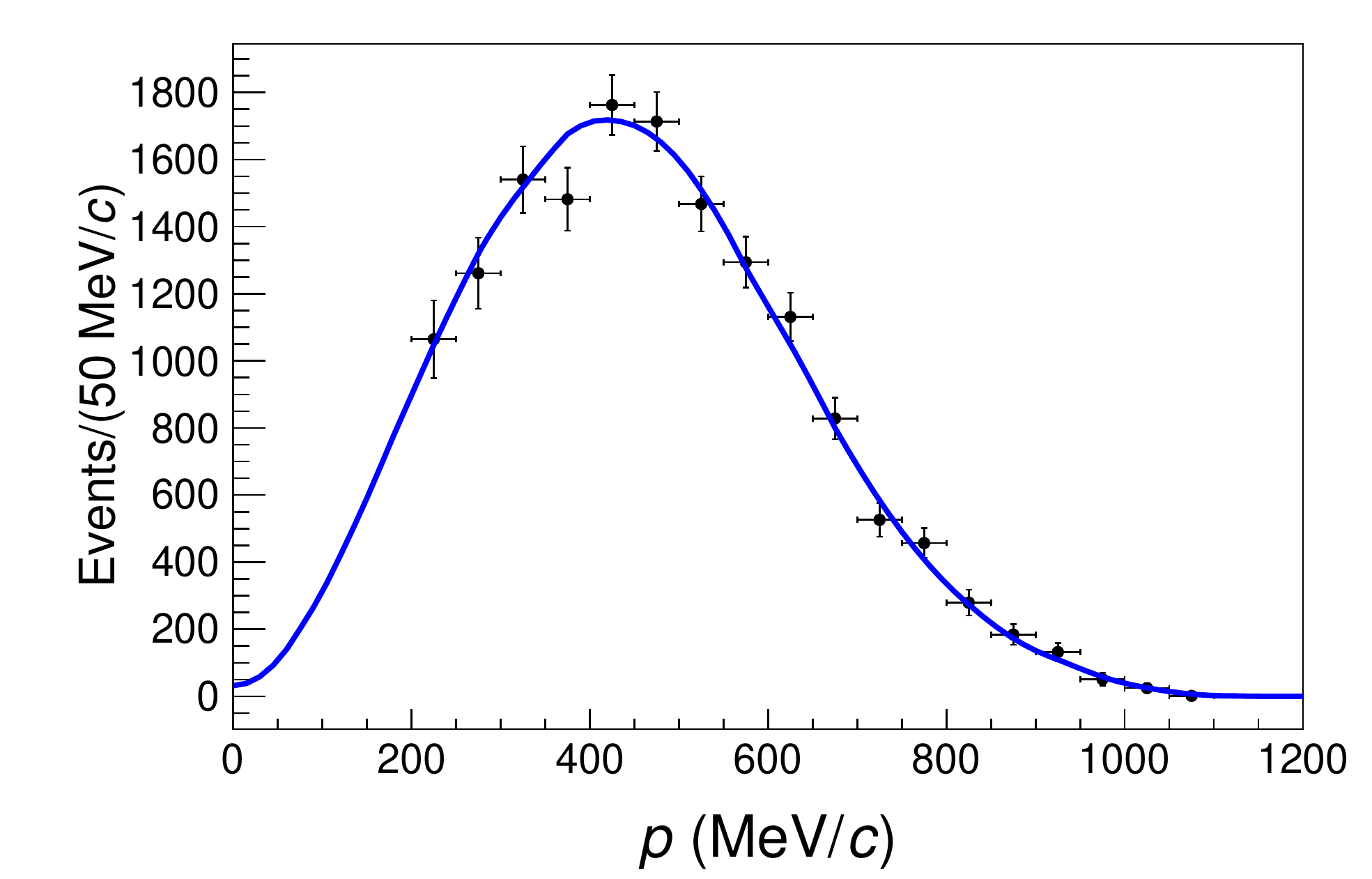}
}
\caption{Momentum spectrum fit used to determine the $\Ds$ semielectronic decay yield below $200$ MeV/$c$.  The black points are sums of the final measured $\semilep$ yields (black circles) for the three data sets from Fig.~\ref{fig:TauSub}, and the solid blue line is result of the fit described in the text.}
\label{fig:Data_extrapolation}
\end{figure}

The integral of the fitted spectrum from $p_{e}=0 \text{ MeV}/c$ to
$p_{e}=200 \text{ MeV}/c$ is added to the number of
signal-efficiency-corrected $\semilep$ events with $p_{e}> 200 \text{
  MeV}/c$ as described in Eq.~\ref{eq:extrap}, and the statistical
uncertainty is scaled by the ratio by which the total yield increases due
to this correction. The fit to data with the assumed momentum spectrum can be found in
Fig.~\ref{fig:Data_extrapolation}. As a crosscheck, we also fit to the
data sets separately, and find consistent results. The yields in data
both without and with the correction for the data with $p_{e}<200
\text{ MeV}/c$ are shown in Table~\ref{table:extrapolation}.

\begin{table*}[hbt]
\centering{
\begin{tabular}{c|p{4cm}|c}
Sample & \centering{Observed Yields with $p_{e}>200 \text{ MeV}/c$ }&Corrected Yields $(N_\text{DT})$\\ \hline
$\EcmA$ &\centering{} $8793 \pm 218$&$9628 \pm 238$ \\ \hline
$\EcmB$ &\centering{} $4695 \pm 174$&$5142 \pm 191$ \\ \hline
$\EcmC$ &\centering{} $1712 \pm 104$&$1874\pm 114$ \\ \hline \hline
Combined & \centering{}$15201 \pm 298$ & $16648\pm 326$
\end{tabular}
}
\caption{Yields in data before and after momentum-extrapolation correction. The ``Combined'' row shows the results of summing together the yields from the three data sets and fitting to the summed distribution. Shown uncertainties are only statistical.}
\label{table:extrapolation}
\end{table*}

Using Eq.~\eqref{eqn:DTagBRF}, we determine $\semilepBF$ with the momentum-extrapolated number of signal-efficiency-corrected double-tag events from Table~\ref{table:extrapolation}, the observed number of  single-tag events from Table~\ref{table:STYields}, and the tag bias from Table~\ref{table:bias}. The branching fractions determined from each data set independently, their average, and the branching fraction determined from their combination are shown along with the associated statistical uncertainties in Table~\ref{table:SemiLepBF}. The difference of the ``Combined'' result from Table~\ref{table:SemiLepBF}, $\semilepBF=\left(6.30\pm0.13(\text{stat.})\right)\%$, and the sum of the observed exclusive semielectronic branching fractions from Table~\ref{table:BFS} gives the unobserved $\Ds$ semielectronic branching fraction as  $\left(-0.04 \pm 0.21\right)\%$, where the stated uncertainty includes the total uncertainty from the exclusive measurements, but only the statistical uncertainty from the inclusive measurement presented in this article.

\begin{table}[hbt]
\centering{
\begin{tabular}{c|c}
Sample & $\semilepBF$\\ \hline
$\EcmA$ &$\left(6.38\pm0.16\right)\%$ \\ \hline
$\EcmB$ &$\left(5.96\pm0.23\right)\%$ \\ \hline
$\EcmC$ & $\left(6.38\pm0.40\right)\%$ \\ \hline \hline
Combined & $\left(6.30\pm0.13\right)\%$
\end{tabular}
}

\caption{$\semilepBF$ determined from data. The ``Combined'' row shows the results of summing together the yields from the three data sets and fitting to the summed distribution. Shown uncertainties are only statistical.}
\label{table:SemiLepBF}
\end{table}

\section{SYSTEMATIC UNCERTAINTY}\label{sec:sys}

Our methods to determine the relative systematic uncertainty on our measured $\semilepBF$ are described below.

\subsection{MC Simulation Statistics and Matrix-Inversion Stability}
We probe the effects of finite MC sample statistics and the stability of the matrix-inversion algorithm by creating a toy ensemble of variations for each efficiency matrix used in the analysis (54 PID matrices and three tracking matrices). For each matrix, we create this ensemble by sampling each entry within the MC sample's statistical uncertainty. For each matrix in each ensemble, we perform the matrix inversion and reperform the analysis with the new inverted matrix. All variations produce a negligible change in our final result, which indicates a negligible systematic uncertainty from the statistical uncertainty of the MC samples as well as the stability of the algorithm for inverting our efficiency matrices.

\subsection{Tracking}\label{sec:TrkSys}

Simulation of our tracking efficiency is studied with a control sample of radiative Bhabha events. Tracking efficiencies as a function of momentum are measured in each data set, as well as in MC samples produced with the BabayagaNLO package~\cite{ref:BabayagaNLO}. The ratios of the measured efficiencies in data and MC samples are weighted by the predicted momentum distribution from signal MC simulation and the number of single-tag events in each data set to determine the systematic uncertainty. This results in a relative systematic uncertainty of $0.7\%$.

In addition, we investigate the systematic uncertainty in the individual tracking efficiency matrix entries. As we assign a systematic uncertainty for the total tracking efficiency, we probe the uncertainty in the individual entries by keeping the sum of a row of the matrix constant while varying the individual entries. The specific variation is as follows:
\begin{itemize}
\item $\left(A_\text{trk}\right)_{i,i-1}\Rightarrow 1.5\left(A_\text{trk}\right)_{i,i-1}$
\item $\left(A_\text{trk}\right)_{i,i+1}\Rightarrow 1.25\left(A_\text{trk}\right)_{i,i+1}$
\item $\left(A_\text{trk}\right)_{i,i}$ decreases to keep the sum of the column constant.
\end{itemize}
This variation is chosen as a conservative estimation of the uncertainties from FSR and detector resolution. We see negligible change when we perform such a variation, so we  only assign the previously stated uncertainty for tracking.
\subsection{PID}\label{sec:PIDSys}

Similar to our procedure in assessing the systematic uncertainty in
our tracking efficiencies, we measure $e$ ID efficiencies as a
function of momentum and track angle in radiative Bhabha control
samples for each data set.

We also probe the accuracy of the pion-faking-electron rates from MC simulation via a control sample of pions collected in each data sample through the decay chains $D^{*+}\to\pi^+ D^0,$  $D^0\to K^-\pi^+,K^-\pi^+\pi^+\pi^-$. To determine the total uncertainty from PID rates, we simultaneously vary the $e$ ID efficiencies and the pion-faking-electron rates using the central values of the measured data-to-MC efficiency ratios and reperform our analysis. This yields a $0.8\%$ change in the final branching fraction, which we assign as the relative systematic uncertainty due to PID.

As our sensitivity to kaon-faking-electron rates is small due to the relatively few number of kaons, the systematic uncertainty in kaon-faking-electron rates is neglected.

\subsection{Tag Bias}
We follow the procedure laid out in~\cite{ref:TagBiasRef}, which assigns a fraction of $1-\btag$ as the systematic uncertainty based on the particles in the final state of the single-tag $D_s^-$ decay. The specific guidelines for variation of detector-response parameters are as follows: $1.0\%$ per kaon for tracking, $0.5\%$ per pion for tracking, and $0.5\%$ per kaon or pion for PID. For $D_s^-\to K^+K^-\pi^-$, with two kaons and one pion, the quadrature sum is $2.9\%$. With $\btag$ from Table~\ref{table:bias} (including the contribution from $\taucont$), taking $2.9\%$ of $1-\btag$ yields a $0.03\%$ relative systematic uncertainty. We additionally propagate the uncertainties in the branching fractions (Table~\ref{table:BFS}) through the calculation of the weighted-average $\btag$. This yields a $0.07\%$ relative systematic uncertainty in the bias. We add these in quadrature and assign the relative systematic uncertainty due to tag bias as $0.1\%$.

\subsection{Number of Single Tags}
We investigate the systematic uncertainty in the invariant-mass fitting procedure used to determine the number of single tags by varying the choice of background distribution from the nominal second-order Chebyshev polynomial. We use both first-order and third-order Chebyshev polynomials as variations in fitting to each data set. Using the first-order Chebyshev polynomial gives a larger difference in the yields in all cases, while not significantly degrading the quality of the fit. We take an average of the changes for each data set weighted by the single-tag yields to determine the systematic uncertainty, which results in a 0.6\%  relative systematic uncertainty in the number of single tags.

\subsection{Background Shapes}
To assess the uncertainty due to our chosen background shapes in our signal-side fits of the tag invariant mass, we use background distributions based on MC simulation instead of the nominal first-order Chebyshev polynomial to model backgrounds in each of the signal-side fits.  We then reperform the analysis with the yields determined from these alternative fits. The relative difference in $N_\text{DT}$ is 0.4\%, which we assign as the relative systematic uncertainty due to this source. 

\subsection{$\BF(\Ds\to\tau^+\nu_\tau)$}
As systematic uncertainty in the kinematic distributions of $\taucont$ events contribute negligible uncertainty, we account for uncertainty in the contribution of $\taucont$ events by propagating the absolute uncertainty for $\BF(\taucont)$ from Table~\ref{table:BFS}, $0.04\%$, which yields a relative systematic uncertainty of $0.6\%$ on our measurement of $\semilepBF$.
 
\subsection{Spectrum Extrapolation}\label{sec:SpectrumSys}

We assess the uncertainty due to the momentum spectrum extrapolation
by generating an ensemble of alternative momentum spectra and fitting
these to the data. Each spectrum in this ensemble is created by
Gaussian sampling the branching fractions of the six observed
exclusive semielectronic modes and adding spectra for unobserved decay
modes. We then add these spectra in proportion based on the sampled
branching fractions. 

We consider effects from the combination of three unobserved decay modes: $\Ds\to h_1\left(1415\right)e^+\nu_e$, $\Ds\to f_1\left(1510\right)e^+\nu_e$, and $\Ds\to\gamma e^+\nu_e$. MC samples for  $\Ds\to h_1\left(1415\right)e^+\nu_e$ and $\Ds\to f_1\left(1510\right)e^+\nu_e$ are generated based on the ISGW2 model's predictions~\cite{ref:ISGW2}. We generate MC samples for $\Ds\to\gamma e^+\nu_e$ based on the model of Yang and Yang~\cite{ref:GammaENuModel}. We determine the normalization of the $\Ds\to h_1\left(1415\right)e^+\nu_e$ and $\Ds\to f_1\left(1510\right)e^+\nu_e$ spectra by fixing the relative branching fraction of these decays to the ISGW2 predictions and fitting them in addition to our nominal momentum spectrum. The $\Ds\to\gamma e^+\nu_e$ spectrum is fixed to its measured $90\%$ confidence level upper limit, $1.3\times 10^{-4}$~\cite{ref:GammaENu}. From this fit, we determine  an upper limit at the $90\%$ confidence-level for the sum of the branching fractions of $\Ds\to h_1\left(1415\right)e^+\nu_e$ and $\Ds\to f_1\left(1510\right)e^+\nu_e$. In our toy ensemble, their summed spectra are fixed based on this upper limit. We performed this same procedure excluding the $\Ds\to\gamma e^+\nu_e$ spectrum, but found the combination of modes to produce the largest variation.

The resulting systematic uncertainty is determined by filling a distribution of the relative change in $\semilepBF$ between the alternative momentum spectra and our nominal spectrum. The linear sum of the mean ($0.33\%$) and RMS ($0.34\%$) of this distribution is taken as the uncertainty, which gives a $0.7\%$ relative systematic uncertainty.

We also probe uncertainty in the models we employ by using the ISGW2-predicted momentum spectra for the $\Ds\to\phi e^+\nu_e$ and $\Ds\to\eta e^+\nu_e$ modes. Using these alternative spectra instead of the nominal spectra gives a less than $0.1\%$ difference in the measured branching fraction, so we conclude that any uncertainty due to model dependence in our analysis is negligible.

\subsection{Summary of Systematic Uncertainties}
The assigned relative systematic uncertainties for all sources are listed in Table~\ref{table:systematics}. Our systematic uncertainty is not dominated by any single source, but the largest contributions come from the momentum spectrum extrapolation and imperfect simulation of PID and tracking efficiencies. The total relative systematic uncertainty is obtained from the quadrature sum of the assigned relative uncertainties. This gives a total relative systematic uncertainty of  $1.6\%$.

\begin{table}[hbt]
\centering{
\begin{tabular}{c|c}
Source&Relative Uncertainty\\ \hline
Tracking & 0.7\%\\ \hline
PID & 0.8\%\\ \hline
Spectrum Extrapolation & 0.7\%\\ \hline
Background Shapes & 0.4\%\\ \hline
Number of Tags & 0.6\%\\ \hline
Tag Bias & 0.1\%\\ \hline
$\BF(\Ds\to\tau^+\nu_\tau)$ & 0.6\%\\ \hline\hline
Total & 1.6\%
\end{tabular}
}
\caption{Systematic uncertainties in the measurement of $\semilepBF$.}
\label{table:systematics}
\end{table}

\section{Summary and Discussion}\label{sec:conc}

Using data collected by the BESIII detector in the center-of-mass-energy range of $\Ecm=4.178-4.230\text{ GeV}$, we measure the inclusive semielectronic branching fraction of the $\Ds$ meson to be
\[
\semilepBF=\left(6.30\pm0.13\;(\text{stat.})\pm 0.10\;(\text{syst.})\right)\%.
\]
We also measure the lab-frame momentum spectrum of the positrons produced in this decay, which can be seen in Fig.~\ref{fig:Data_extrapolation}.

Our result is consistent with the measurement from the CLEO-c experiment\cite{ref:CLEOMeas},
\[
\semilepBF =(6.52\pm 0.39\;(\text{stat.}) \pm 0.15\;(\text{syst.}))\%,
\]
with a factor of $3$ reduction of the statistical uncertainty and a factor of $1.5$ reduction of the systematic uncertainty. The total precision of our measurement is $2.6\%$,  which corresponds to approximately a $2.5$ times improvement in the total precision compared to the measurement from CLEO-c. 

By taking the difference between our measurement of $\semilepBF$ and the sum of the best available measurements for the exclusive semielectronic modes in Table~\ref{table:BFS}, we calculate the unobserved semielectronic branching fraction to be
\[
\begin{aligned}
\semilepBF-\sum_i \mathcal B\left(\Ds\to X_i e^+\nu_e\right)=\\\left(-0.04\pm 0.13\;(\text{stat.})\pm 0.20\;(\text{syst.})\right)\%,
\end{aligned}
\]
where the systematic uncertainty includes the total uncertainty on the measured exclusive branching fractions. Our measurement provides no evidence for the existence of unobserved $\Ds$ semileptonic decay modes and constrains the branching fractions of all unobserved decay modes. In addition, the measured momentum spectrum can be used to further constrain the decay rates of modes with characteristic momentum spectra. The spectrum is included in tabular form in the supplemental material~\cite{ref:supplement}.

With our updated measurement of the $\Ds$ semielectronic branching fraction, the CLEO-c measurement of the $\D0$ semielectronic branching fraction~\cite{ref:CLEOMeas}, and the 2020 PDG values for the $\Ds$ and $\D0$ lifetimes~\cite{ref:PDG}, we find
\[
\frac{\Gamma(D_{s}^{+}\rightarrow Xe^+\nu_e)}{\Gamma(D^0\rightarrow Xe^+\nu_e)}=0.790\pm 0.016\;(\text{stat.})\pm0.020\;(\text{syst.}),
\]
where the systematic uncertainty includes the total uncertainty from $\mathcal B \left(\D0\to X e^+\nu_e\right)$. This result is in agreement with the prediction of $\frac{\Gamma(\Ds \to Xe^+\nu)}{\Gamma(\D0 \to Xe^+\nu) }=0.813$ from~\cite{ref:Rosner}, supporting the conclusion that the difference in the semileptonic decay widths of $\Ds$ and $\D0$ mesons can be accounted for within the Standard Model by non-spectator interactions. Further theoretical analysis of our measured spectrum, similar to those of~\cite{ref:Manohar} and \cite{ref:Gambino}, can constrain specific processes like WA of the constituent $c$ and $\overline s$ quarks of the $\Ds$, with potential extensions to determinations of $|V_{ub}|$ in semileptonic $B$ decays~\cite{ref:Vos}.

\vspace{3ex}
\section{Acknowledgments}\label{sec:acknowledgements}
 
The BESIII collaboration thanks the staff of BEPCII and the IHEP computing center for their strong support. This work is supported in part by National Key Research and Development Program of China under Contracts Nos. 2020YFA0406400, 2020YFA0406300; National Natural Science Foundation of China (NSFC) under Contracts Nos. 11625523, 11635010, 11735014, 11822506, 11835012, 11935015, 11935016, 11935018, 11961141012; the Chinese Academy of Sciences (CAS) Large-Scale Scientific Facility Program; Joint Large-Scale Scientific Facility Funds of the NSFC and CAS under Contracts Nos. U1732263, U1832207; CAS Key Research Program of Frontier Sciences under Contract No. QYZDJ-SSW-SLH040; 100 Talents Program of CAS; INPAC and Shanghai Key Laboratory for Particle Physics and Cosmology; ERC under Contract No. 758462; European Union Horizon 2020 research and innovation programme under Contract No. Marie Sklodowska-Curie grant agreement No 894790; German Research Foundation DFG under Contracts Nos. 443159800, Collaborative Research Center CRC 1044, FOR 2359, FOR 2359, GRK 214; Istituto Nazionale di Fisica Nucleare, Italy; Ministry of Development of Turkey under Contract No. DPT2006K-120470; National Science and Technology fund; Olle Engkvist Foundation under Contract No. 200-0605; STFC (United Kingdom); The Knut and Alice Wallenberg Foundation (Sweden) under Contract No. 2016.0157; The Royal Society, UK under Contracts Nos. DH140054, DH160214; The Swedish Research Council; U. S. Department of Energy under Contracts Nos. DE-FG02-05ER41374, DE-SC-0012069.



\pagebreak
\widetext
\clearpage
\begin{center}
\textbf{\large Supplemental Material}
\end{center}
\setcounter{equation}{0}
\setcounter{page}{1}
\makeatletter

\begin{table*}[h!]
\centering{
\begin{tabular}{c|c}
$p_e$ (MeV/$c$) & $N_{DT,Xe^+\nu_e}$\tabularnewline \hline
200 - 250    & $1064 \pm 115$ \tabularnewline\hline
250 - 300    & $1261 \pm 106$ \tabularnewline\hline
300 - 350    & $1540 \pm 99$ \tabularnewline\hline
350 - 400    & $1482 \pm 93$ \tabularnewline\hline
400 - 450    & $1762 \pm 90$ \tabularnewline\hline
450 - 500    & $1713 \pm 88$ \tabularnewline\hline
500 - 550    & $1468 \pm 82$ \tabularnewline\hline
550 - 600    & $1294 \pm 76$ \tabularnewline\hline
600 - 650    & $1130 \pm 72$ \tabularnewline\hline
650 - 700    & $829 \pm 62$ \tabularnewline\hline
700 - 750    & $526 \pm 51$ \tabularnewline\hline
750 - 800    & $457 \pm 44$ \tabularnewline\hline
800 - 850    & $280 \pm 38$  \tabularnewline\hline
850 - 900    & $184 \pm 31$  \tabularnewline\hline
900 - 950    & $132 \pm 27$  \tabularnewline\hline
950 - 1000   & $52 \pm 19$  \tabularnewline\hline
1000 - 1050  & $24 \pm 14$    \tabularnewline\hline
1050 - 1100  & $2 \pm 6$    \tabularnewline\hline
\end{tabular}
}
 \caption{Efficiency-corrected and background subtracted yields ($N_{DT,Xe^+\nu_e}$) with statistical errors as a function of $e^+$ momentum.}
\label{table:Yields}
\end{table*}

\section{Double-tag Fits}
\label{sec:DataFits}
The following section contains the results of the double-tag fits in which the observed $\Ds\to X e^+$,$\Ds\to X \pi^+$, and $\Ds\to X K^+$  yields are determined for each sign hypothesis and momentum bin. In each plot, the solid blue line is the result of the total fit, the dashed red line is the fitted distribution of non-$D_s^-\to K^+K^-\pi^-$ backgrounds, the dotted black line is the fitted signal distribution, the filled red histogram is the MC simulation-predicted contributions from backgrounds, and the black points are data. Binning is arbitrary and used solely for display.
\subsection{$\EcmA$ Double-tag Fits}
\label{subsec:4180DataFits}
\subsubsection{$\EcmA$ Data $e$ ID Fits}
\label{subsubsec:4180DataEIDFits}
\centering
\begin{tabular}{cc}
\textbf{RS 200-250 MeV/$c$} & \textbf{WS 200-250 MeV/$c$}\\
\includegraphics[width=3in]{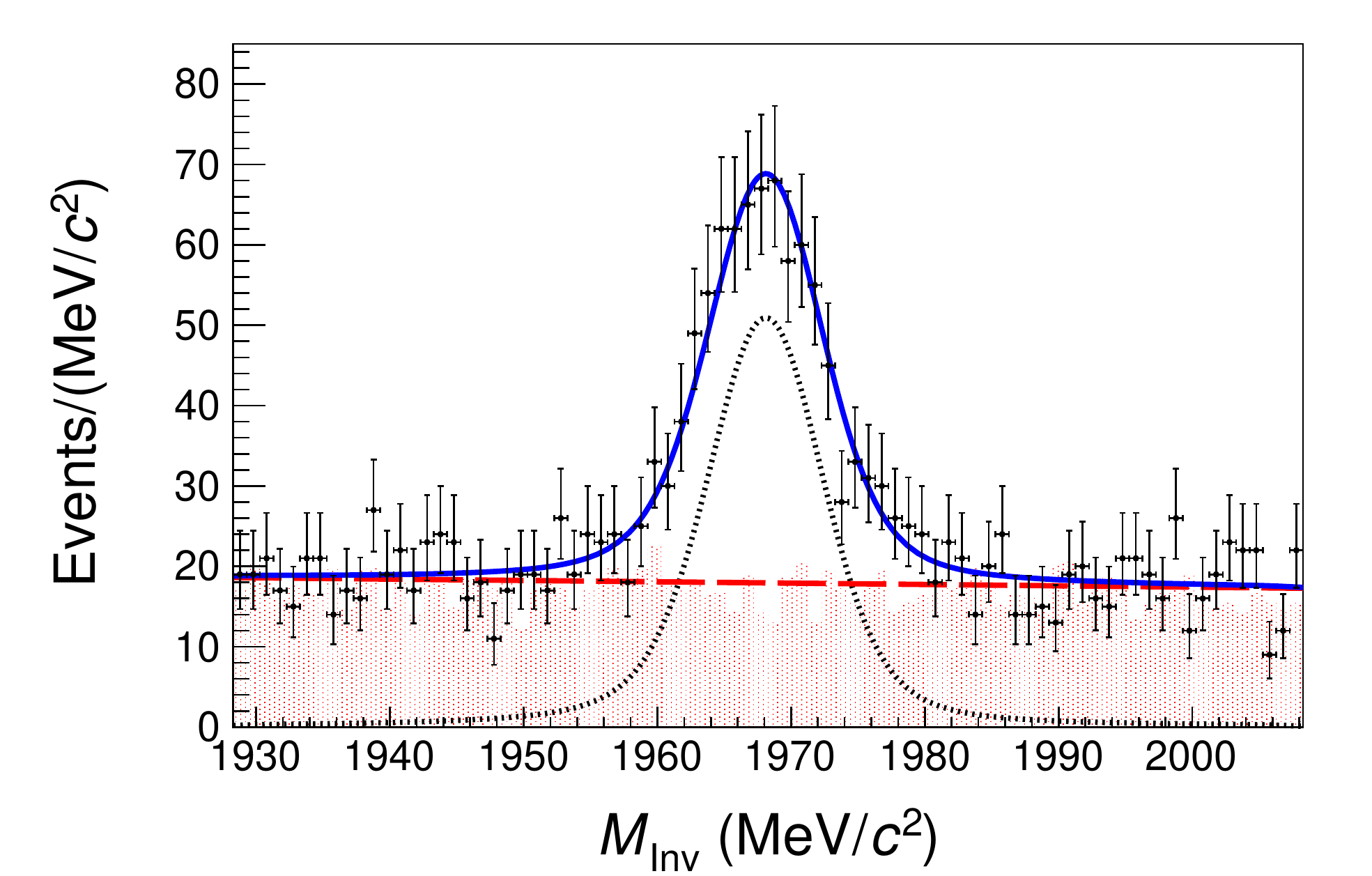} & \includegraphics[width=3in]{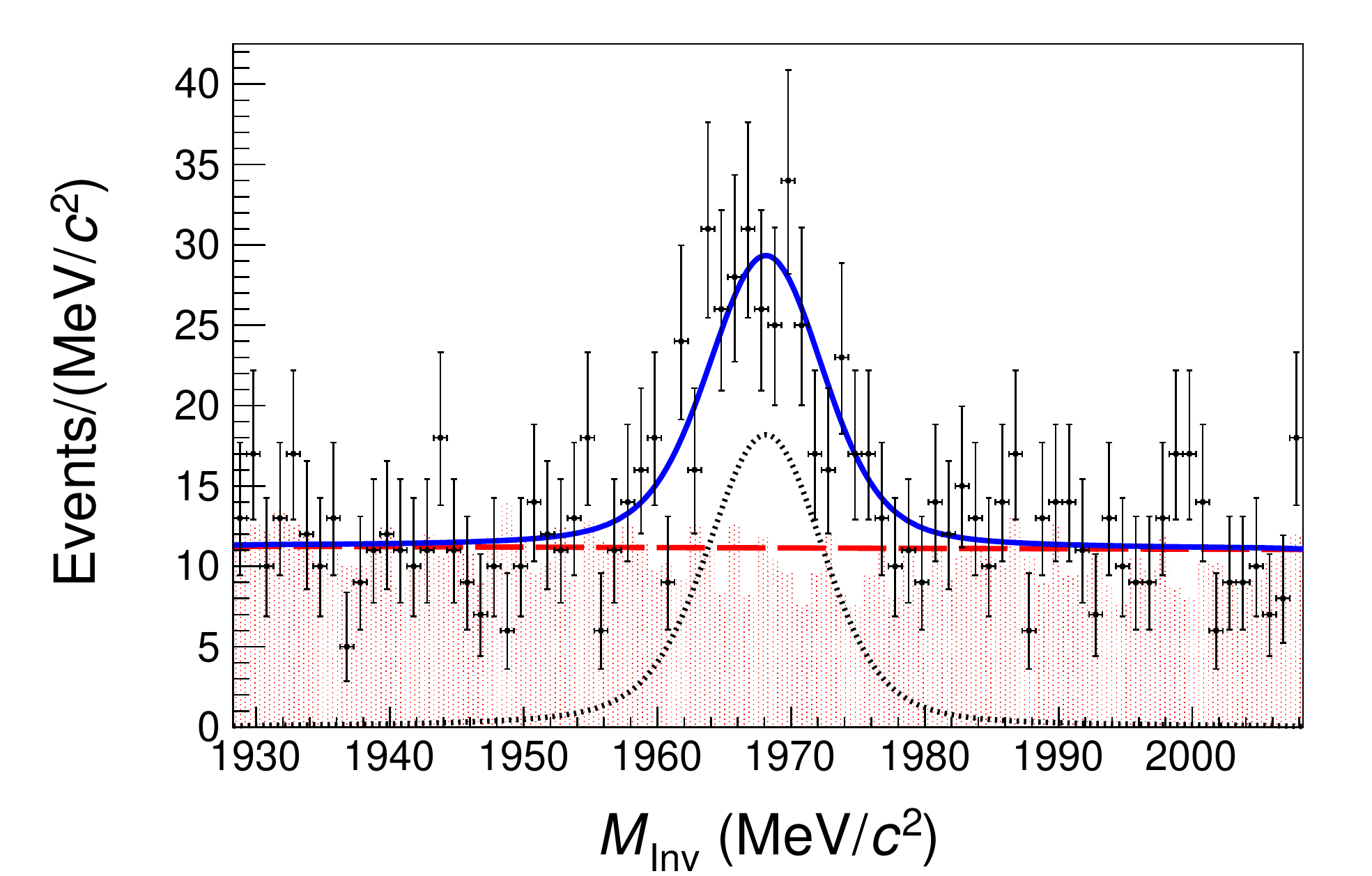}\\
\textbf{RS 250-300 MeV/$c$} & \textbf{WS 250-300 MeV/$c$}\\
\includegraphics[width=3in]{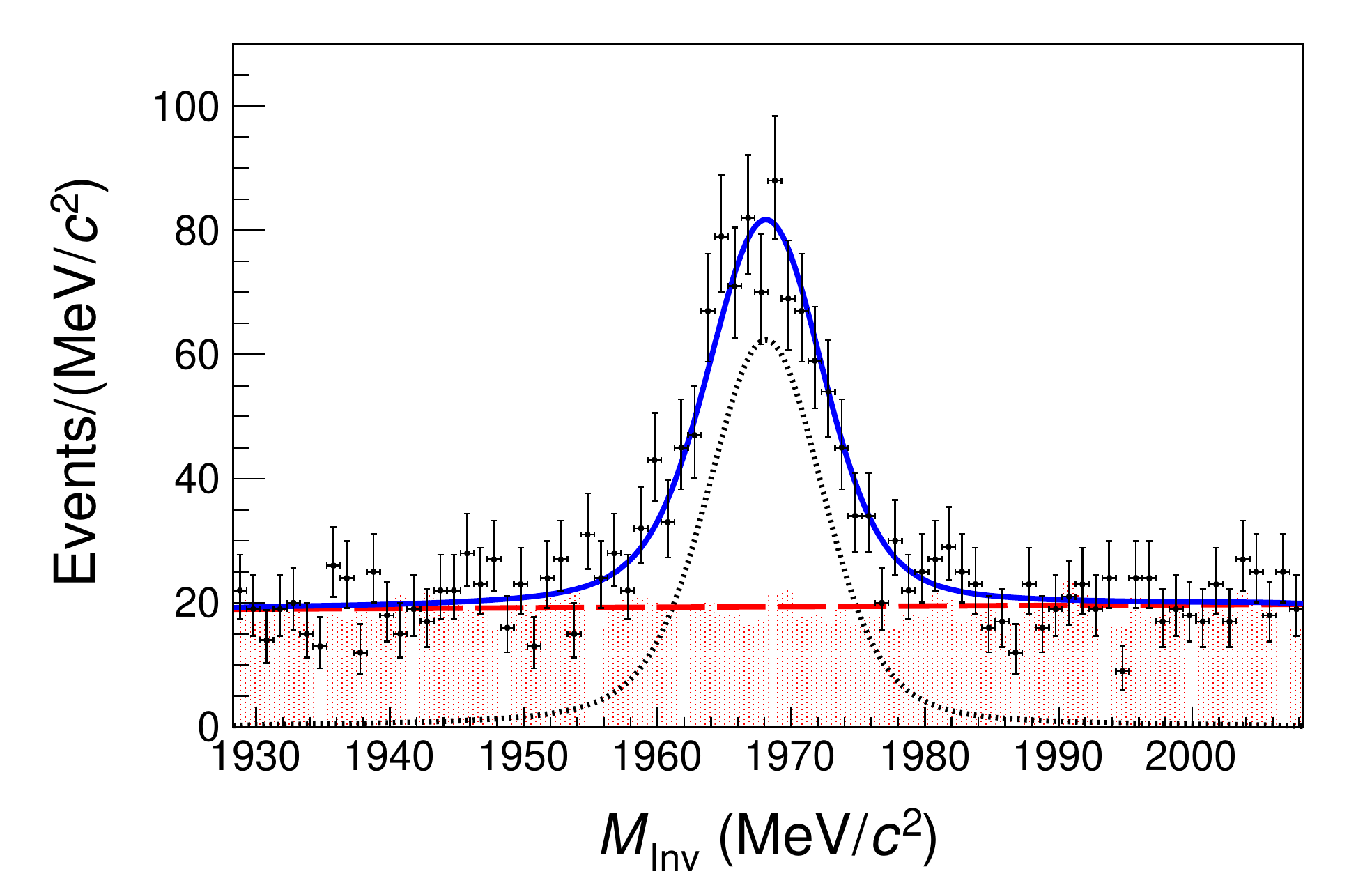} & \includegraphics[width=3in]{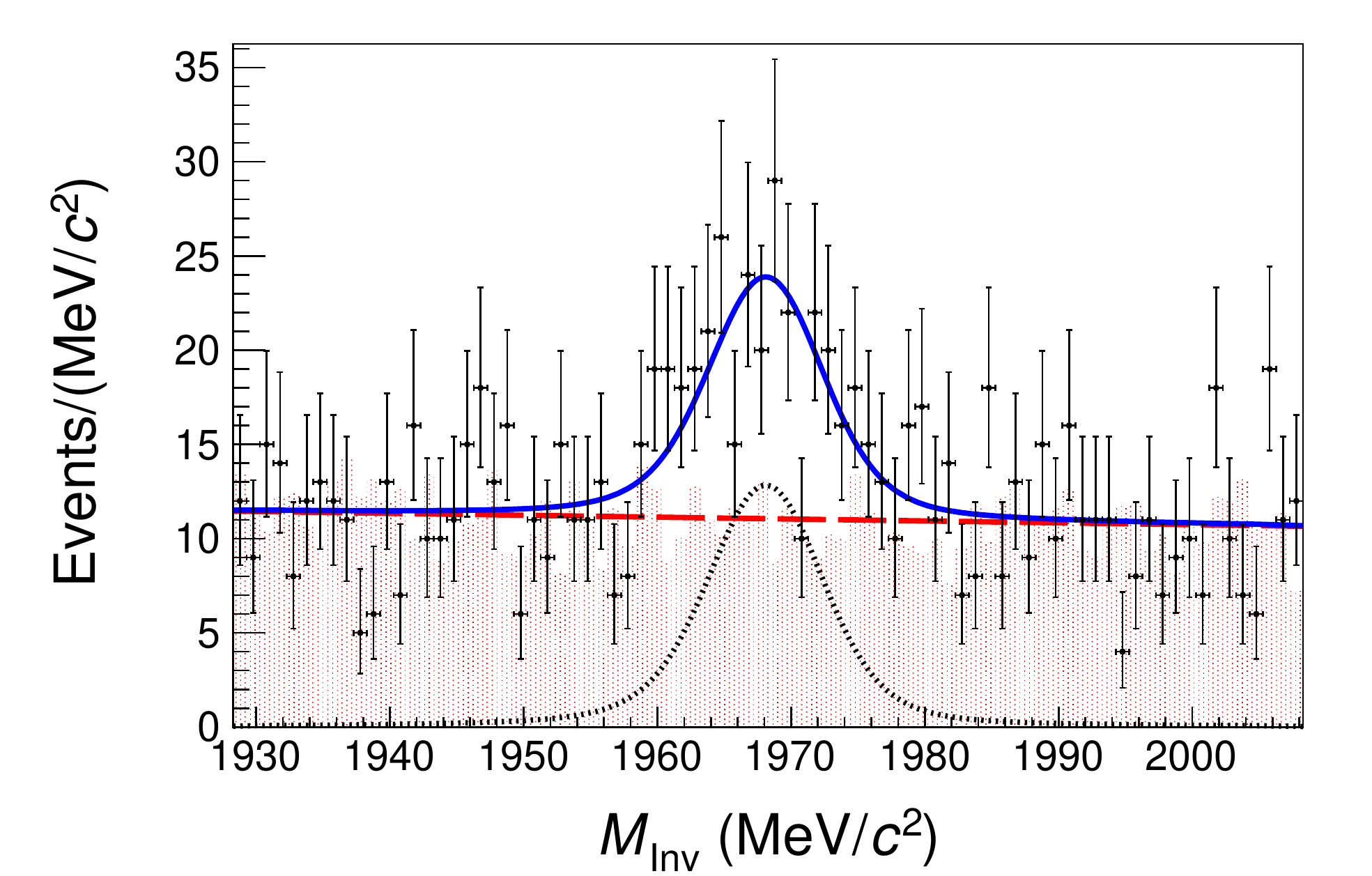}\\
\textbf{RS 300-350 MeV/$c$} & \textbf{WS 300-350 MeV/$c$}\\
\includegraphics[width=3in]{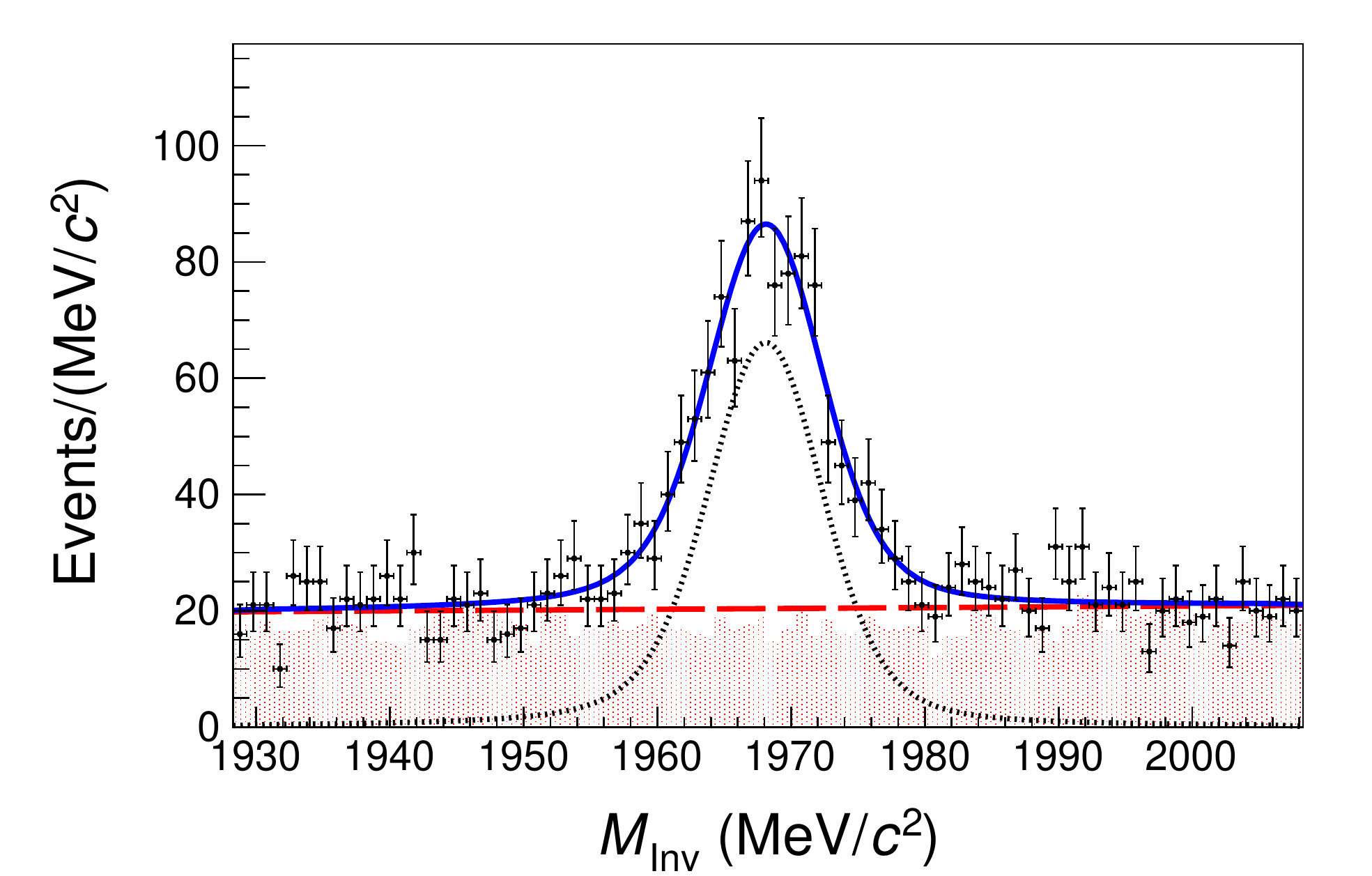} & \includegraphics[width=3in]{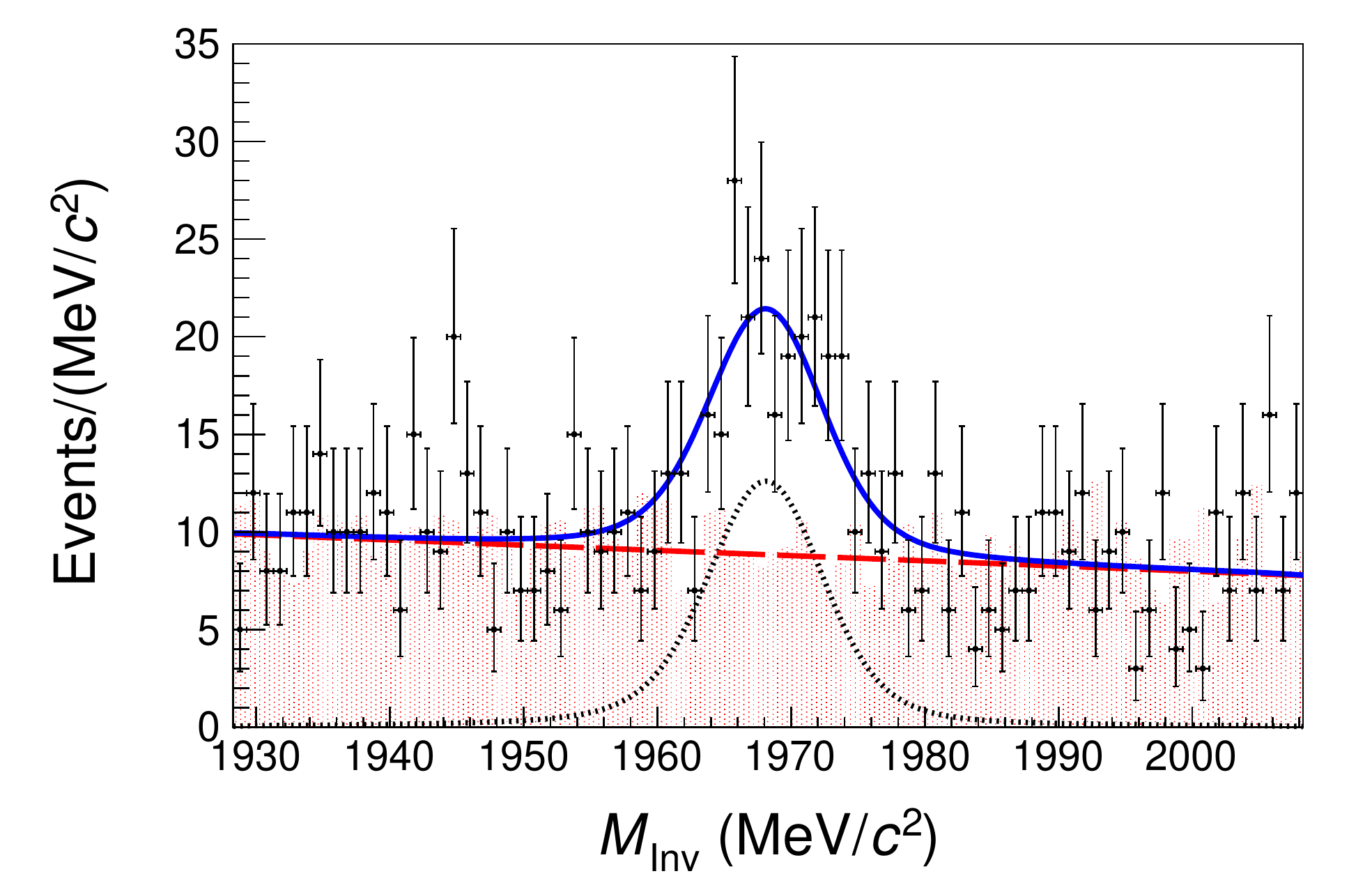}
\end{tabular}
\begin{tabular}{cc}
\textbf{RS 350-400 MeV/$c$} & \textbf{WS 350-400 MeV/$c$}\\
\includegraphics[width=3in]{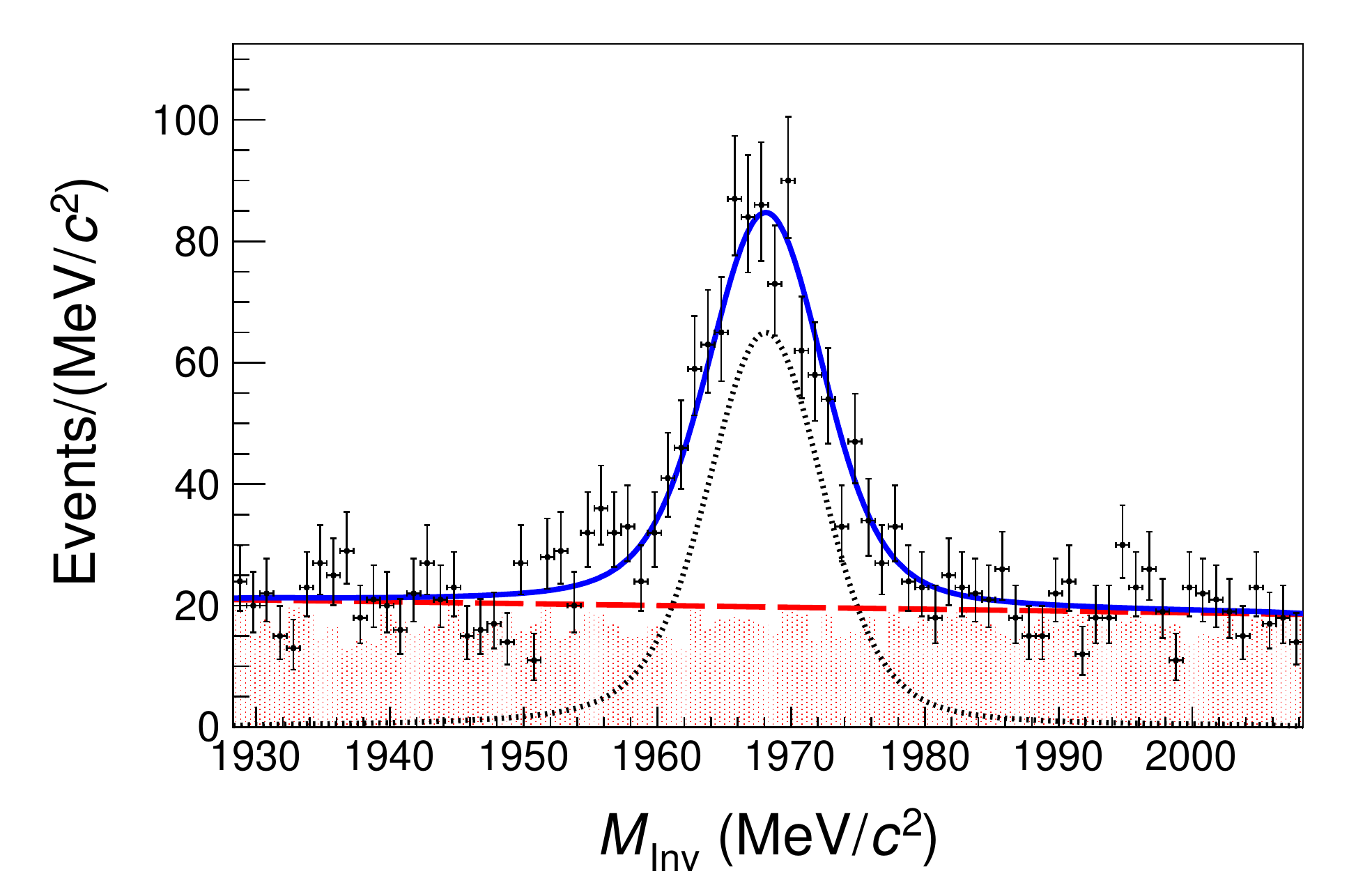} & \includegraphics[width=3in]{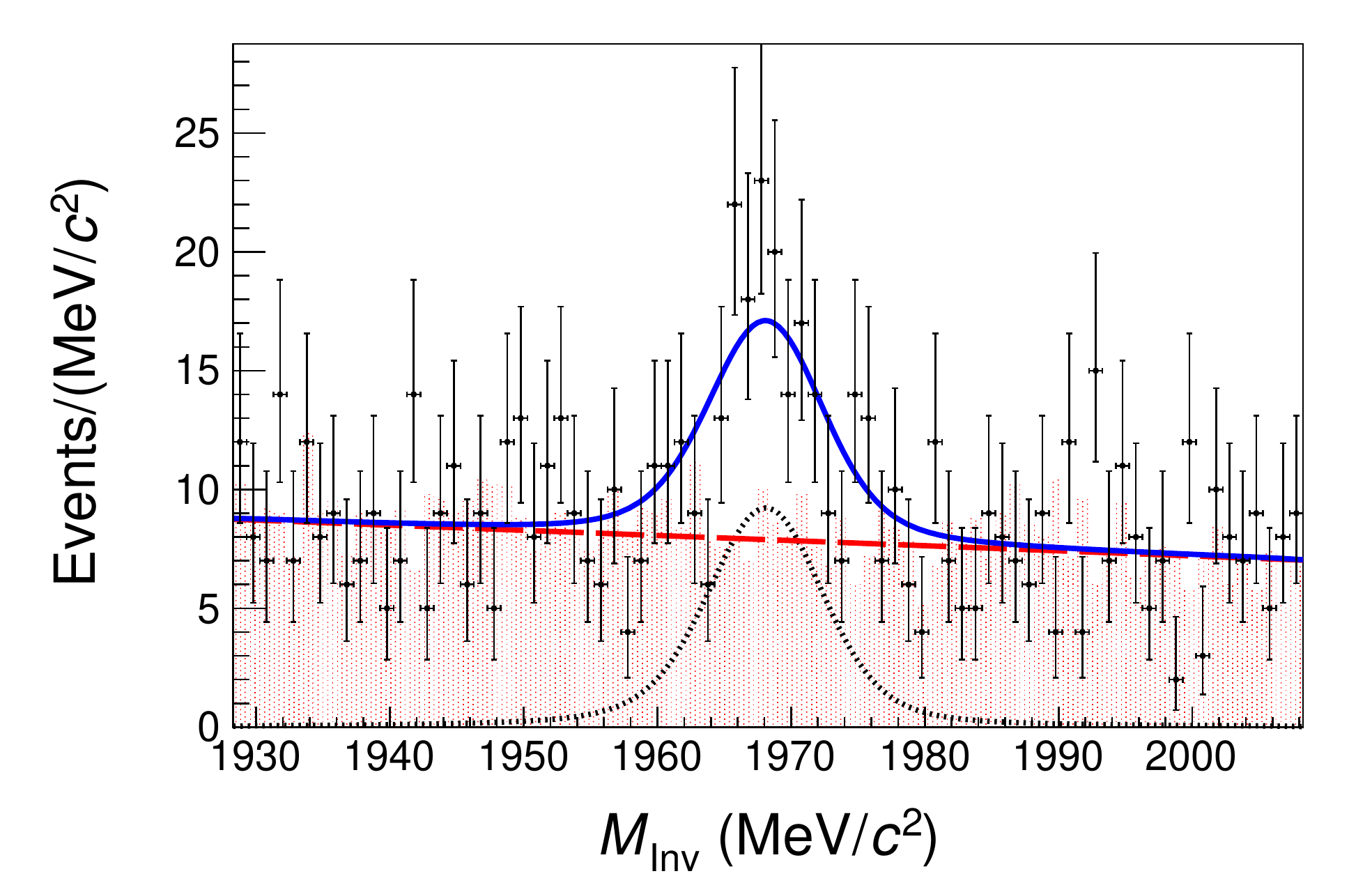}\\
\end{tabular}
\begin{tabular}{cc}
\textbf{RS 400-450 MeV/$c$} & \textbf{WS 400-450 MeV/$c$}\\
\includegraphics[width=3in]{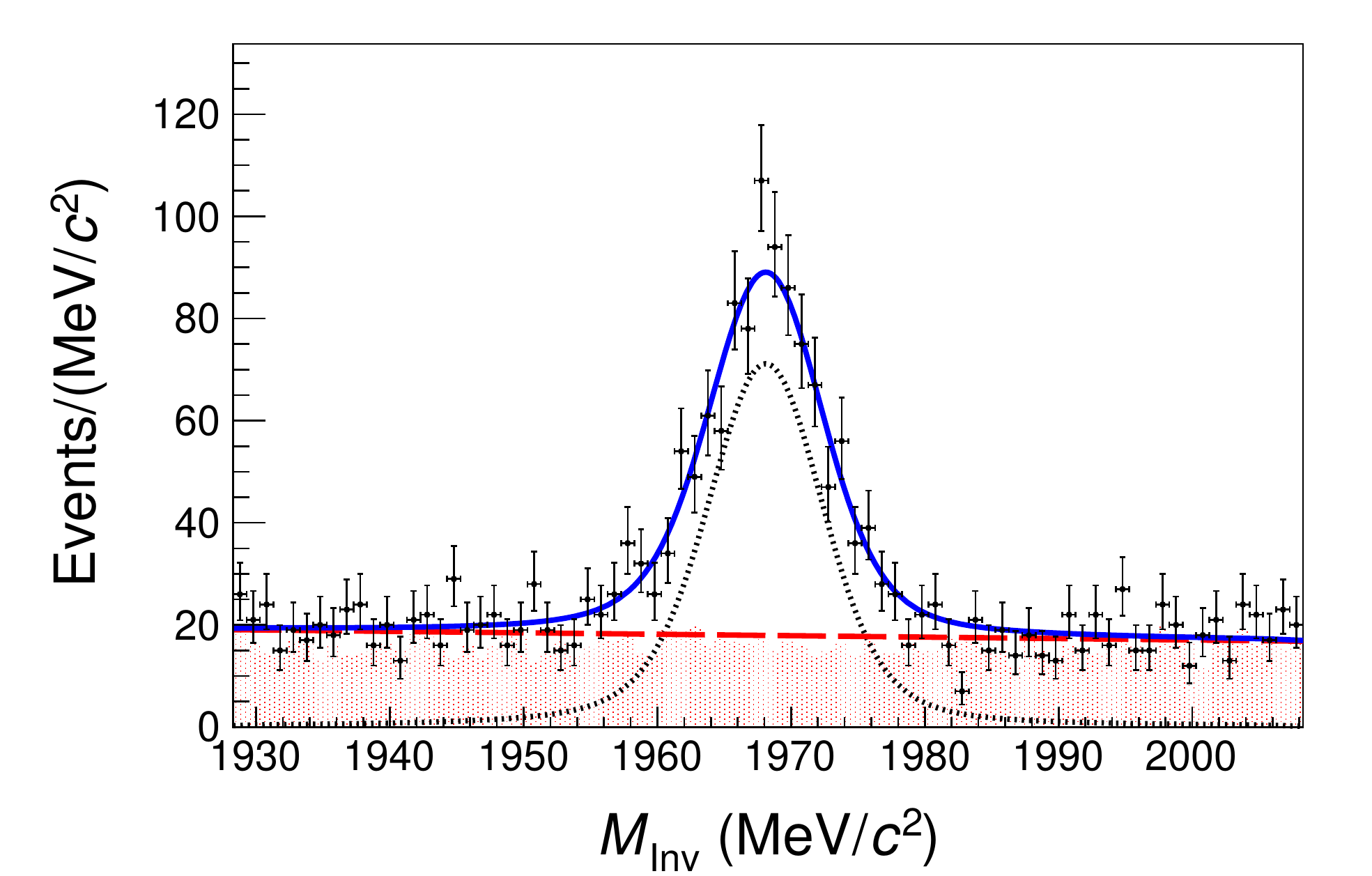} & \includegraphics[width=3in]{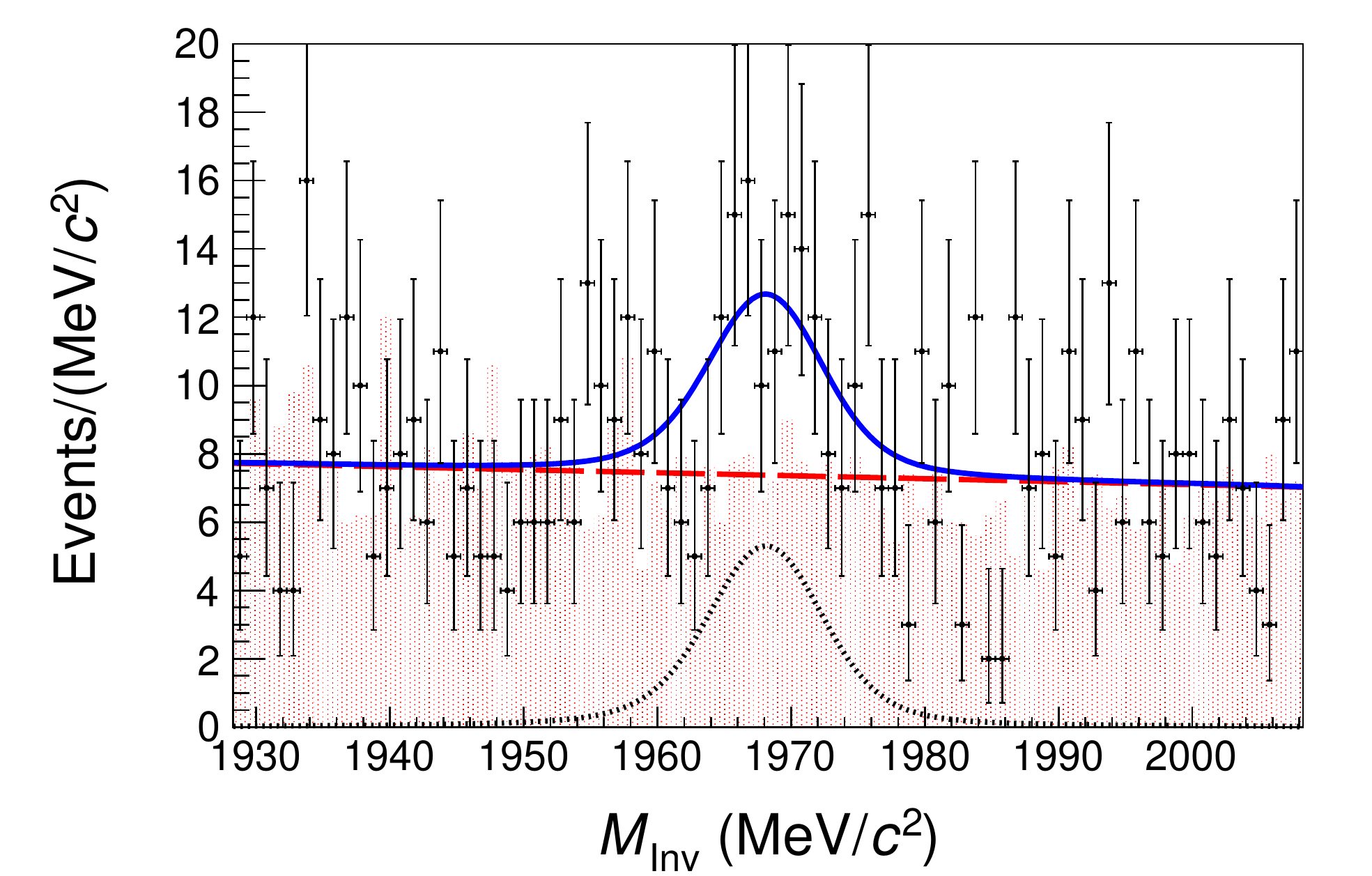}\\
\textbf{RS 450-500 MeV/$c$} & \textbf{WS 450-500 MeV/$c$}\\
\includegraphics[width=3in]{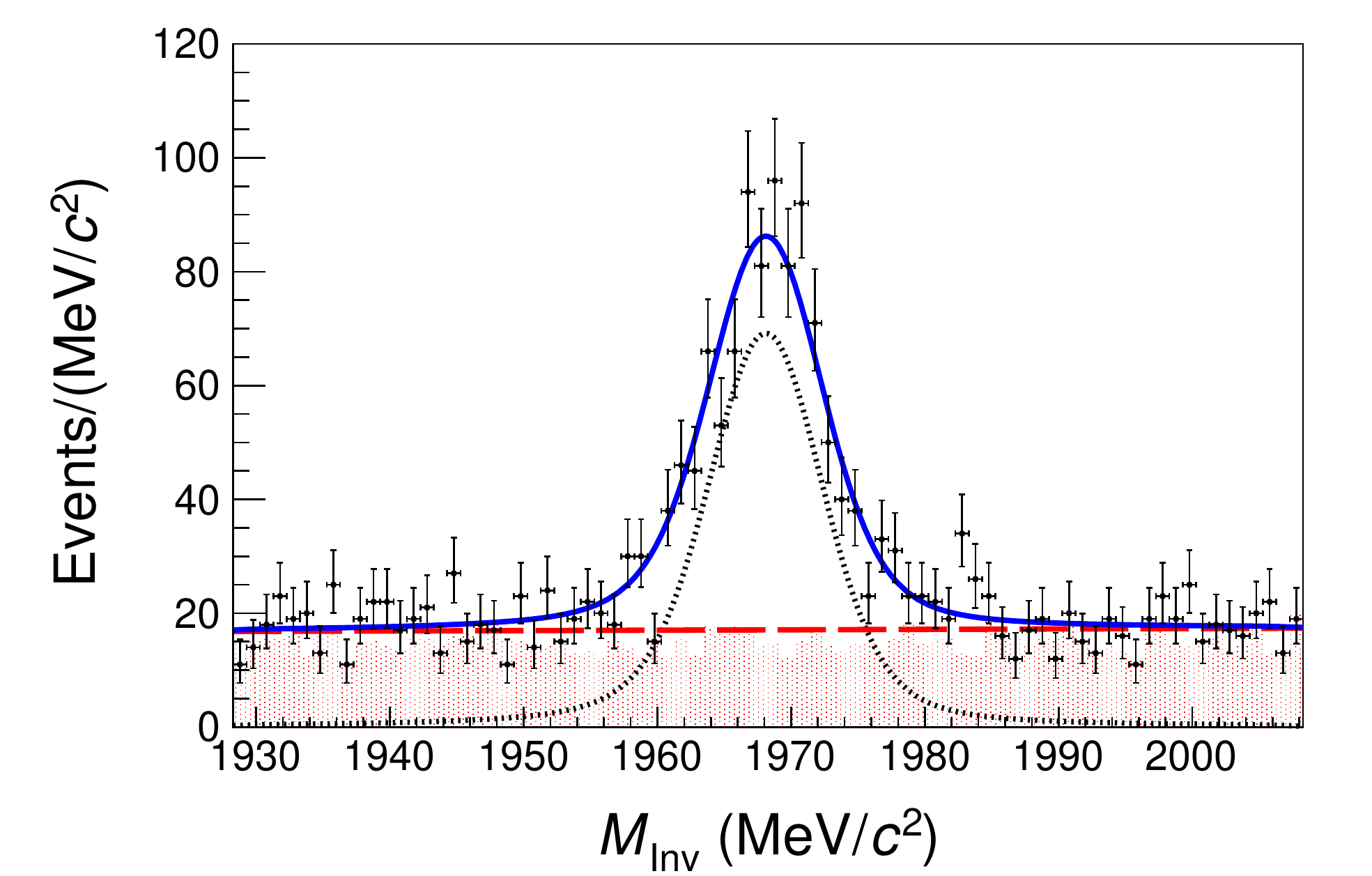} & \includegraphics[width=3in]{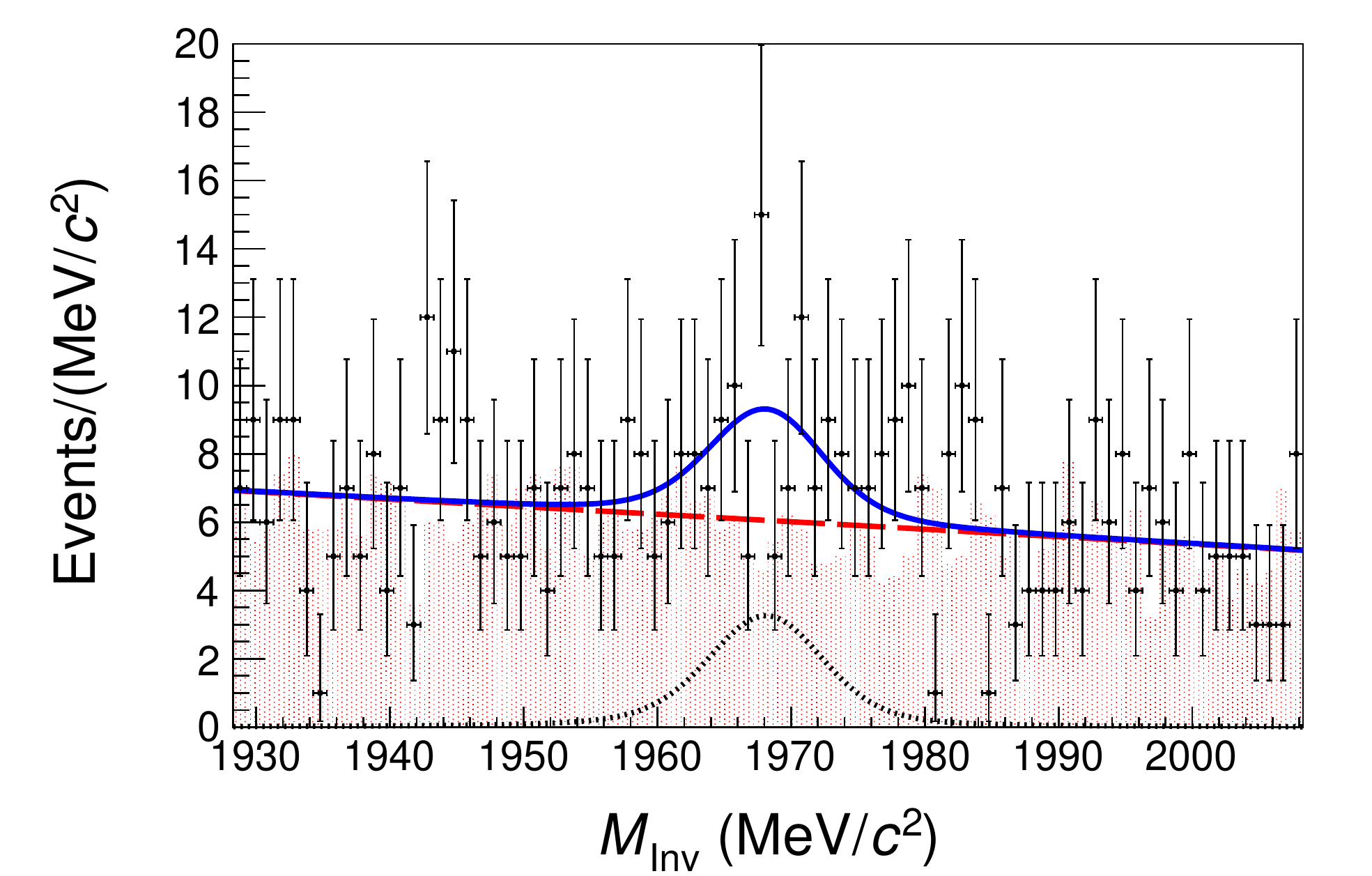}\\
\textbf{RS 500-550 MeV/$c$} & \textbf{WS 500-550 MeV/$c$}\\
\includegraphics[width=3in]{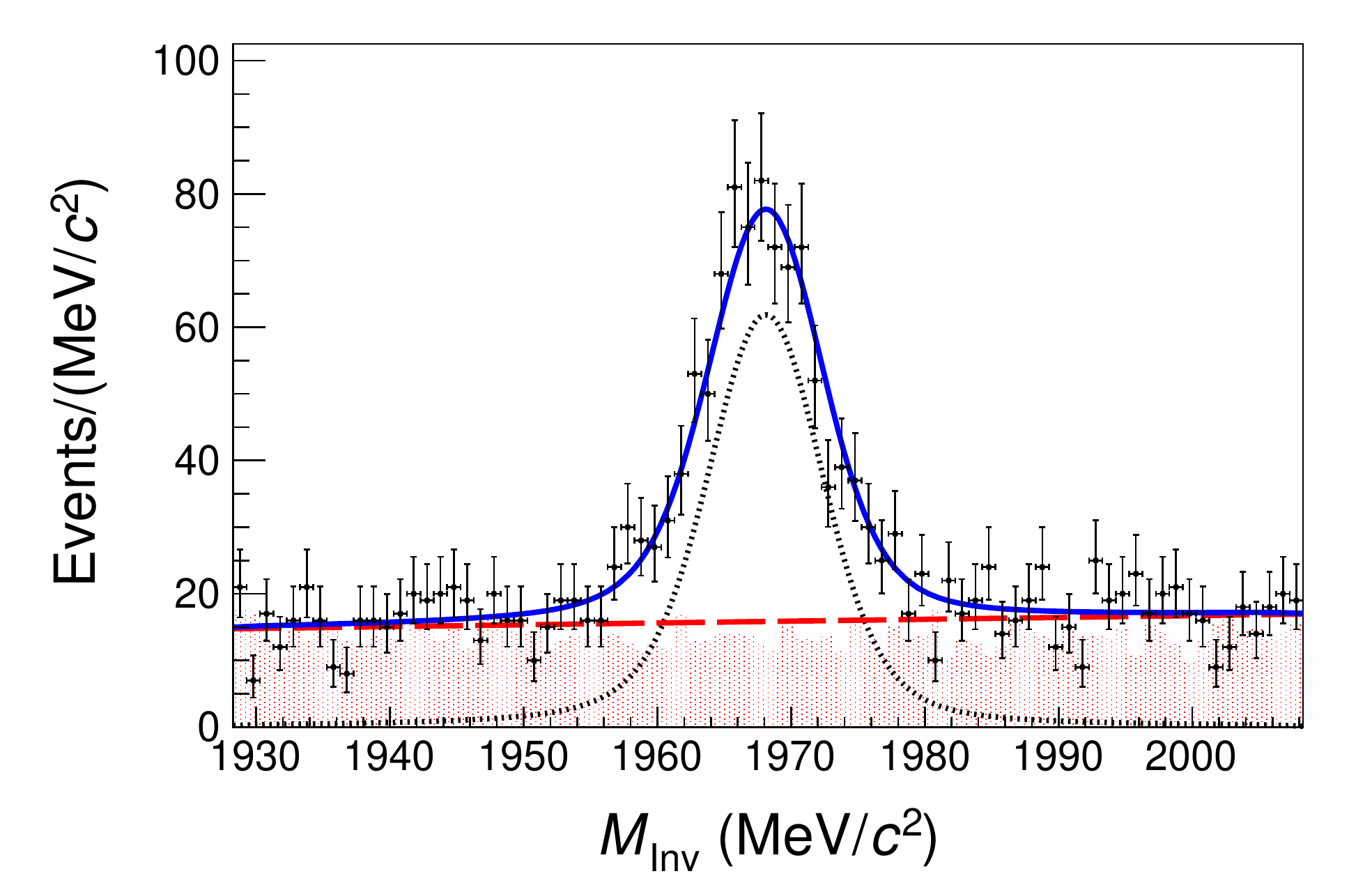} & \includegraphics[width=3in]{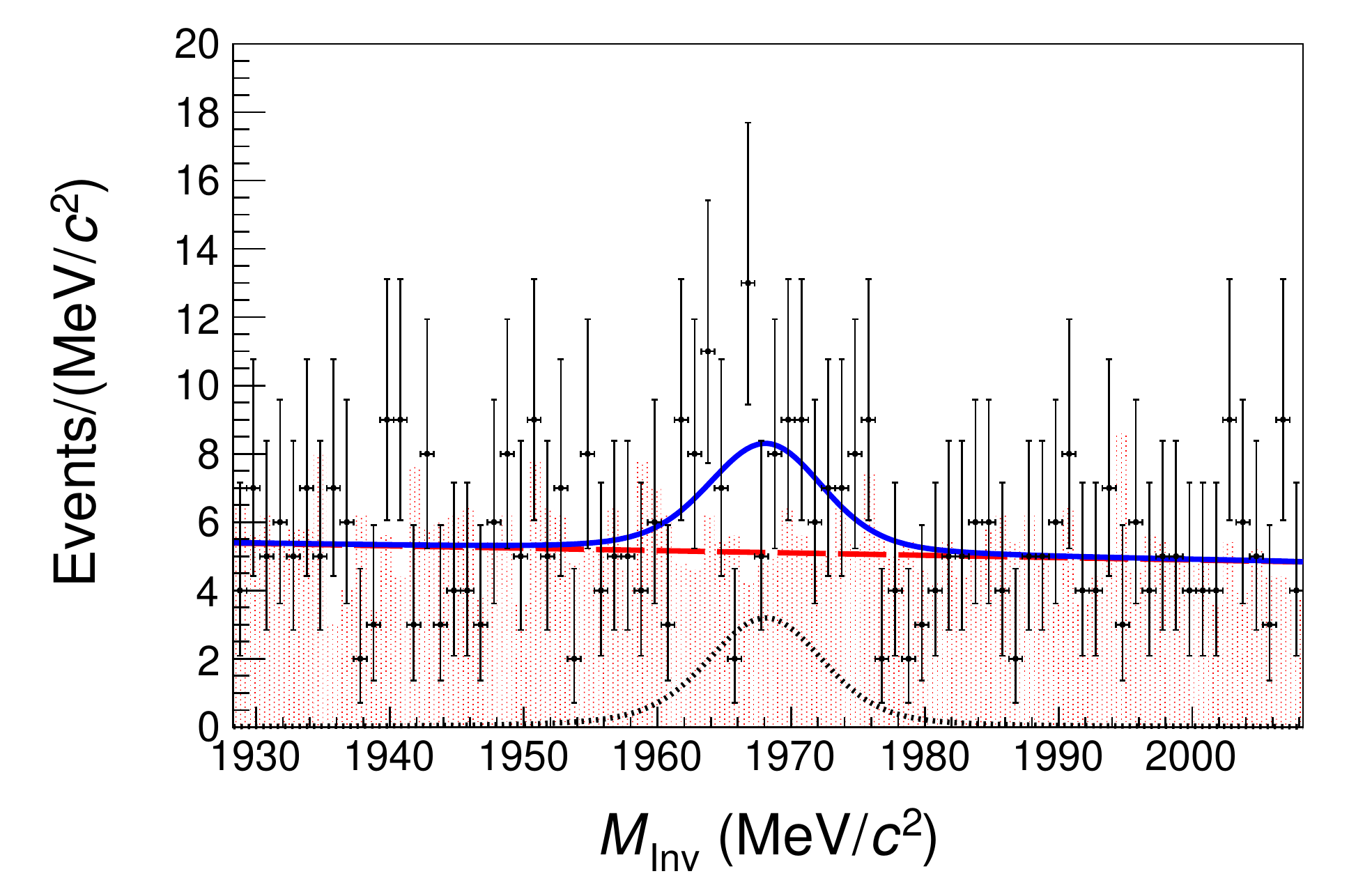}\\
\textbf{RS 550-600 MeV/$c$} & \textbf{WS 550-600 MeV/$c$}\\
\includegraphics[width=3in]{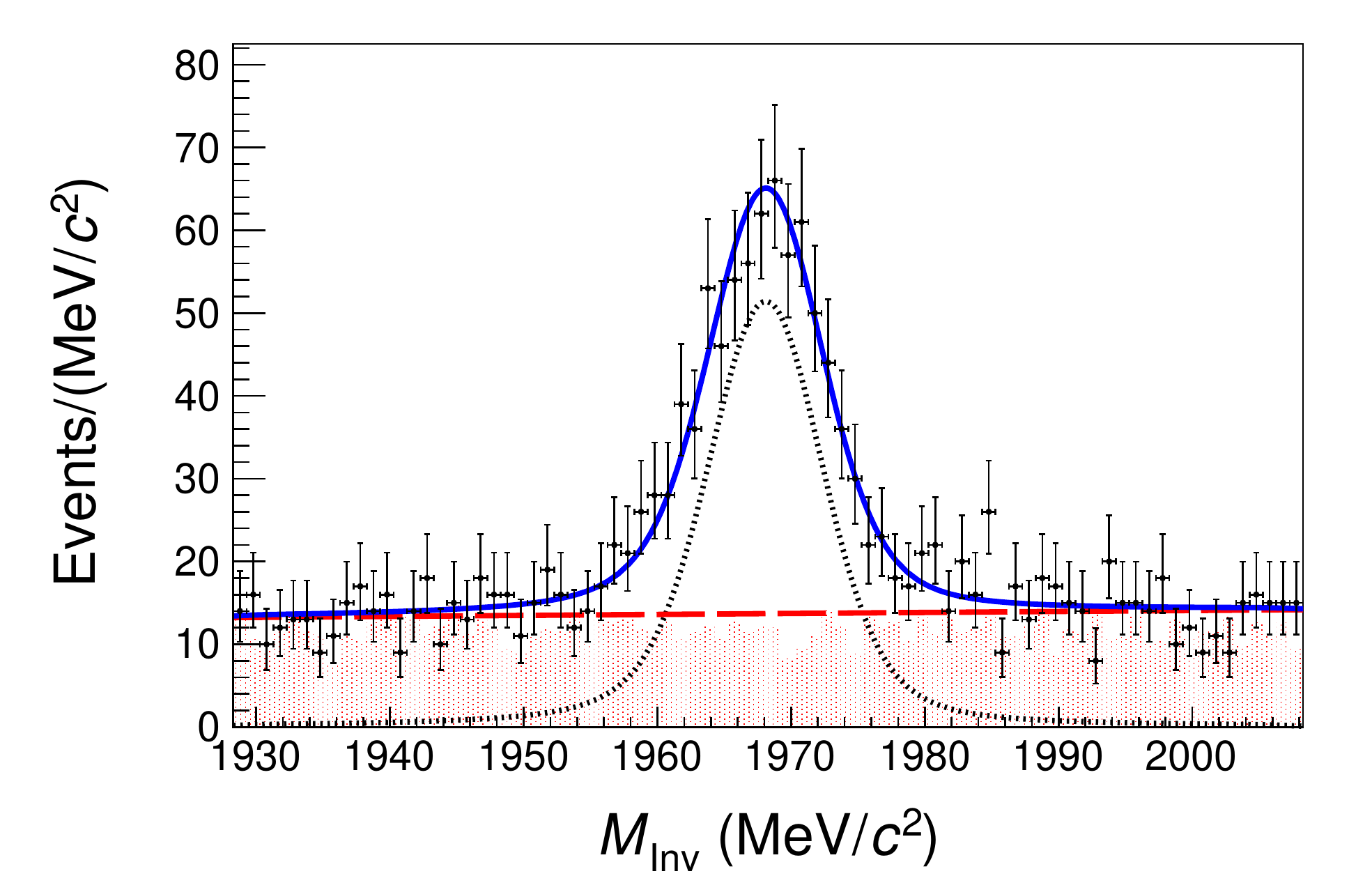} & \includegraphics[width=3in]{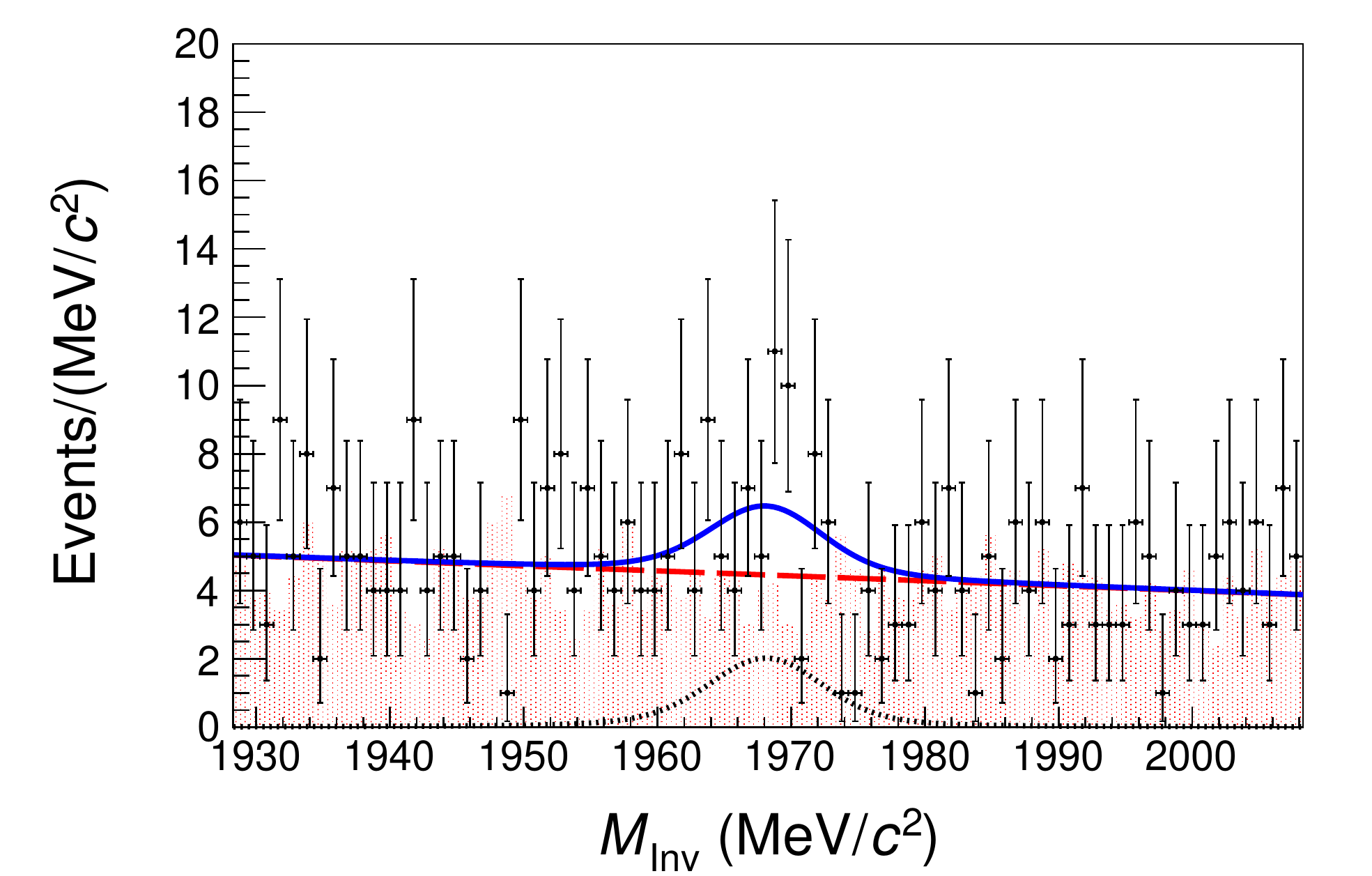}\\
\end{tabular}
\begin{tabular}{cc}
\textbf{RS 600-650 MeV/$c$} & \textbf{WS 600-650 MeV/$c$}\\
\includegraphics[width=3in]{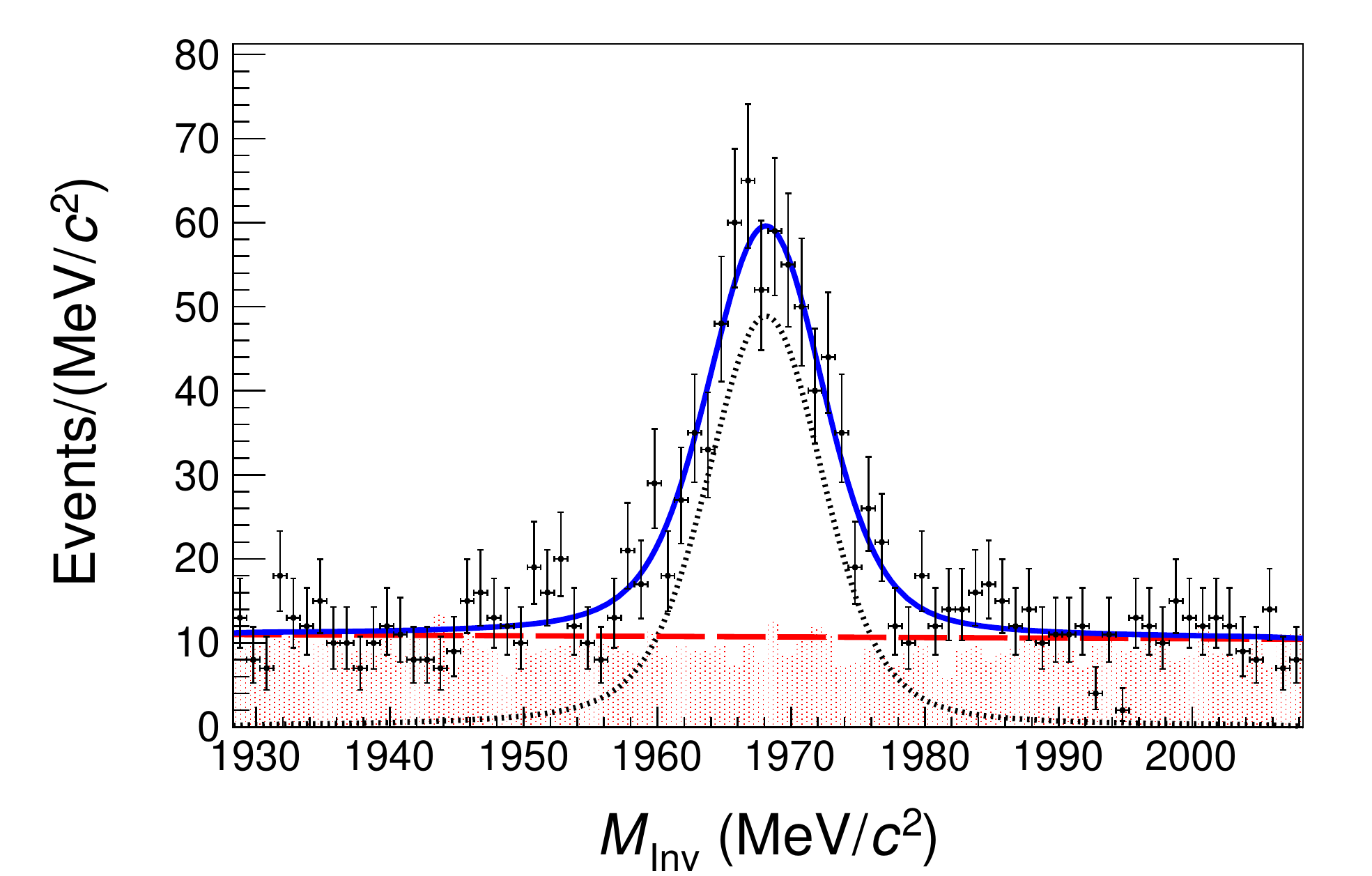} & \includegraphics[width=3in]{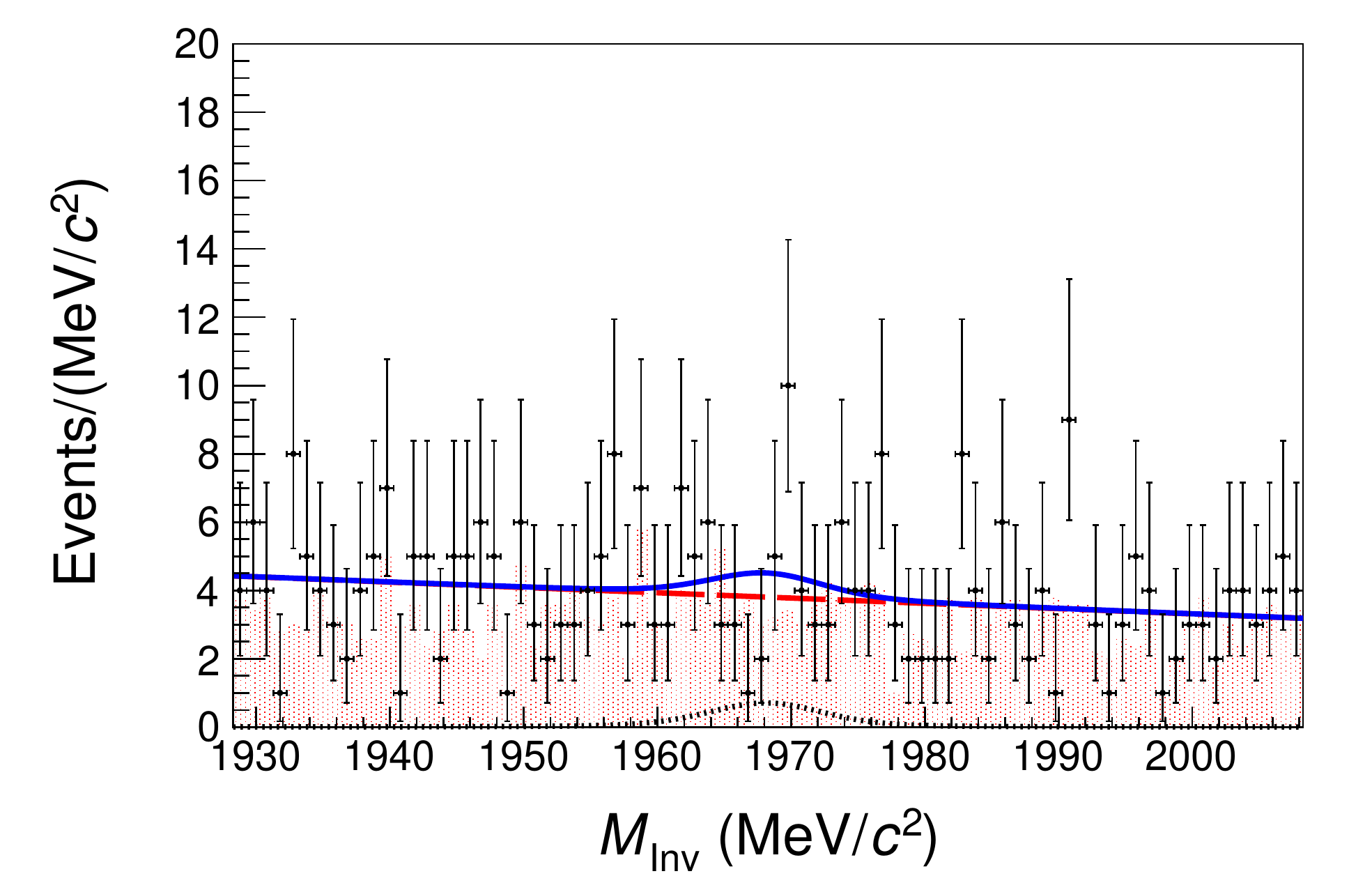}\\
\textbf{RS 650-700 MeV/$c$} & \textbf{WS 650-700 MeV/$c$}\\
\includegraphics[width=3in]{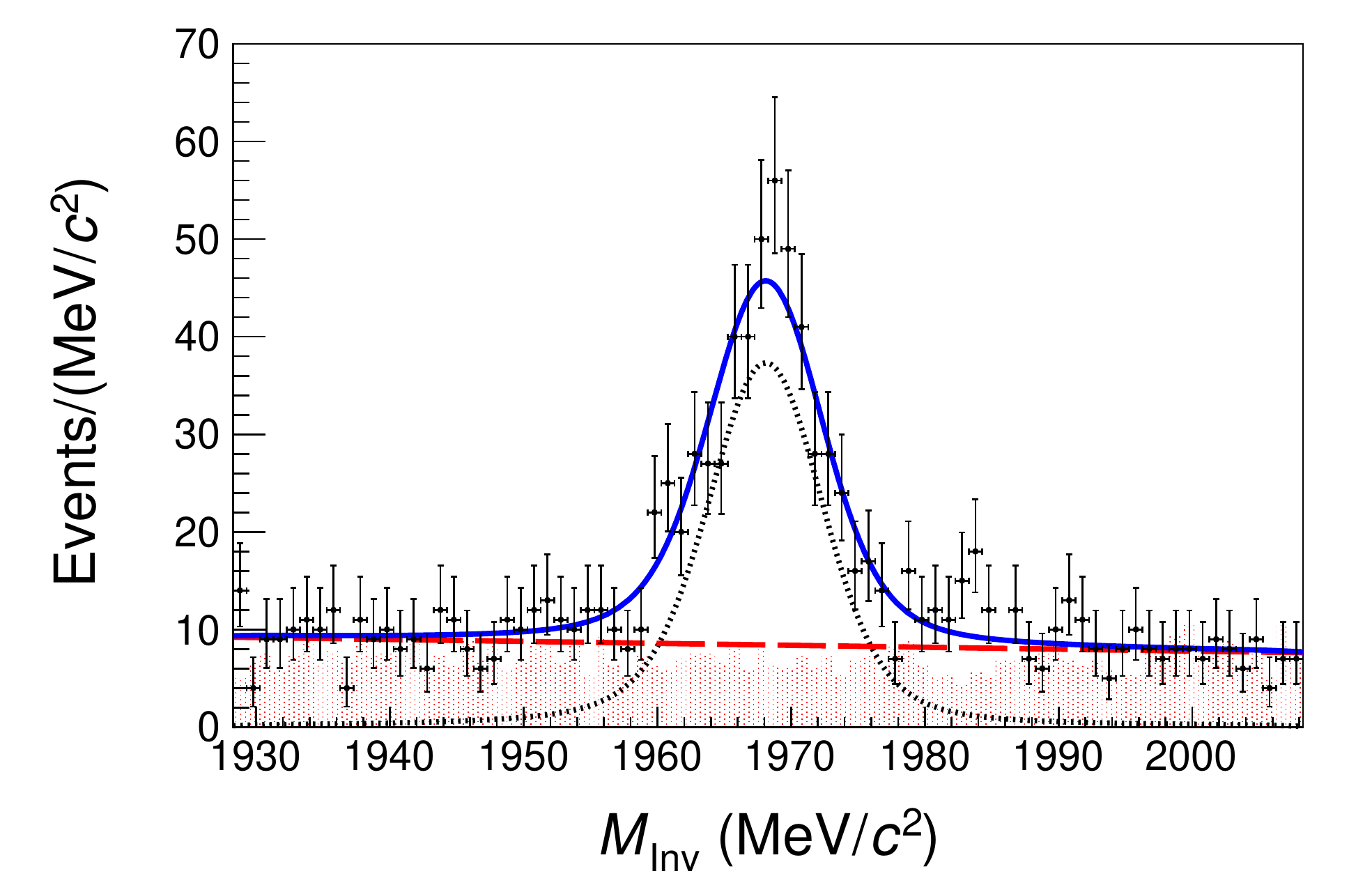} & \includegraphics[width=3in]{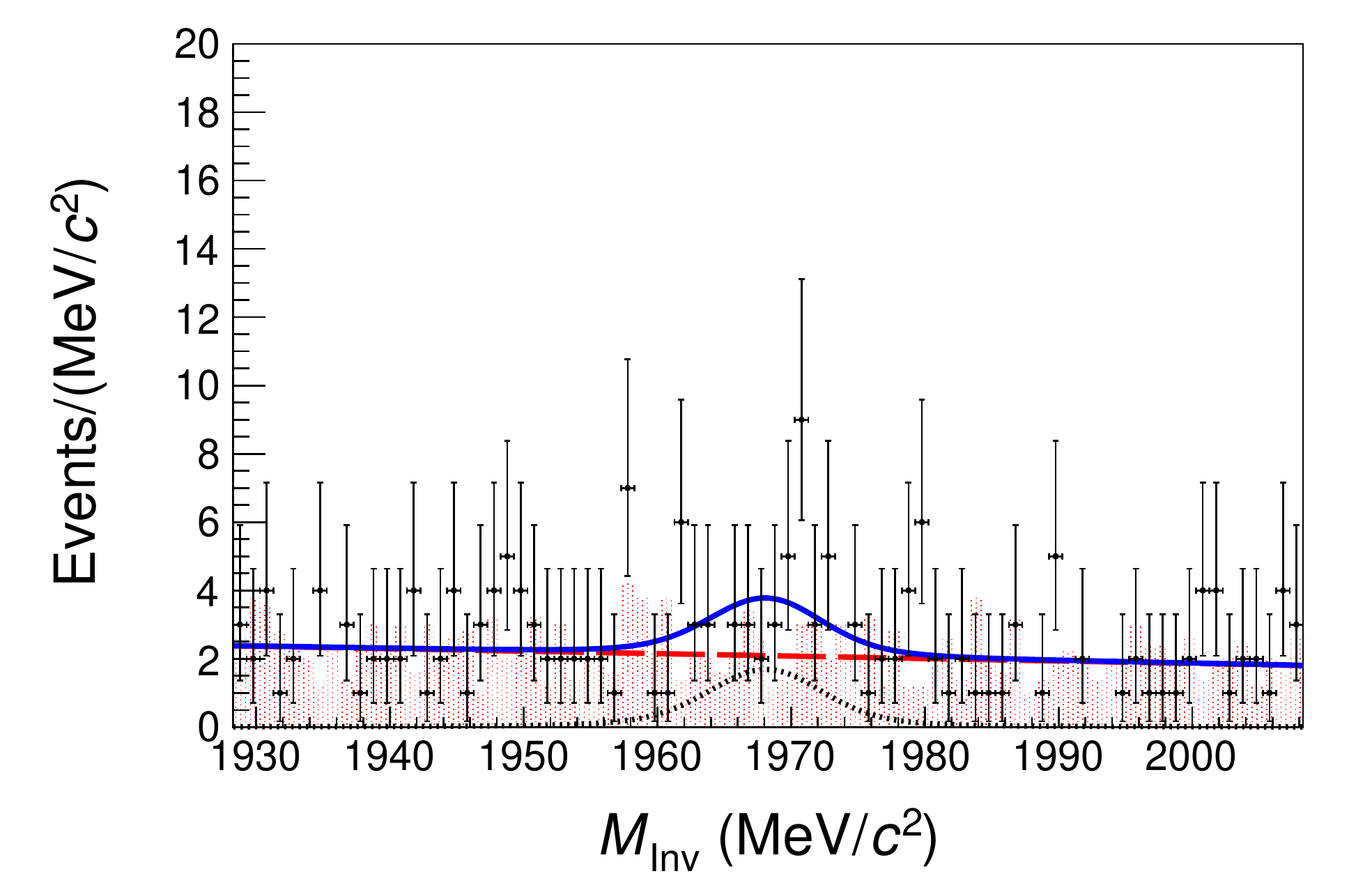}\\
\textbf{RS 700-750 MeV/$c$} & \textbf{WS 700-750 MeV/$c$}\\
\includegraphics[width=3in]{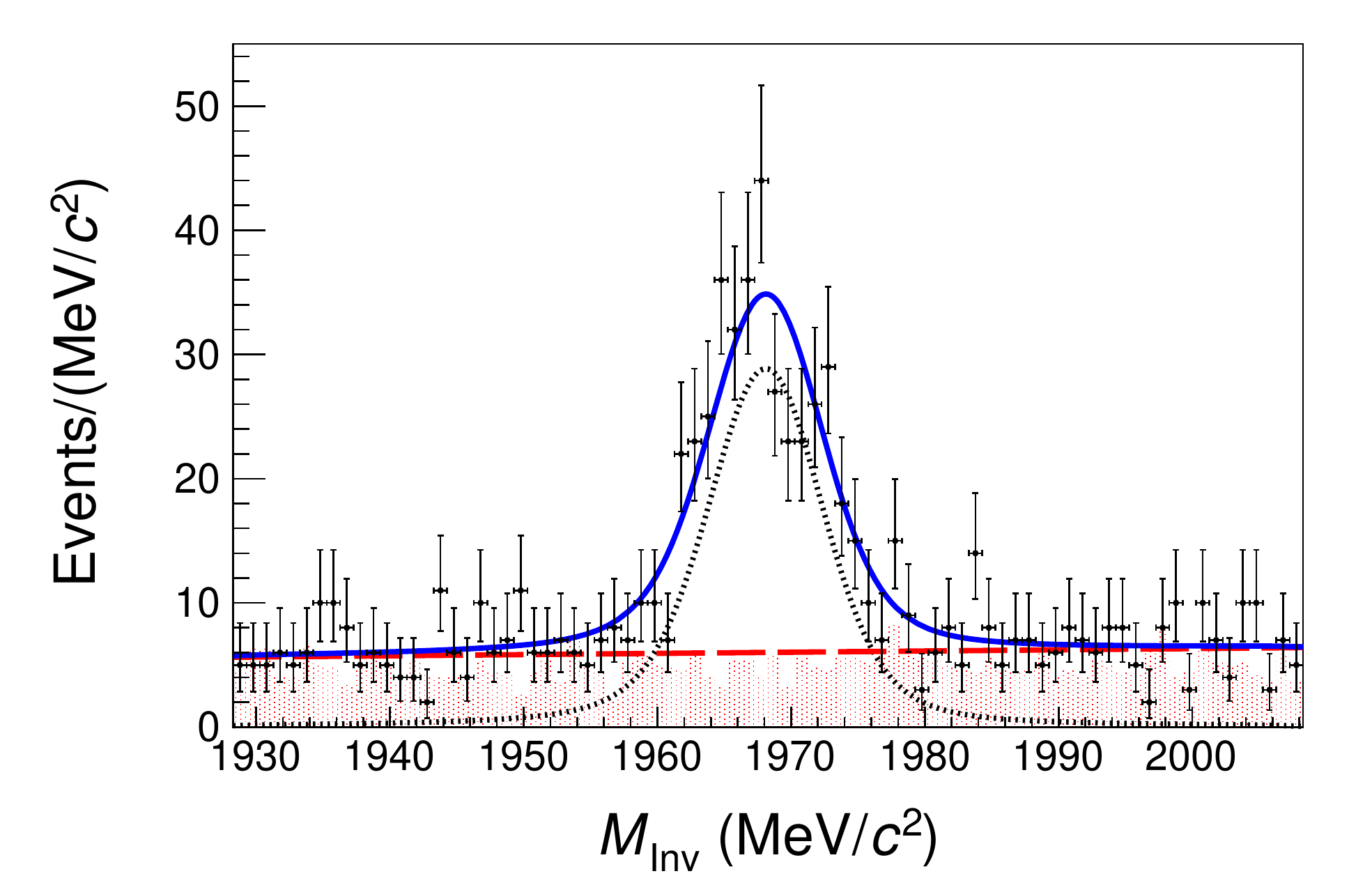} & \includegraphics[width=3in]{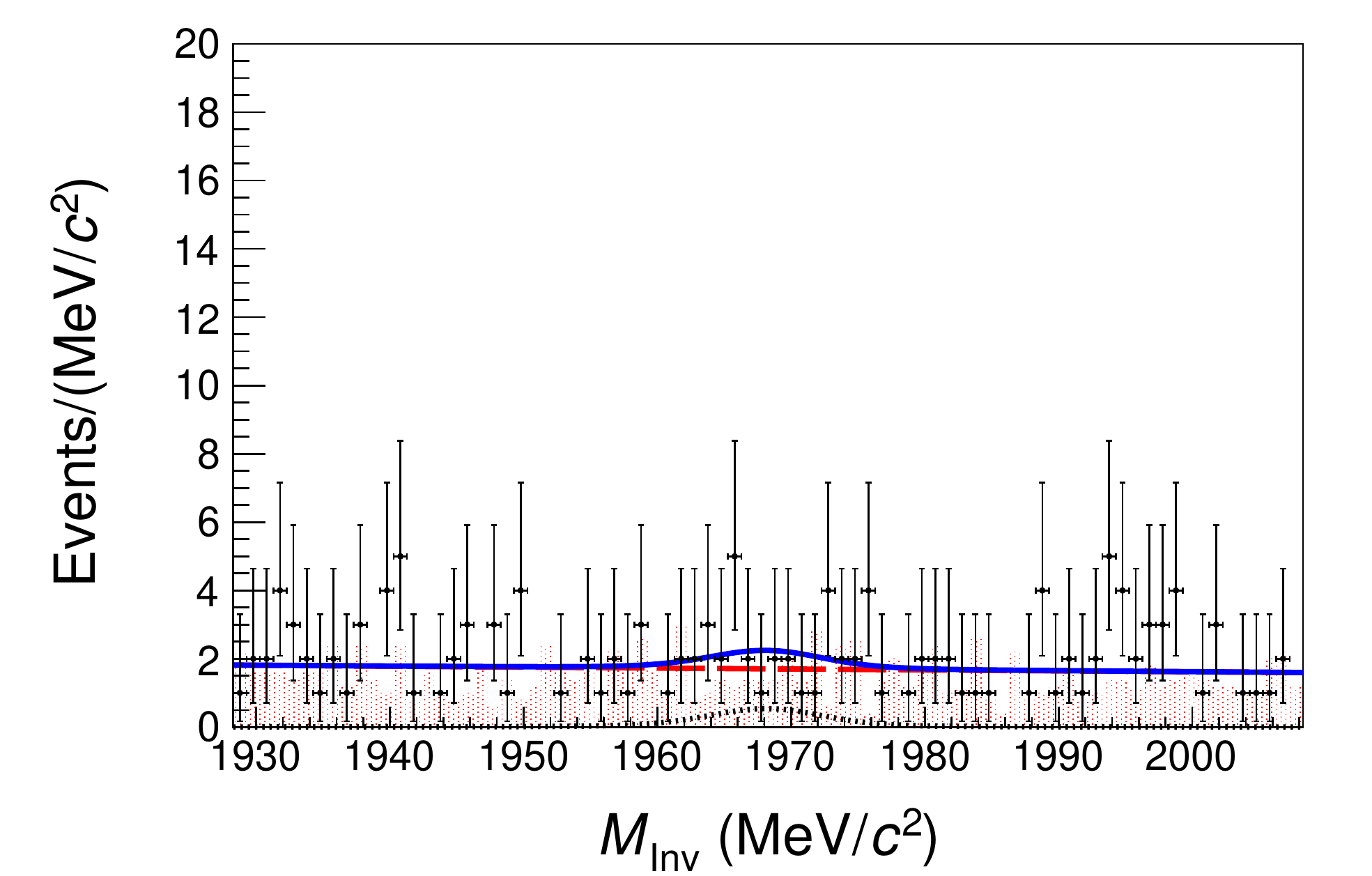}\\
\textbf{RS 750-800 MeV/$c$} & \textbf{WS 750-800 MeV/$c$}\\
\includegraphics[width=3in]{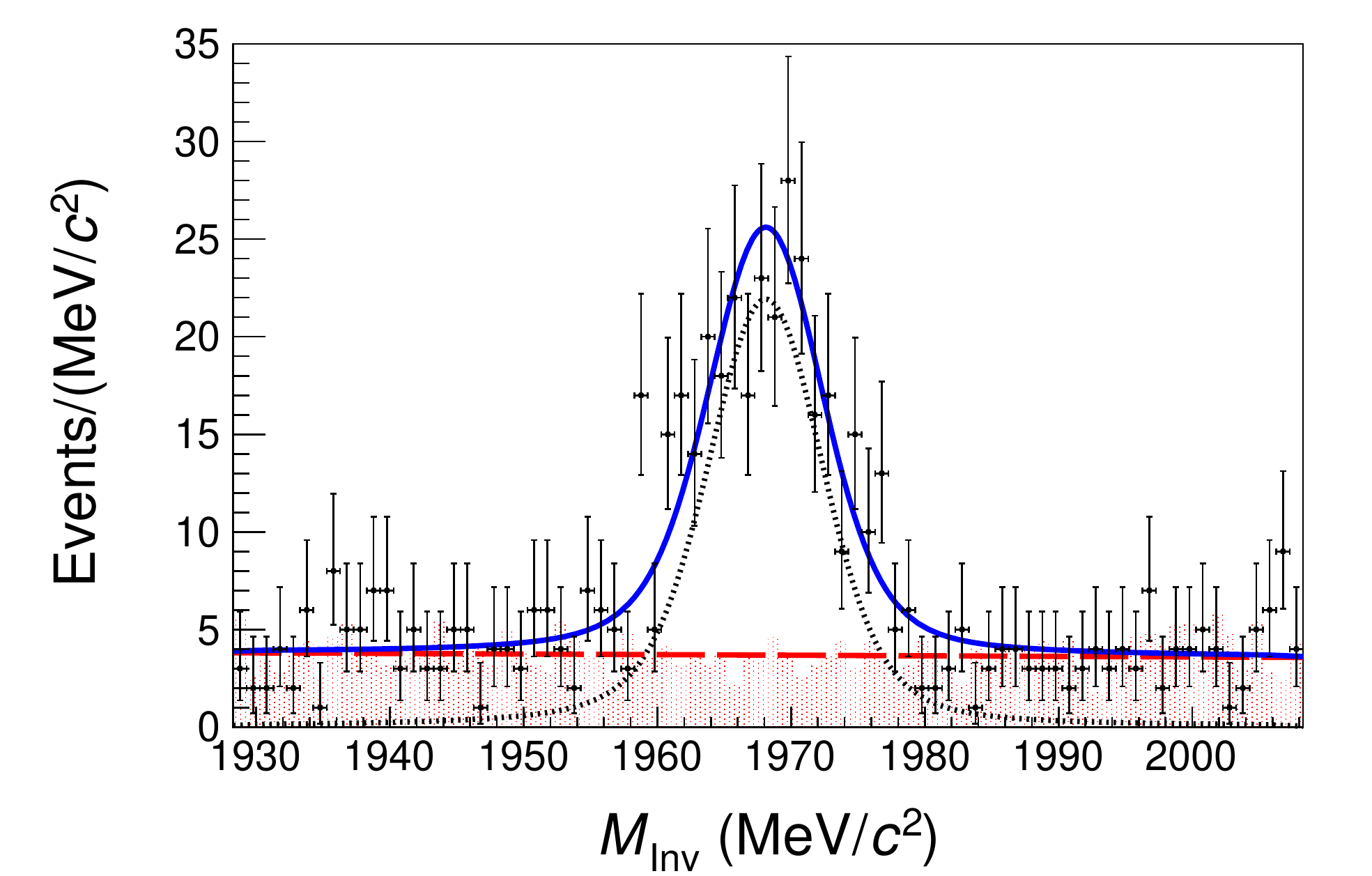} & \includegraphics[width=3in]{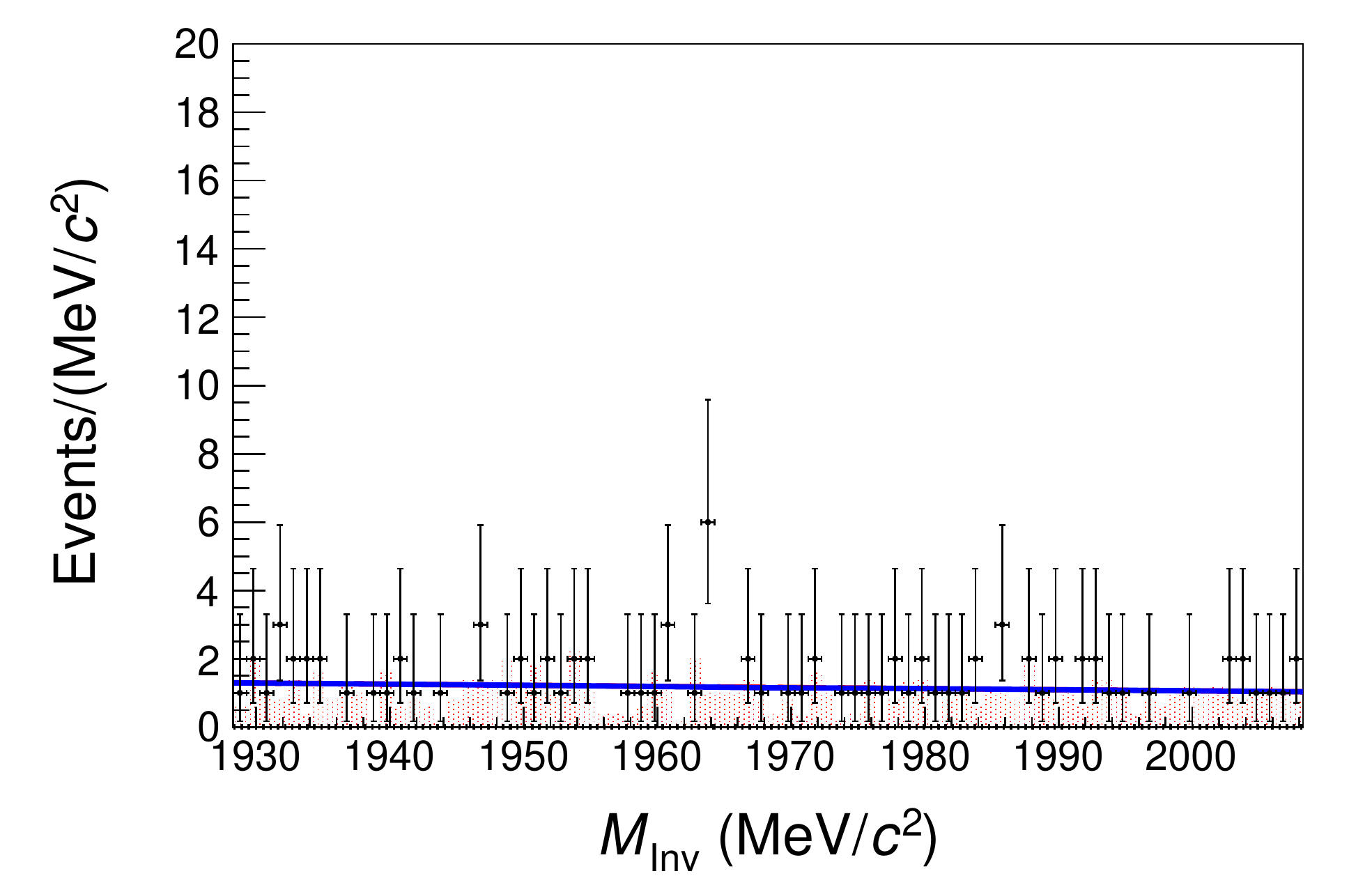}\\
\end{tabular}
\begin{tabular}{cc}
\textbf{RS 800-850 MeV/$c$} & \textbf{WS 800-850 MeV/$c$}\\
\includegraphics[width=3in]{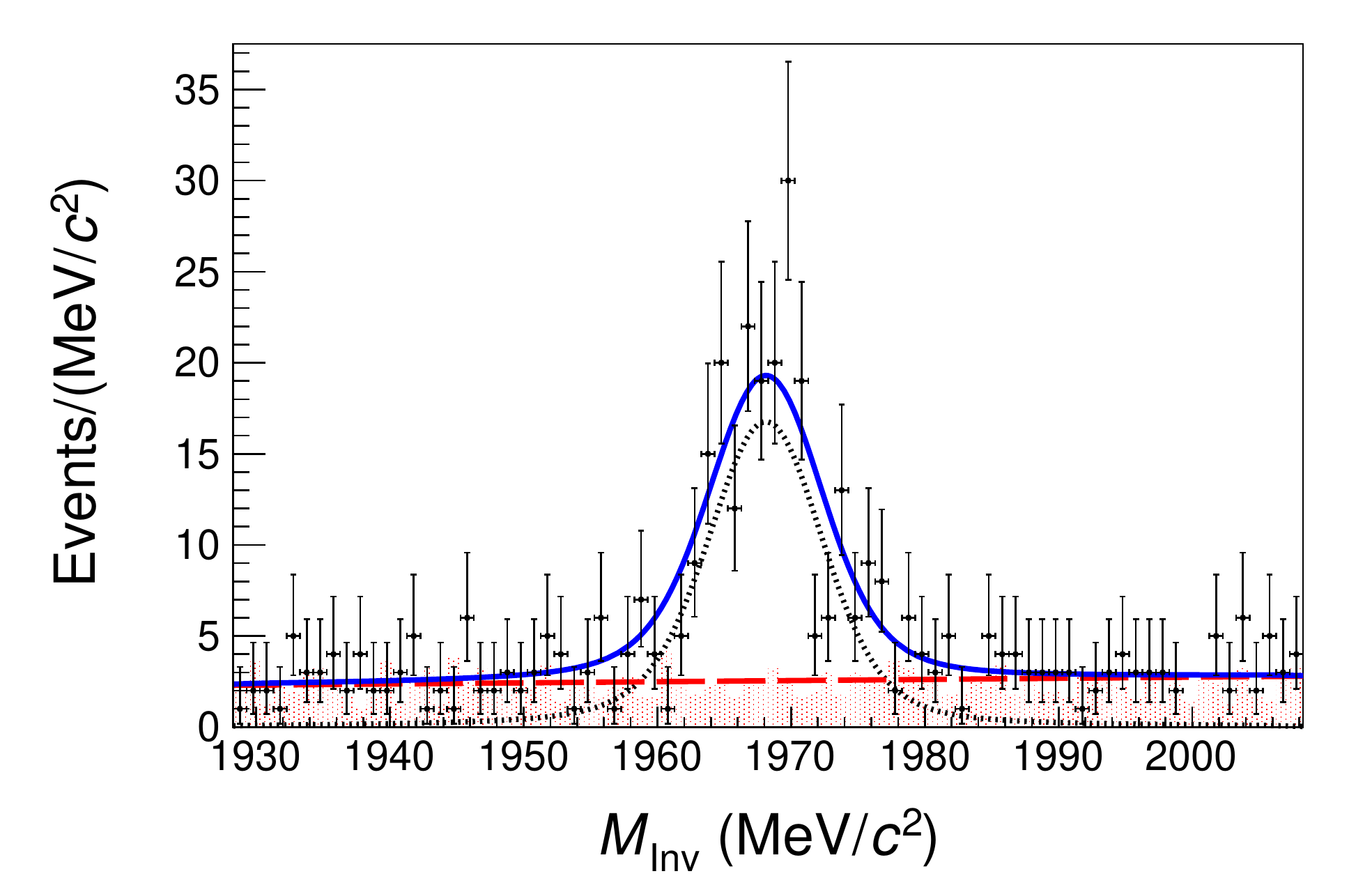} & \includegraphics[width=3in]{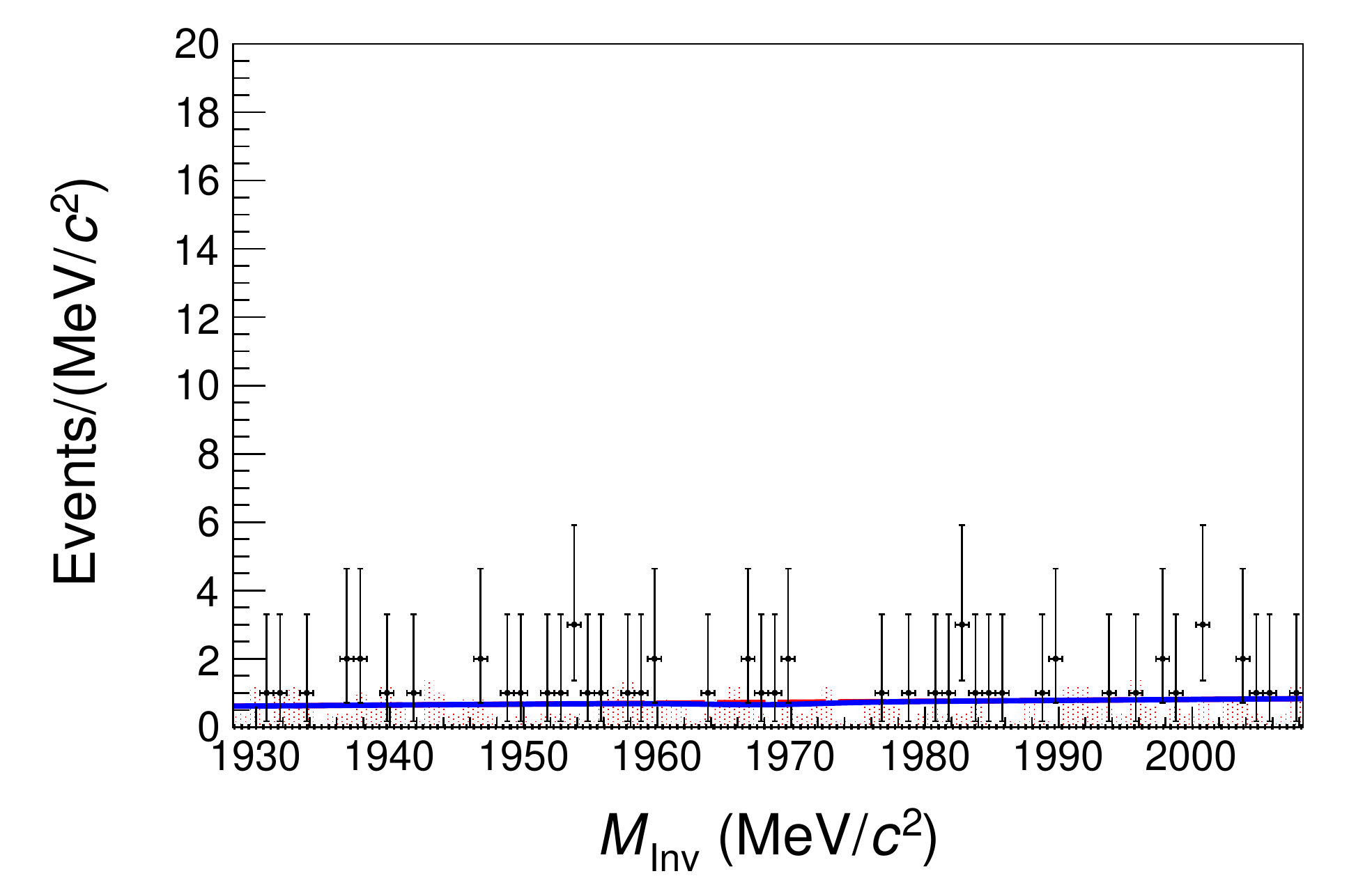}\\
\textbf{RS 850-900 MeV/$c$} & \textbf{WS 850-900 MeV/$c$}\\
\includegraphics[width=3in]{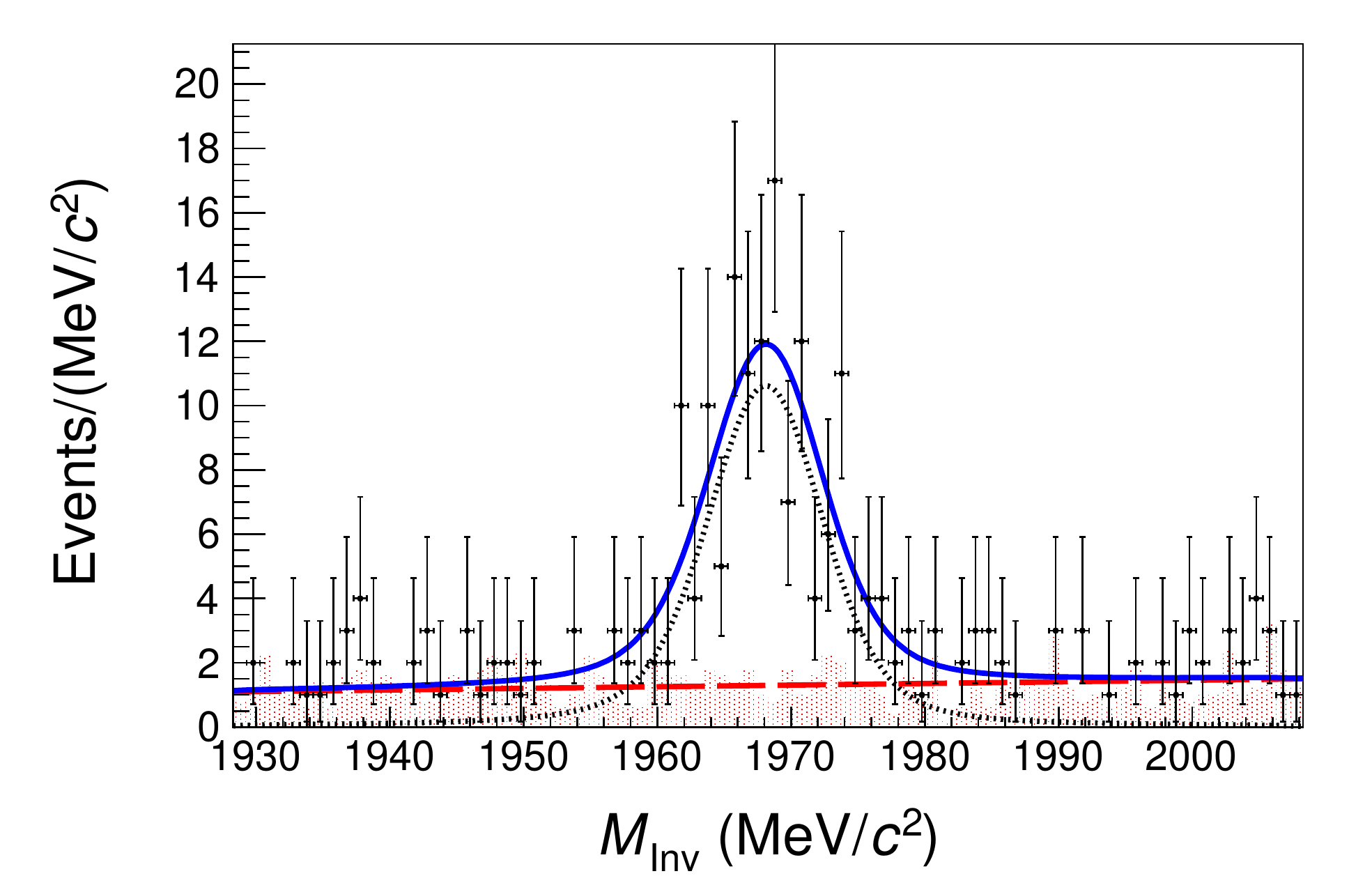} & \includegraphics[width=3in]{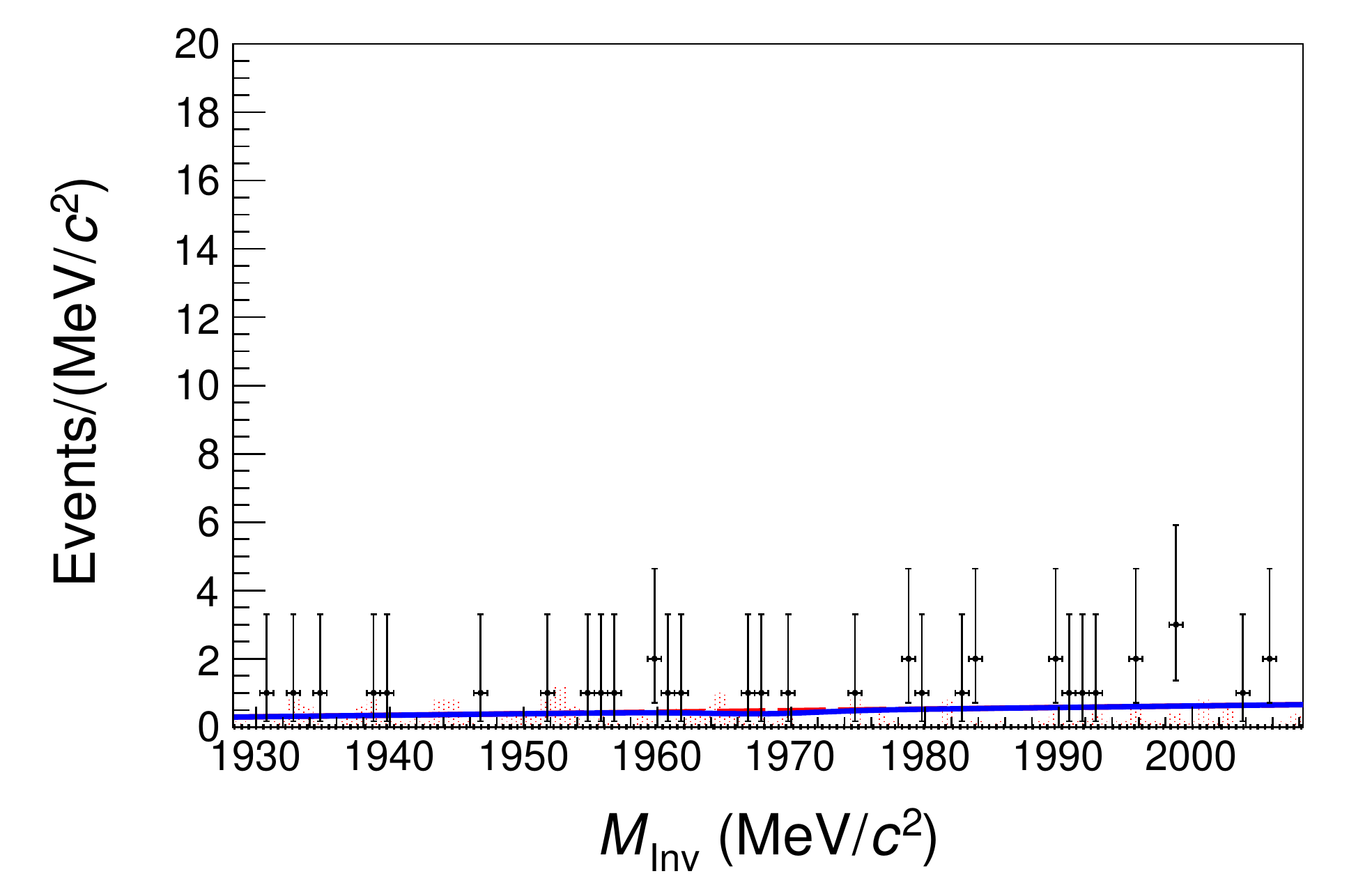}\\
\textbf{RS 900-950 MeV/$c$} & \textbf{WS 900-950 MeV/$c$}\\
\includegraphics[width=3in]{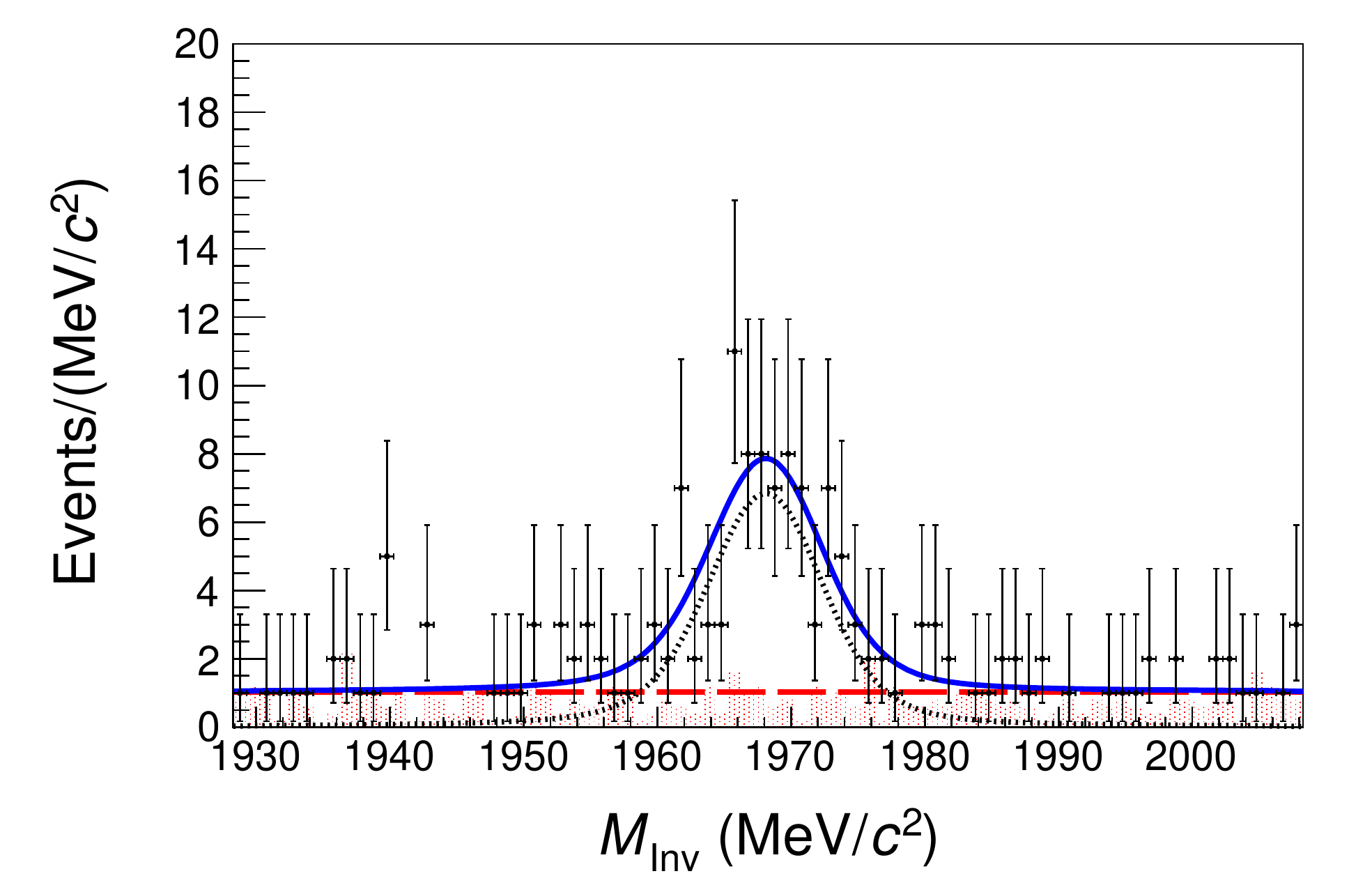} & \includegraphics[width=3in]{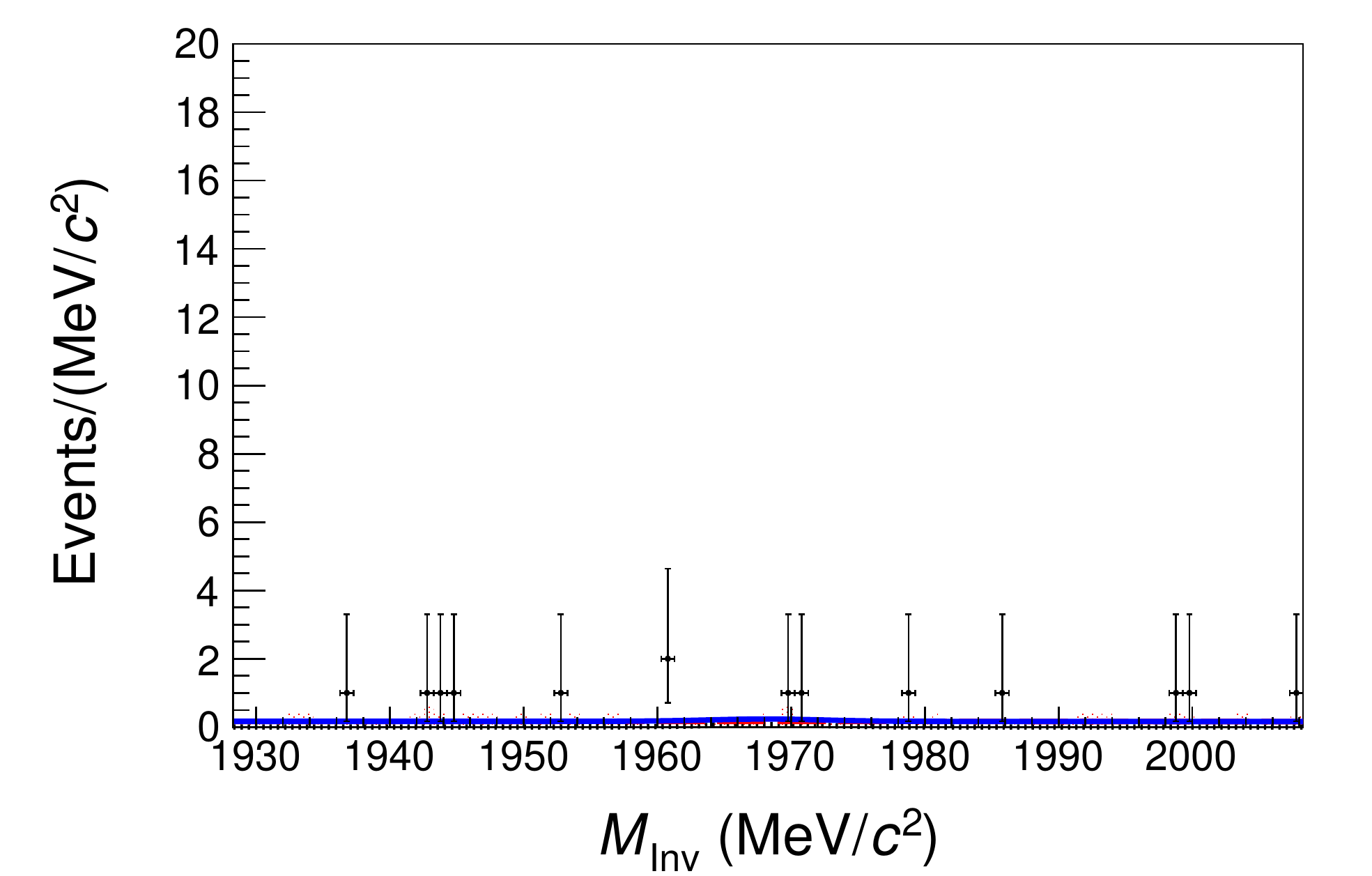}\\
\textbf{RS 950-1000 MeV/$c$} & \textbf{WS 950-1000 MeV/$c$}\\
\includegraphics[width=3in]{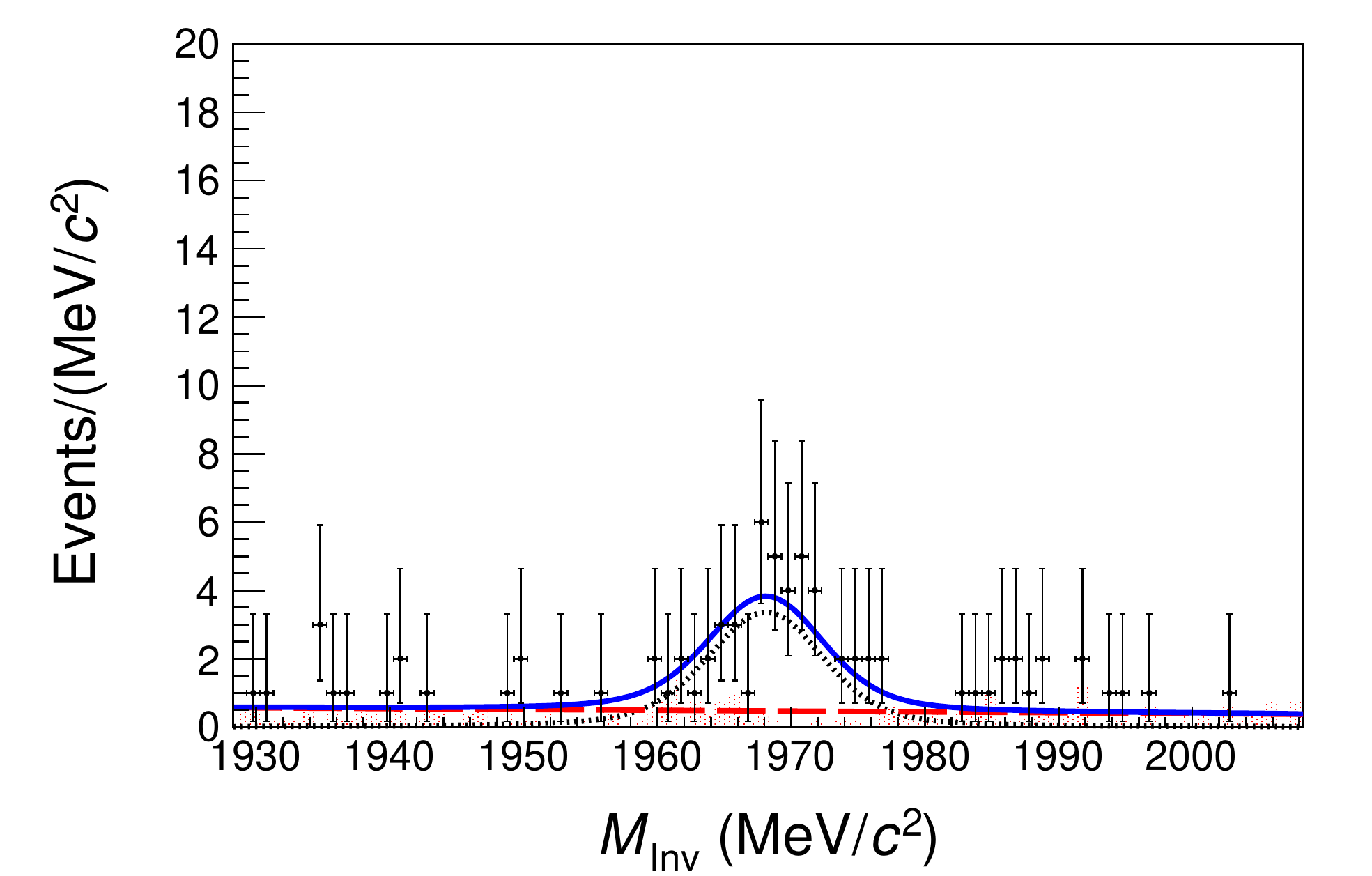} & \includegraphics[width=3in]{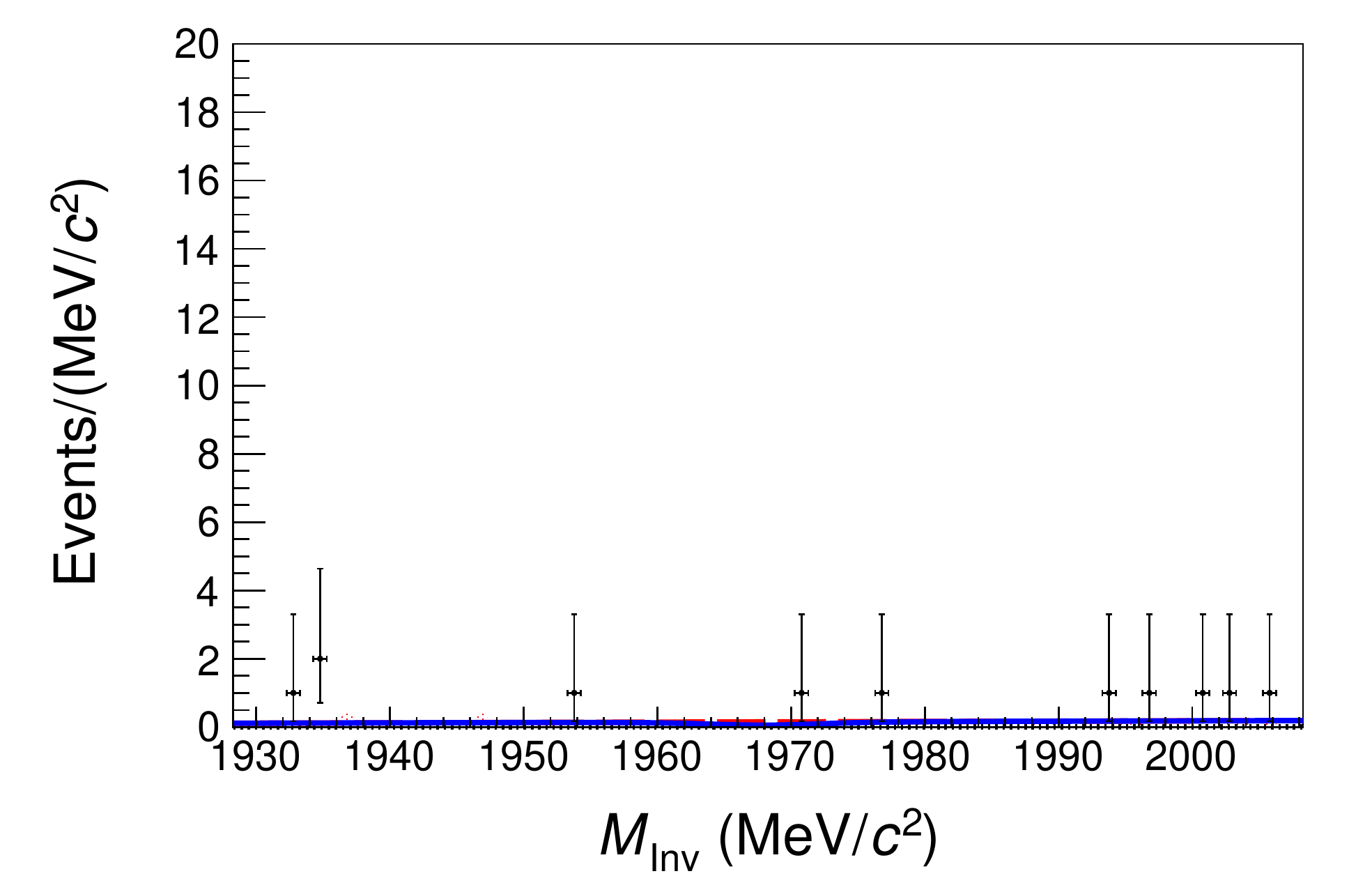}\\
\end{tabular}

\begin{tabular}{cc}
\textbf{RS 1000-1050 MeV/$c$} & \textbf{WS 1000-1050 MeV/$c$}\\
\includegraphics[width=3in,valign=m]{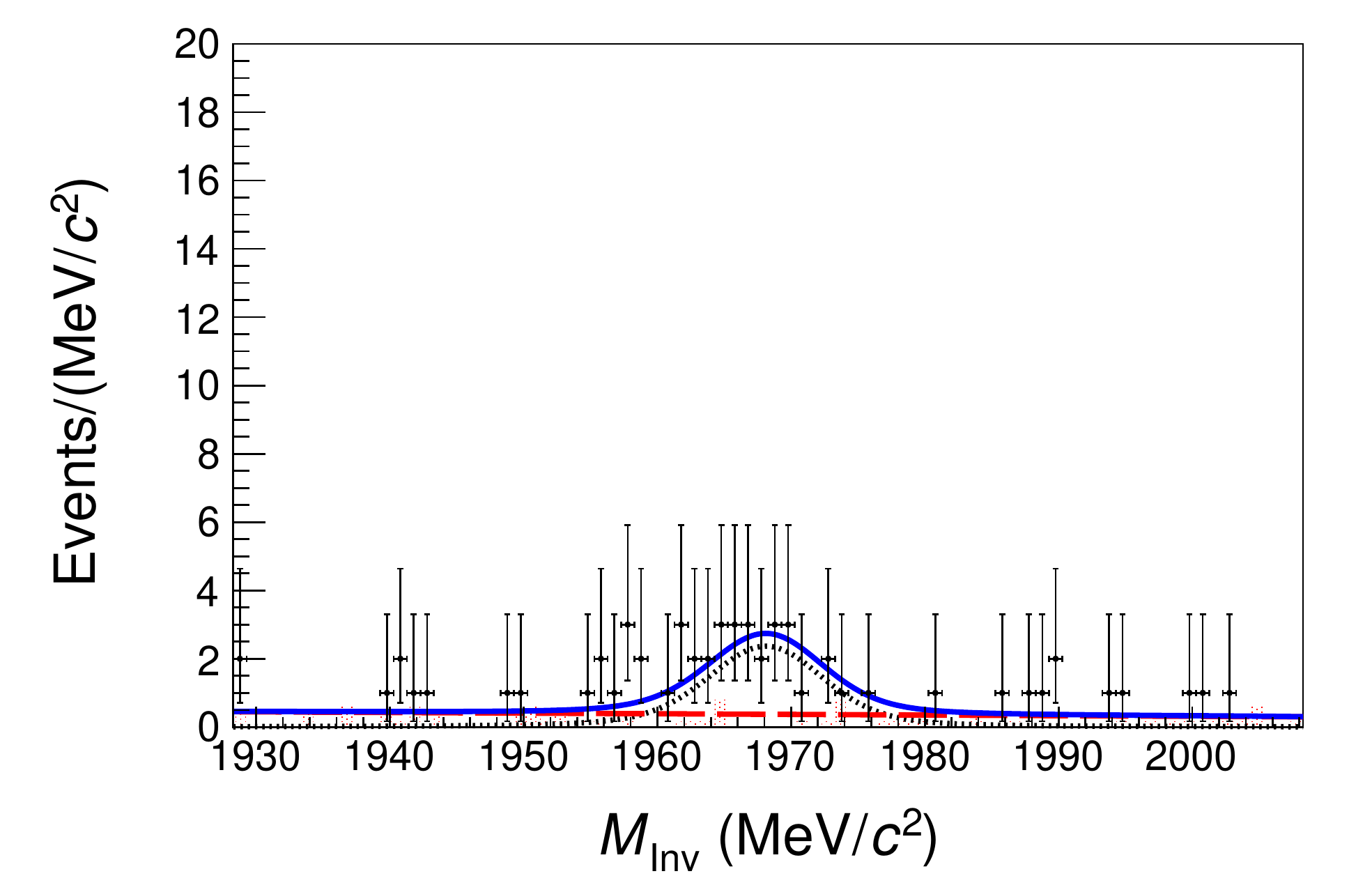} & No entries are seen in the signal region\\
\textbf{RS 1050-1100 MeV/$c$} & \textbf{WS 1050-1100 MeV/$c$}\\
\includegraphics[width=3in,valign=m]{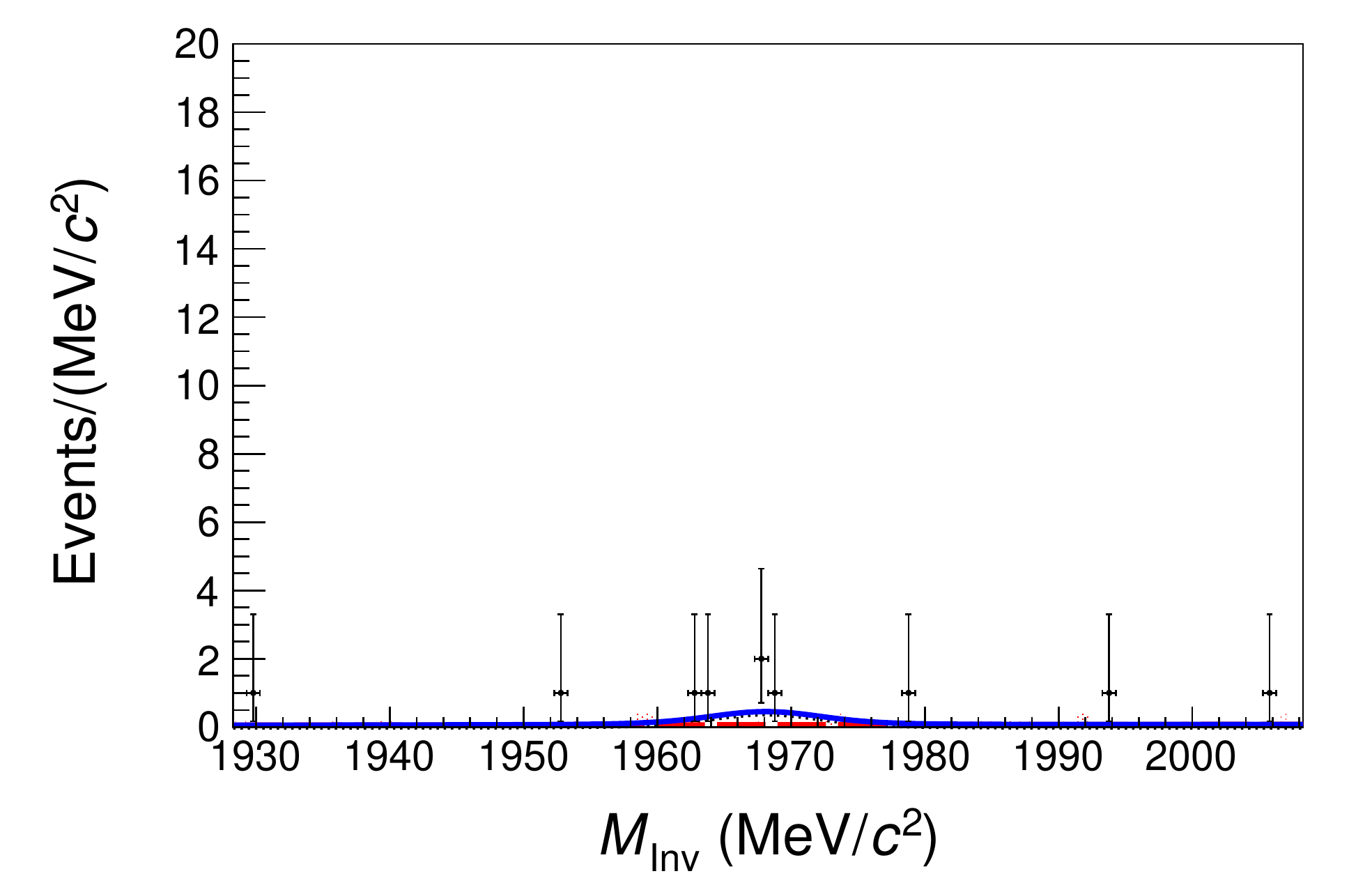} & No entries are seen in the signal region
\end{tabular}
\pagebreak
\raggedright

\subsubsection{$\EcmA$ Data $\pi$ ID Fits}
\label{subsubsec:4180DataPiIDFits}

\centering
\begin{tabular}{cc}
\textbf{RS 200-250 MeV/$c$} & \textbf{WS 200-250 MeV/$c$}\\
\includegraphics[width=3in]{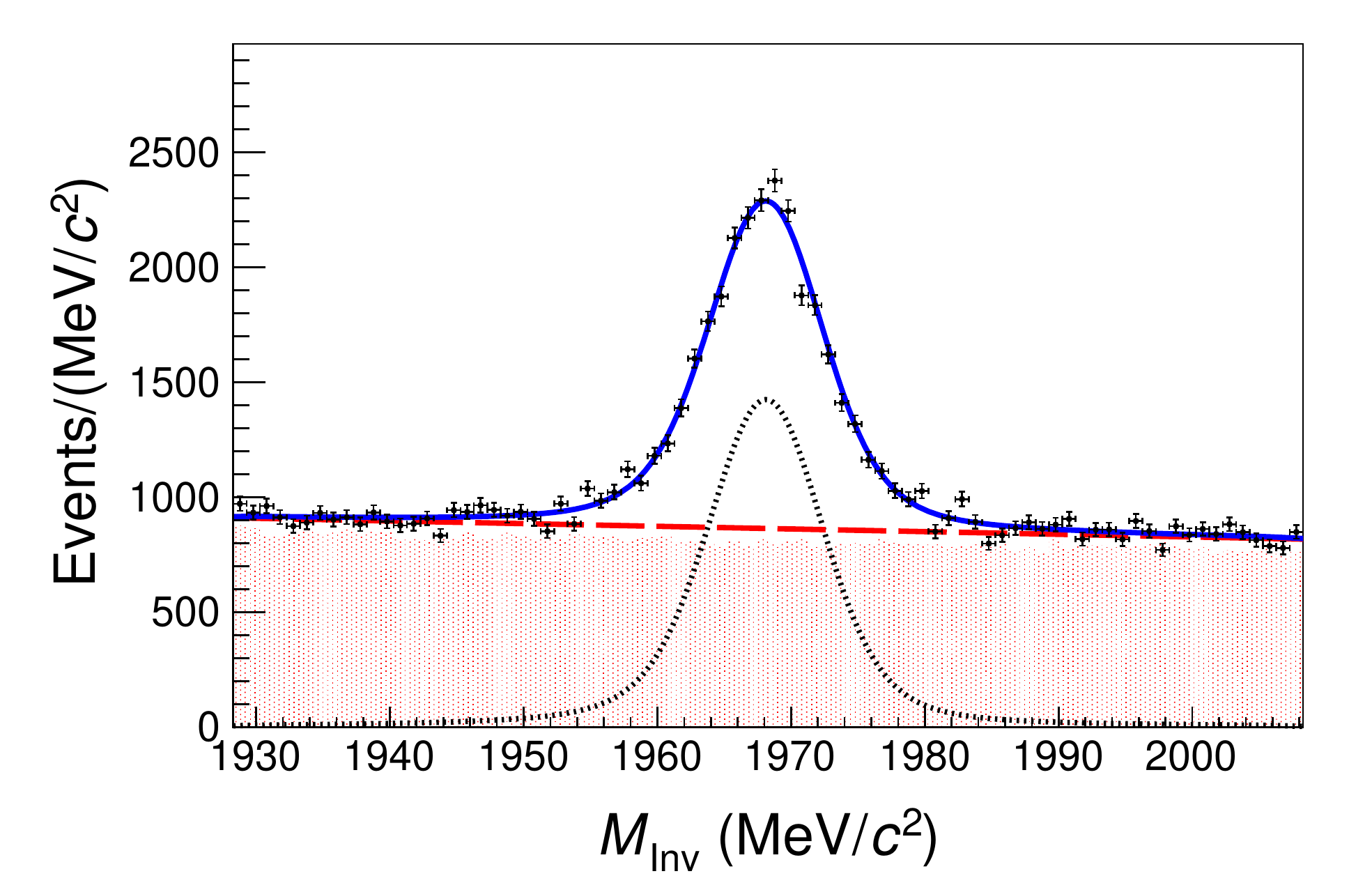} & \includegraphics[width=3in]{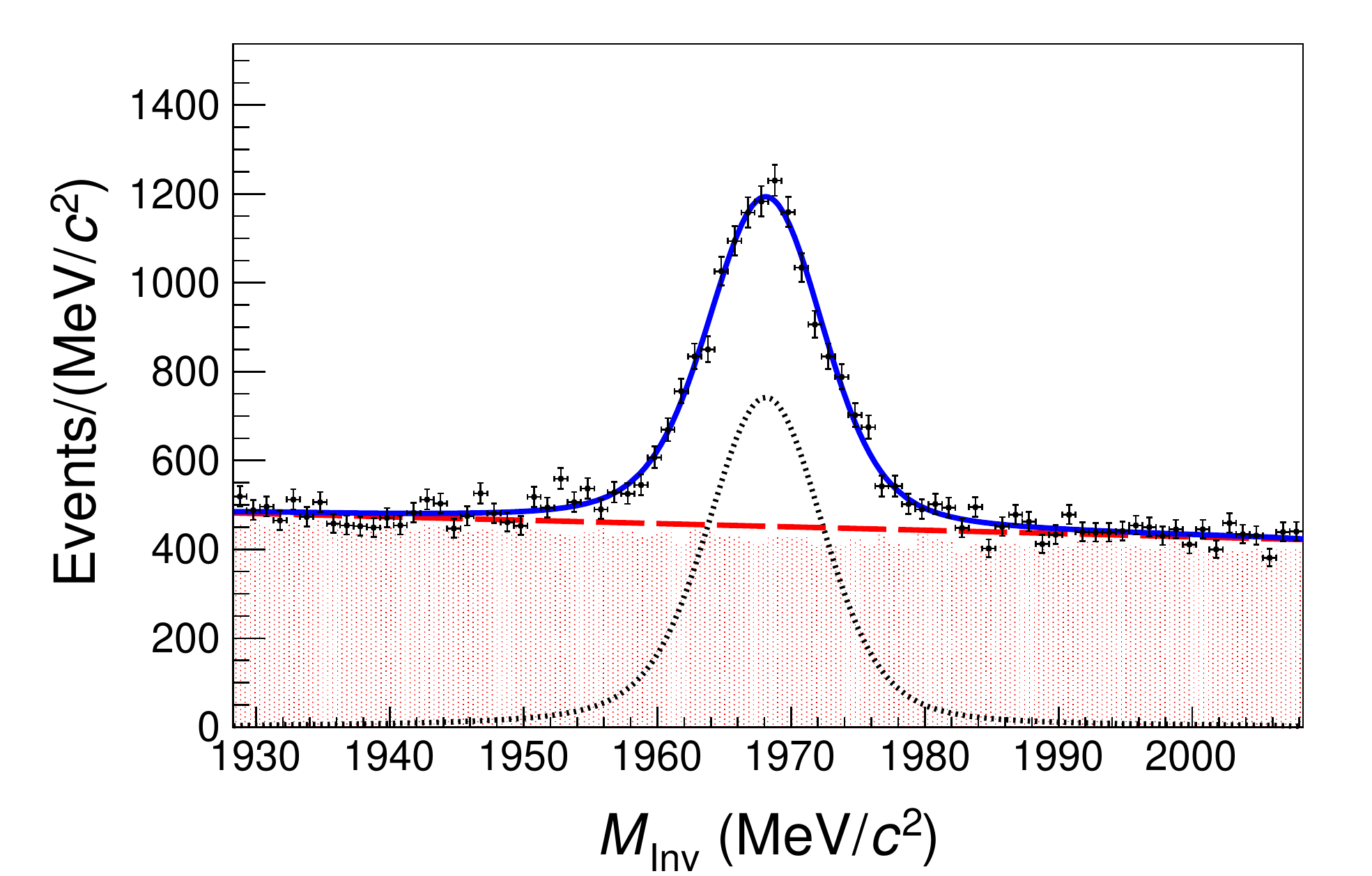}\\
\textbf{RS 250-300 MeV/$c$} & \textbf{WS 250-300 MeV/$c$}\\
\includegraphics[width=3in]{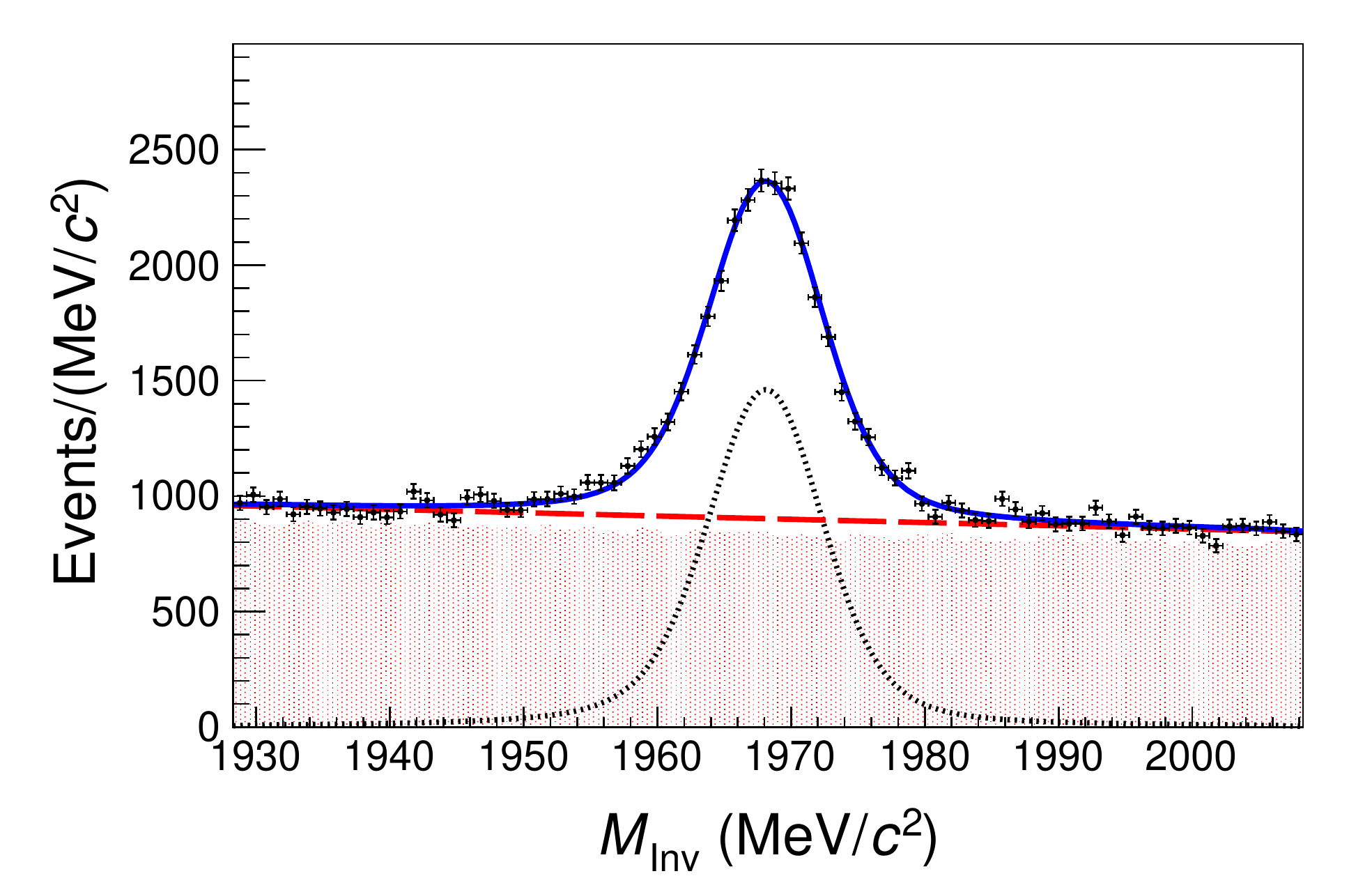} & \includegraphics[width=3in]{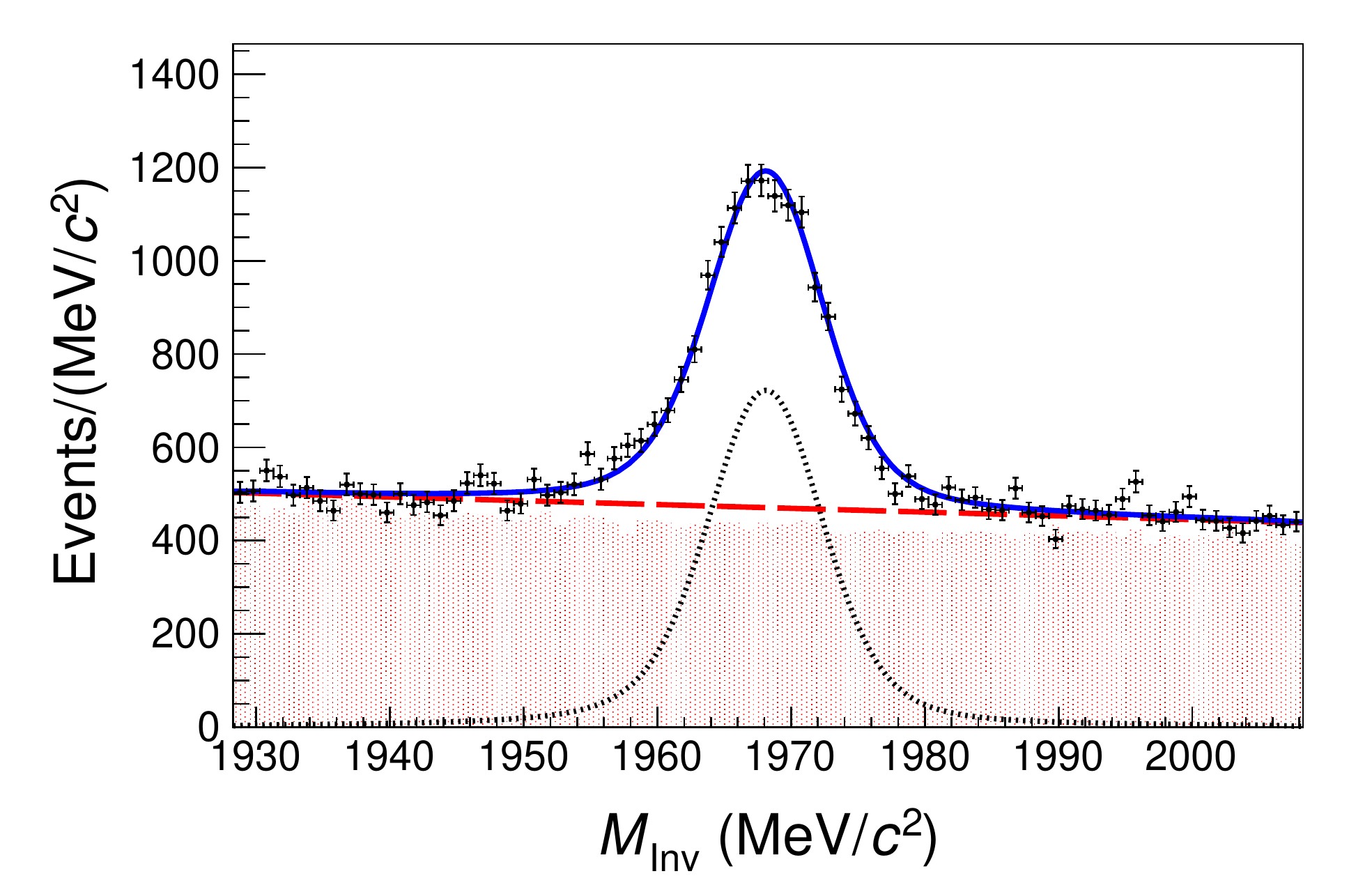}\\
\textbf{RS 300-350 MeV/$c$} & \textbf{WS 300-350 MeV/$c$}\\
\includegraphics[width=3in]{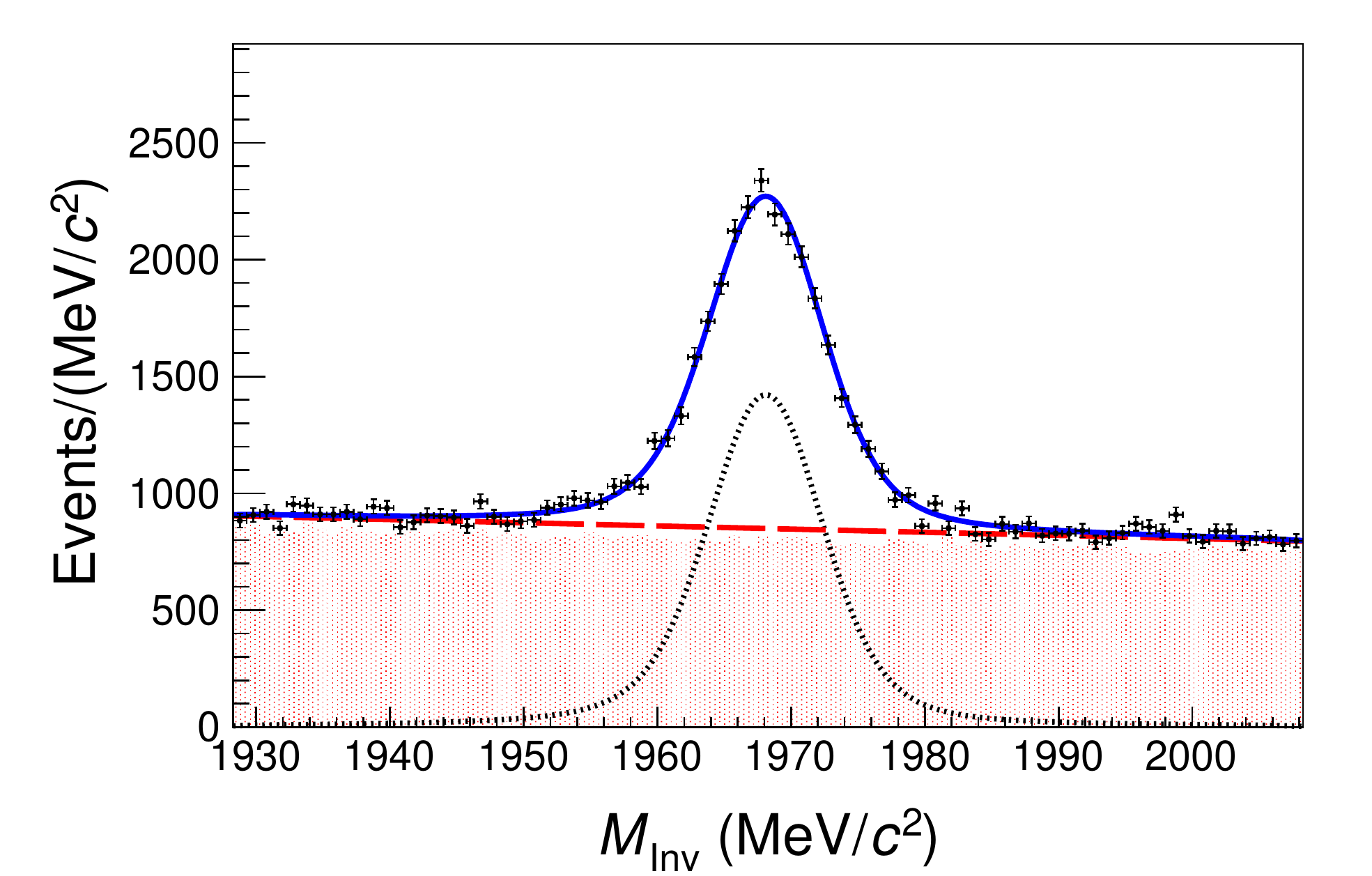} & \includegraphics[width=3in]{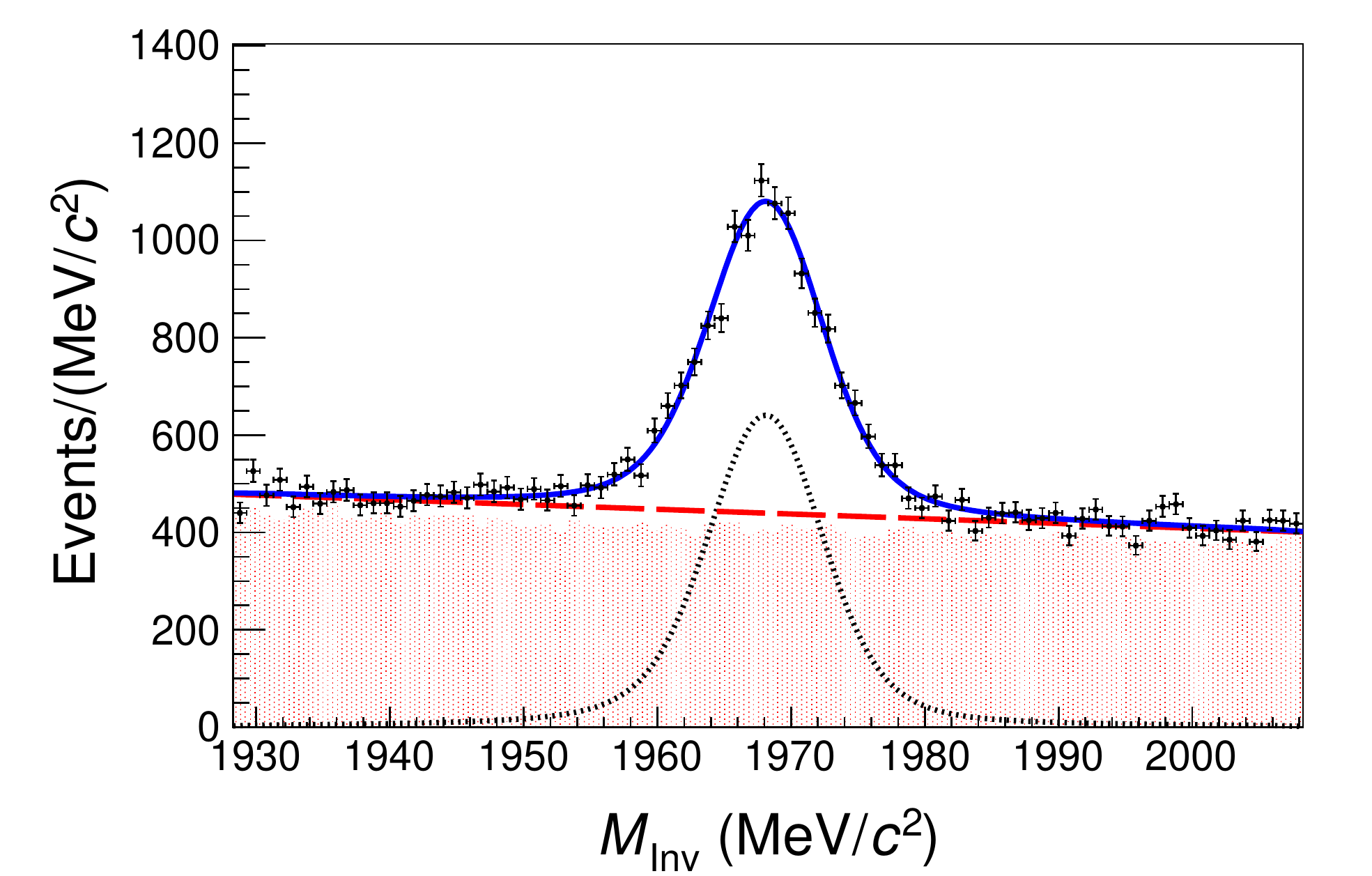}\\
\textbf{RS 350-400 MeV/$c$} & \textbf{WS 350-400 MeV/$c$}\\
\includegraphics[width=3in]{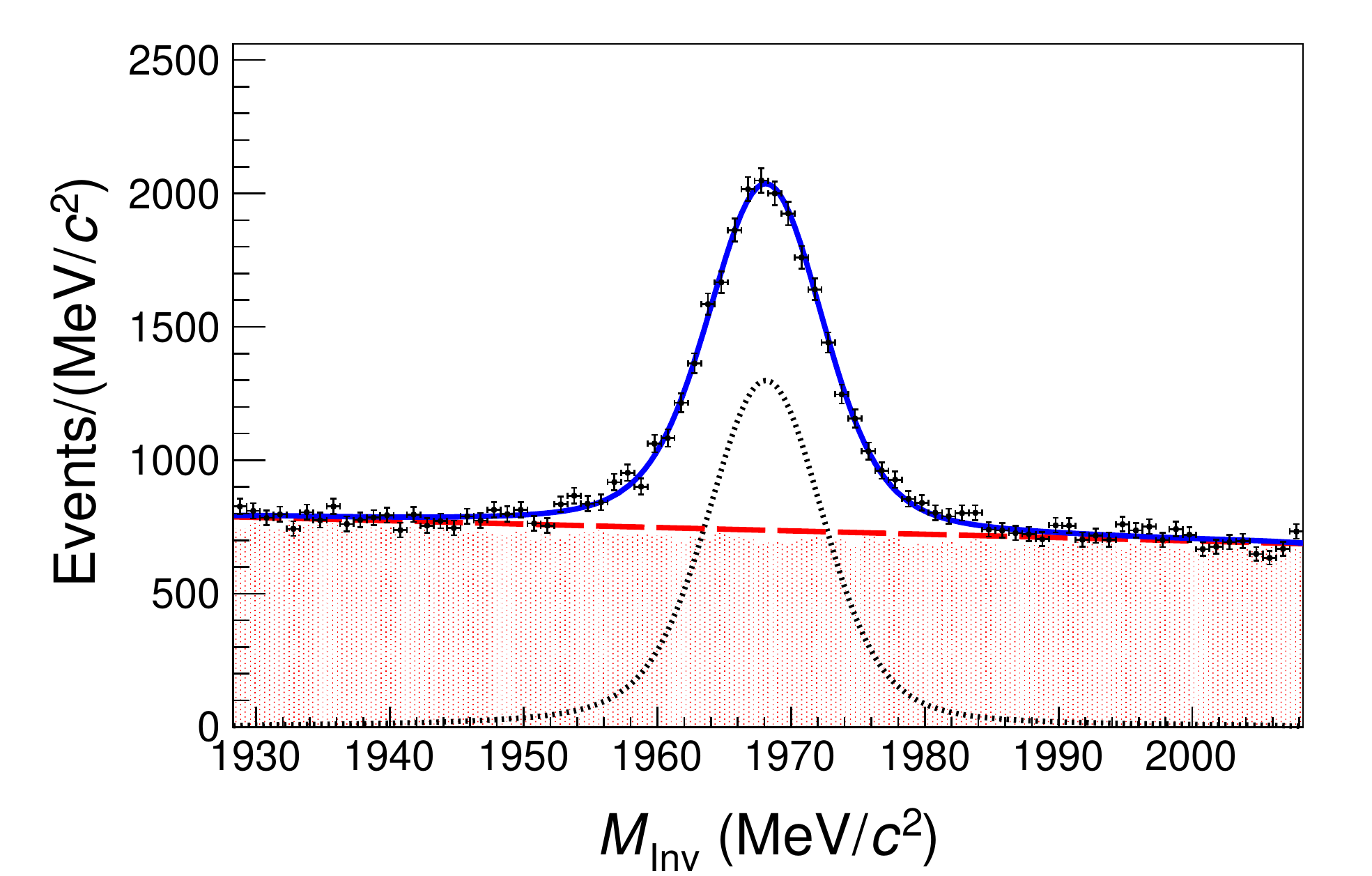} & \includegraphics[width=3in]{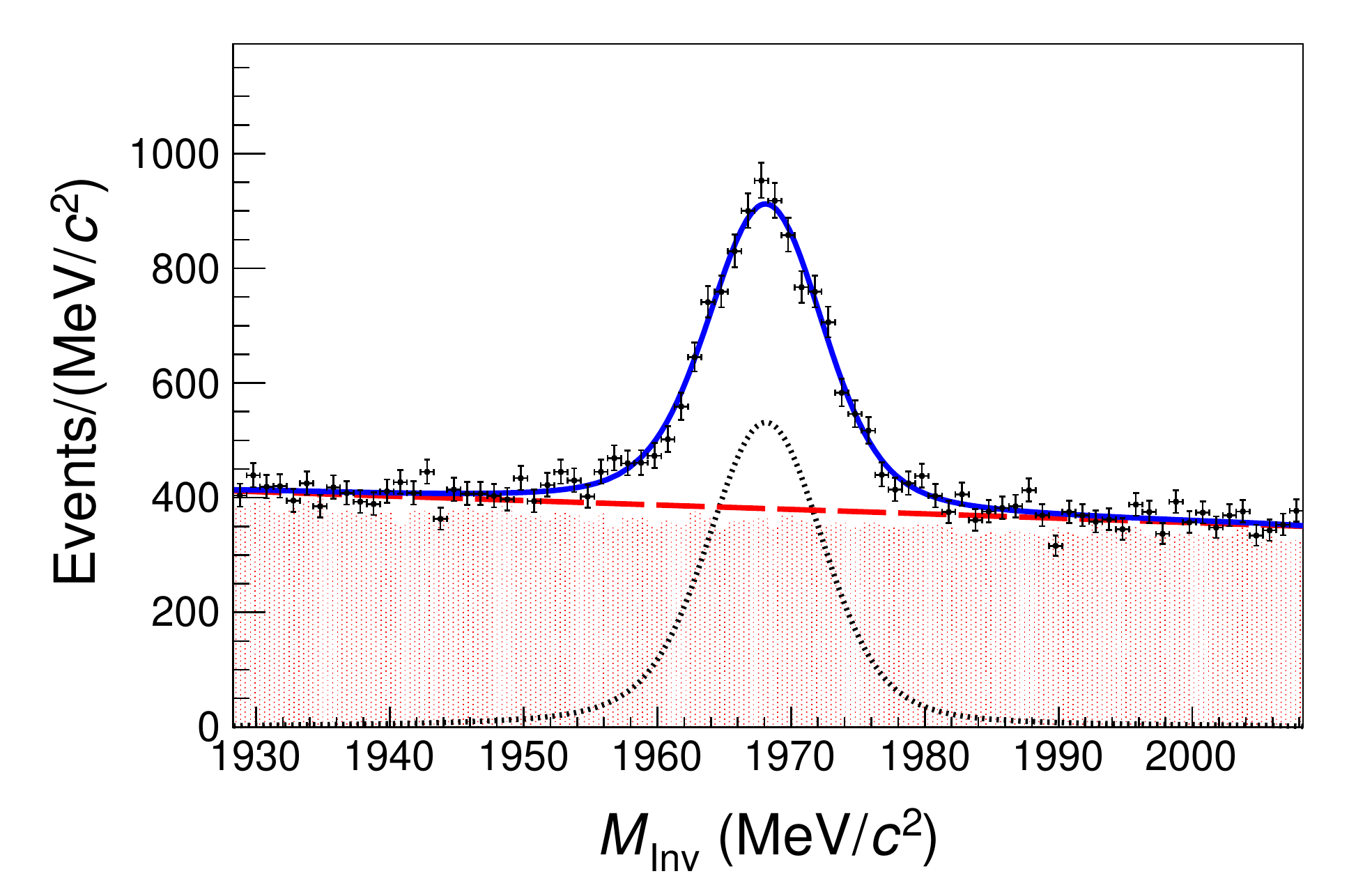}
\end{tabular}

\begin{tabular}{cc}
\textbf{RS 400-450 MeV/$c$} & \textbf{WS 400-450 MeV/$c$}\\
\includegraphics[width=3in]{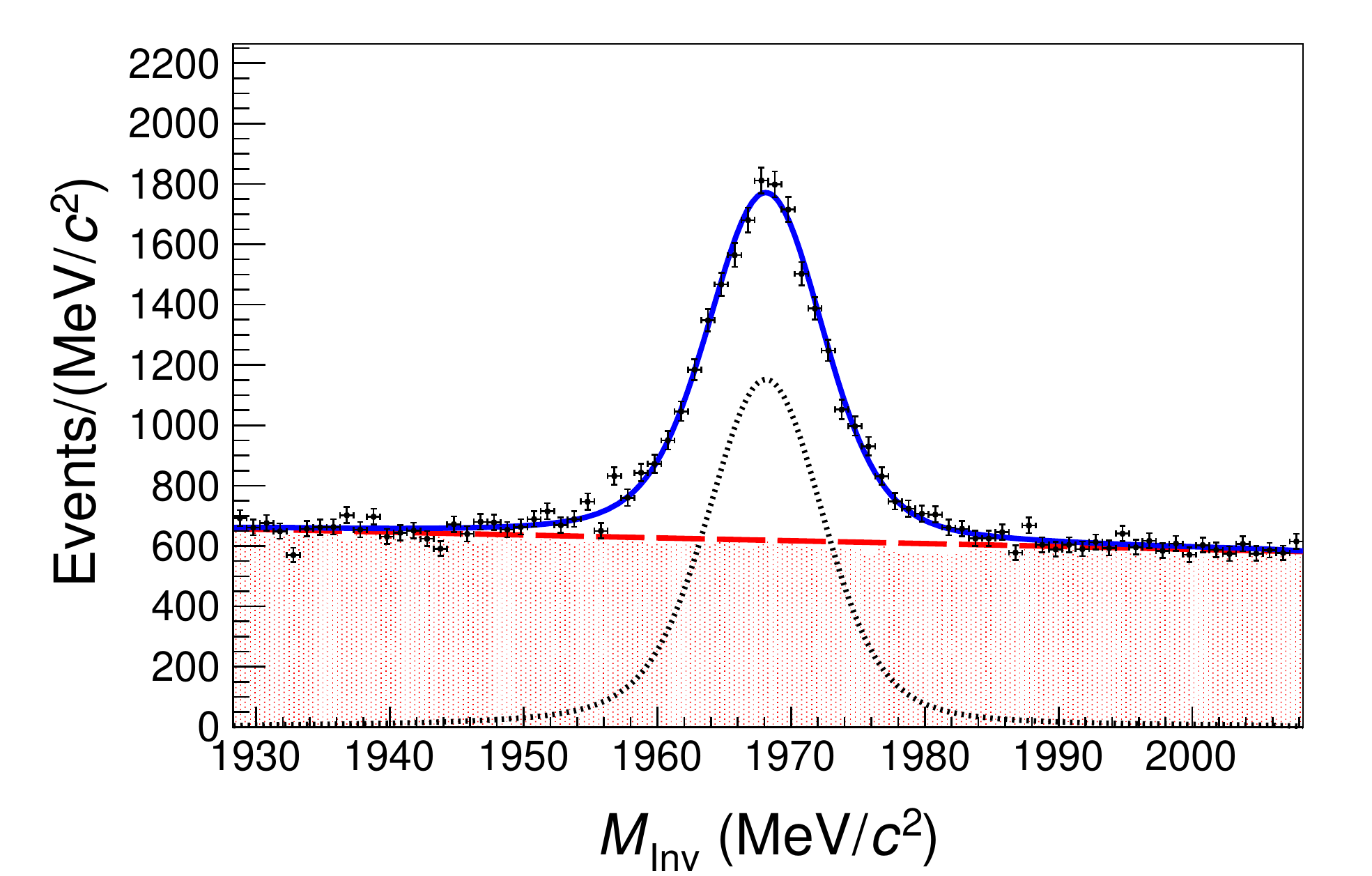} & \includegraphics[width=3in]{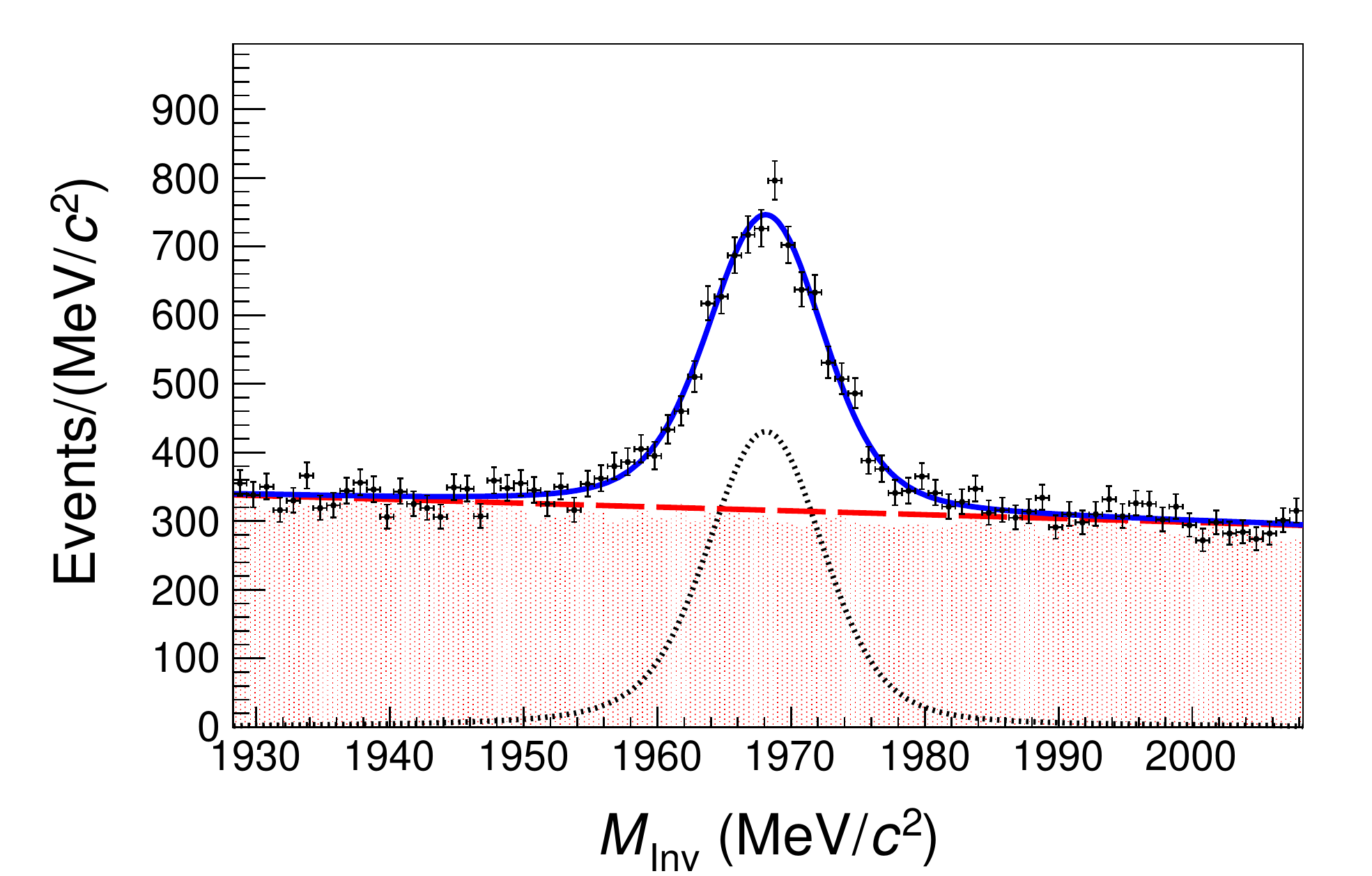}\\
\textbf{RS 450-500 MeV/$c$} & \textbf{WS 450-500 MeV/$c$}\\
\includegraphics[width=3in]{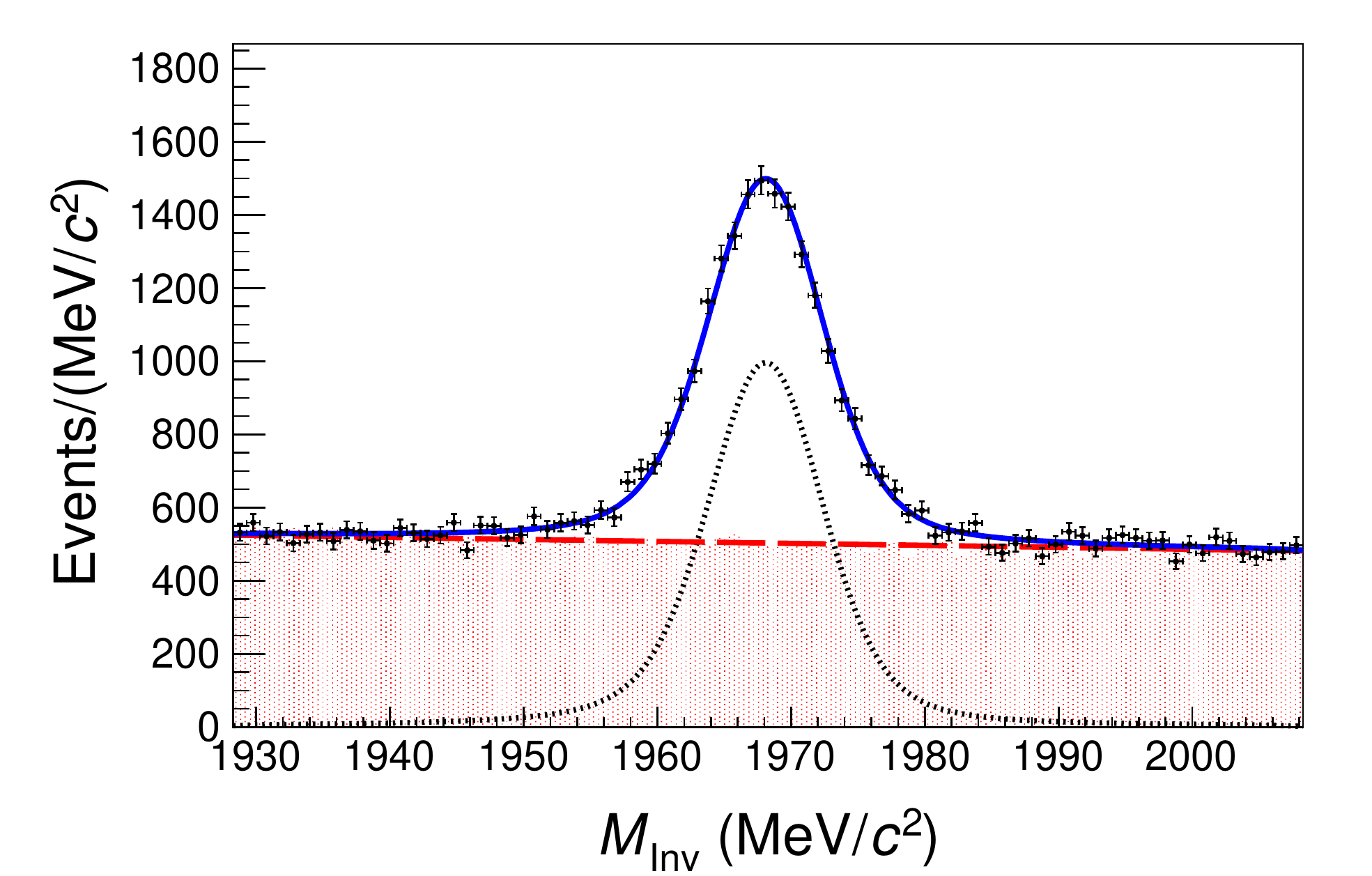} & \includegraphics[width=3in]{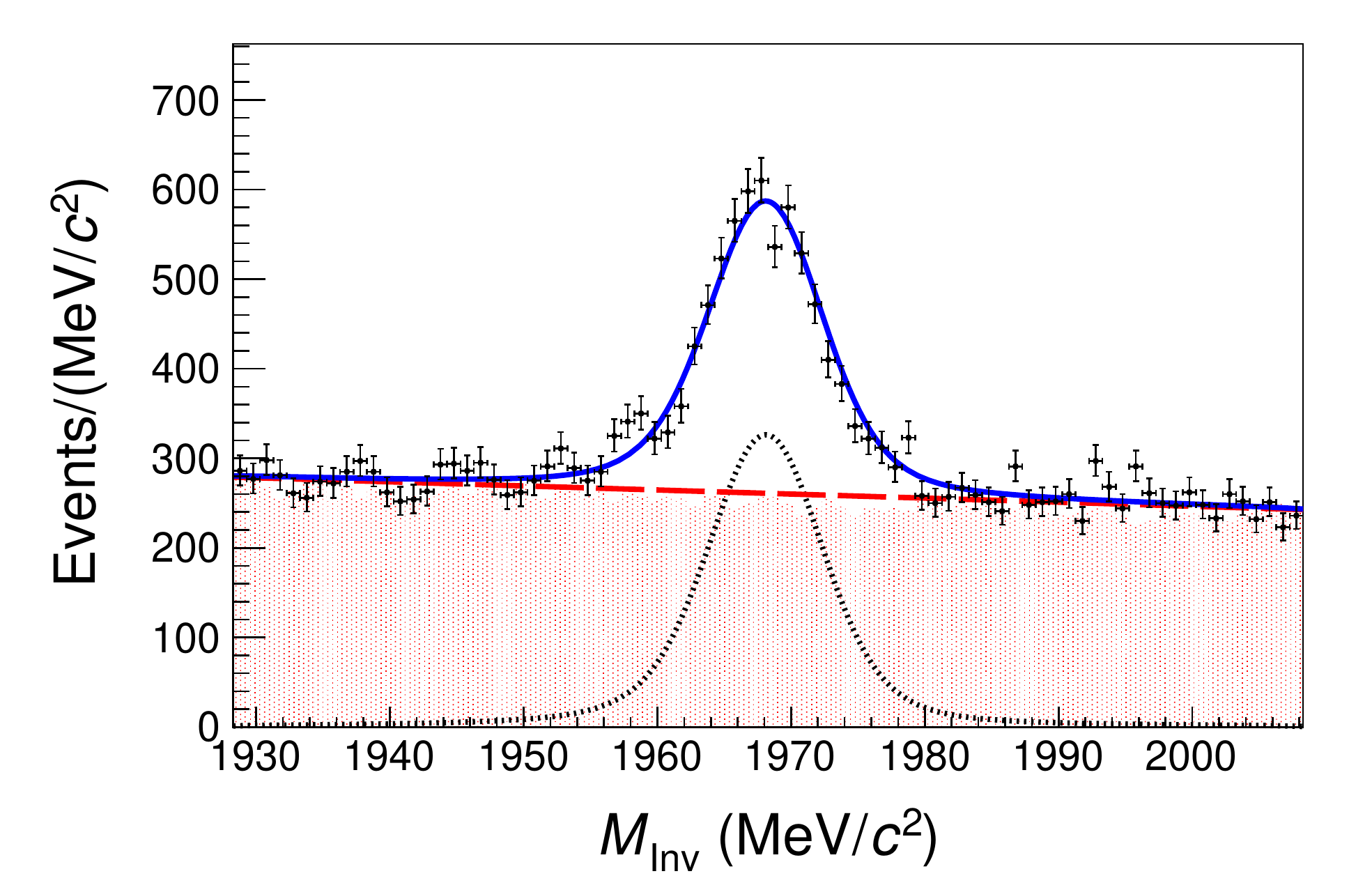}\\
\textbf{RS 500-550 MeV/$c$} & \textbf{WS 500-550 MeV/$c$}\\
\includegraphics[width=3in]{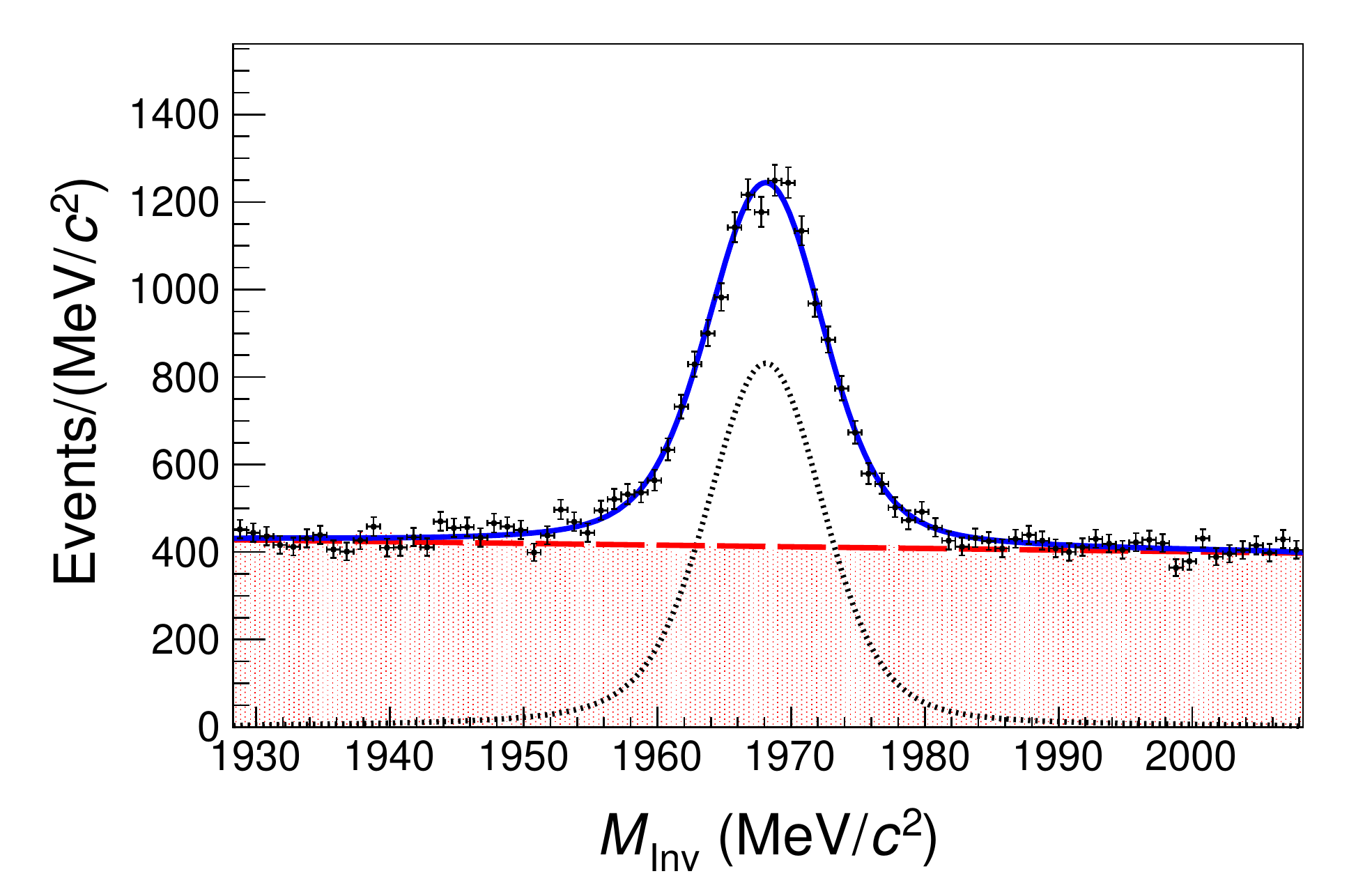} & \includegraphics[width=3in]{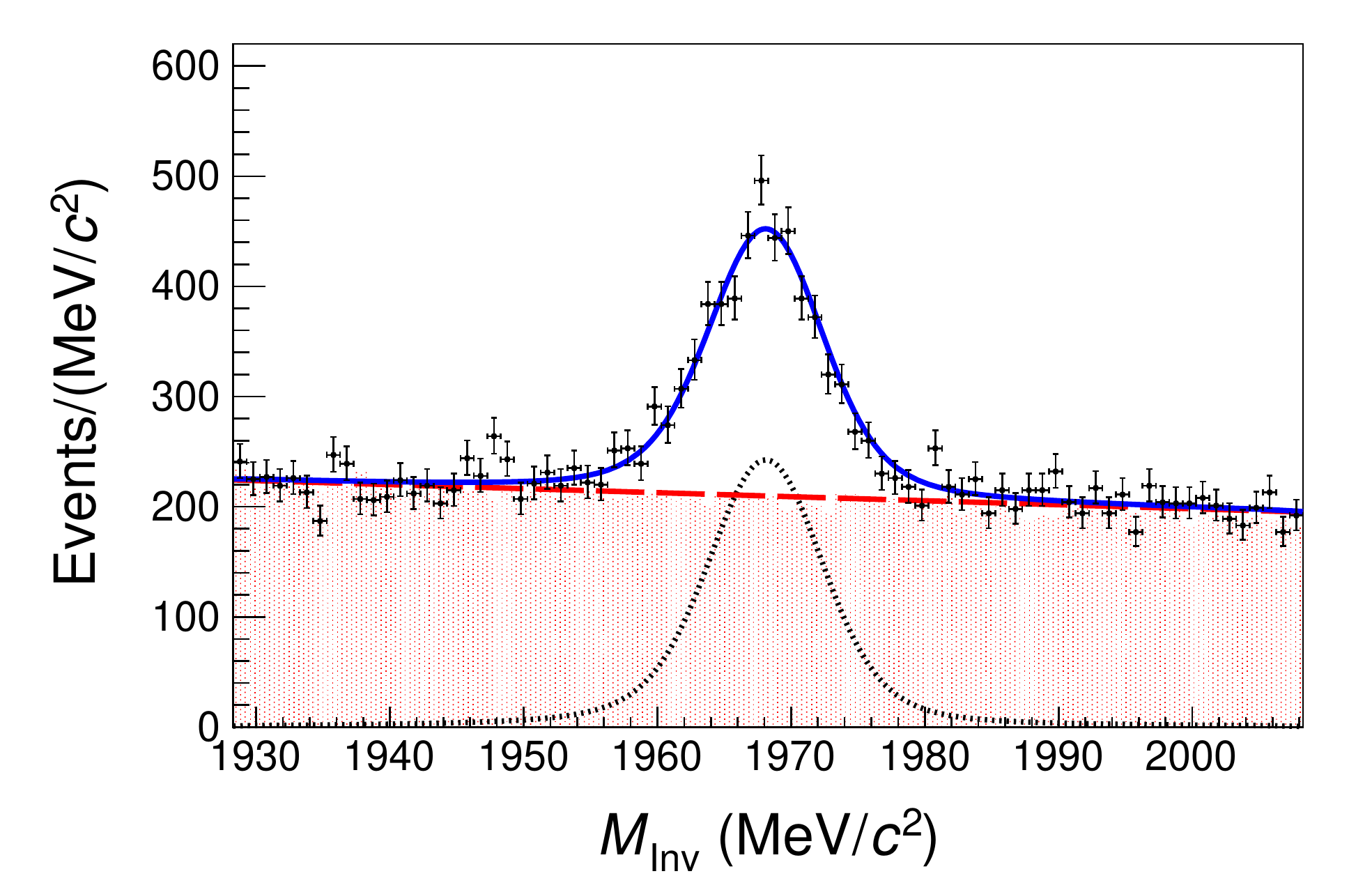}\\
\textbf{RS 550-600 MeV/$c$} & \textbf{WS 550-600 MeV/$c$}\\
\includegraphics[width=3in]{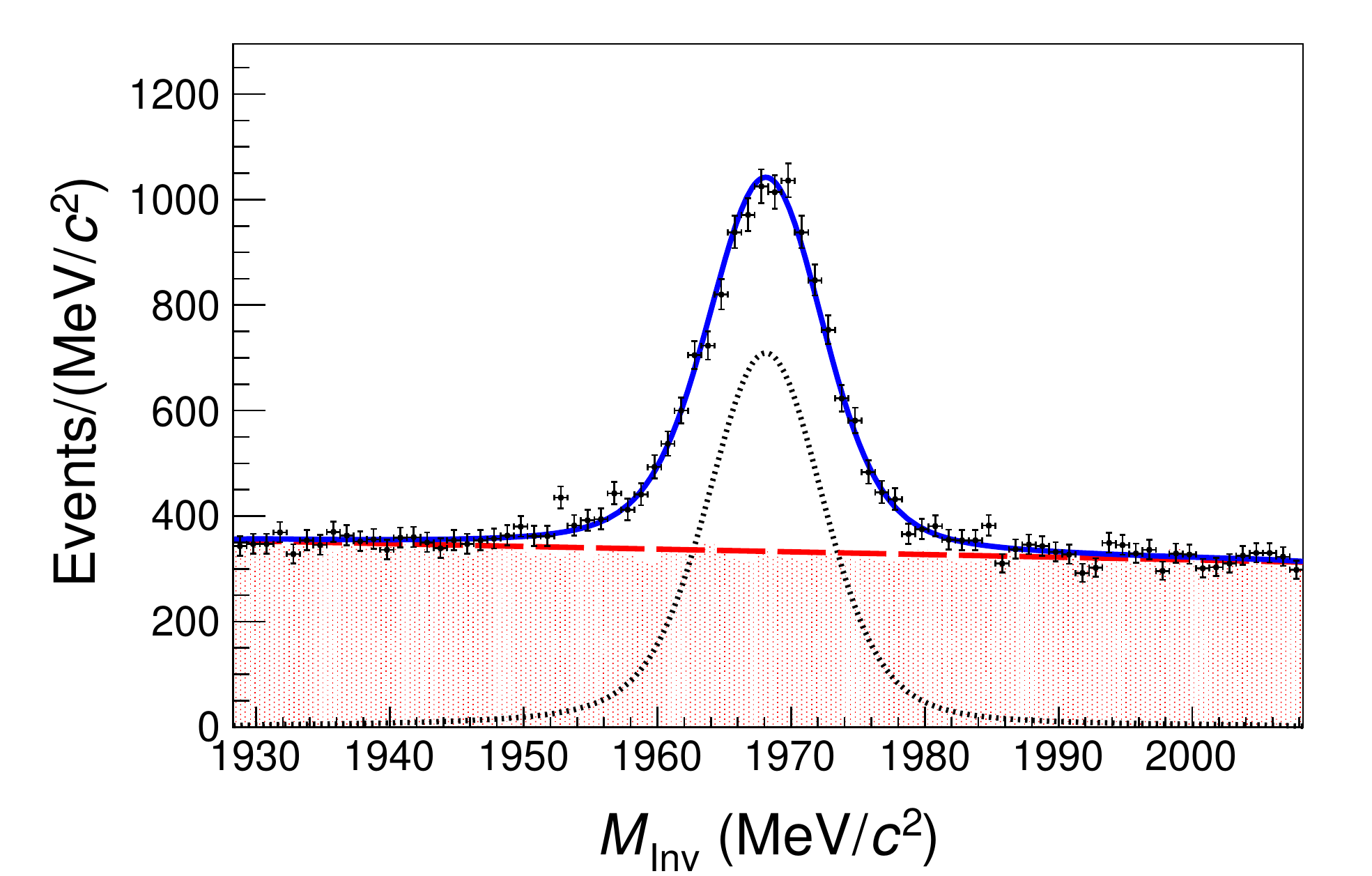} & \includegraphics[width=3in]{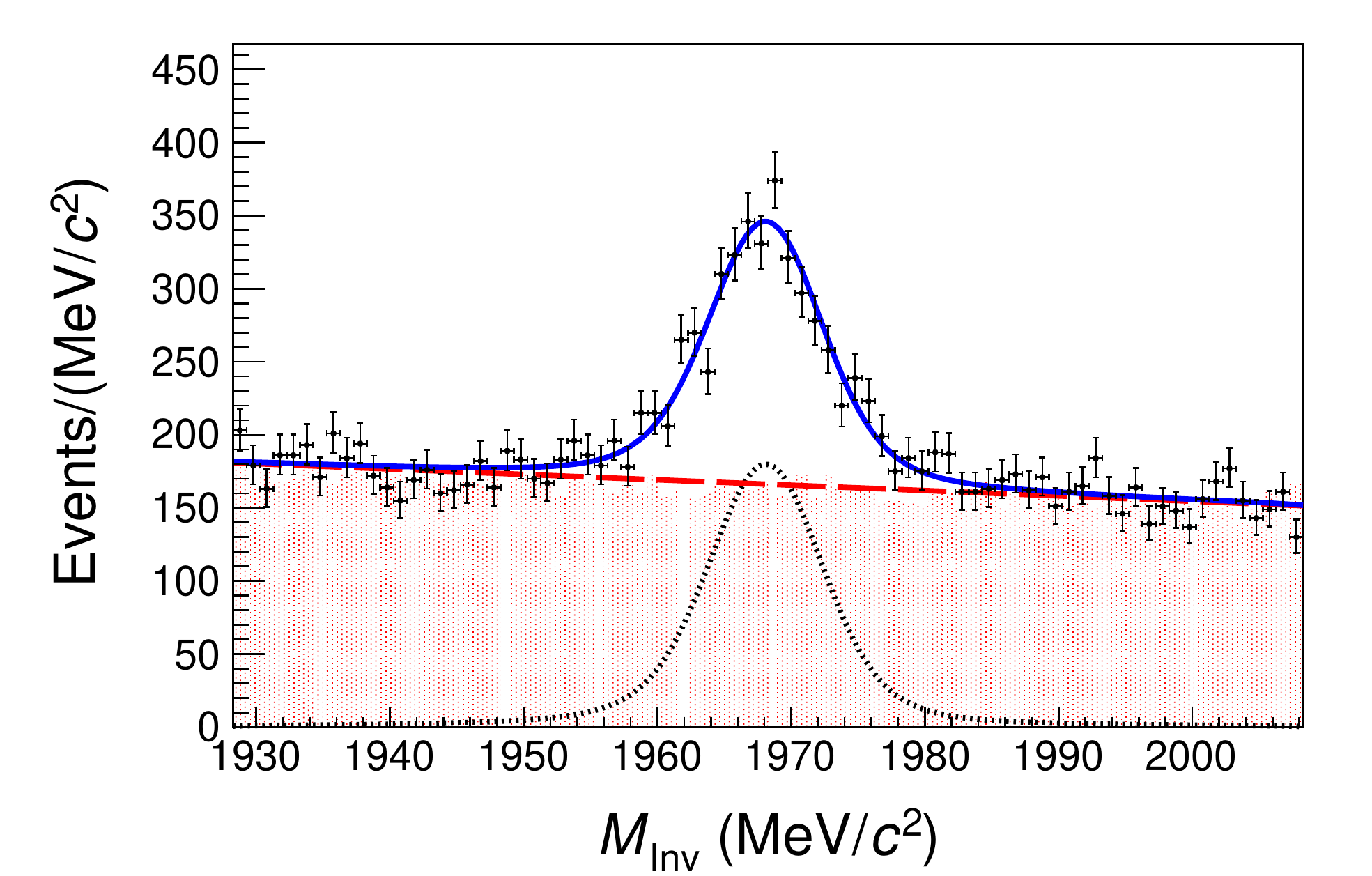}
\end{tabular}

\begin{tabular}{cc}
\textbf{RS 600-650 MeV/$c$} & \textbf{WS 600-650 MeV/$c$}\\
\includegraphics[width=3in]{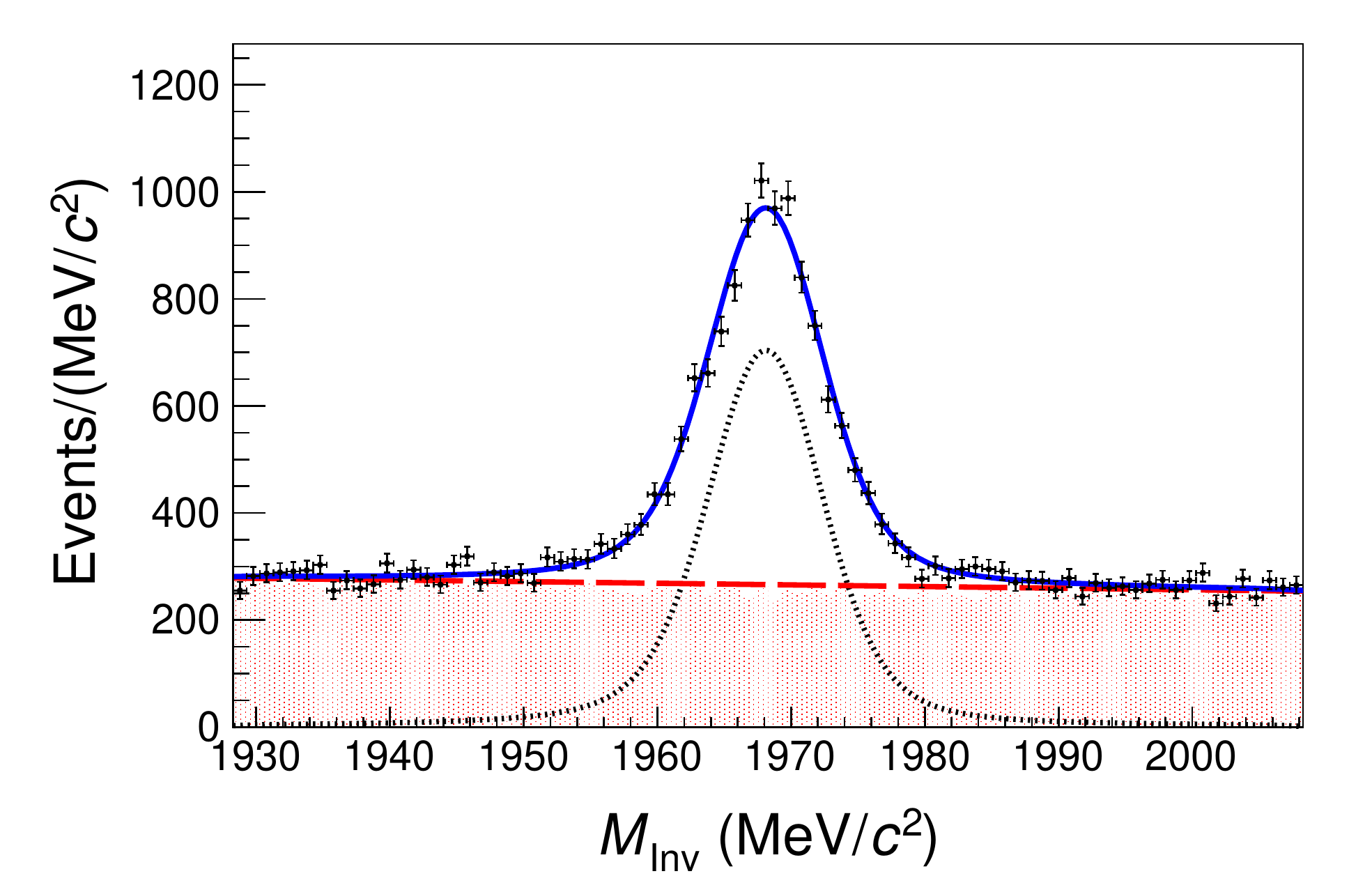} & \includegraphics[width=3in]{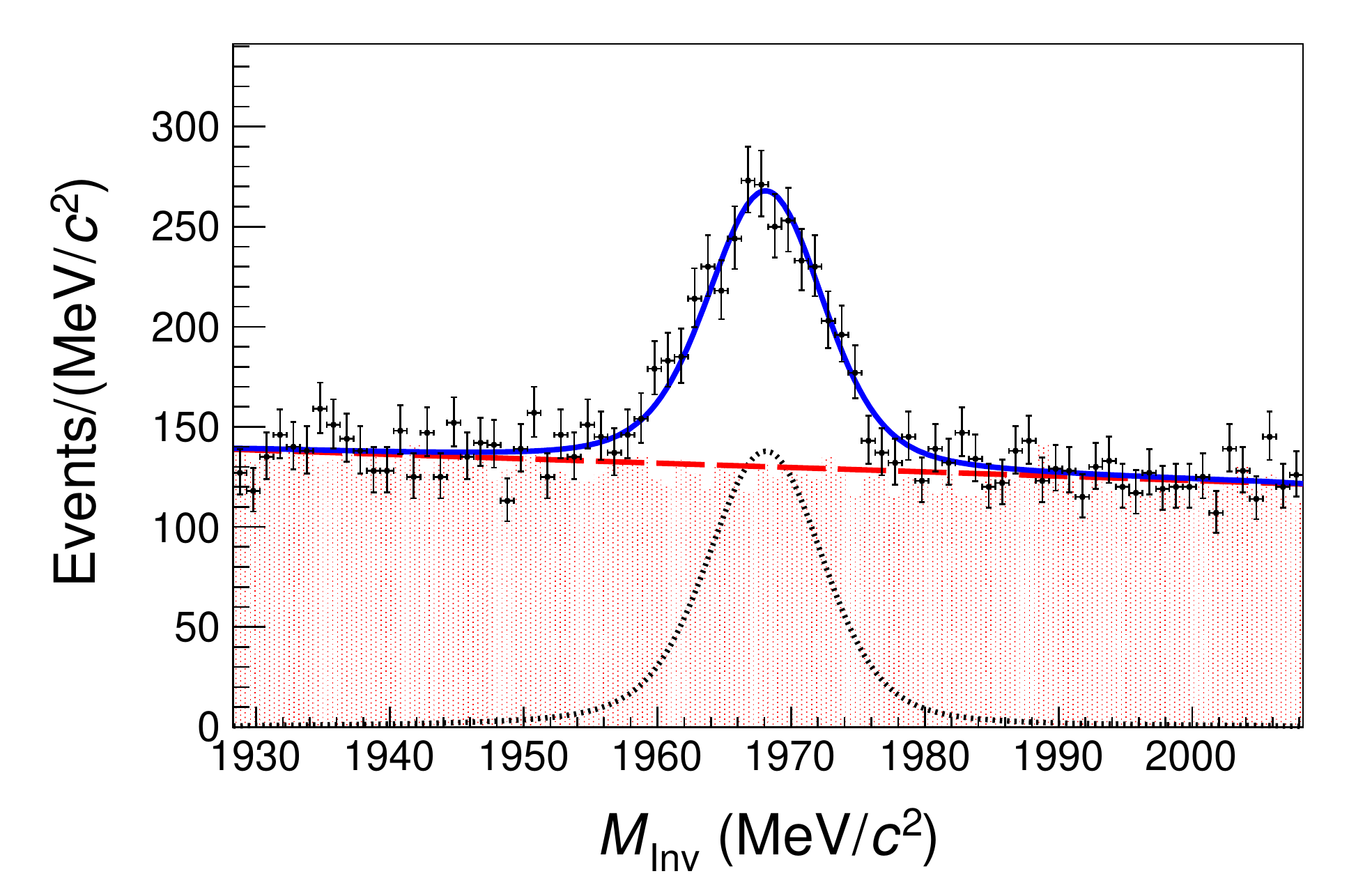}\\
\textbf{RS 650-700 MeV/$c$} & \textbf{WS 650-700 MeV/$c$}\\
\includegraphics[width=3in]{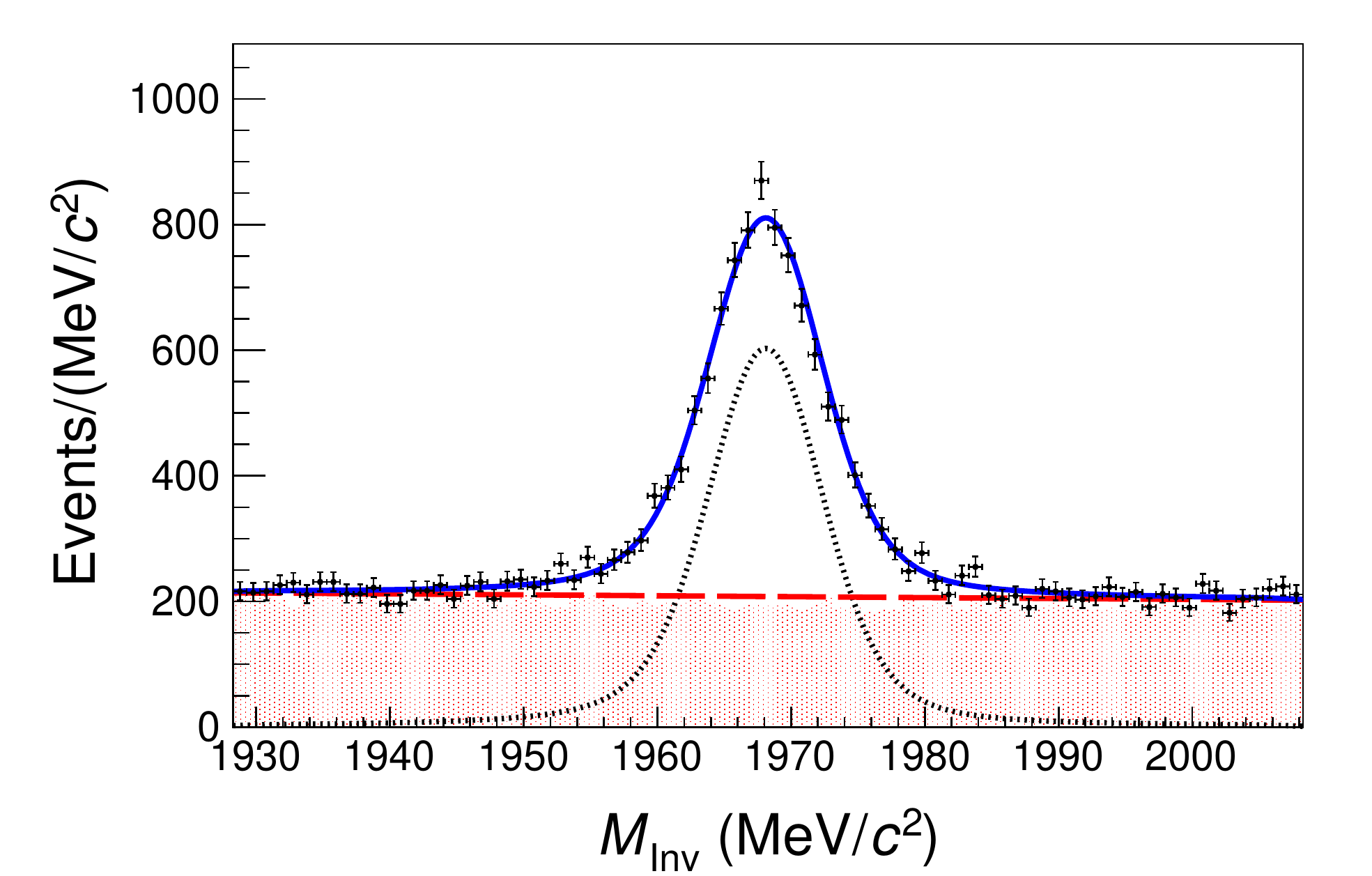} & \includegraphics[width=3in]{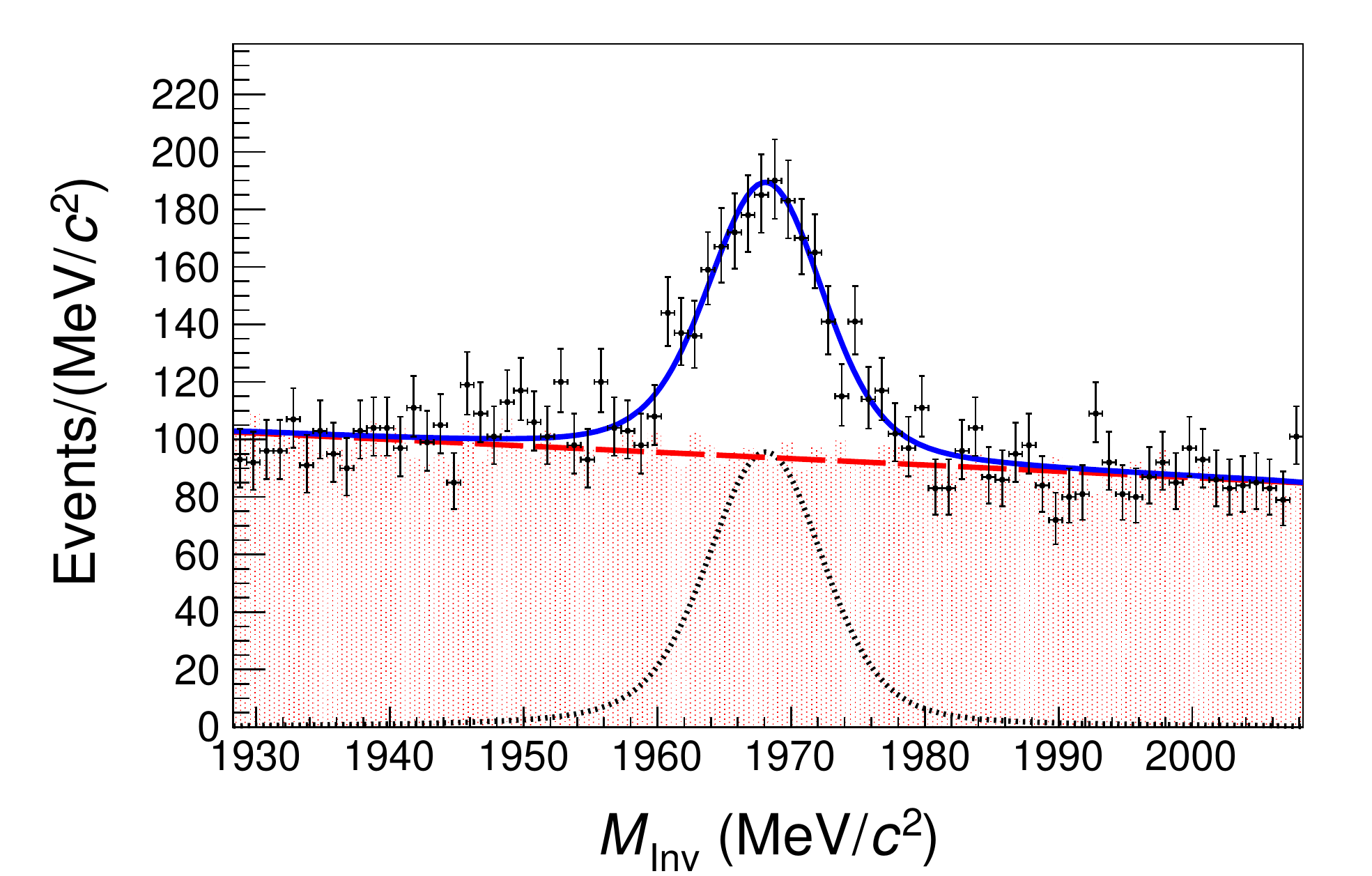}\\
\textbf{RS 700-750 MeV/$c$} & \textbf{WS 700-750 MeV/$c$}\\
\includegraphics[width=3in]{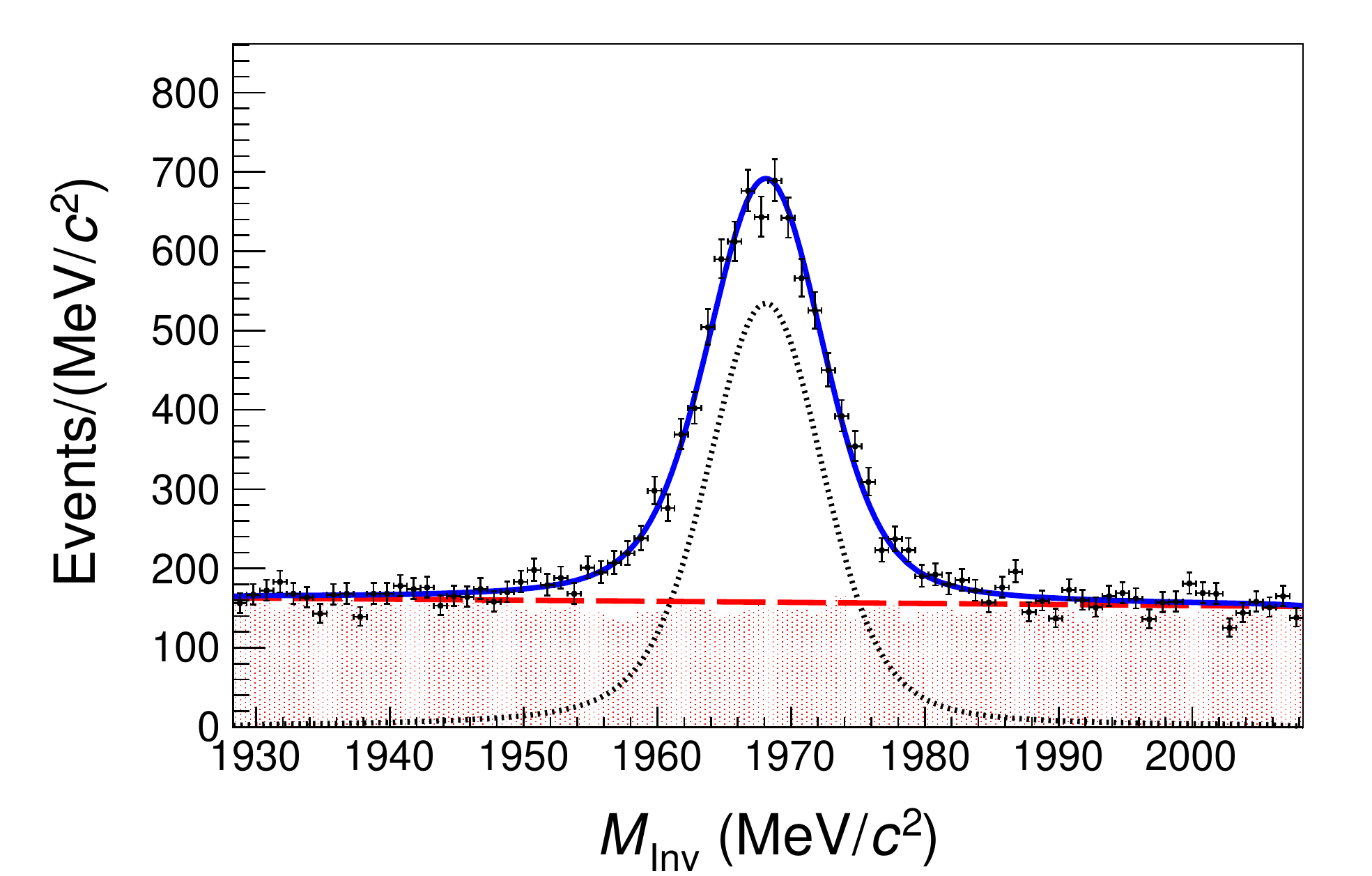} & \includegraphics[width=3in]{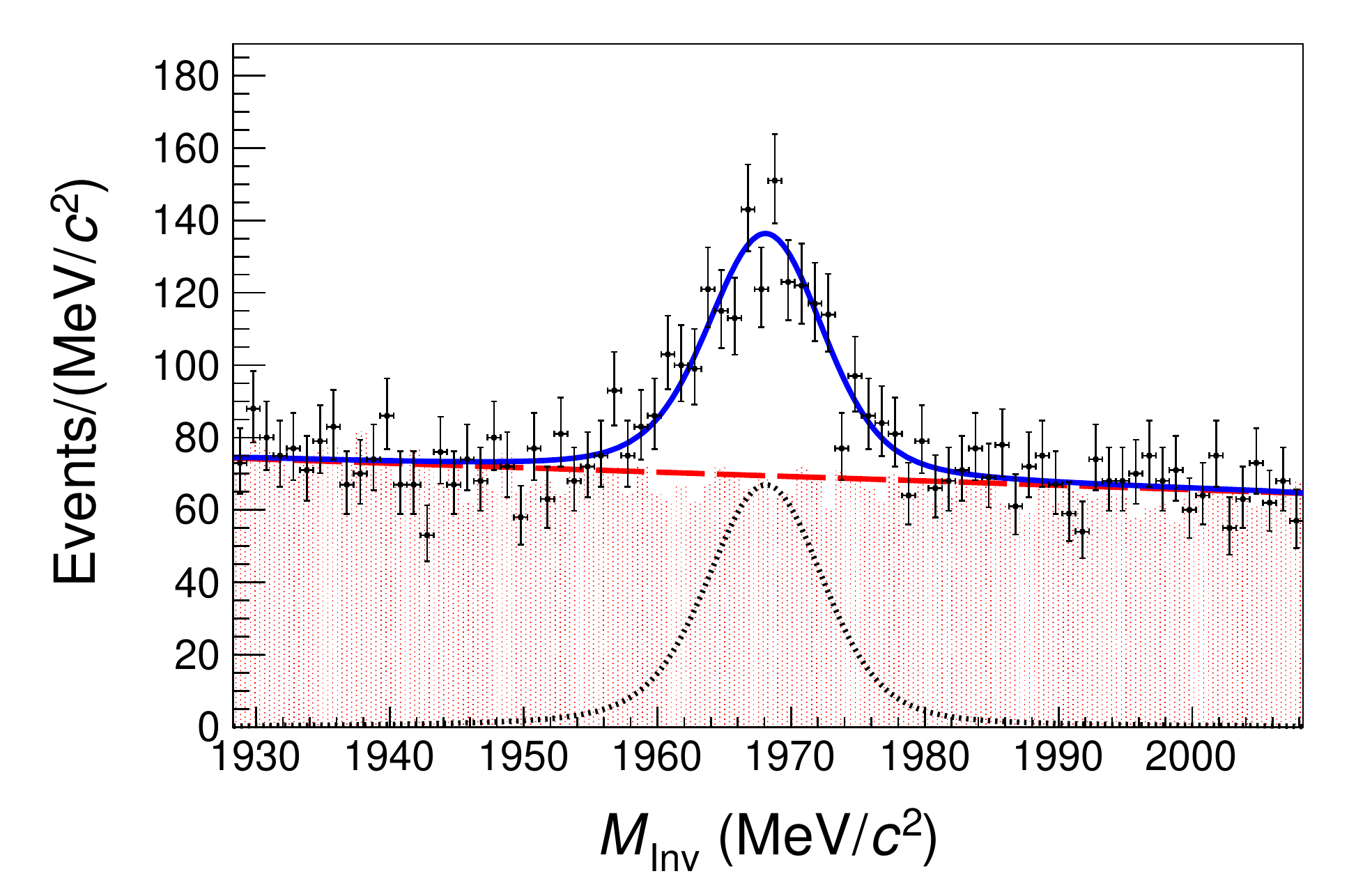}\\
\textbf{RS 750-800 MeV/$c$} & \textbf{WS 750-800 MeV/$c$}\\
\includegraphics[width=3in]{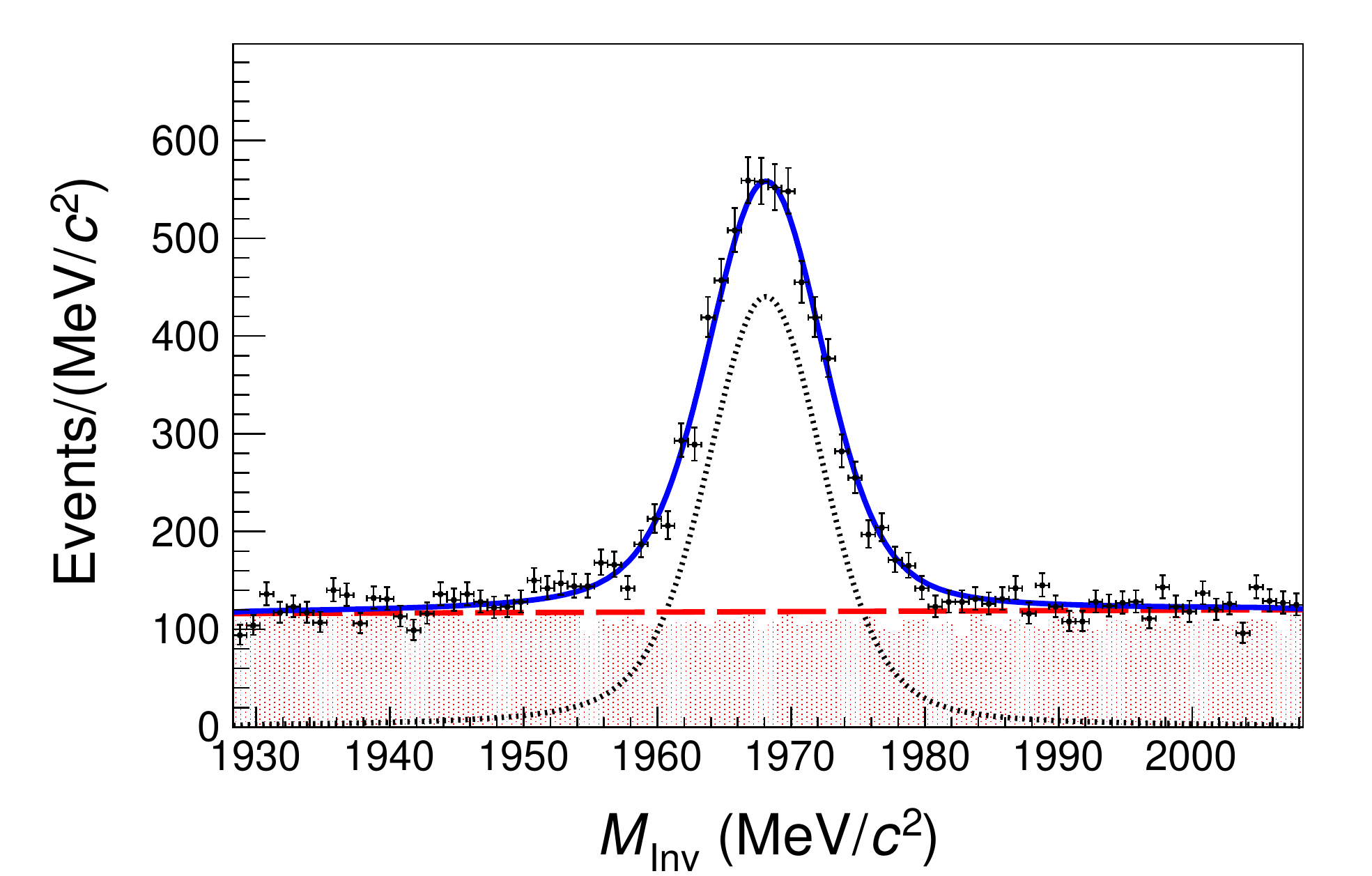} & \includegraphics[width=3in]{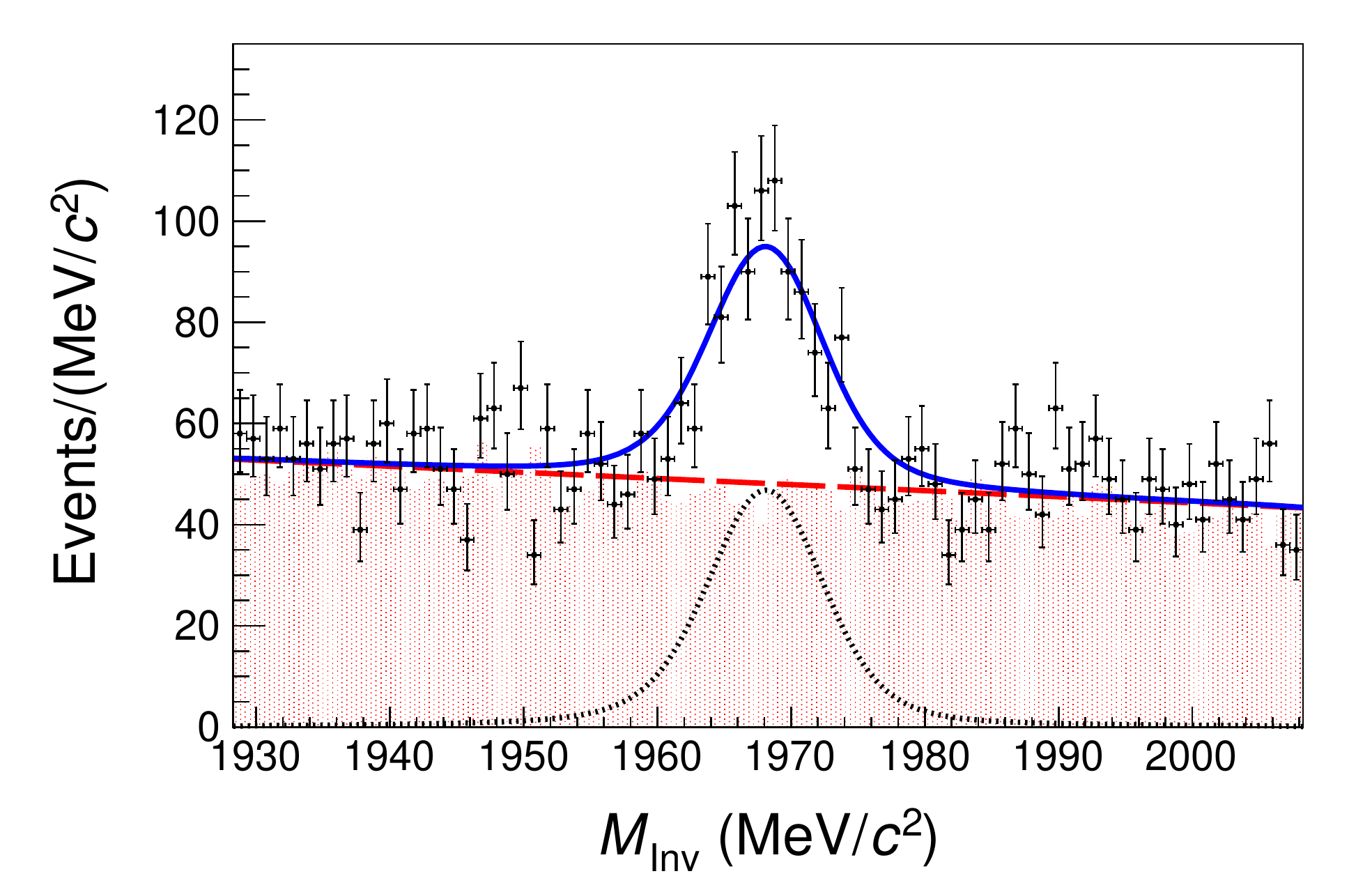}\\
\end{tabular}

\begin{tabular}{cc}
\textbf{RS 800-850 MeV/$c$} & \textbf{WS 800-850 MeV/$c$}\\
\includegraphics[width=3in]{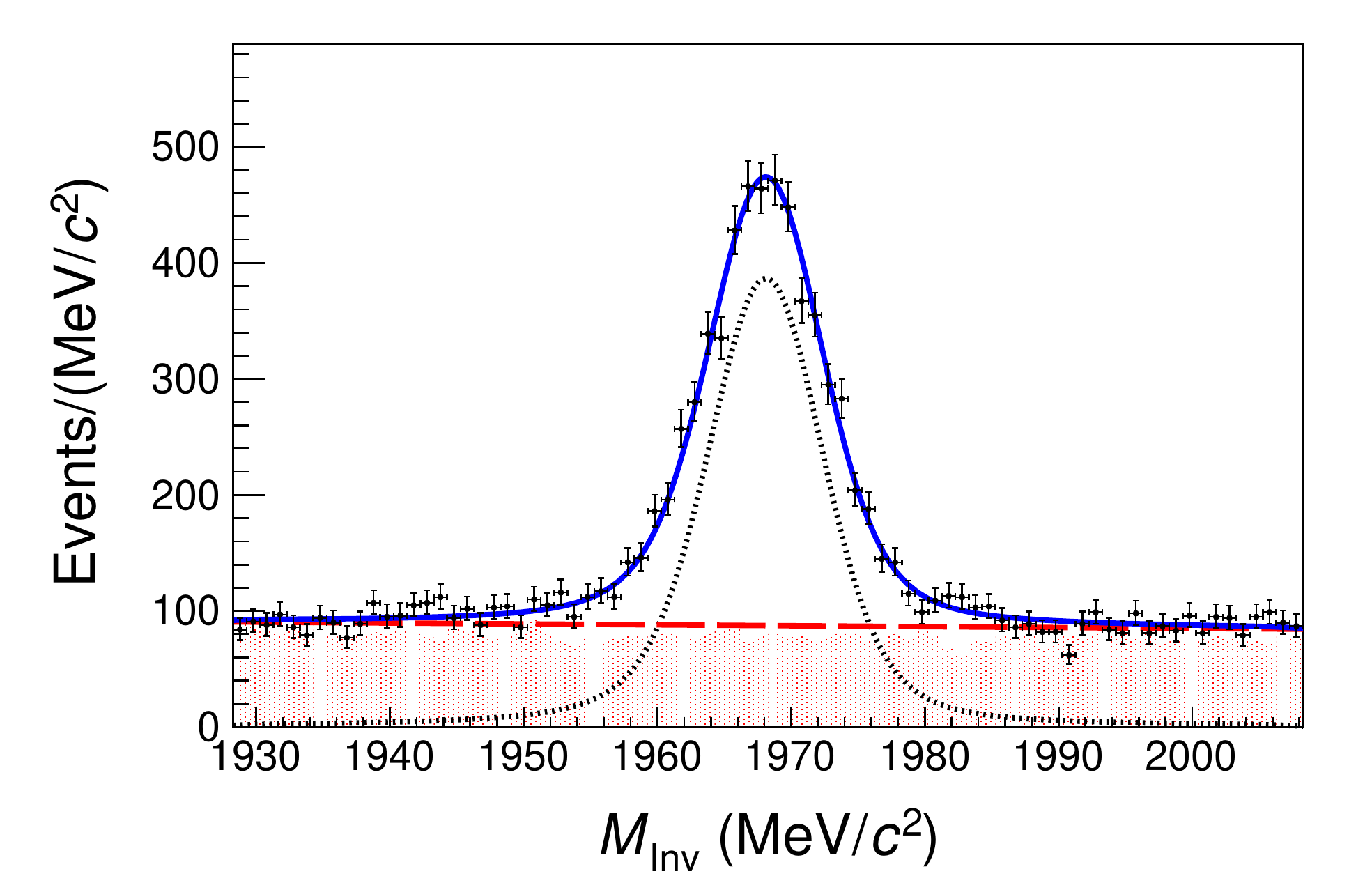} & \includegraphics[width=3in]{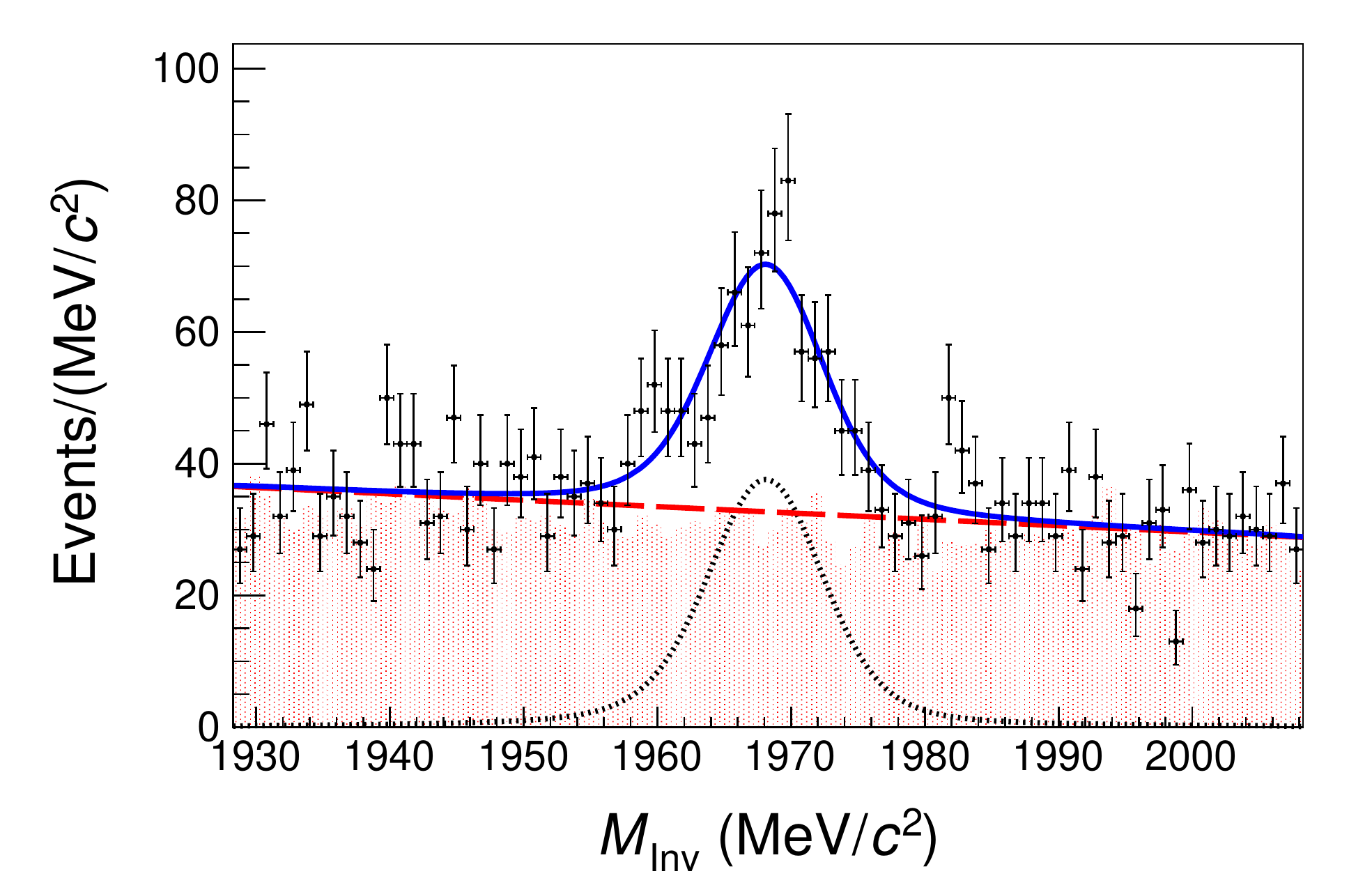}\\
\textbf{RS 850-900 MeV/$c$} & \textbf{WS 850-900 MeV/$c$}\\
\includegraphics[width=3in]{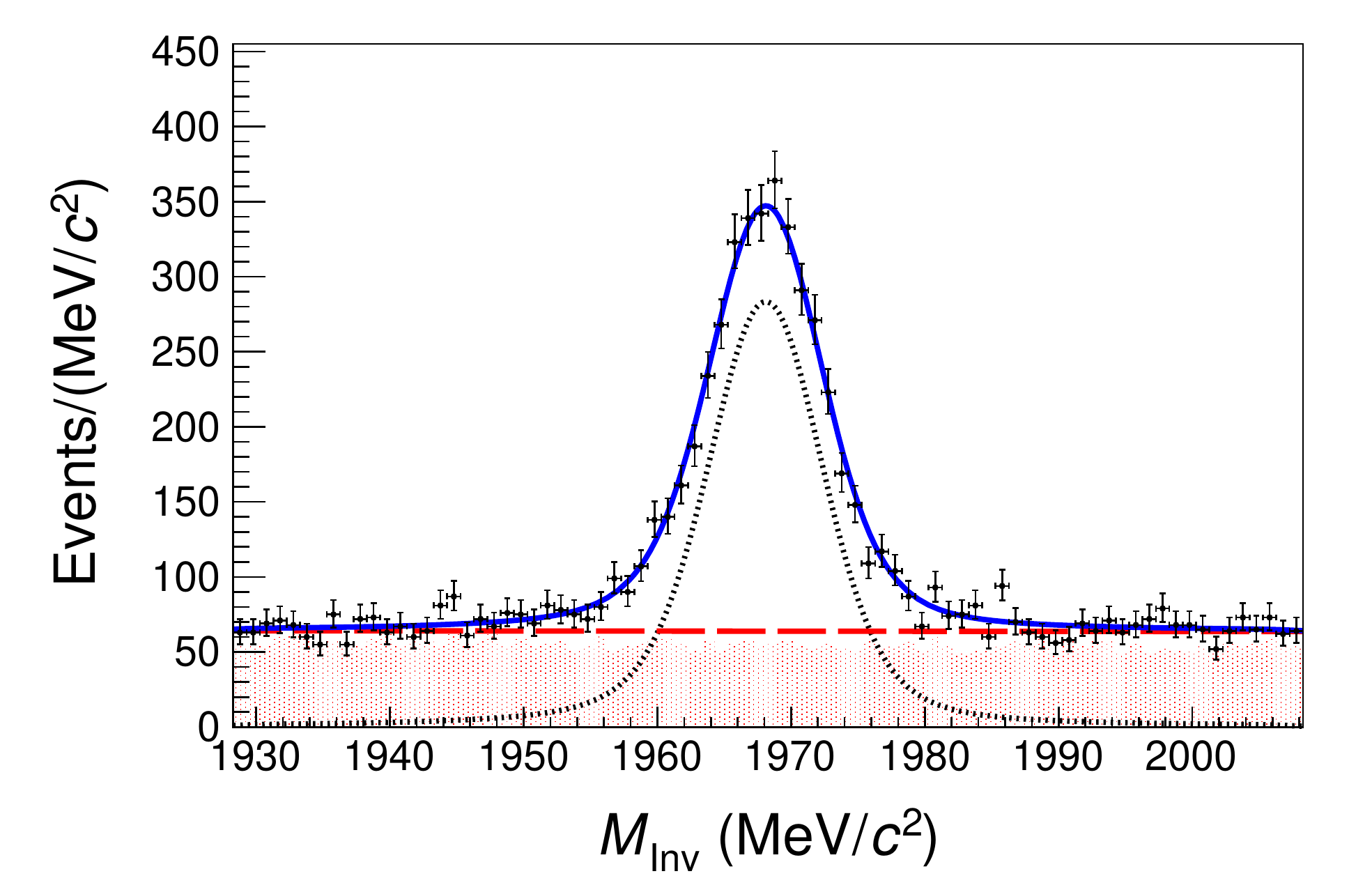} & \includegraphics[width=3in]{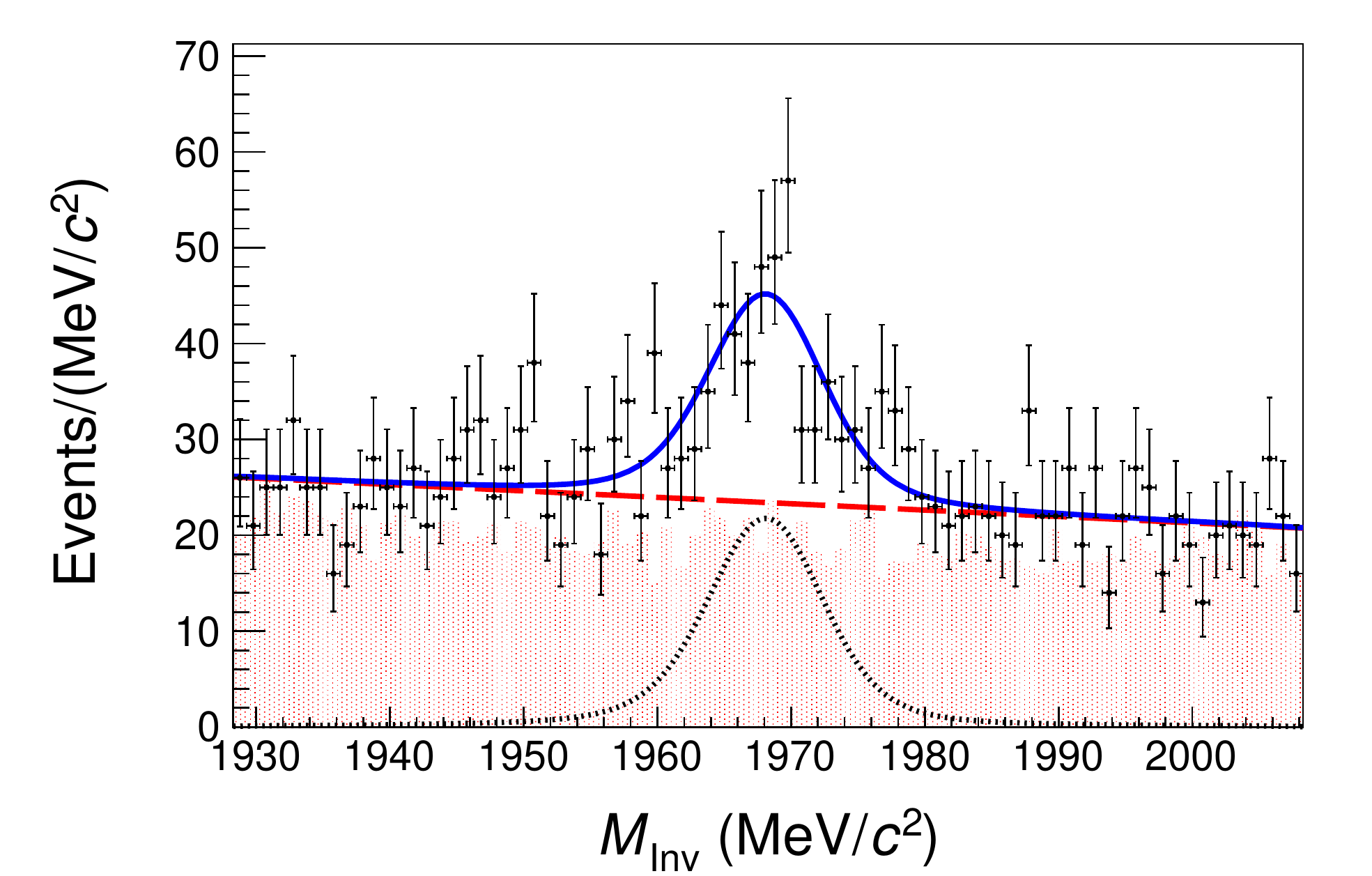}\\
\textbf{RS 900-950 MeV/$c$} & \textbf{WS 900-950 MeV/$c$}\\
\includegraphics[width=3in]{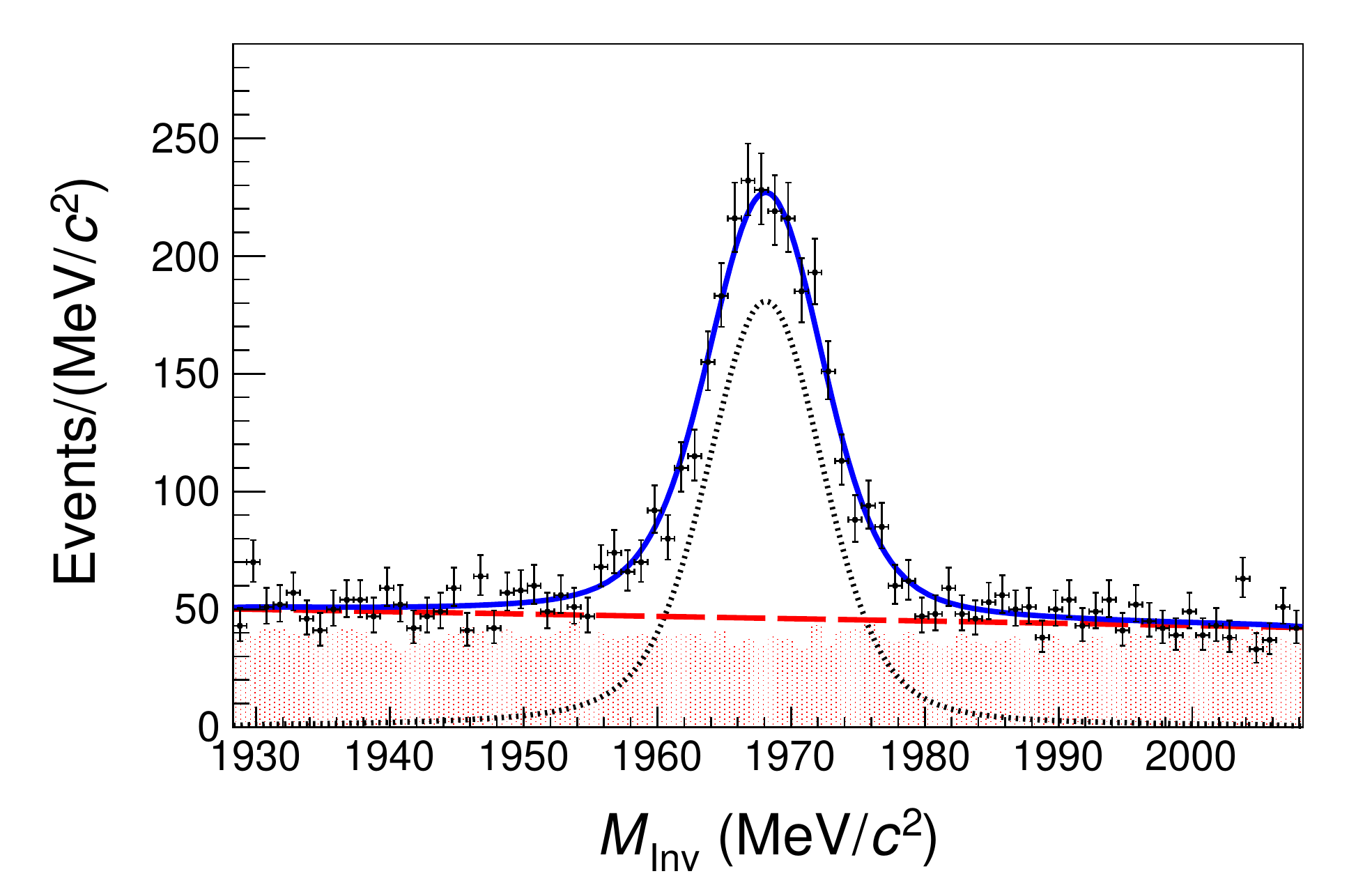} & \includegraphics[width=3in]{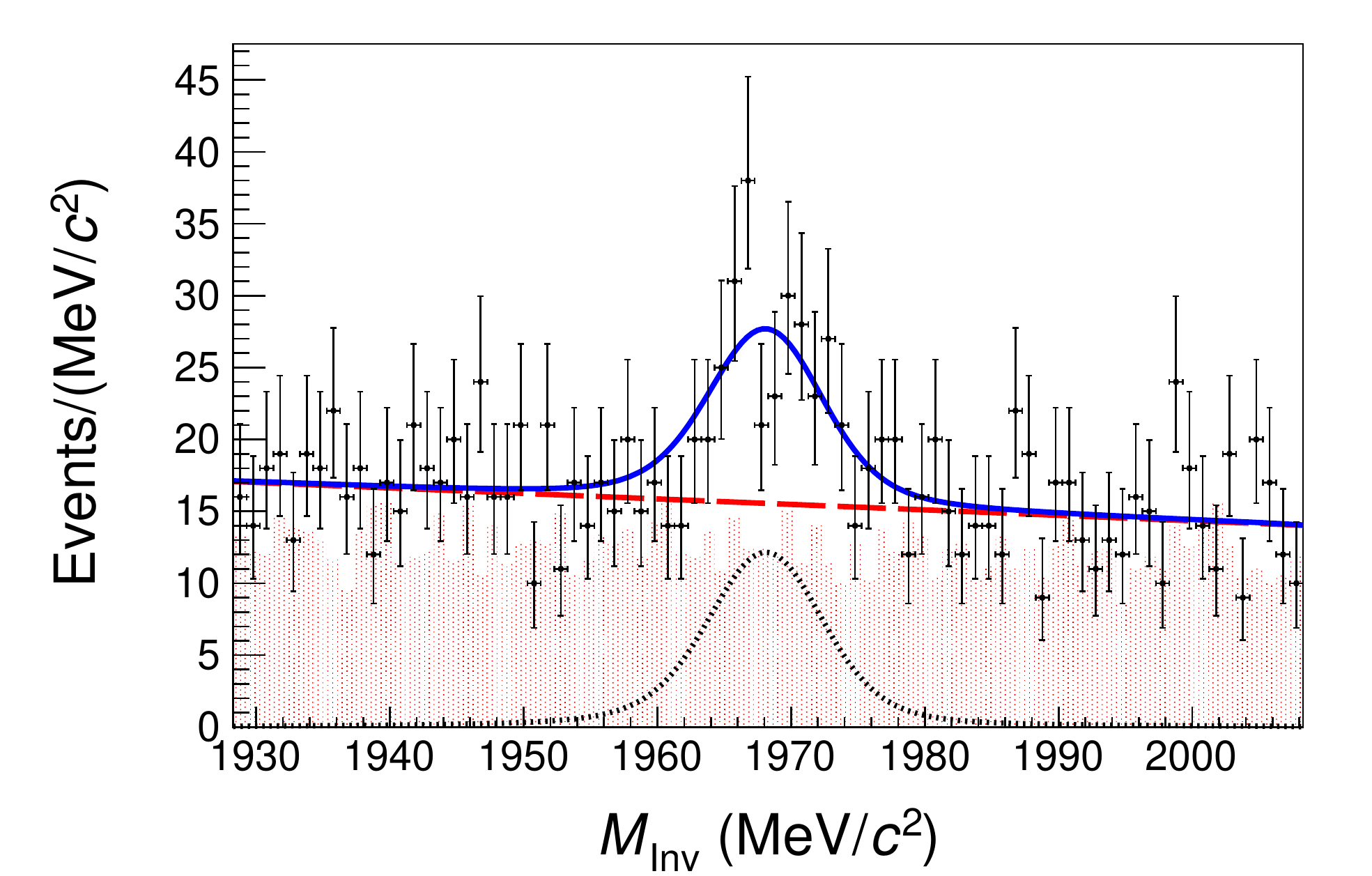}\\
\textbf{RS 950-1000 MeV/$c$} & \textbf{WS 950-1000 MeV/$c$}\\
\includegraphics[width=3in]{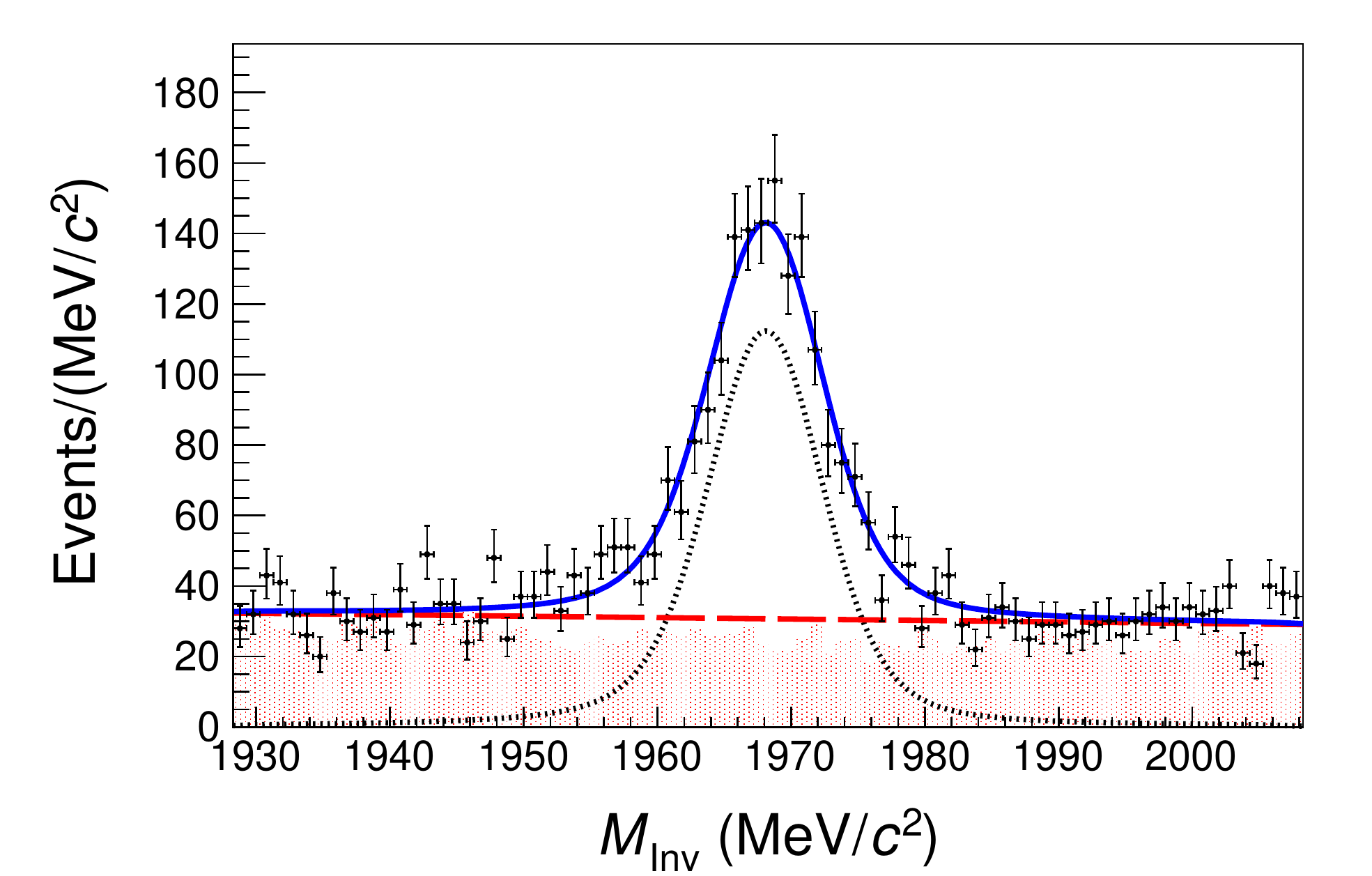} & \includegraphics[width=3in]{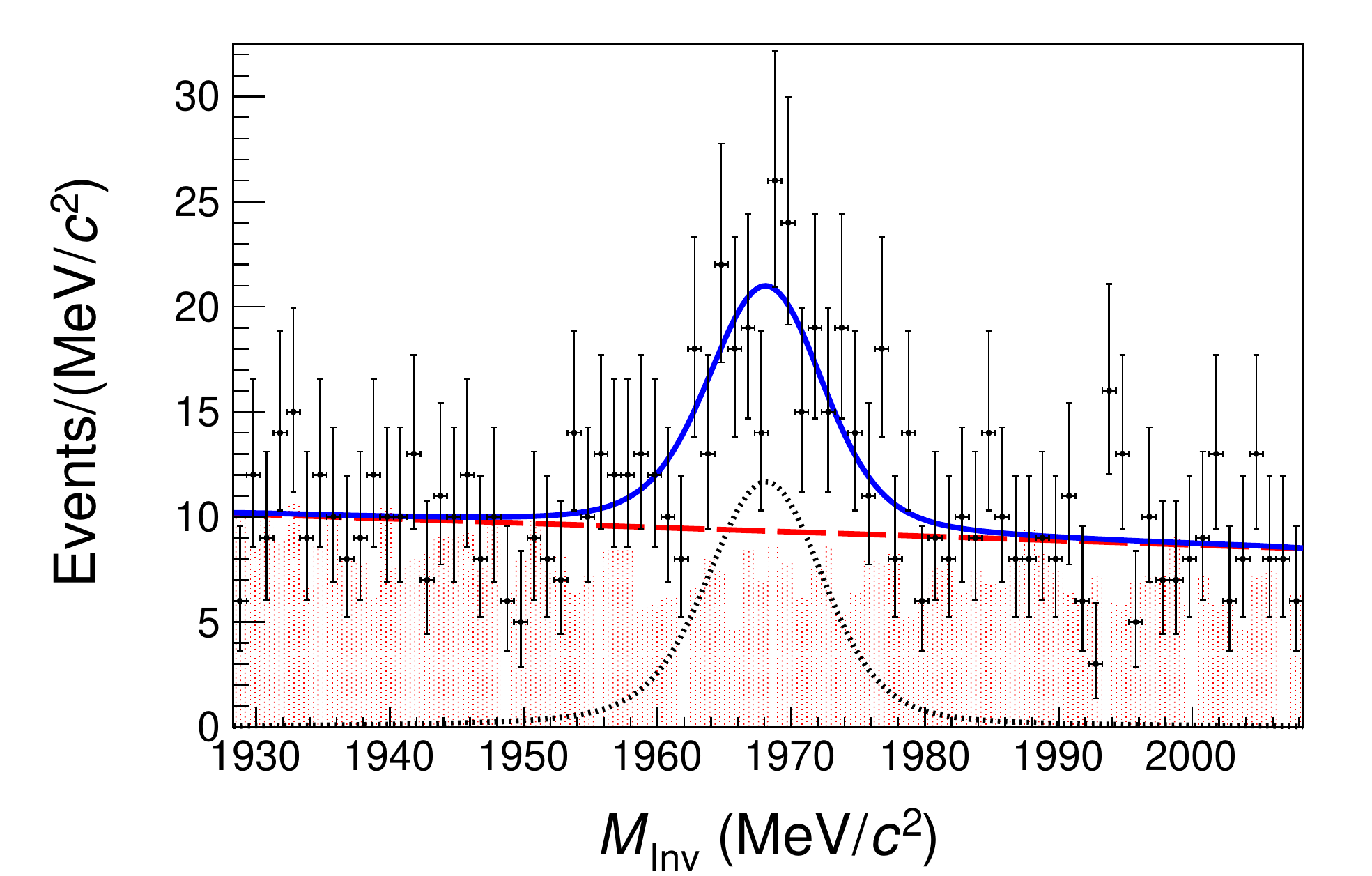}\\
\end{tabular}

\begin{tabular}{cc}
\textbf{RS 1000-1050 MeV/$c$} & \textbf{WS 1000-1050 MeV/$c$}\\
\includegraphics[width=3in]{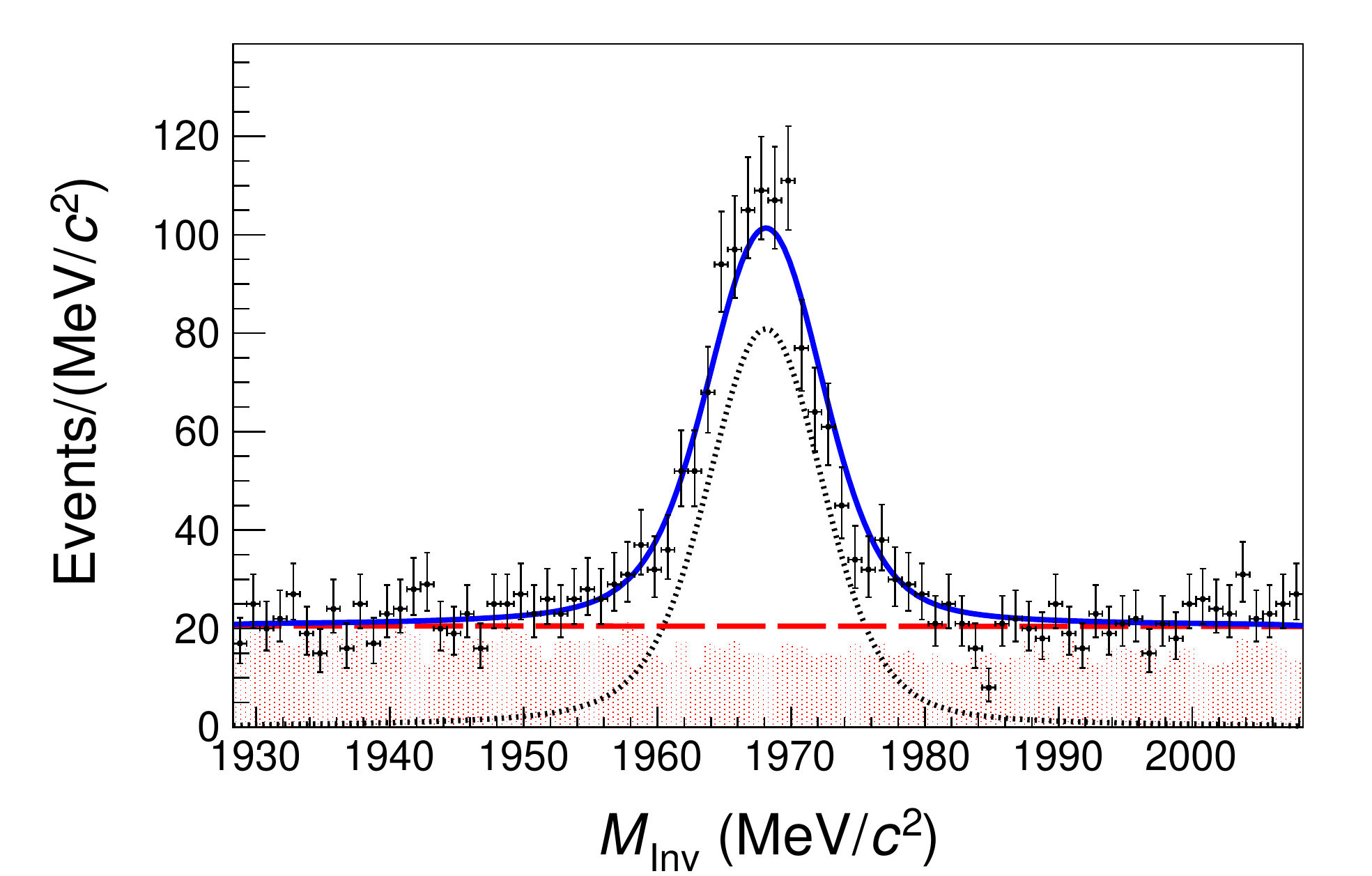} & \includegraphics[width=3in]{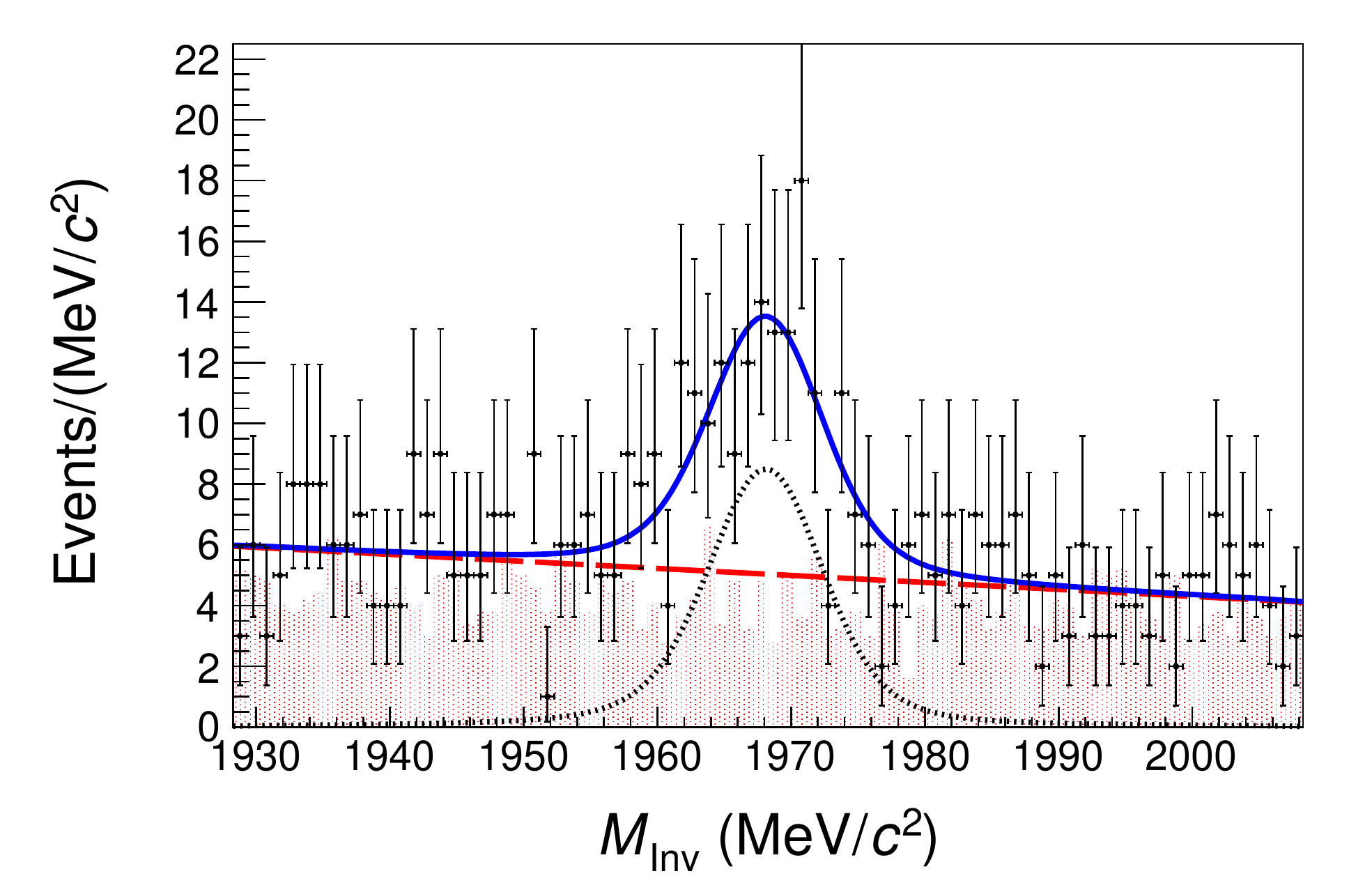}\\
\textbf{RS 1050-1100 MeV/$c$} & \textbf{WS 1050-1100 MeV/$c$}\\
\includegraphics[width=3in]{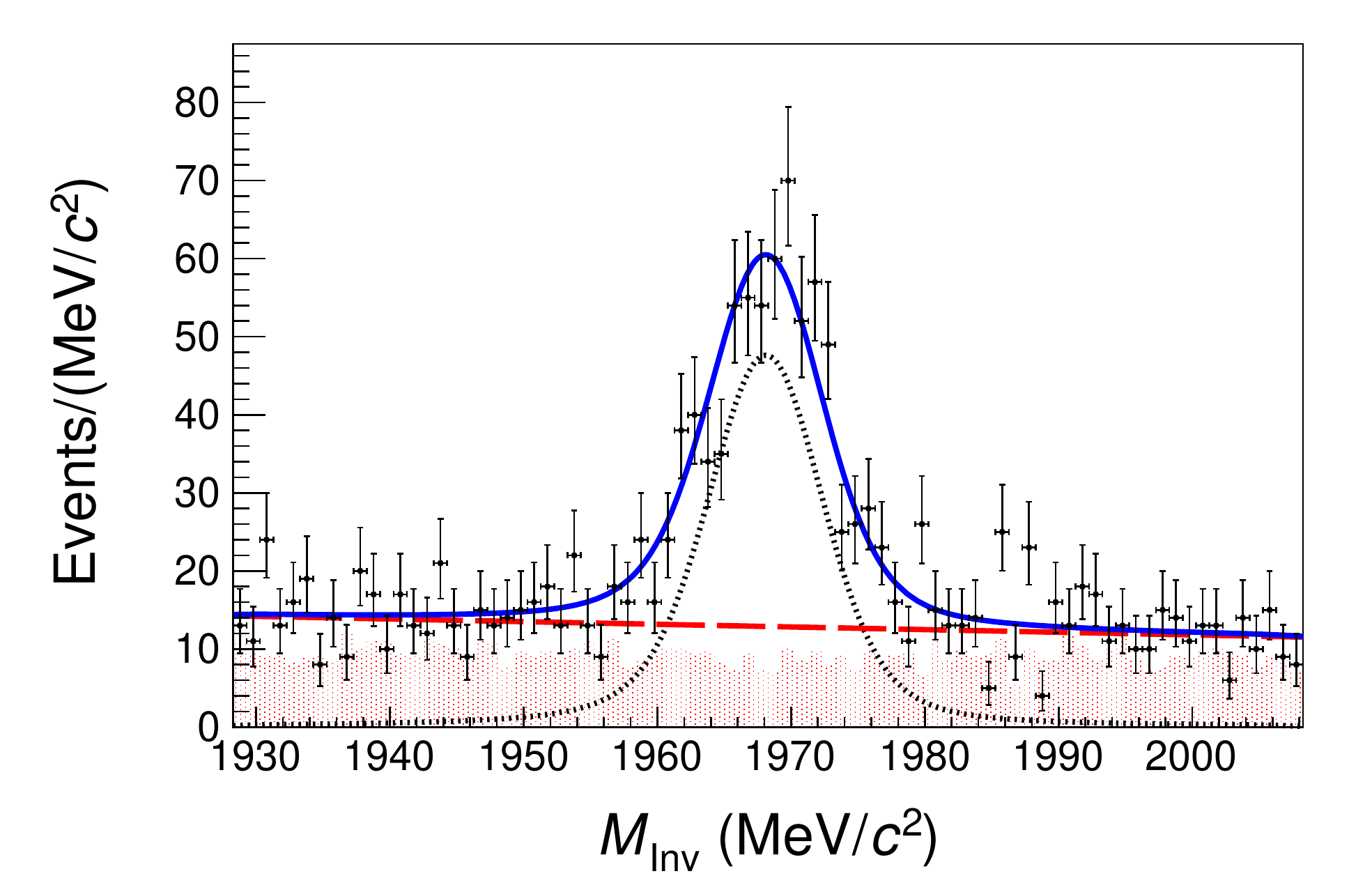} & \includegraphics[width=3in]{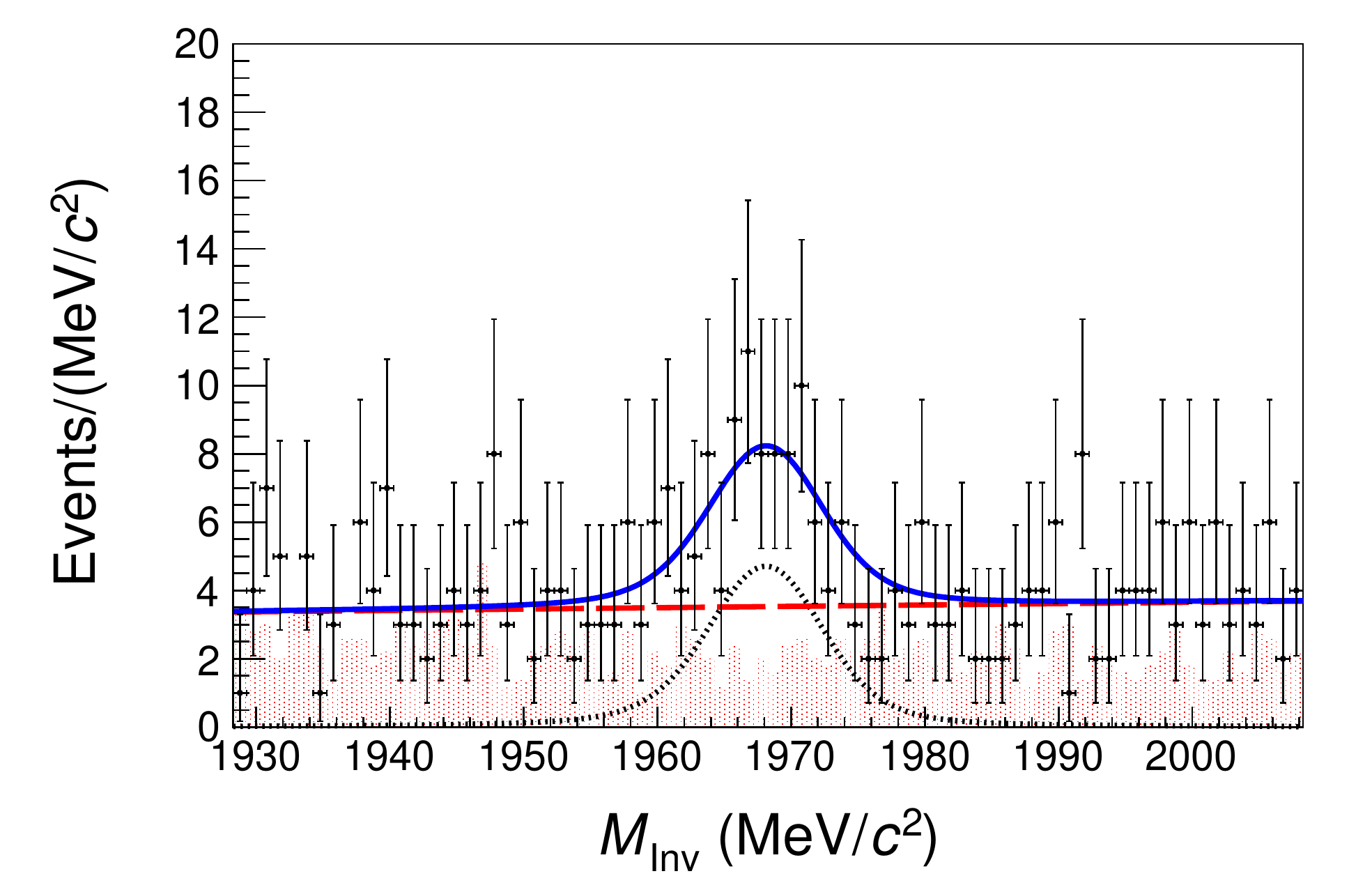}
\end{tabular}

\pagebreak
\raggedright

\subsubsection{$\EcmA$ Data $K$ ID Fits}
\label{subsubsec:4180DataKIDFits}
\centering
\begin{tabular}{cc}
\textbf{RS 200-250 MeV/$c$} & \textbf{WS 200-250 MeV/$c$}\\
\includegraphics[width=3in]{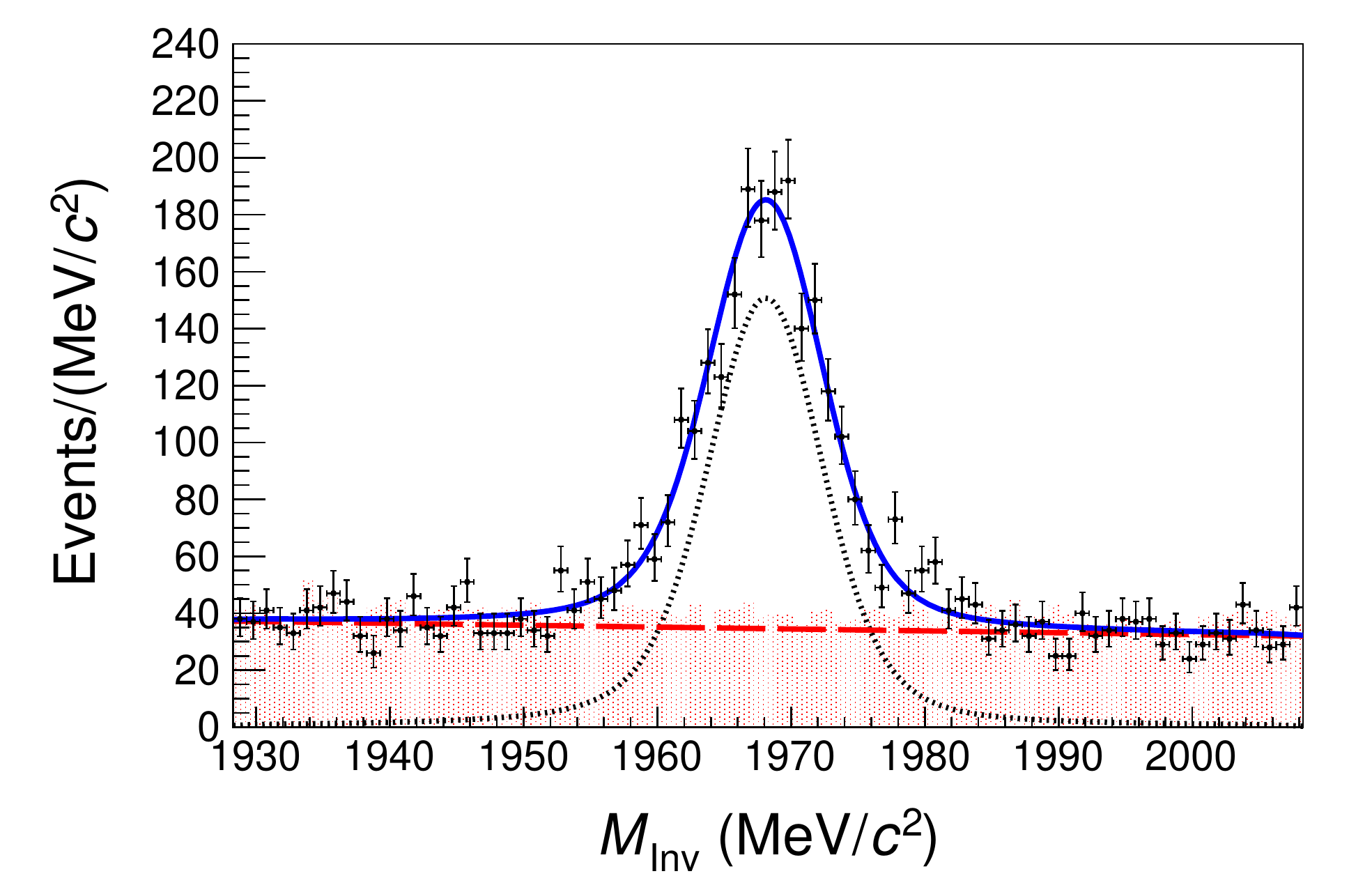} & \includegraphics[width=3in]{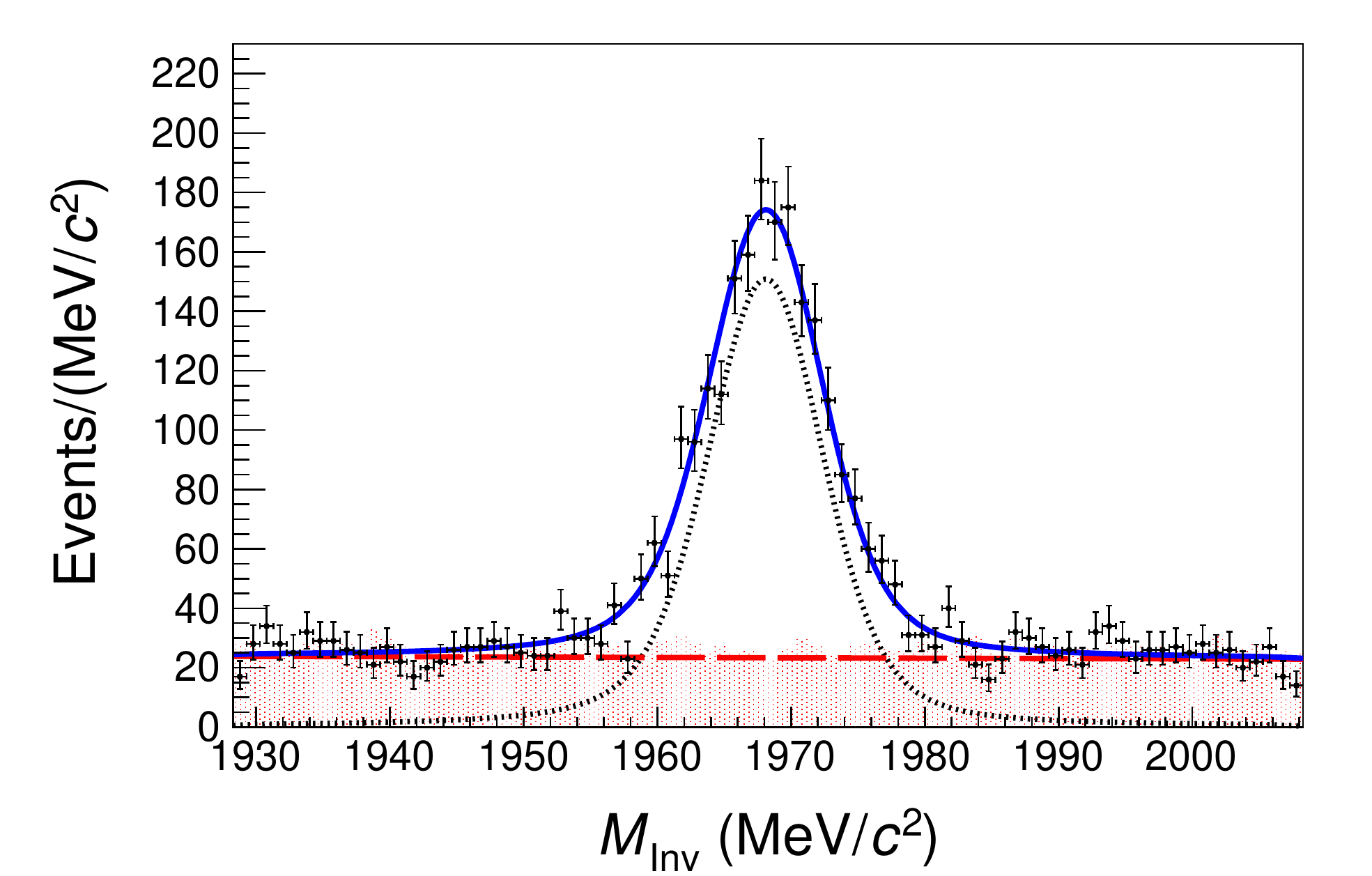}\\
\textbf{RS 250-300 MeV/$c$} & \textbf{WS 250-300 MeV/$c$}\\
\includegraphics[width=3in]{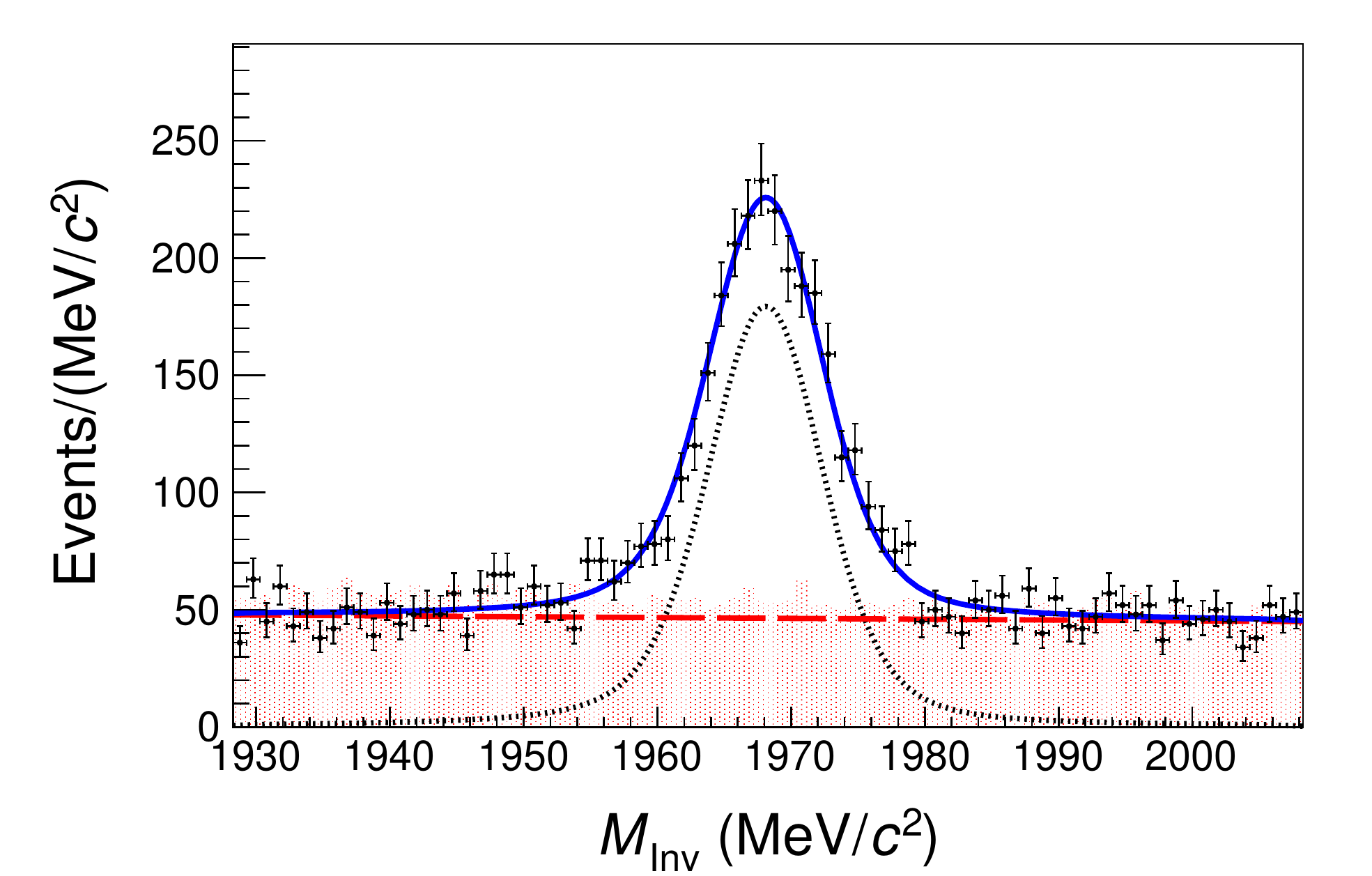} & \includegraphics[width=3in]{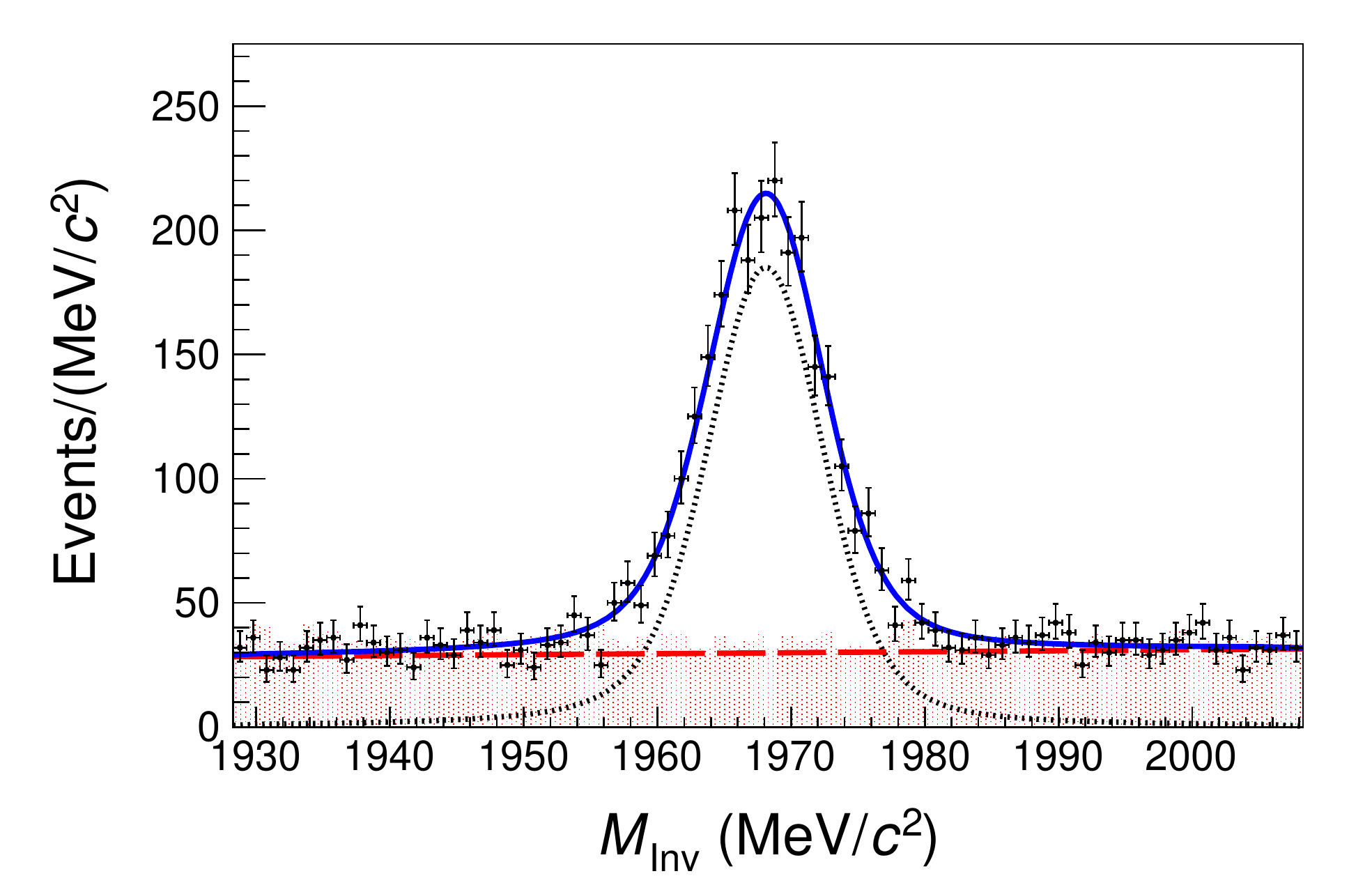}\\
\textbf{RS 300-350 MeV/$c$} & \textbf{WS 300-350 MeV/$c$}\\
\includegraphics[width=3in]{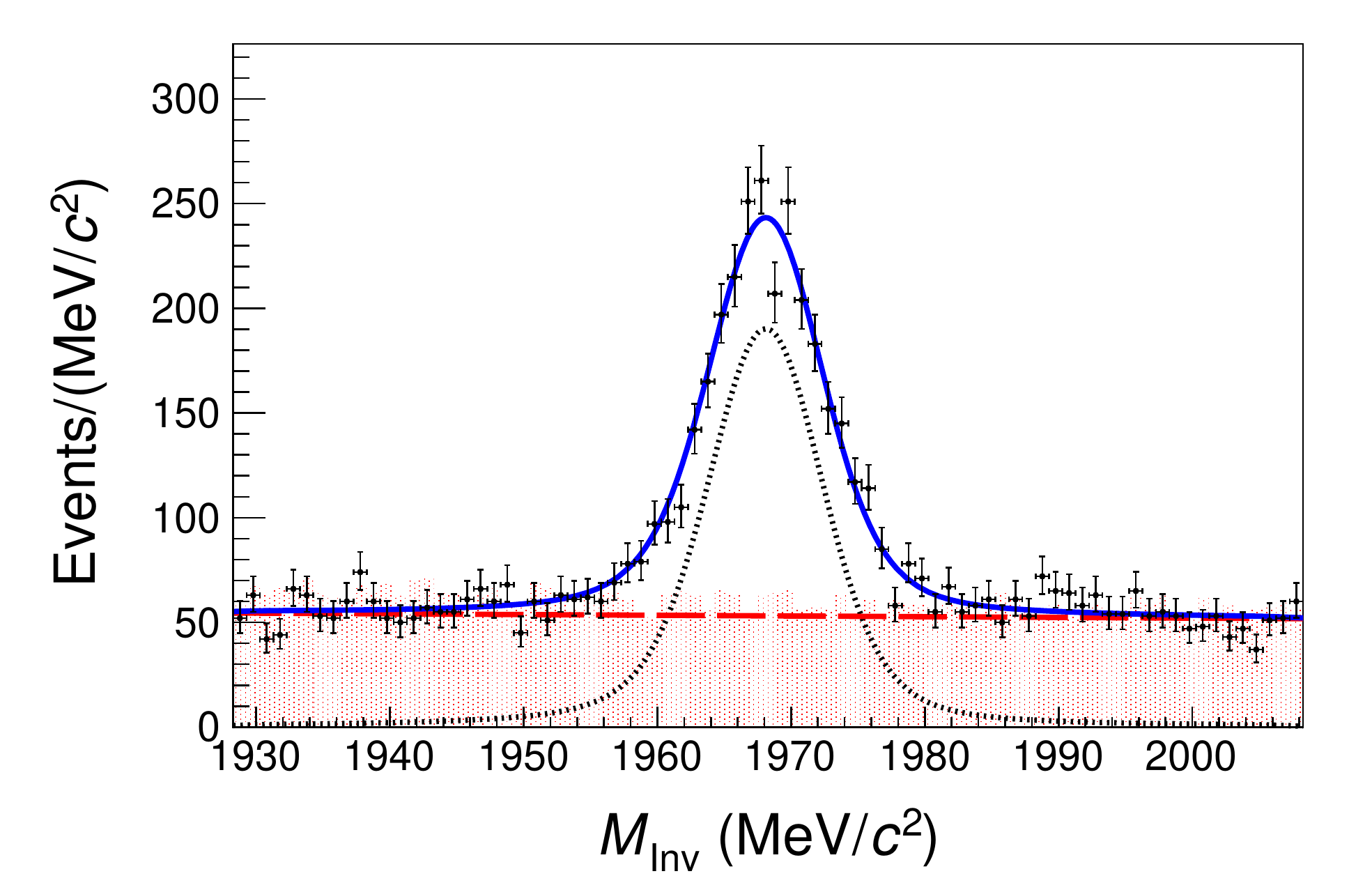} & \includegraphics[width=3in]{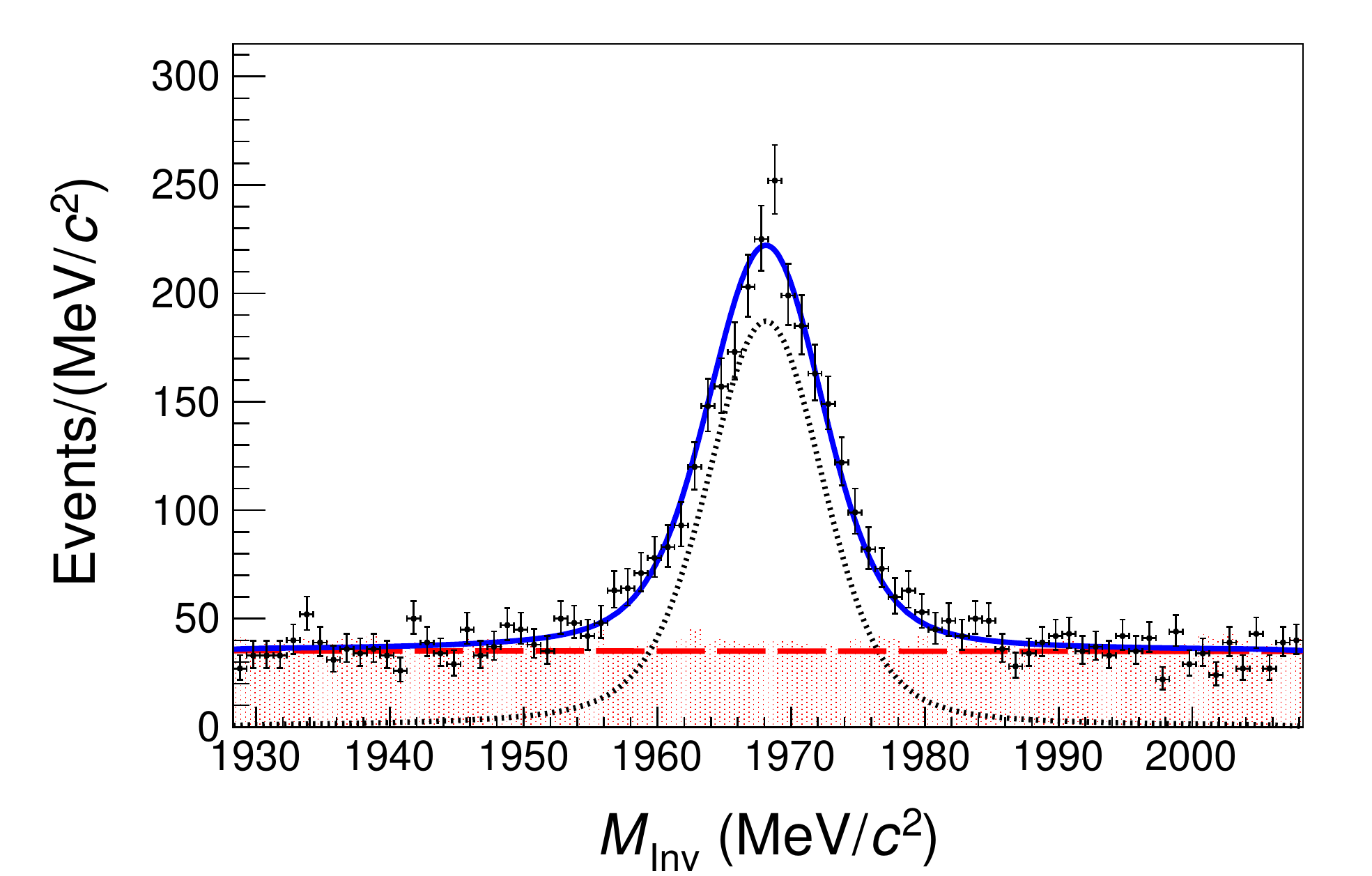}\\
\textbf{RS 350-400 MeV/$c$} & \textbf{WS 350-400 MeV/$c$}\\
\includegraphics[width=3in]{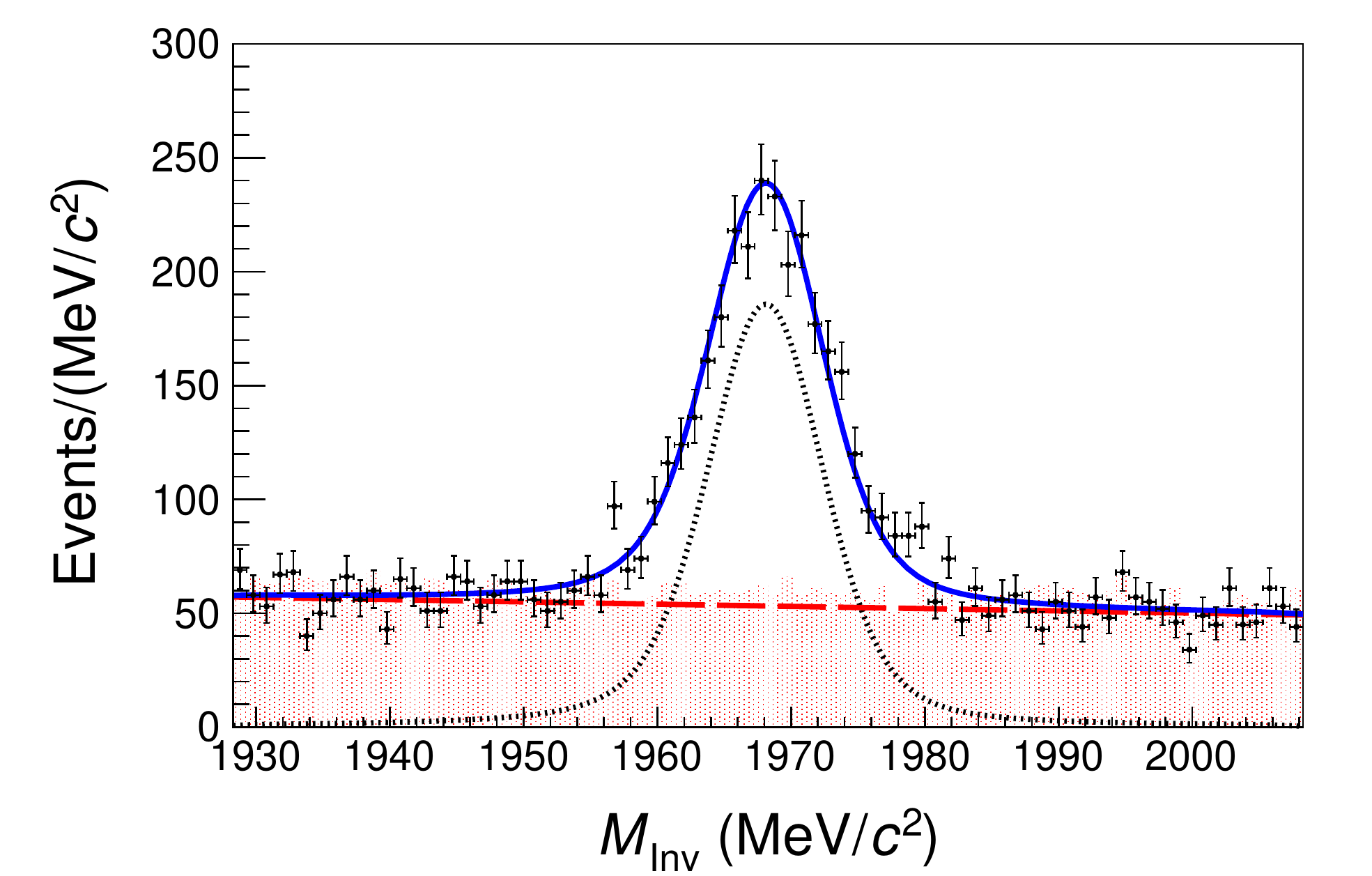} & \includegraphics[width=3in]{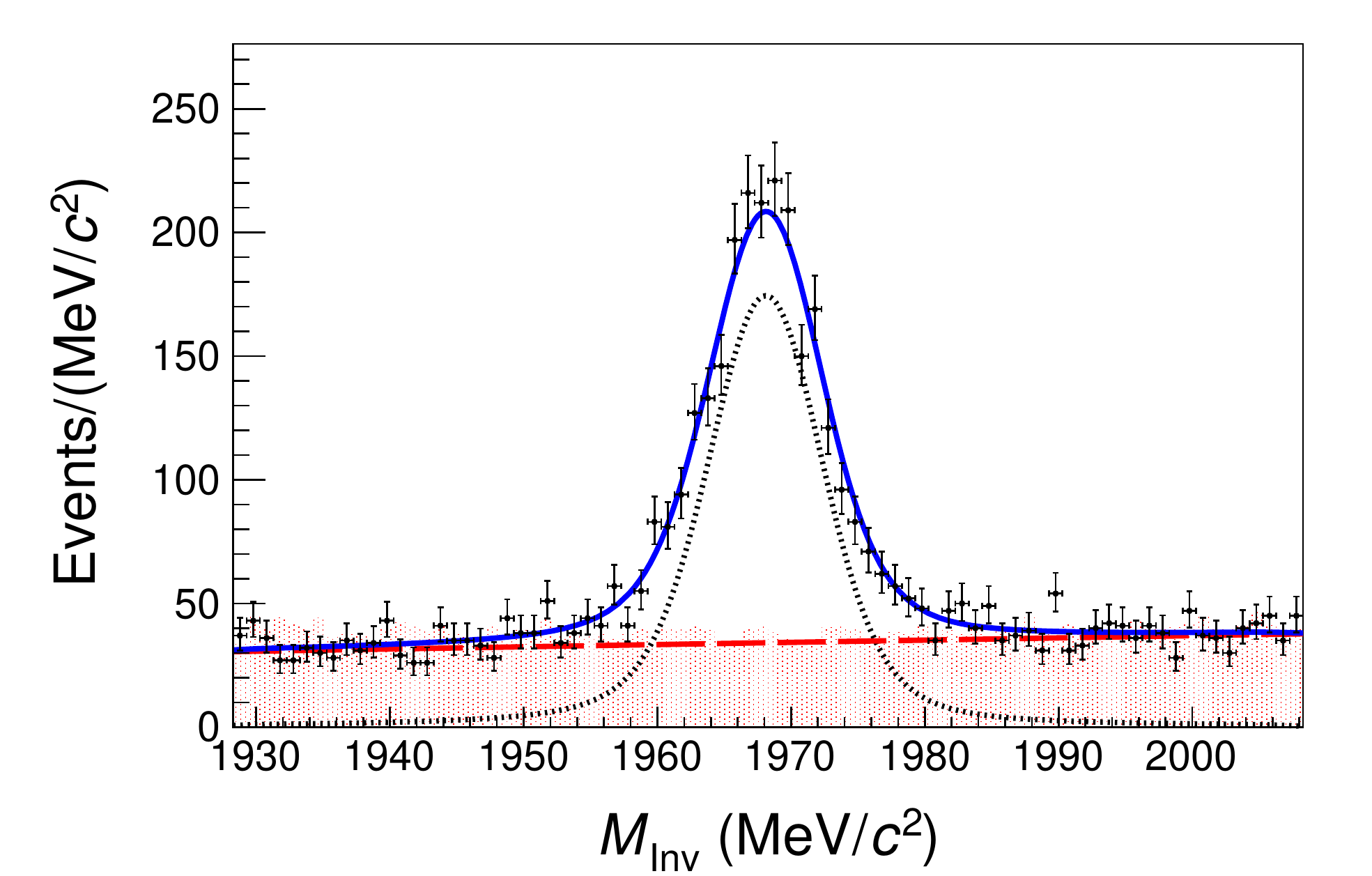}
\end{tabular}
\pagebreak

\begin{tabular}{cc}
\textbf{RS 400-450 MeV/$c$} & \textbf{WS 400-450 MeV/$c$}\\
\includegraphics[width=3in]{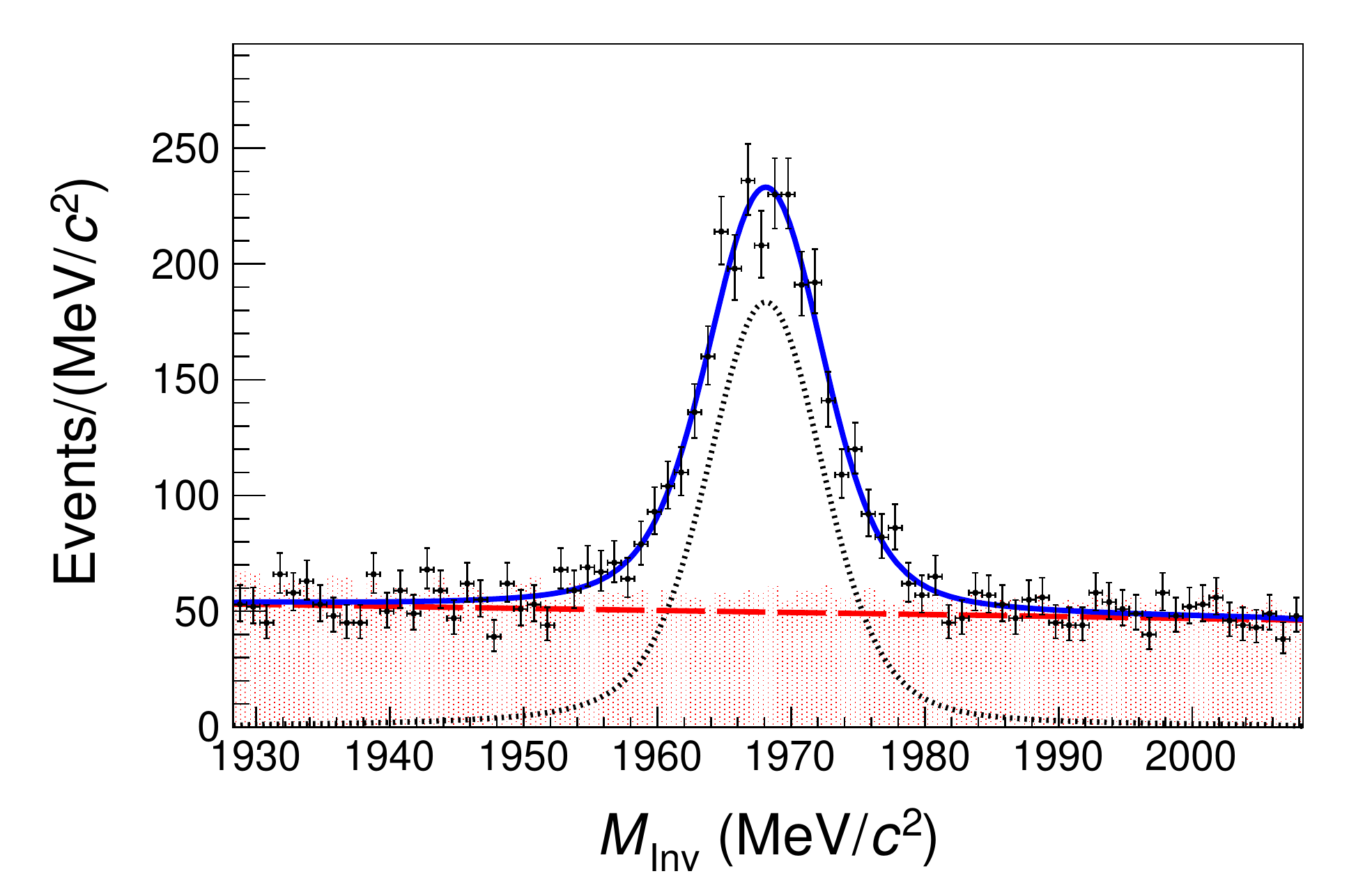} & \includegraphics[width=3in]{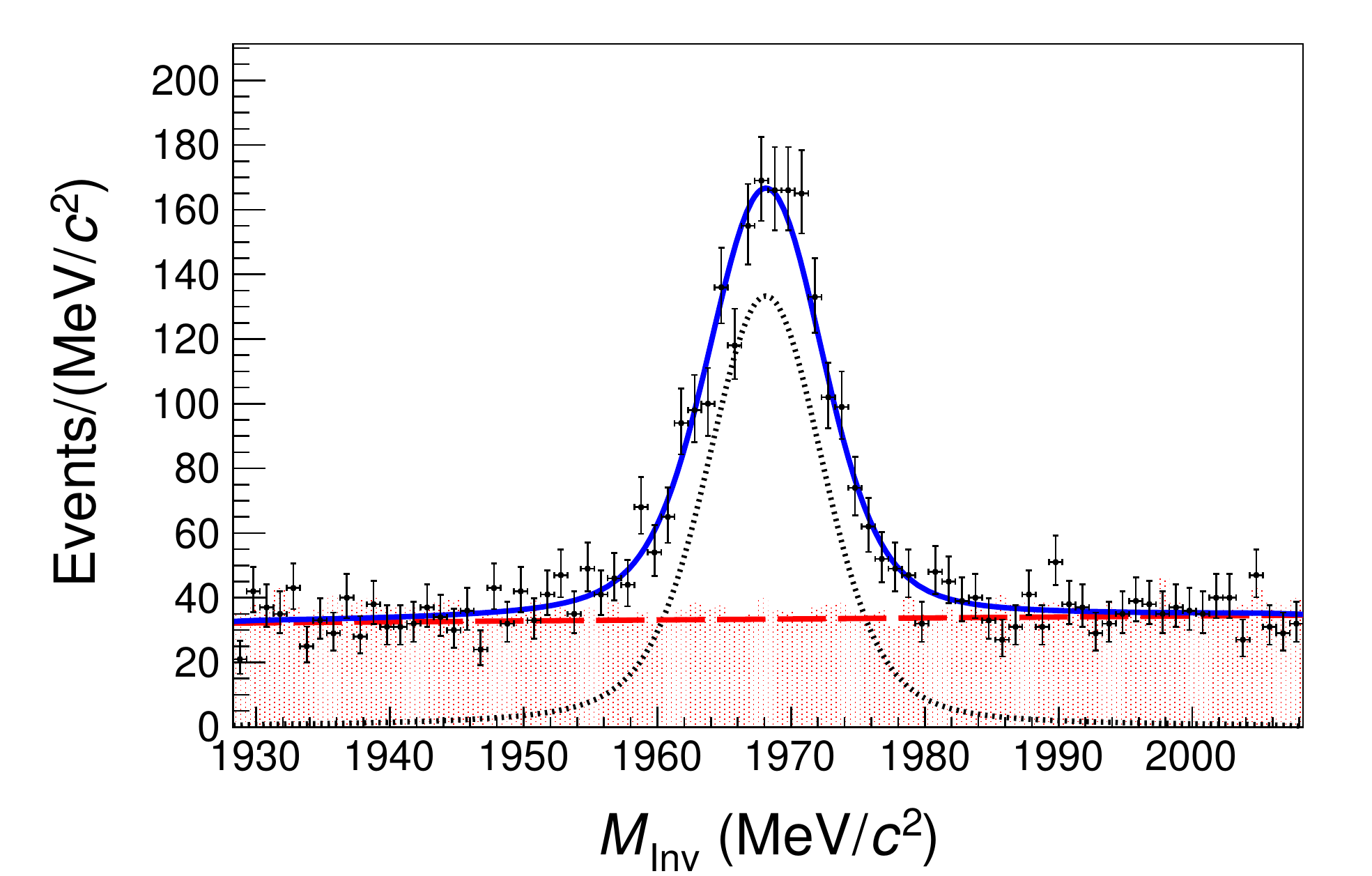}\\
\textbf{RS 450-500 MeV/$c$} & \textbf{WS 450-500 MeV/$c$}\\
\includegraphics[width=3in]{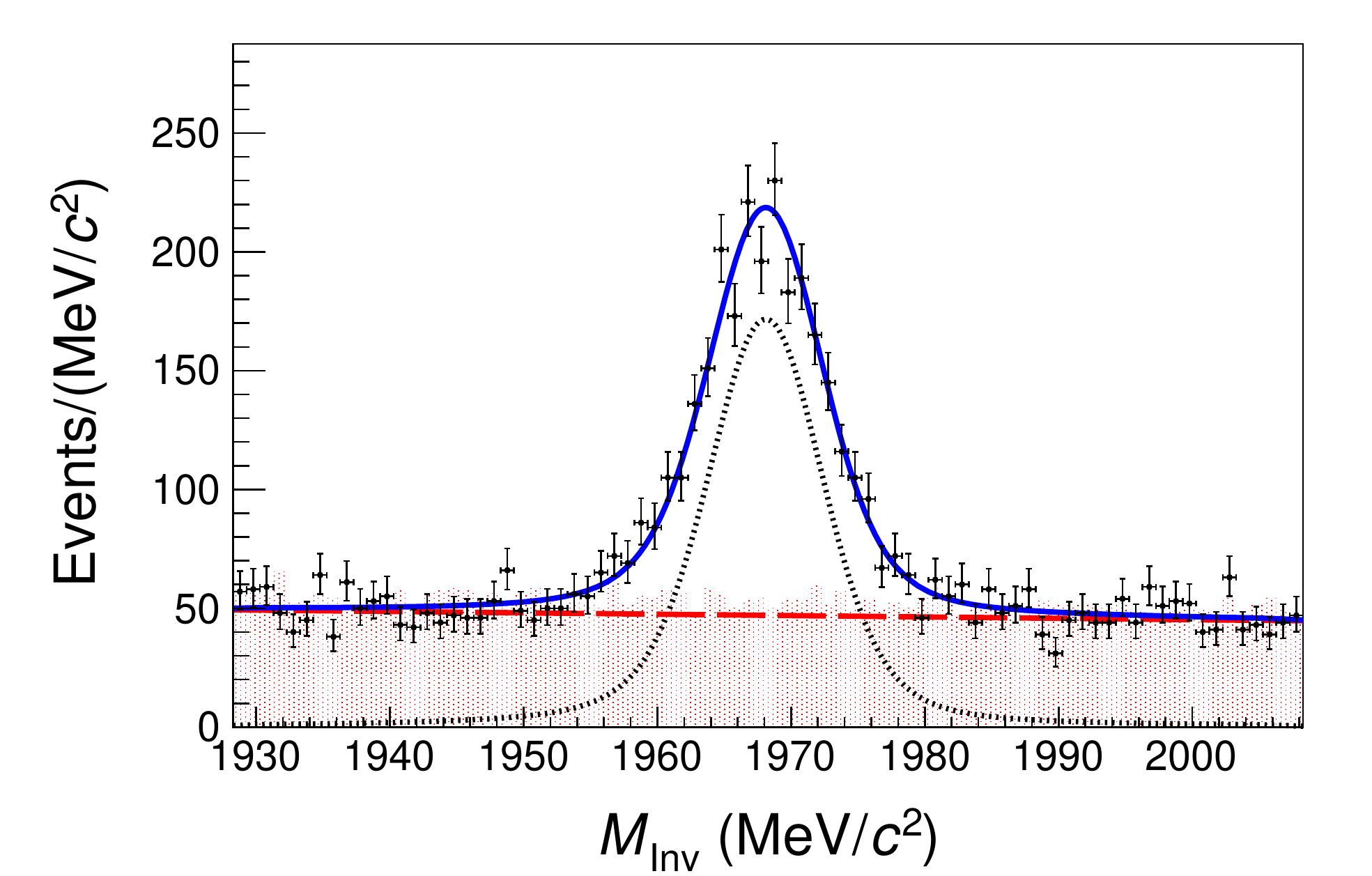} & \includegraphics[width=3in]{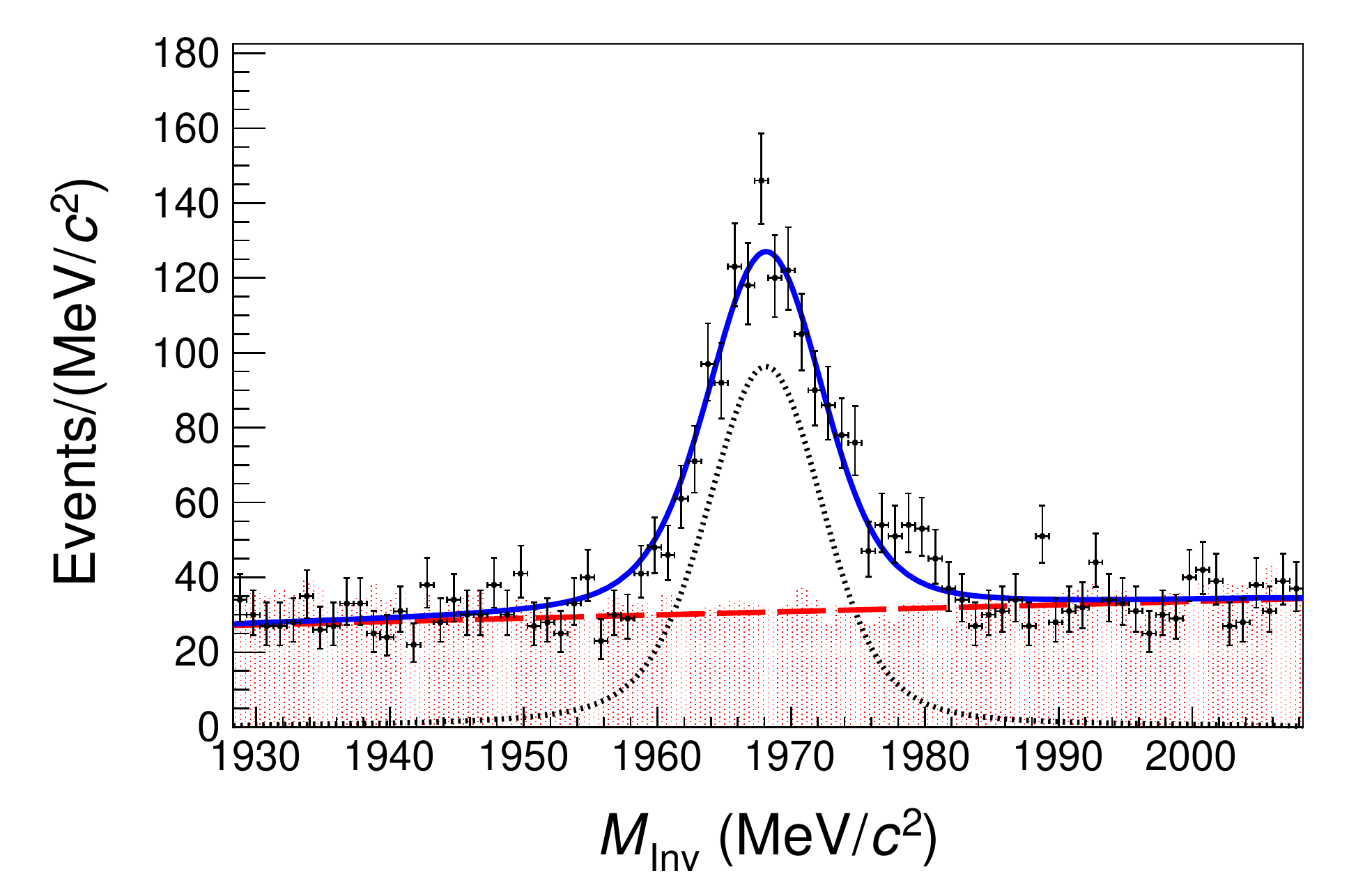}\\
\textbf{RS 500-550 MeV/$c$} & \textbf{WS 500-550 MeV/$c$}\\
\includegraphics[width=3in]{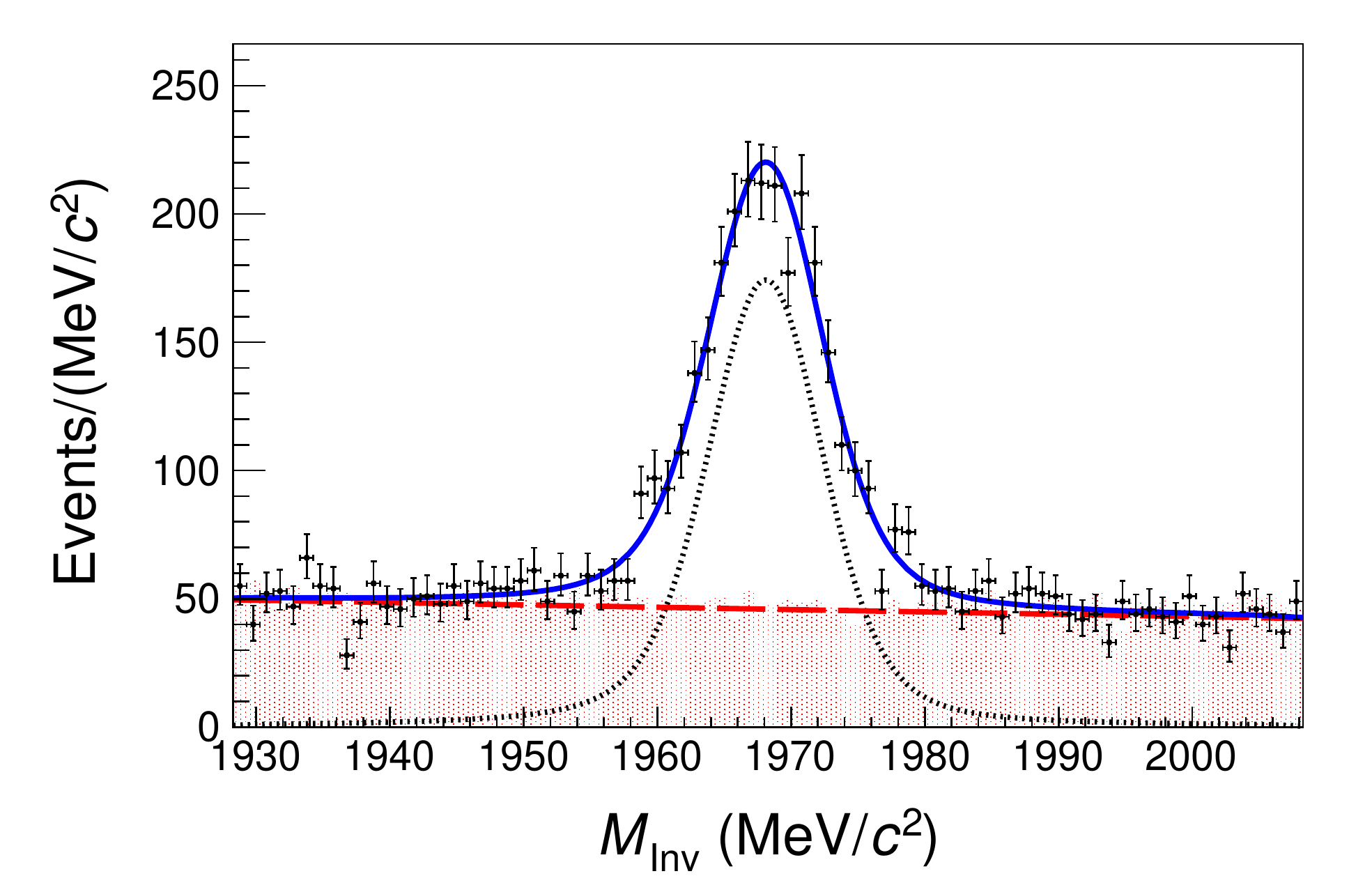} & \includegraphics[width=3in]{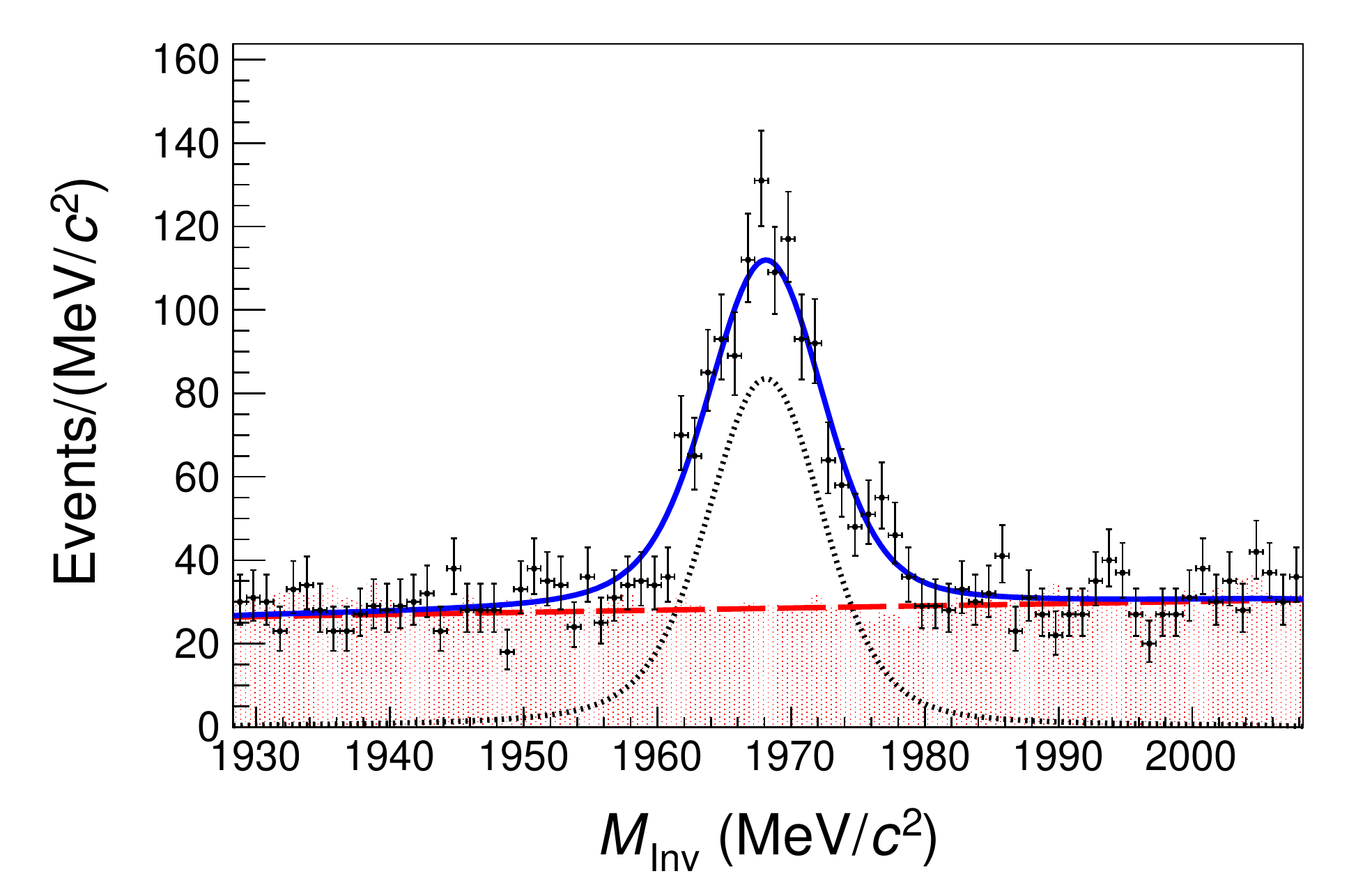}\\
\textbf{RS 550-600 MeV/$c$} & \textbf{WS 550-600 MeV/$c$}\\
\includegraphics[width=3in]{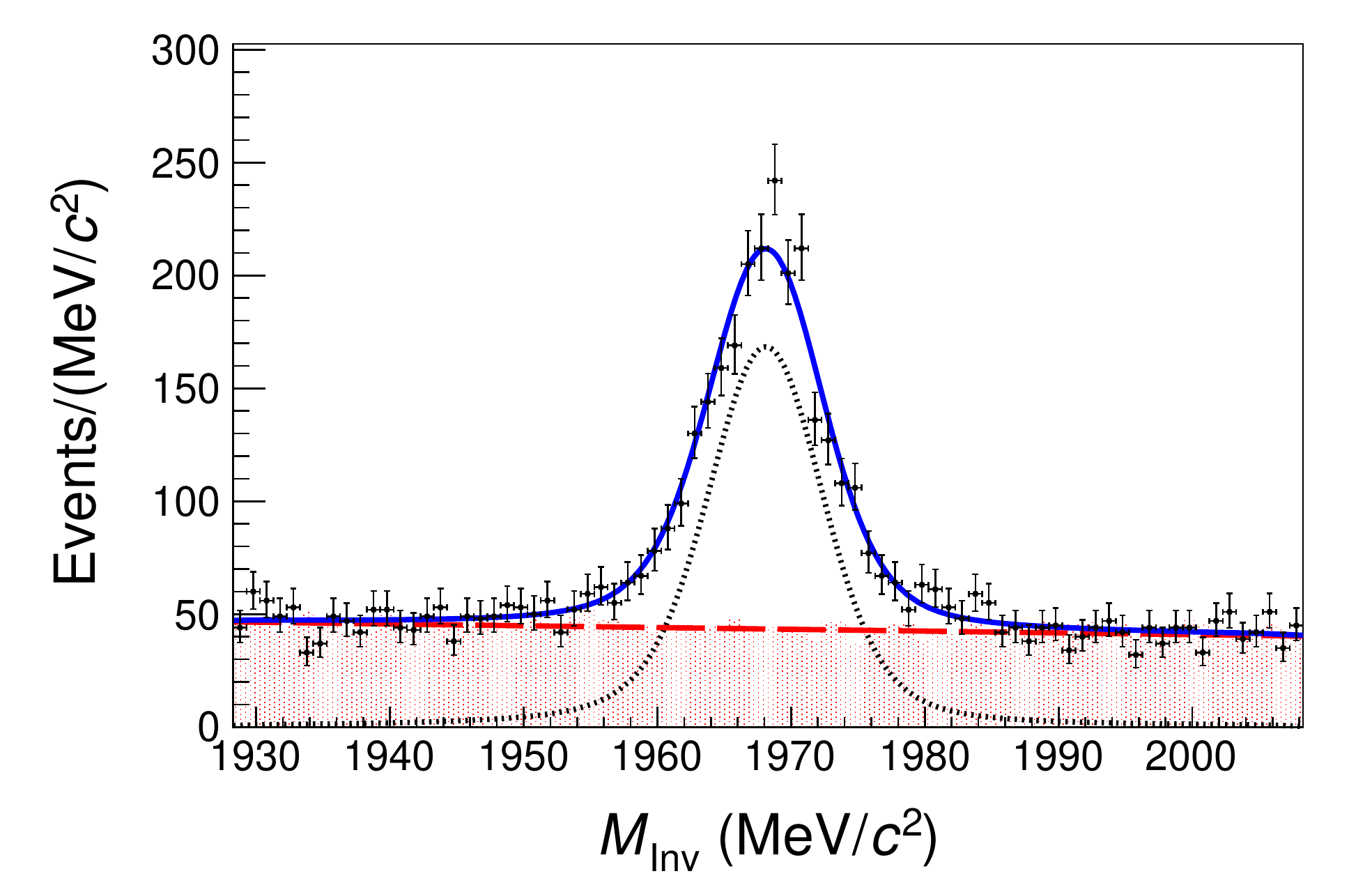} & \includegraphics[width=3in]{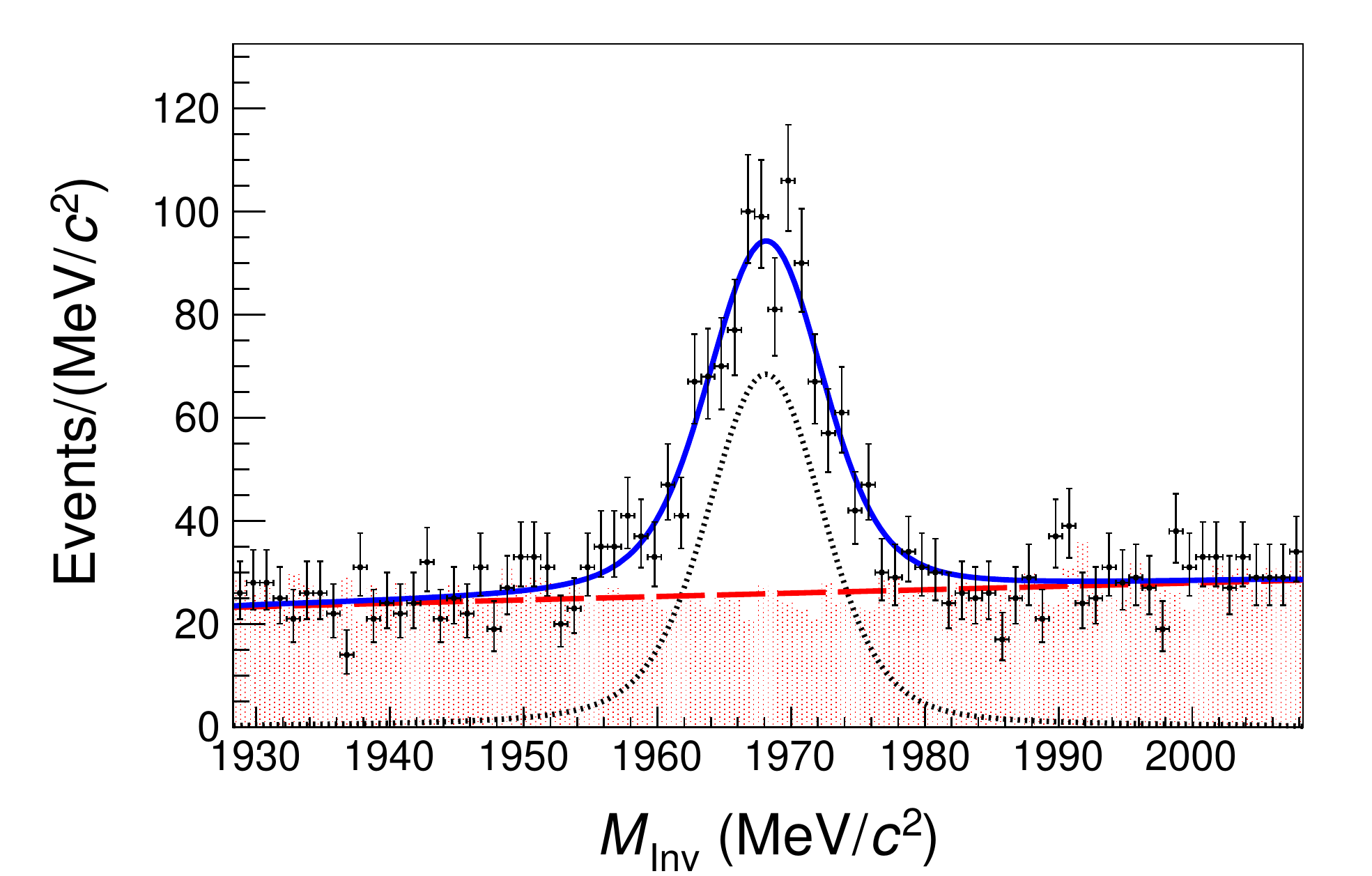}
\end{tabular}

\begin{tabular}{cc}
\textbf{RS 600-650 MeV/$c$} & \textbf{WS 600-650 MeV/$c$}\\
\includegraphics[width=3in]{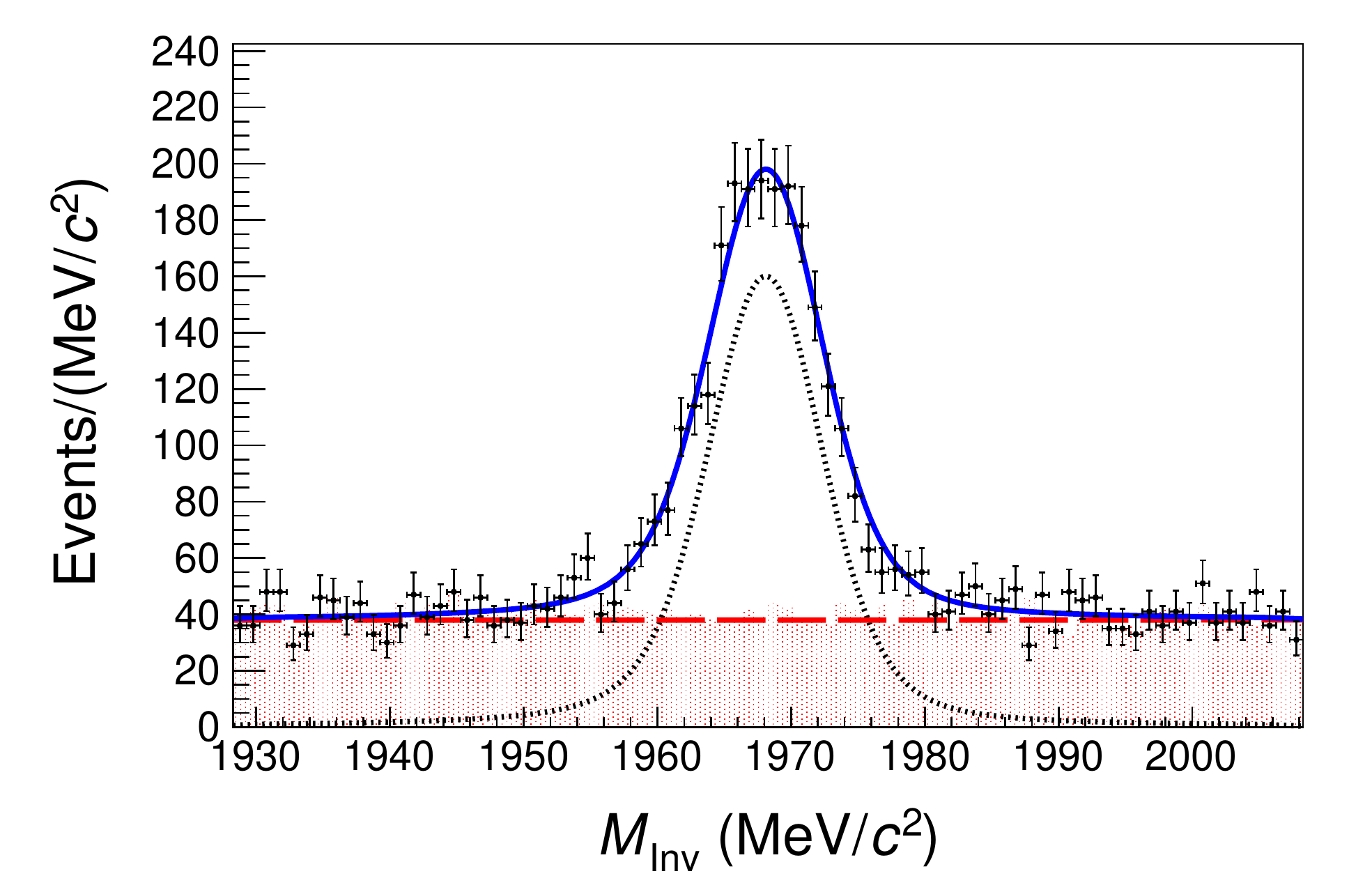} & \includegraphics[width=3in]{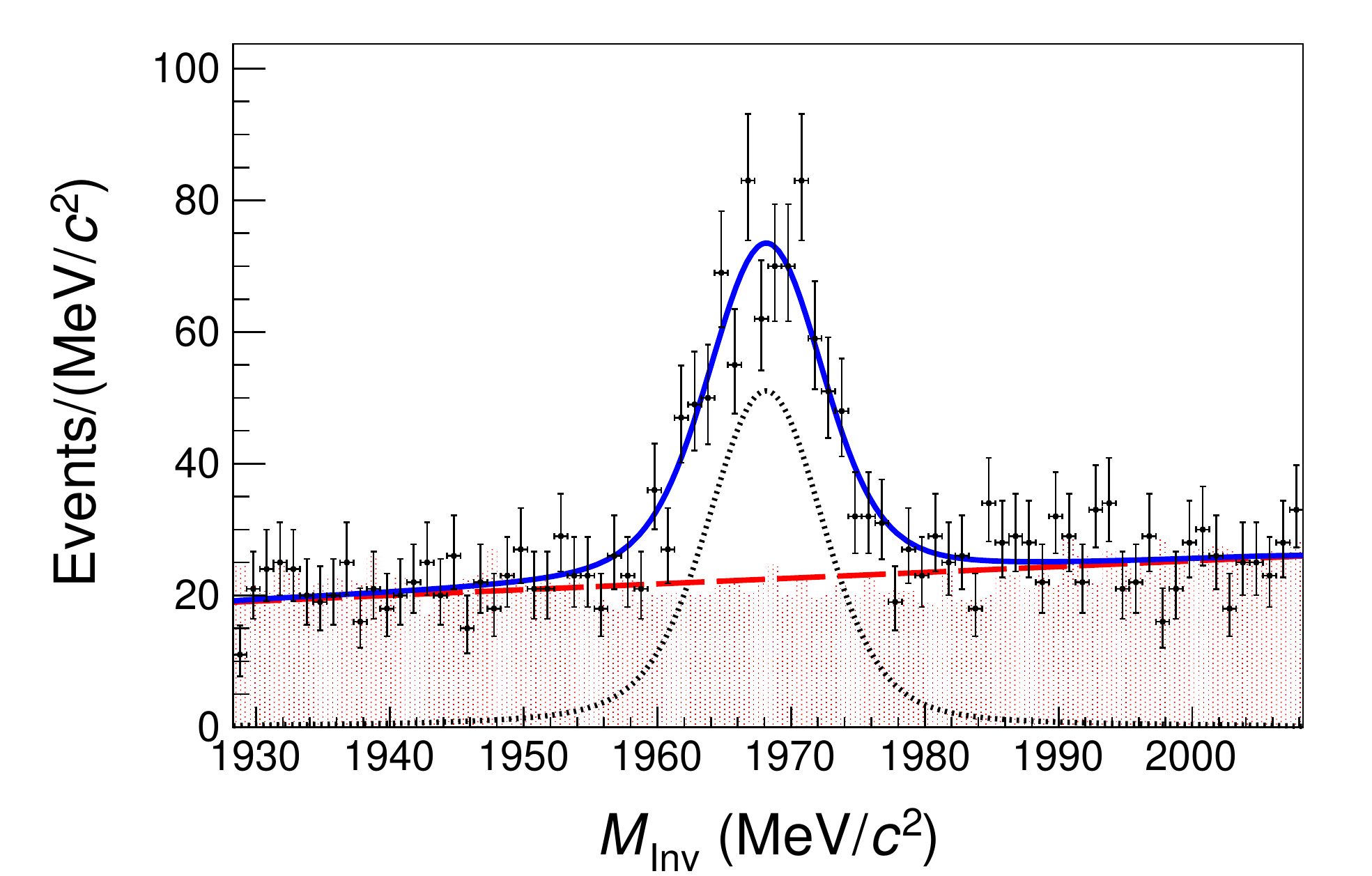}\\
\textbf{RS 650-700 MeV/$c$} & \textbf{WS 650-700 MeV/$c$}\\
\includegraphics[width=3in]{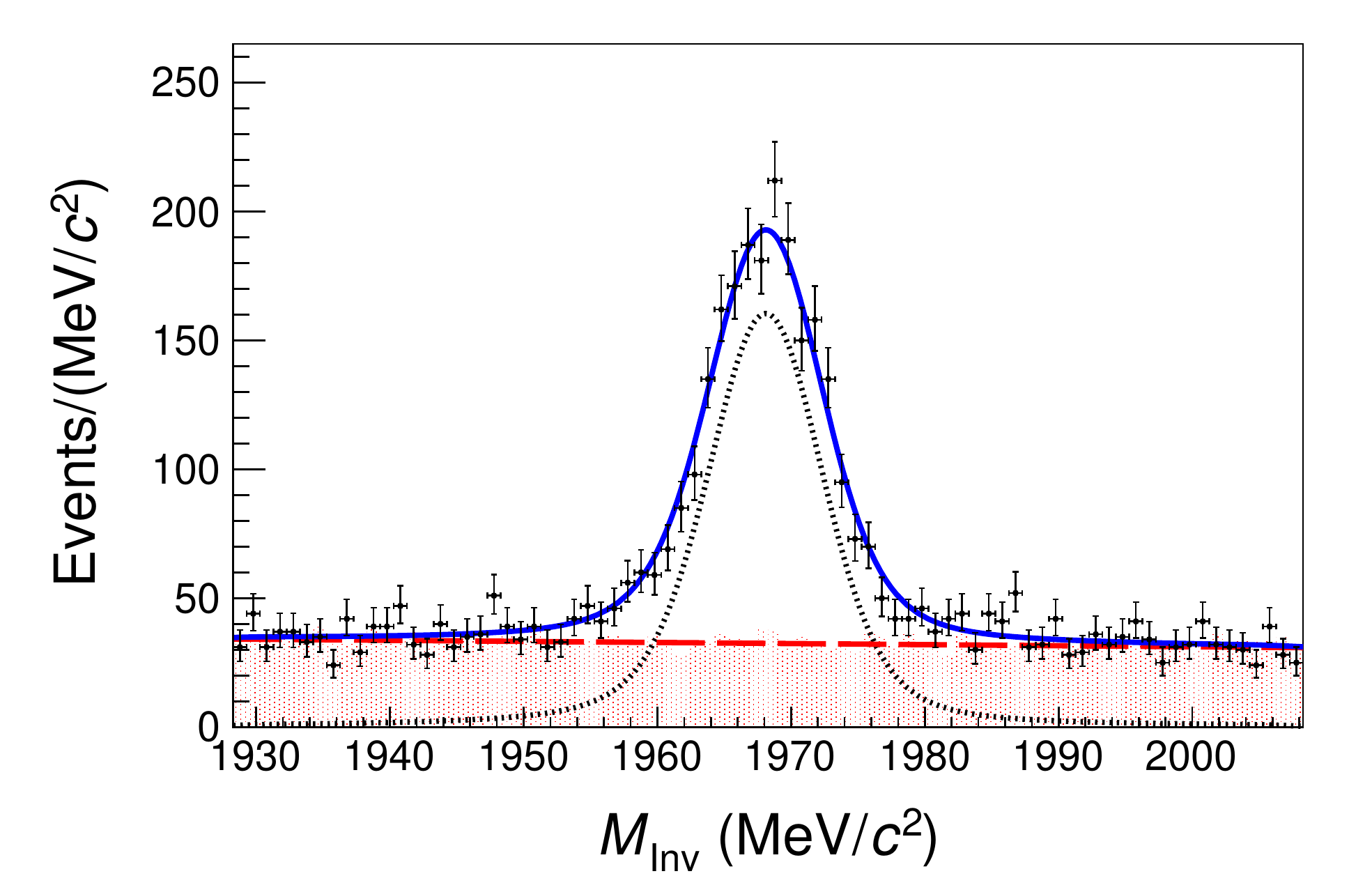} & \includegraphics[width=3in]{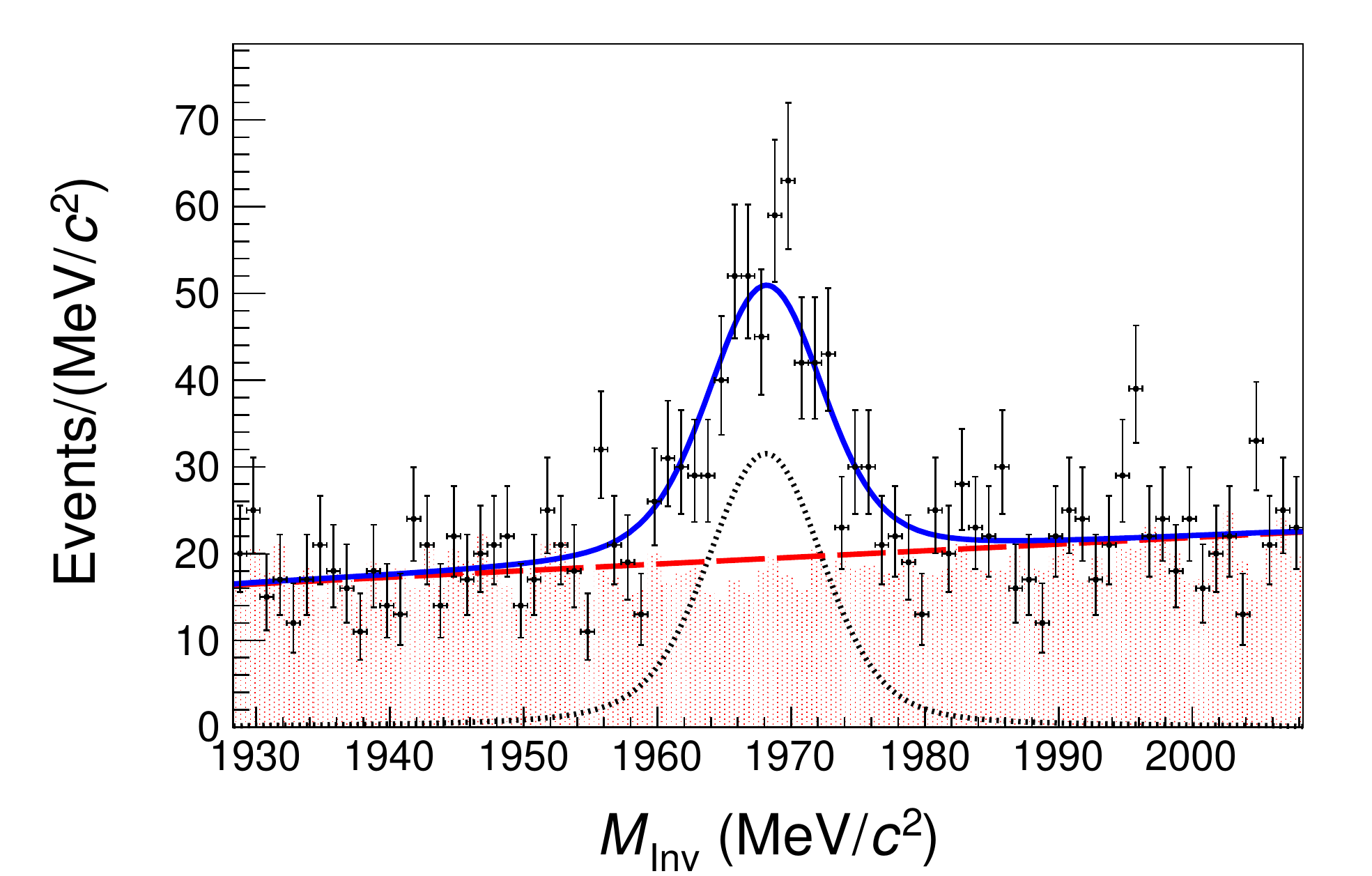}\\
\textbf{RS 700-750 MeV/$c$} & \textbf{WS 700-750 MeV/$c$}\\
\includegraphics[width=3in]{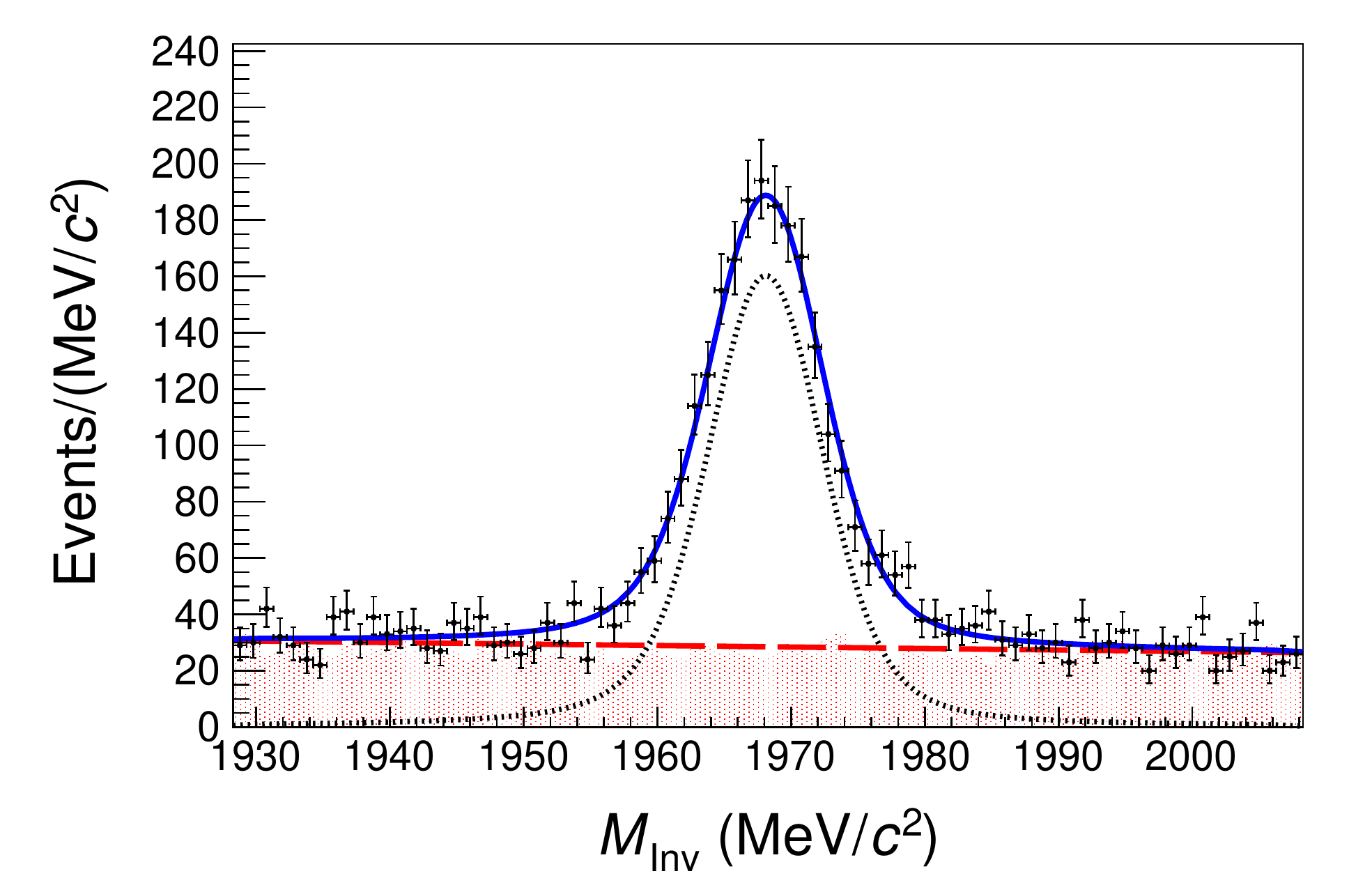} & \includegraphics[width=3in]{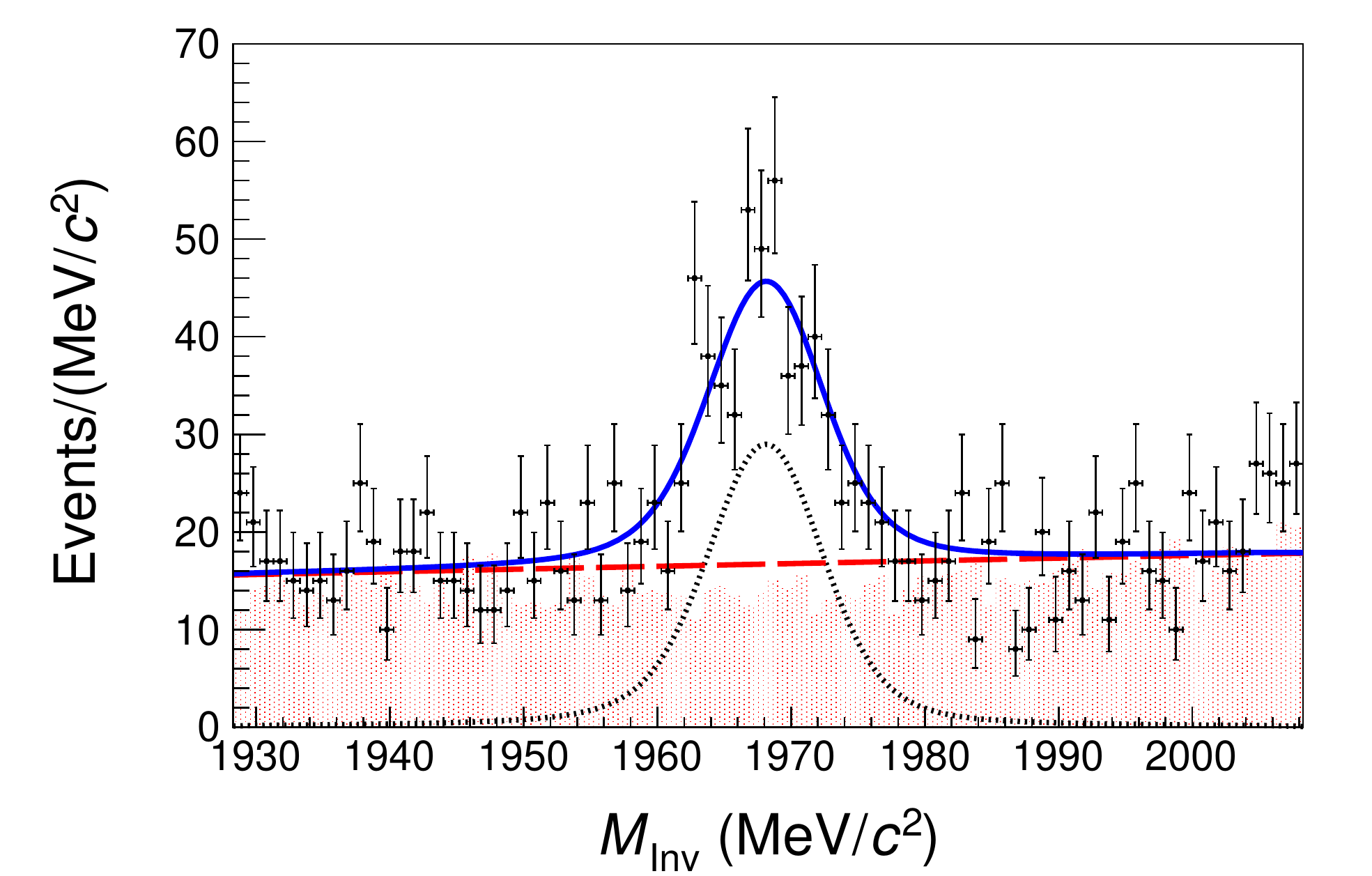}\\
\textbf{RS 750-800 MeV/$c$} & \textbf{WS 750-800 MeV/$c$}\\
\includegraphics[width=3in]{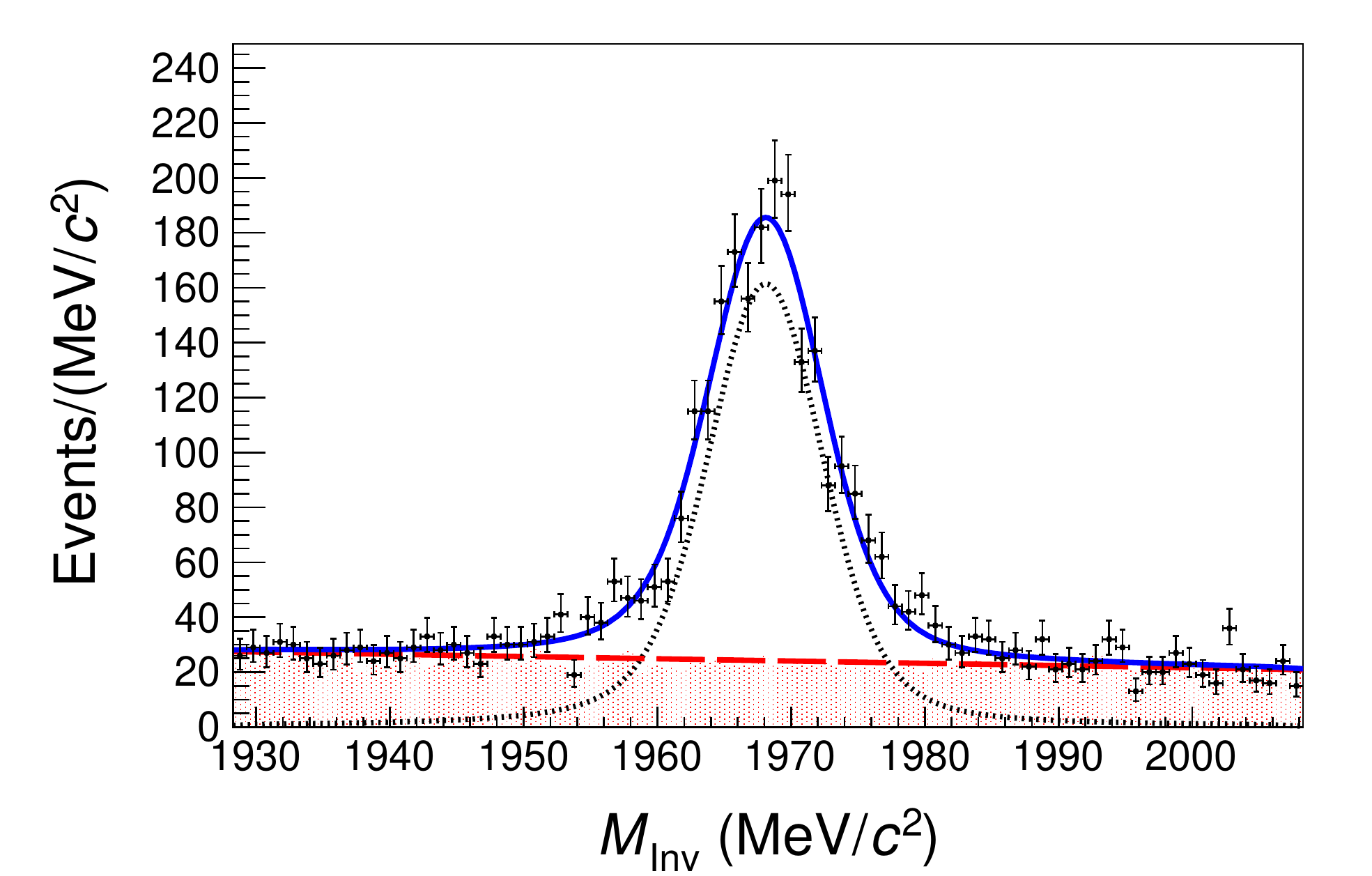} & \includegraphics[width=3in]{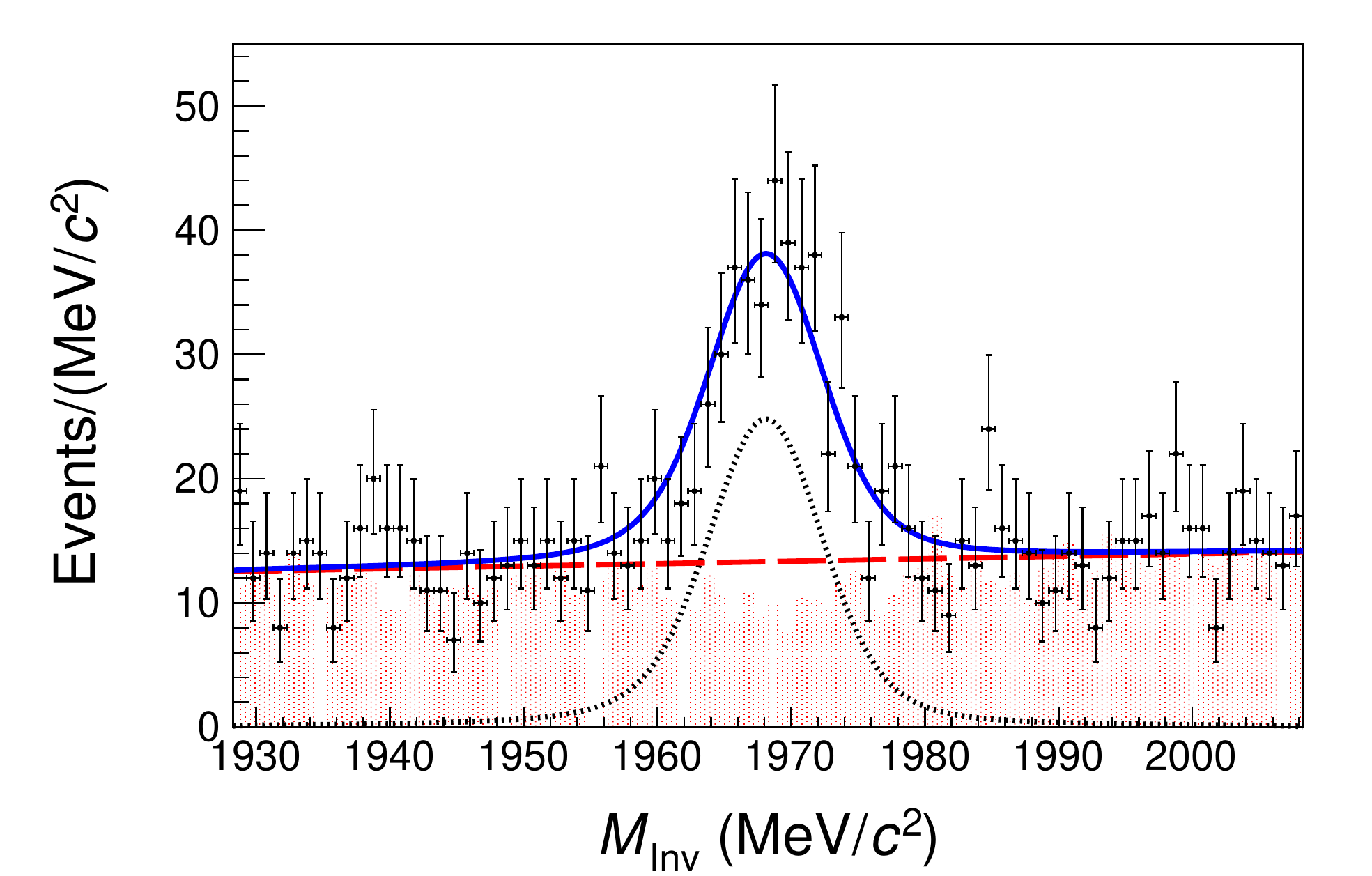}\\
\end{tabular}

\begin{tabular}{cc}
\textbf{RS 800-850 MeV/$c$} & \textbf{WS 800-850 MeV/$c$}\\
\includegraphics[width=3in]{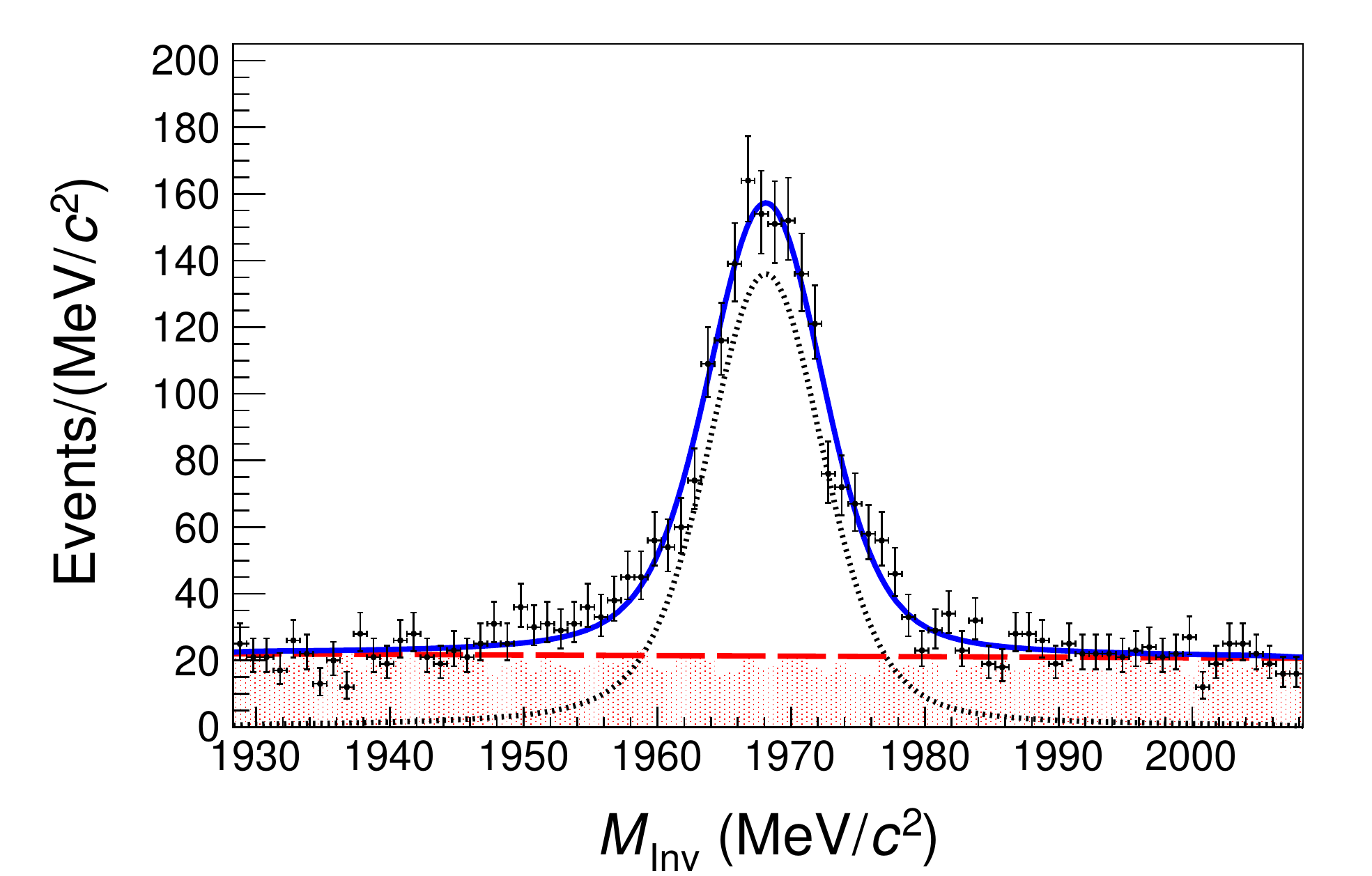} & \includegraphics[width=3in]{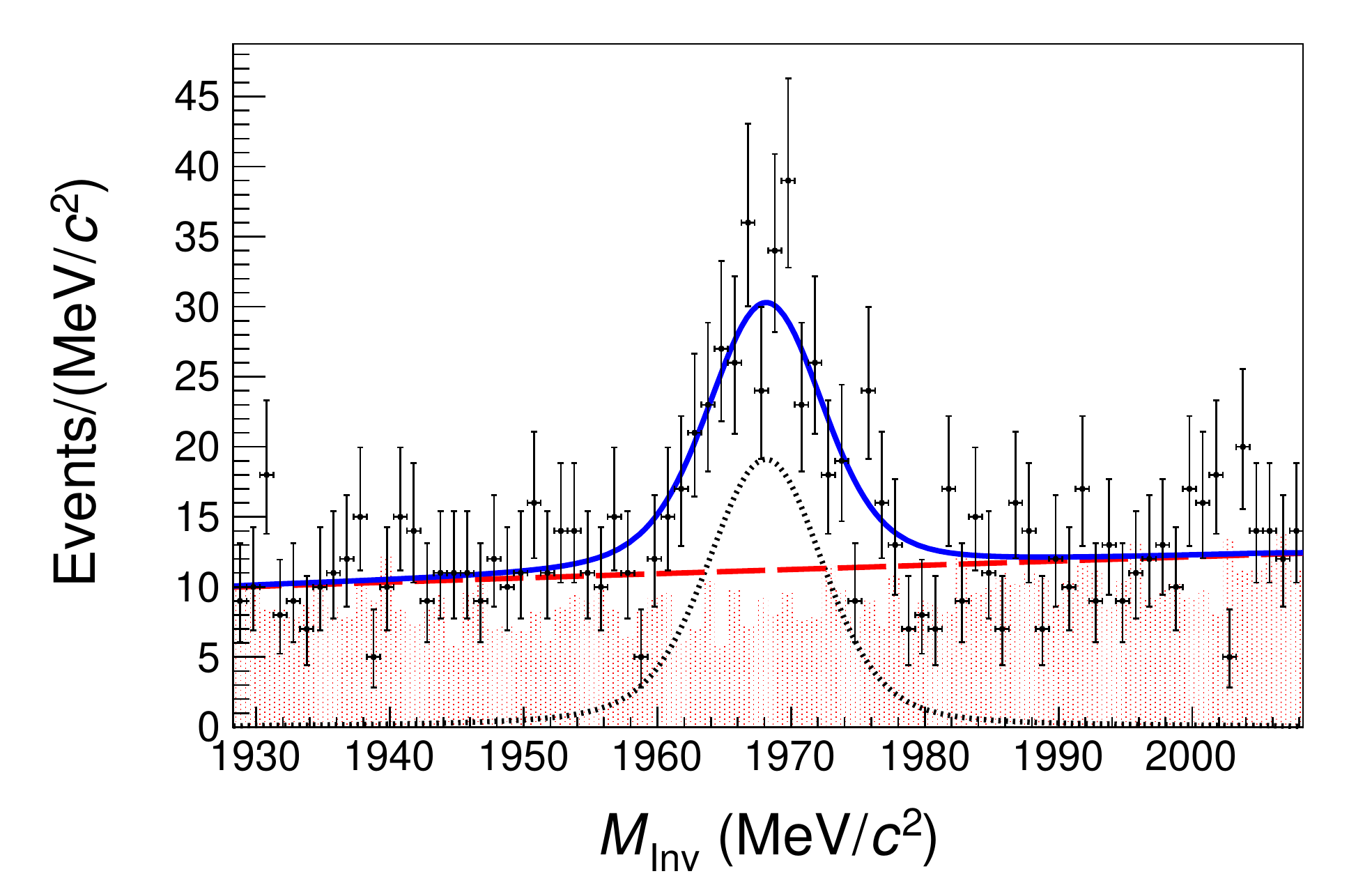}\\
\textbf{RS 850-900 MeV/$c$} & \textbf{WS 850-900 MeV/$c$}\\
\includegraphics[width=3in]{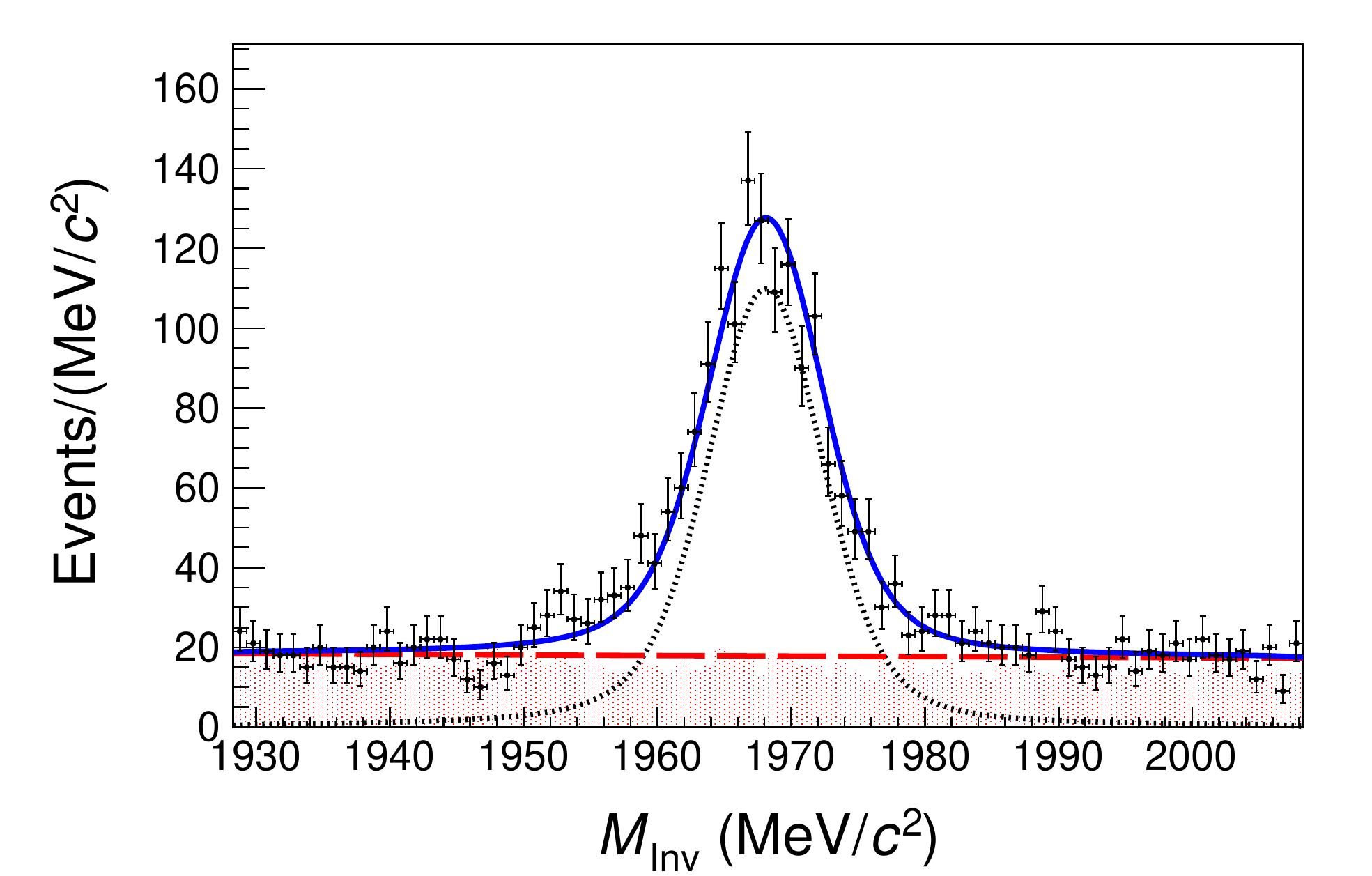} & \includegraphics[width=3in]{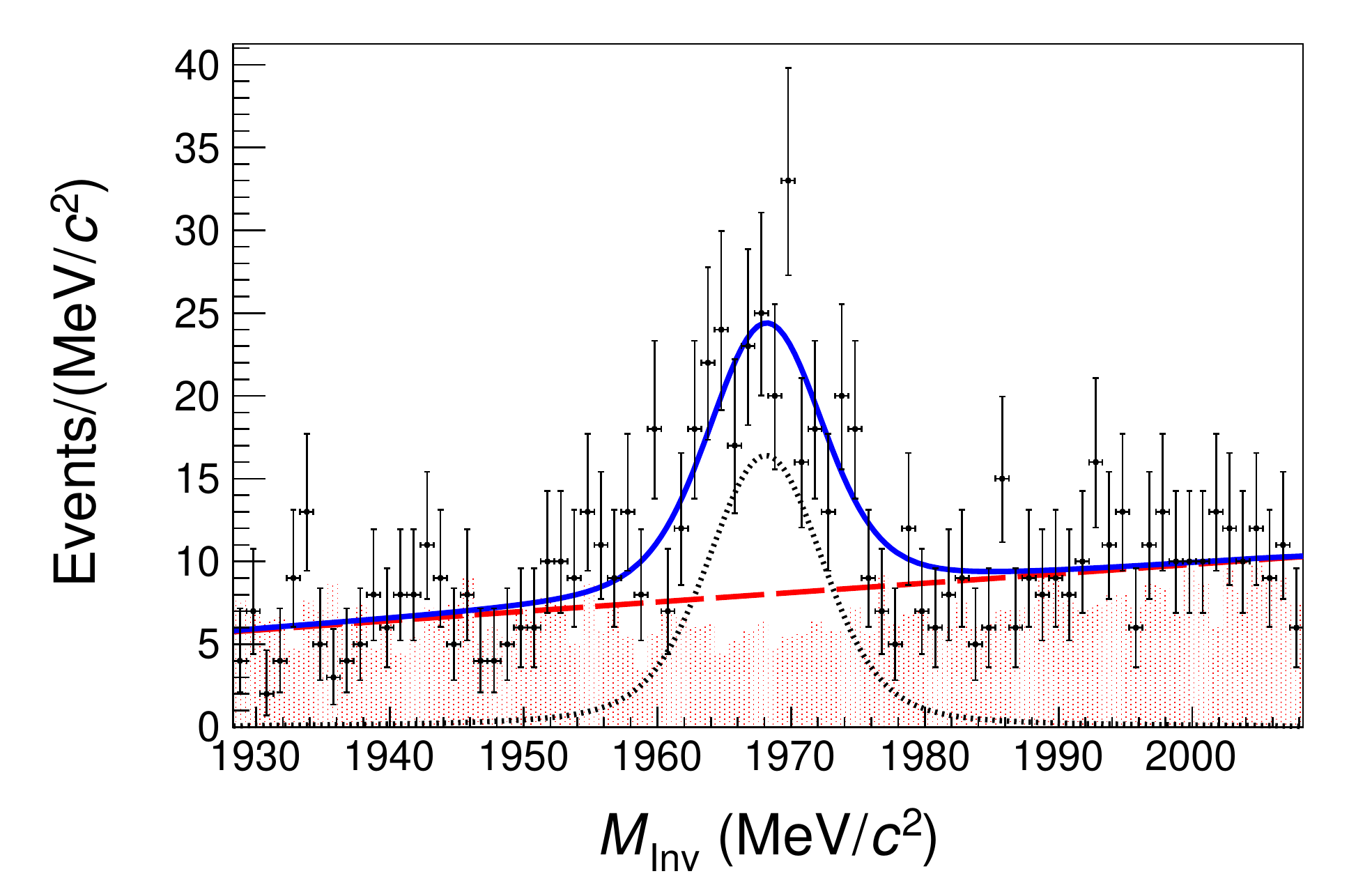}\\
\textbf{RS 900-950 MeV/$c$} & \textbf{WS 900-950 MeV/$c$}\\
\includegraphics[width=3in]{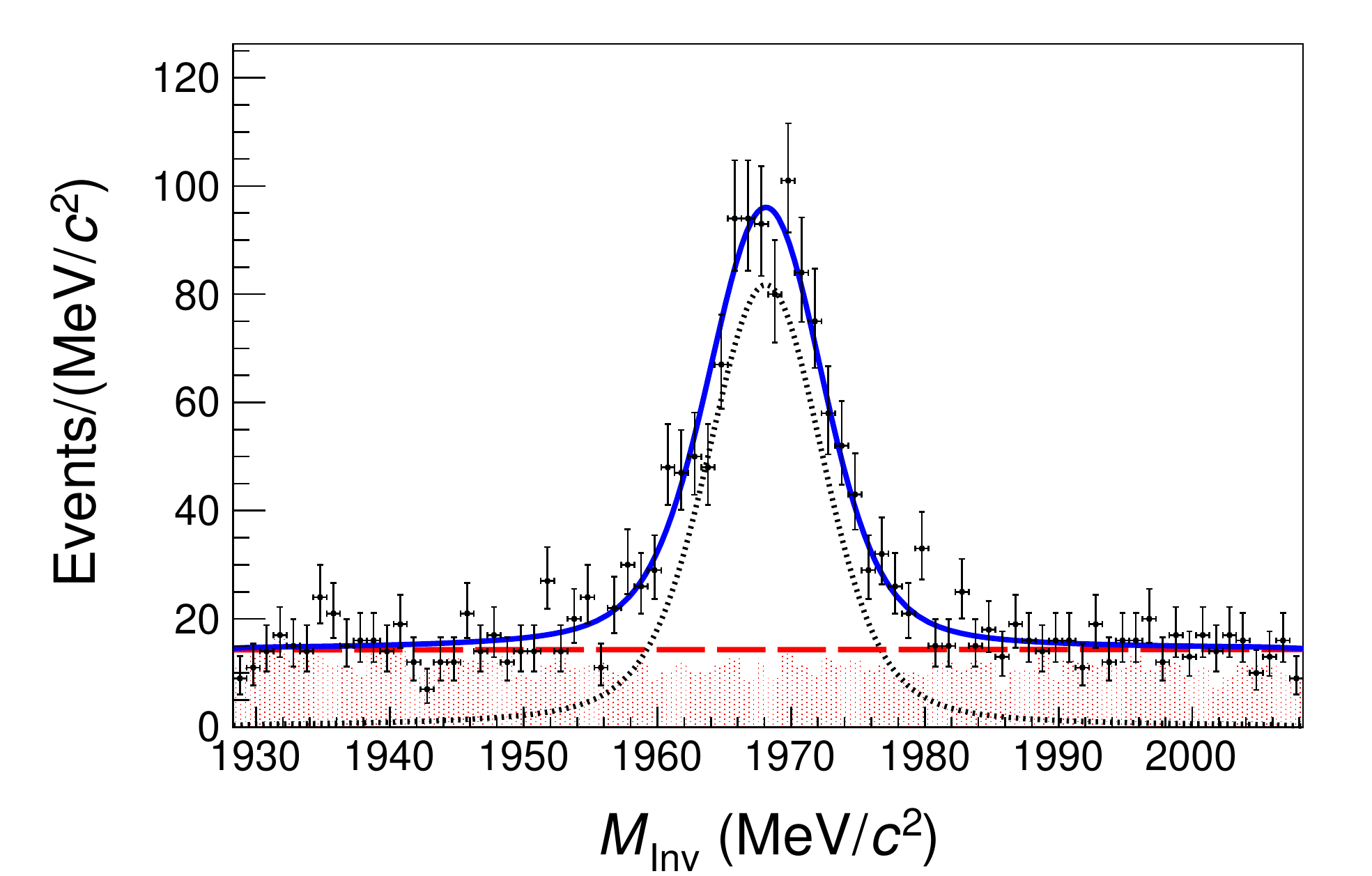} & \includegraphics[width=3in]{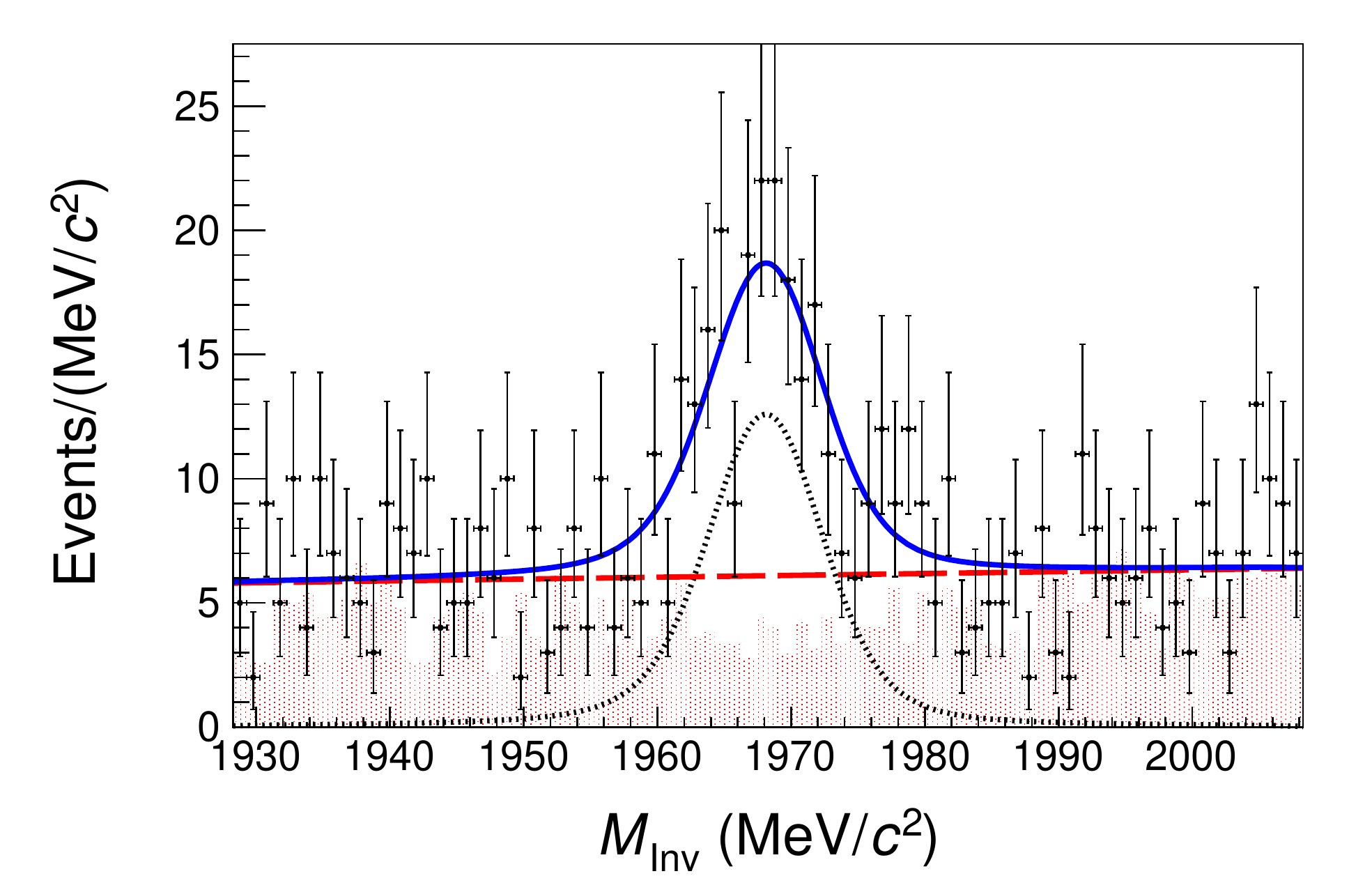}\\
\textbf{RS 950-1000 MeV/$c$} & \textbf{WS 950-1000 MeV/$c$}\\
\includegraphics[width=3in]{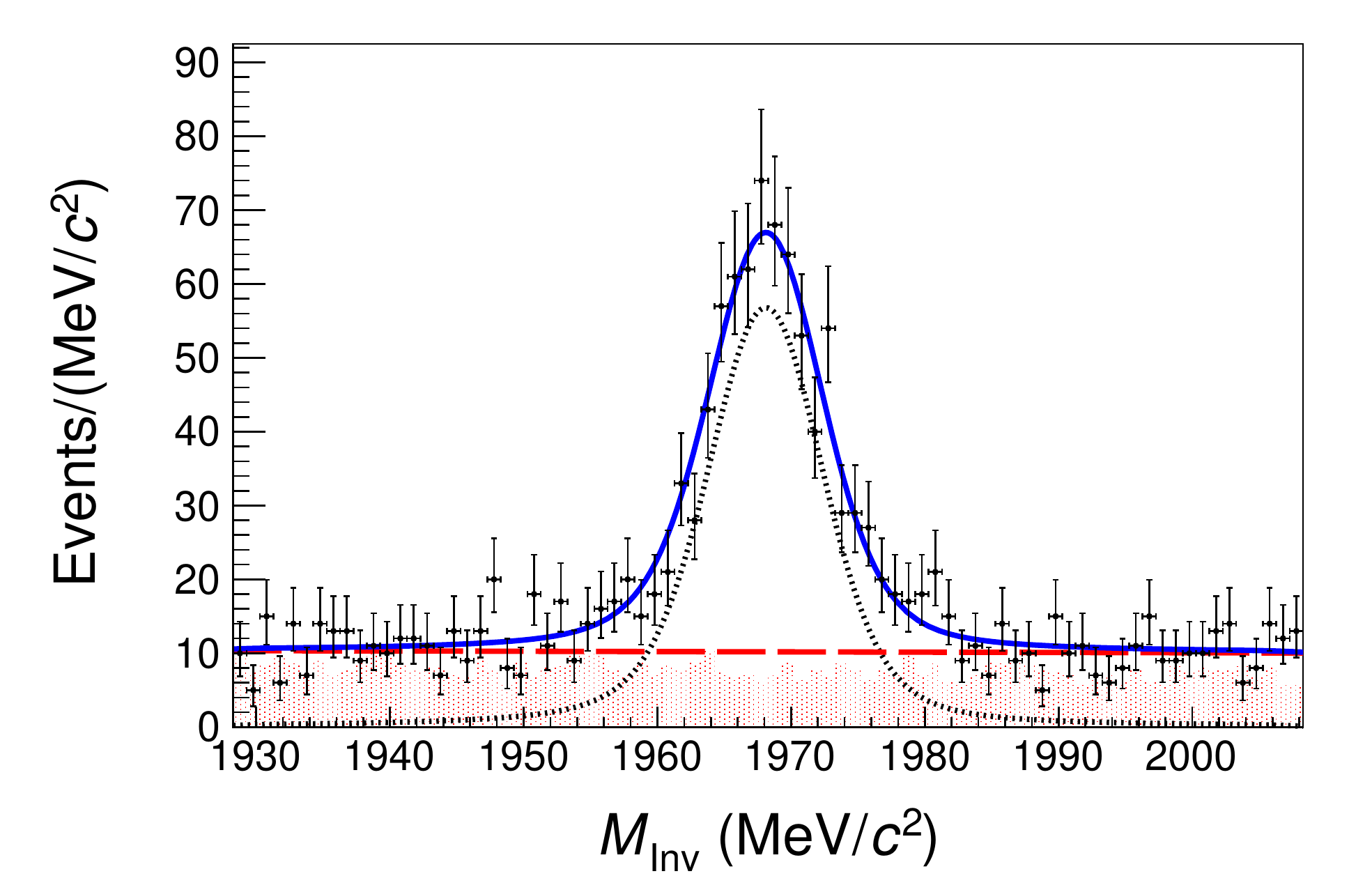} & \includegraphics[width=3in]{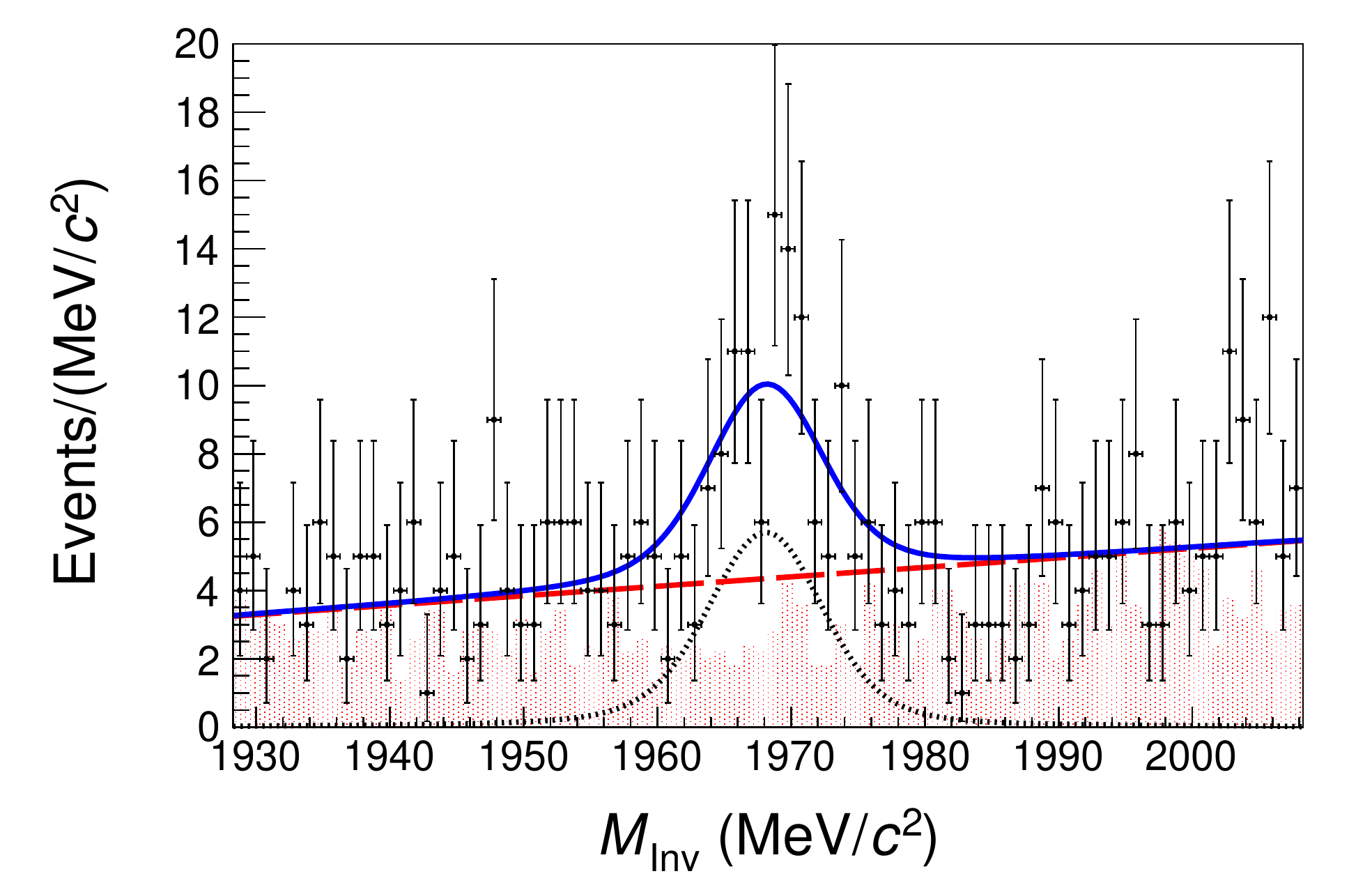}\\
\end{tabular}

\begin{tabular}{cc}
\textbf{RS 1000-1050 MeV/$c$} & \textbf{WS 1000-1050 MeV/$c$}\\
\includegraphics[width=3in]{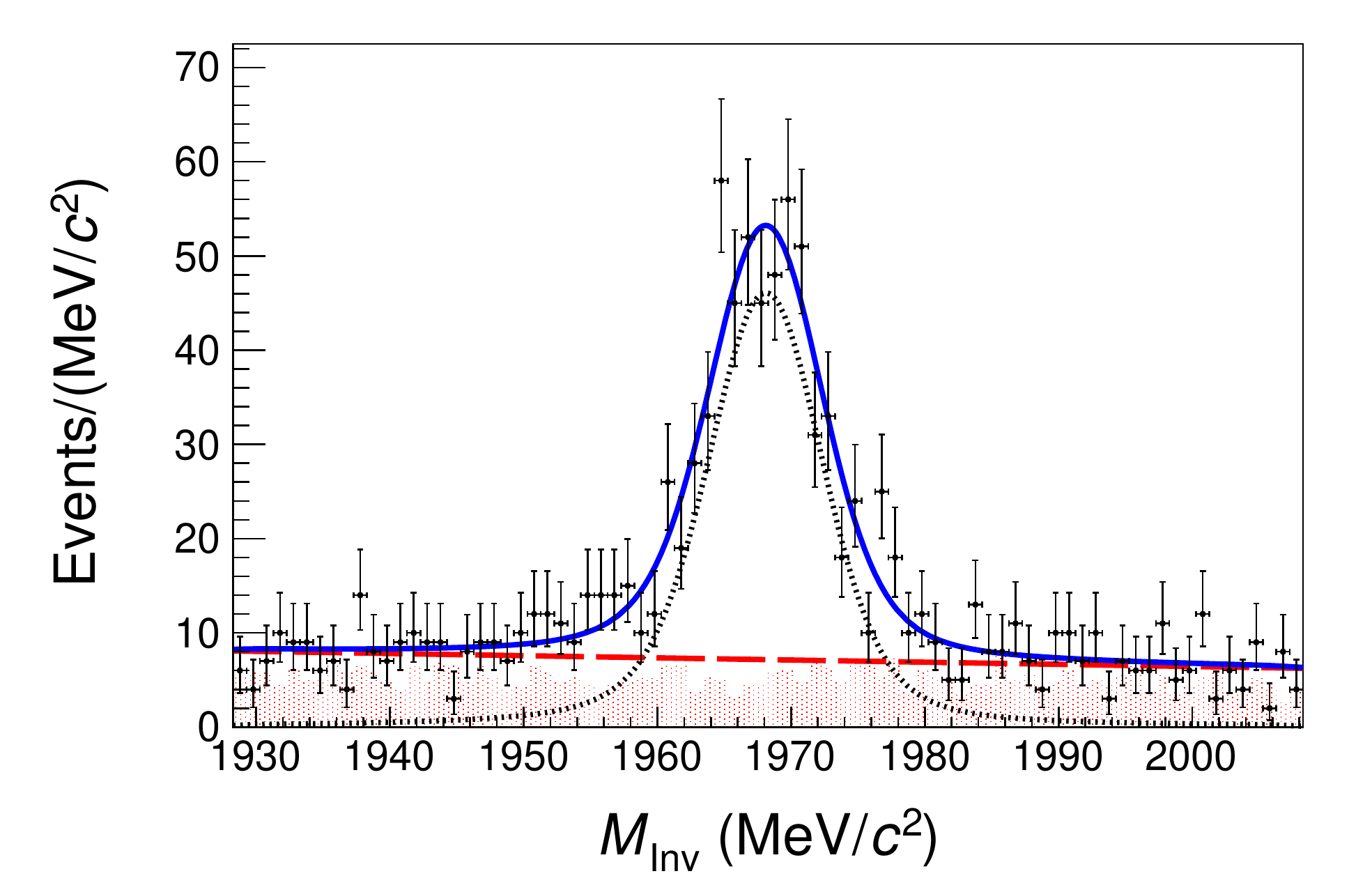} & \includegraphics[width=3in]{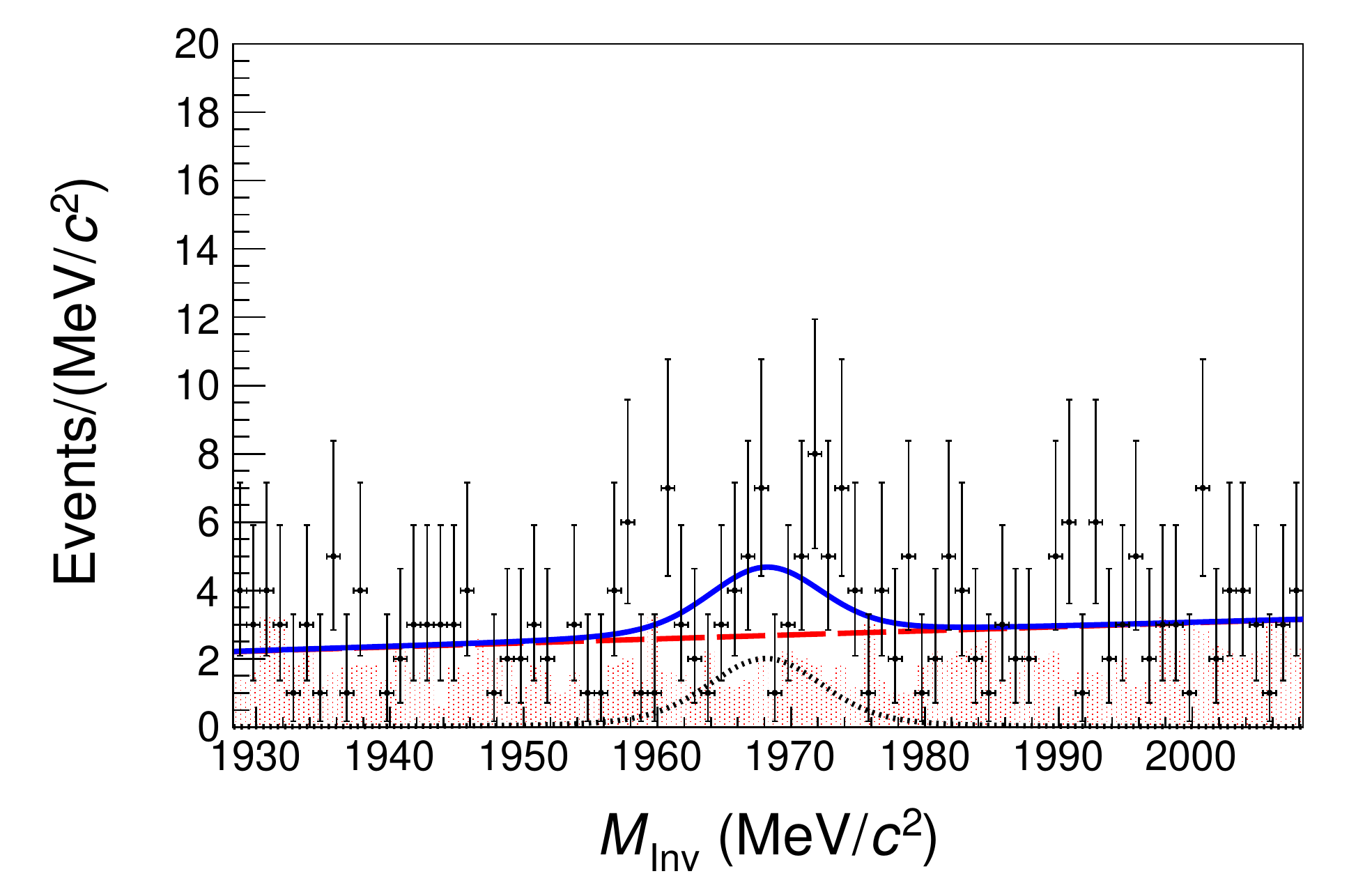}\\
\textbf{RS 1050-1100 MeV/$c$} & \textbf{WS 1050-1100 MeV/$c$}\\
\includegraphics[width=3in]{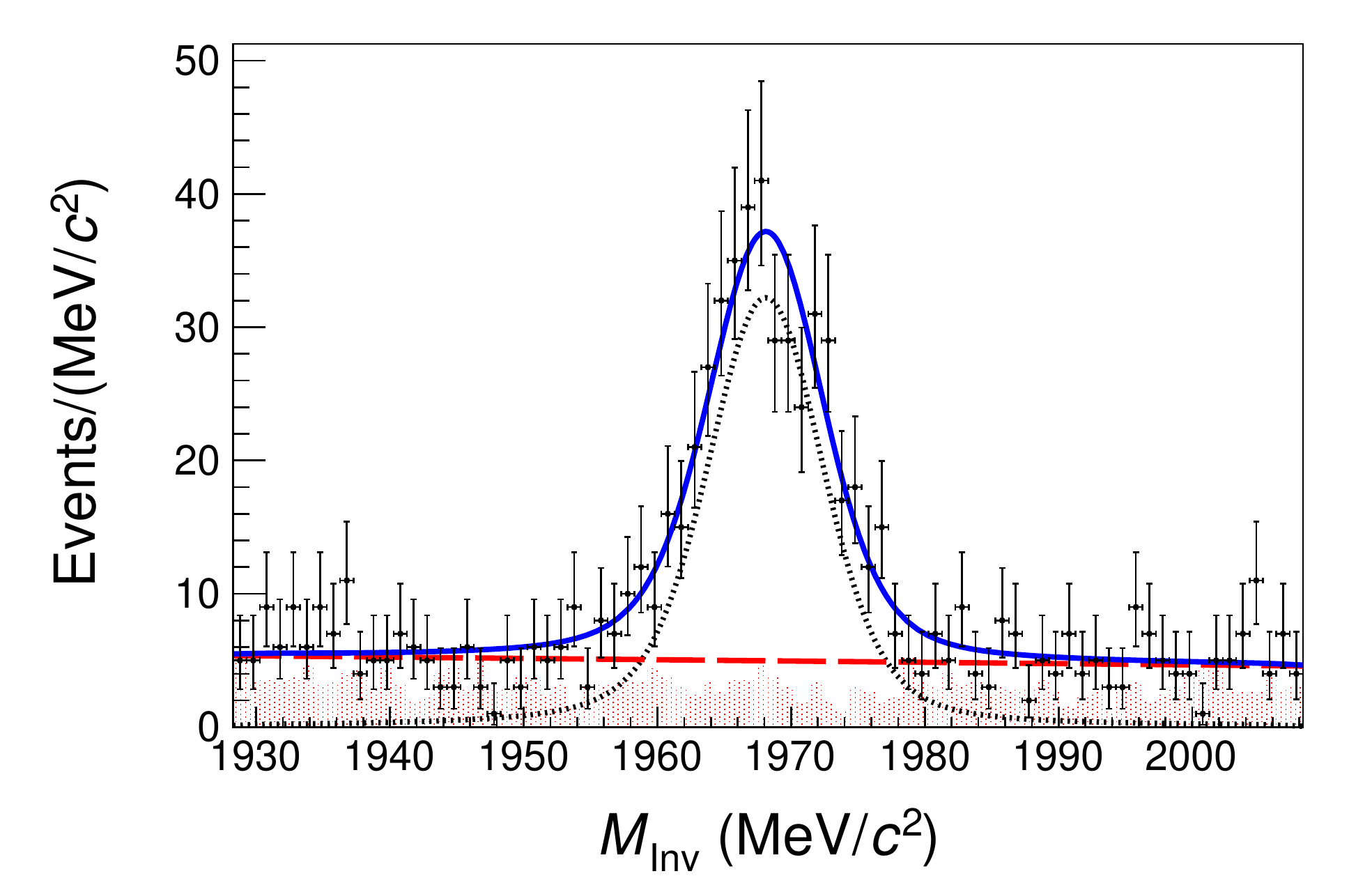} & \includegraphics[width=3in]{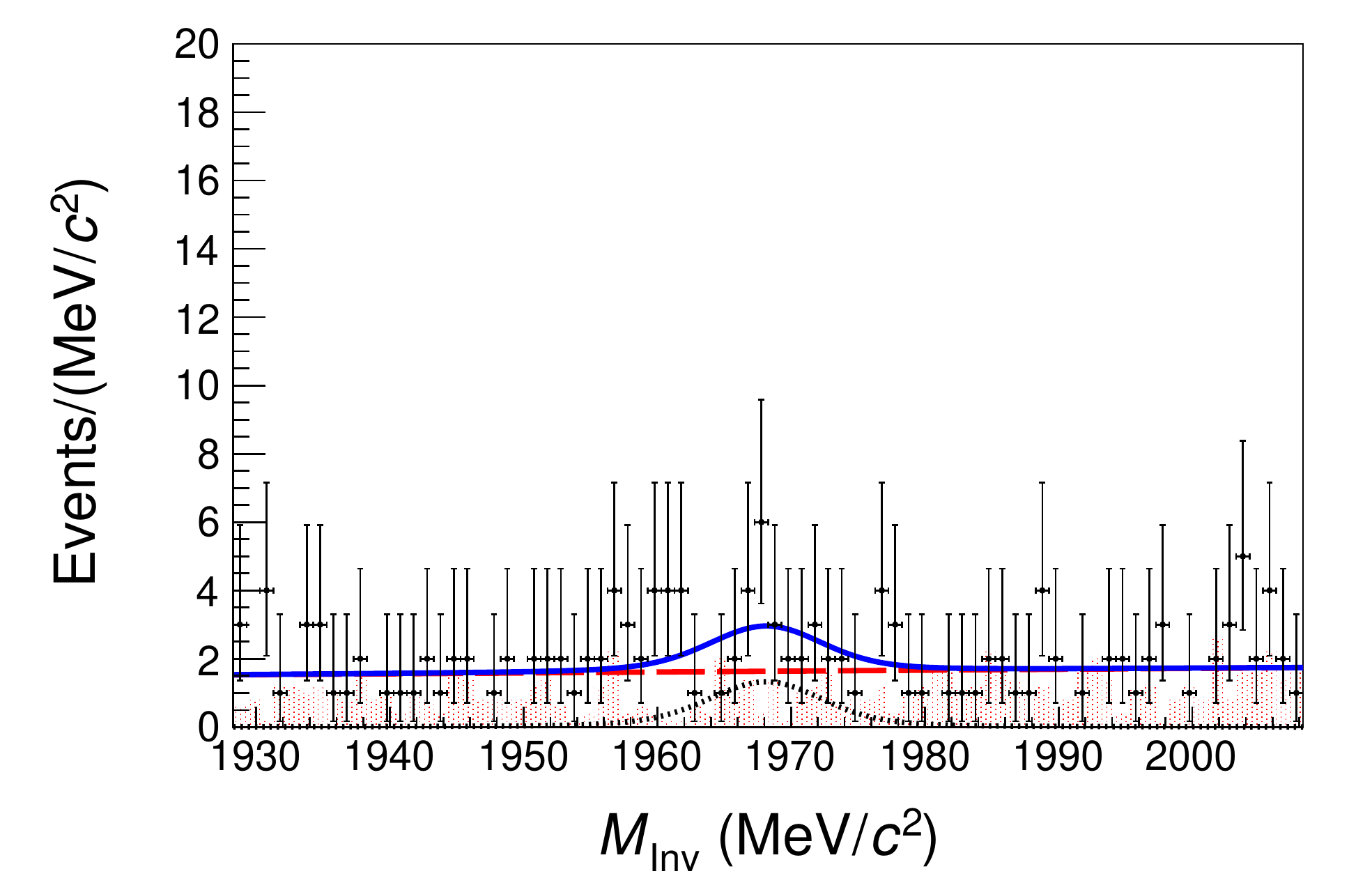}
\end{tabular}

\pagebreak
\raggedright

\subsection{$\EcmB$ Double-tag Fits }
\label{subsec:XYZDataFits}
\subsubsection{$\EcmB$ Data $e$ ID Fits}
\label{subsubsec:XYZDataEIDFits}
\centering
\begin{tabular}{cc}
\textbf{RS 200-250 MeV/$c$} & \textbf{WS 200-250 MeV/$c$}\\
\includegraphics[width=3in]{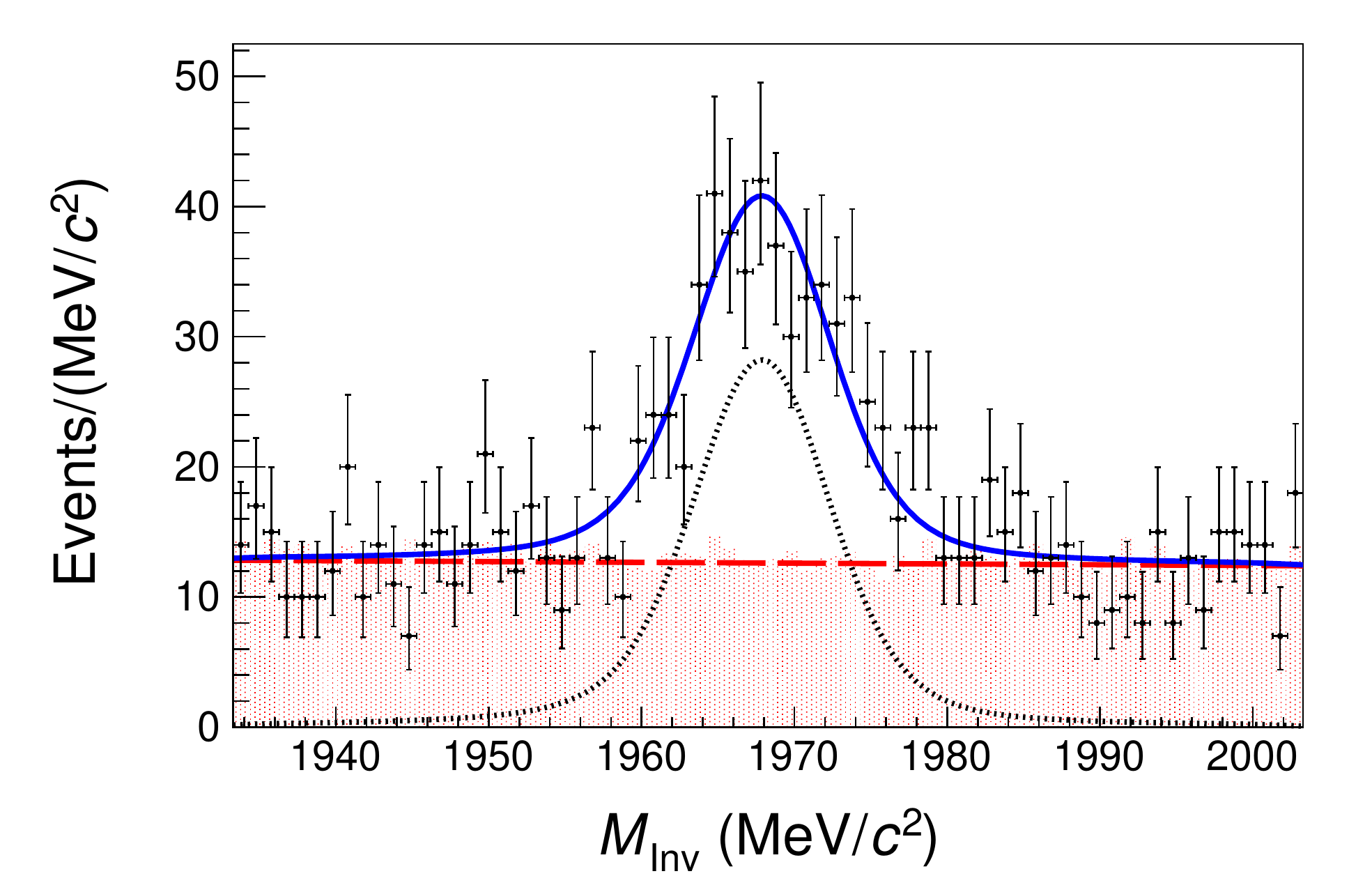} & \includegraphics[width=3in]{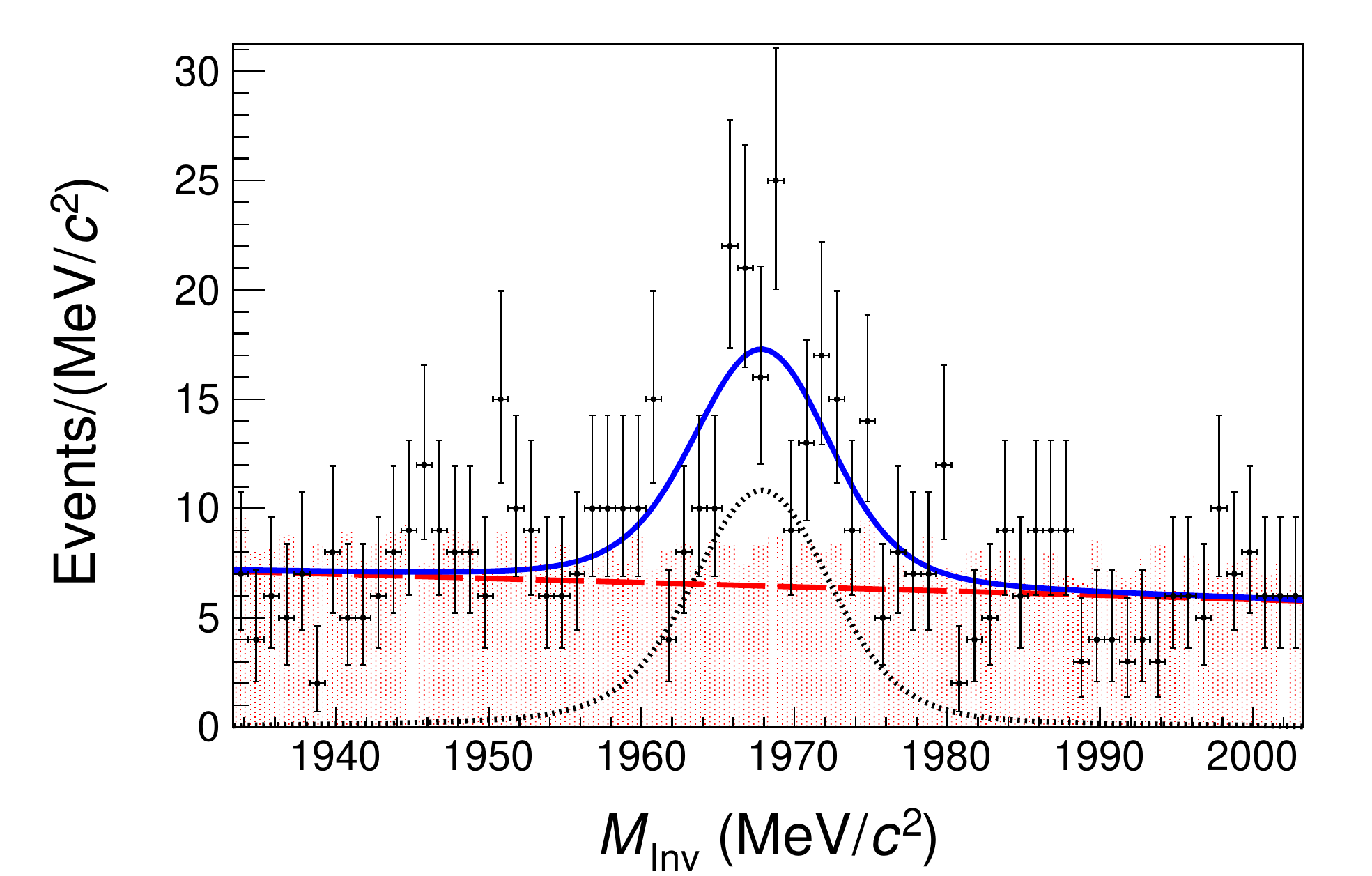}\\
\textbf{RS 250-300 MeV/$c$} & \textbf{WS 250-300 MeV/$c$}\\
\includegraphics[width=3in]{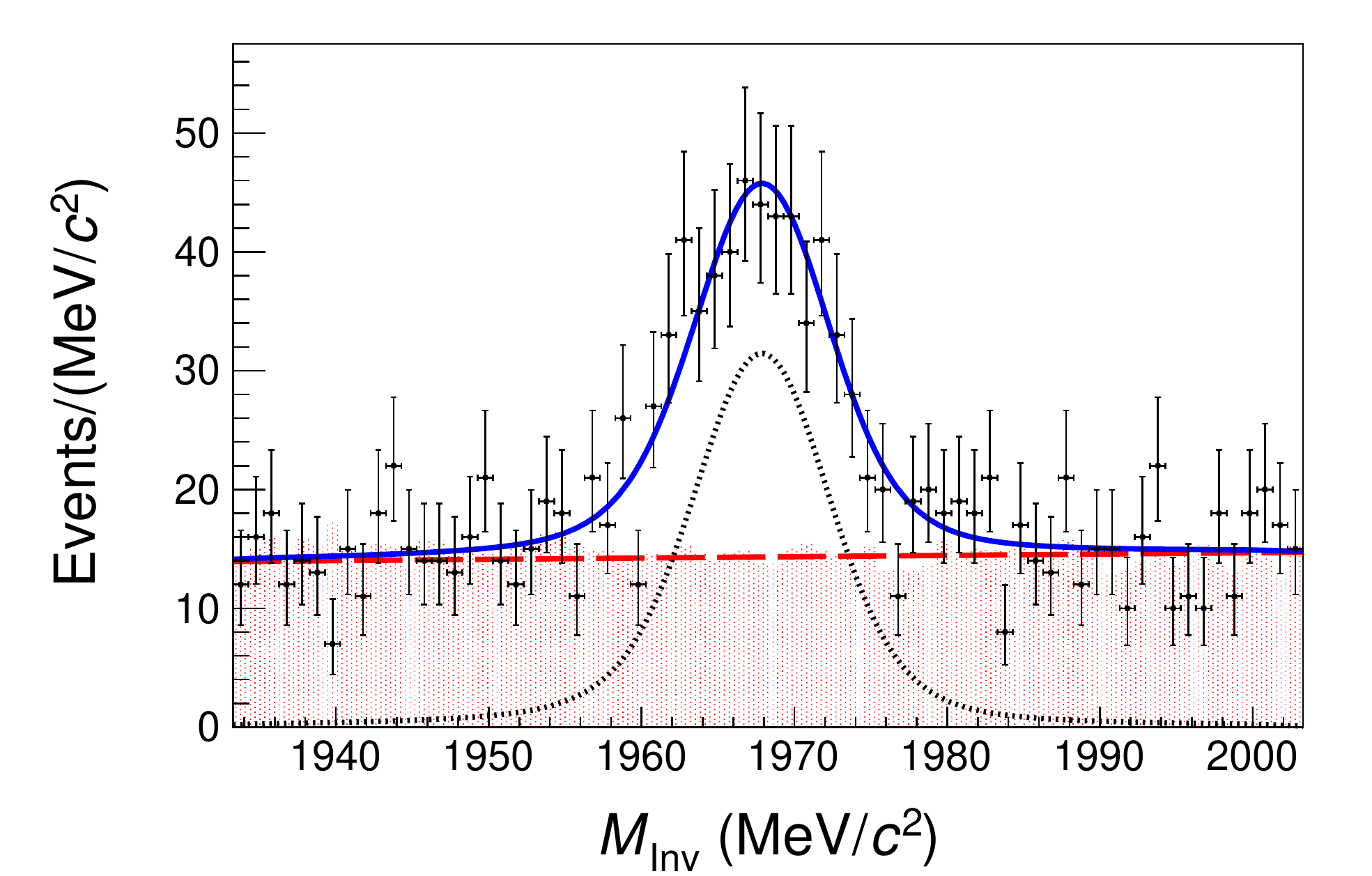} & \includegraphics[width=3in]{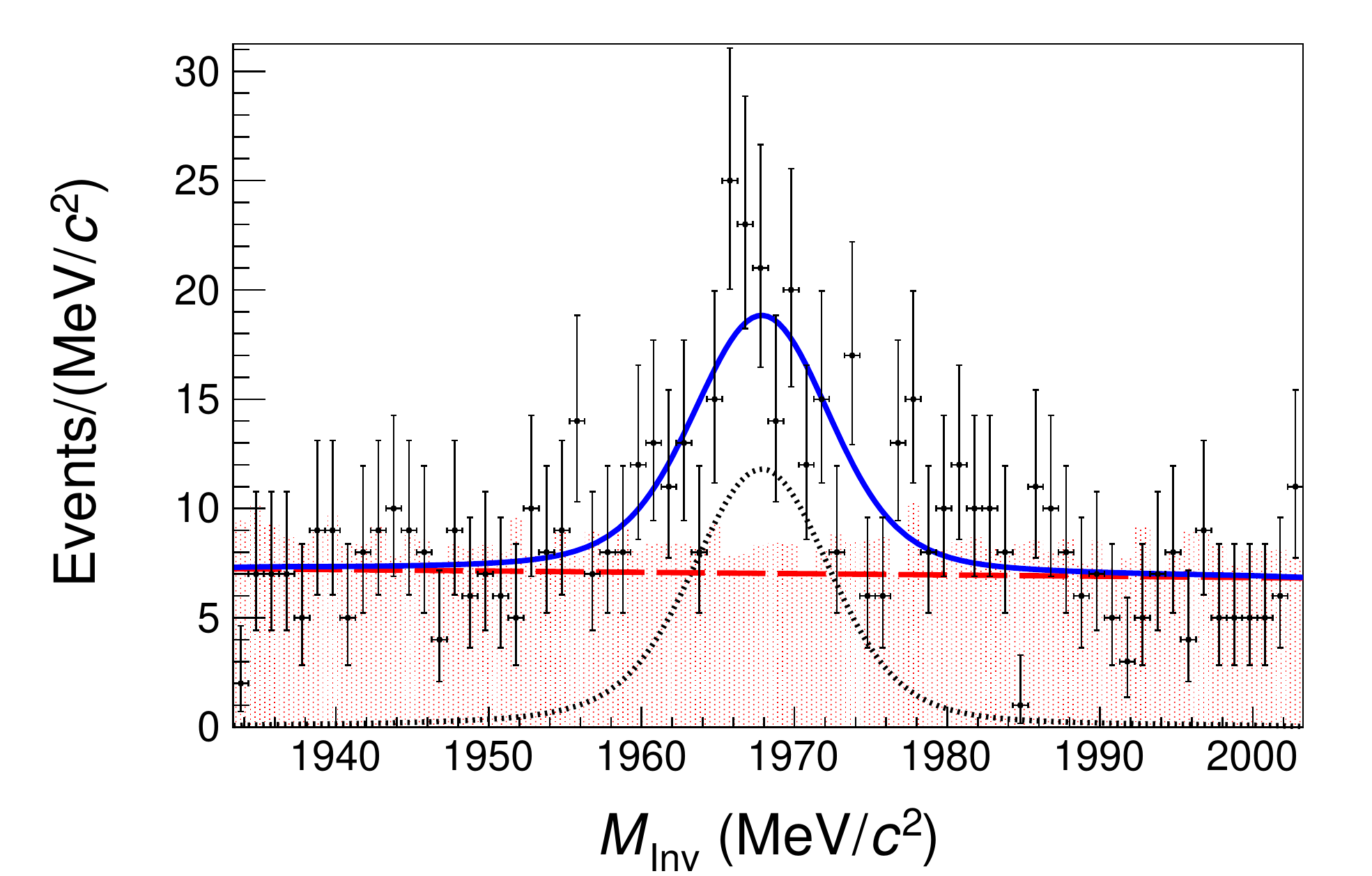}\\
\textbf{RS 300-350 MeV/$c$} & \textbf{WS 300-350 MeV/$c$}\\
\includegraphics[width=3in]{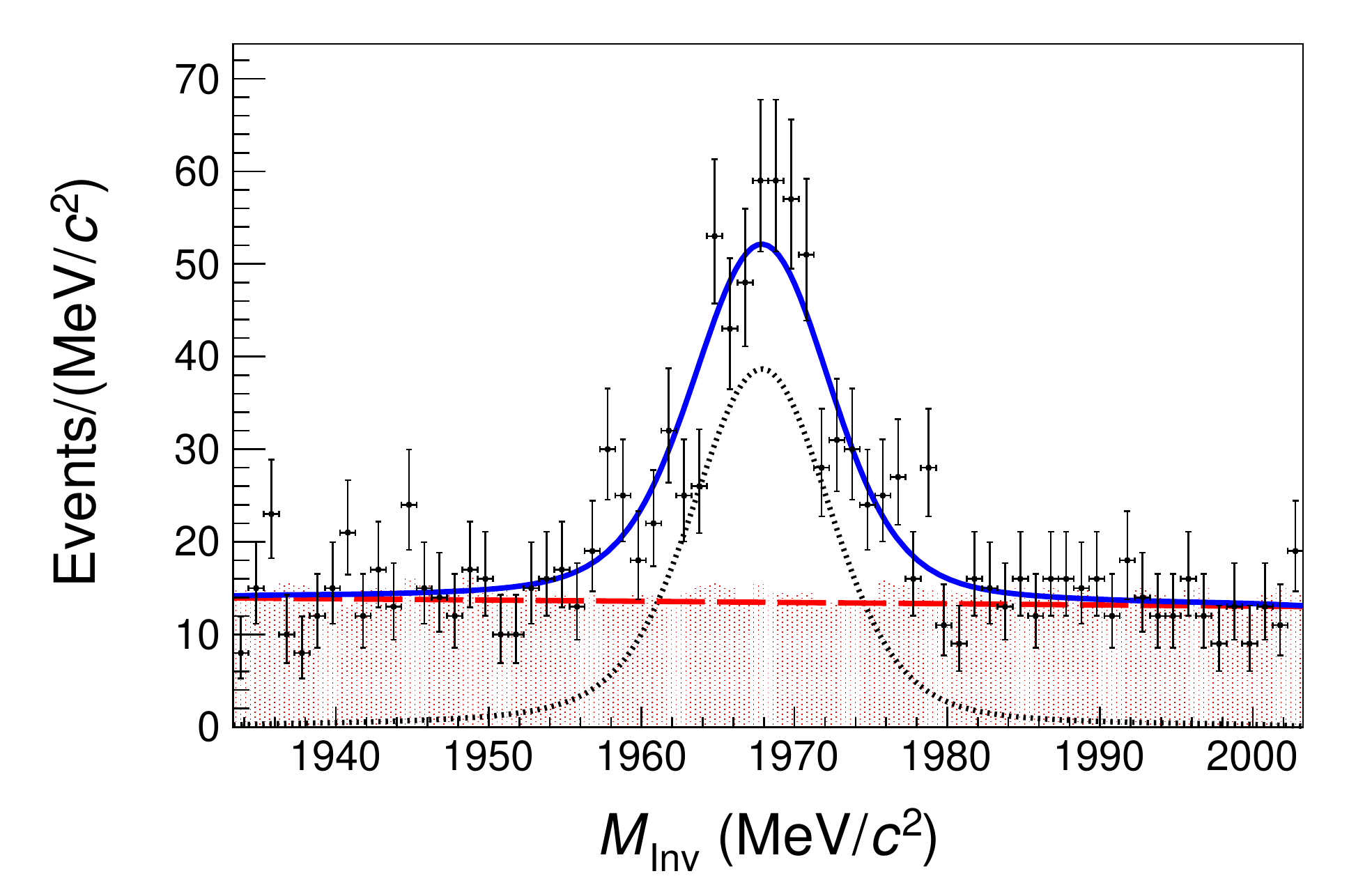} & \includegraphics[width=3in]{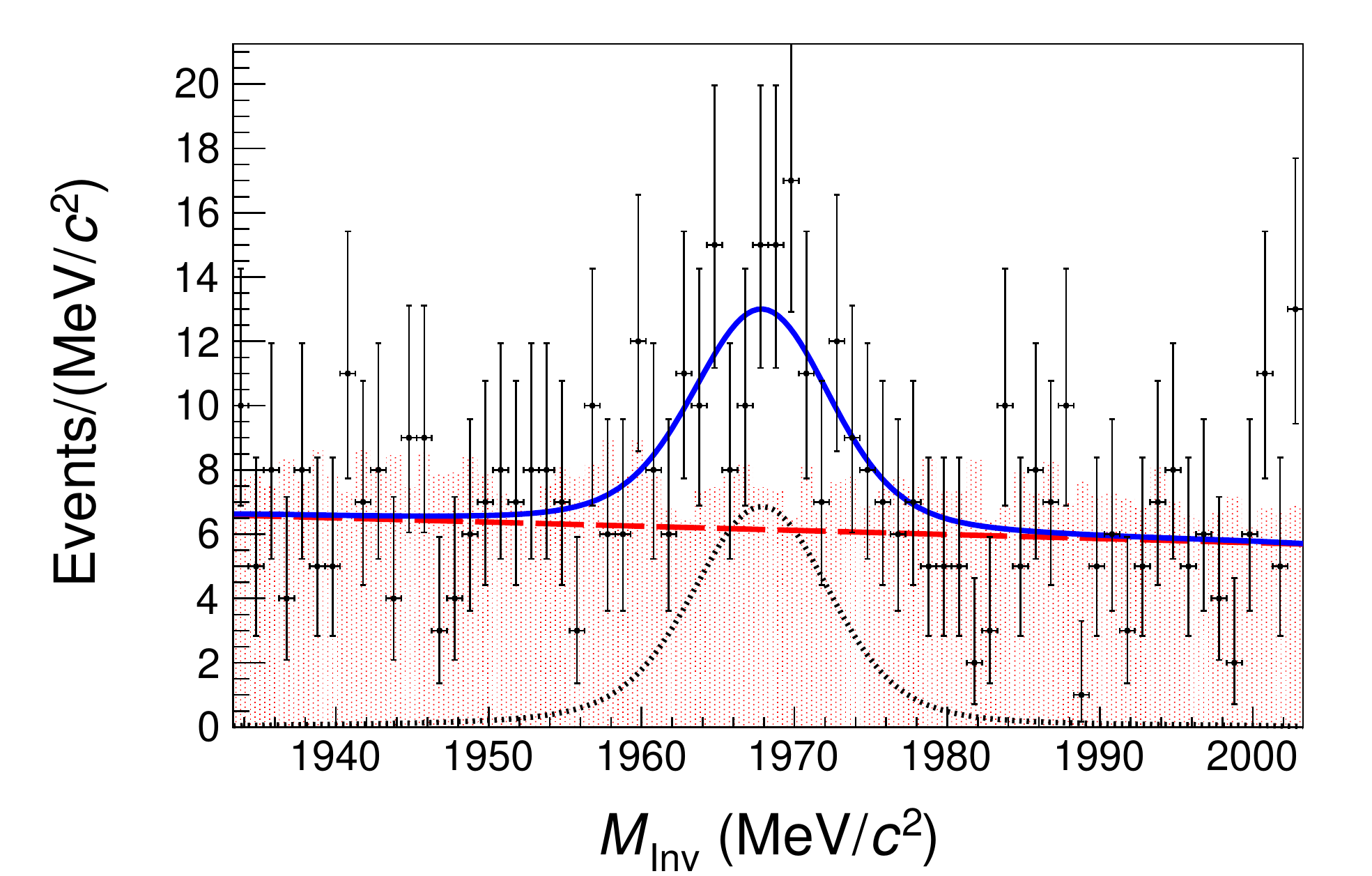}\\
\textbf{RS 350-400 MeV/$c$} & \textbf{WS 350-400 MeV/$c$}\\
\includegraphics[width=3in]{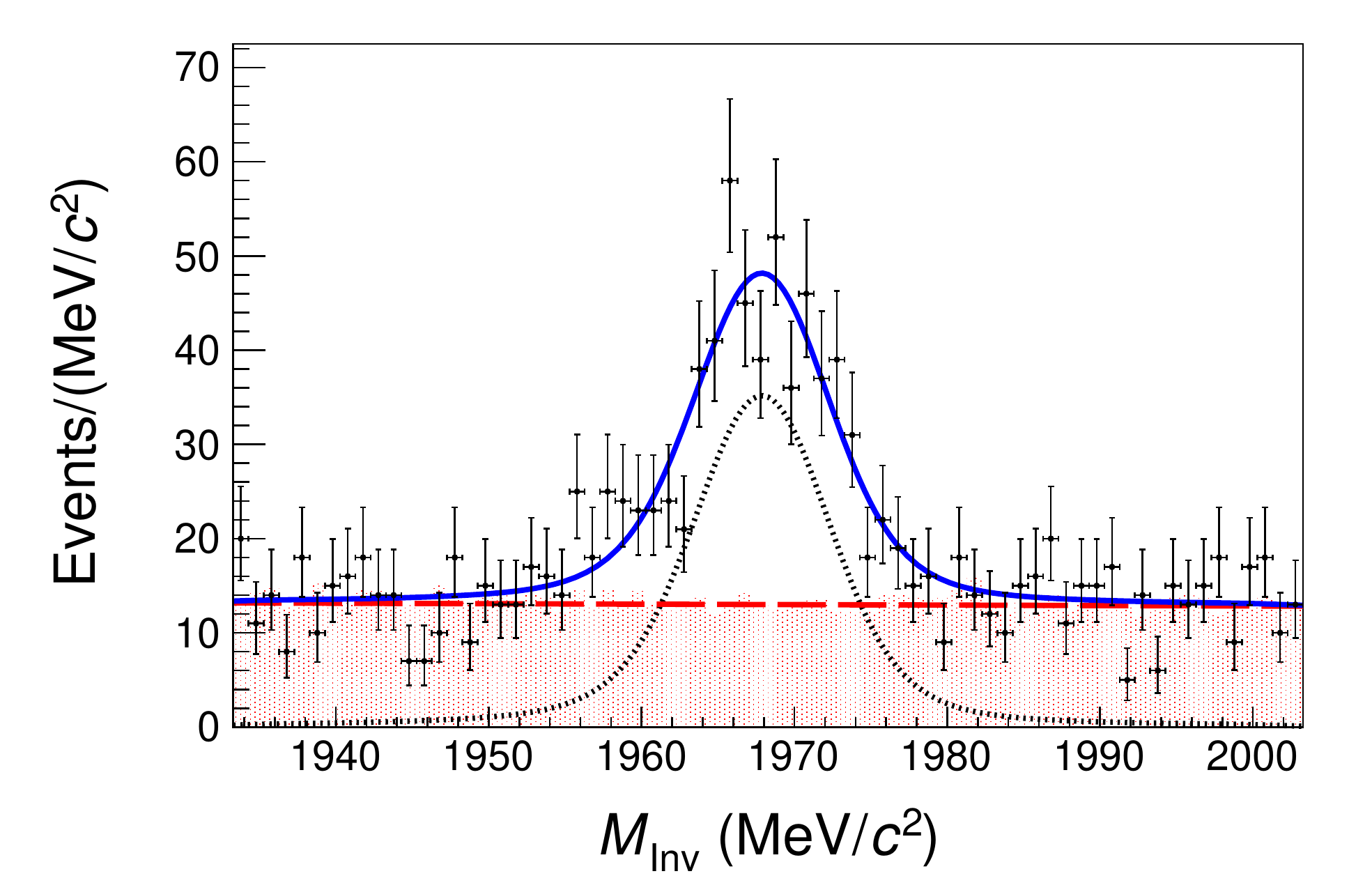} & \includegraphics[width=3in]{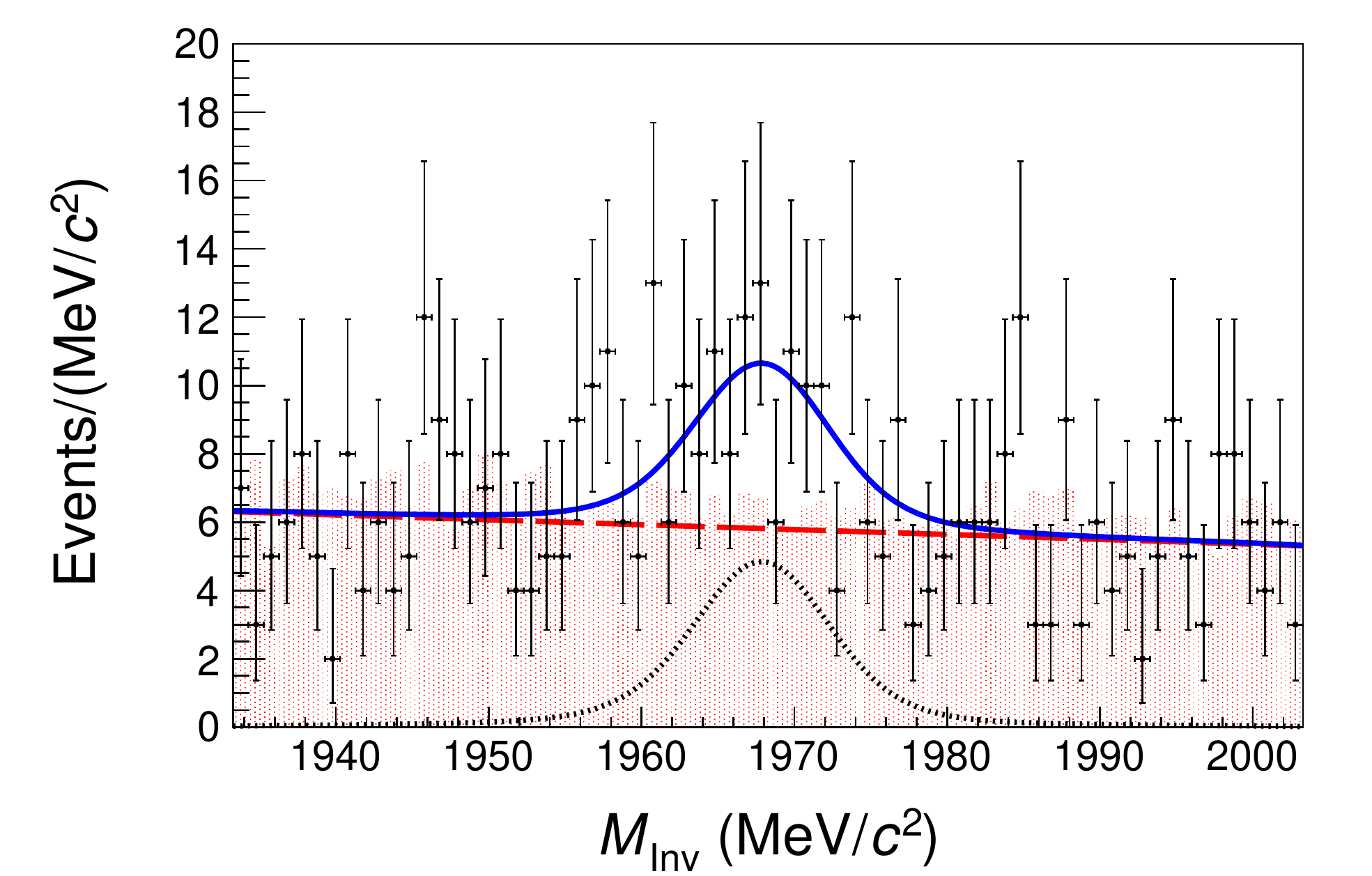}
\end{tabular}

\begin{tabular}{cc}
\textbf{RS 400-450 MeV/$c$} & \textbf{WS 400-450 MeV/$c$}\\
\includegraphics[width=3in]{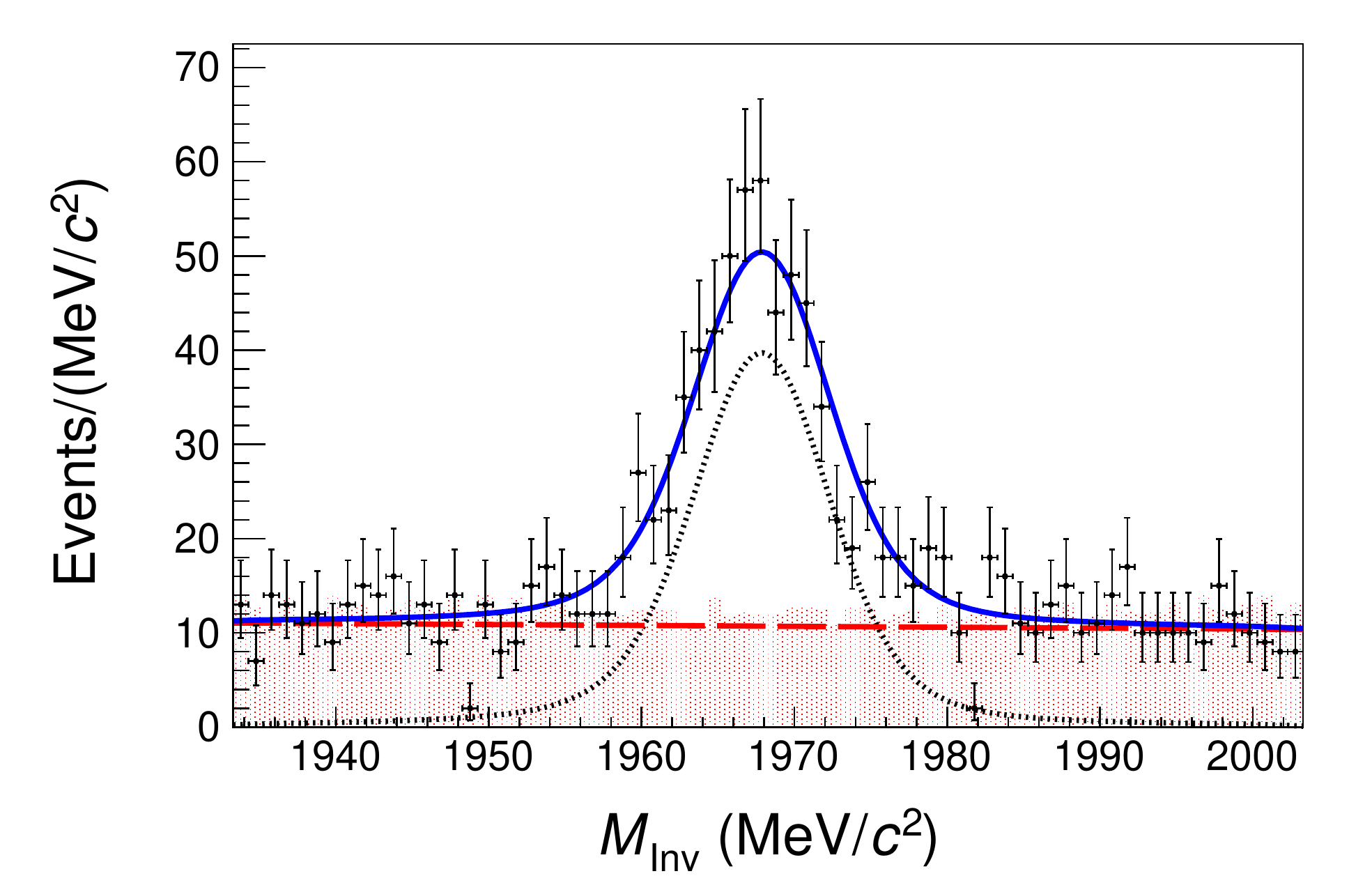} & \includegraphics[width=3in]{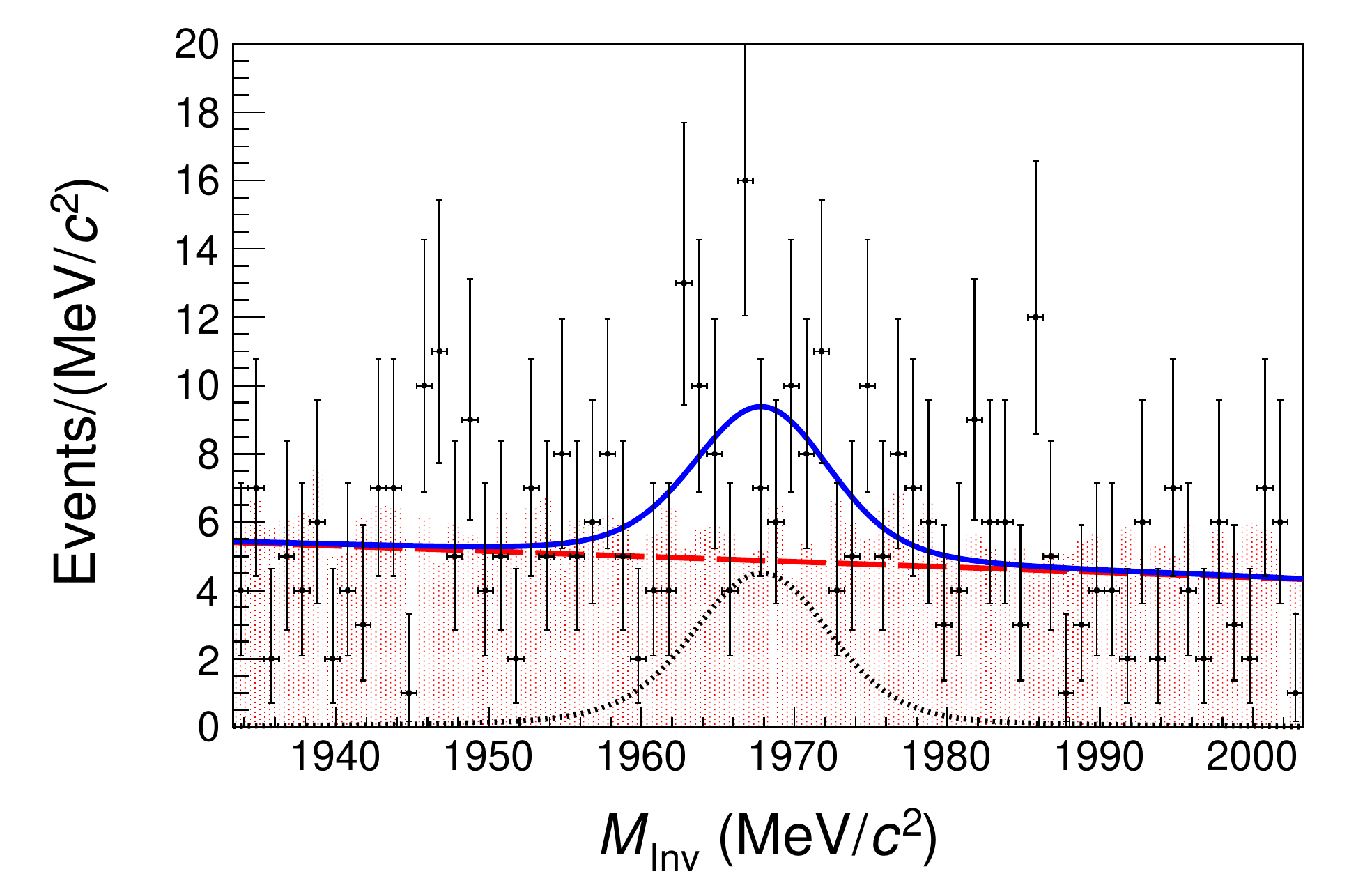}\\
\textbf{RS 450-500 MeV/$c$} & \textbf{WS 450-500 MeV/$c$}\\
\includegraphics[width=3in]{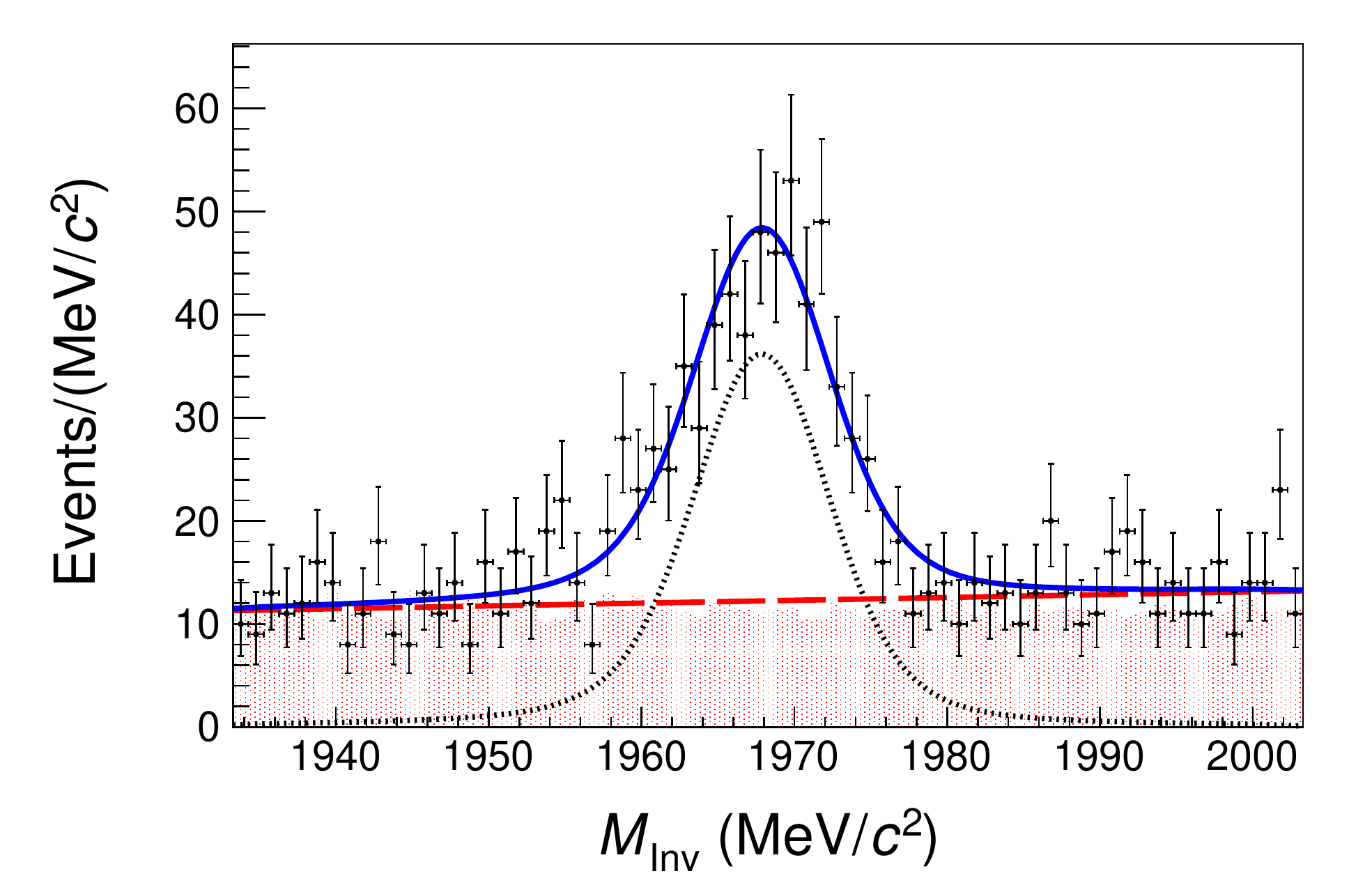} & \includegraphics[width=3in]{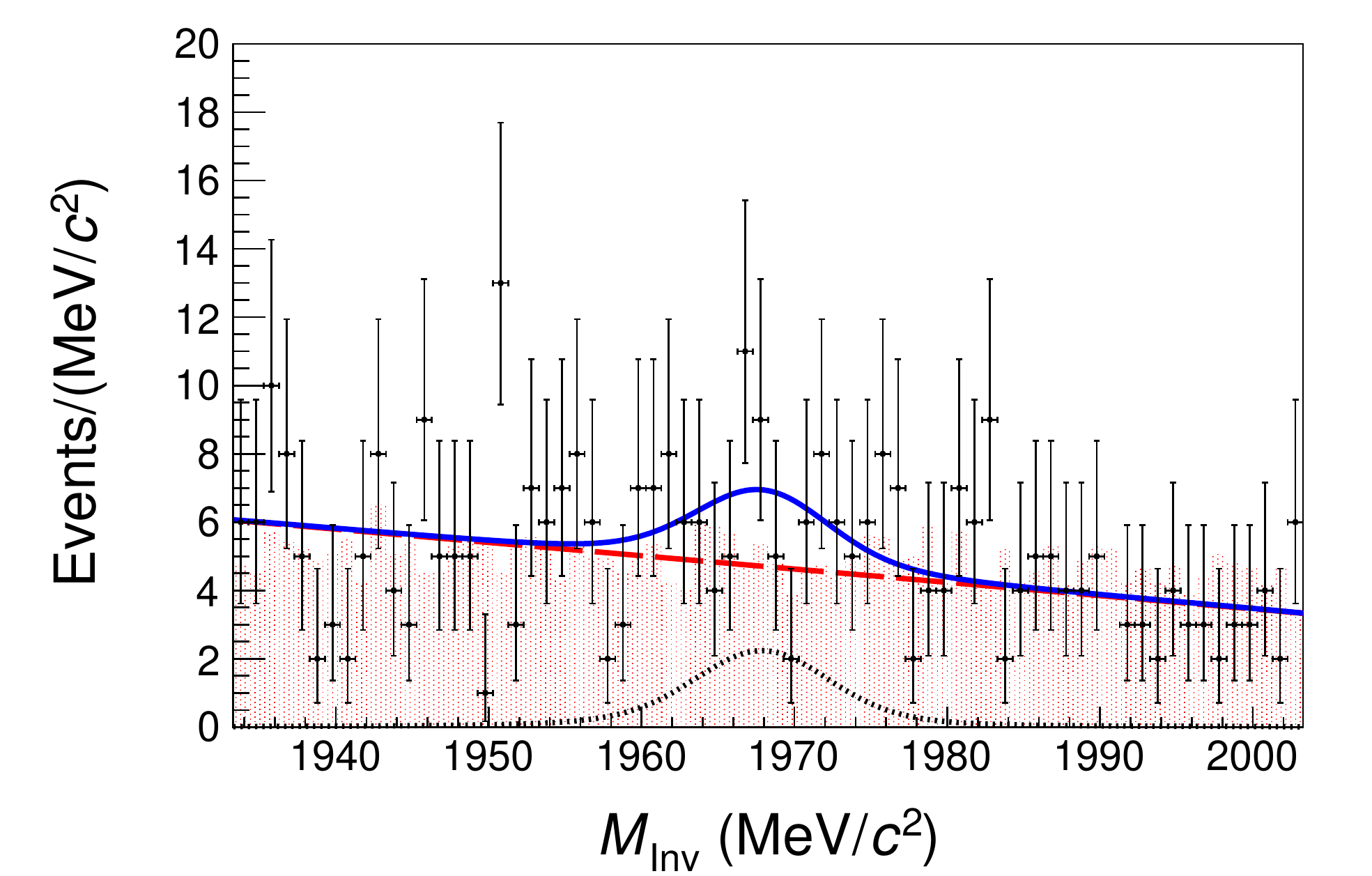}\\
\textbf{RS 500-550 MeV/$c$} & \textbf{WS 500-550 MeV/$c$}\\
\includegraphics[width=3in]{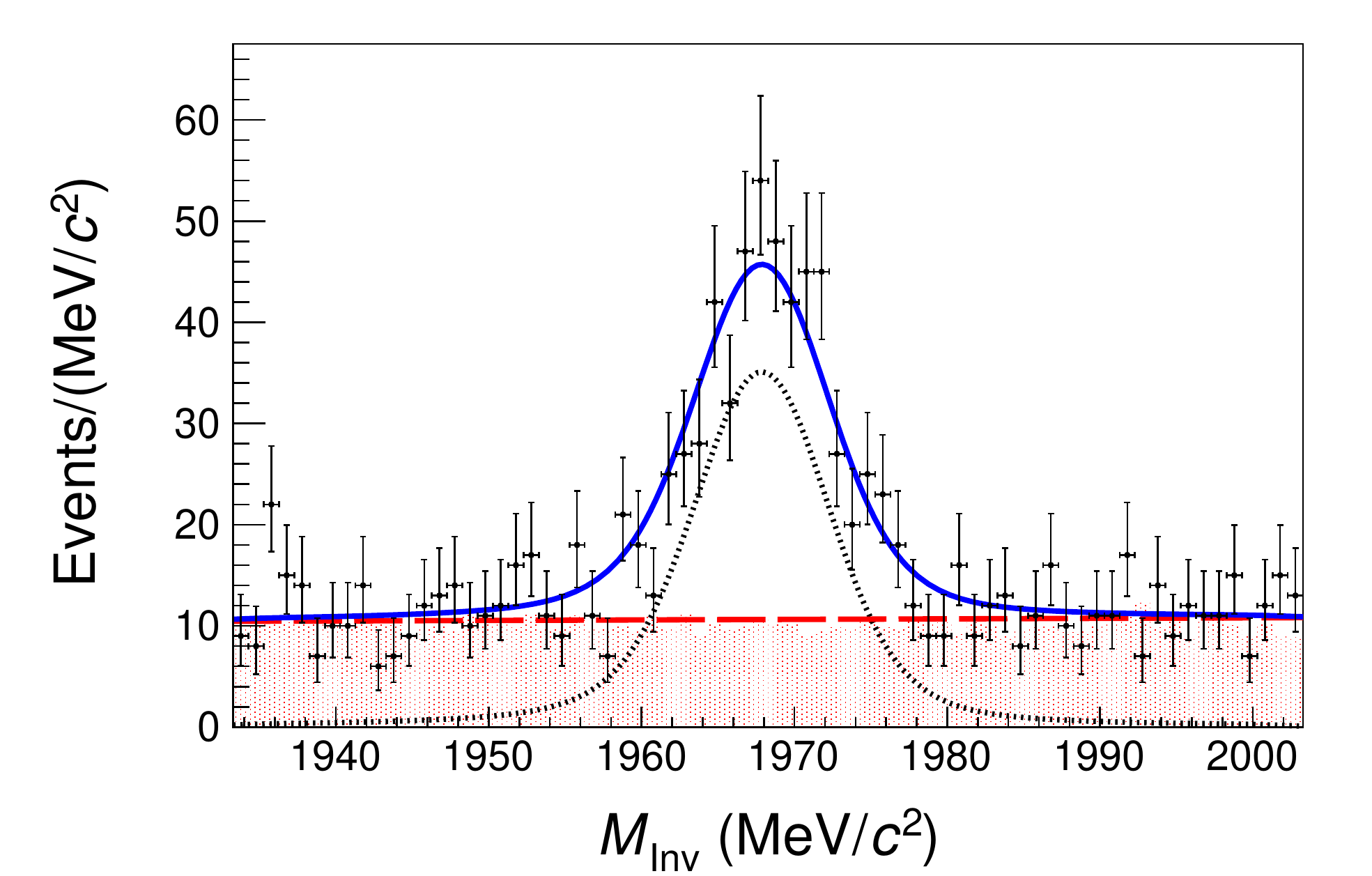} & \includegraphics[width=3in]{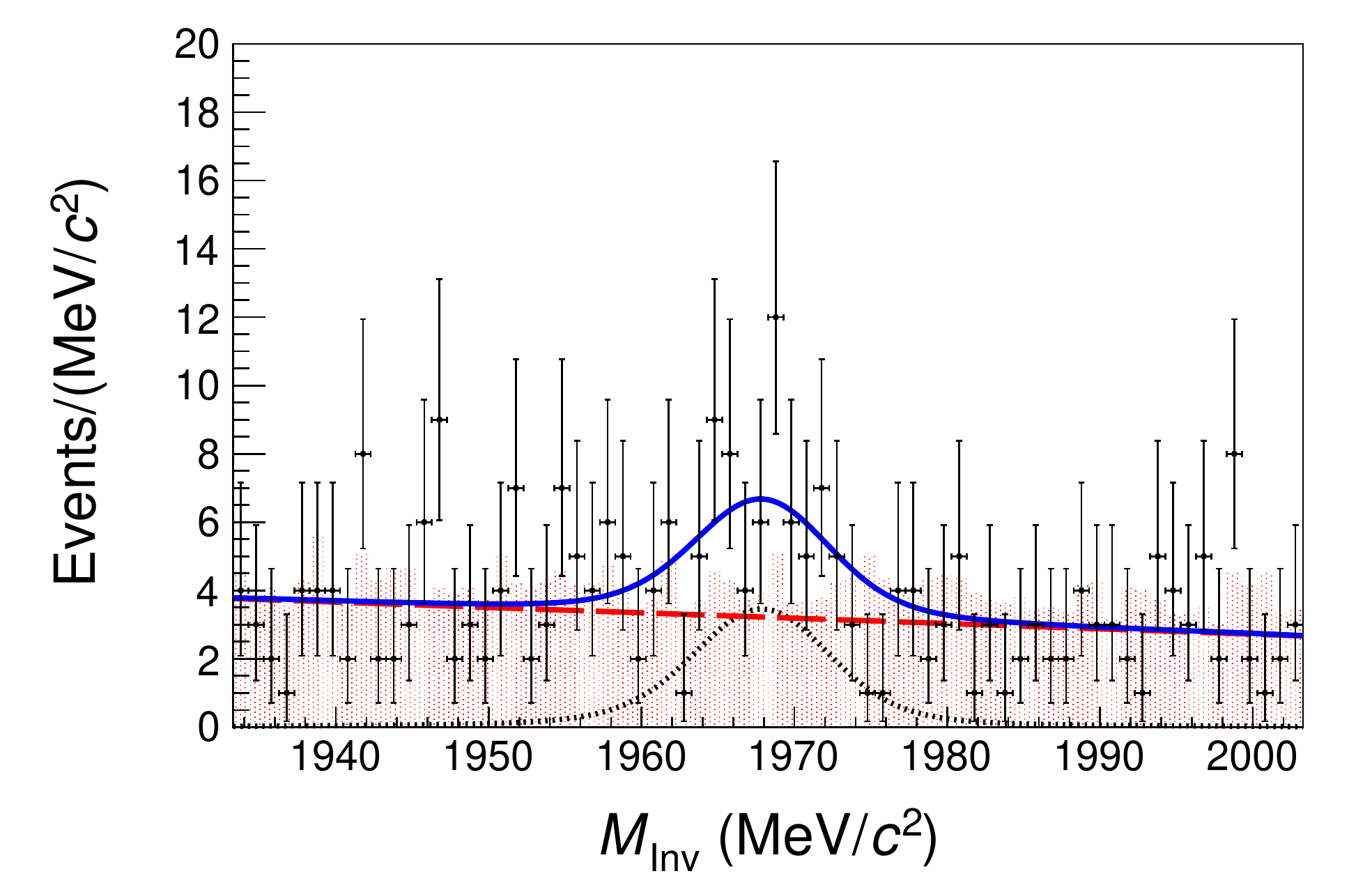}\\
\textbf{RS 550-600 MeV/$c$} & \textbf{WS 550-600 MeV/$c$}\\
\includegraphics[width=3in]{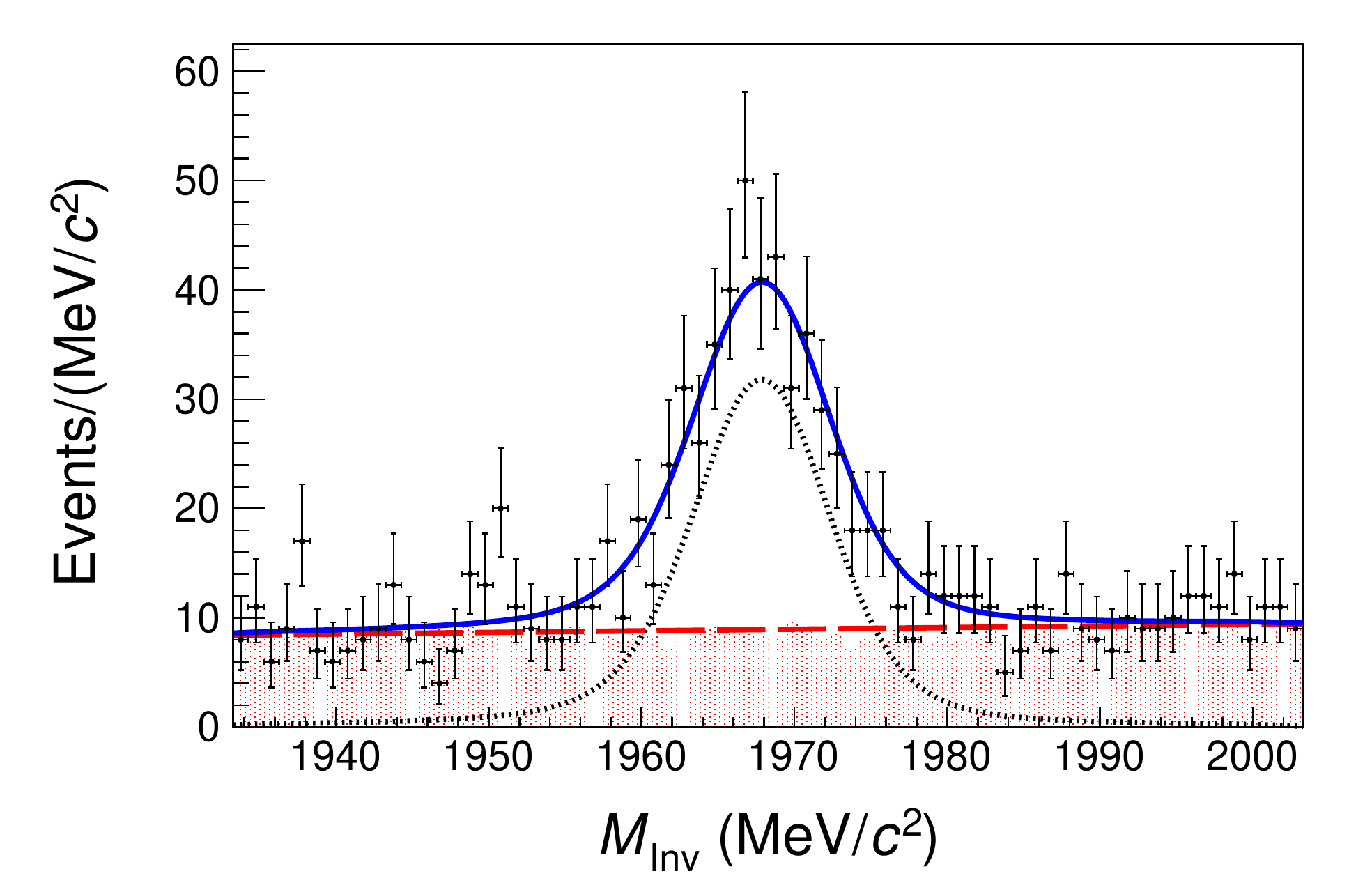} & \includegraphics[width=3in]{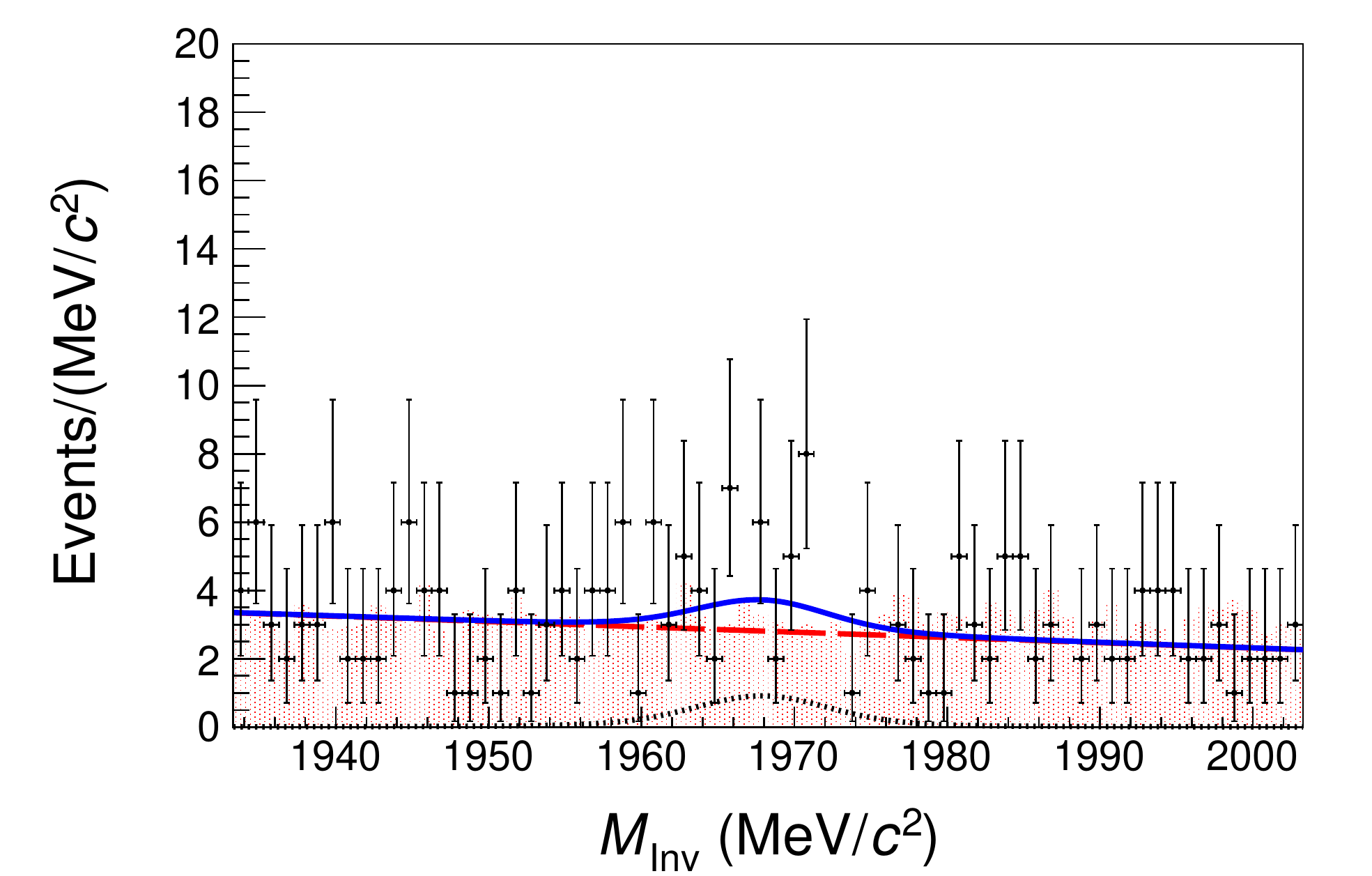}
\end{tabular}

\begin{tabular}{cc}
\textbf{RS 600-650 MeV/$c$} & \textbf{WS 600-650 MeV/$c$}\\
\includegraphics[width=3in]{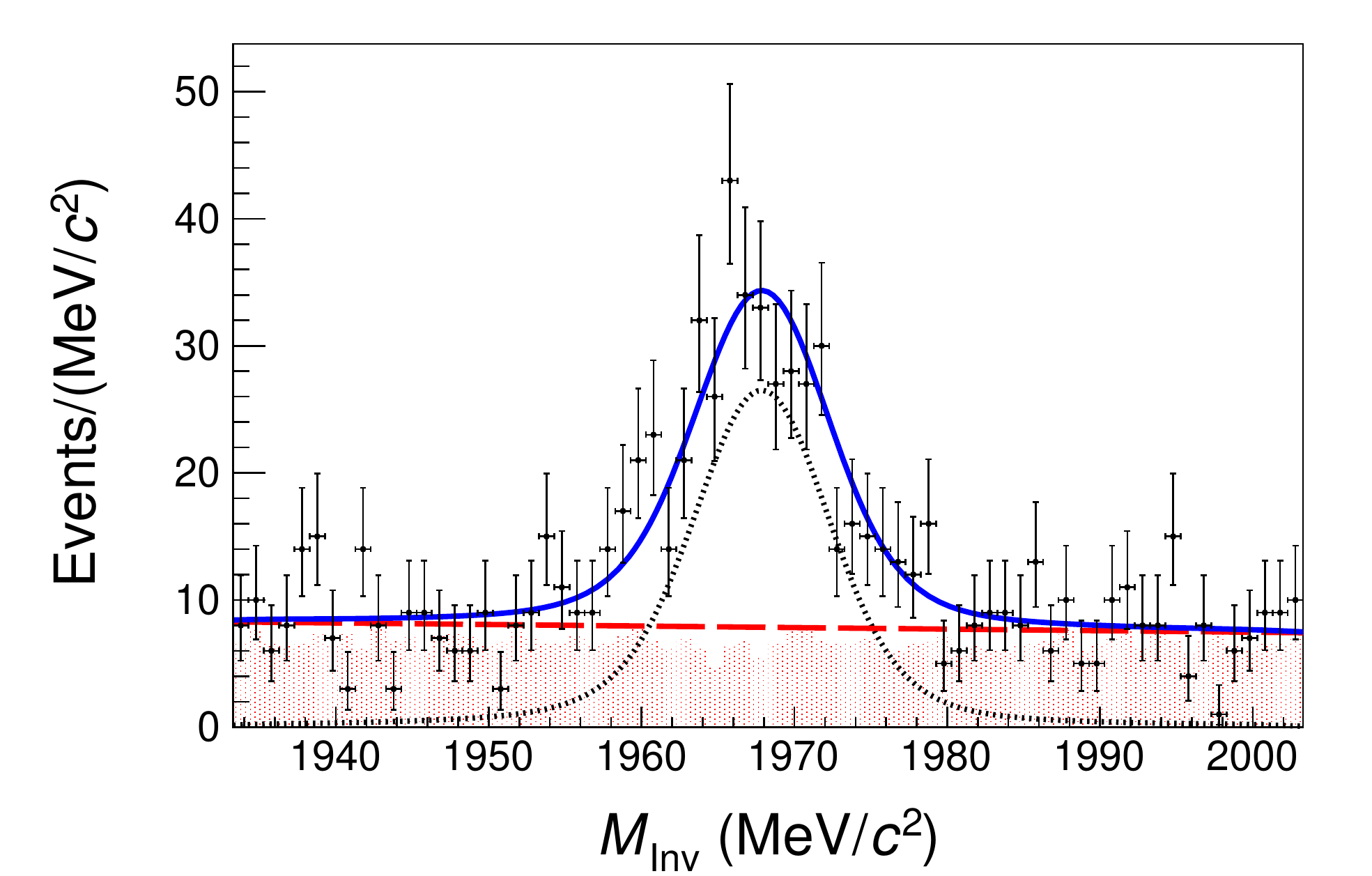} & \includegraphics[width=3in]{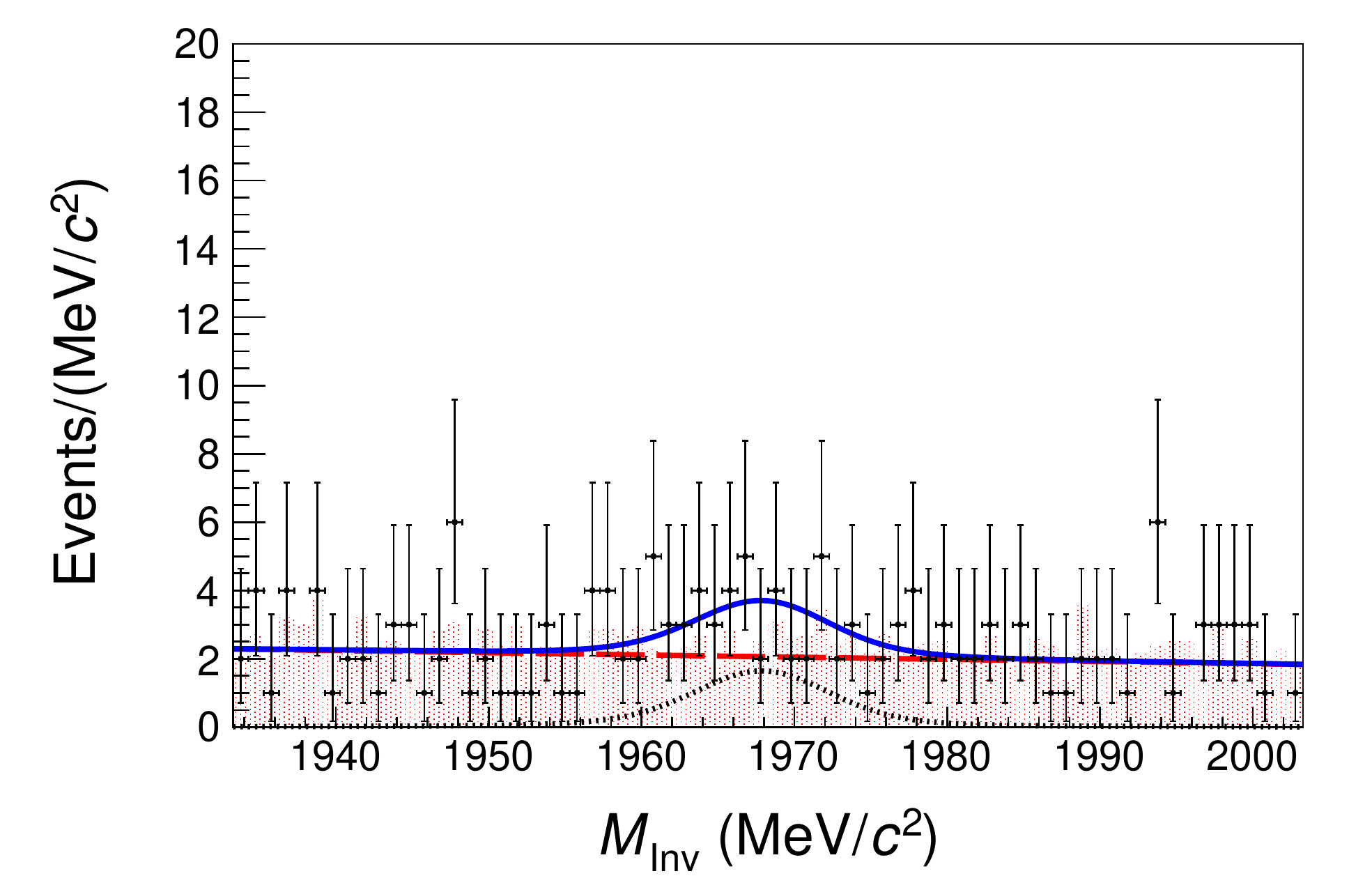}\\
\textbf{RS 650-700 MeV/$c$} & \textbf{WS 650-700 MeV/$c$}\\
\includegraphics[width=3in]{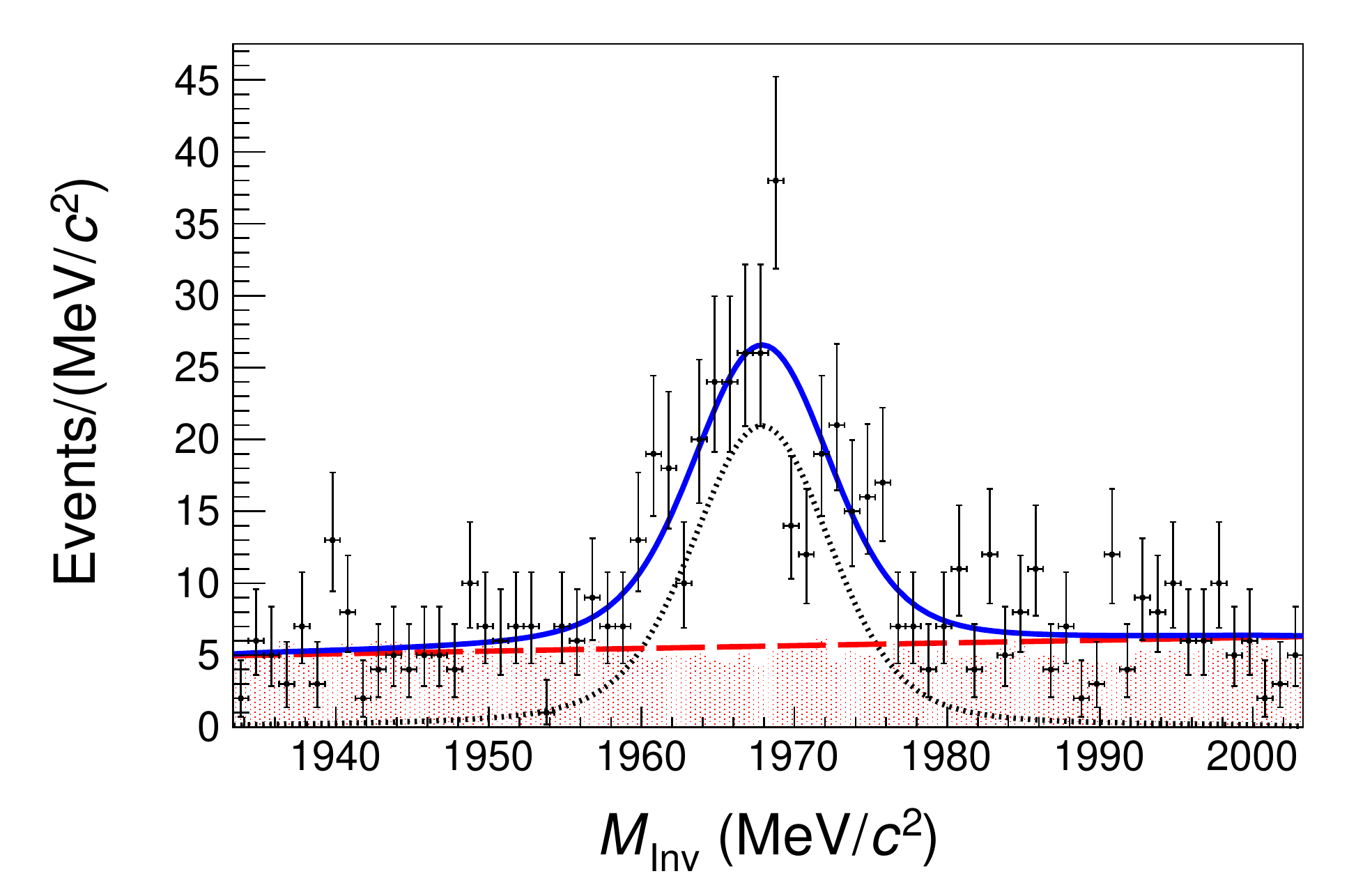} & \includegraphics[width=3in]{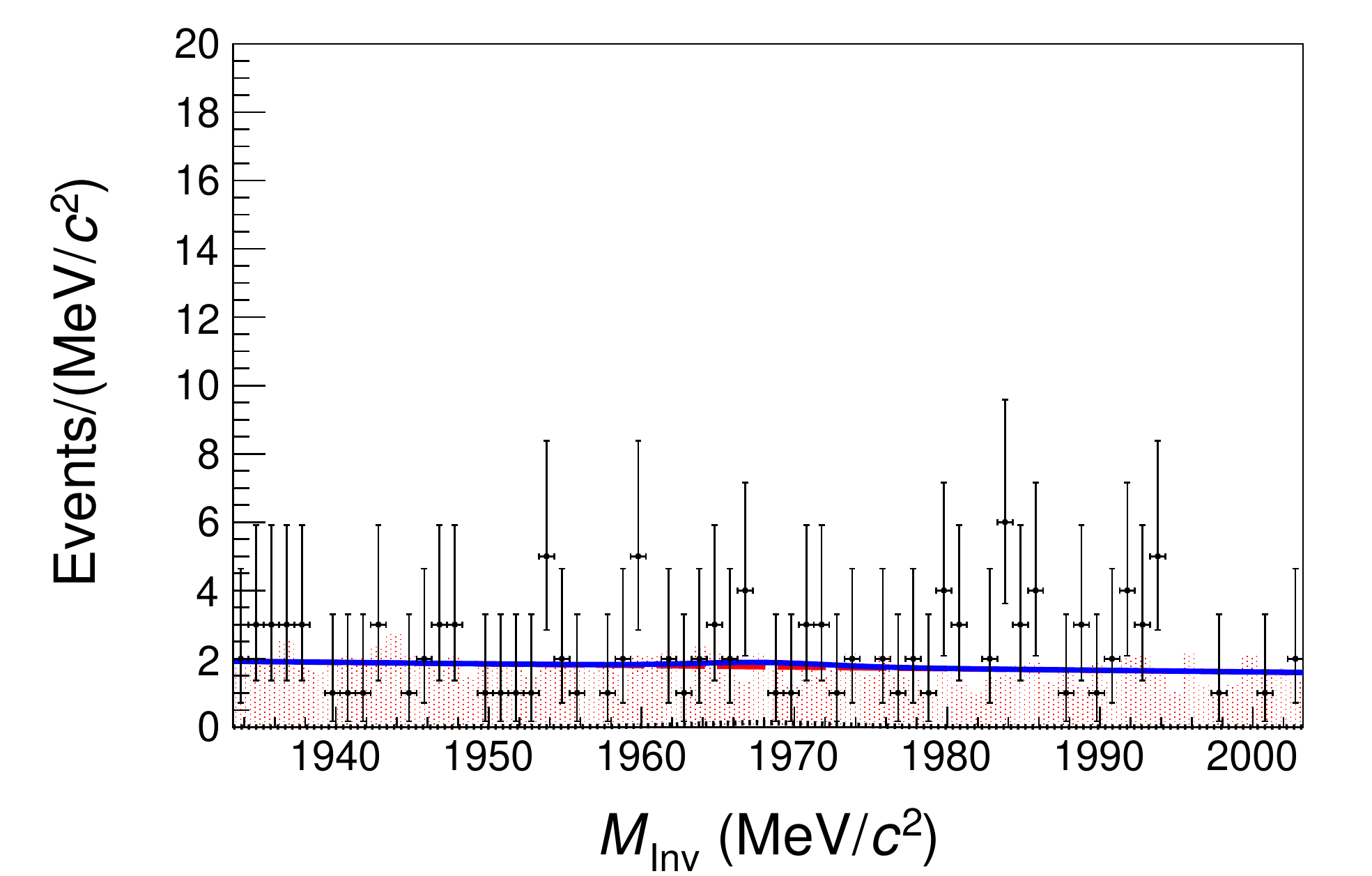}\\
\textbf{RS 700-750 MeV/$c$} & \textbf{WS 700-750 MeV/$c$}\\
\includegraphics[width=3in]{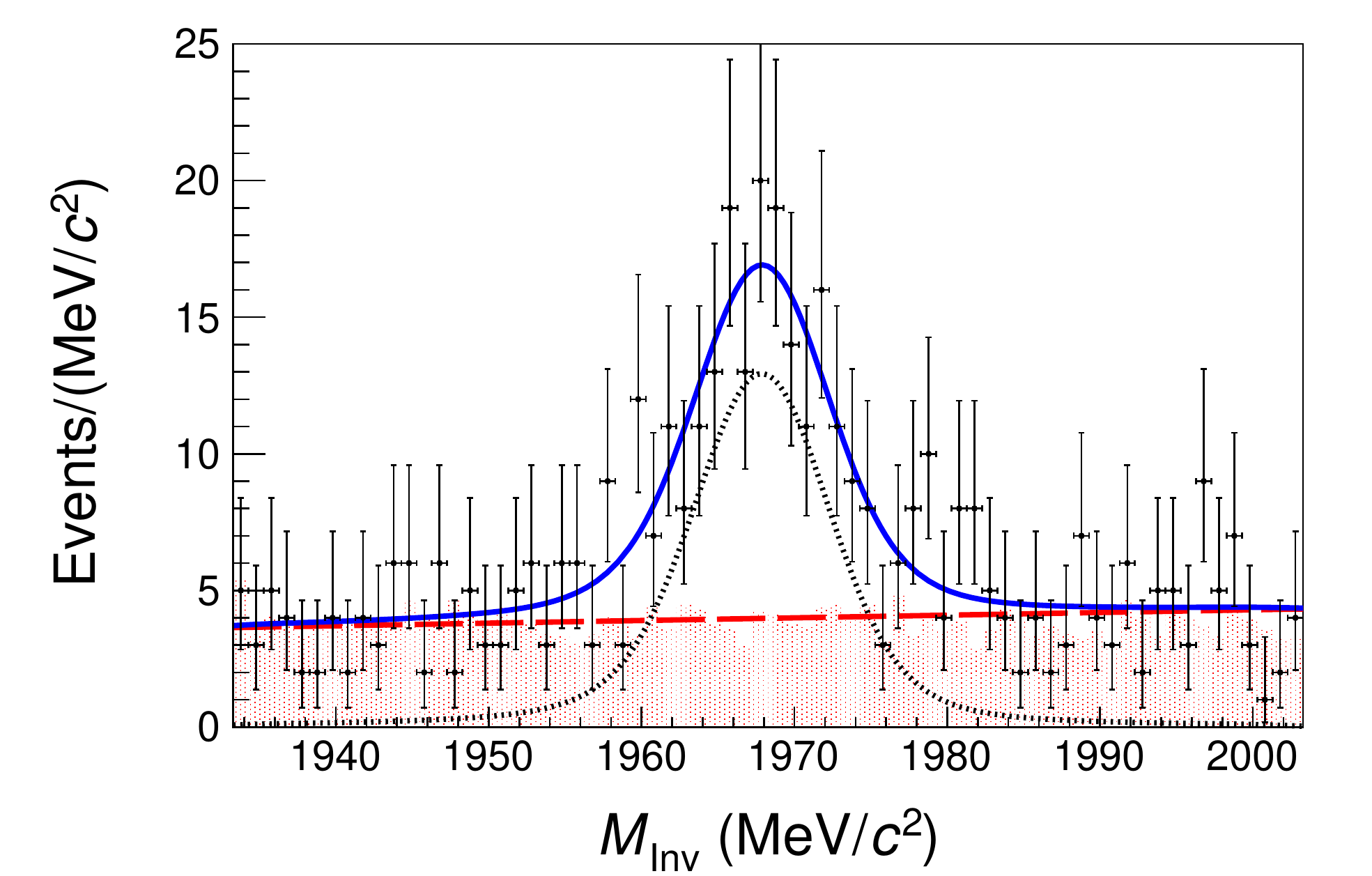} & \includegraphics[width=3in]{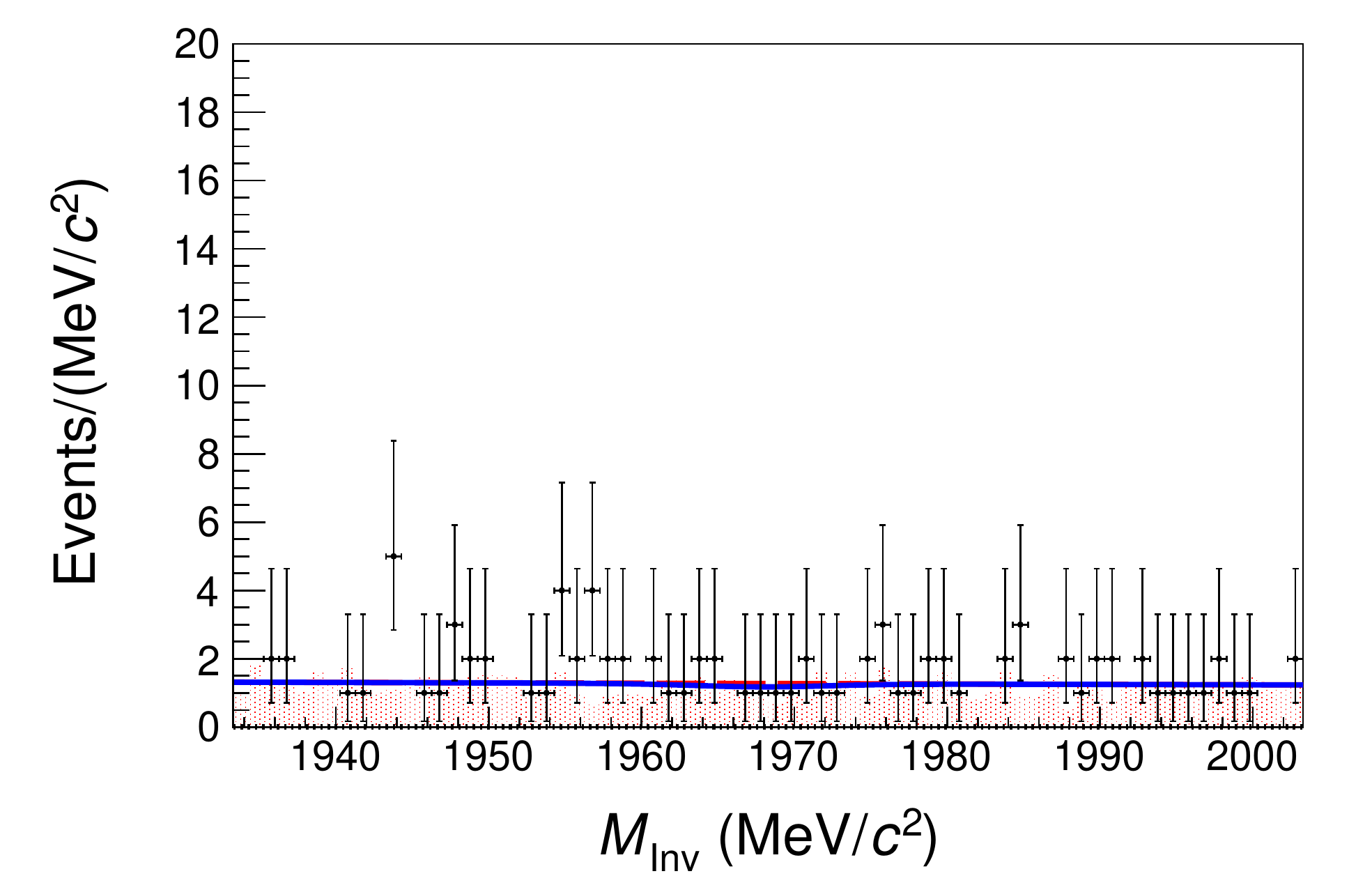}\\
\textbf{RS 750-800 MeV/$c$} & \textbf{WS 750-800 MeV/$c$}\\
\includegraphics[width=3in]{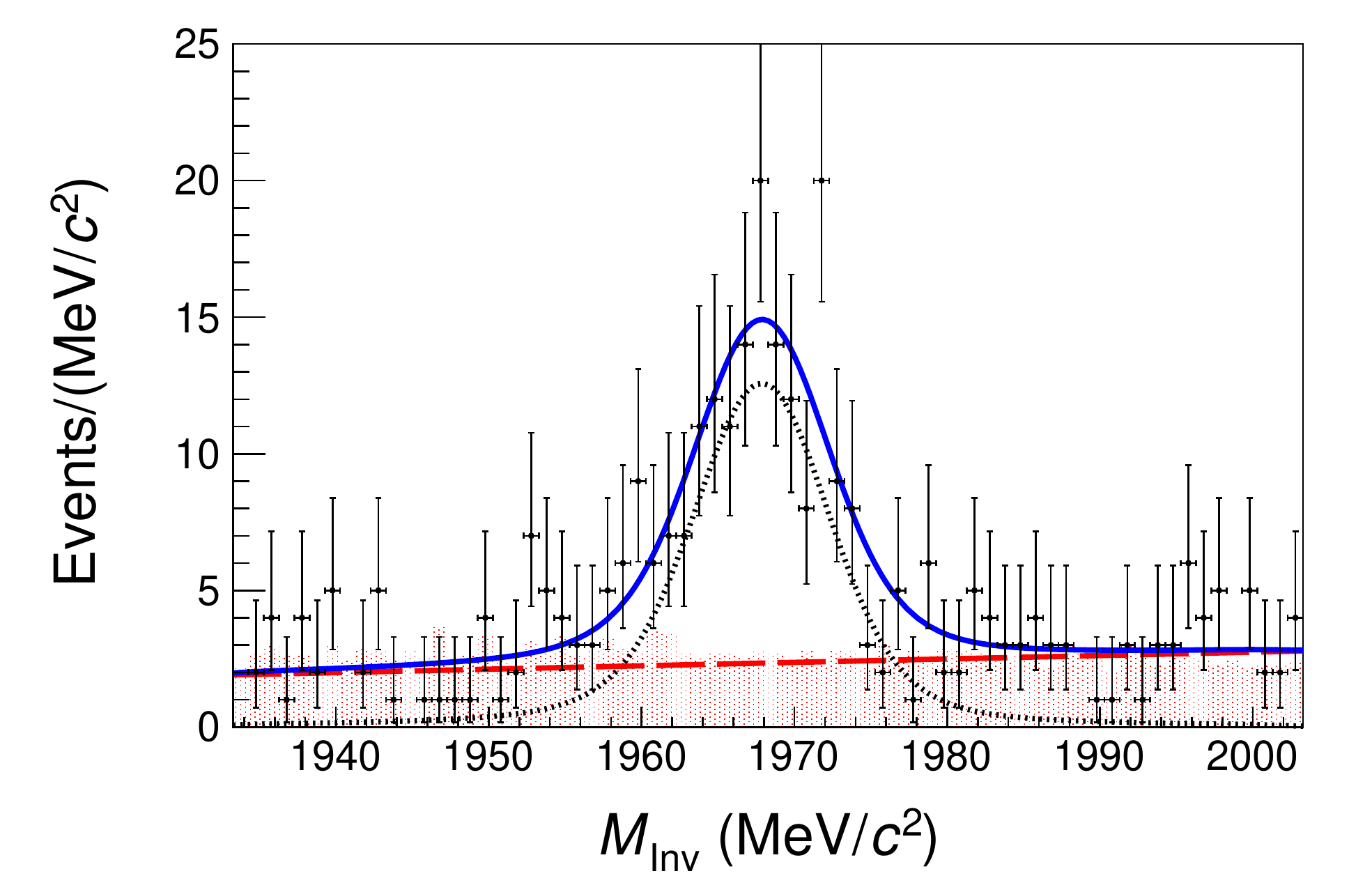} & \includegraphics[width=3in]{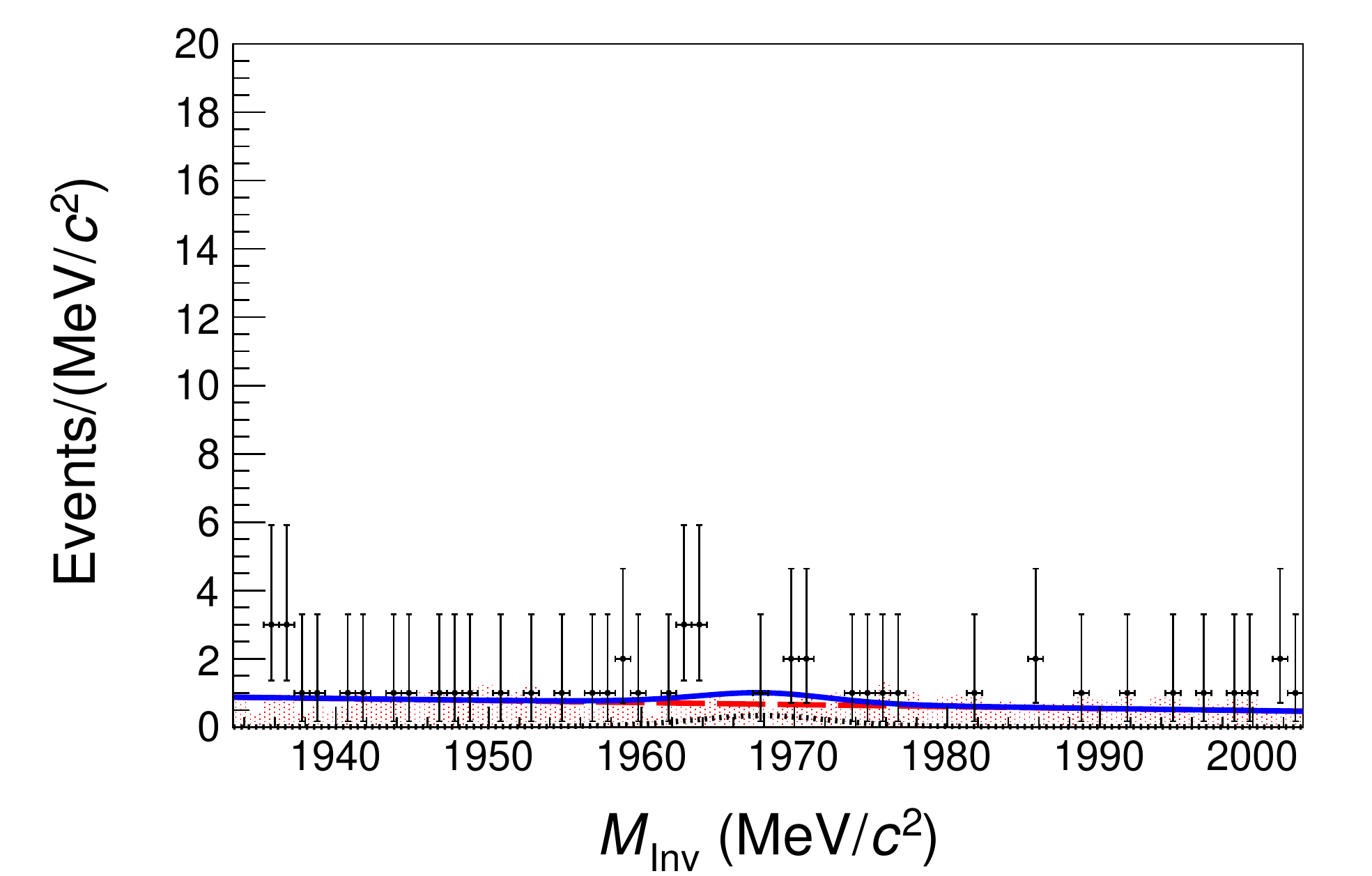}\\
\end{tabular}

\begin{tabular}{cc}
\textbf{RS 800-850 MeV/$c$} & \textbf{WS 800-850 MeV/$c$}\\
\includegraphics[width=3in]{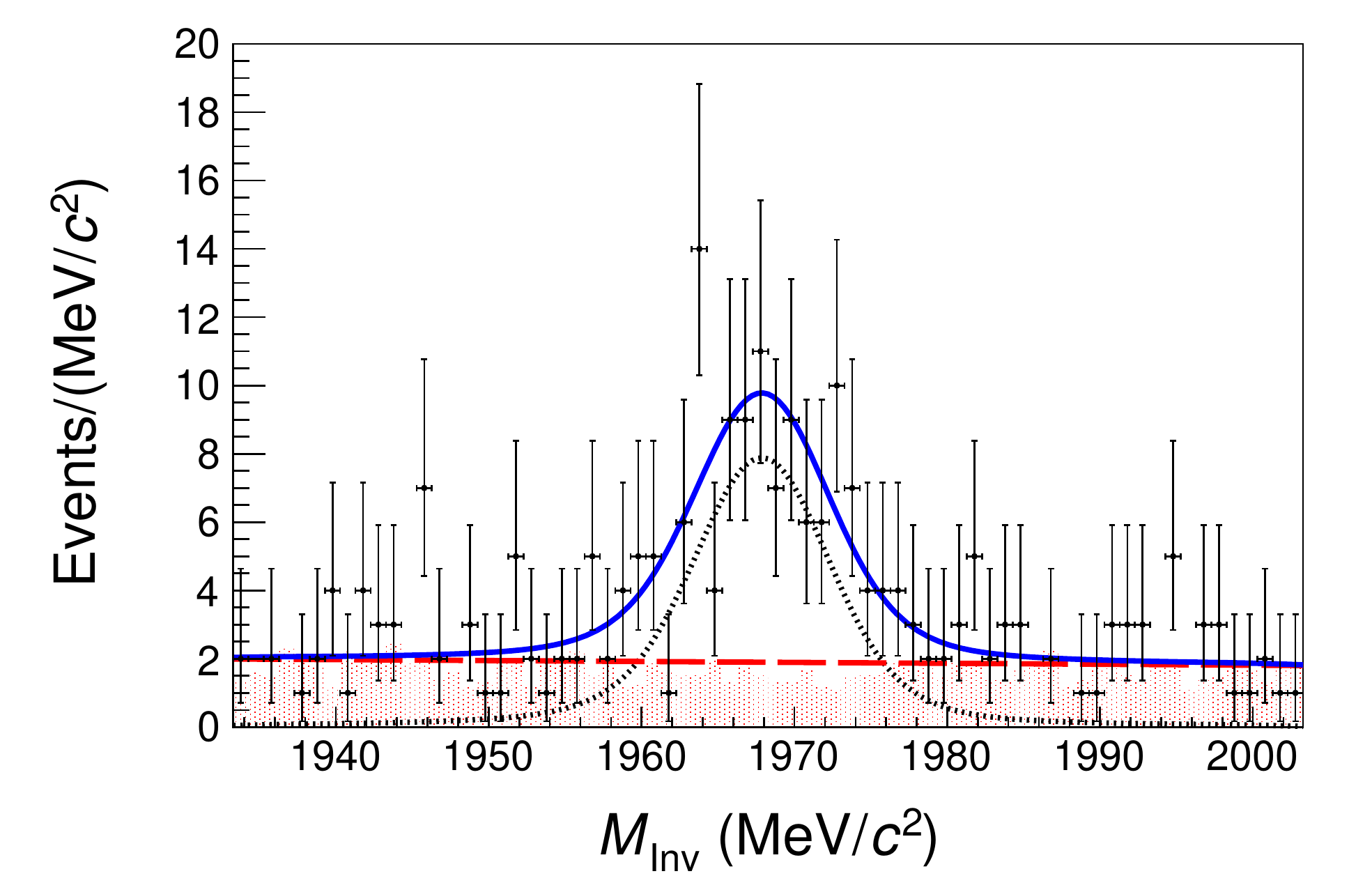} & \includegraphics[width=3in]{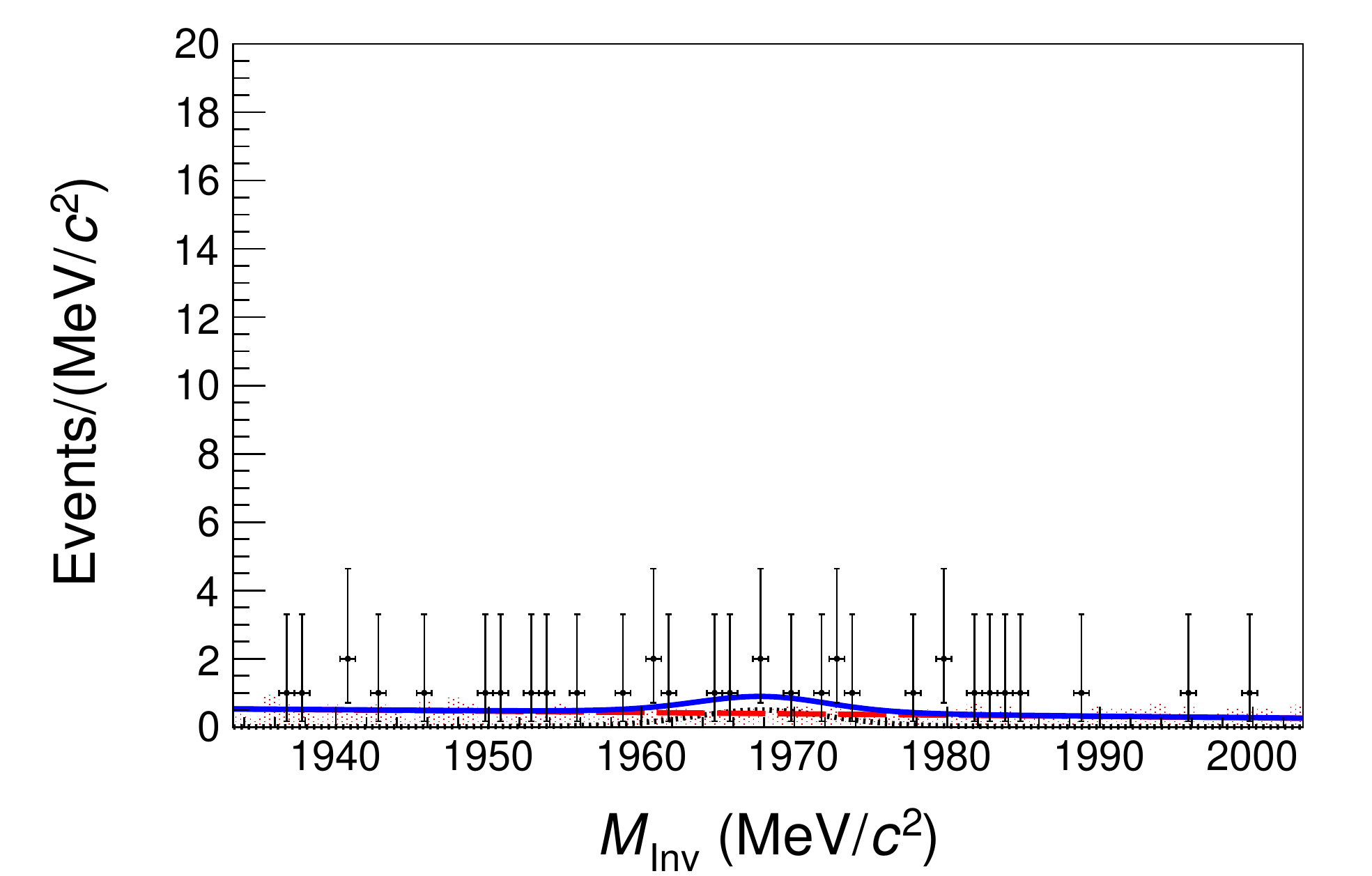}\\
\textbf{RS 850-900 MeV/$c$} & \textbf{WS 850-900 MeV/$c$}\\
\includegraphics[width=3in]{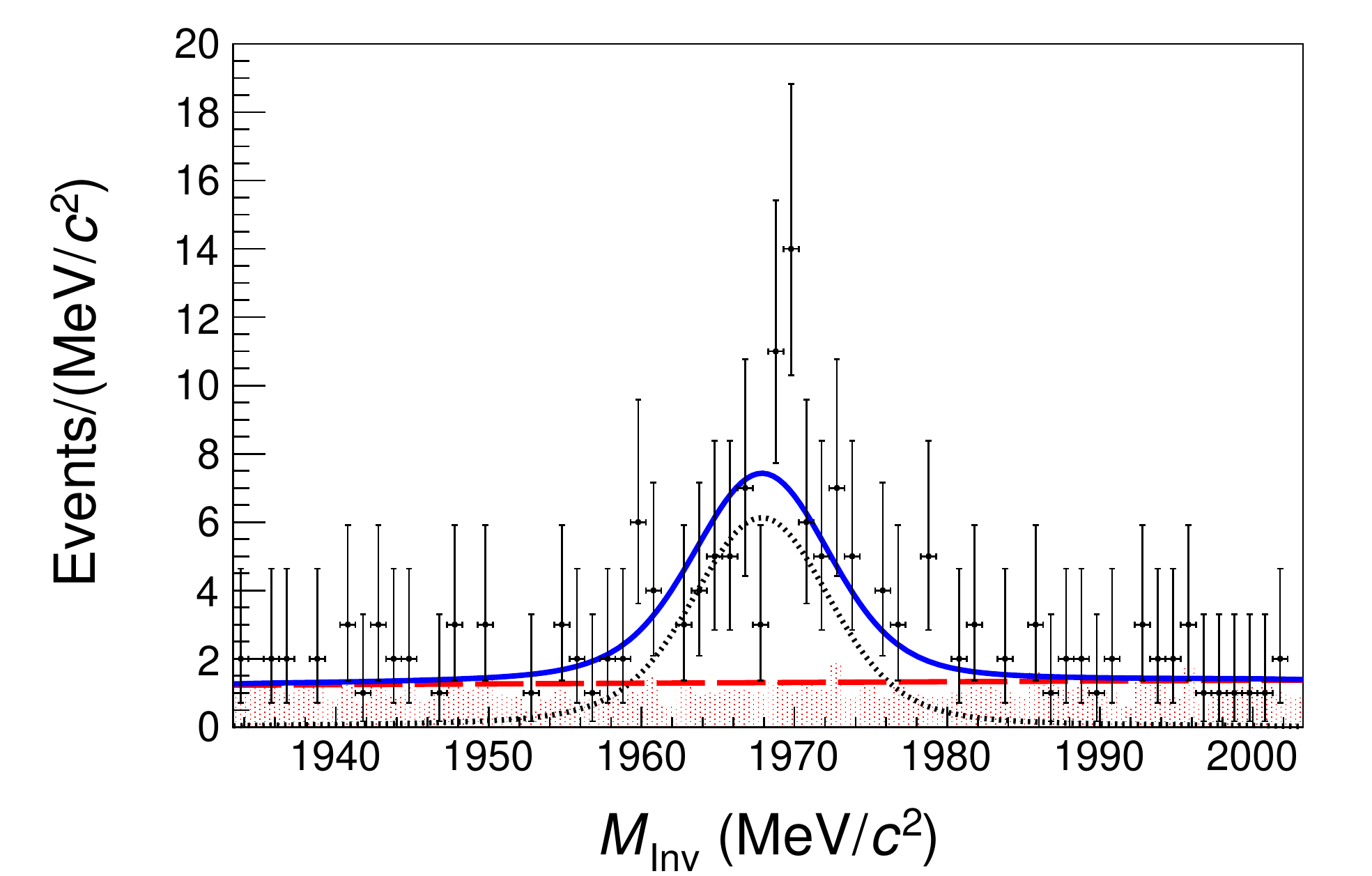} & \includegraphics[width=3in]{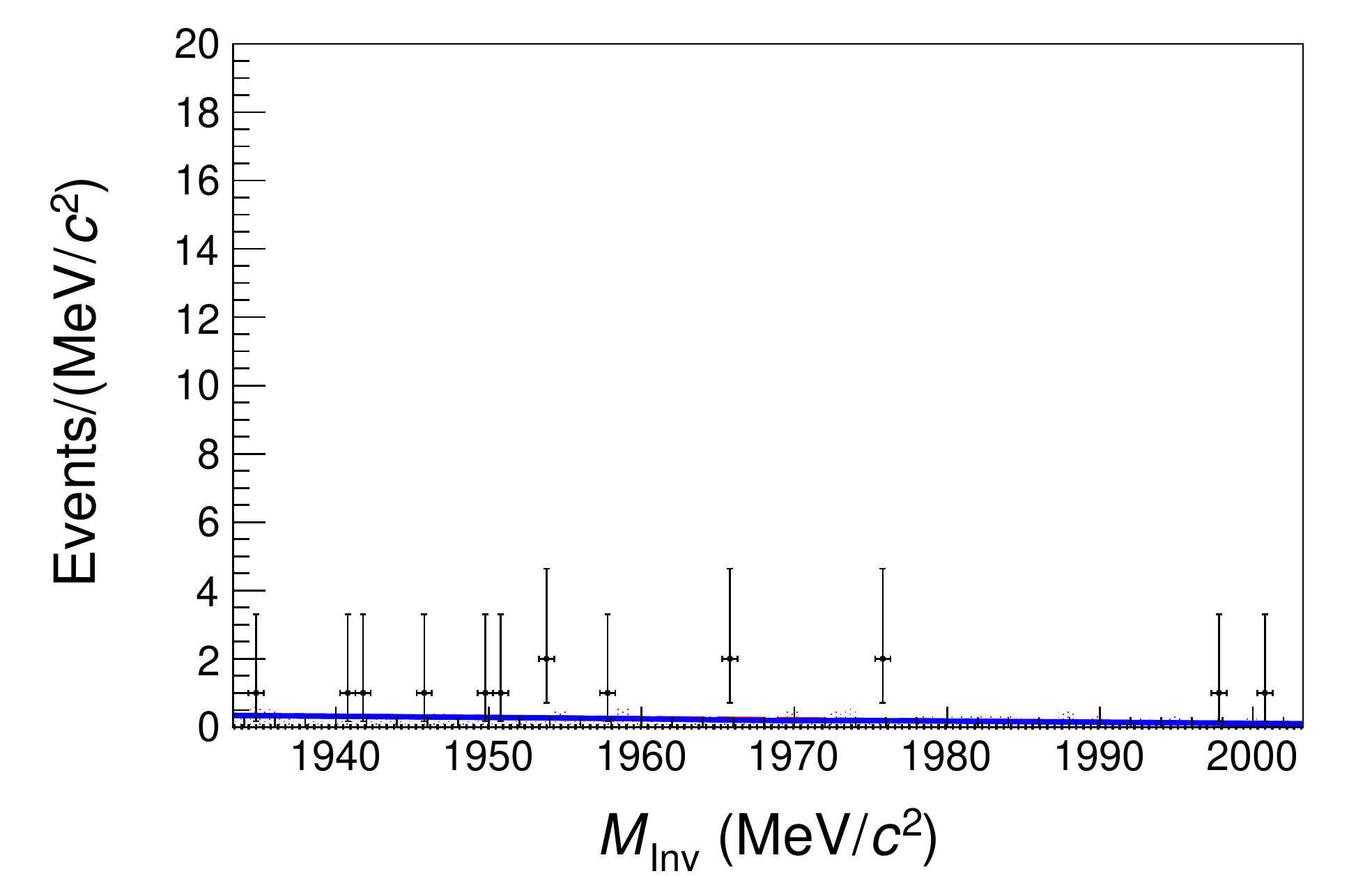}\\
\textbf{RS 900-950 MeV/$c$} & \textbf{WS 900-950 MeV/$c$}\\
\includegraphics[width=3in]{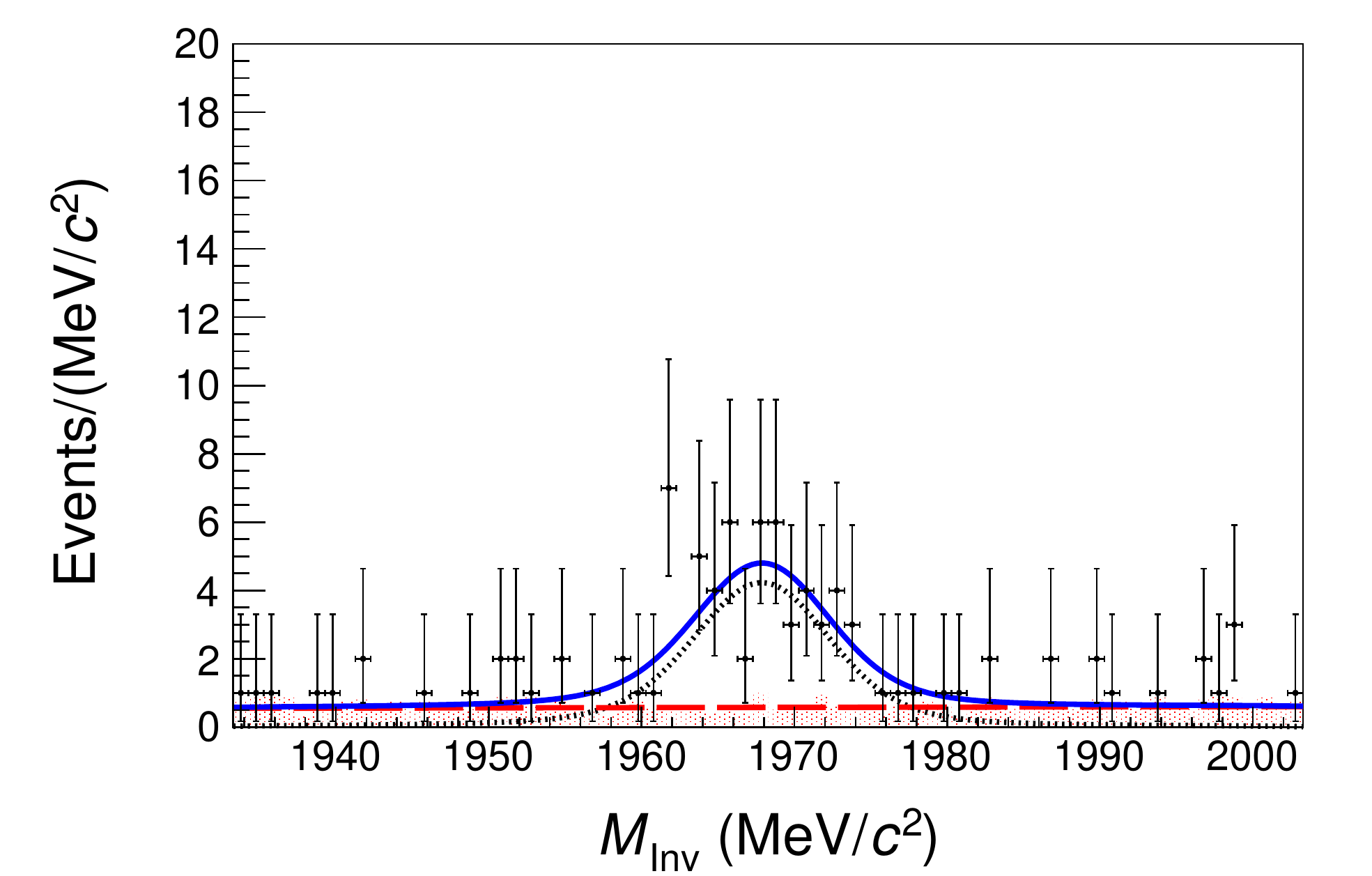} & \includegraphics[width=3in]{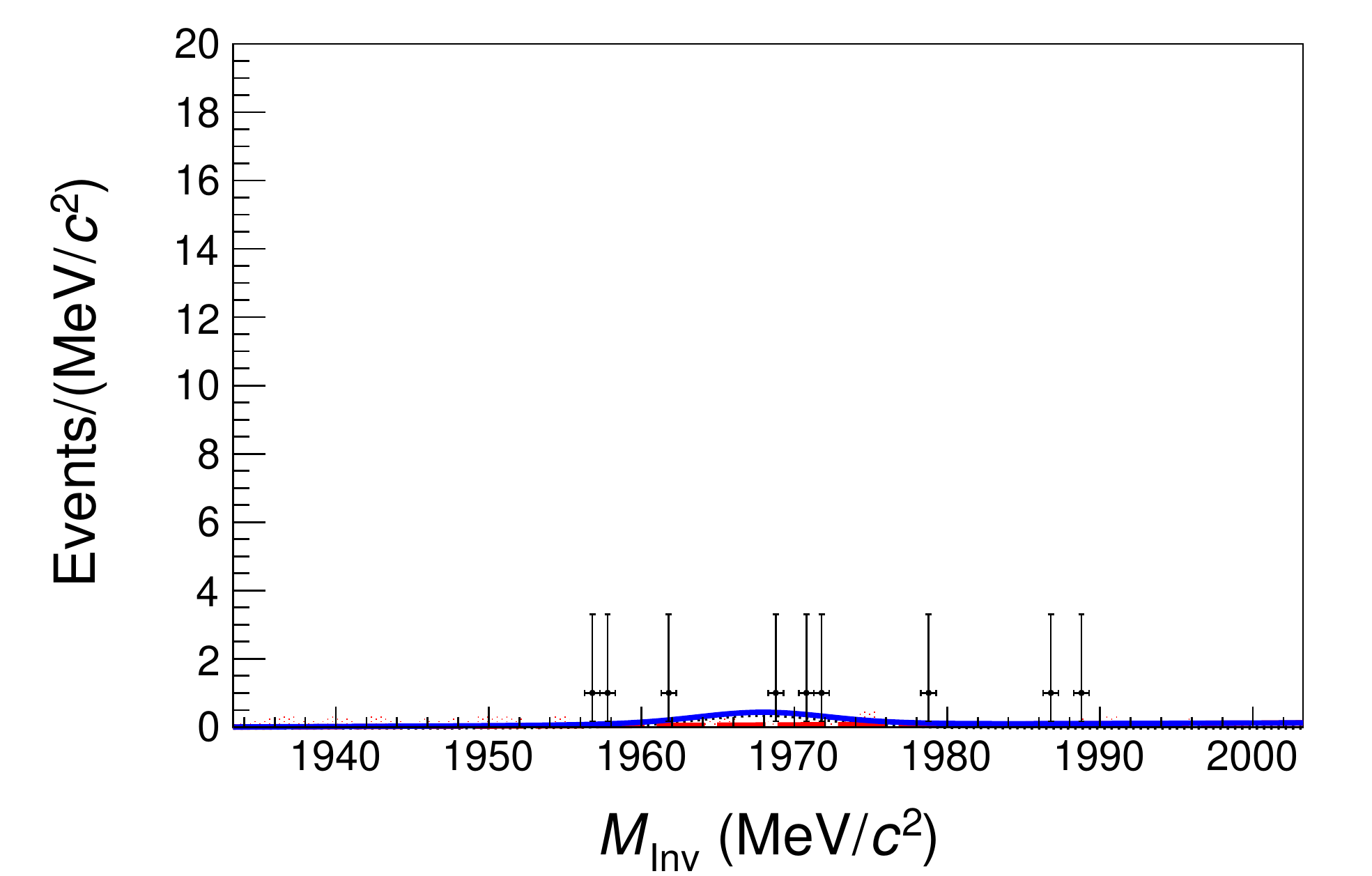}\\
\textbf{RS 950-1000 MeV/$c$} & \textbf{WS 950-1000 MeV/$c$}\\
\includegraphics[width=3in,valign=m]{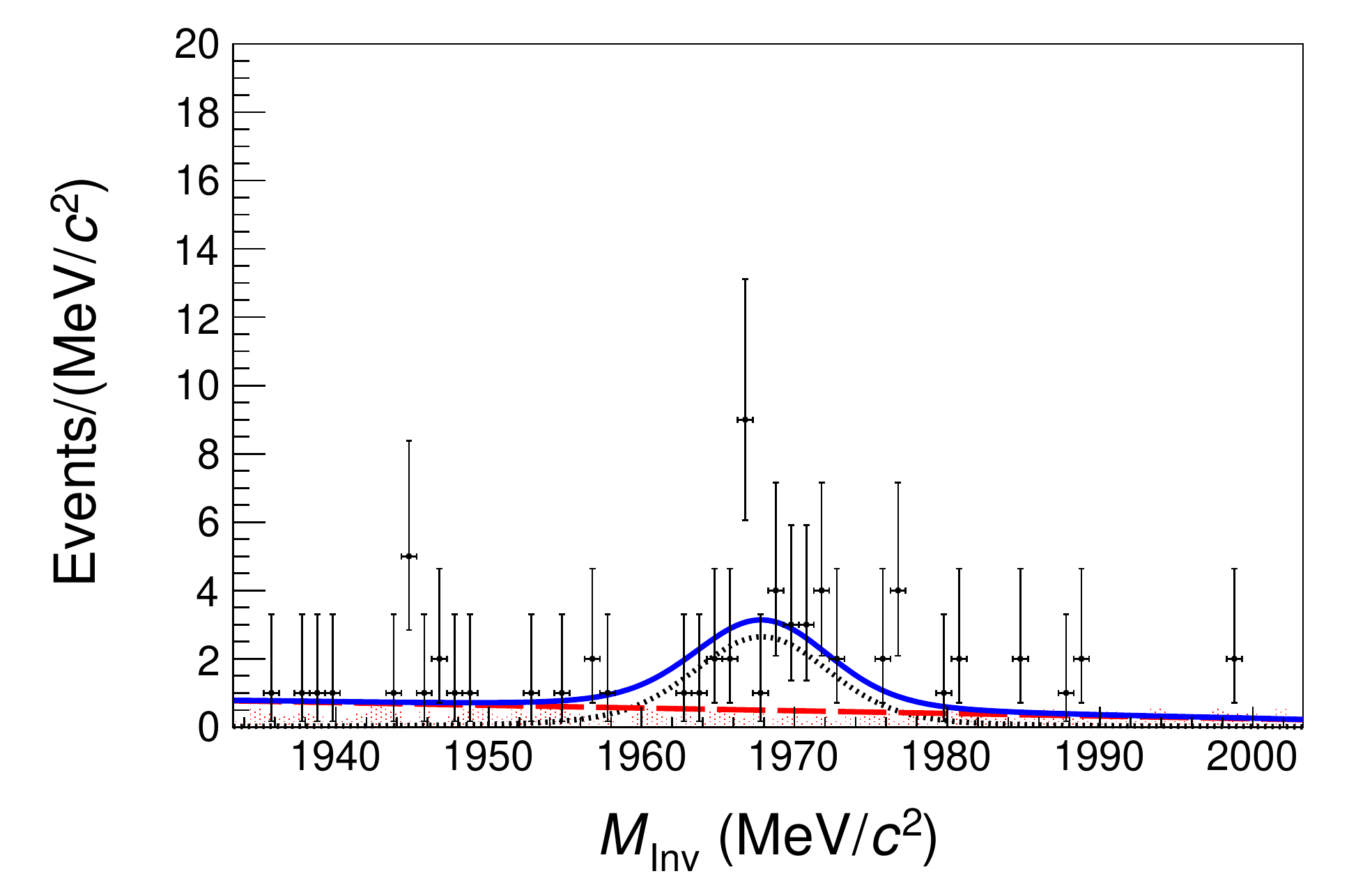} & No entries are seen in the signal region\\
\end{tabular}

\begin{tabular}{cc}
\textbf{RS 1000-1050 MeV/$c$} & \textbf{WS 1000-1050 MeV/$c$}\\
\includegraphics[width=3in,valign=m]{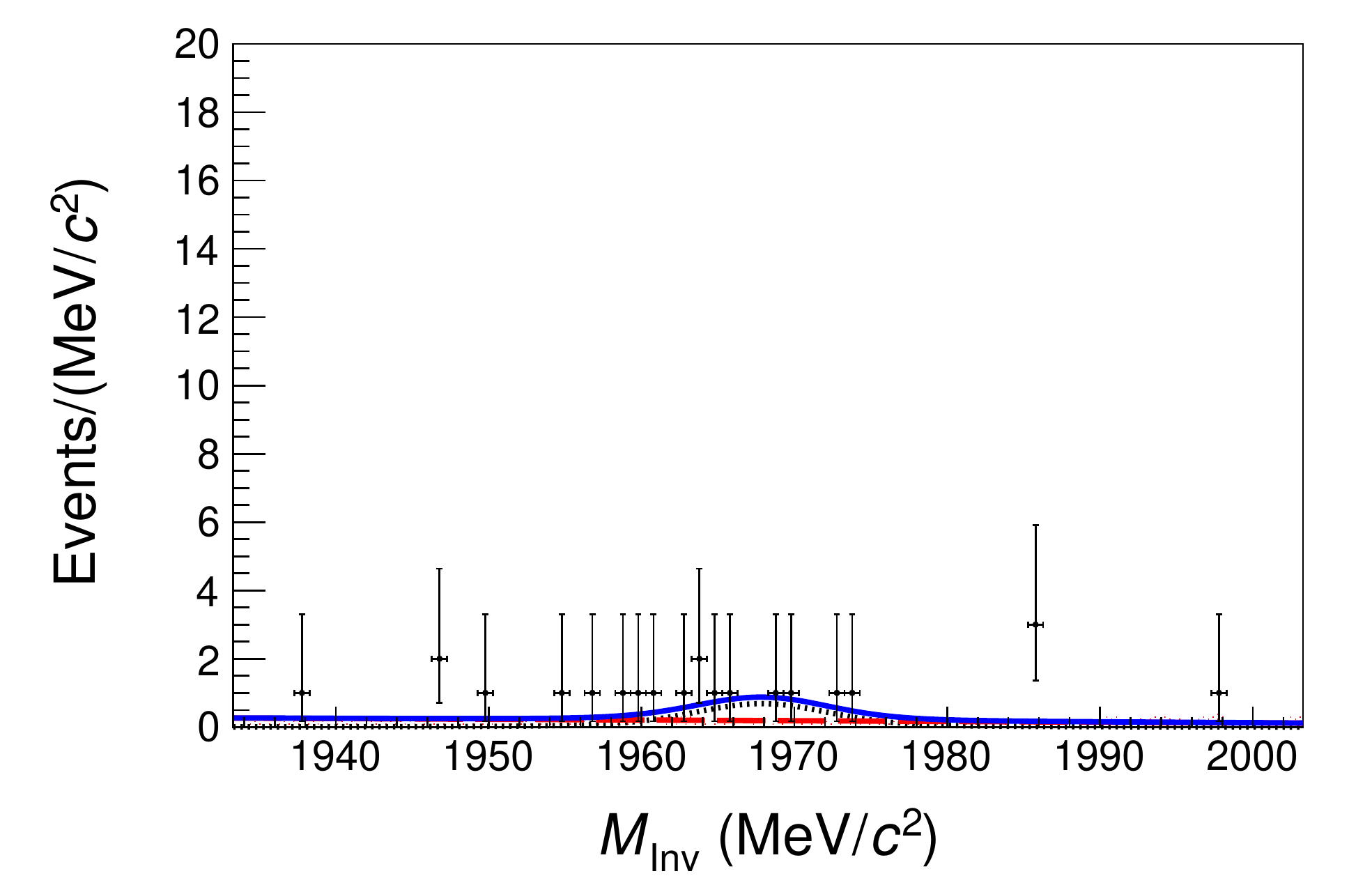} & No entries are seen in the signal region\\
\textbf{RS 1050-1100 MeV/$c$} & \textbf{WS 1050-1100 MeV/$c$}\\
\includegraphics[width=3in,valign=m]{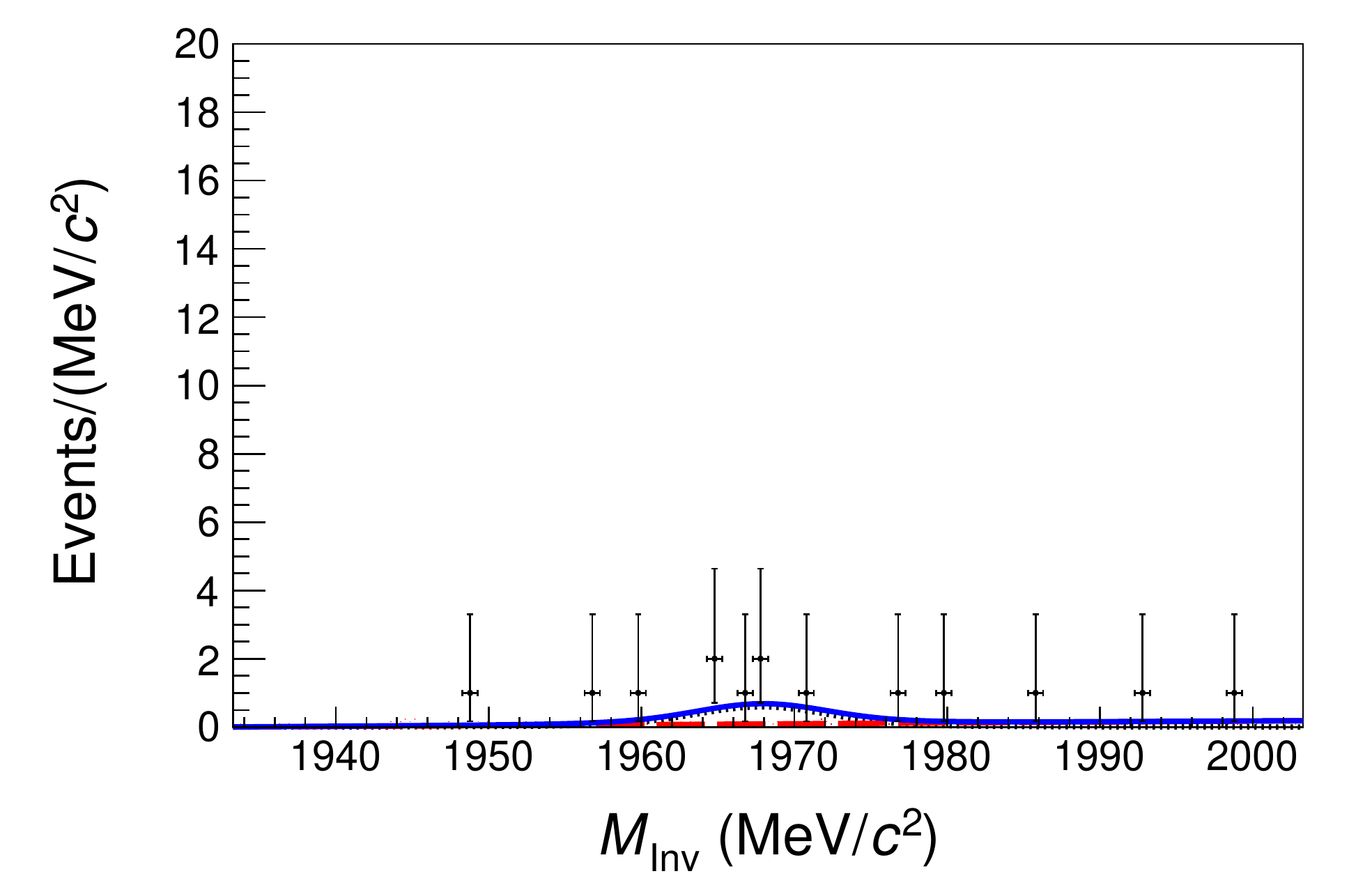} & No entries are seen in the signal region
\end{tabular}

\pagebreak
\raggedright

\subsubsection{$\EcmB$ Data $\pi$ ID Fits}
\label{subsubsec:XYZDataPiIDFits}
\centering
\begin{tabular}{cc}
\textbf{RS 200-250 MeV/$c$} & \textbf{WS 200-250 MeV/$c$}\\
\includegraphics[width=3in]{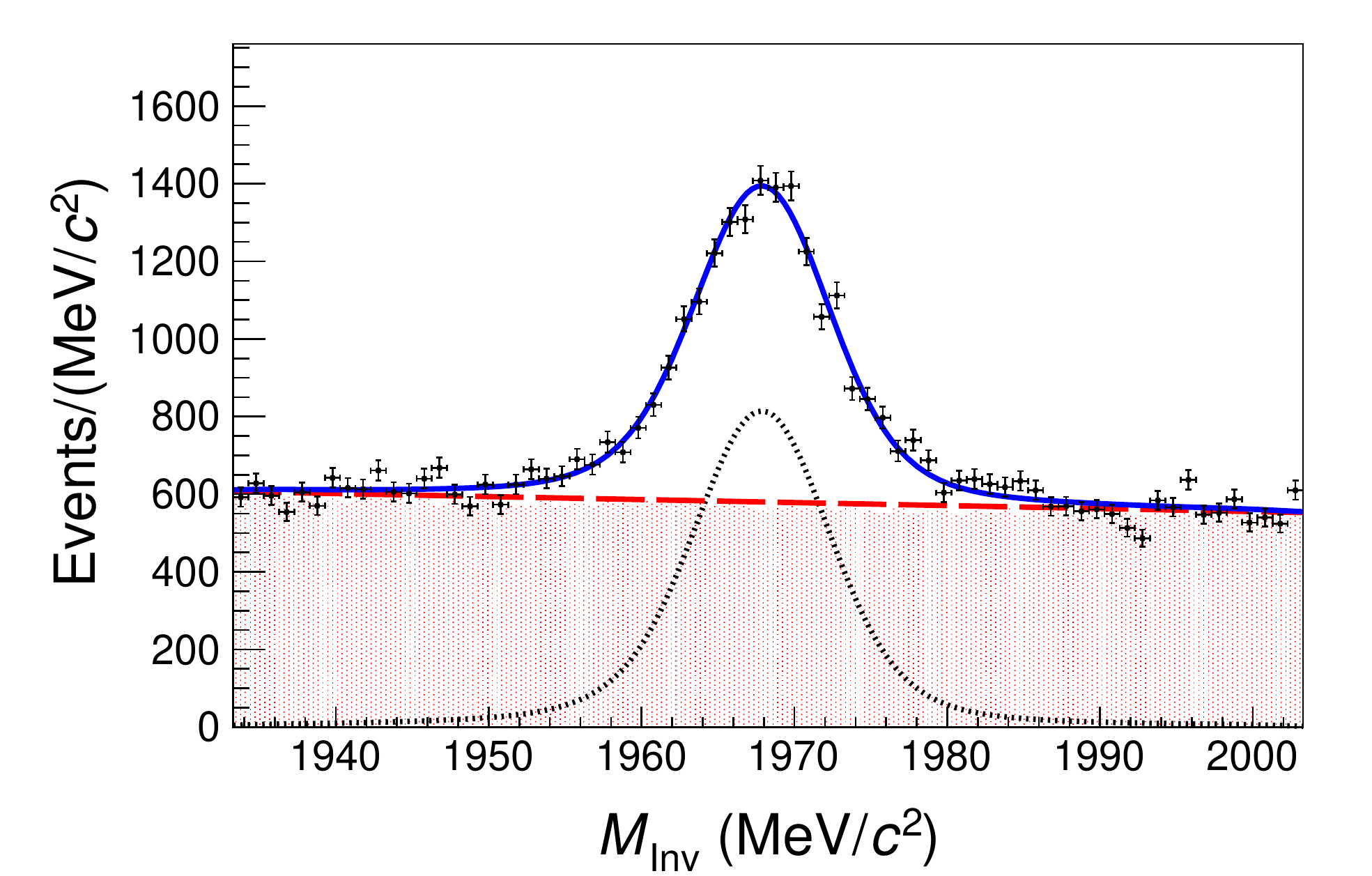} & \includegraphics[width=3in]{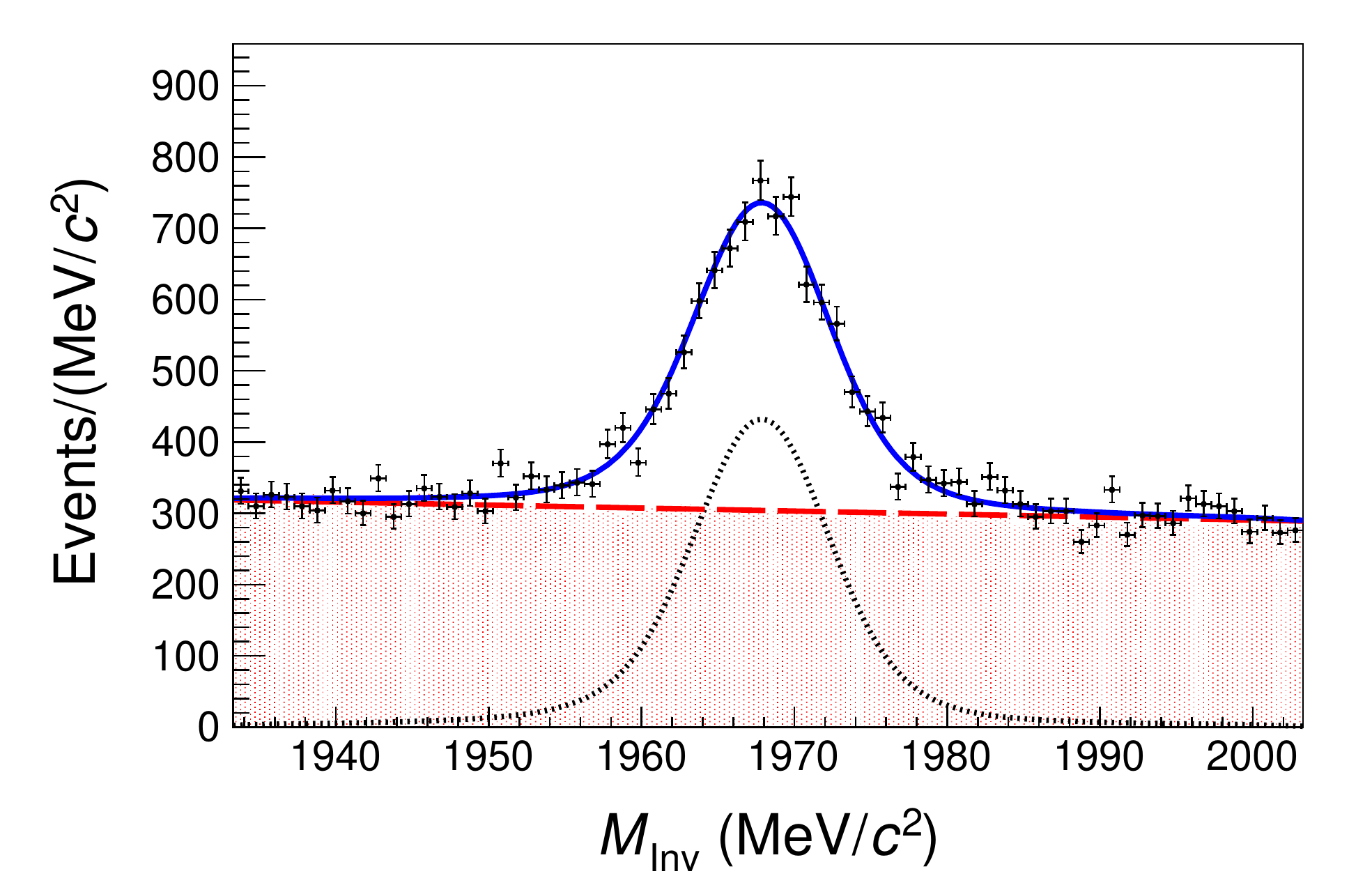}\\
\textbf{RS 250-300 MeV/$c$} & \textbf{WS 250-300 MeV/$c$}\\
\includegraphics[width=3in]{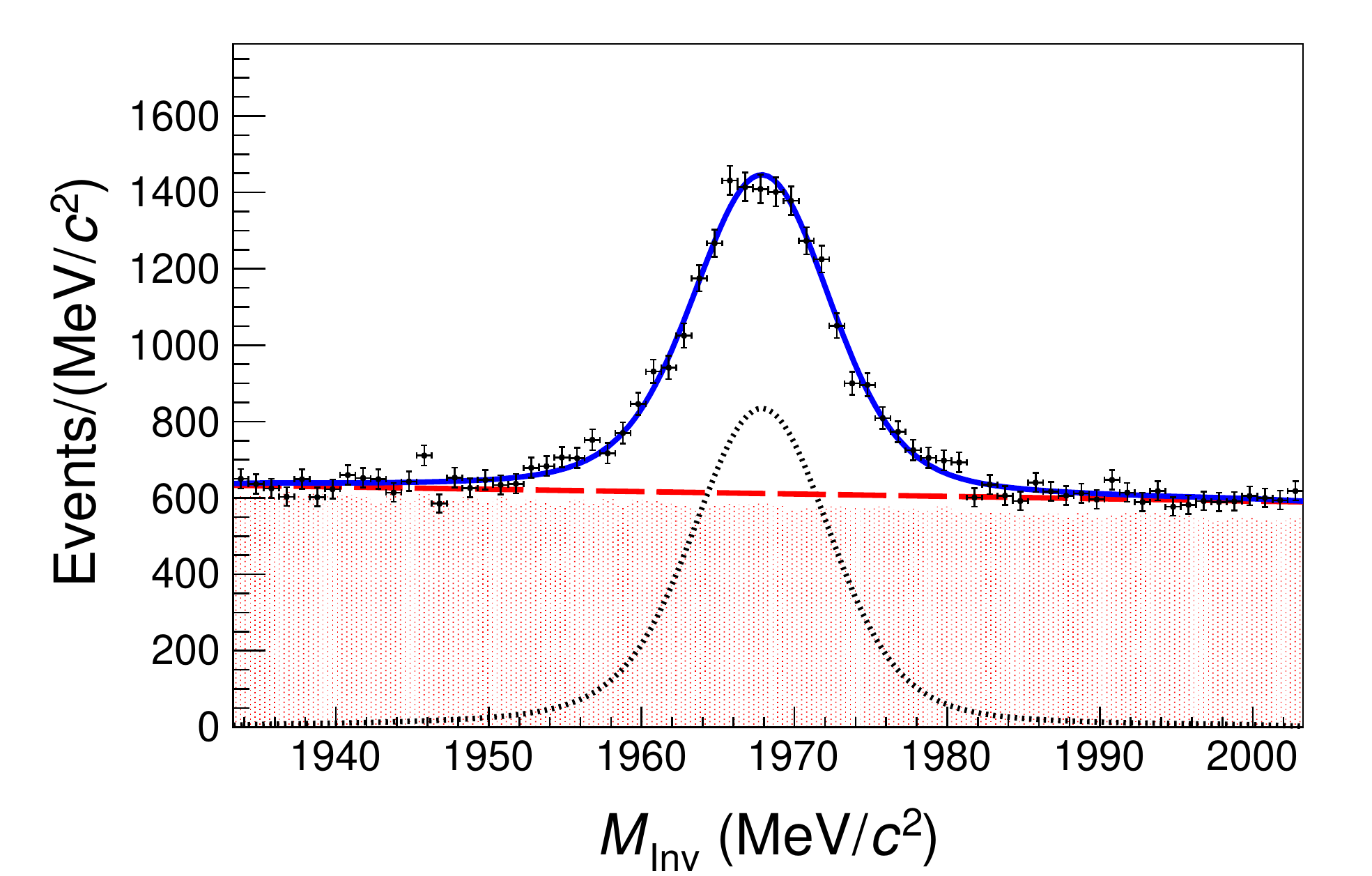} & \includegraphics[width=3in]{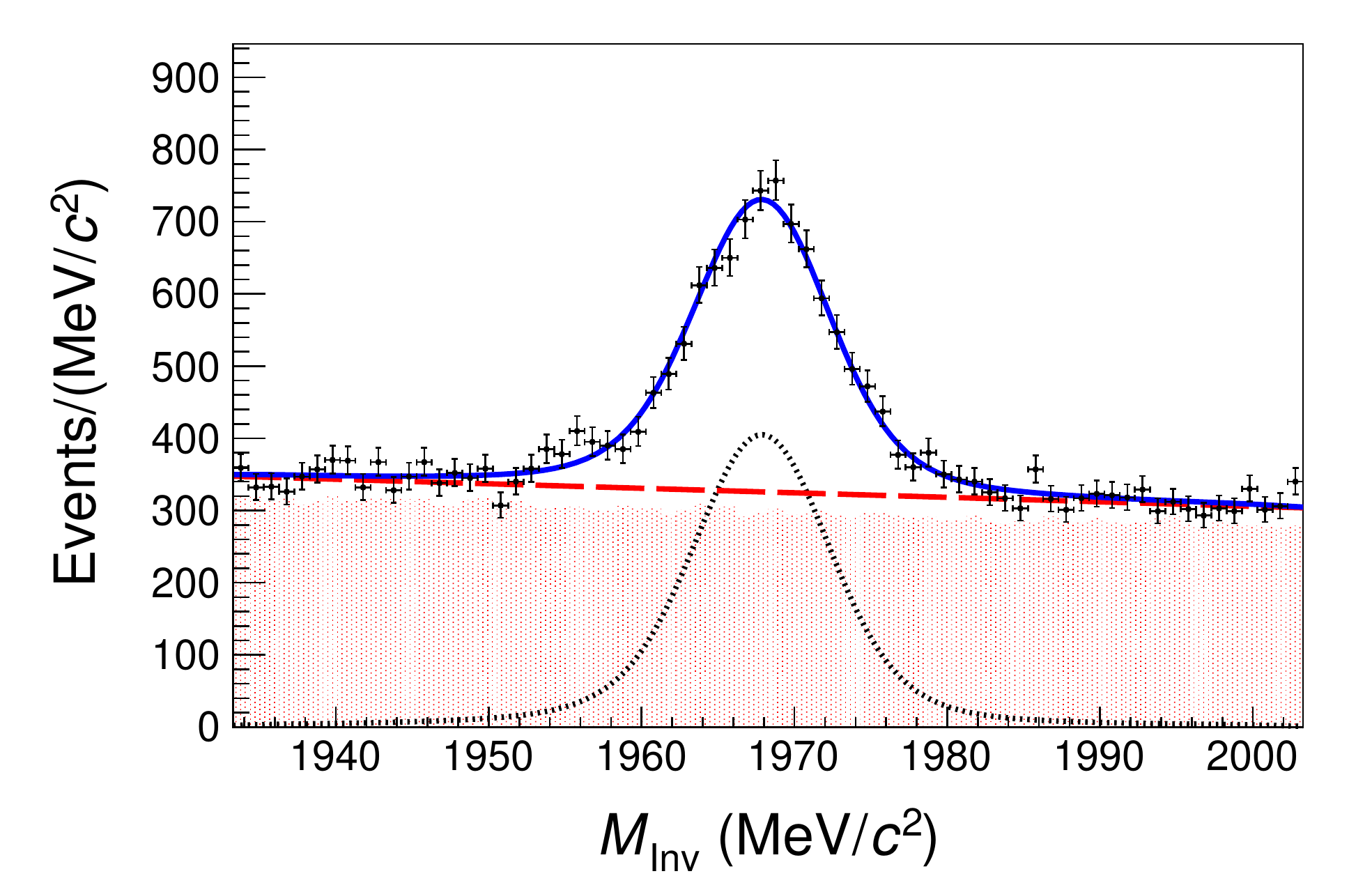}\\
\textbf{RS 300-350 MeV/$c$} & \textbf{WS 300-350 MeV/$c$}\\
\includegraphics[width=3in]{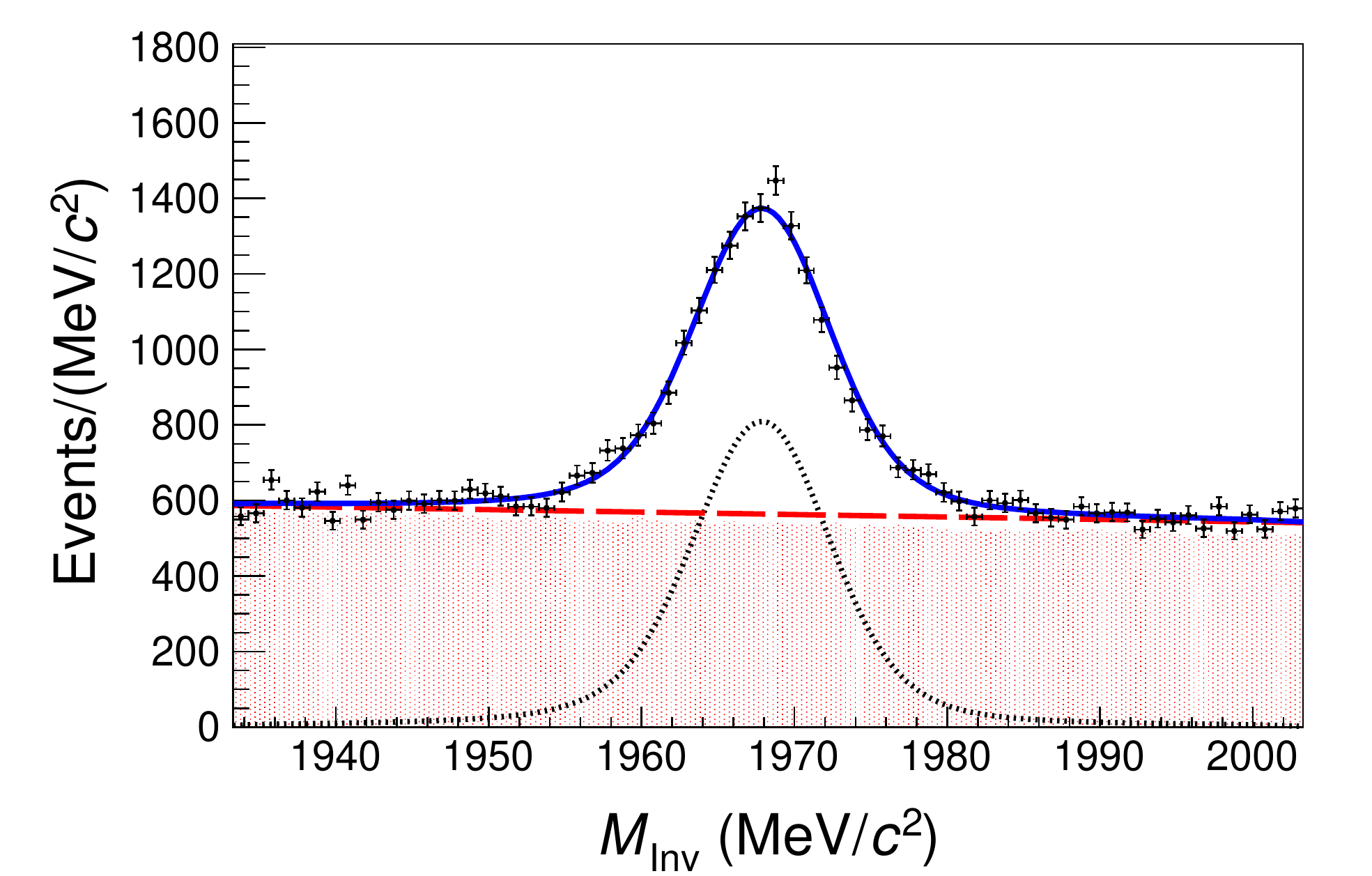} & \includegraphics[width=3in]{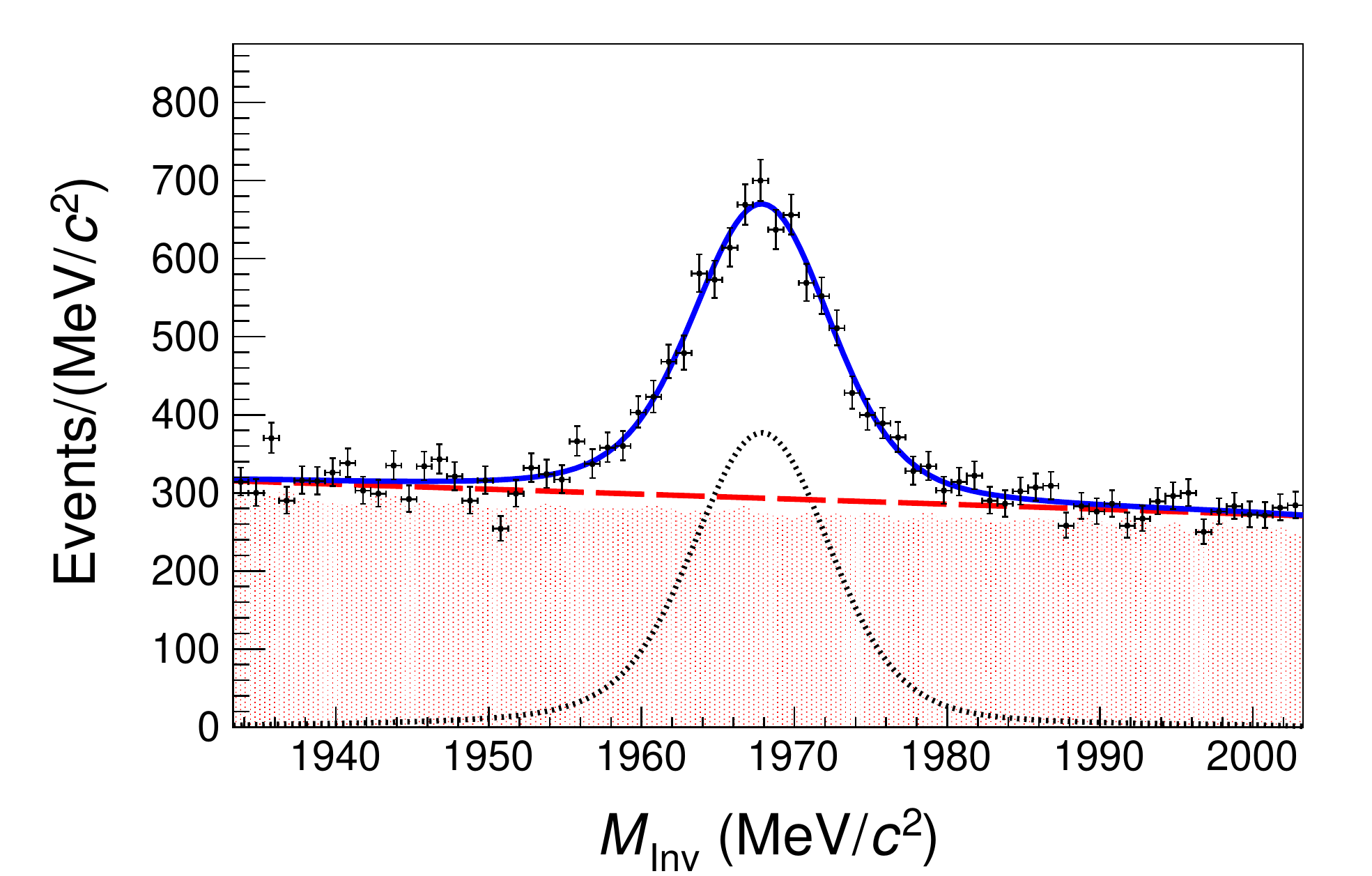}\\
\textbf{RS 350-400 MeV/$c$} & \textbf{WS 350-400 MeV/$c$}\\
\includegraphics[width=3in]{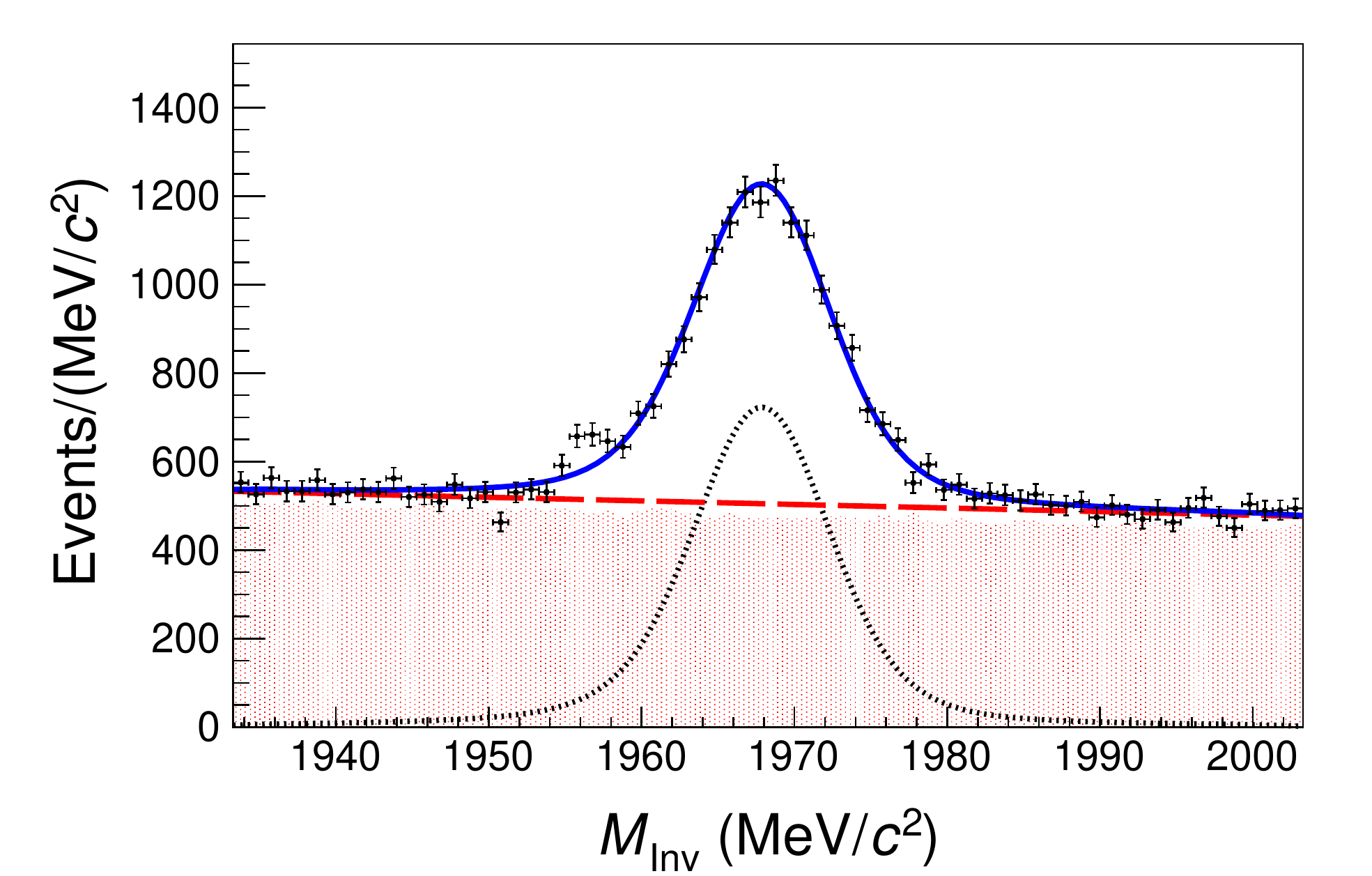} & \includegraphics[width=3in]{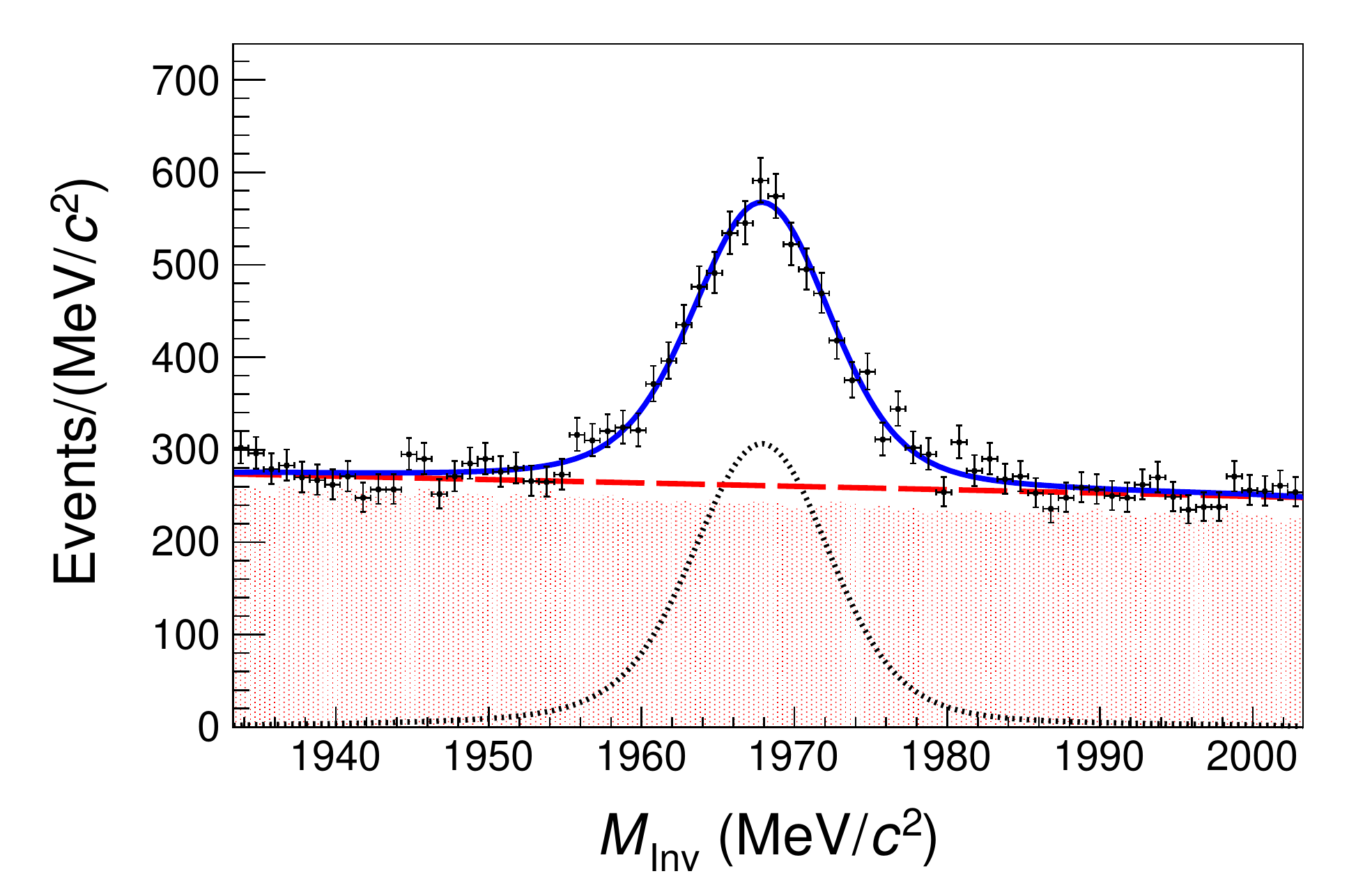}
\end{tabular}
\pagebreak

\begin{tabular}{cc}
\textbf{RS 400-450 MeV/$c$} & \textbf{WS 400-450 MeV/$c$}\\
\includegraphics[width=3in]{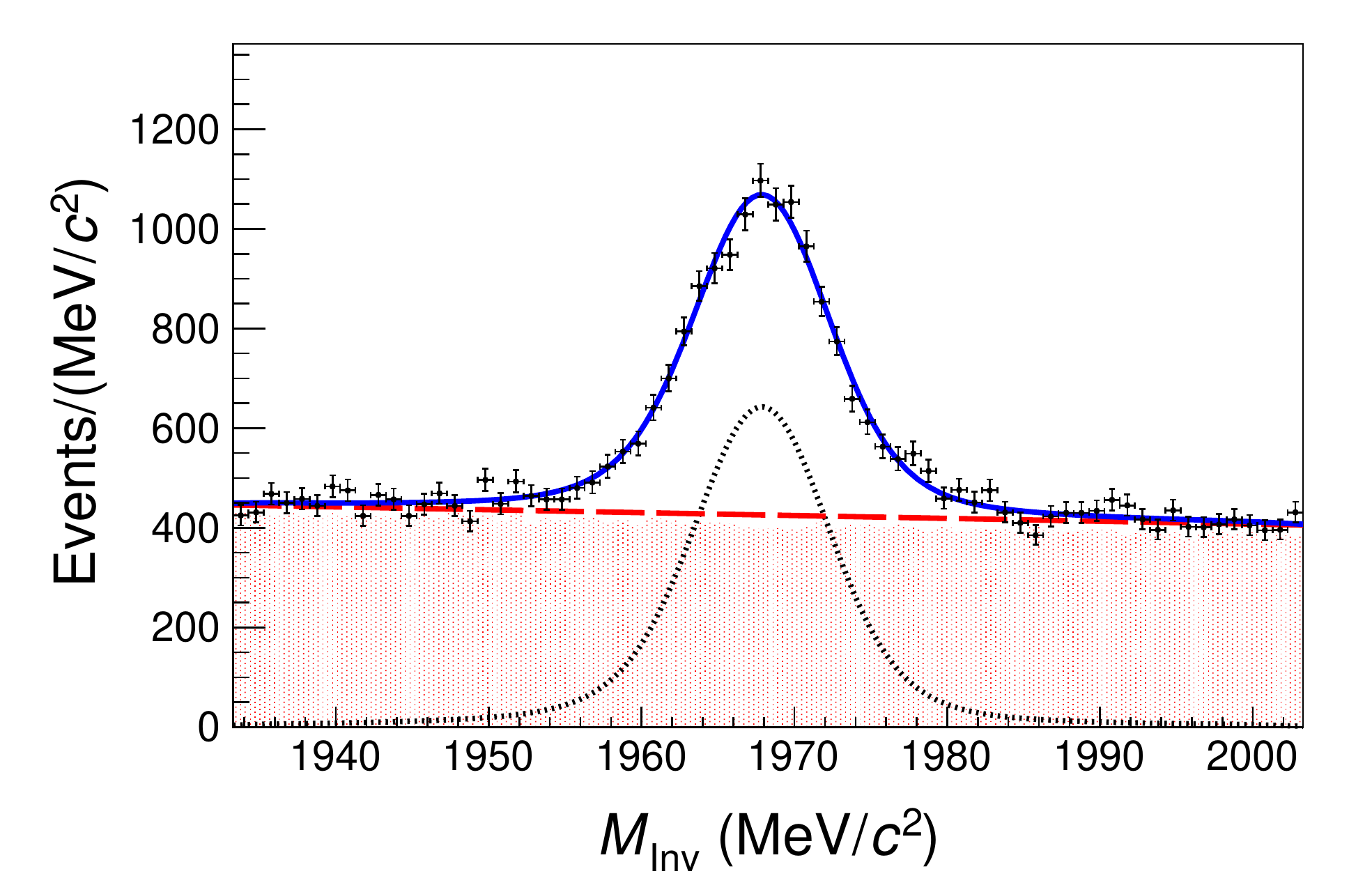} & \includegraphics[width=3in]{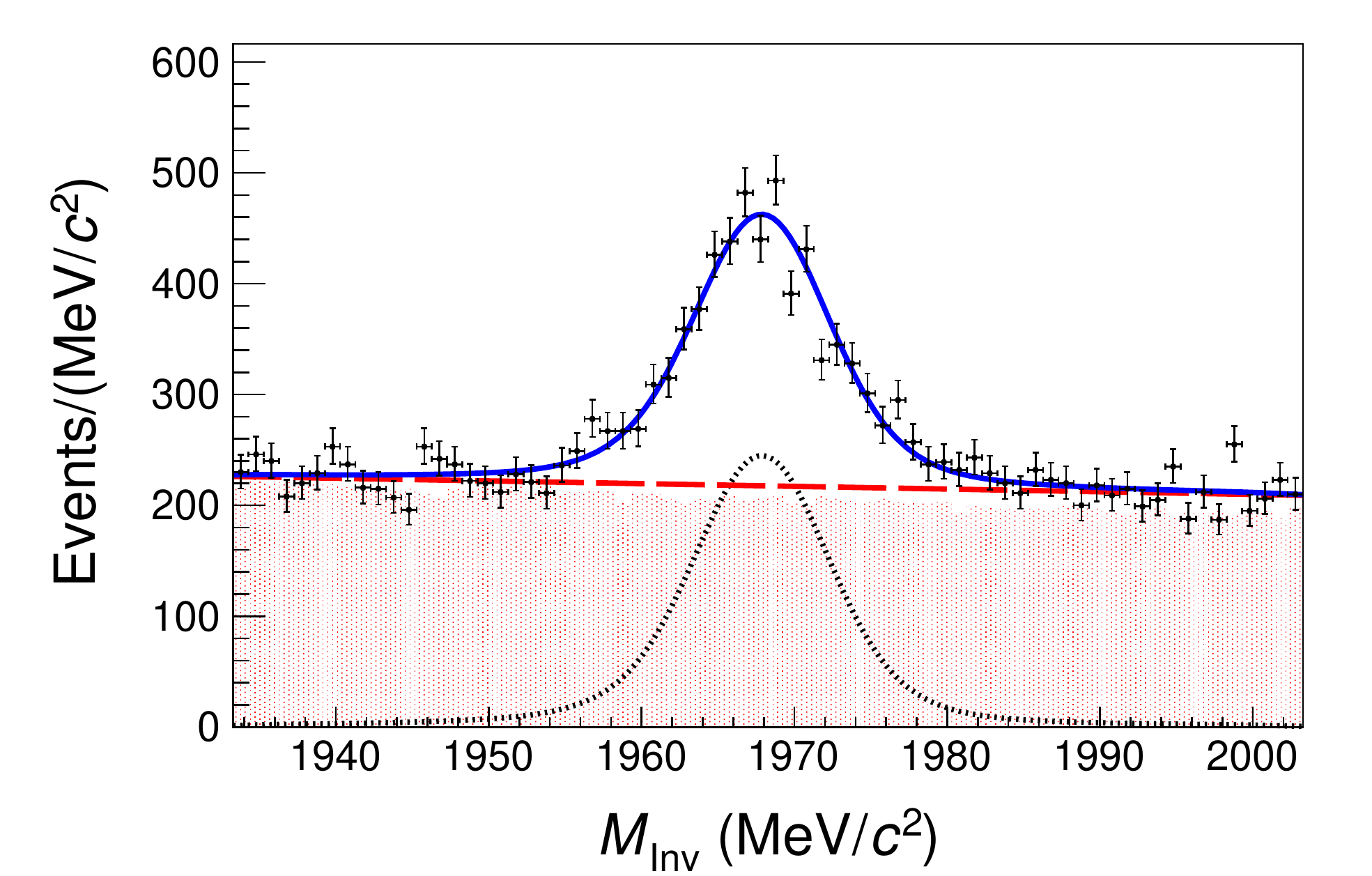}\\
\textbf{RS 450-500 MeV/$c$} & \textbf{WS 450-500 MeV/$c$}\\
\includegraphics[width=3in]{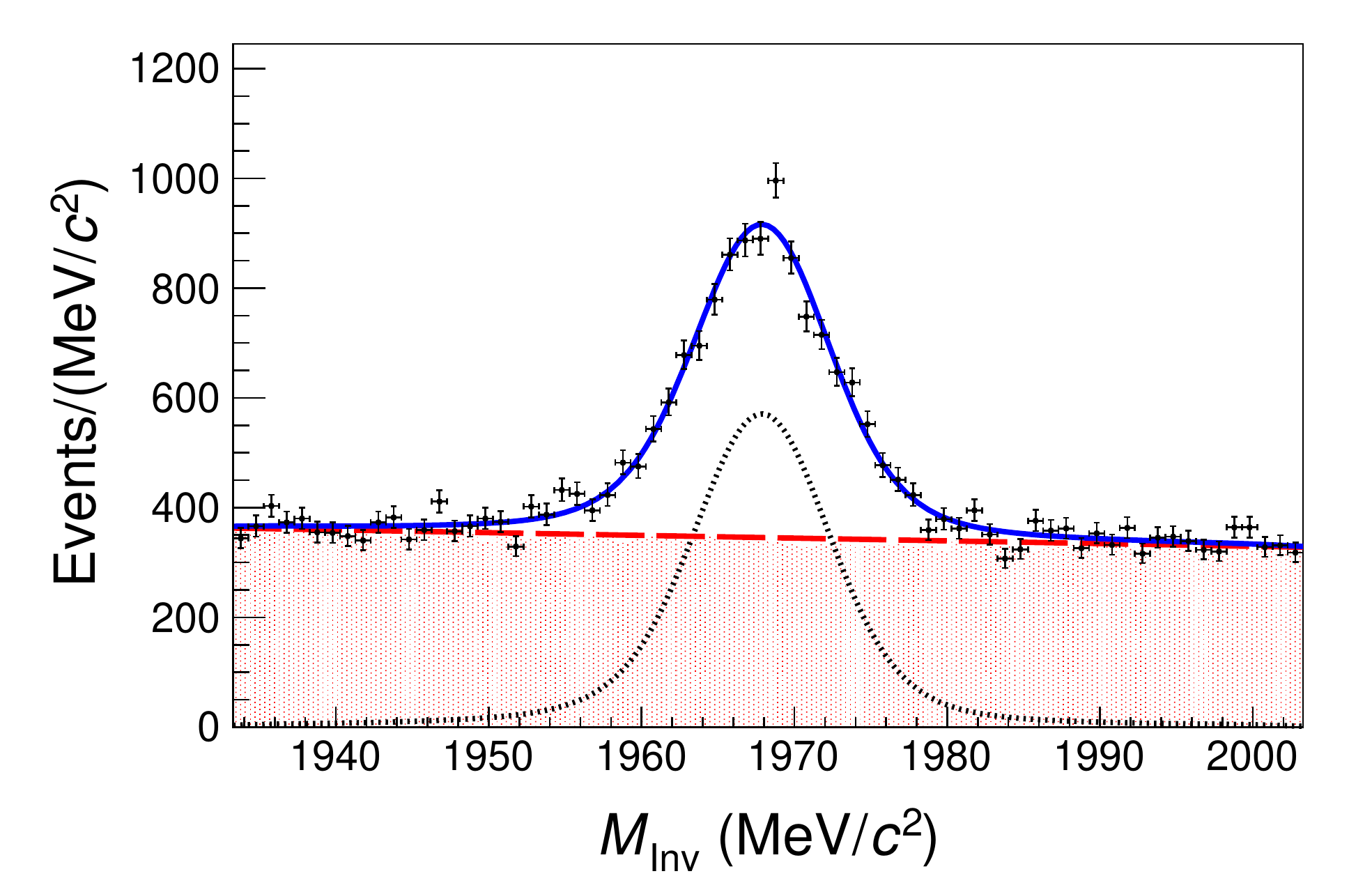} & \includegraphics[width=3in]{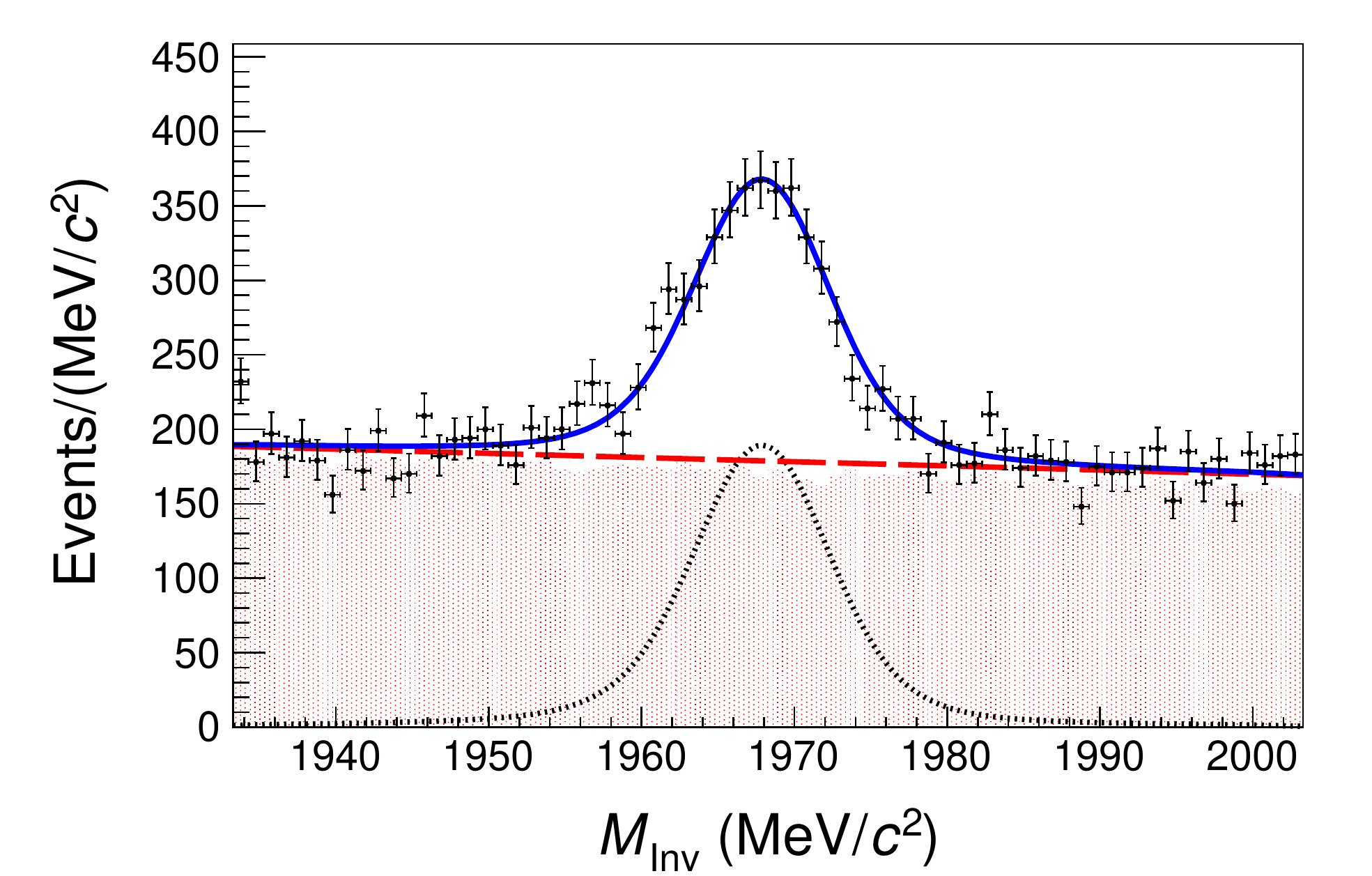}\\
\textbf{RS 500-550 MeV/$c$} & \textbf{WS 500-550 MeV/$c$}\\
\includegraphics[width=3in]{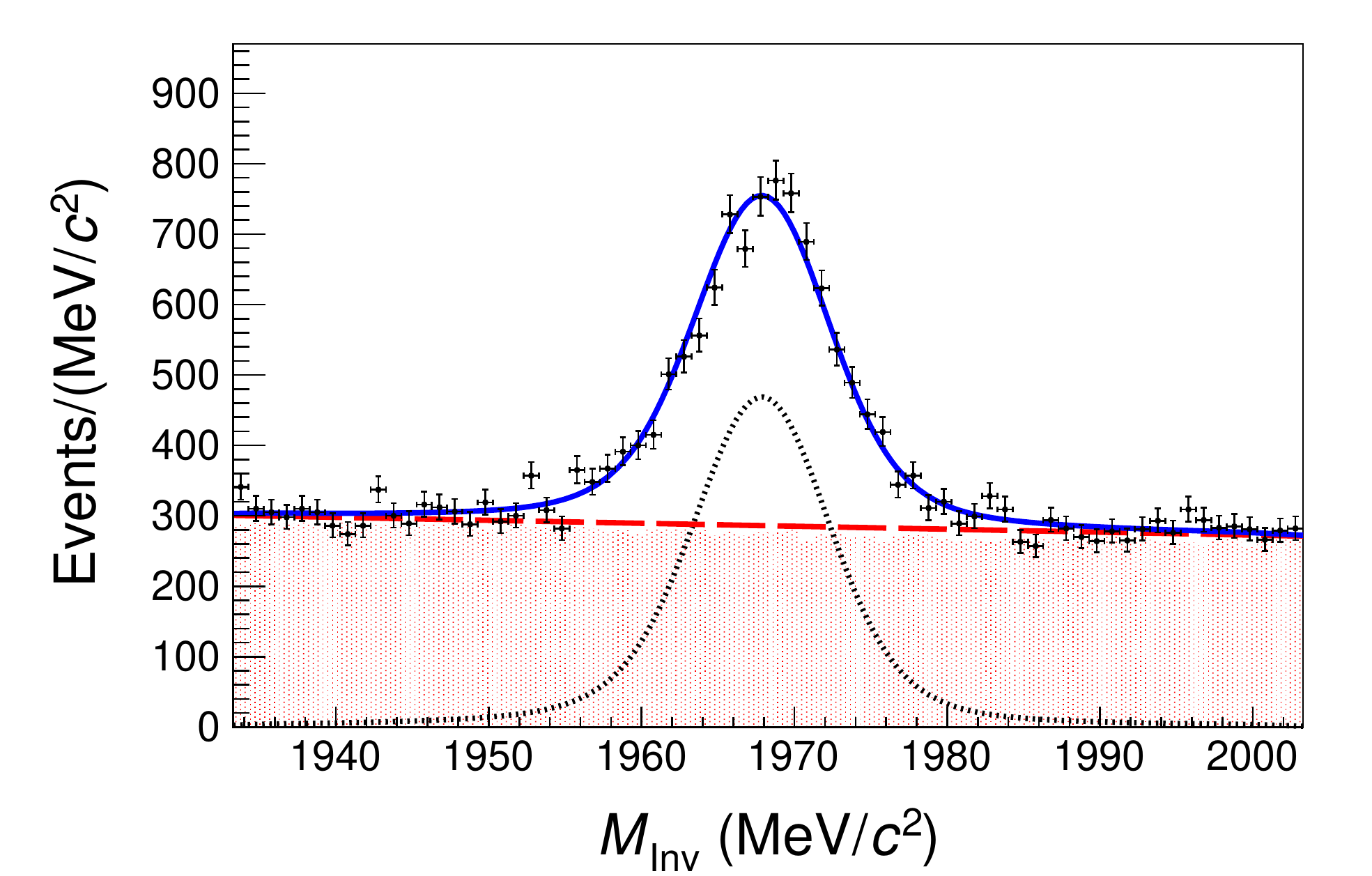} & \includegraphics[width=3in]{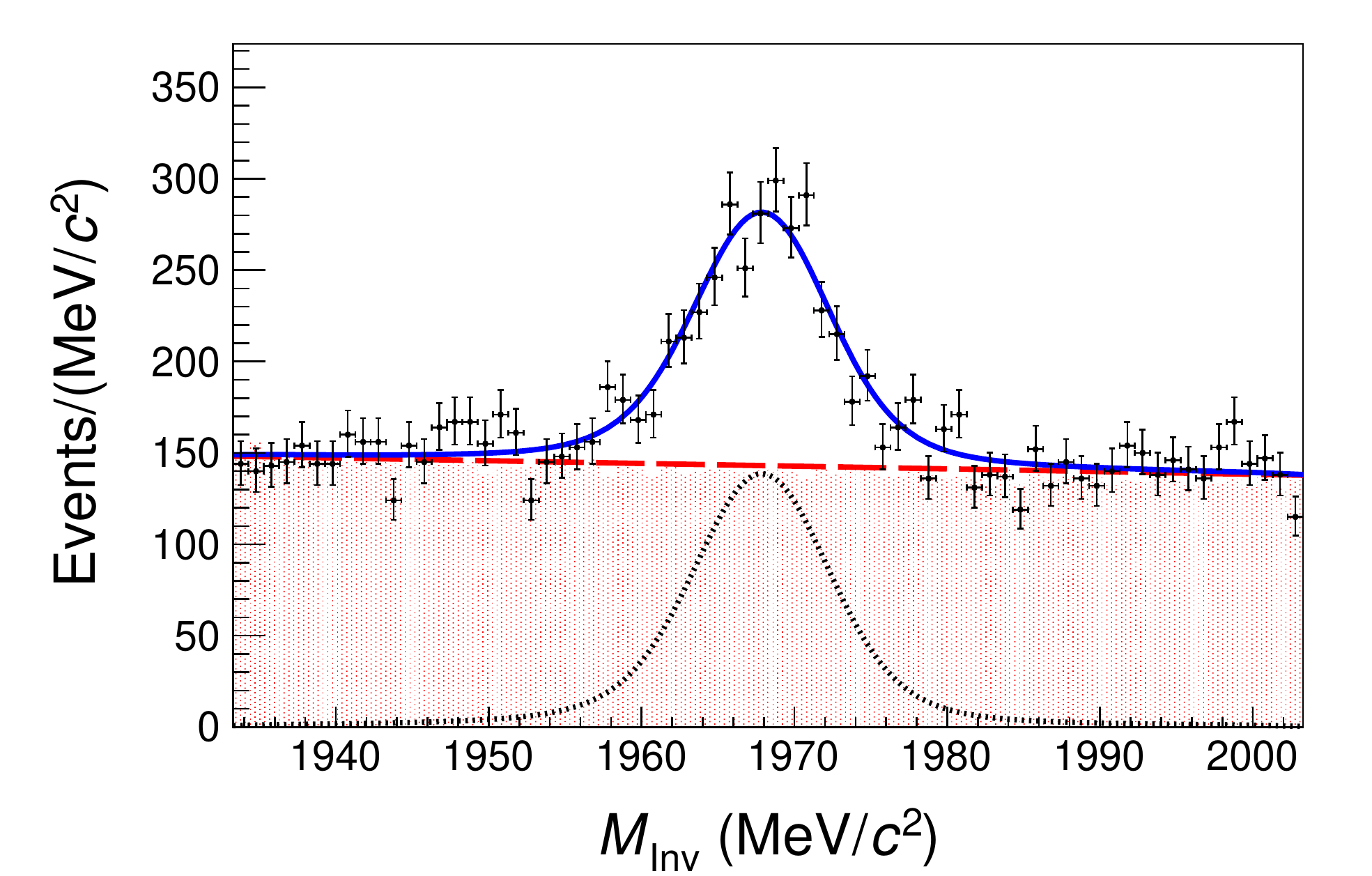}\\
\textbf{RS 550-600 MeV/$c$} & \textbf{WS 550-600 MeV/$c$}\\
\includegraphics[width=3in]{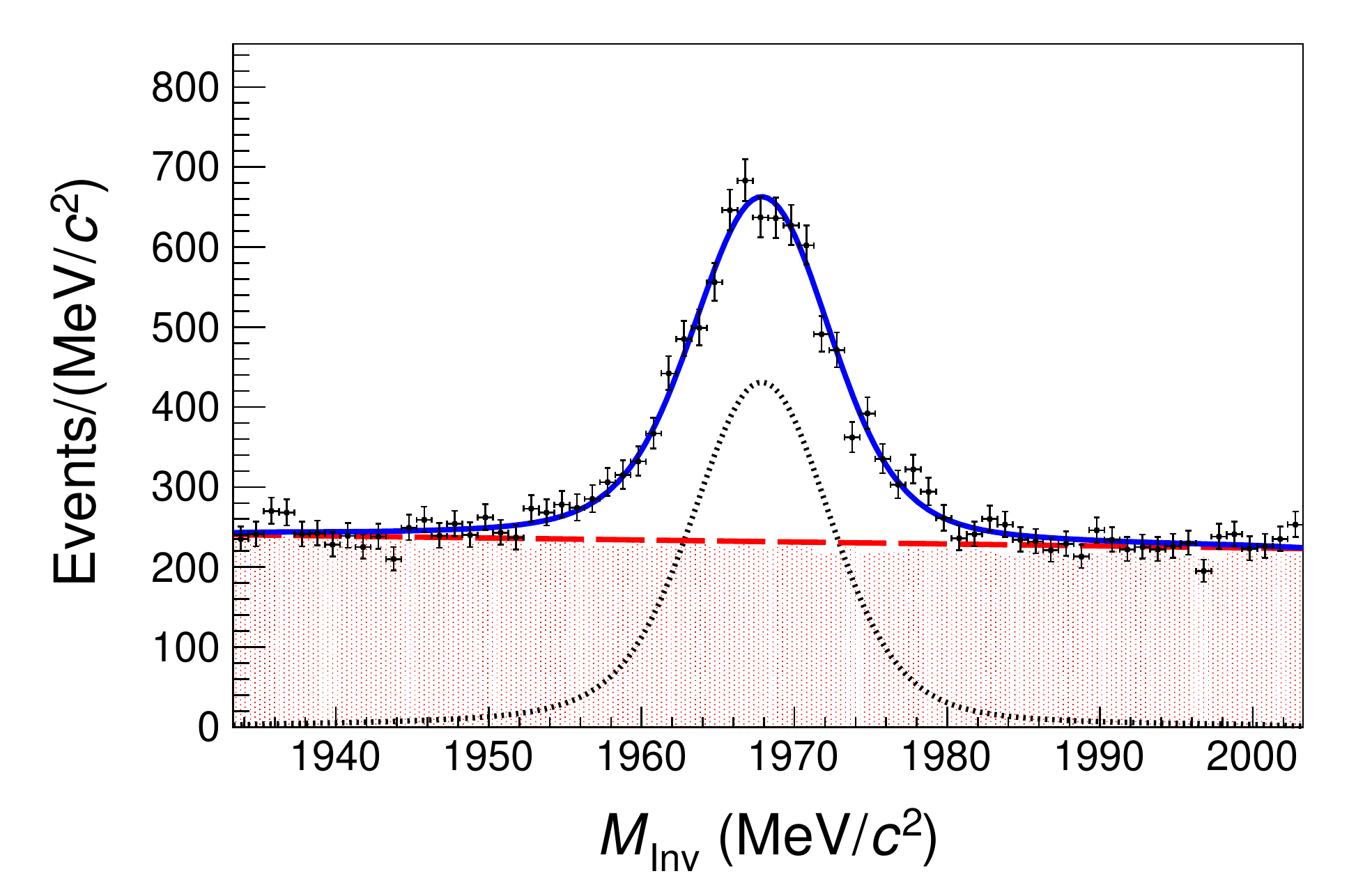} & \includegraphics[width=3in]{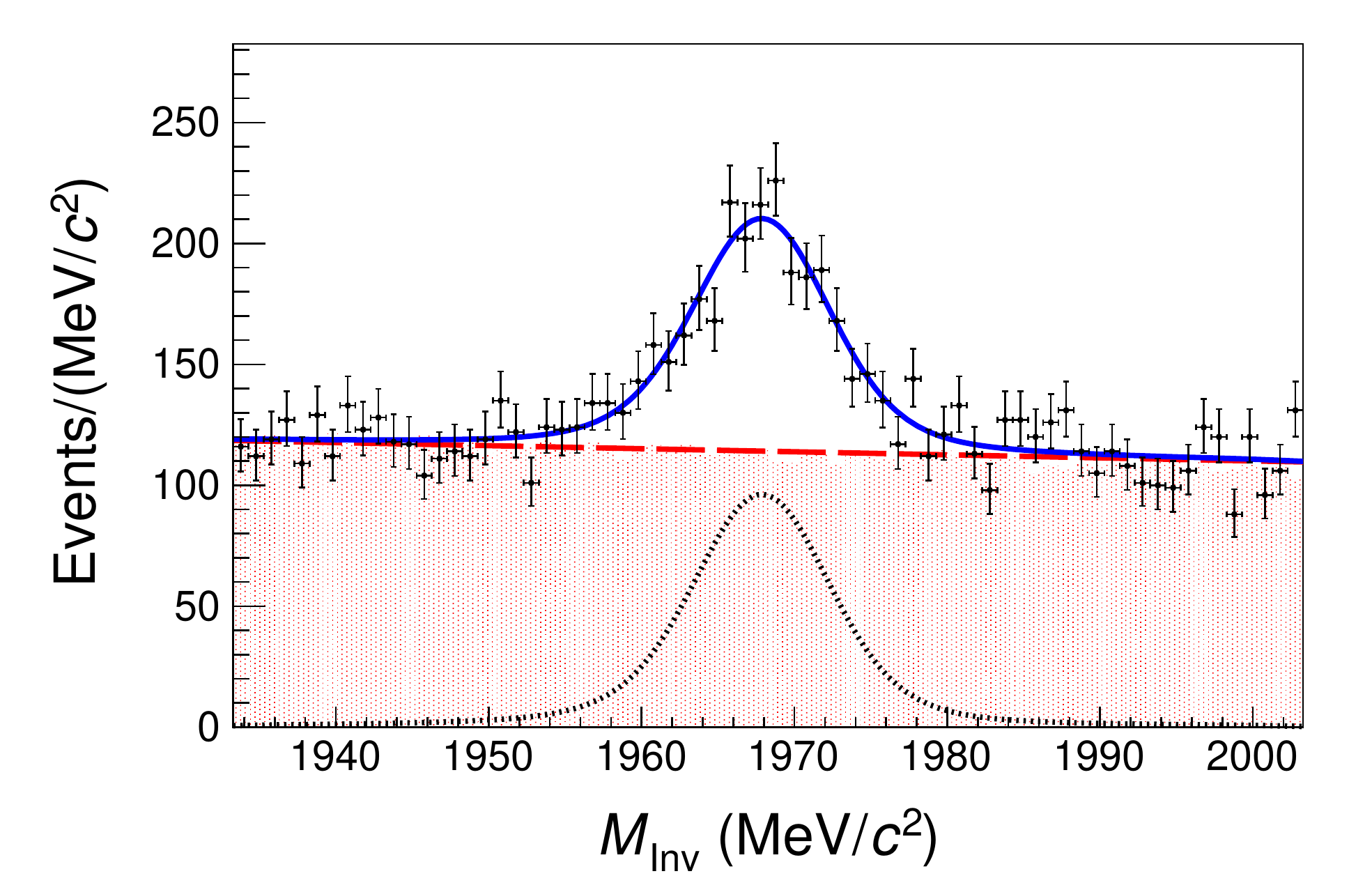}
\end{tabular}

\begin{tabular}{cc}
\textbf{RS 600-650 MeV/$c$} & \textbf{WS 600-650 MeV/$c$}\\
\includegraphics[width=3in]{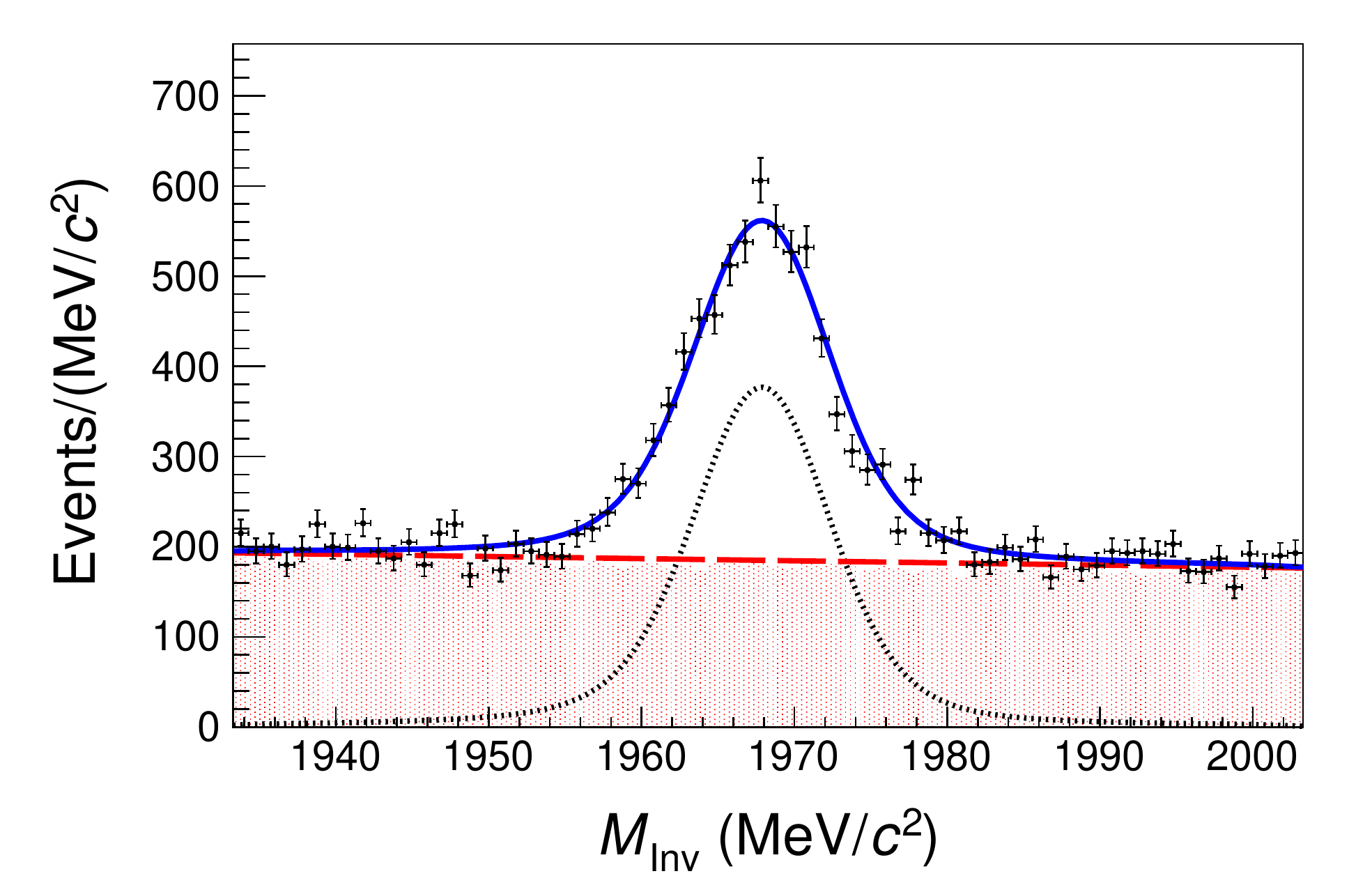} & \includegraphics[width=3in]{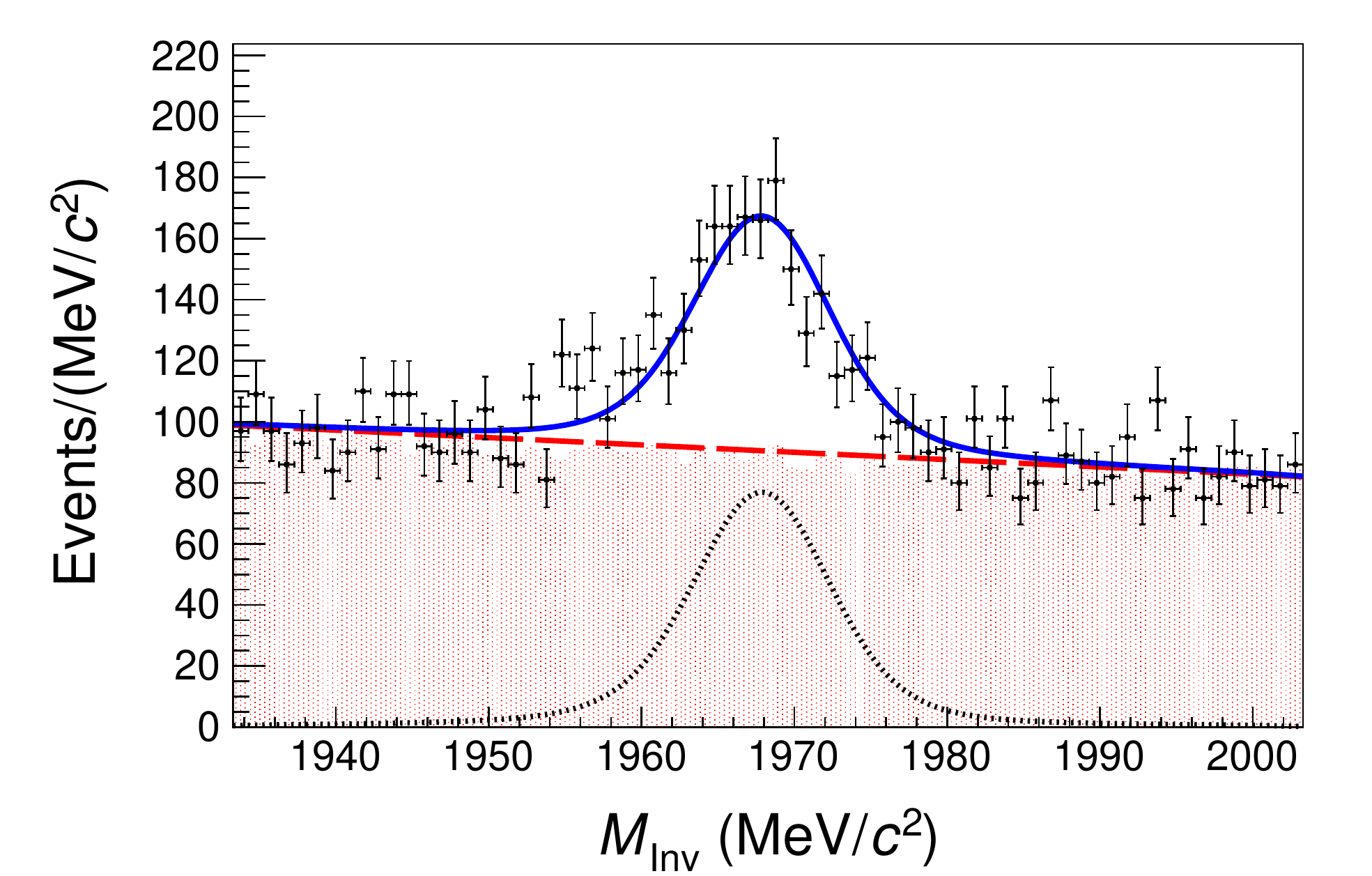}\\
\textbf{RS 650-700 MeV/$c$} & \textbf{WS 650-700 MeV/$c$}\\
\includegraphics[width=3in]{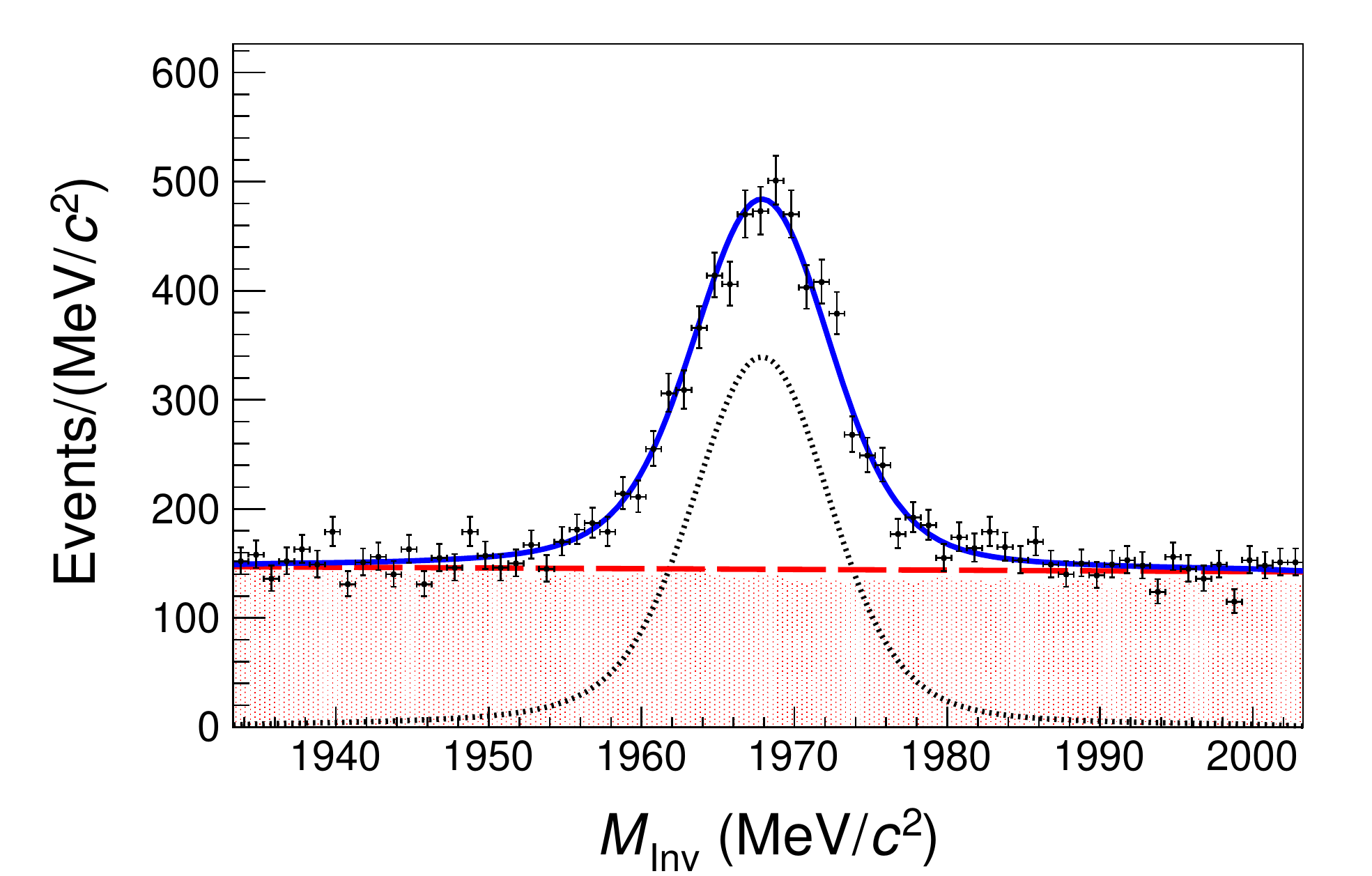} & \includegraphics[width=3in]{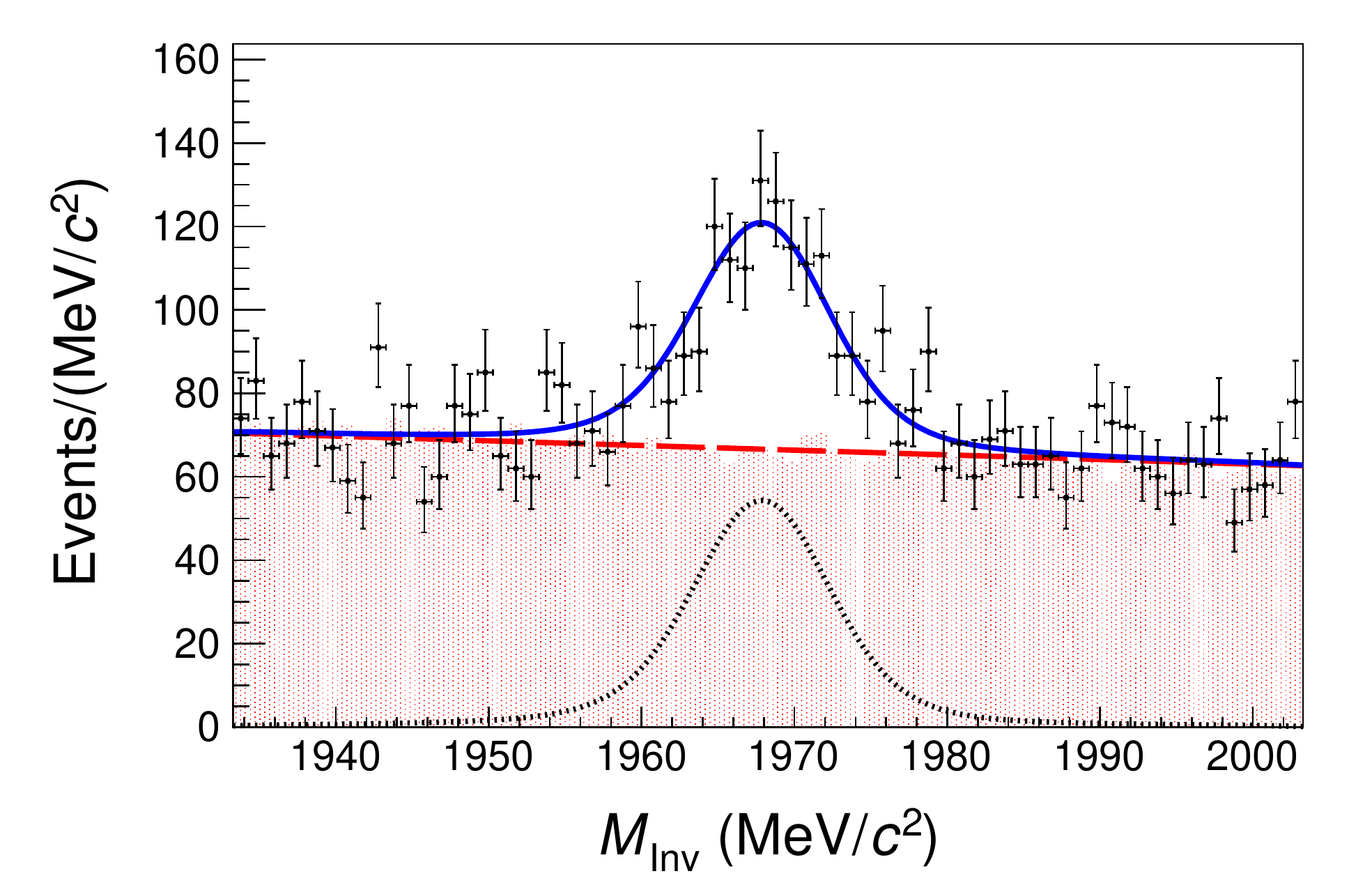}\\
\textbf{RS 700-750 MeV/$c$} & \textbf{WS 700-750 MeV/$c$}\\
\includegraphics[width=3in]{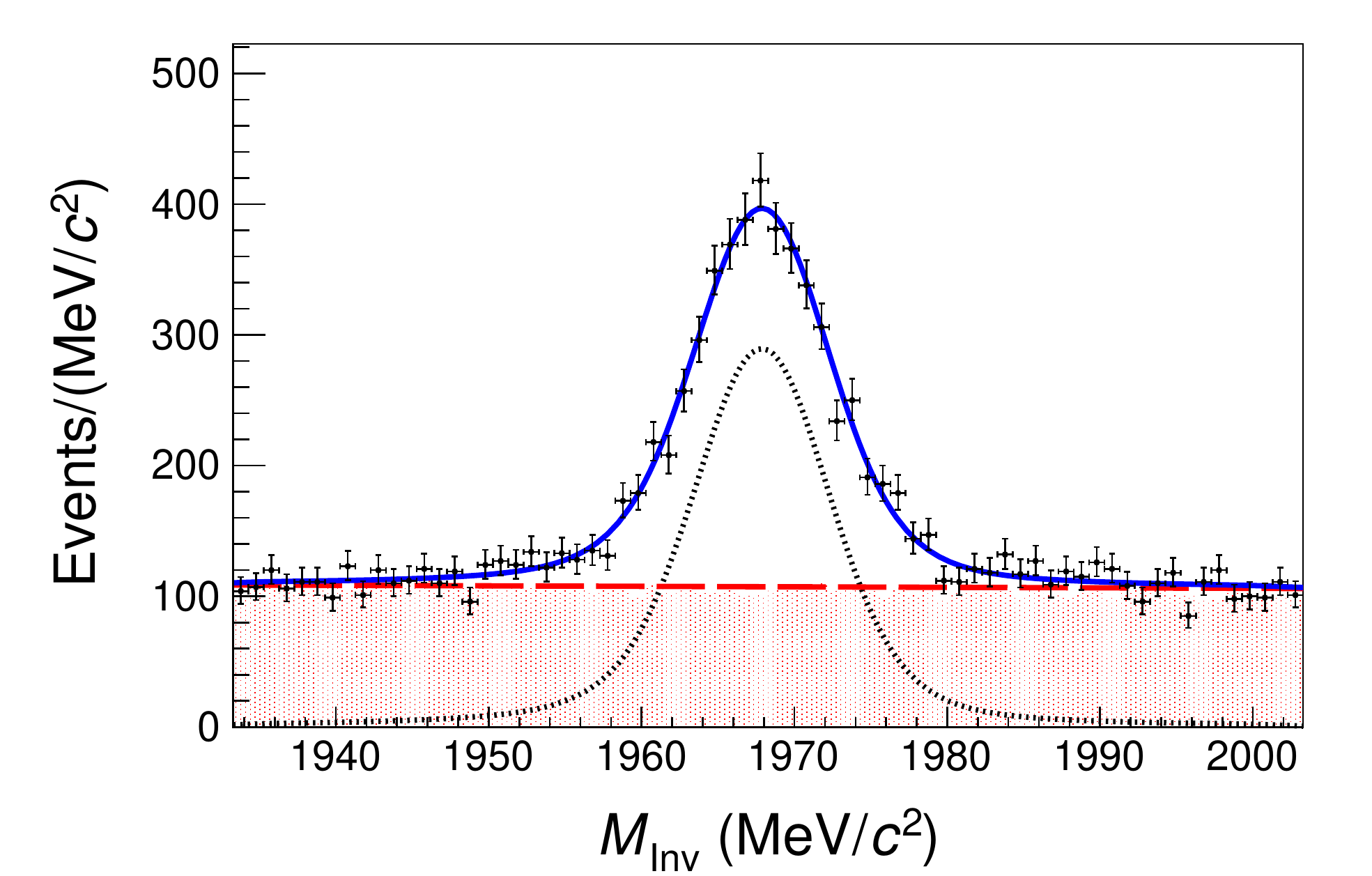} & \includegraphics[width=3in]{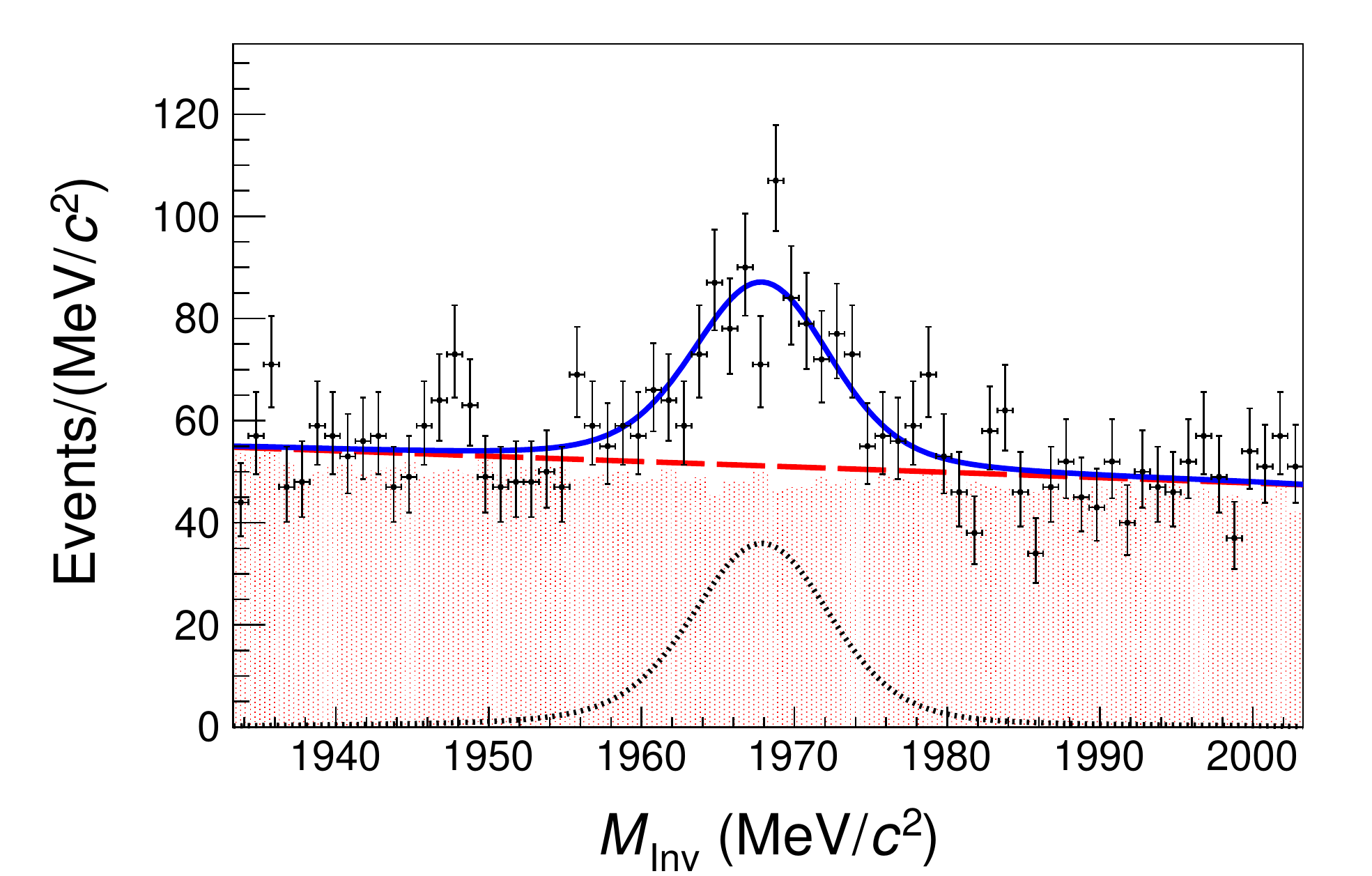}\\
\textbf{RS 750-800 MeV/$c$} & \textbf{WS 750-800 MeV/$c$}\\
\includegraphics[width=3in]{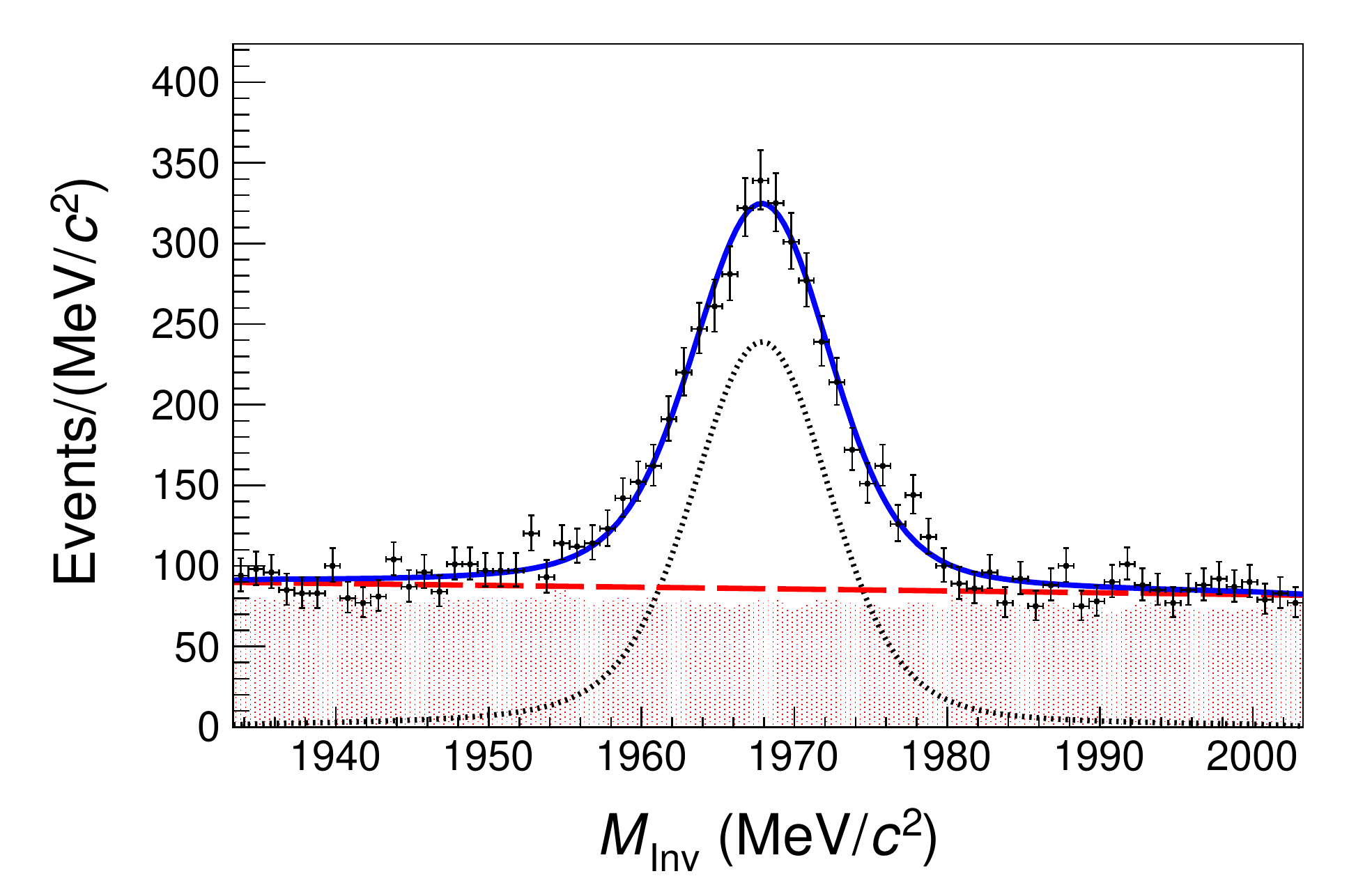} & \includegraphics[width=3in]{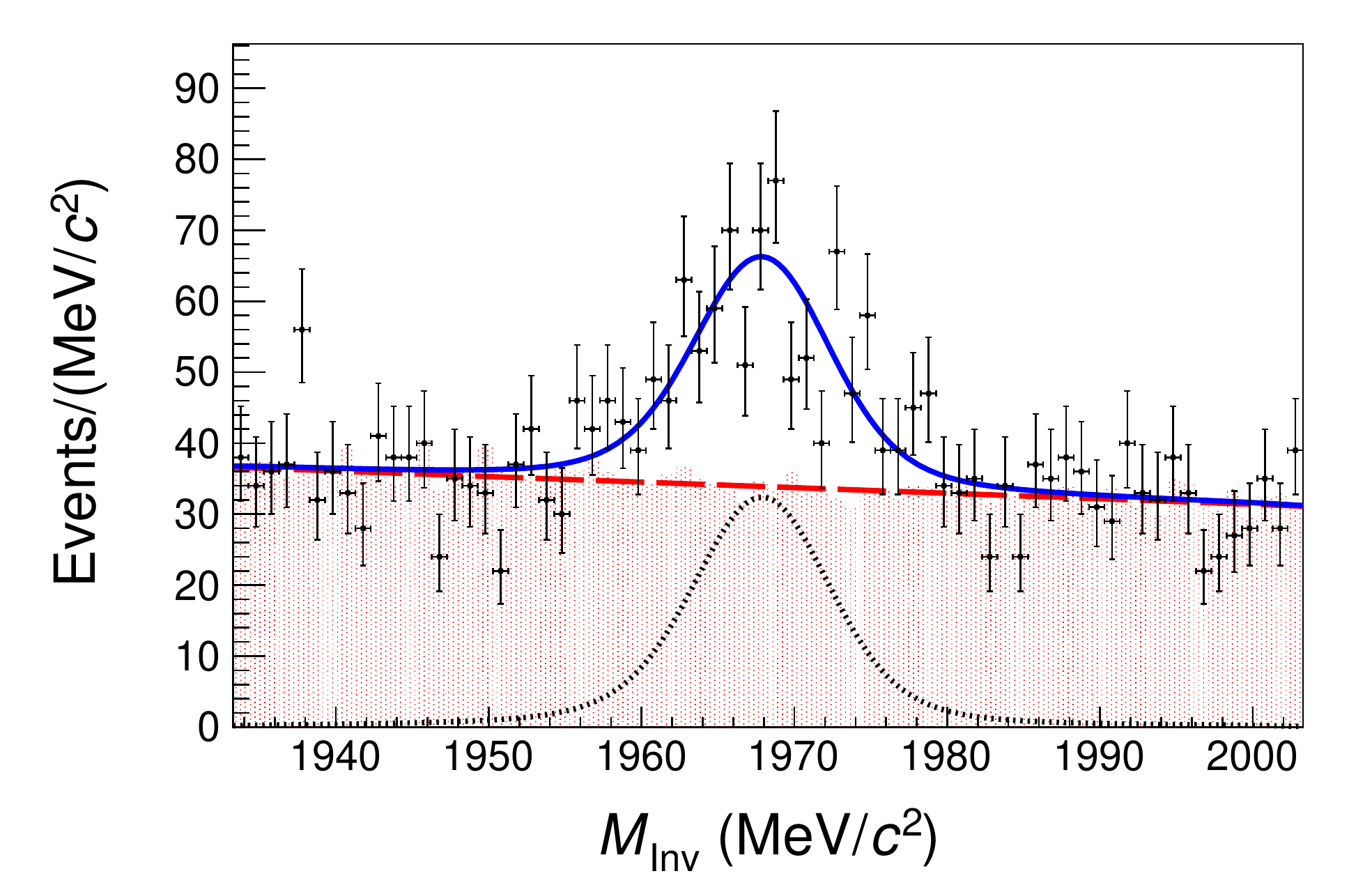}\\
\end{tabular}

\begin{tabular}{cc}
\textbf{RS 800-850 MeV/$c$} & \textbf{WS 800-850 MeV/$c$}\\
\includegraphics[width=3in]{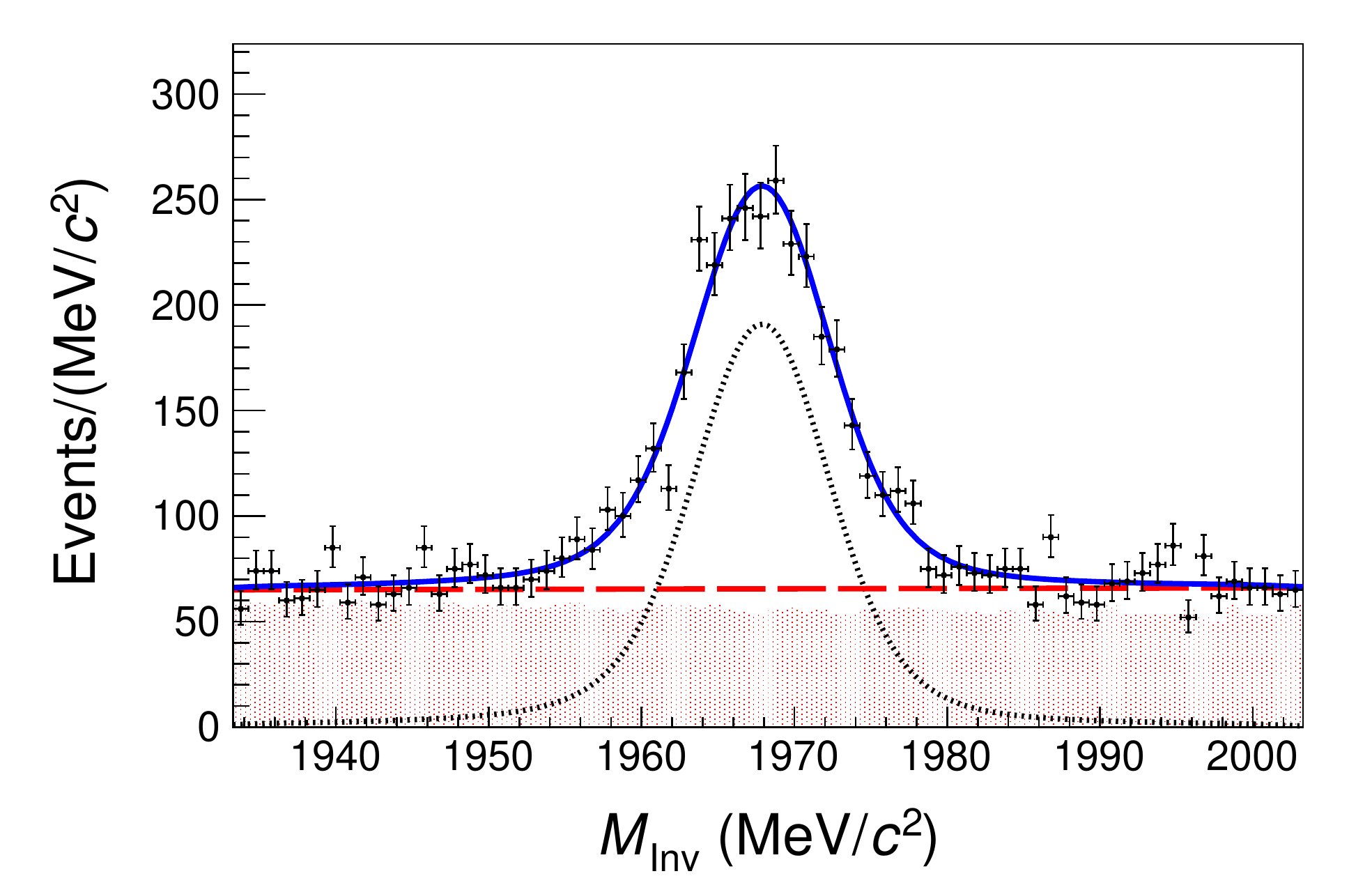} & \includegraphics[width=3in]{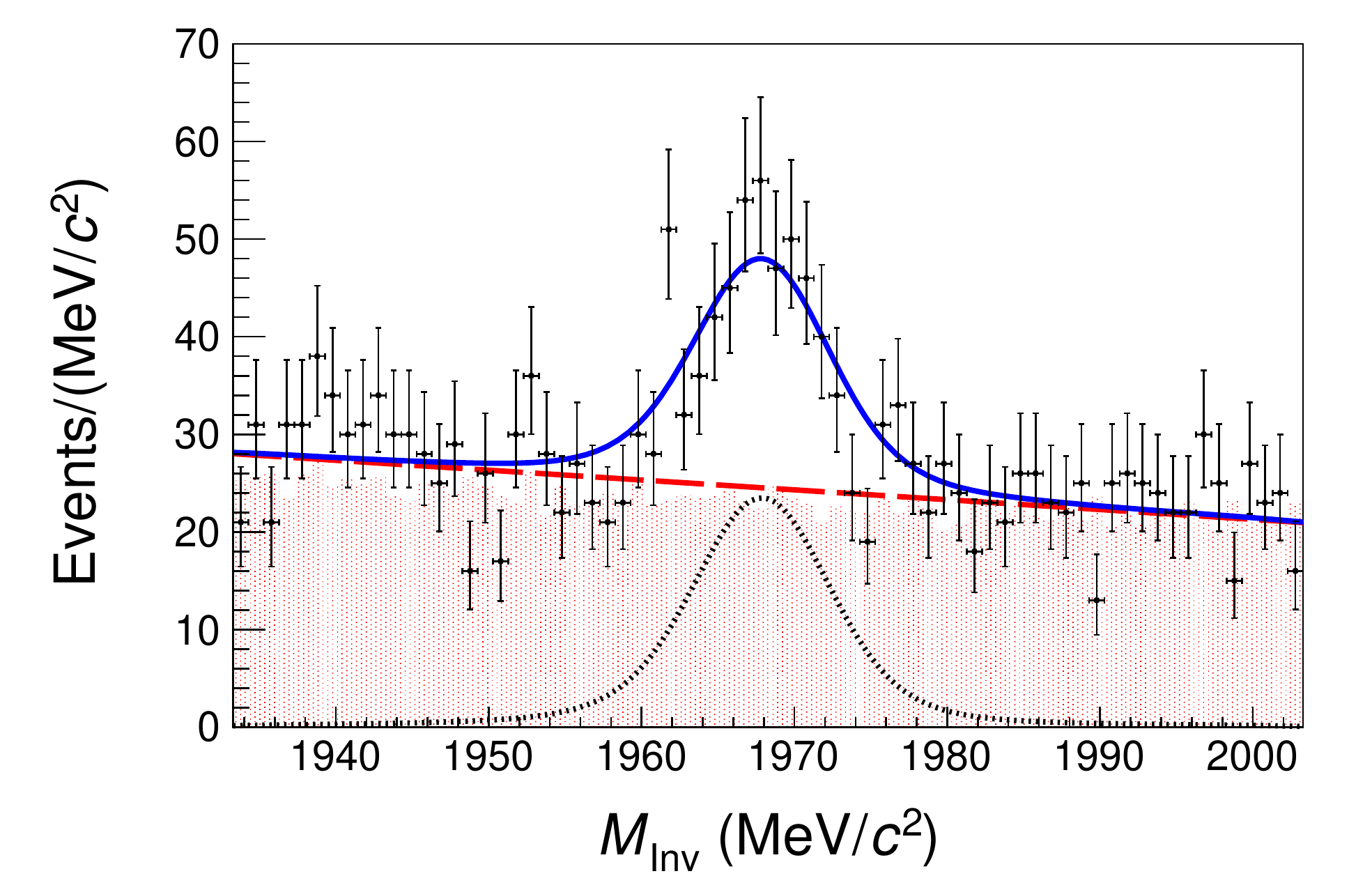}\\
\textbf{RS 850-900 MeV/$c$} & \textbf{WS 850-900 MeV/$c$}\\
\includegraphics[width=3in]{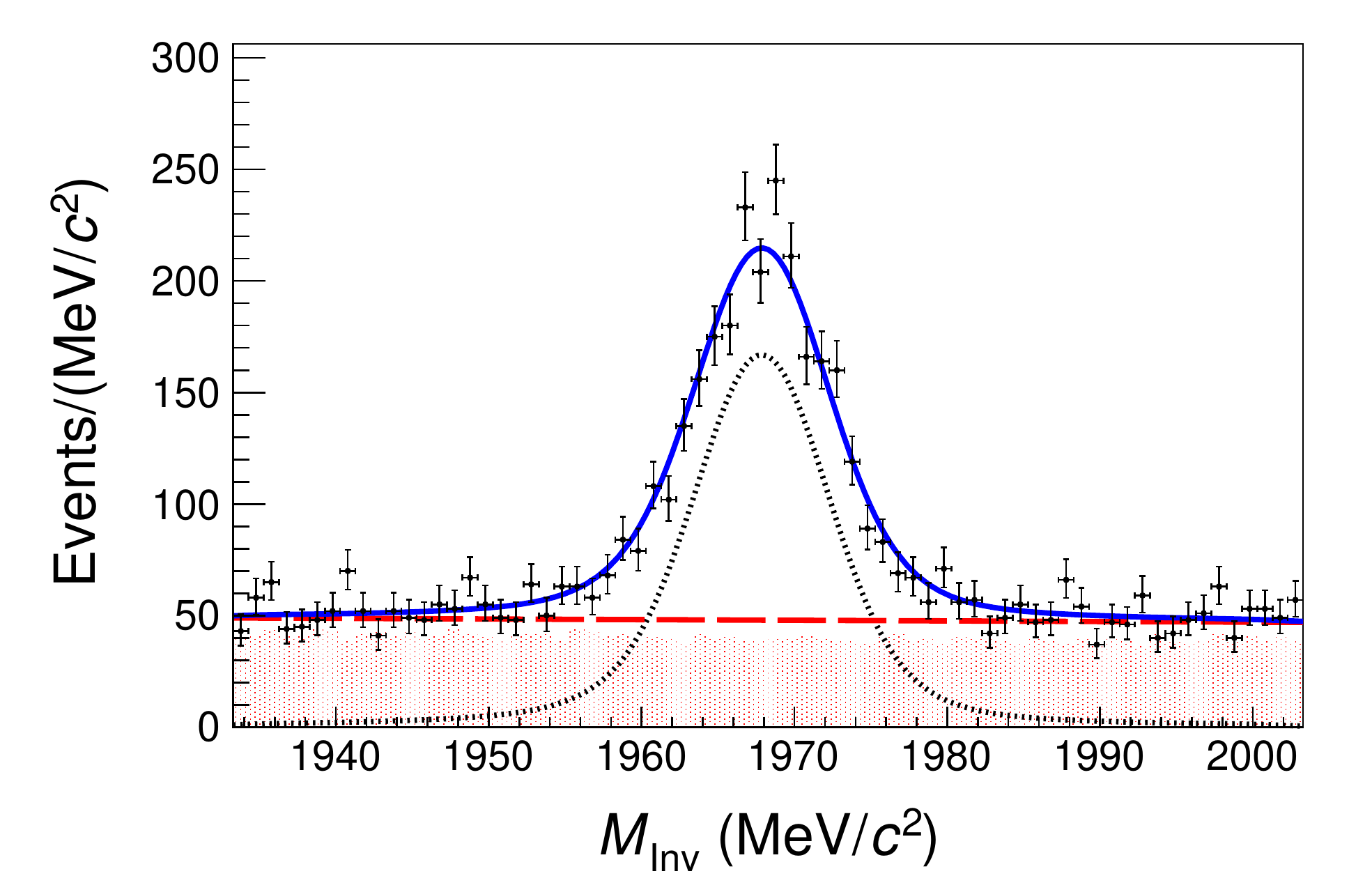} & \includegraphics[width=3in]{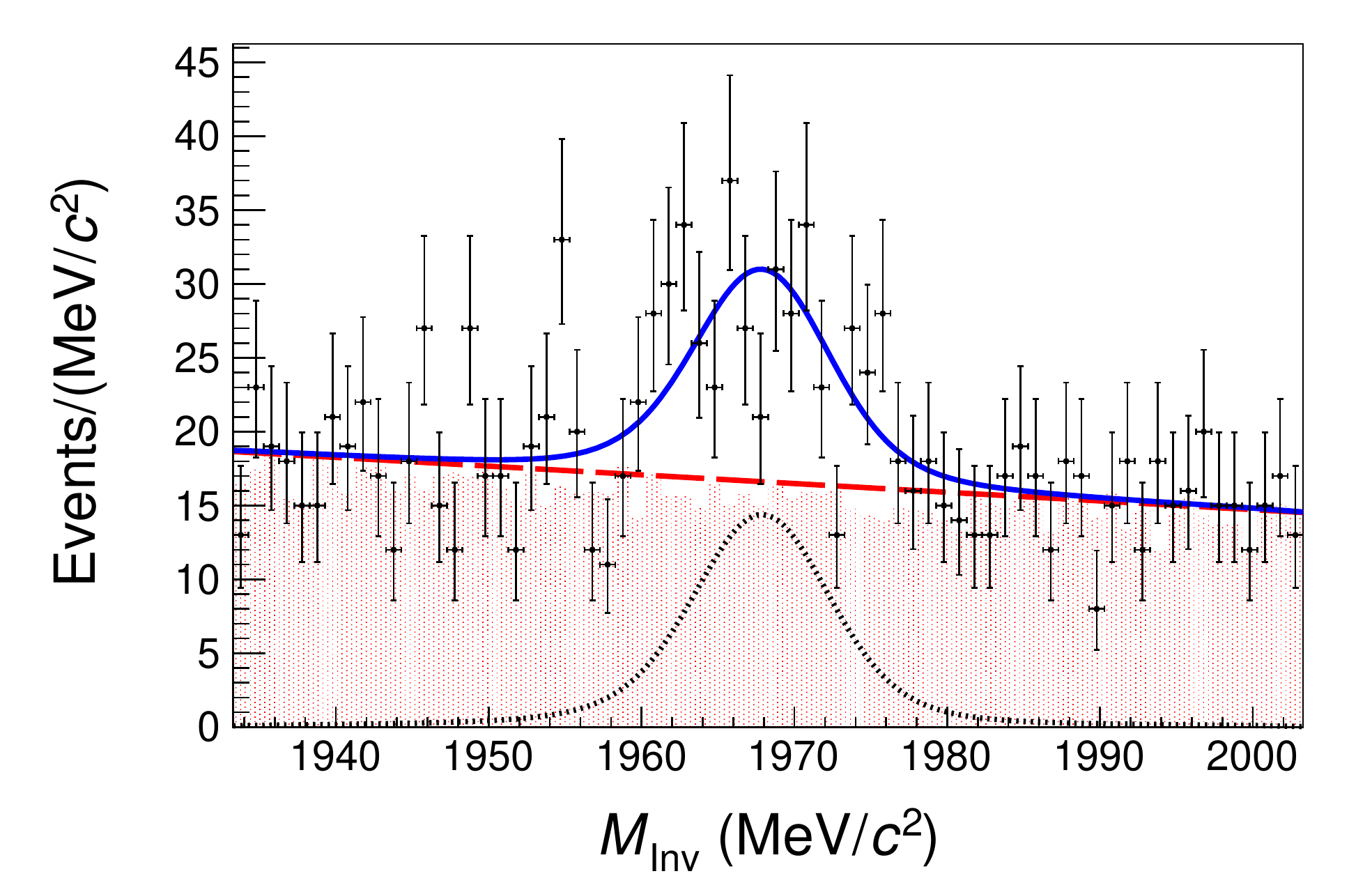}\\
\textbf{RS 900-950 MeV/$c$} & \textbf{WS 900-950 MeV/$c$}\\
\includegraphics[width=3in]{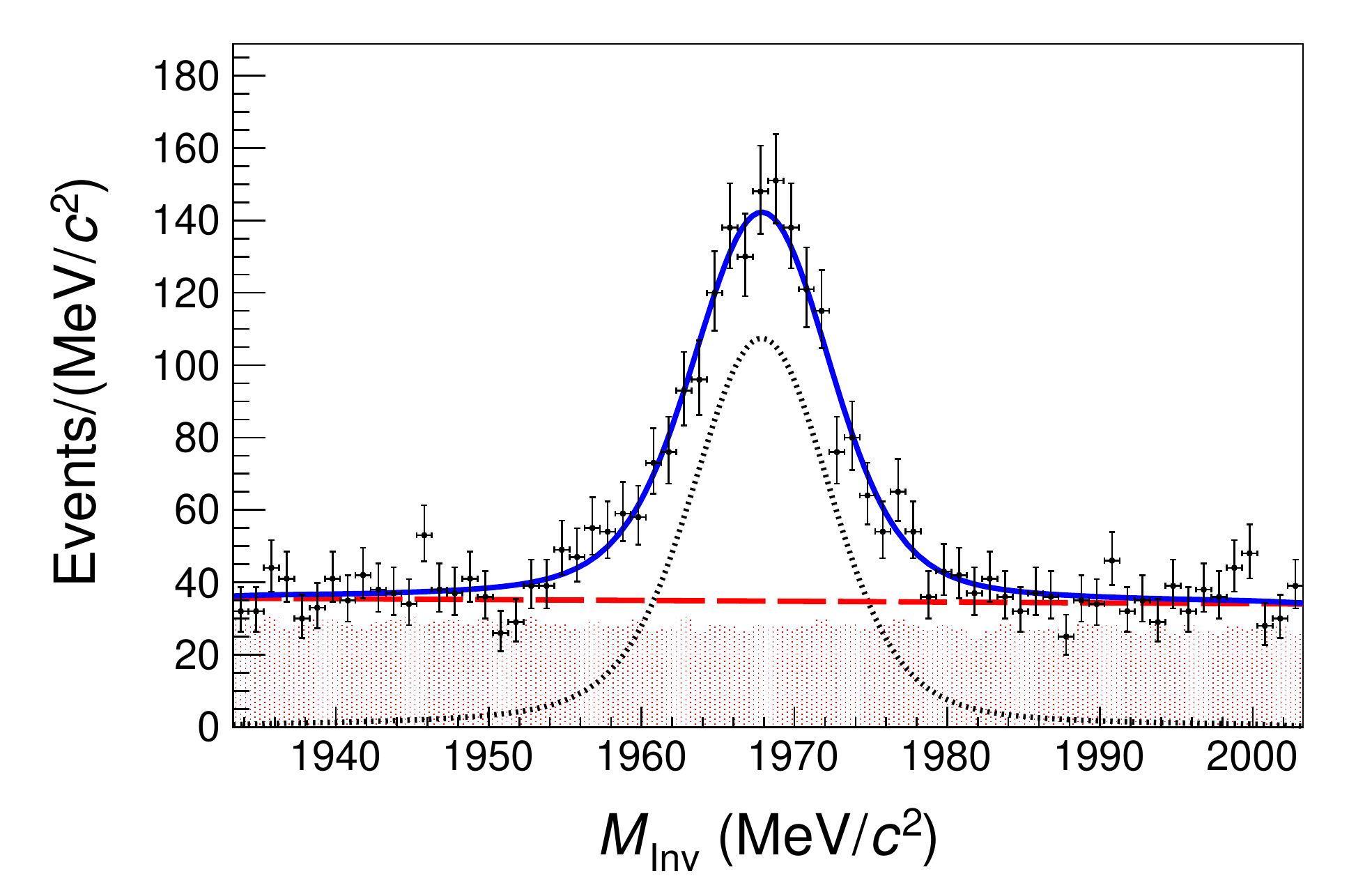} & \includegraphics[width=3in]{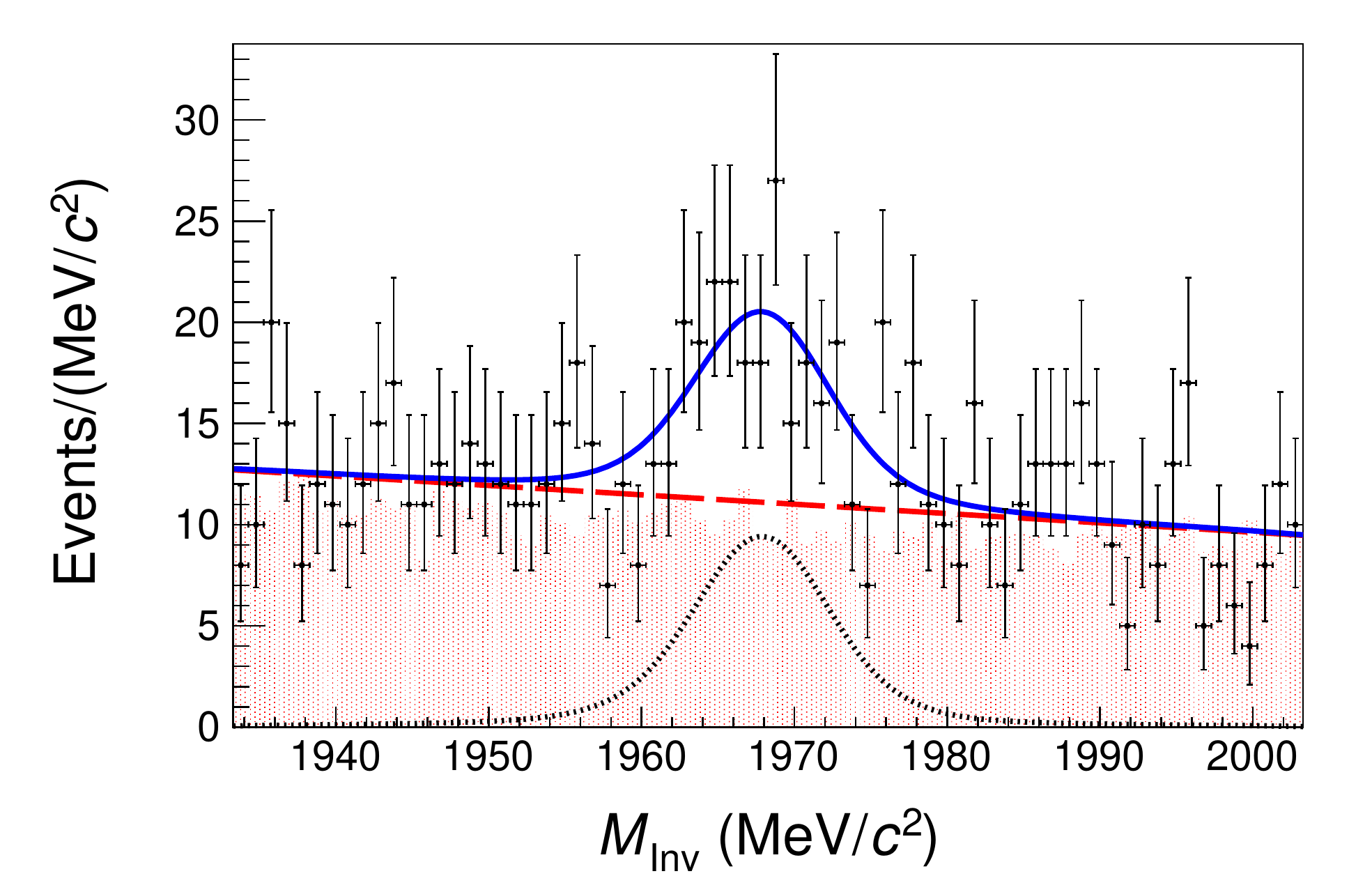}\\
\textbf{RS 950-1000 MeV/$c$} & \textbf{WS 950-1000 MeV/$c$}\\
\includegraphics[width=3in]{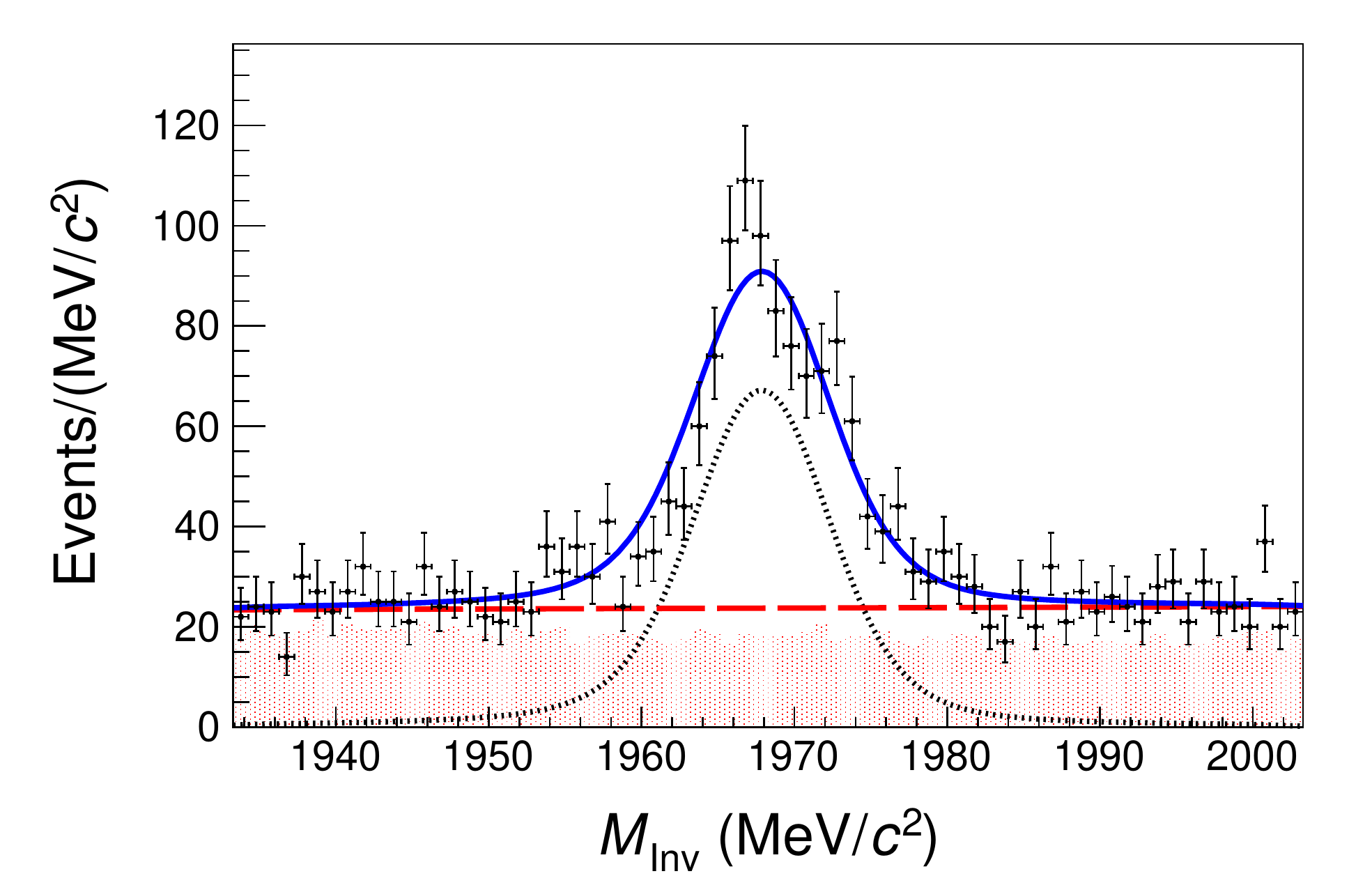} & \includegraphics[width=3in]{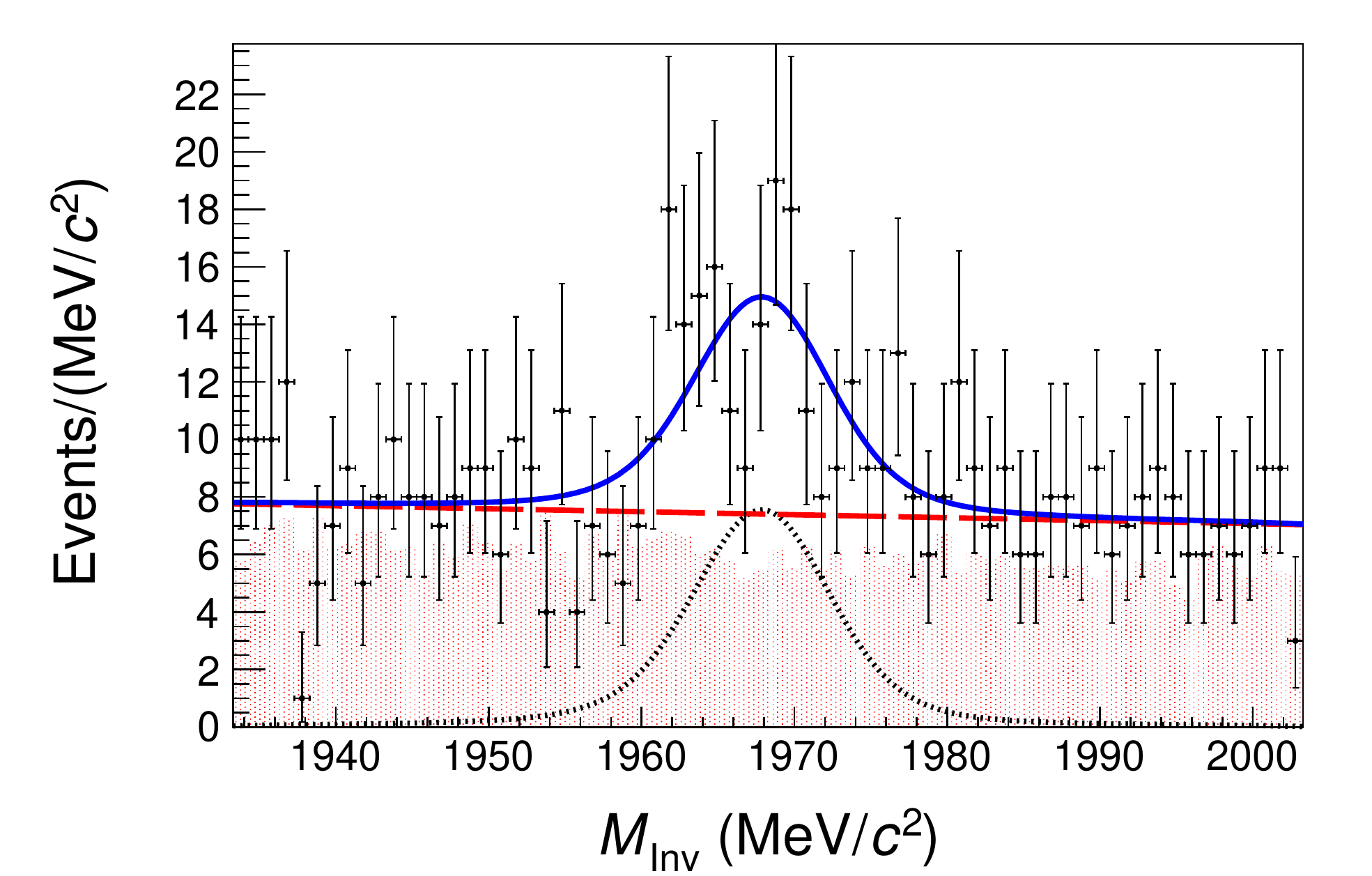}\\
\end{tabular}

\begin{tabular}{cc}
\textbf{RS 1000-1050 MeV/$c$} & \textbf{WS 1000-1050 MeV/$c$}\\
\includegraphics[width=3in]{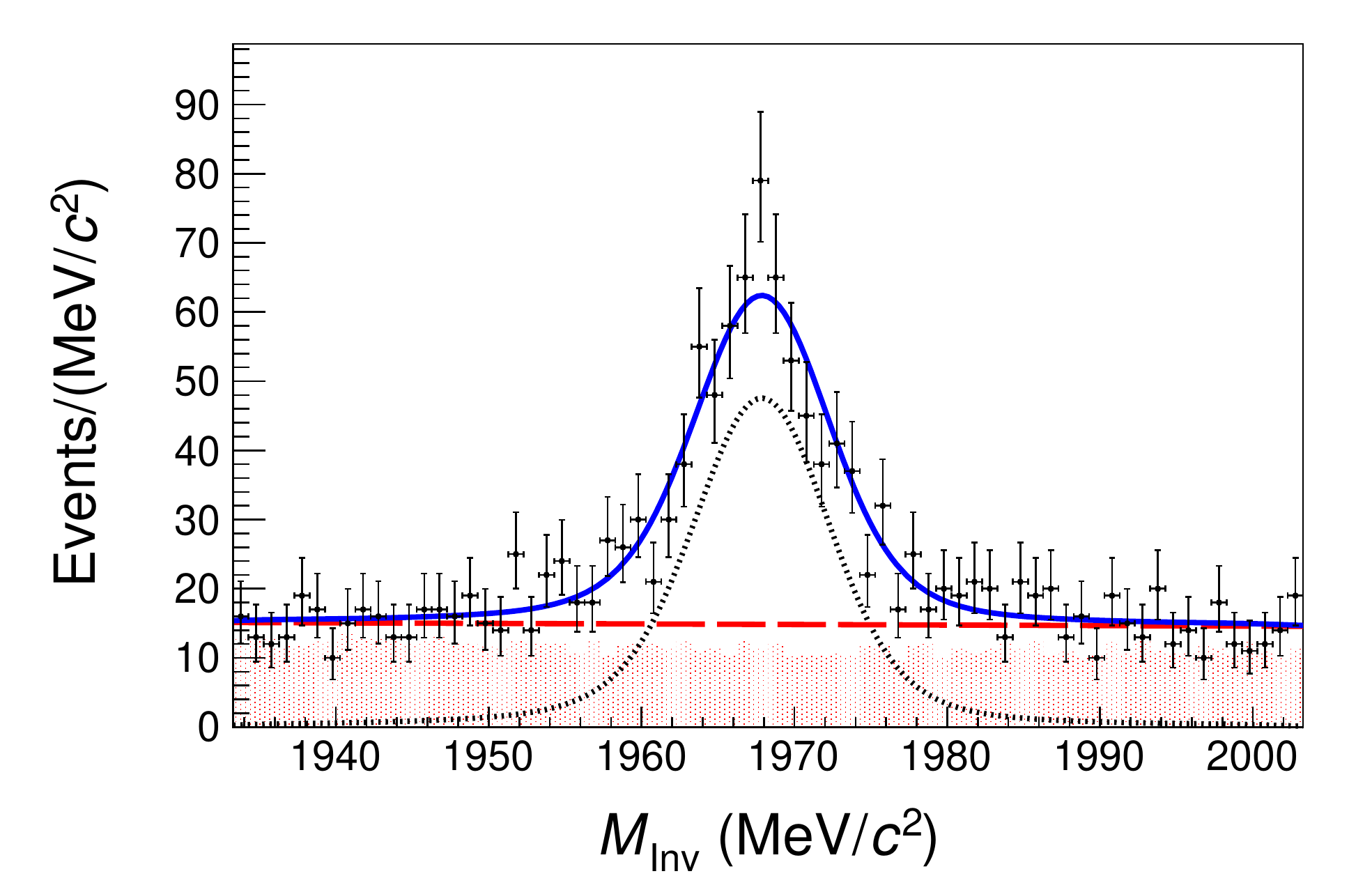} & \includegraphics[width=3in]{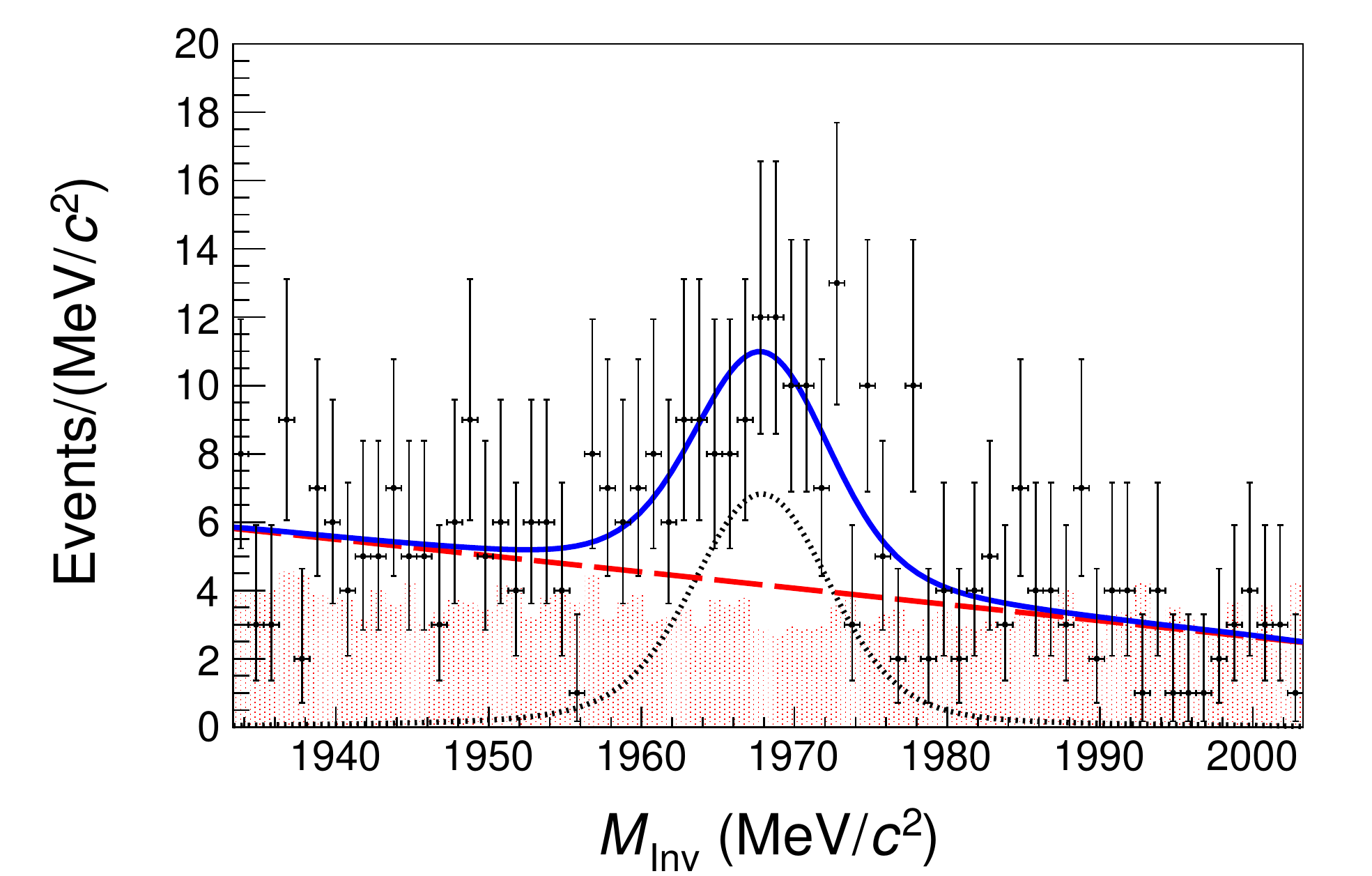}\\
\textbf{RS 1050-1100 MeV/$c$} & \textbf{WS 1050-1100 MeV/$c$}\\
\includegraphics[width=3in]{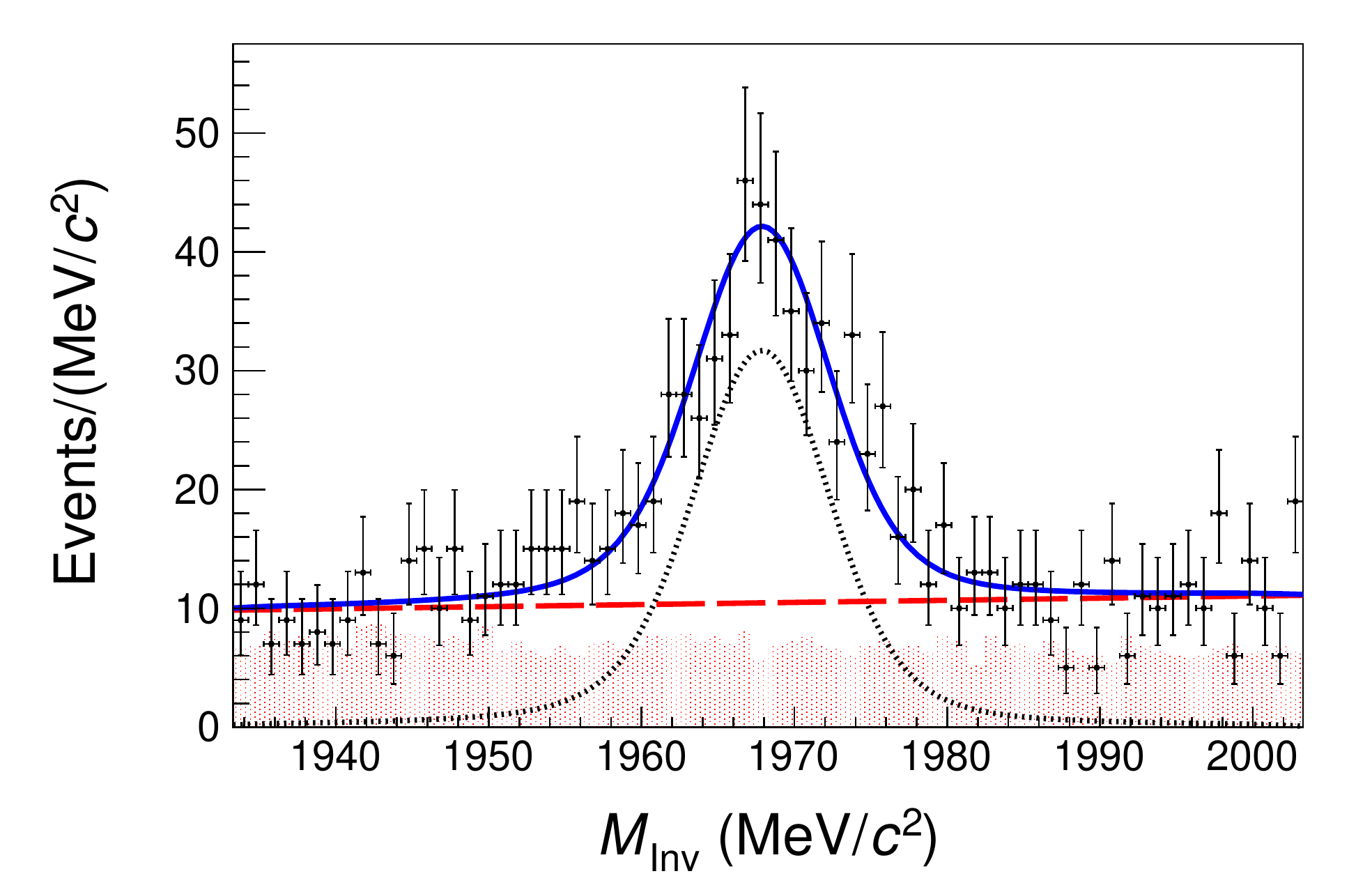} & \includegraphics[width=3in]{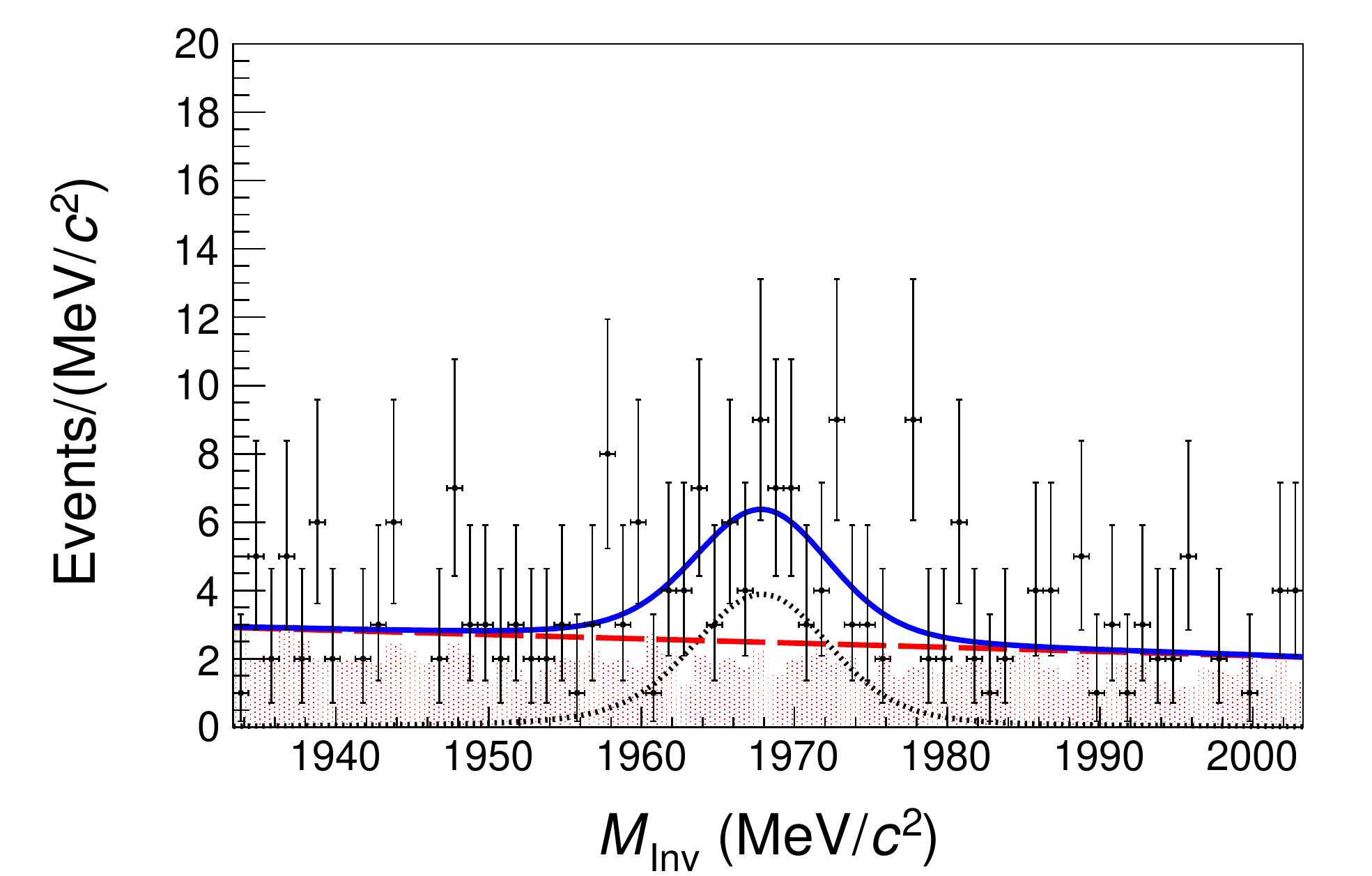}
\end{tabular}

\pagebreak
\raggedright

\subsubsection{$\EcmB$ Data $K$ ID Fits}
\label{subsubsec:XYZDataKIDFits}
\centering
\begin{tabular}{cc}
\textbf{RS 200-250 MeV/$c$} & \textbf{WS 200-250 MeV/$c$}\\
\includegraphics[width=3in]{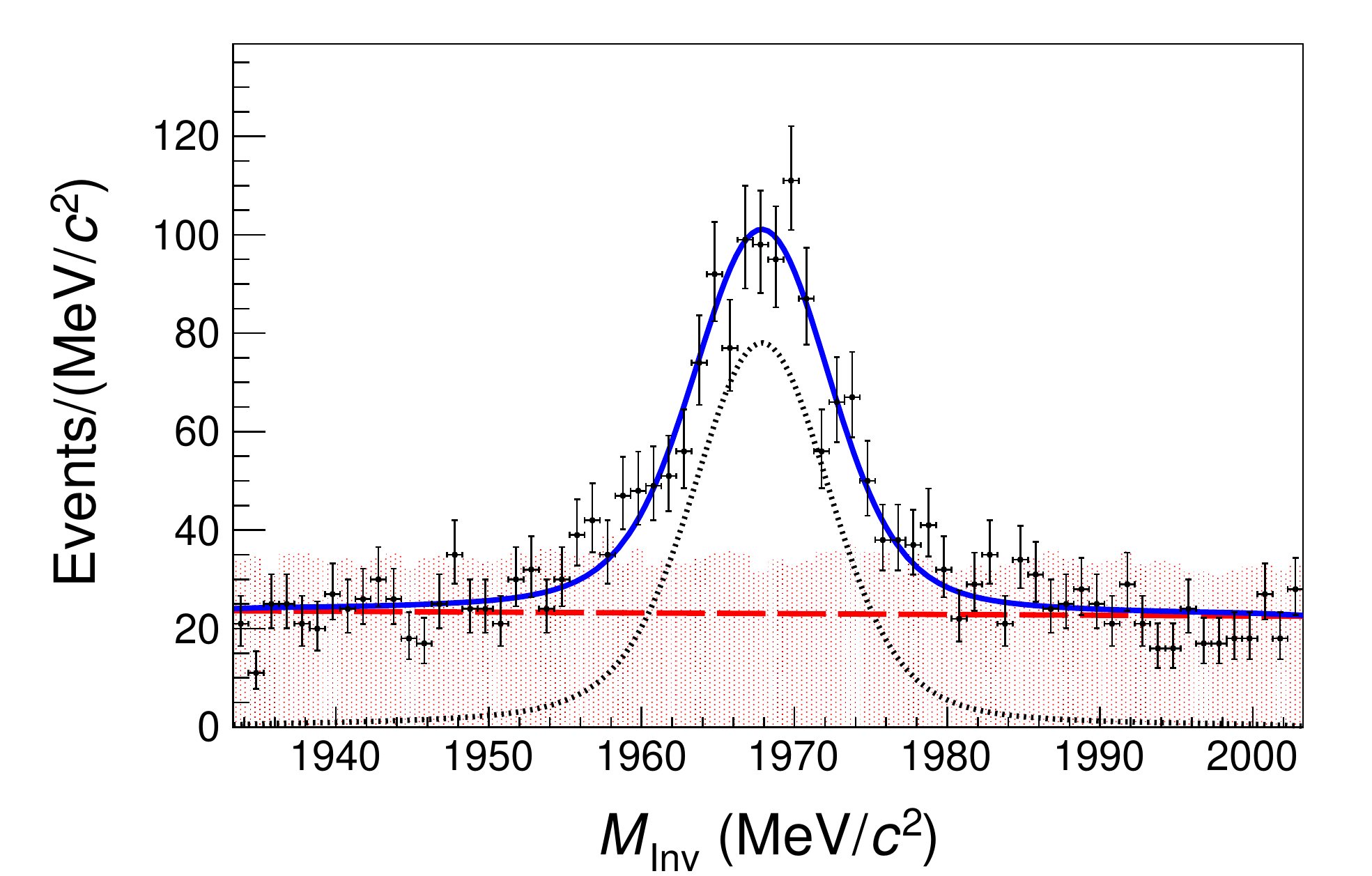} & \includegraphics[width=3in]{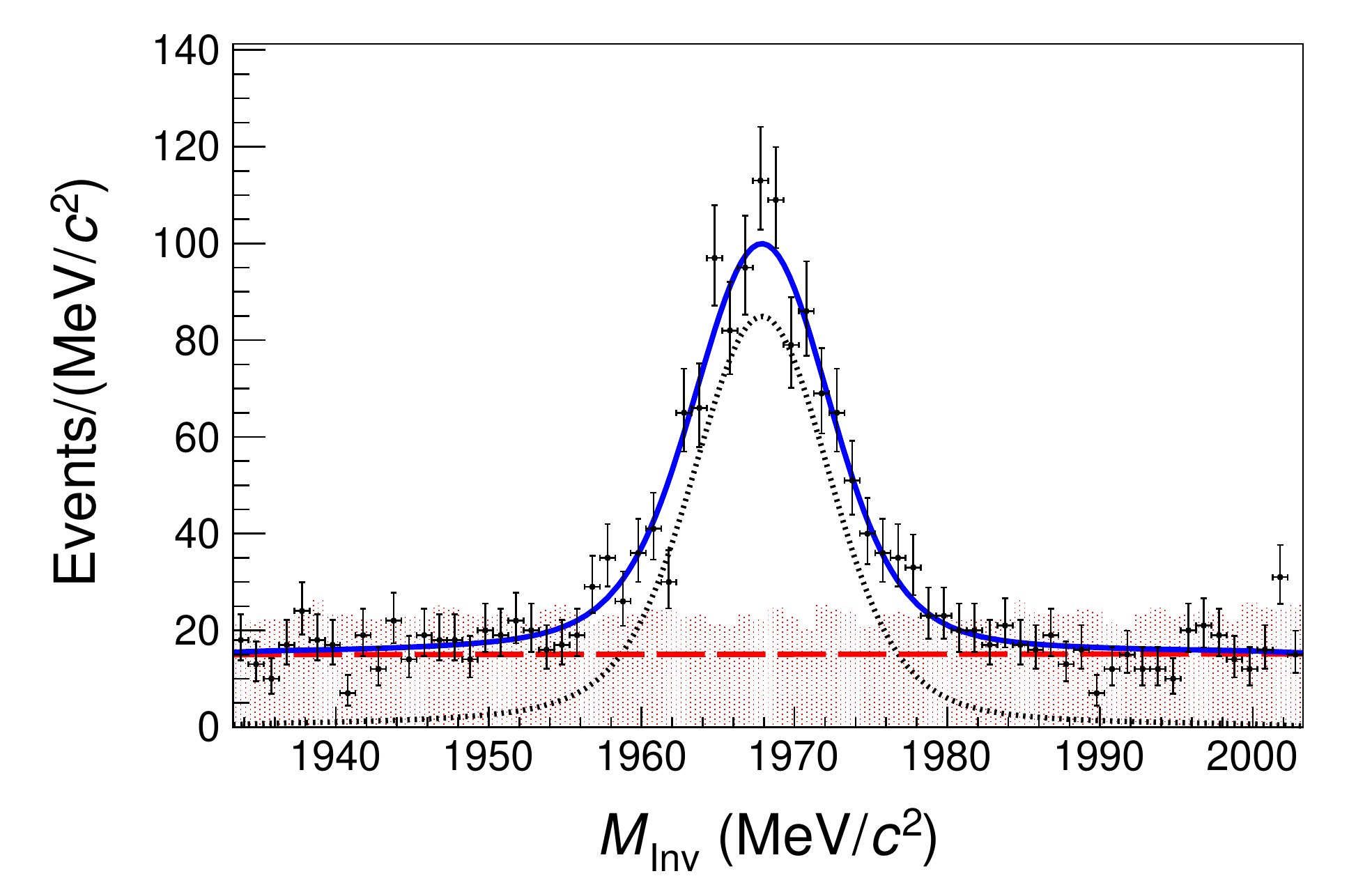}\\
\textbf{RS 250-300 MeV/$c$} & \textbf{WS 250-300 MeV/$c$}\\
\includegraphics[width=3in]{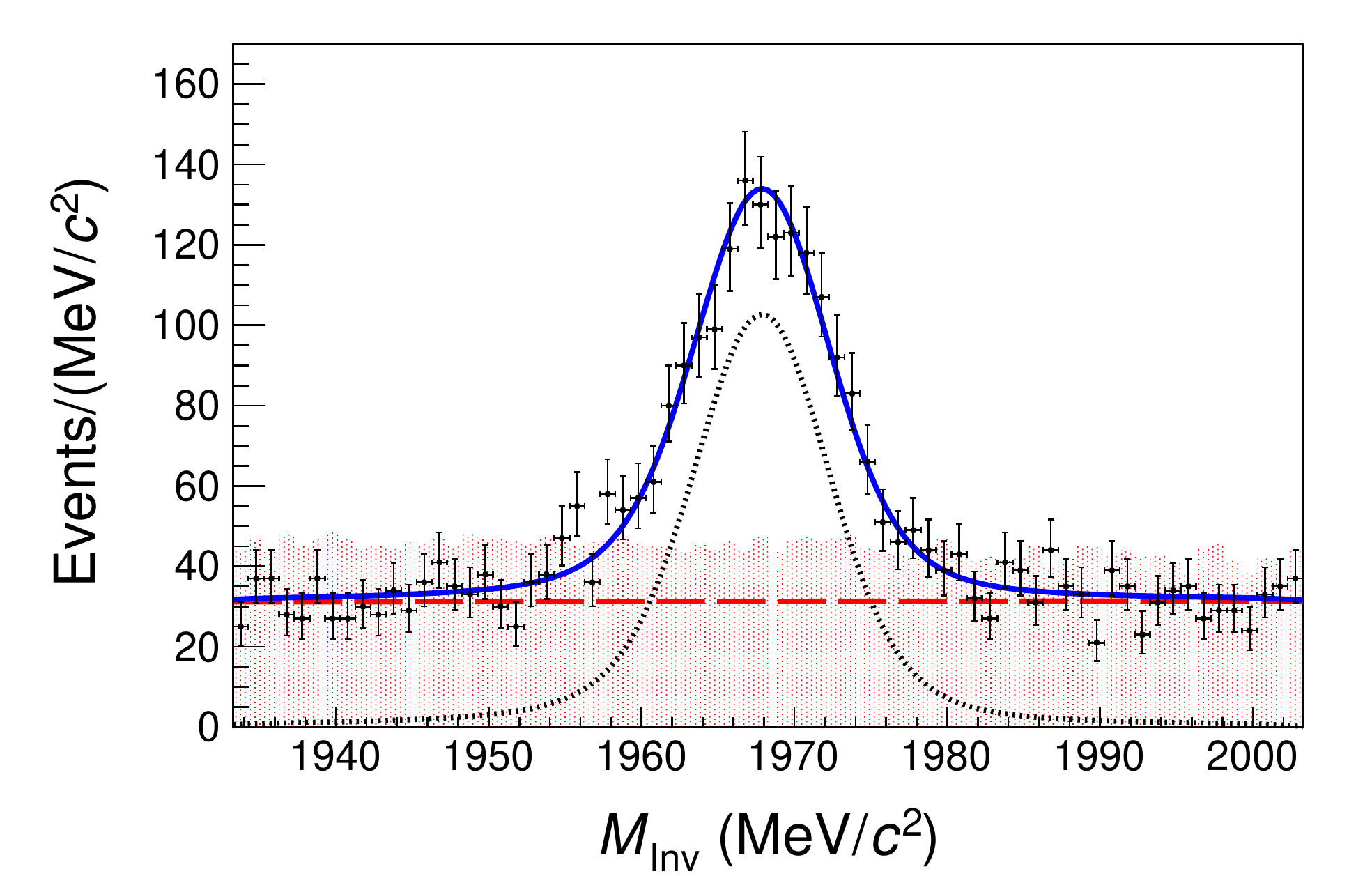} & \includegraphics[width=3in]{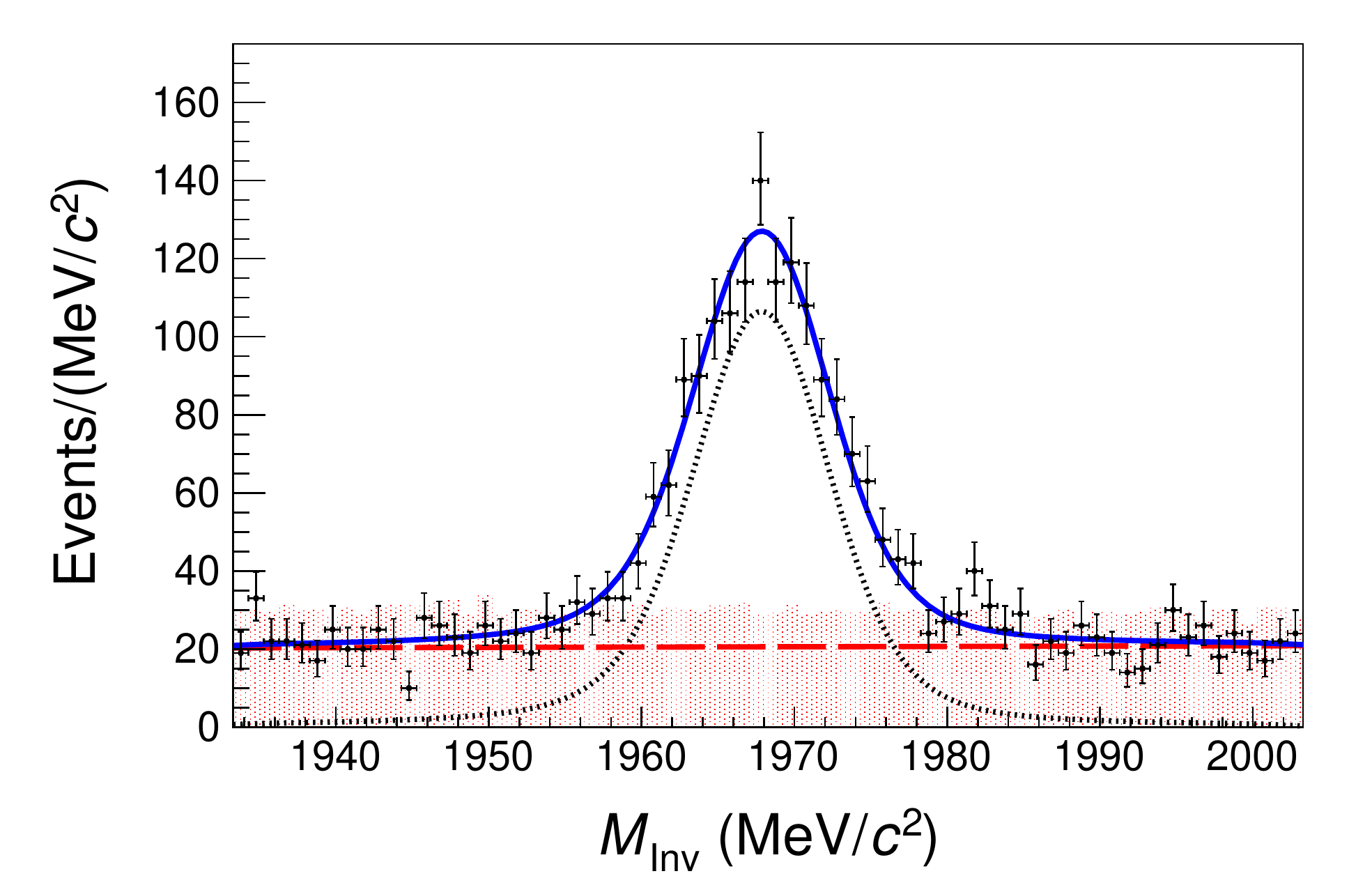}\\
\textbf{RS 300-350 MeV/$c$} & \textbf{WS 300-350 MeV/$c$}\\
\includegraphics[width=3in]{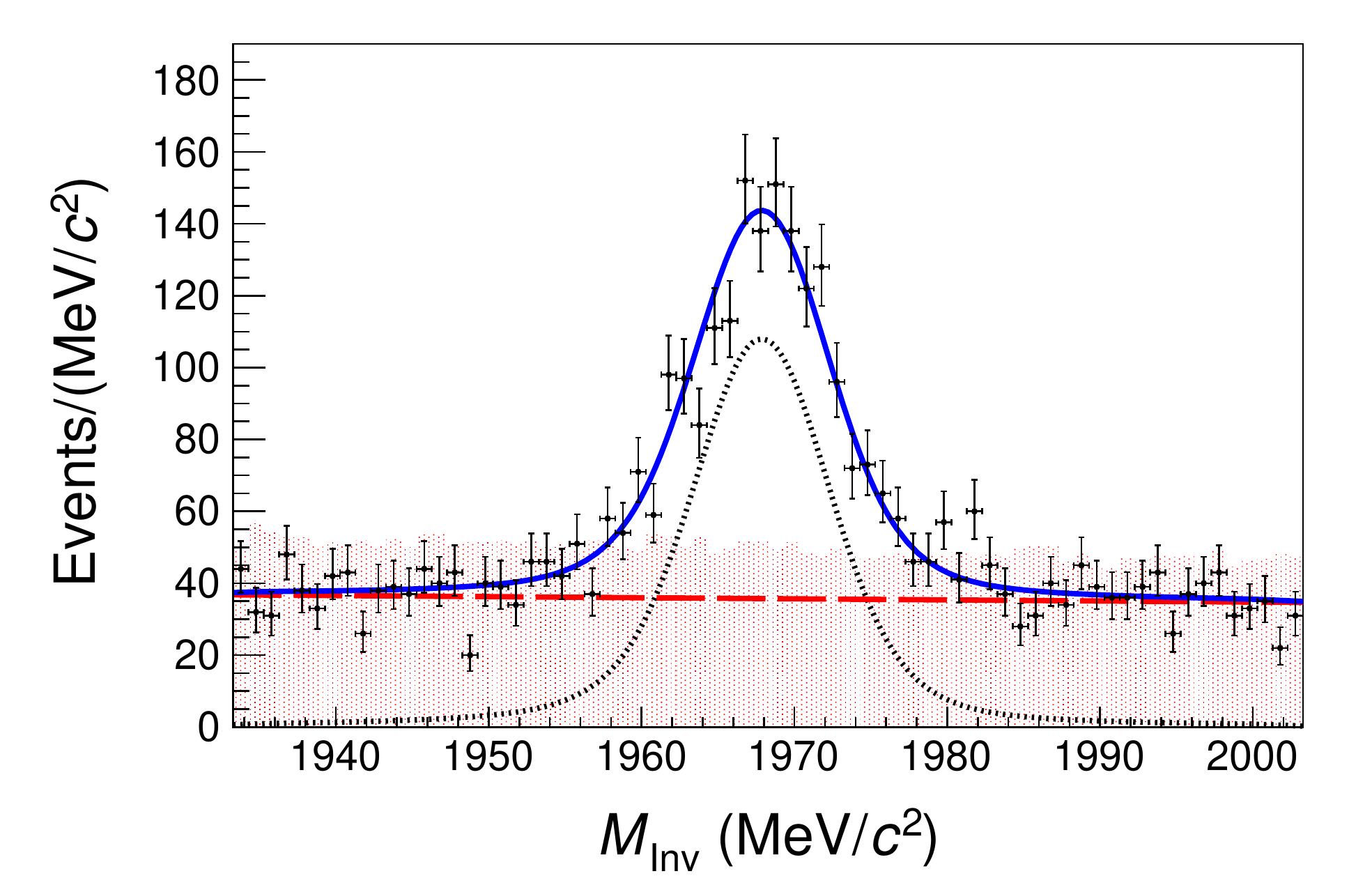} & \includegraphics[width=3in]{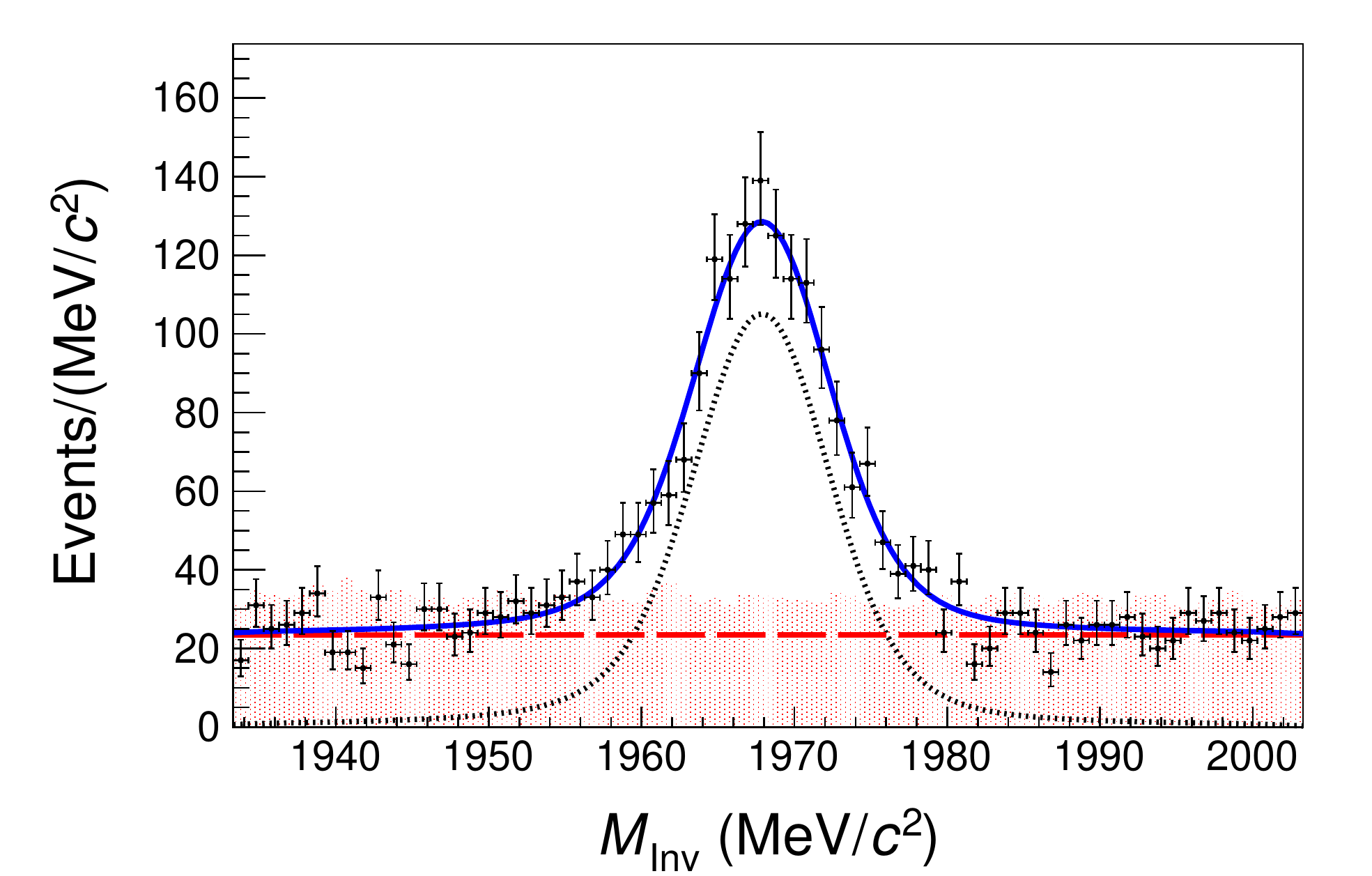}\\
\textbf{RS 350-400 MeV/$c$} & \textbf{WS 350-400 MeV/$c$}\\
\includegraphics[width=3in]{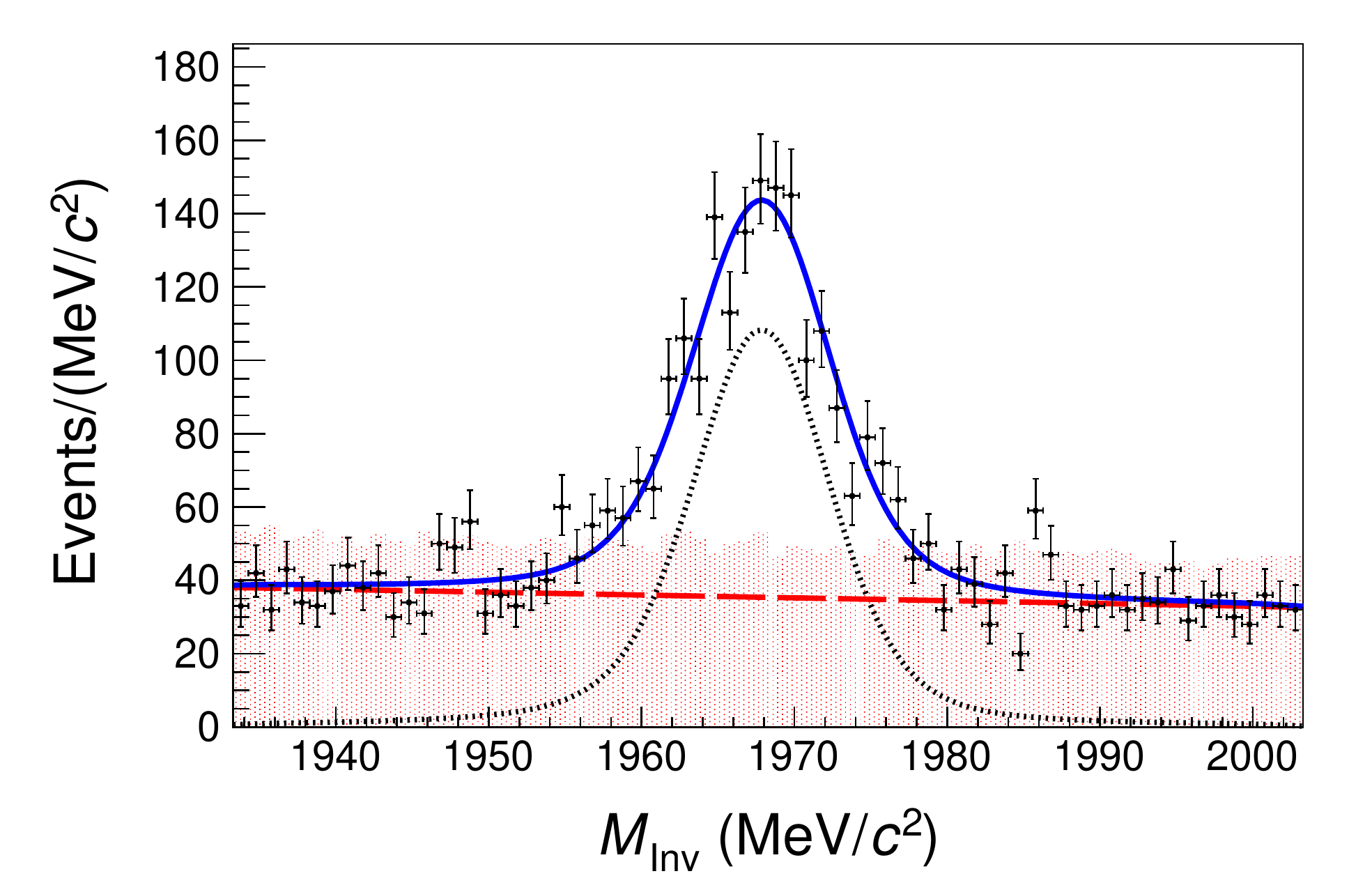} & \includegraphics[width=3in]{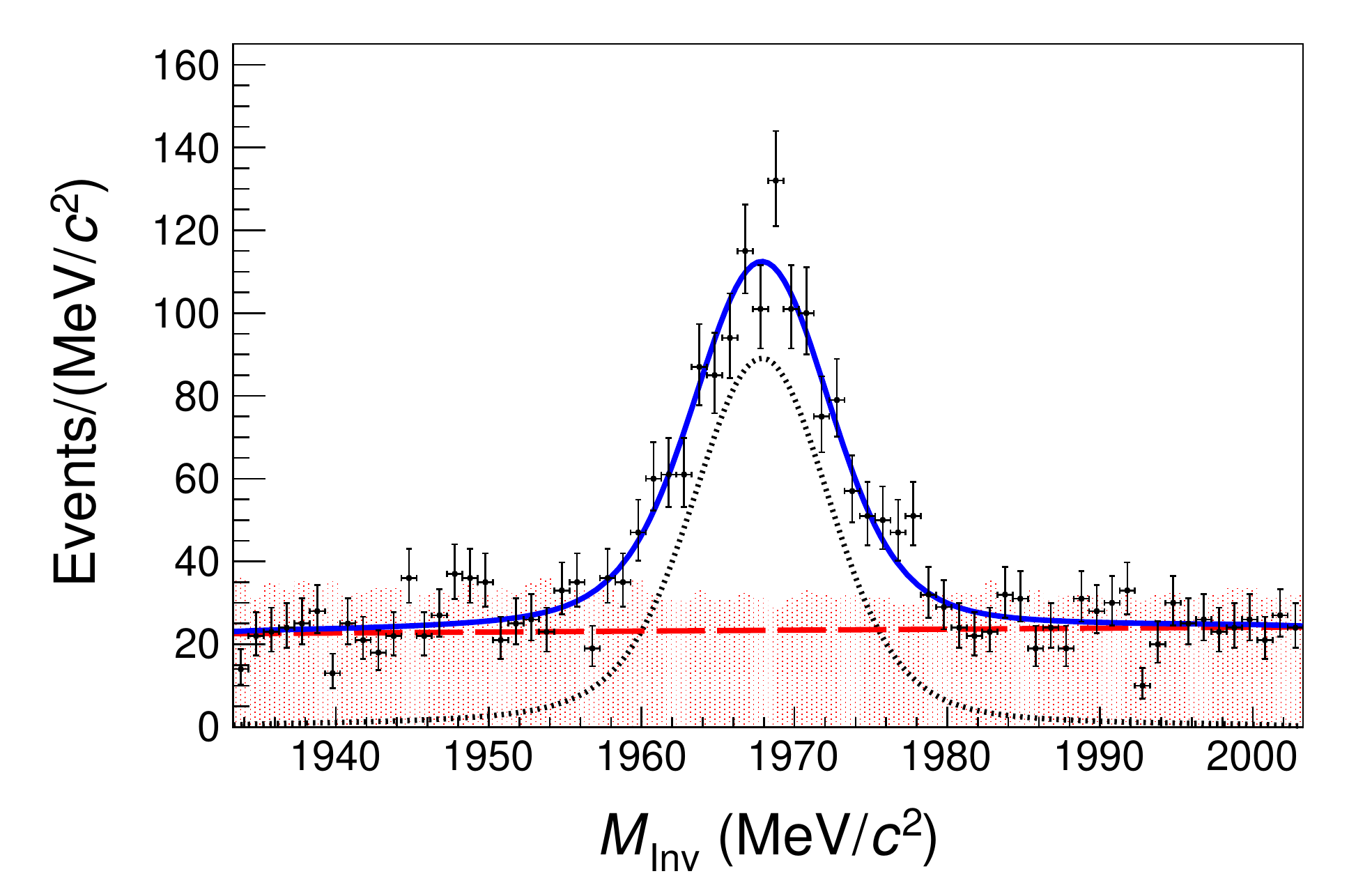}
\end{tabular}
\pagebreak

\begin{tabular}{cc}
\textbf{RS 400-450 MeV/$c$} & \textbf{WS 400-450 MeV/$c$}\\
\includegraphics[width=3in]{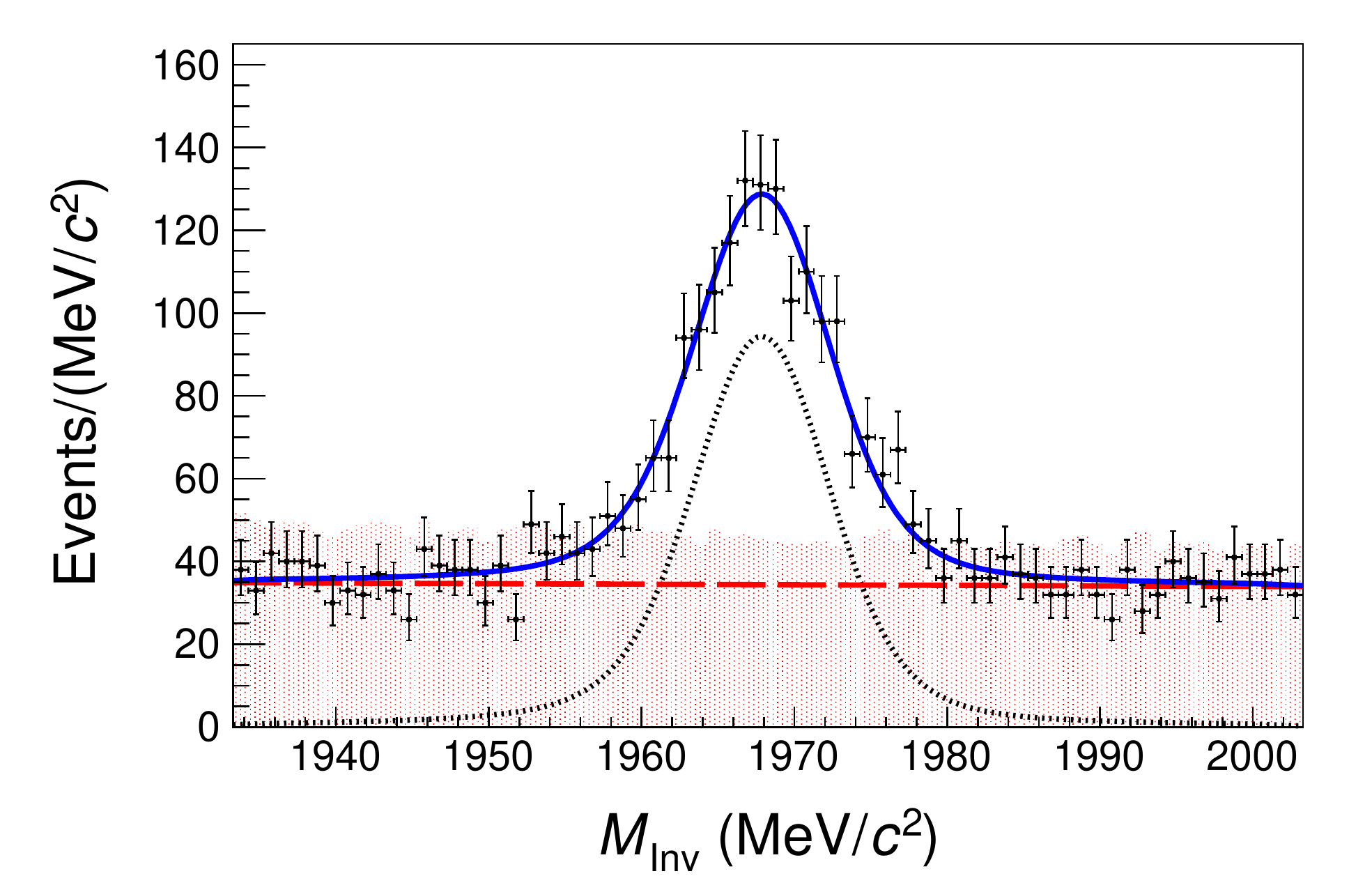} & \includegraphics[width=3in]{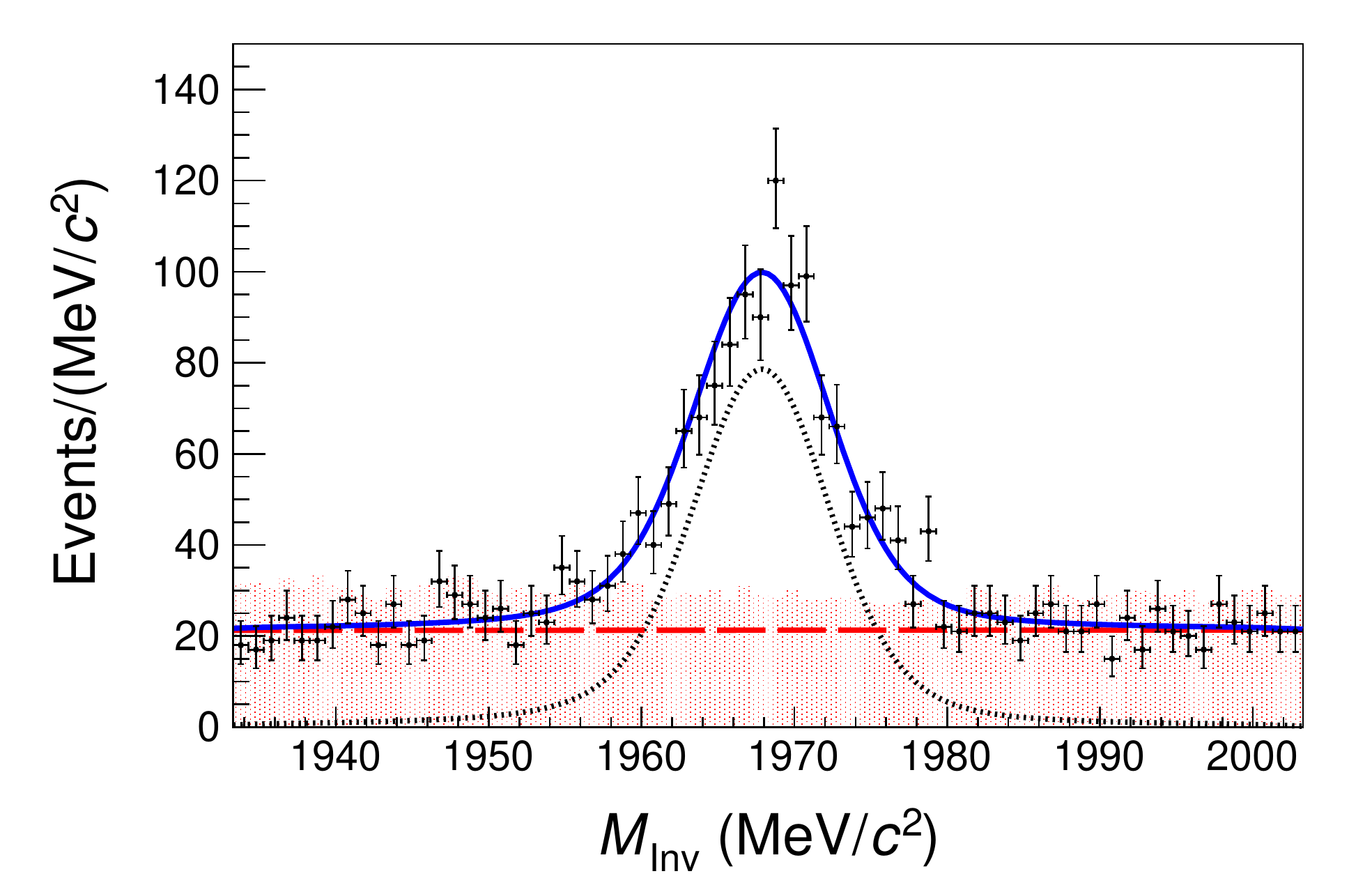}\\
\textbf{RS 450-500 MeV/$c$} & \textbf{WS 450-500 MeV/$c$}\\
\includegraphics[width=3in]{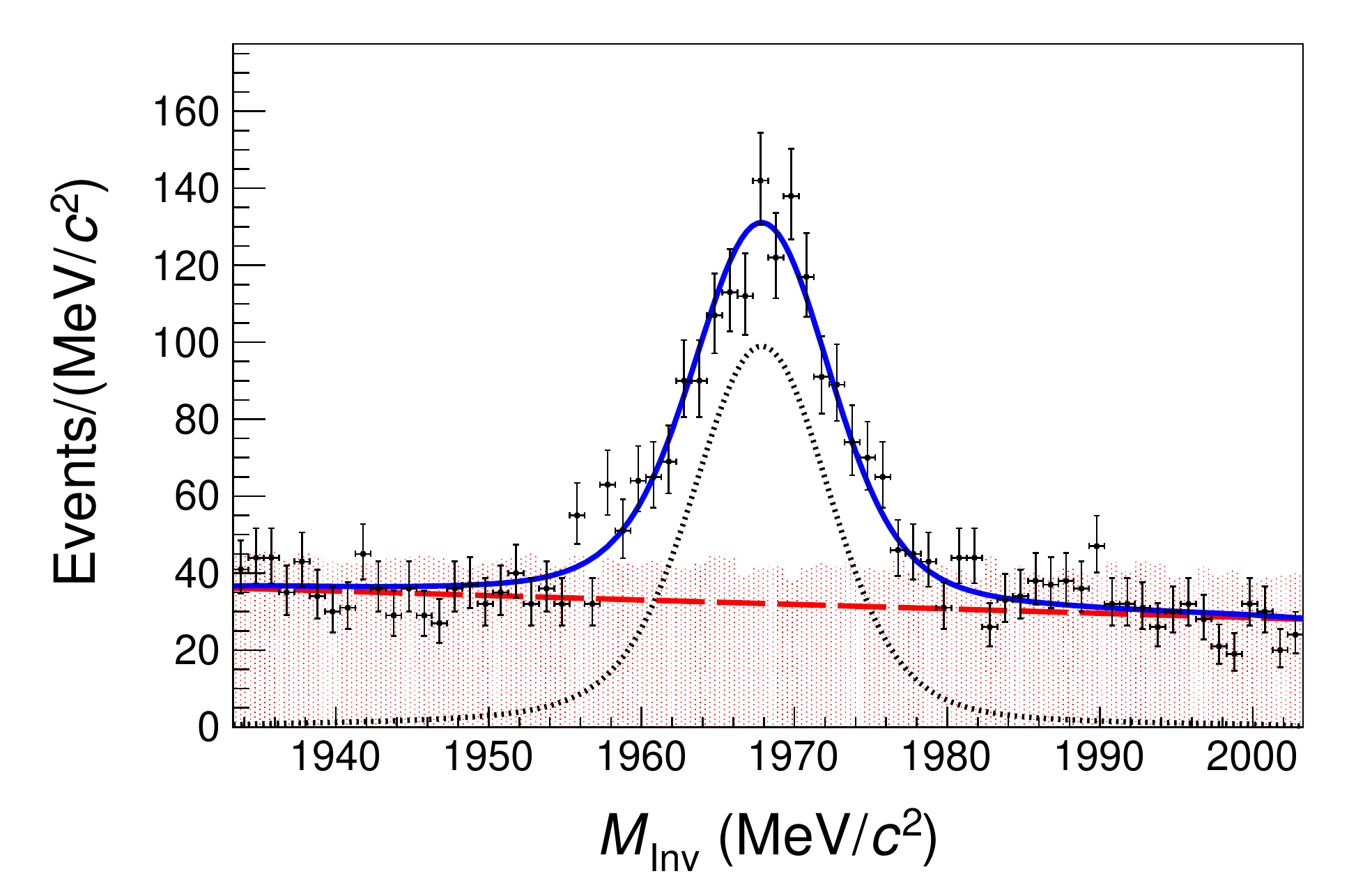} & \includegraphics[width=3in]{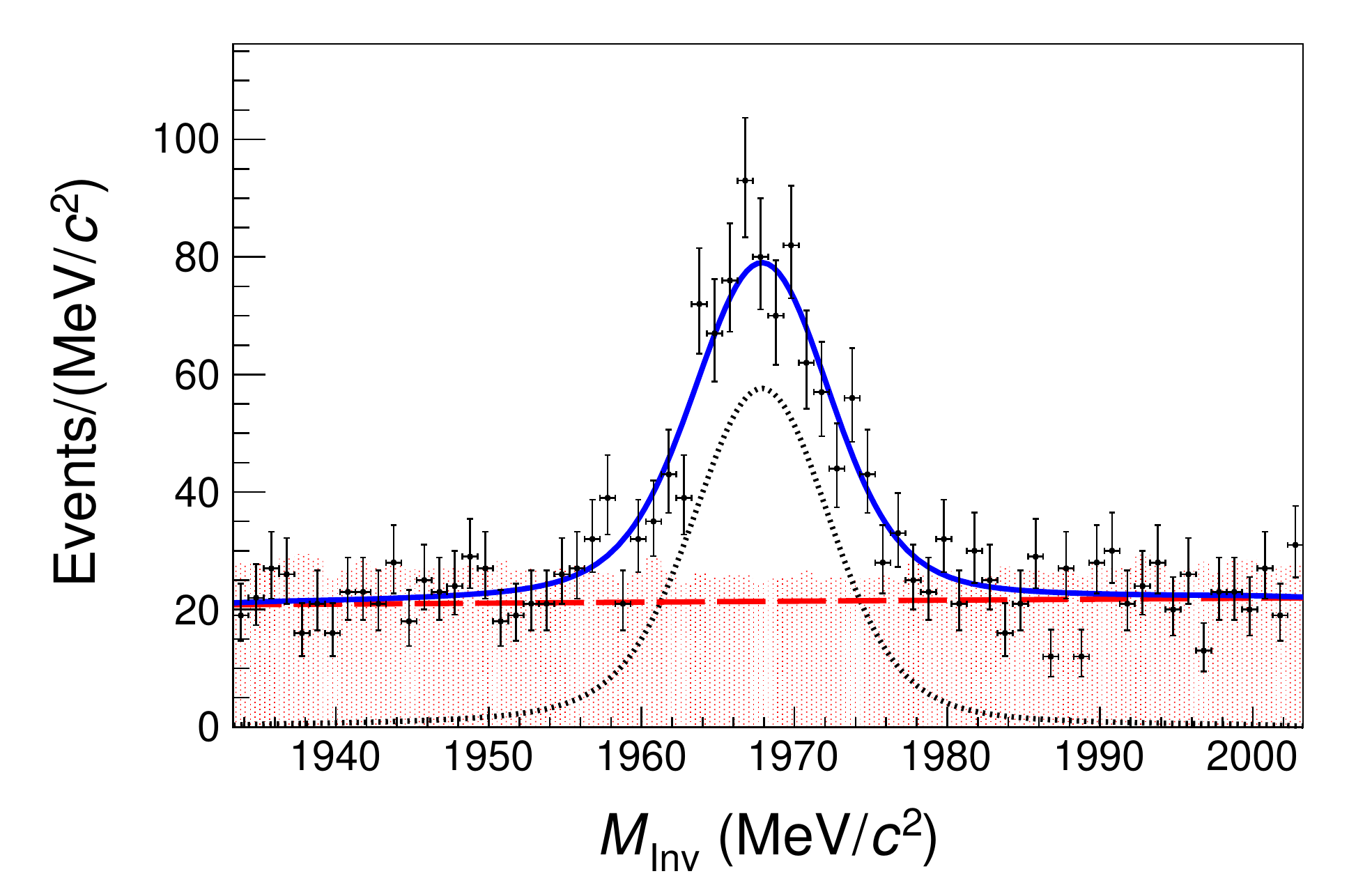}\\
\textbf{RS 500-550 MeV/$c$} & \textbf{WS 500-550 MeV/$c$}\\
\includegraphics[width=3in]{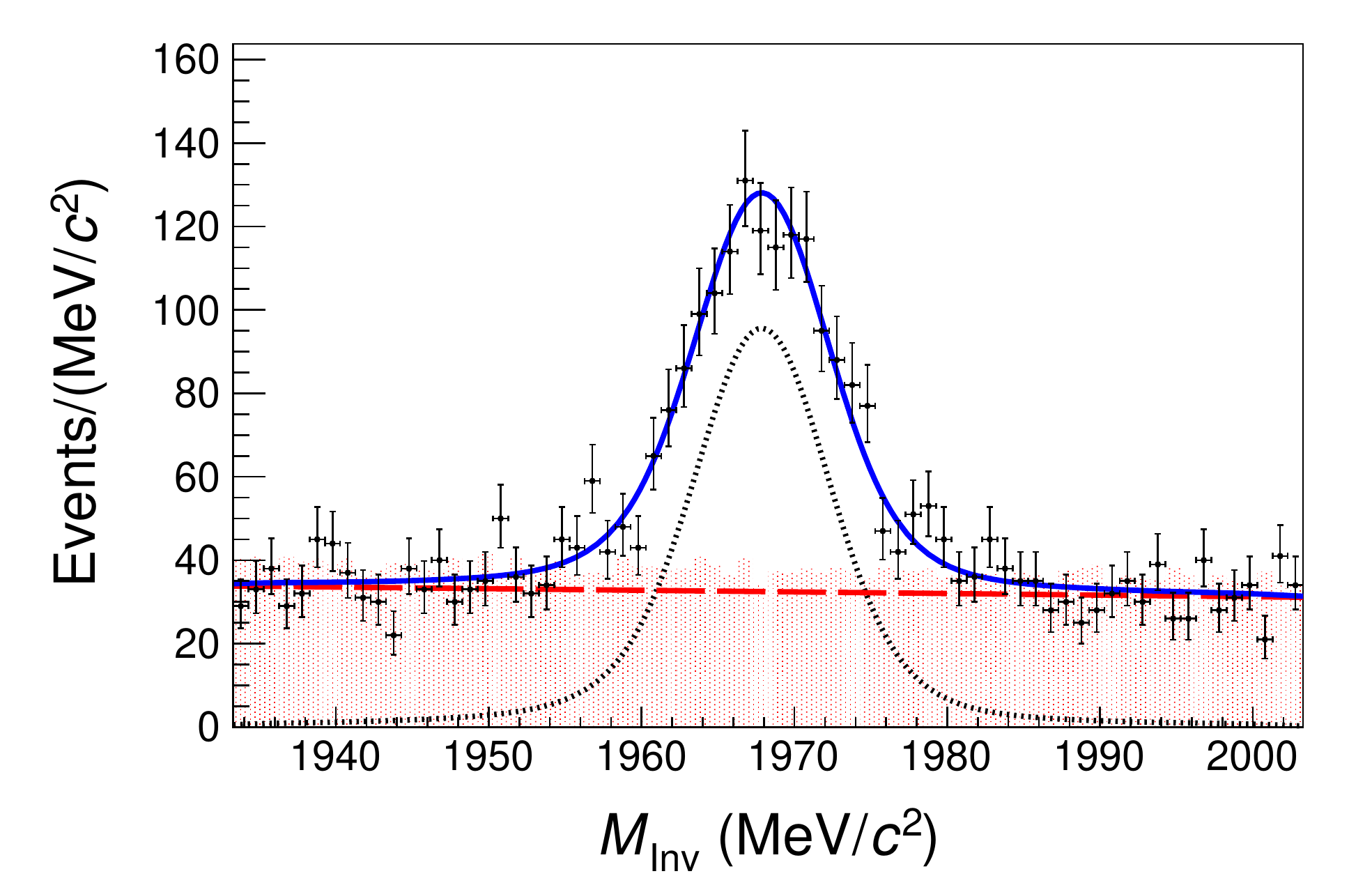} & \includegraphics[width=3in]{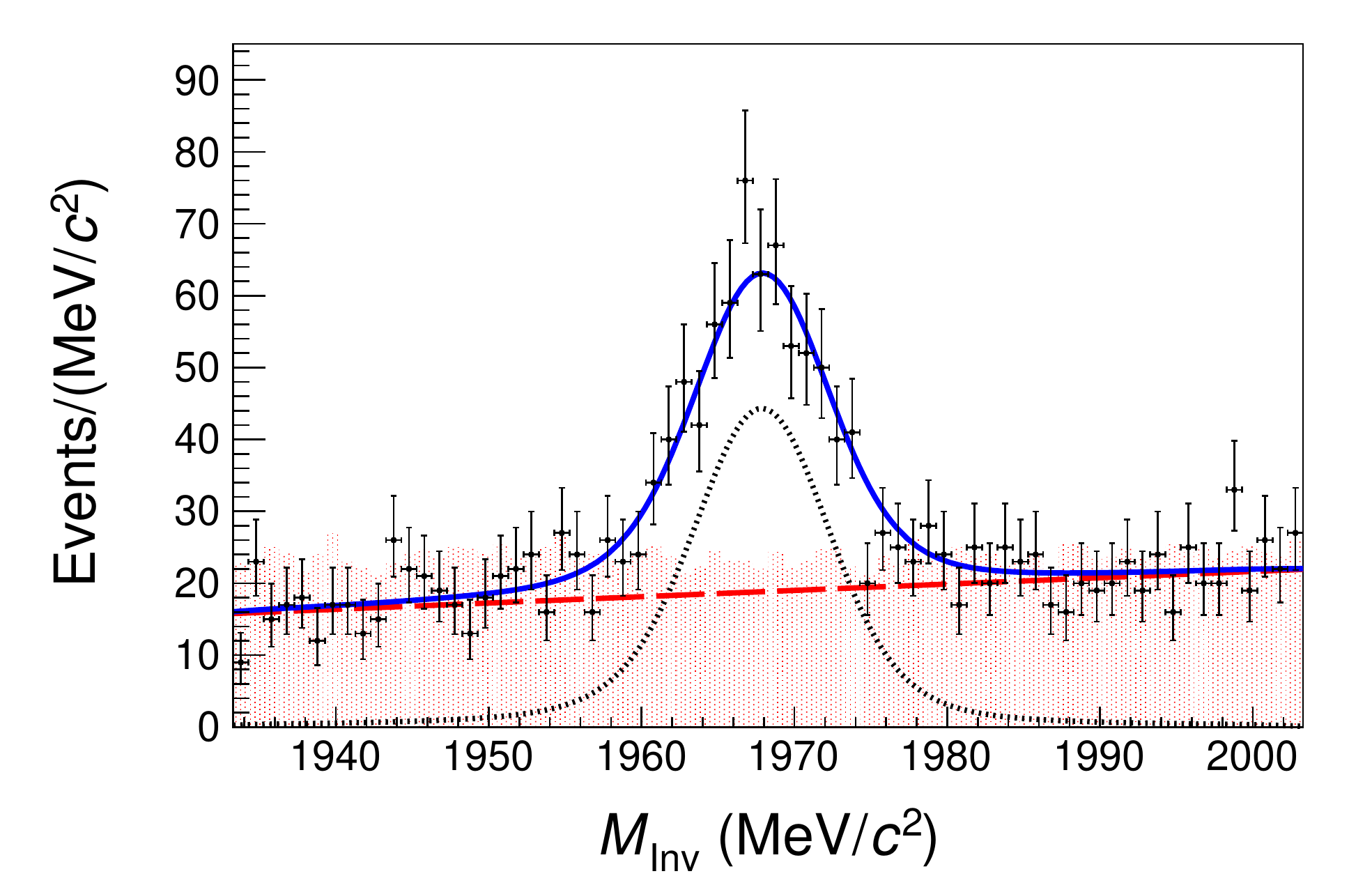}\\
\textbf{RS 550-600 MeV/$c$} & \textbf{WS 550-600 MeV/$c$}\\
\includegraphics[width=3in]{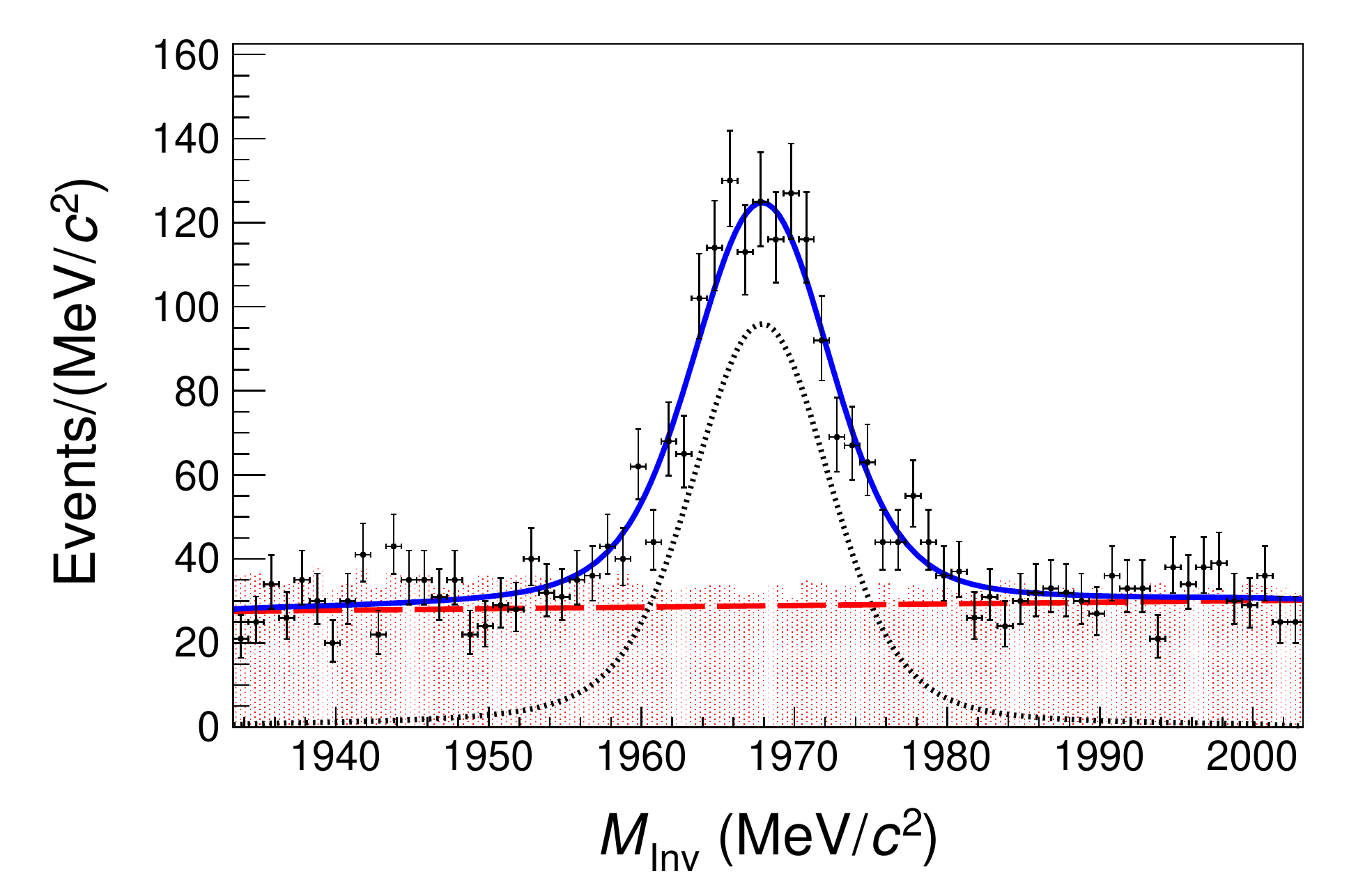} & \includegraphics[width=3in]{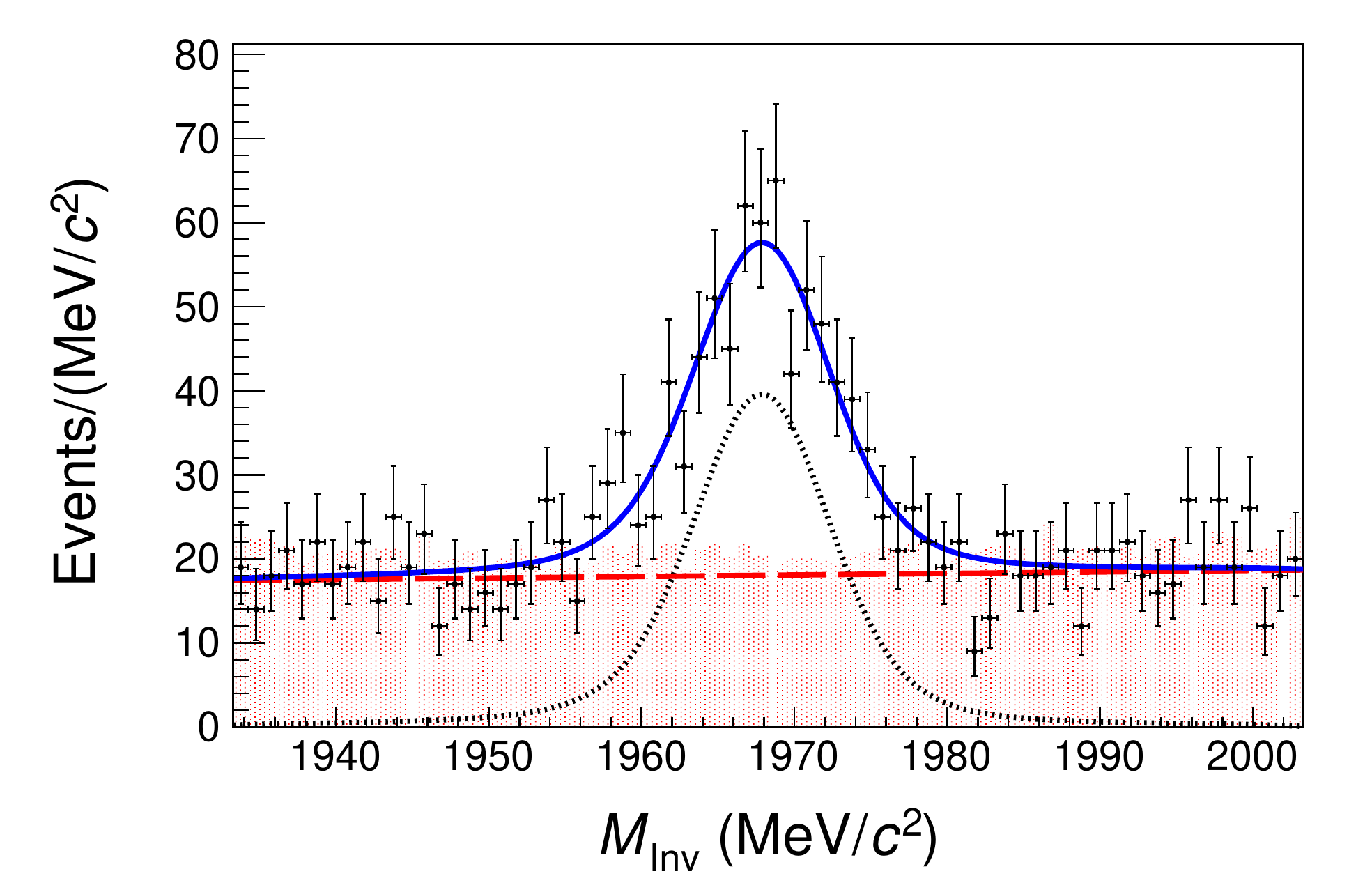}
\end{tabular}

\begin{tabular}{cc}
\textbf{RS 600-650 MeV/$c$} & \textbf{WS 600-650 MeV/$c$}\\
\includegraphics[width=3in]{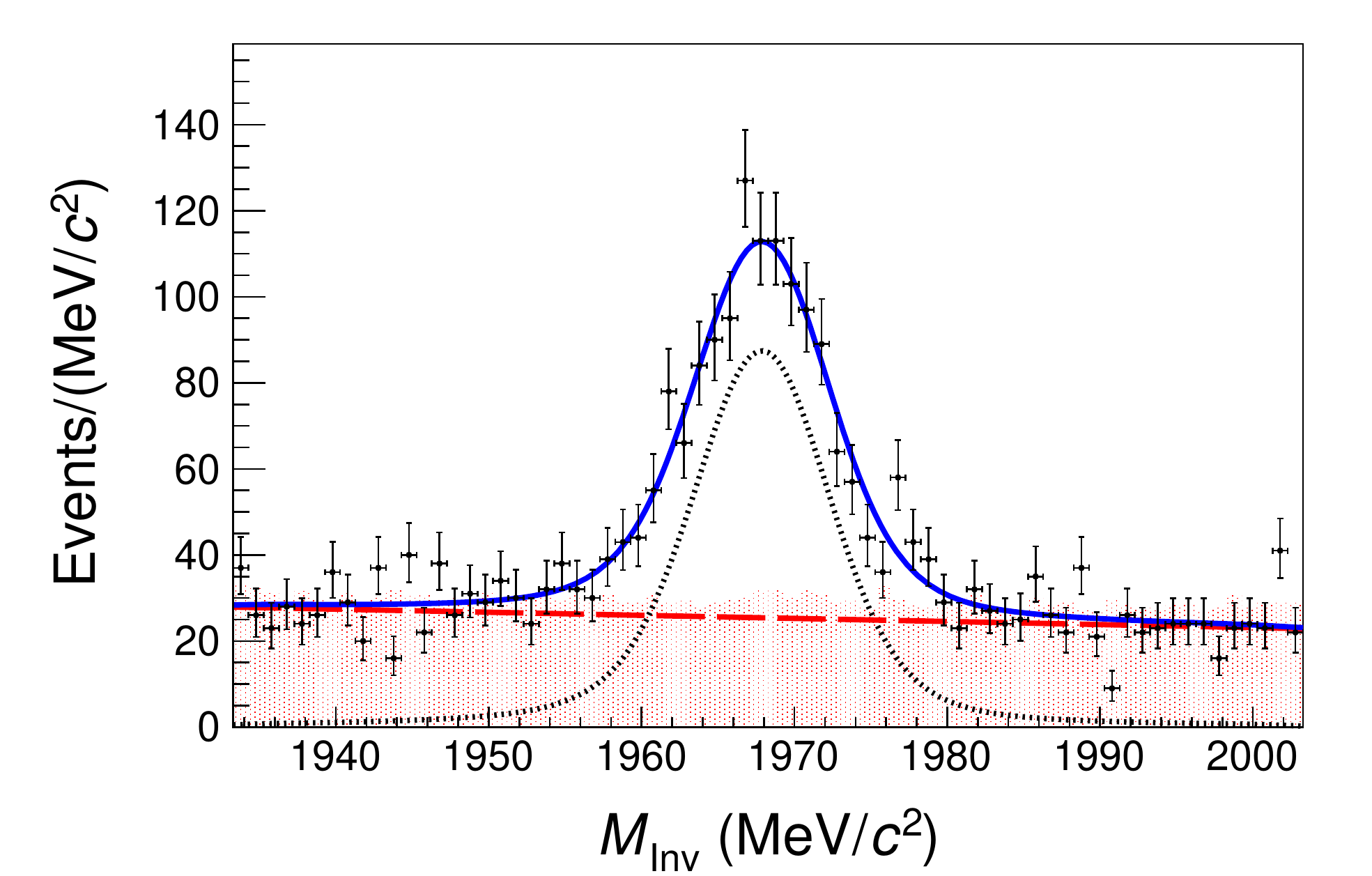} & \includegraphics[width=3in]{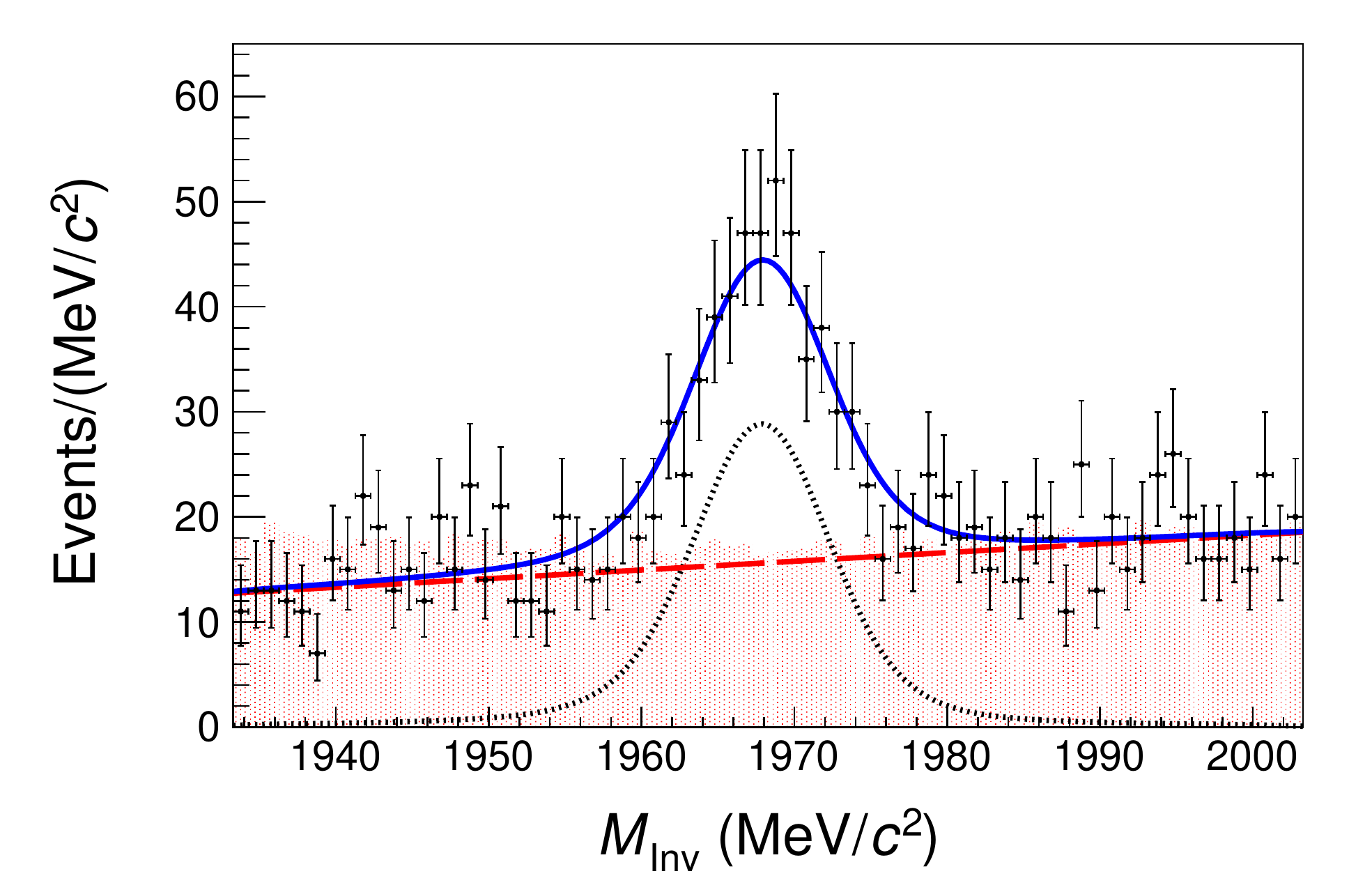}\\
\textbf{RS 650-700 MeV/$c$} & \textbf{WS 650-700 MeV/$c$}\\
\includegraphics[width=3in]{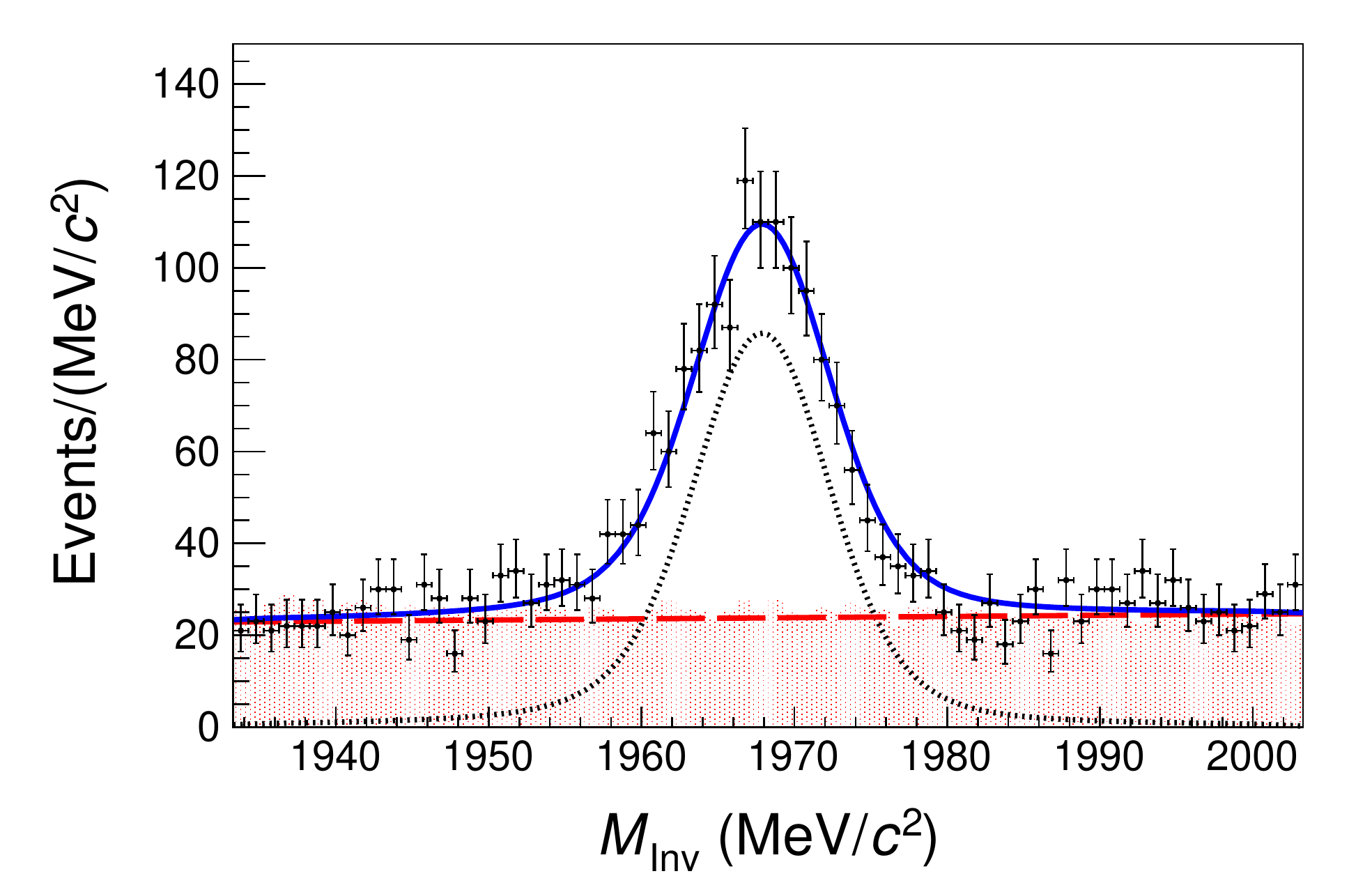} & \includegraphics[width=3in]{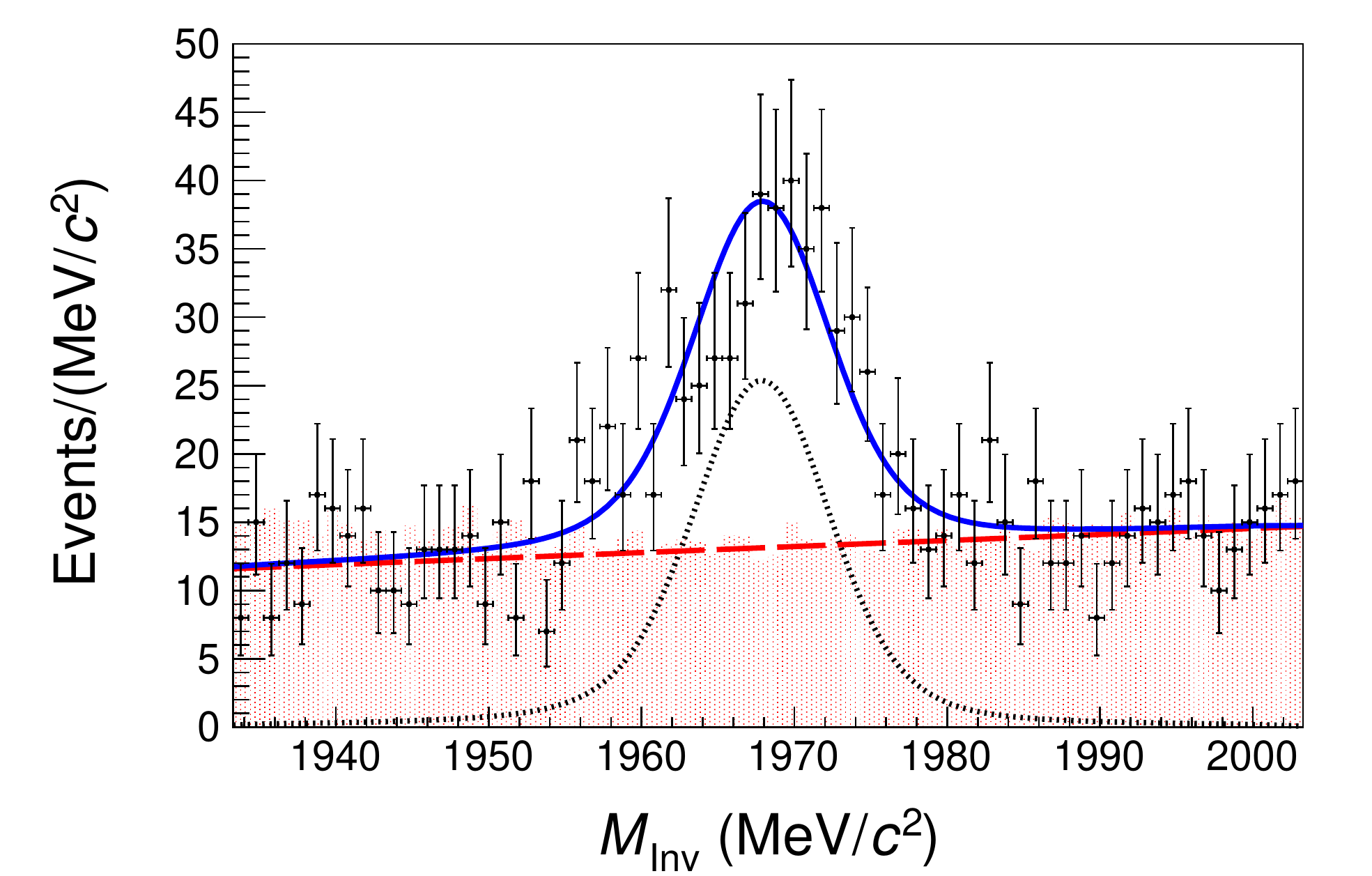}\\
\textbf{RS 700-750 MeV/$c$} & \textbf{WS 700-750 MeV/$c$}\\
\includegraphics[width=3in]{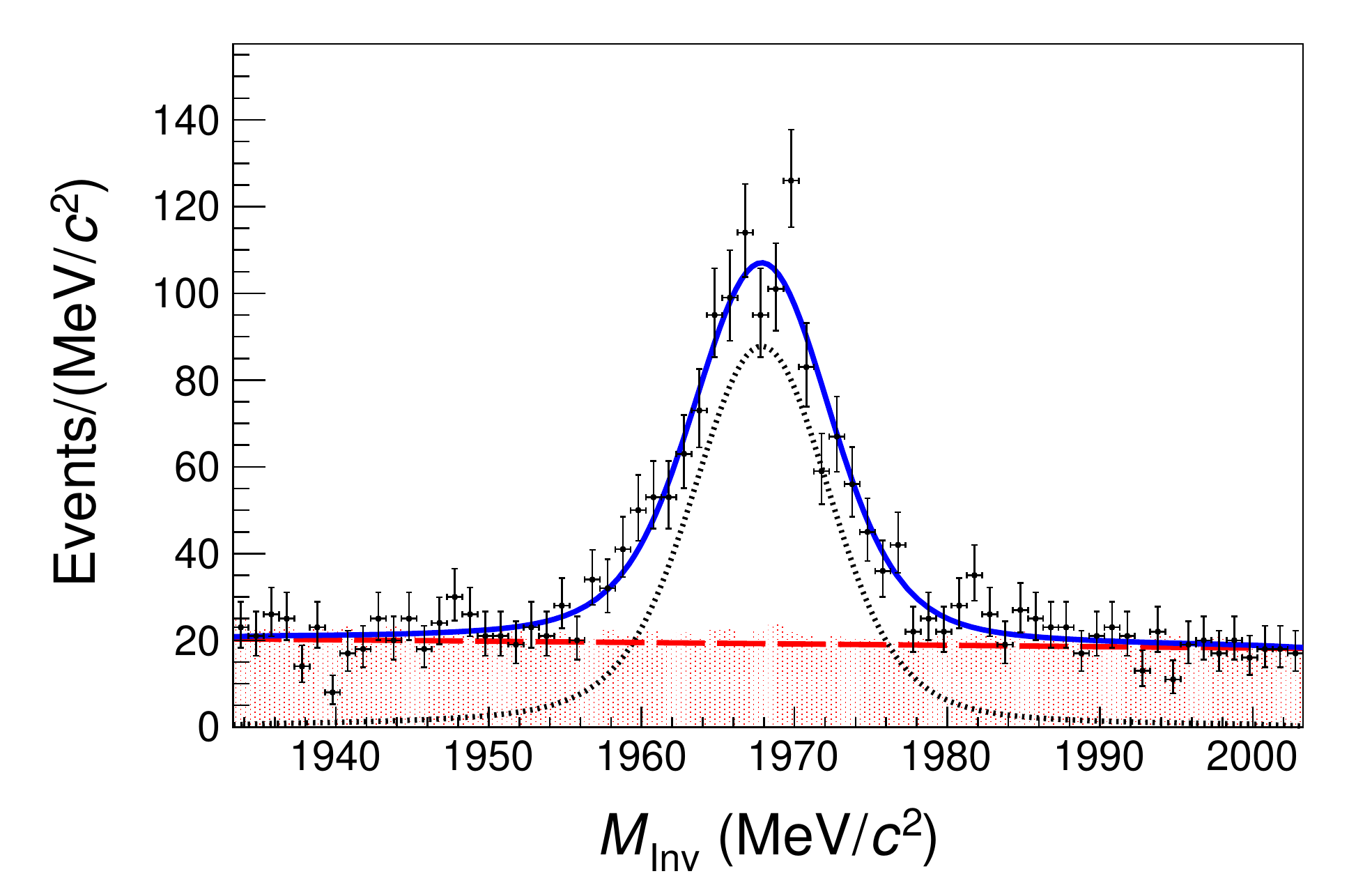} & \includegraphics[width=3in]{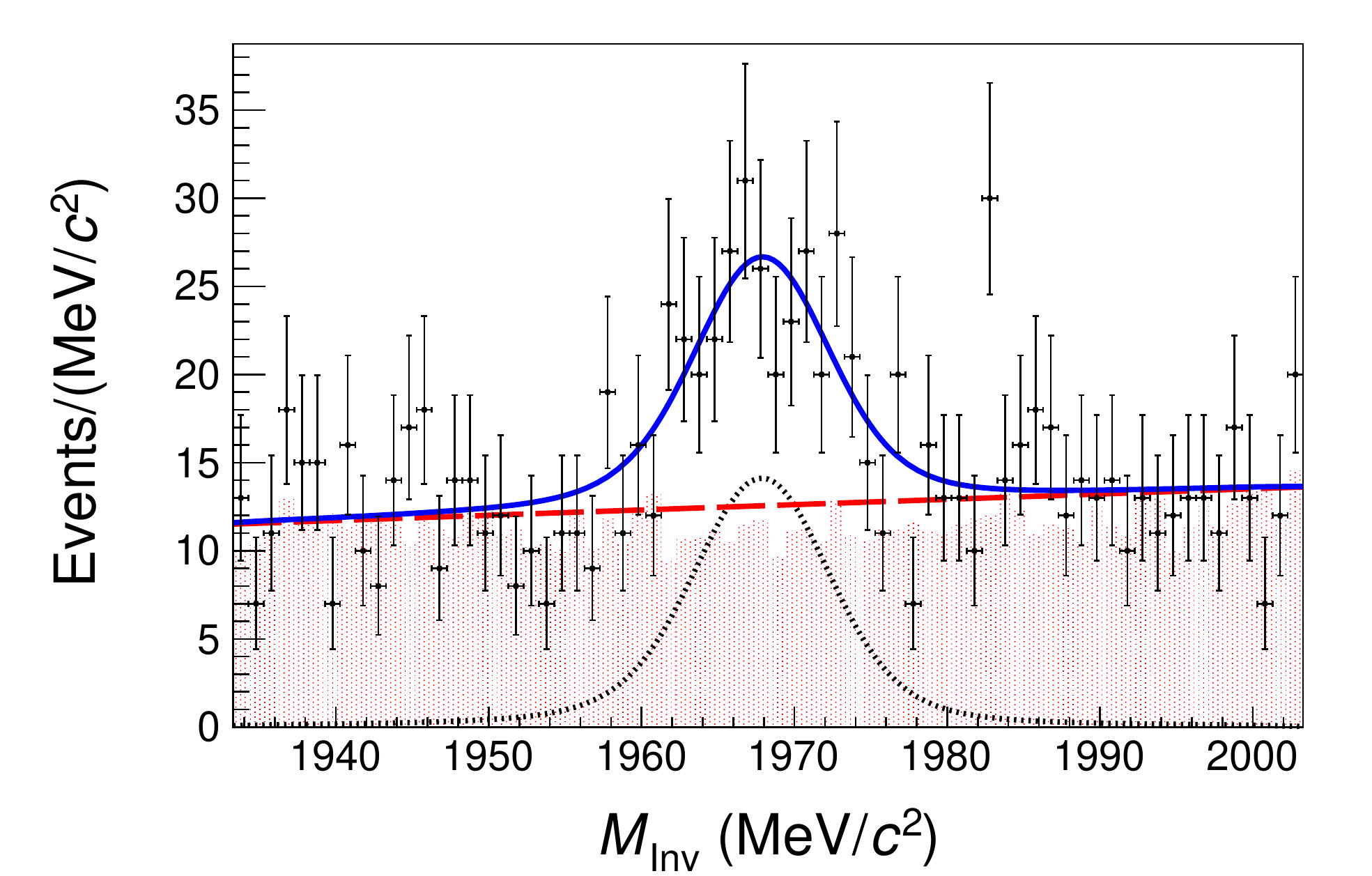}\\
\textbf{RS 750-800 MeV/$c$} & \textbf{WS 750-800 MeV/$c$}\\
\includegraphics[width=3in]{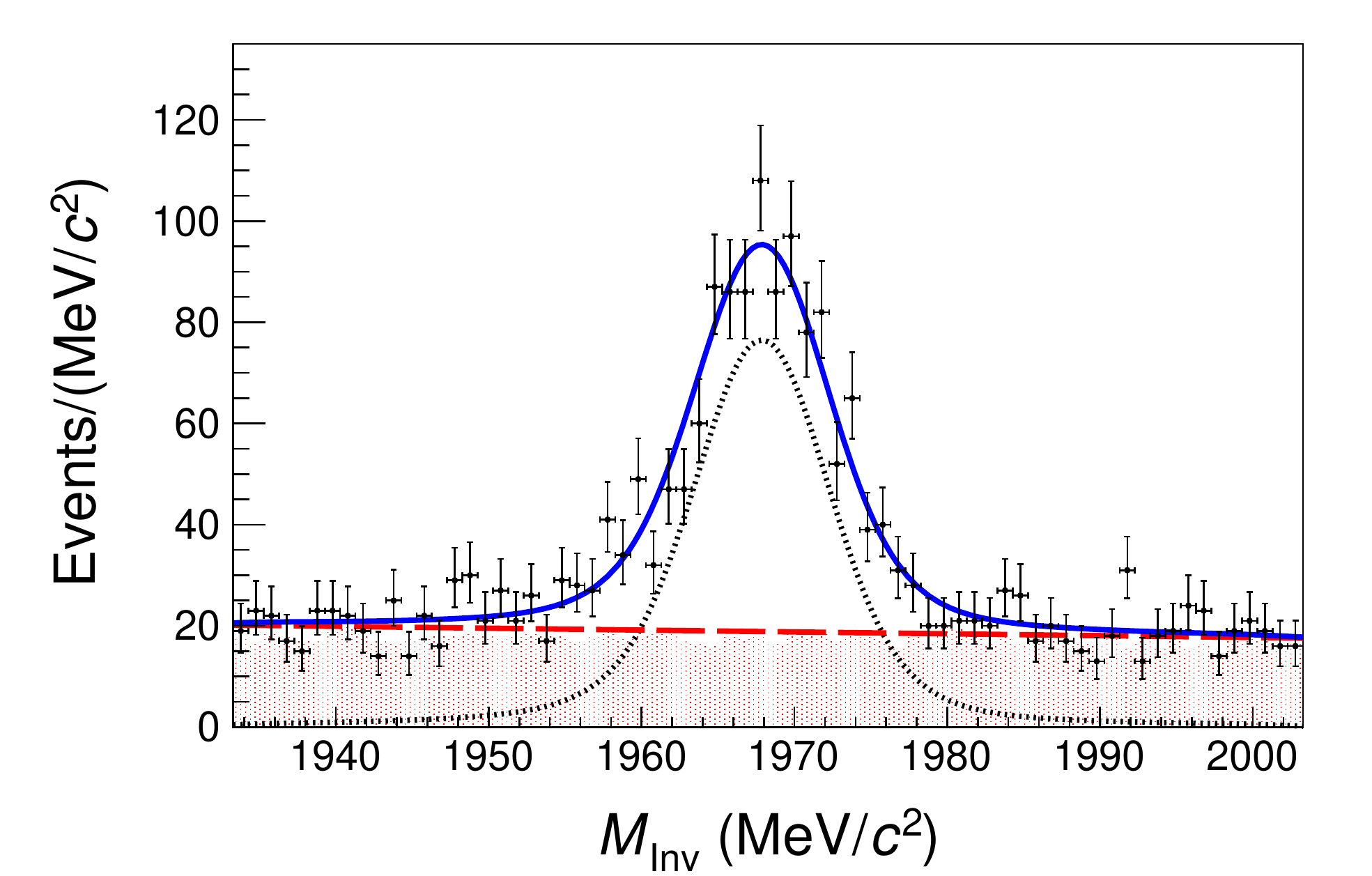} & \includegraphics[width=3in]{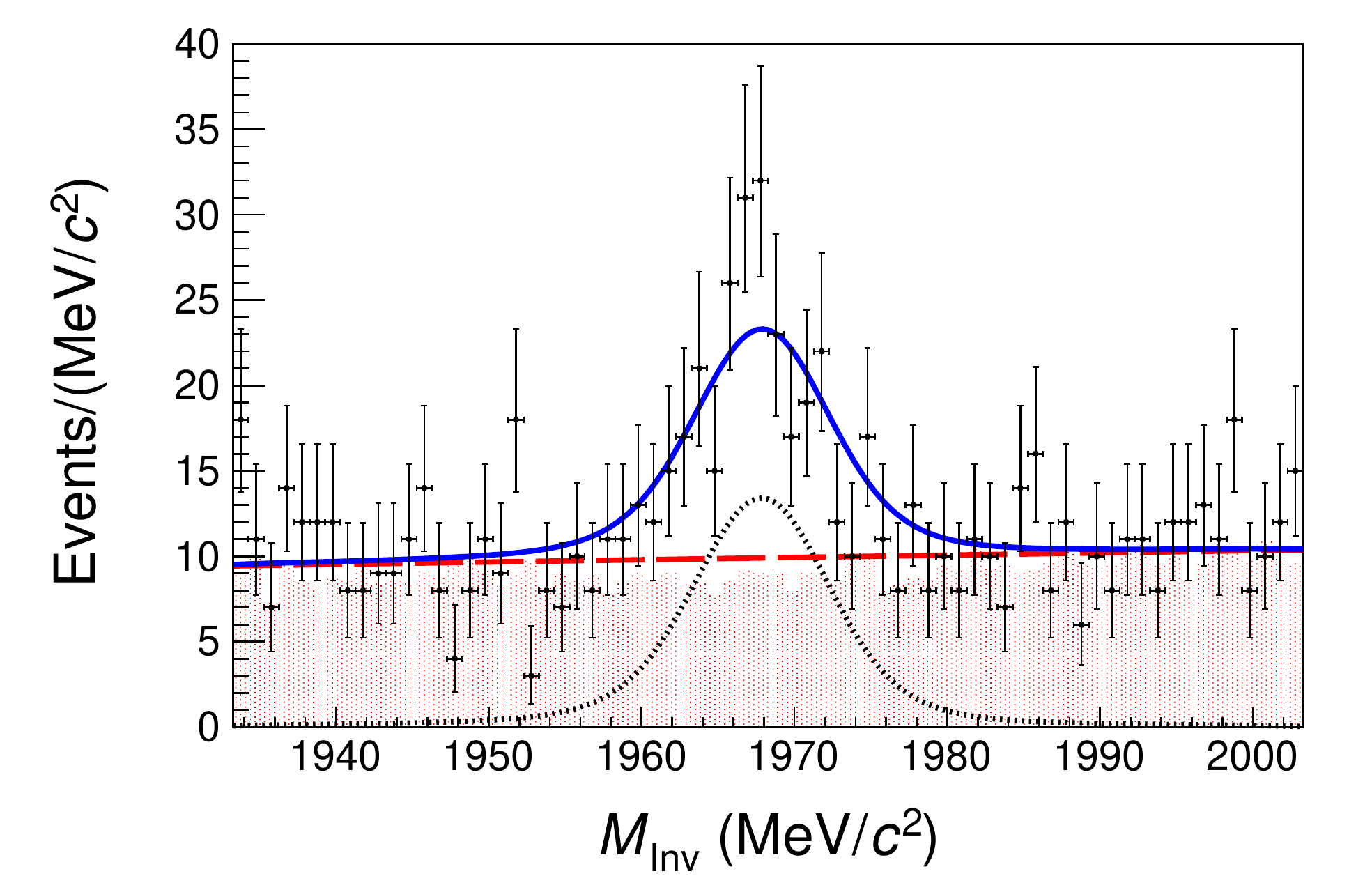}\\
\end{tabular}

\begin{tabular}{cc}
\textbf{RS 800-850 MeV/$c$} & \textbf{WS 800-850 MeV/$c$}\\
\includegraphics[width=3in]{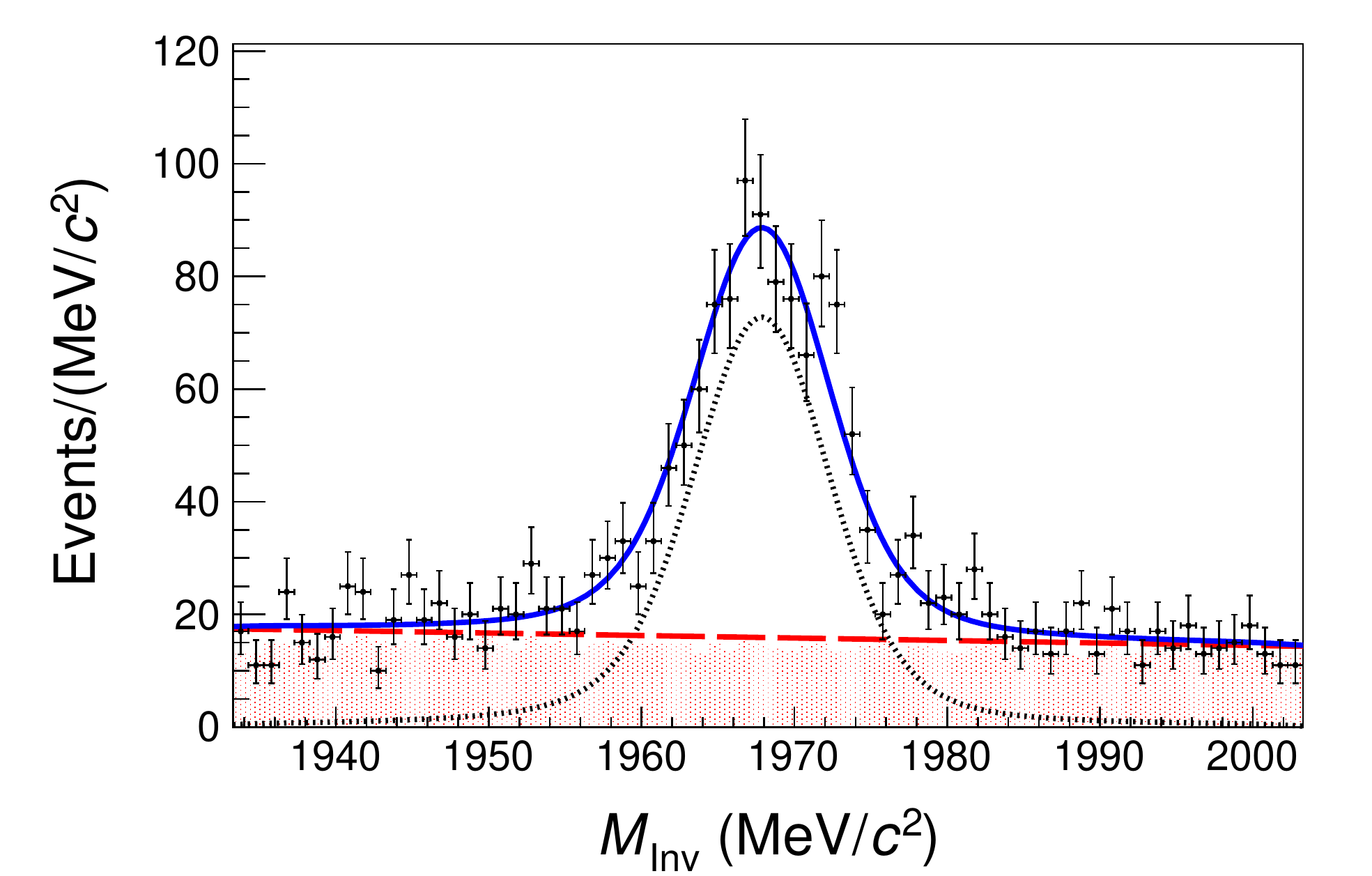} & \includegraphics[width=3in]{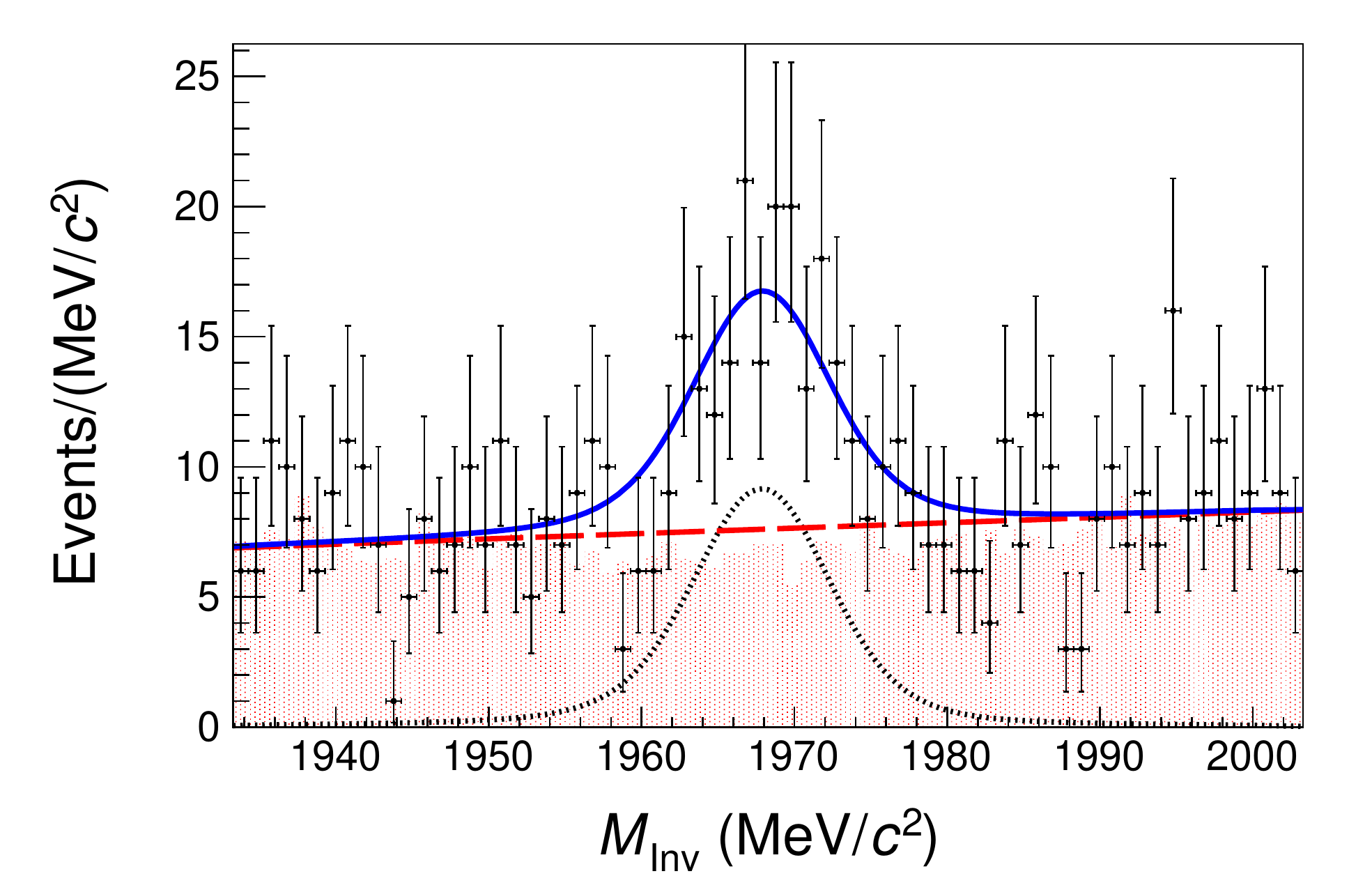}\\
\textbf{RS 850-900 MeV/$c$} & \textbf{WS 850-900 MeV/$c$}\\
\includegraphics[width=3in]{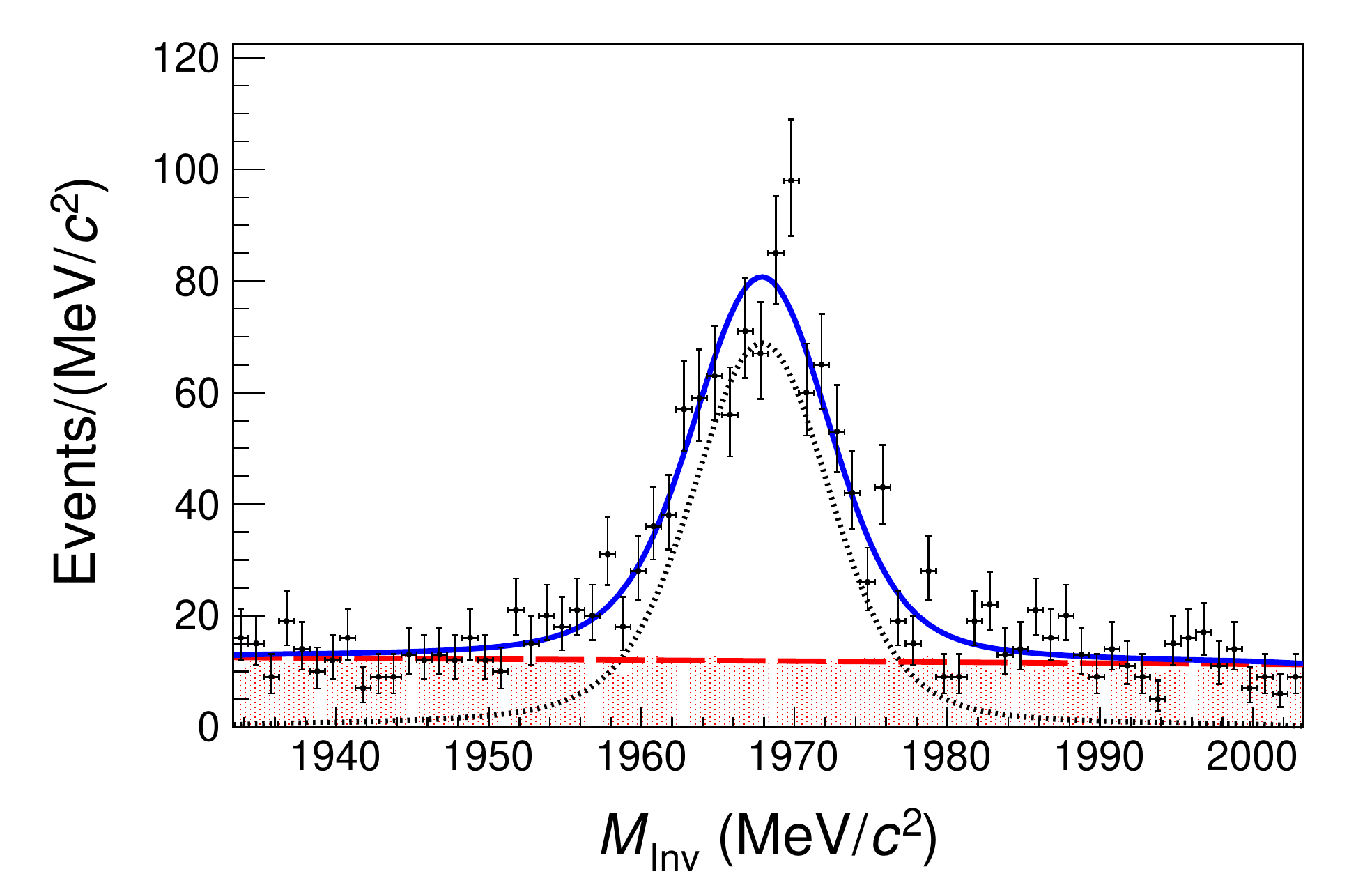} & \includegraphics[width=3in]{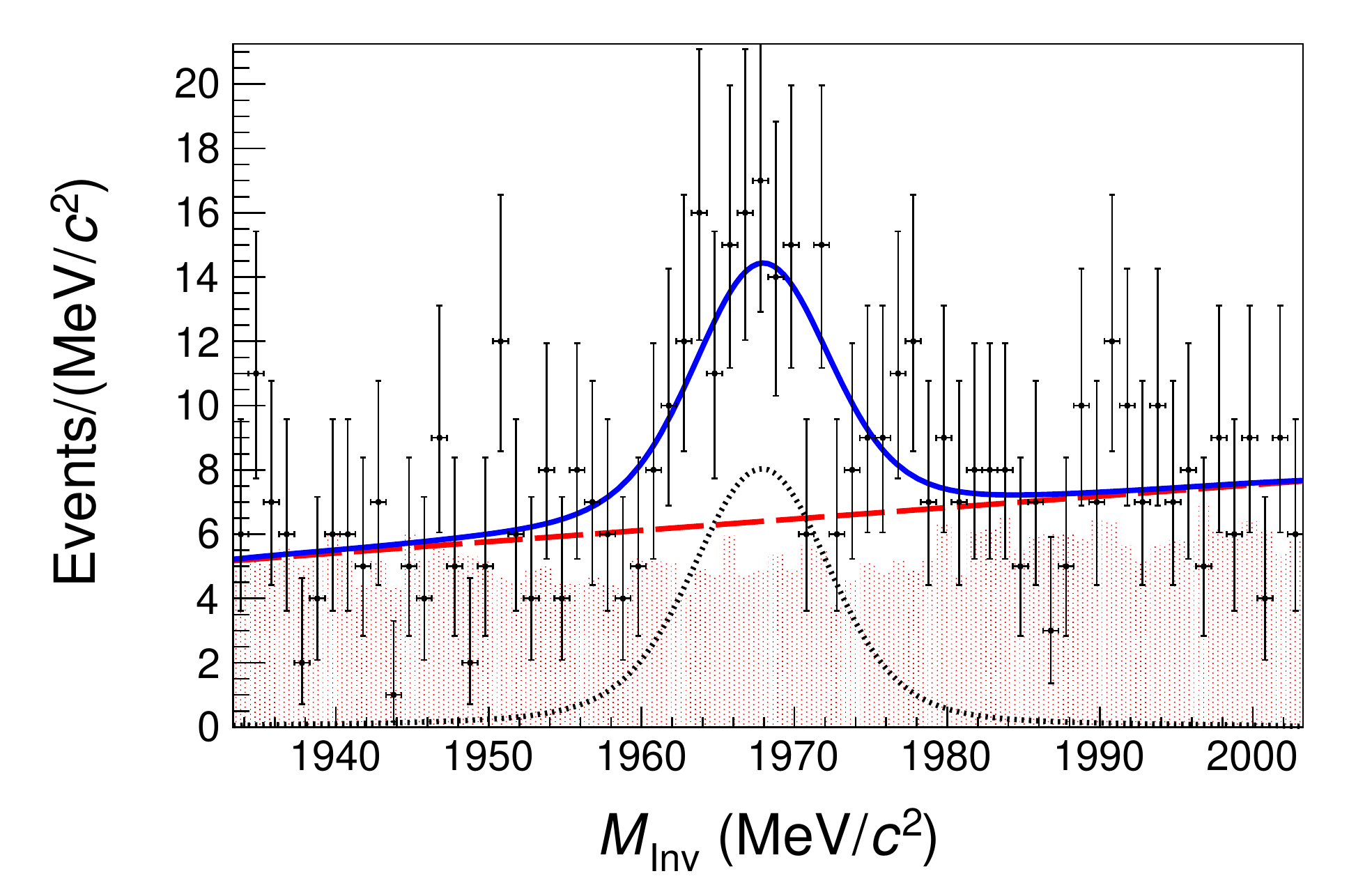}\\
\textbf{RS 900-950 MeV/$c$} & \textbf{WS 900-950 MeV/$c$}\\
\includegraphics[width=3in]{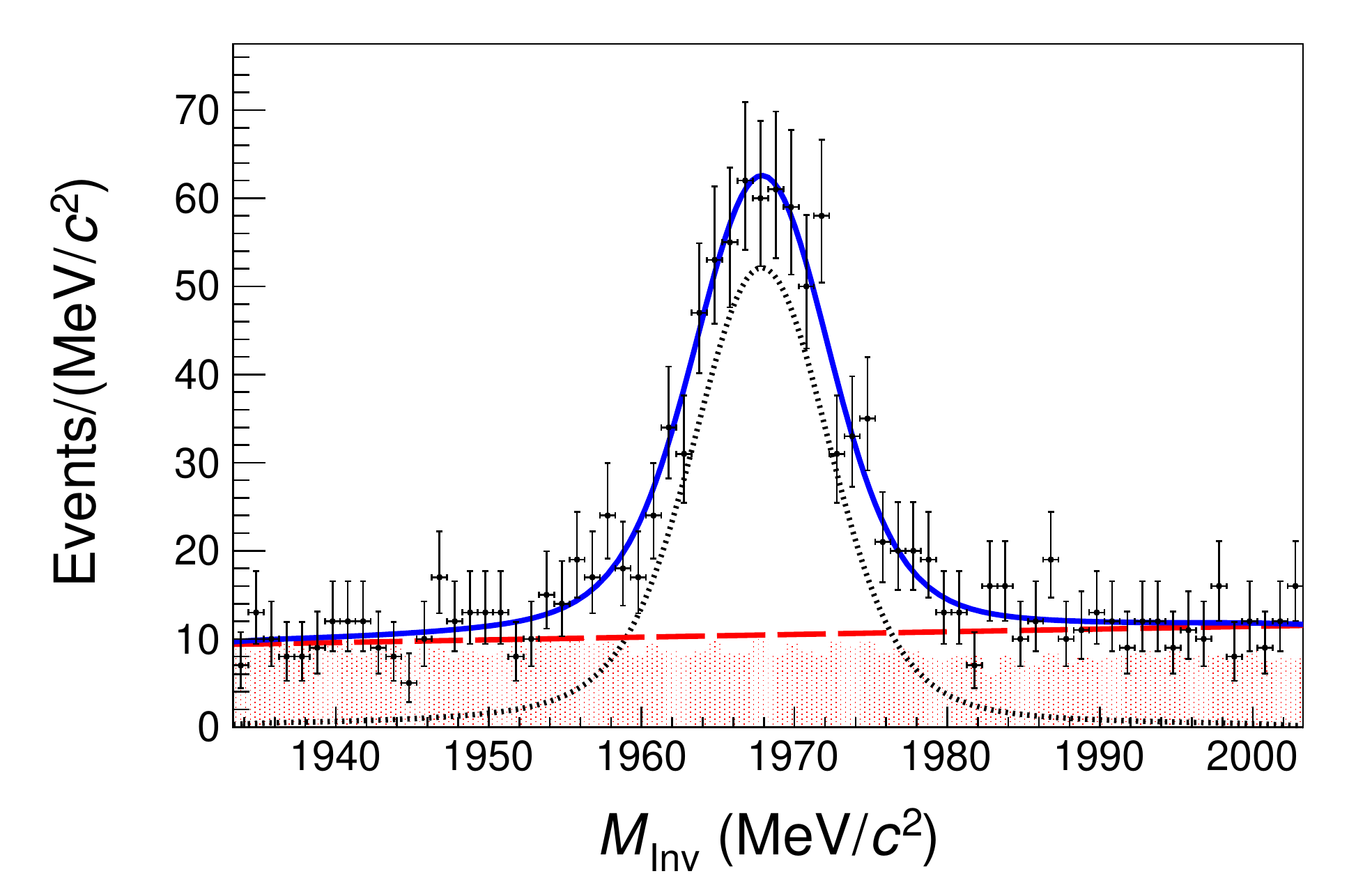} & \includegraphics[width=3in]{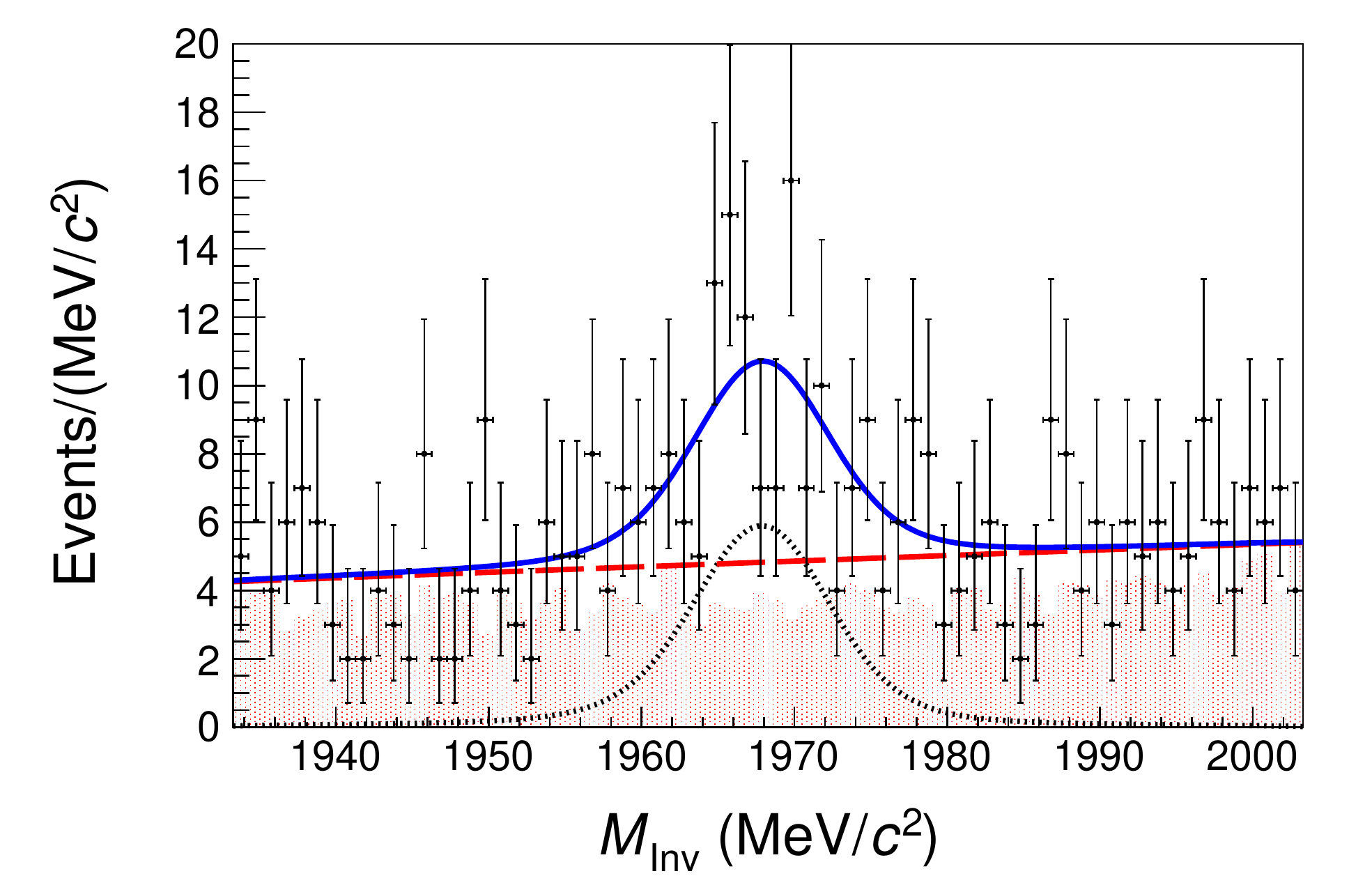}\\
\textbf{RS 950-1000 MeV/$c$} & \textbf{WS 950-1000 MeV/$c$}\\
\includegraphics[width=3in]{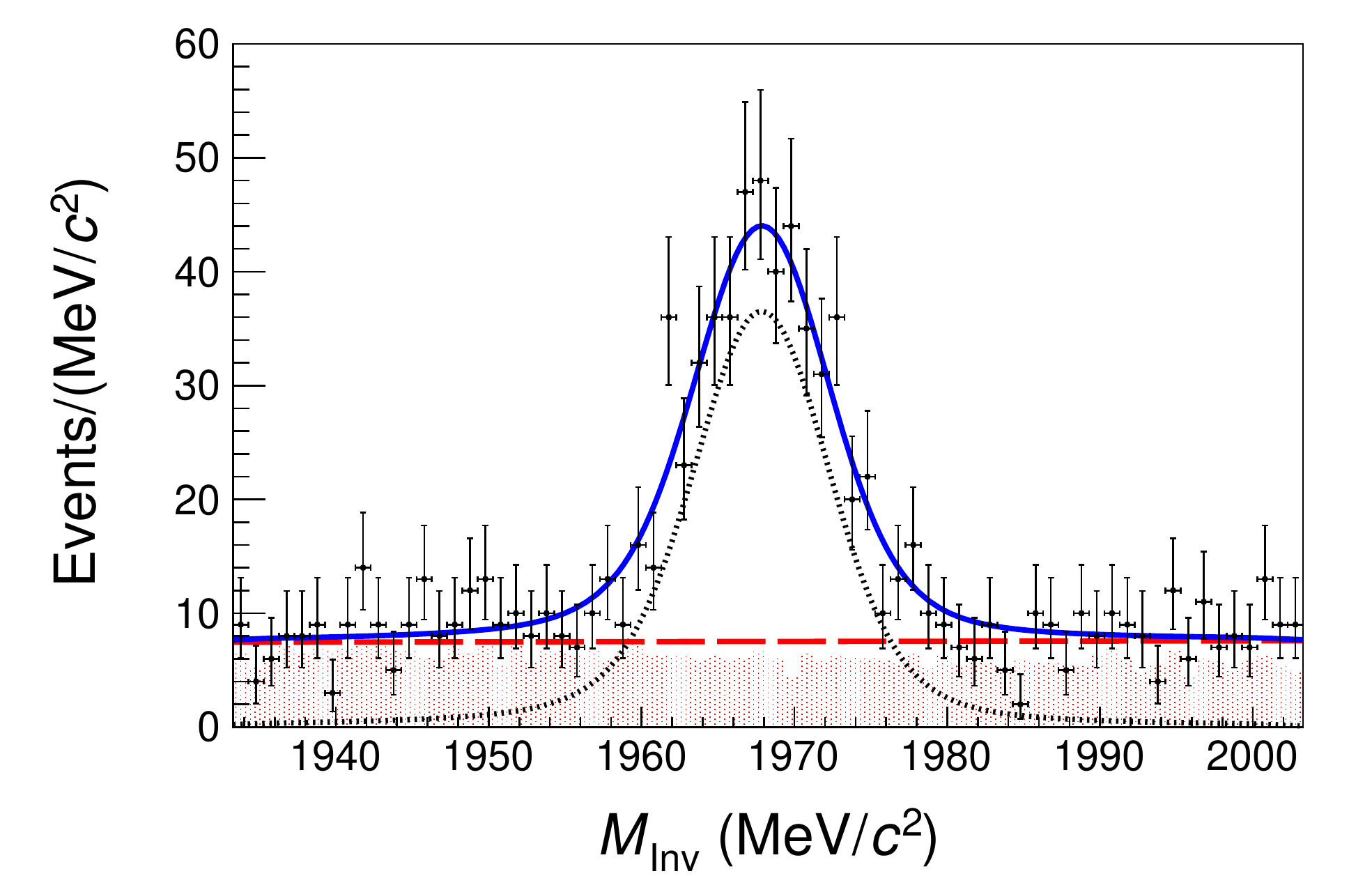} & \includegraphics[width=3in]{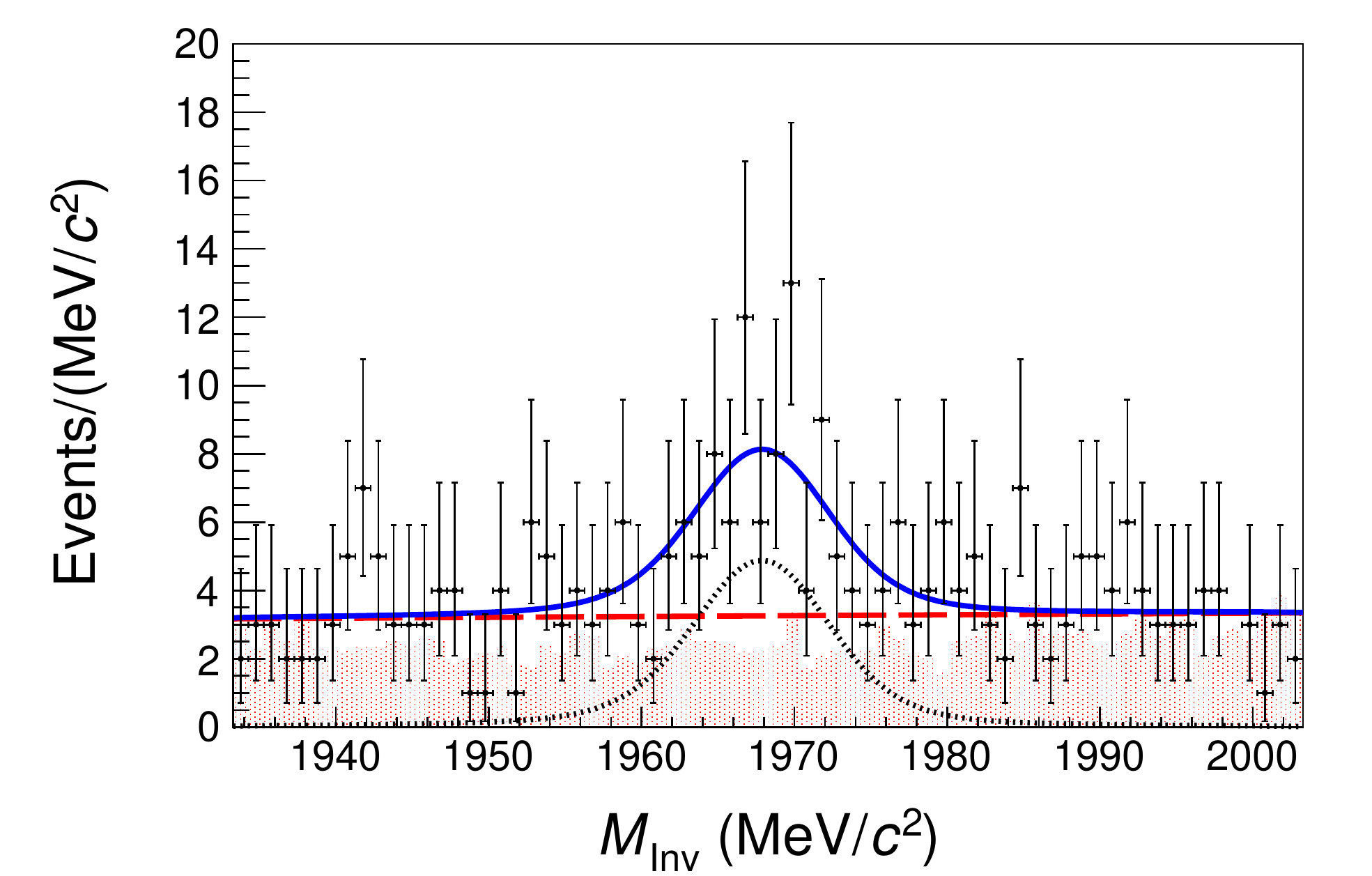}\\
\end{tabular}

\begin{tabular}{cc}
\textbf{RS 1000-1050 MeV/$c$} & \textbf{WS 1000-1050 MeV/$c$}\\
\includegraphics[width=3in]{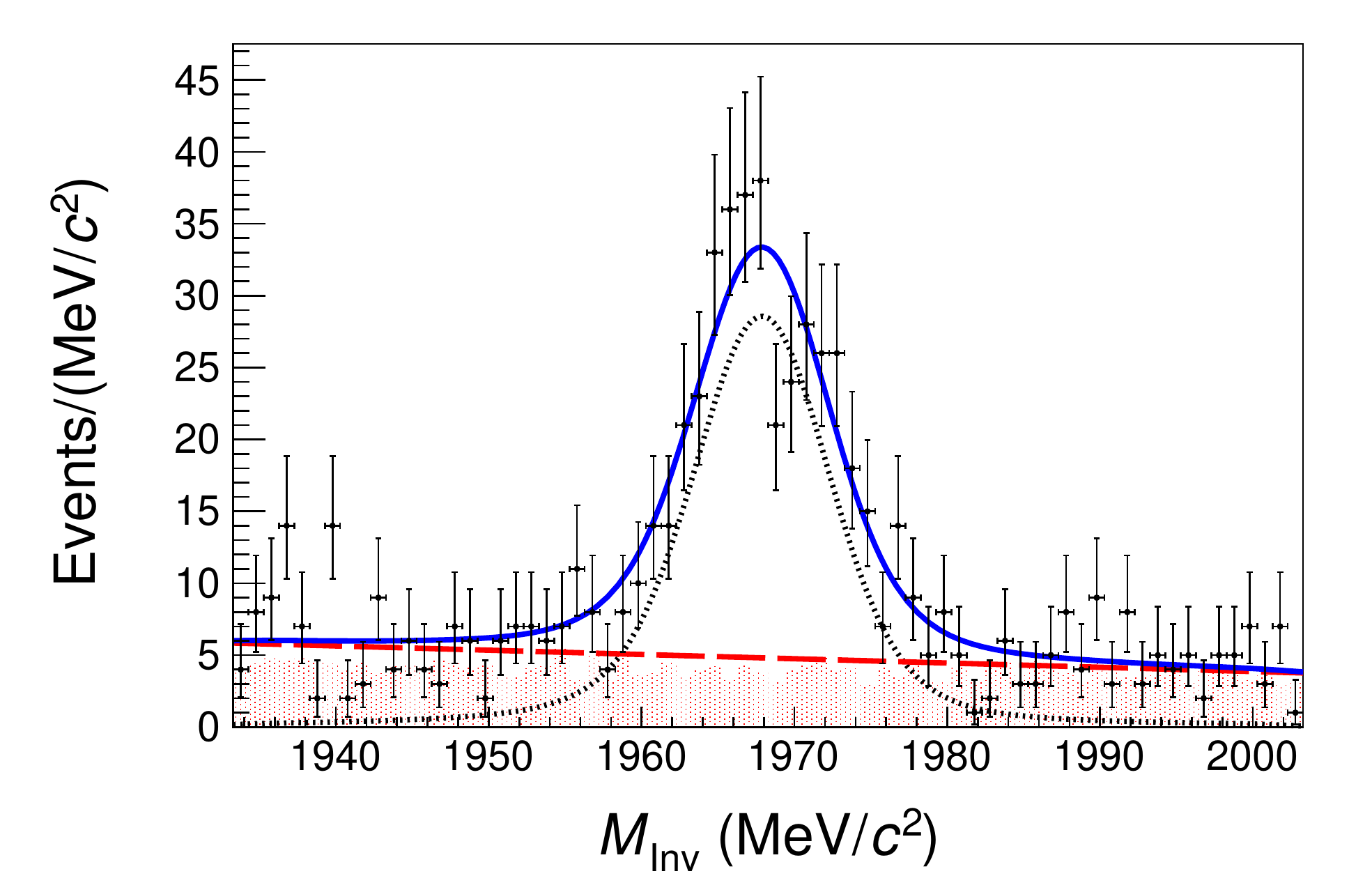} & \includegraphics[width=3in]{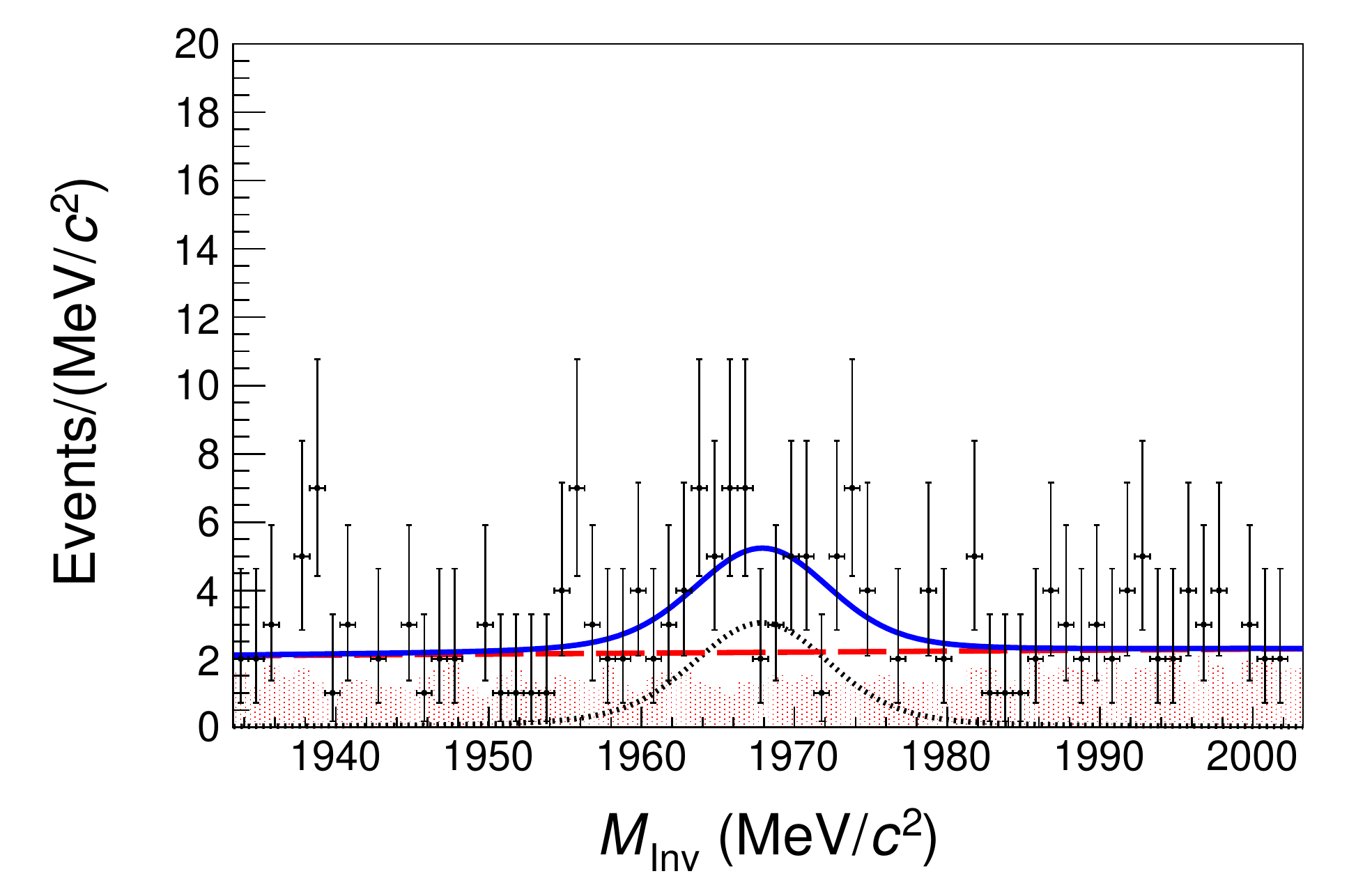}\\
\textbf{RS 1050-1100 MeV/$c$} & \textbf{WS 1050-1100 MeV/$c$}\\
\includegraphics[width=3in]{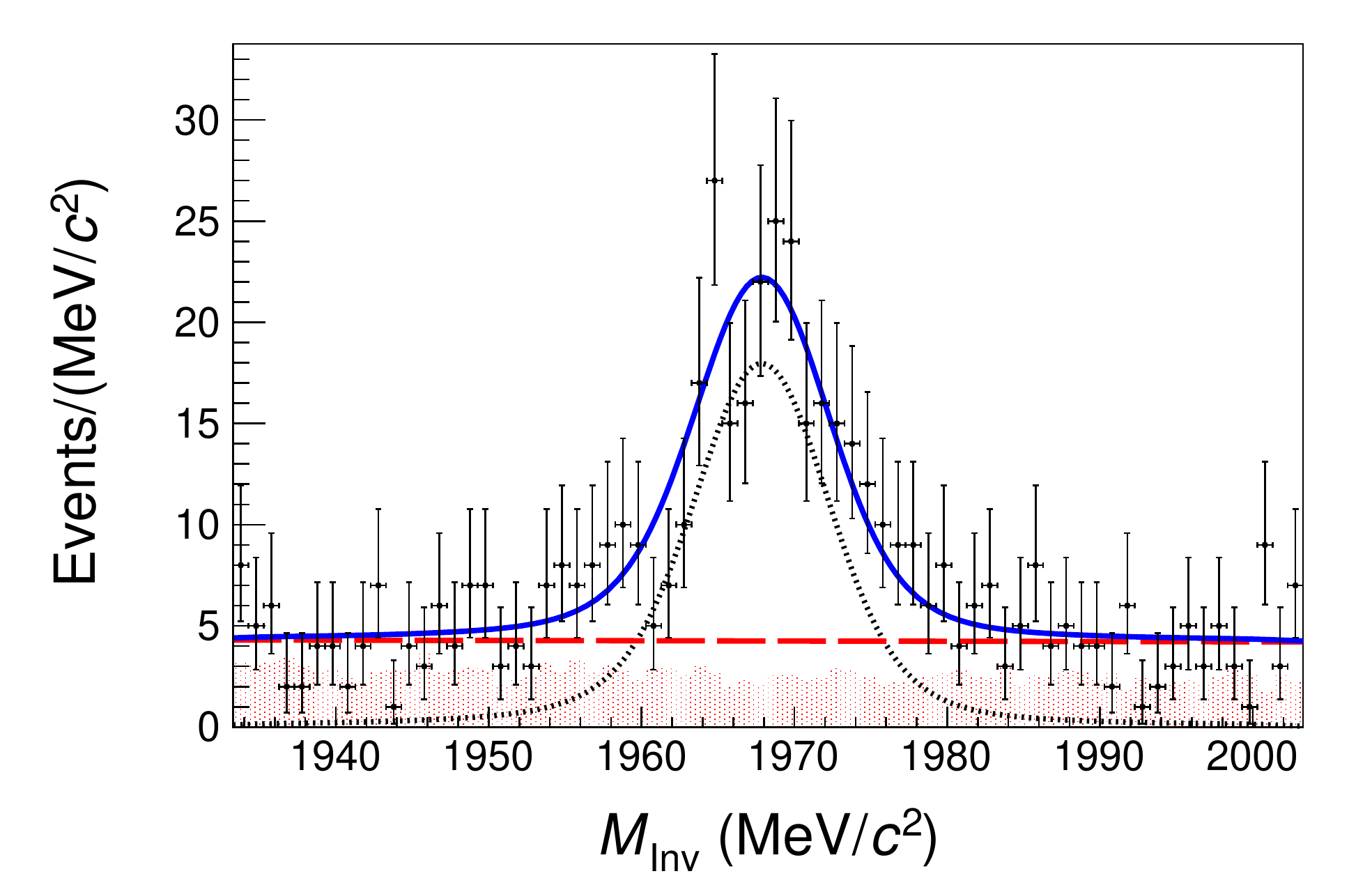} & \includegraphics[width=3in]{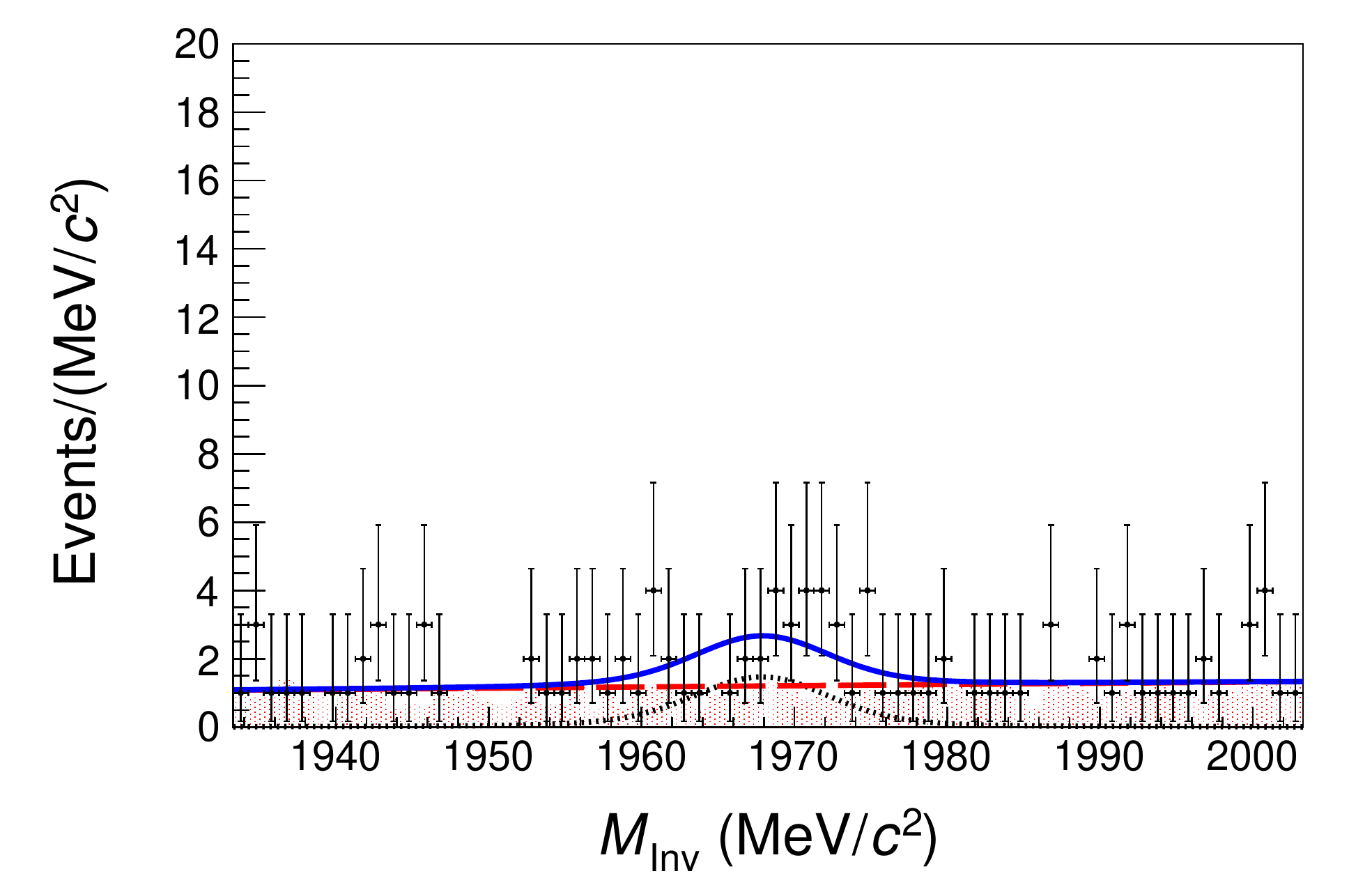}
\end{tabular}

\pagebreak
\raggedright

\subsection{$\EcmC$ Data Sample}
\label{subsec:4230DataFits}
\subsubsection{$\EcmC$ Data $e$ ID Fits}
\label{subsubsec:4230DataEIDFits}
\centering
\begin{tabular}{cc}
\textbf{RS 200-250 MeV/$c$} & \textbf{WS 200-250 MeV/$c$}\\
\includegraphics[width=3in]{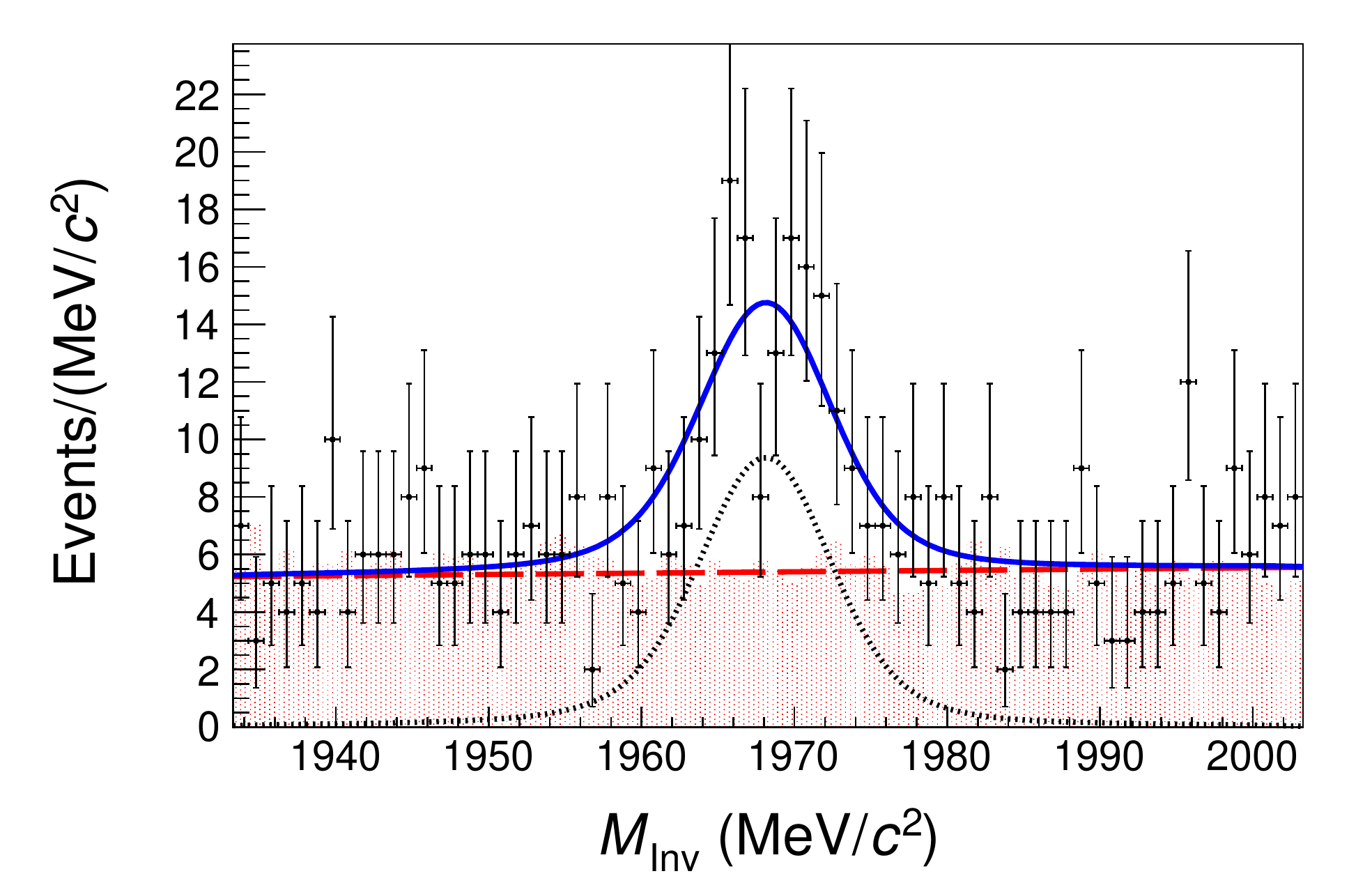} & \includegraphics[width=3in]{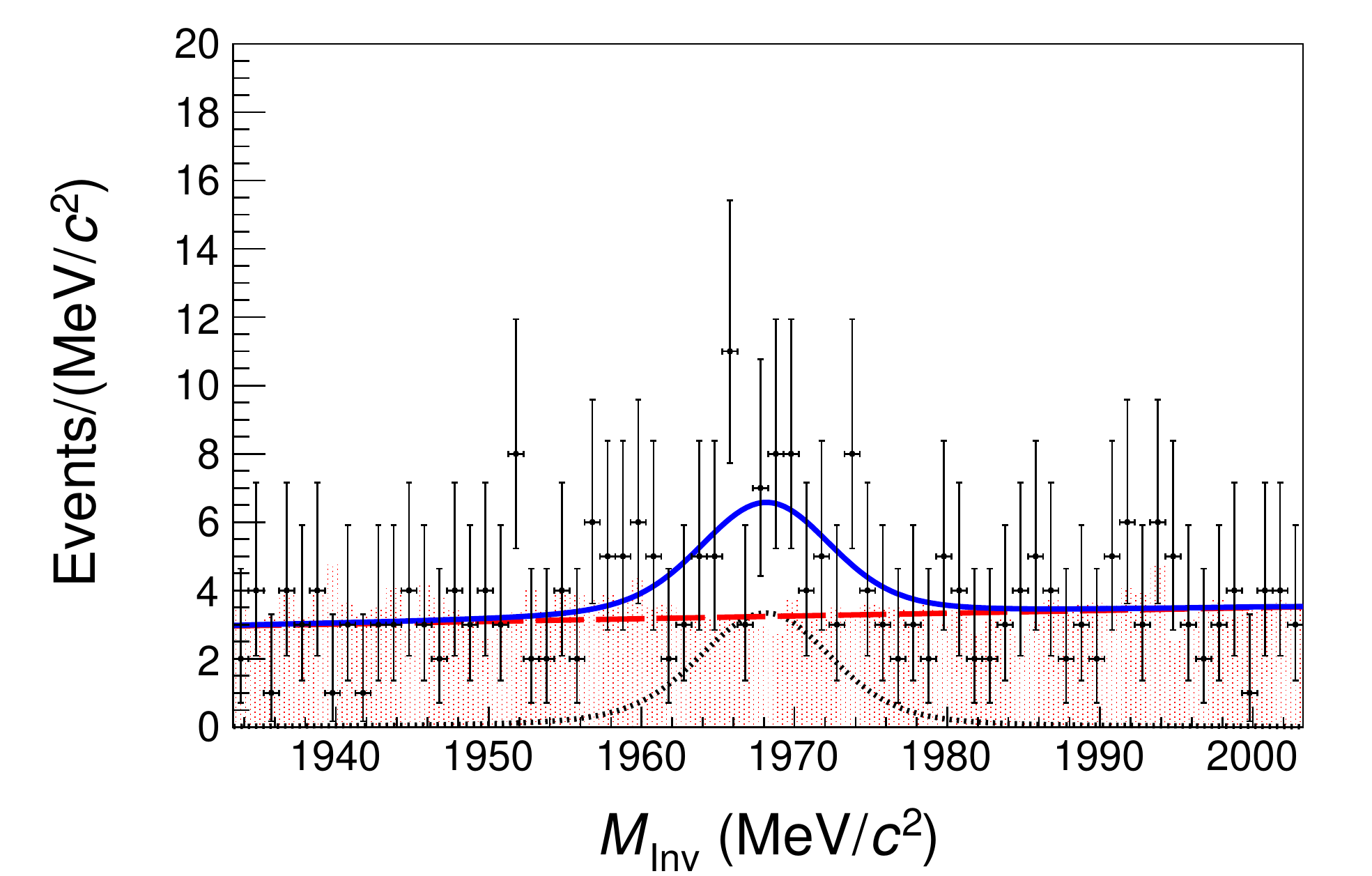}\\
\textbf{RS 250-300 MeV/$c$} & \textbf{WS 250-300 MeV/$c$}\\
\includegraphics[width=3in]{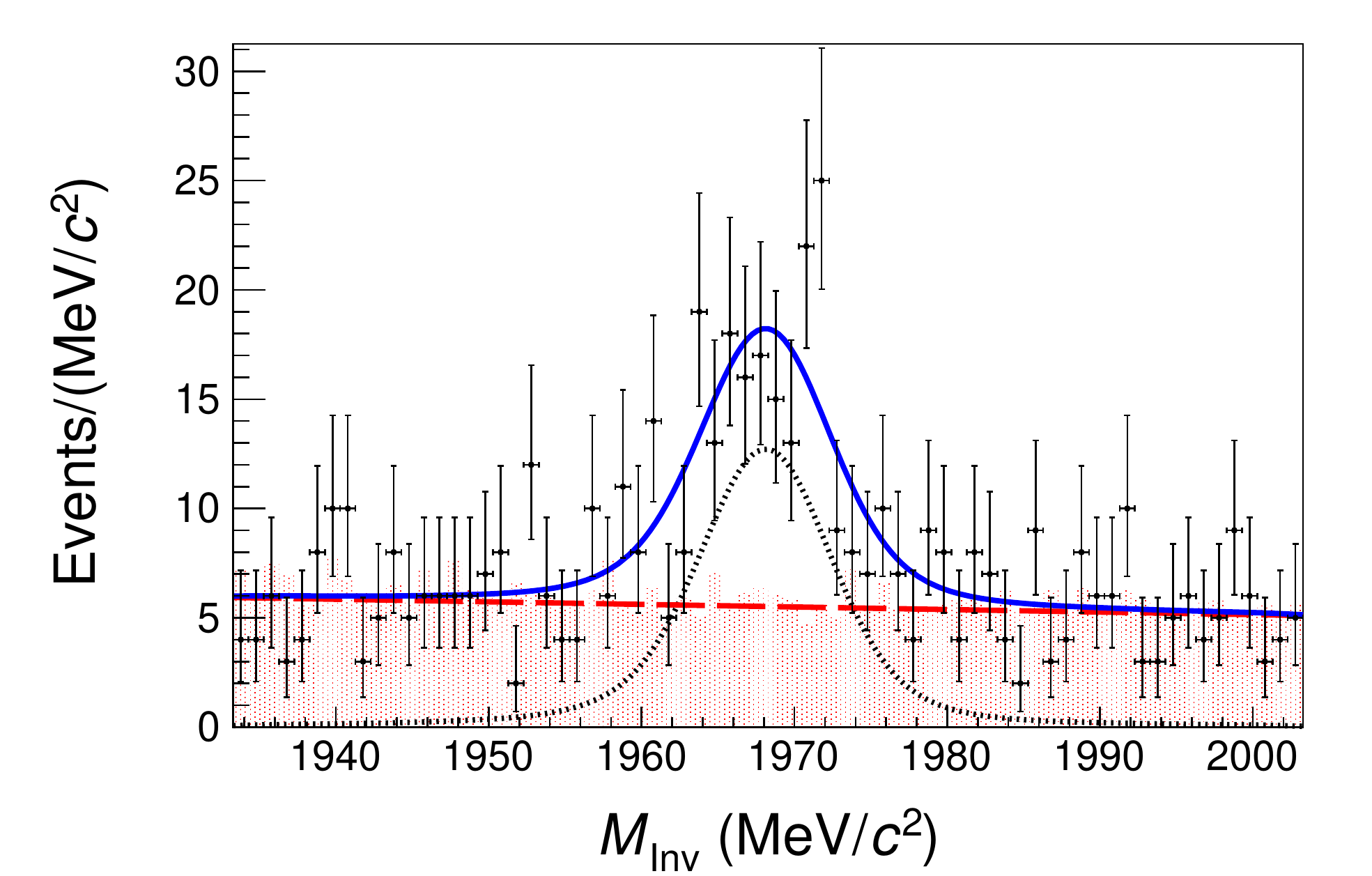} & \includegraphics[width=3in]{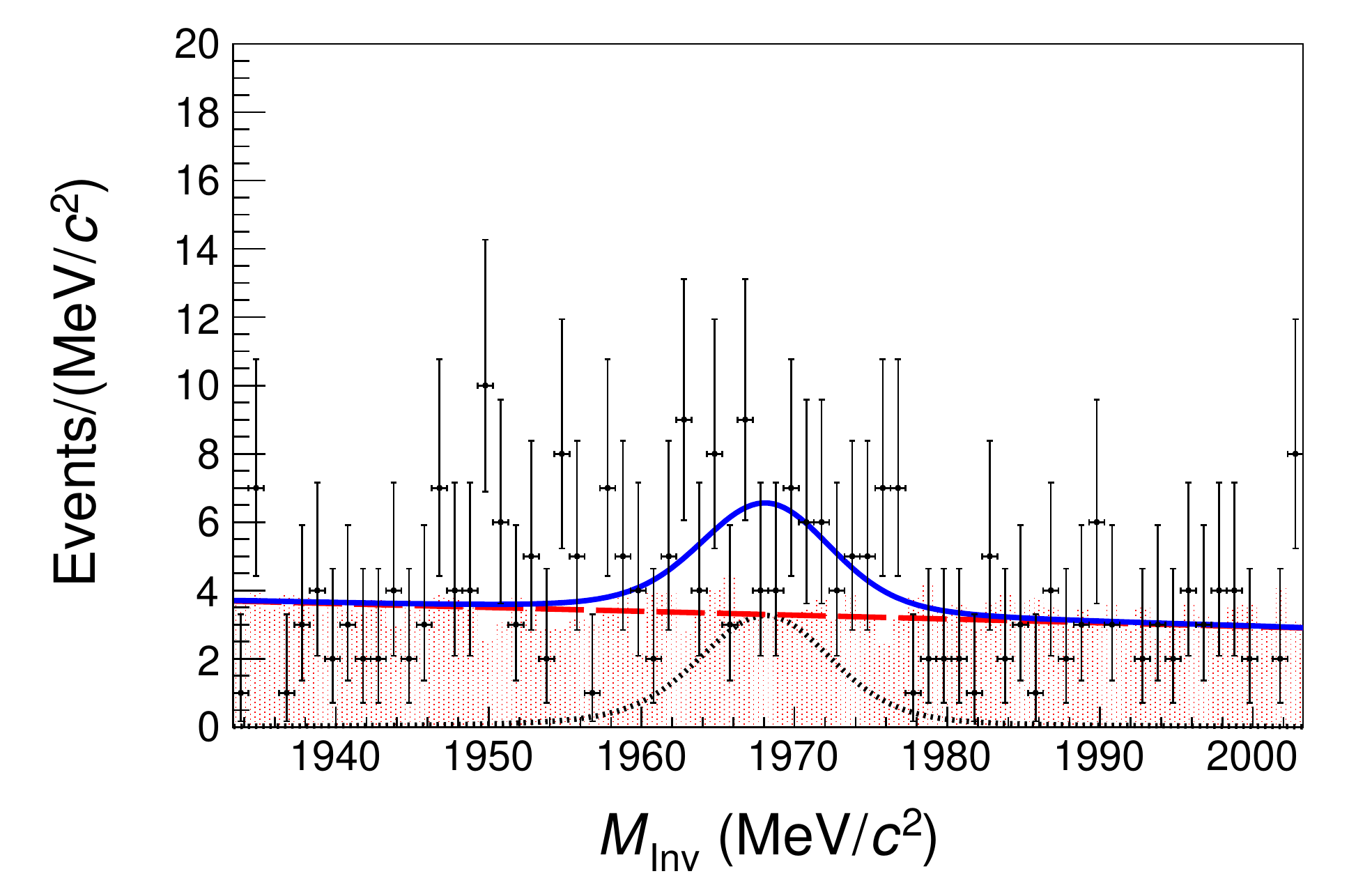}\\
\textbf{RS 300-350 MeV/$c$} & \textbf{WS 300-350 MeV/$c$}\\
\includegraphics[width=3in]{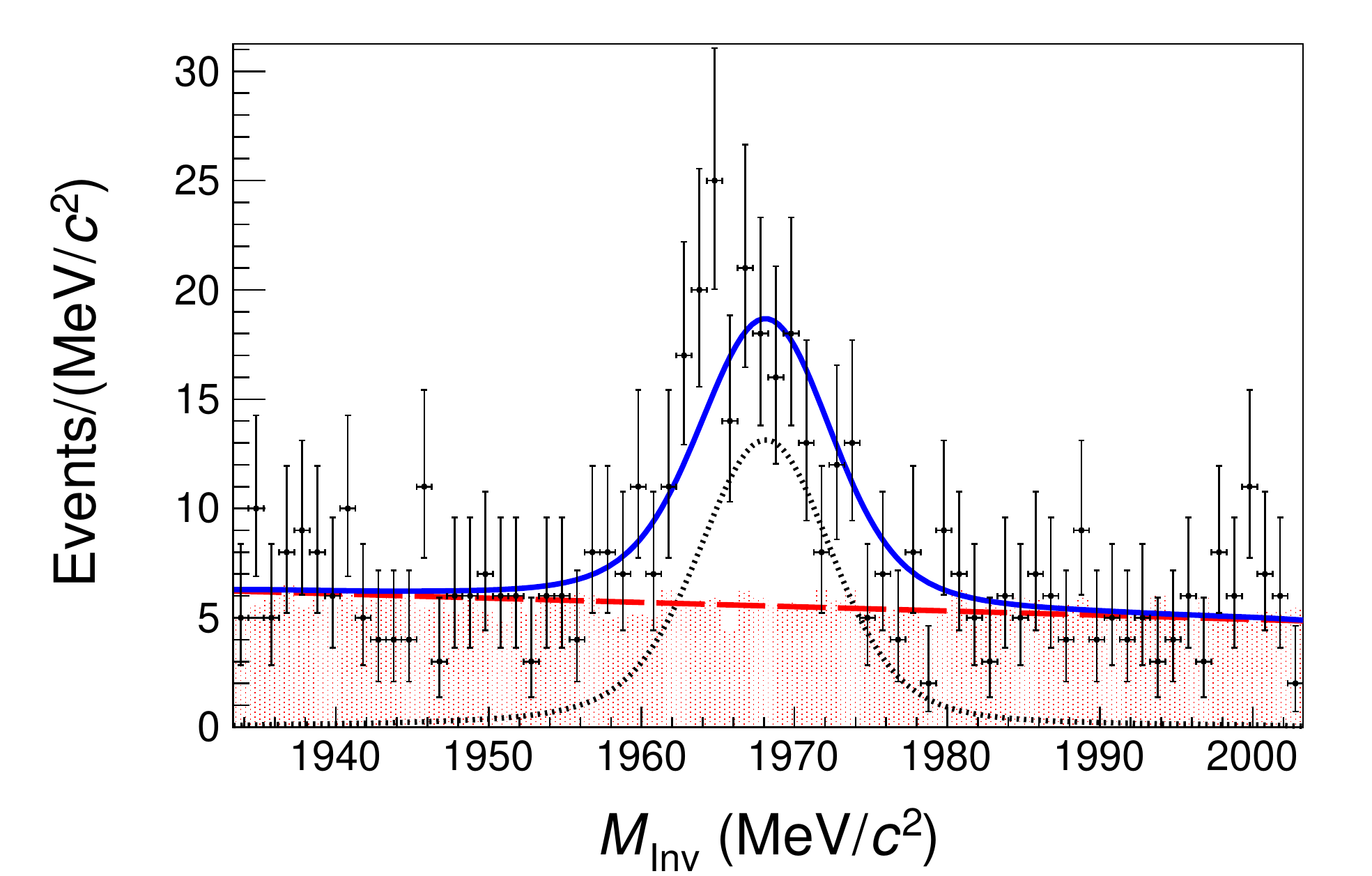} & \includegraphics[width=3in]{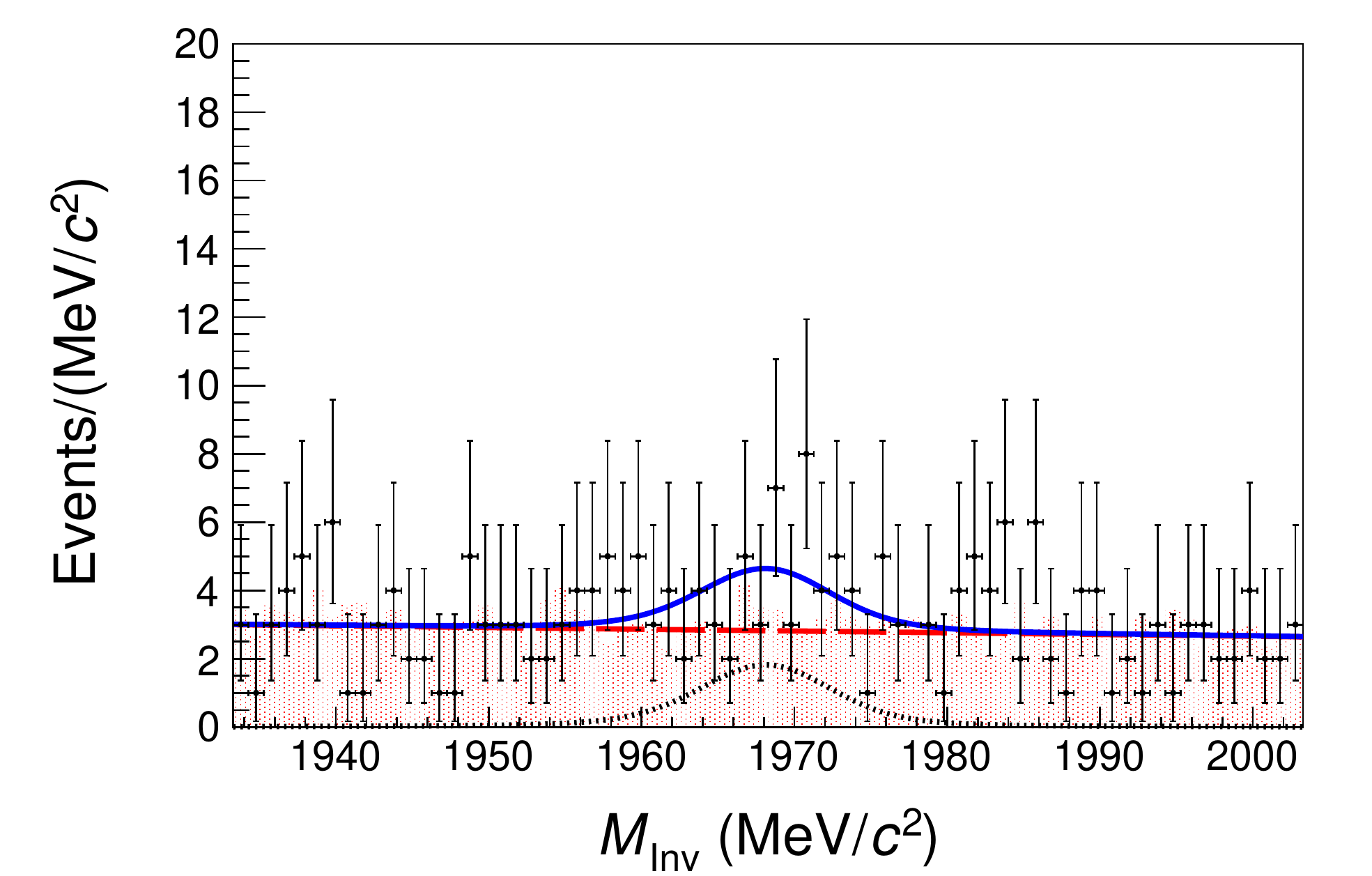}\\
\textbf{RS 350-400 MeV/$c$} & \textbf{WS 350-400 MeV/$c$}\\
\includegraphics[width=3in]{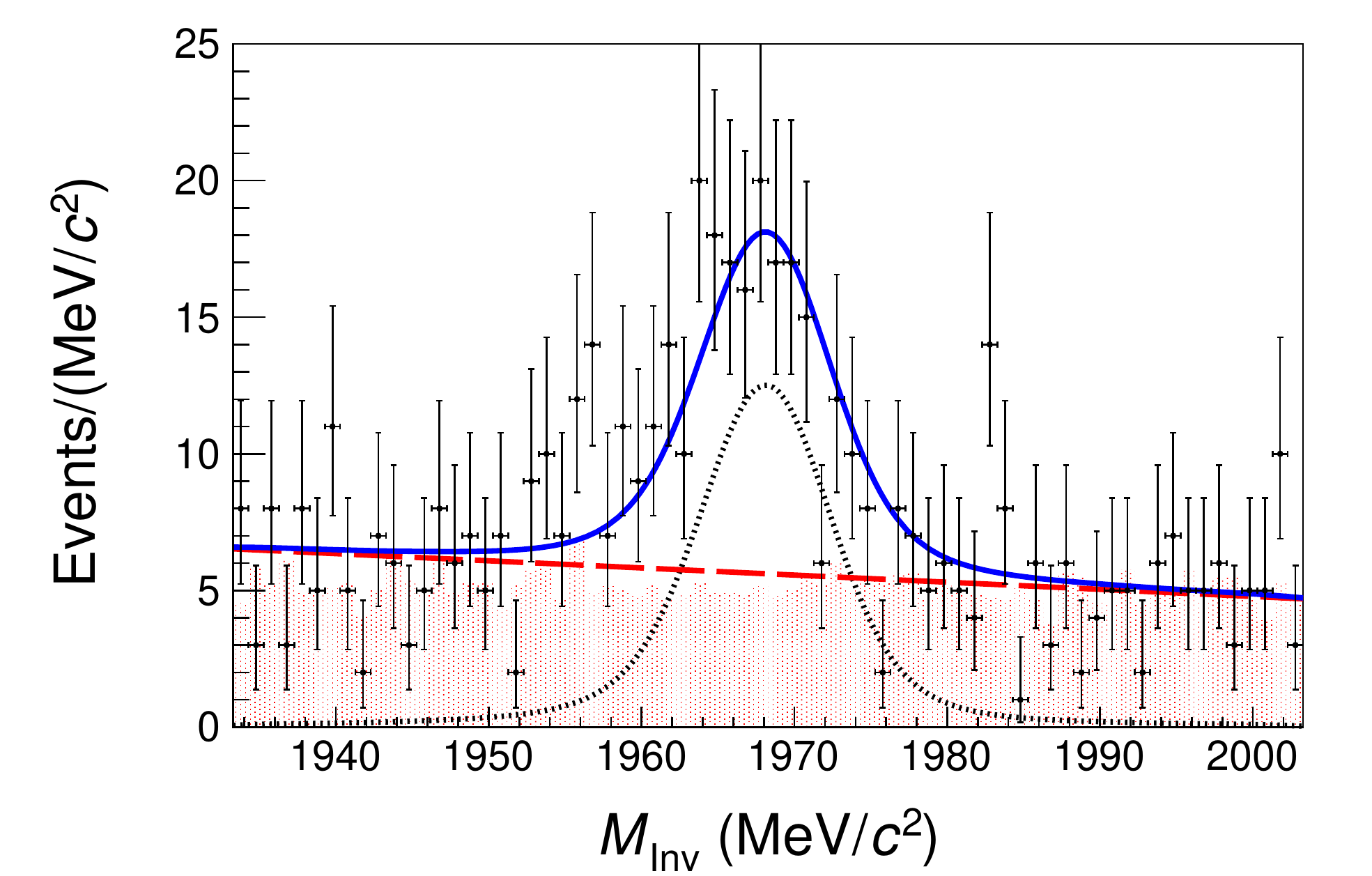} & \includegraphics[width=3in]{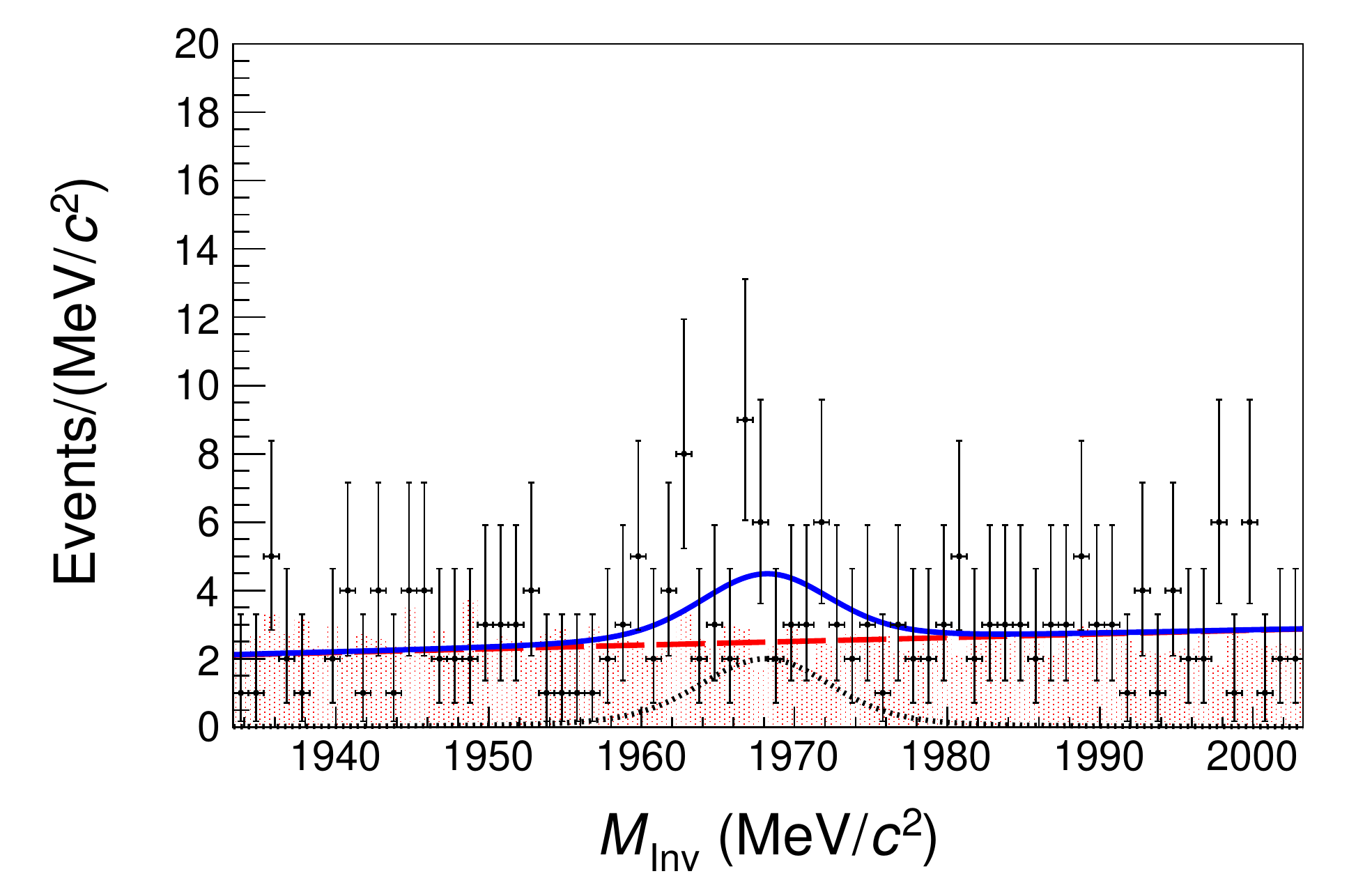}
\end{tabular}
\pagebreak

\begin{tabular}{cc}
\textbf{RS 400-450 MeV/$c$} & \textbf{WS 400-450 MeV/$c$}\\
\includegraphics[width=3in]{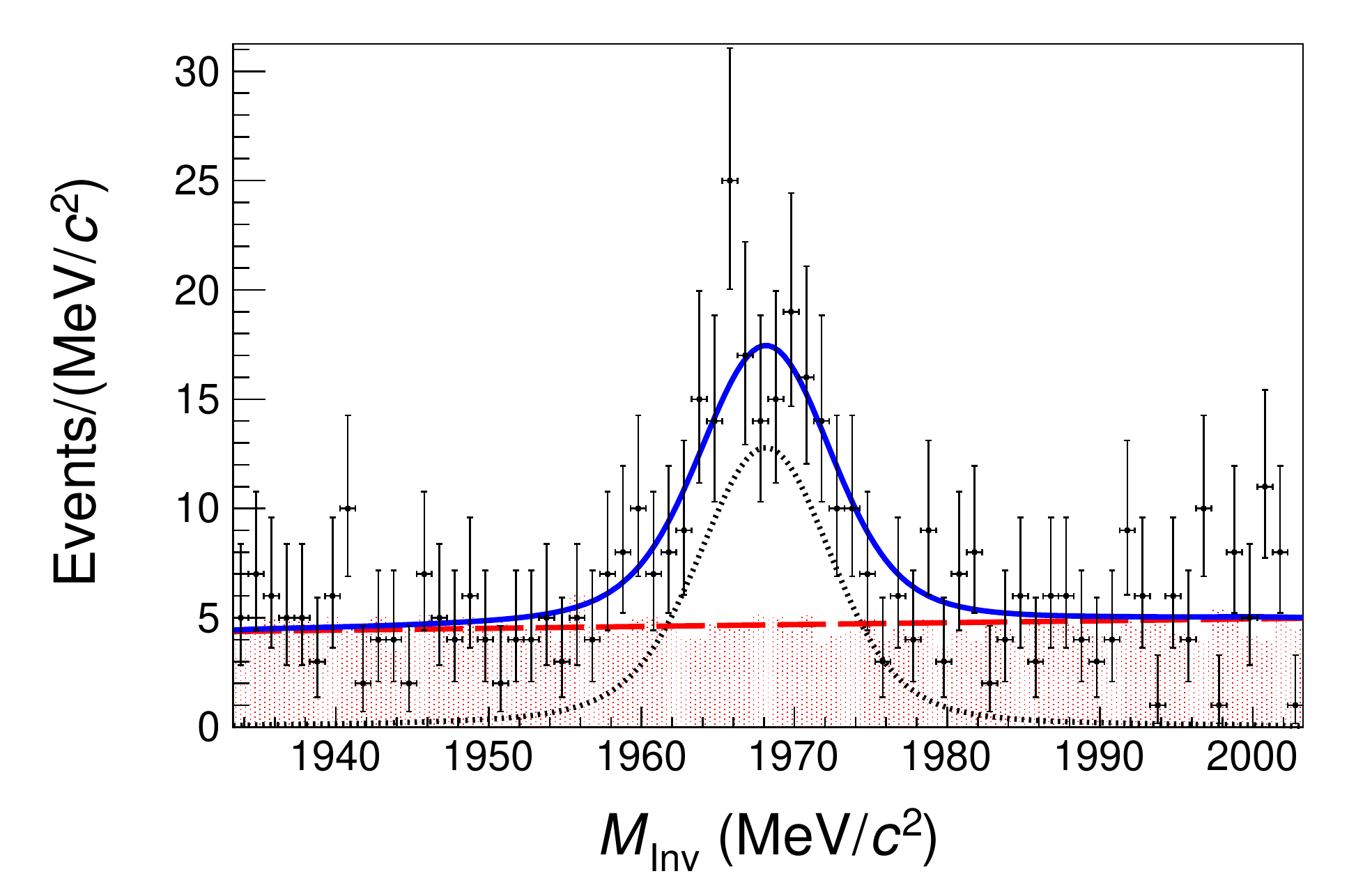} & \includegraphics[width=3in]{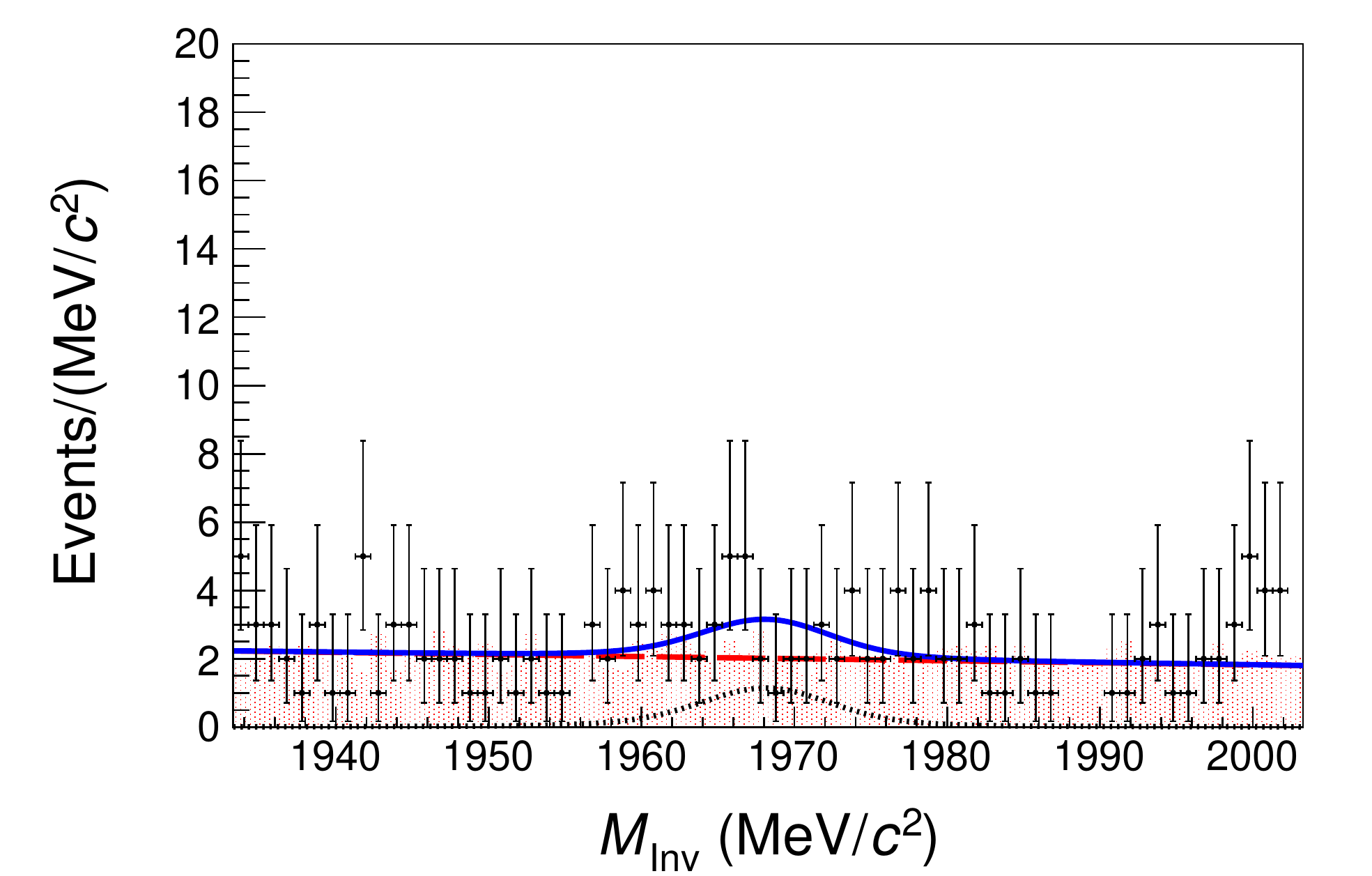}\\
\textbf{RS 450-500 MeV/$c$} & \textbf{WS 450-500 MeV/$c$}\\
\includegraphics[width=3in]{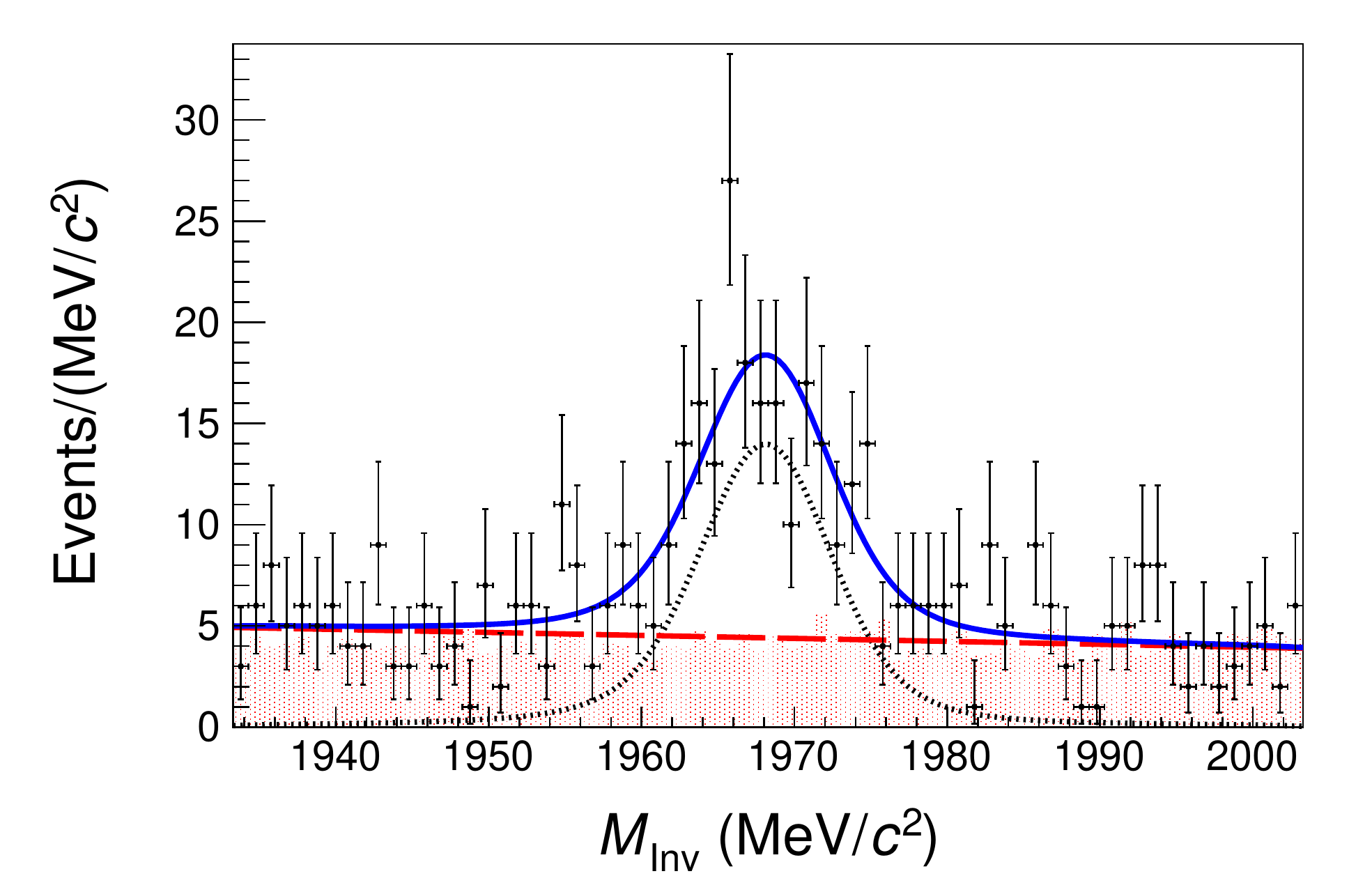} & \includegraphics[width=3in]{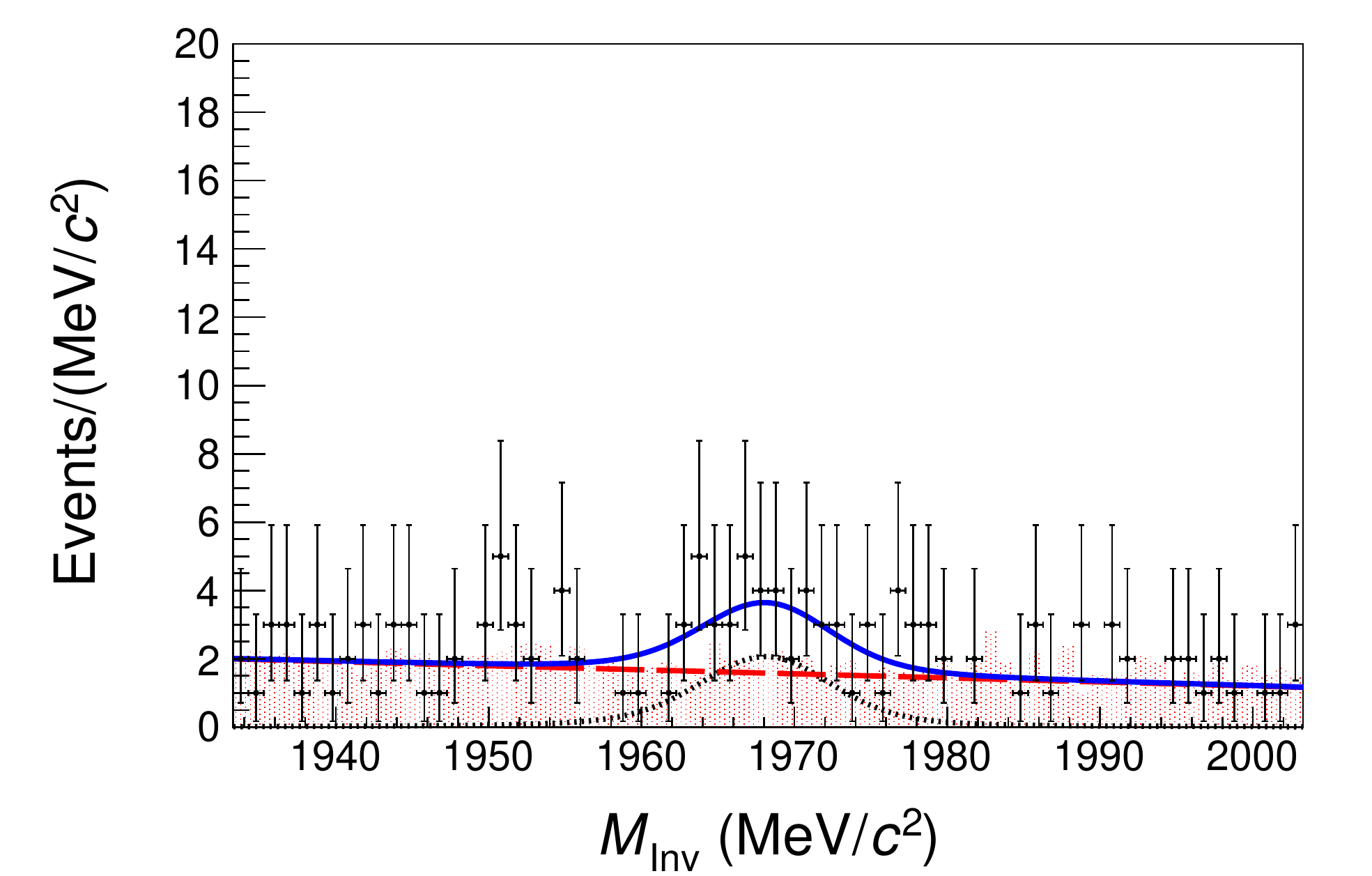}\\
\textbf{RS 500-550 MeV/$c$} & \textbf{WS 500-550 MeV/$c$}\\
\includegraphics[width=3in]{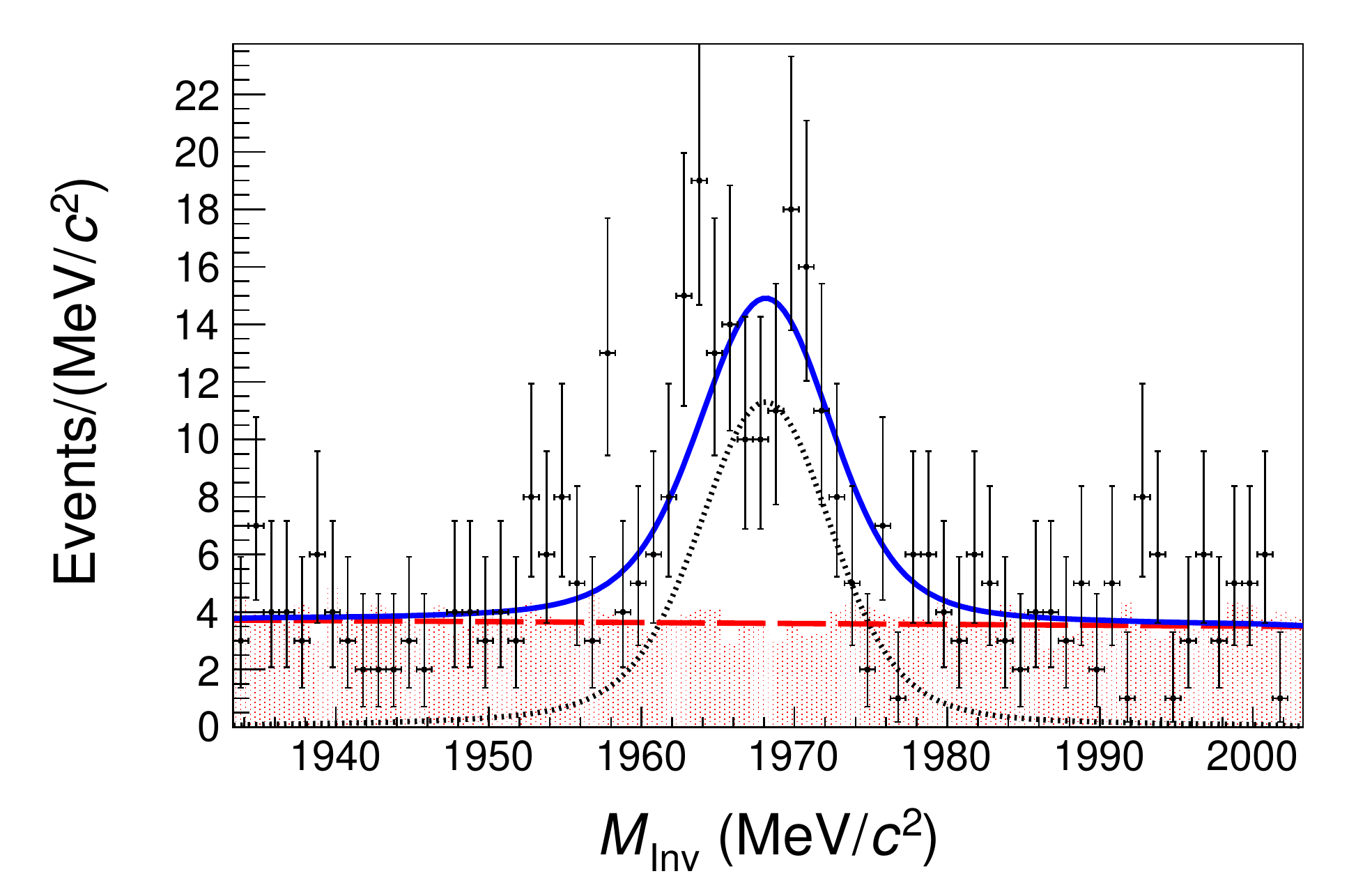} & \includegraphics[width=3in]{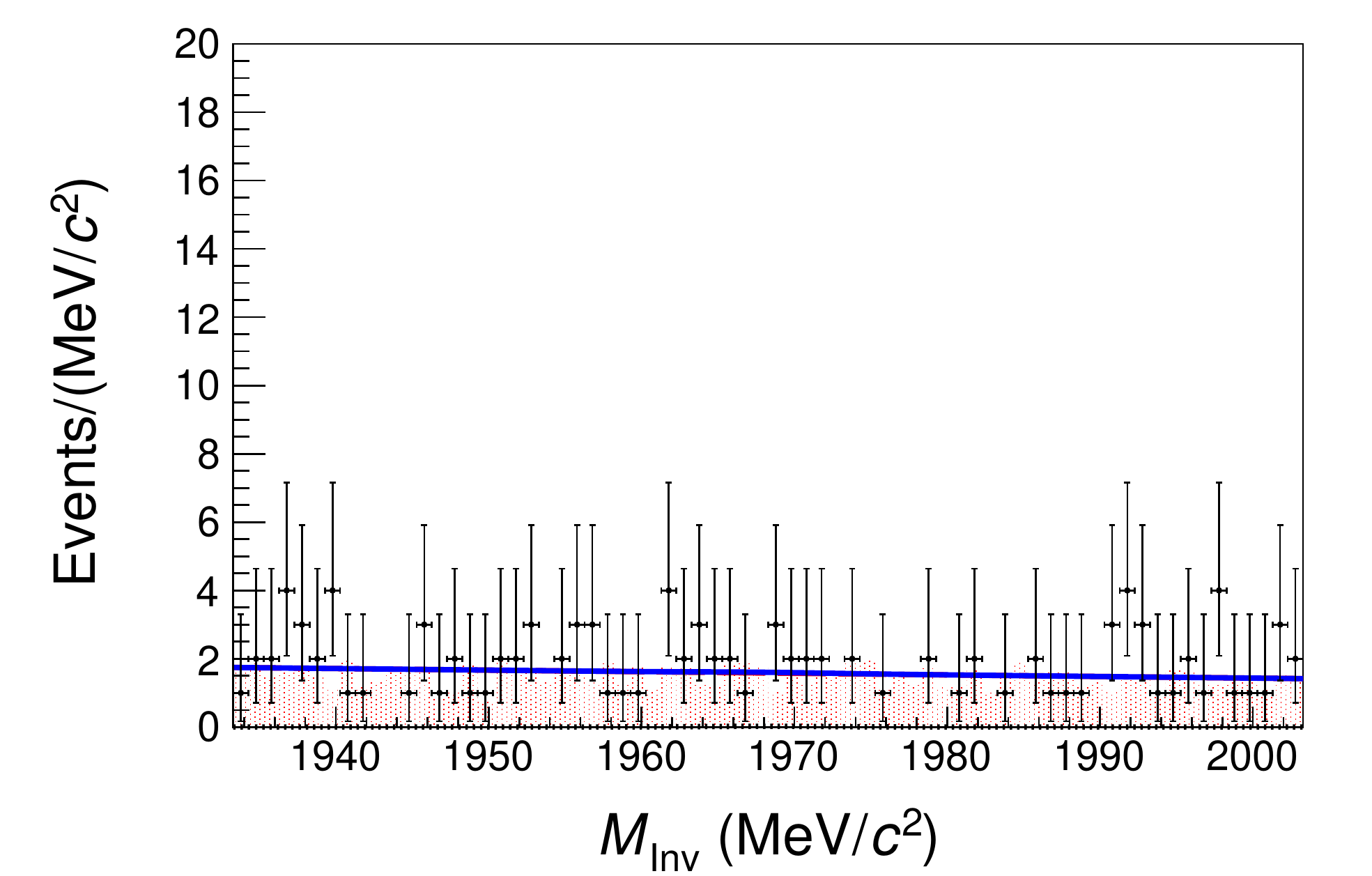}\\
\textbf{RS 550-600 MeV/$c$} & \textbf{WS 550-600 MeV/$c$}\\
\includegraphics[width=3in]{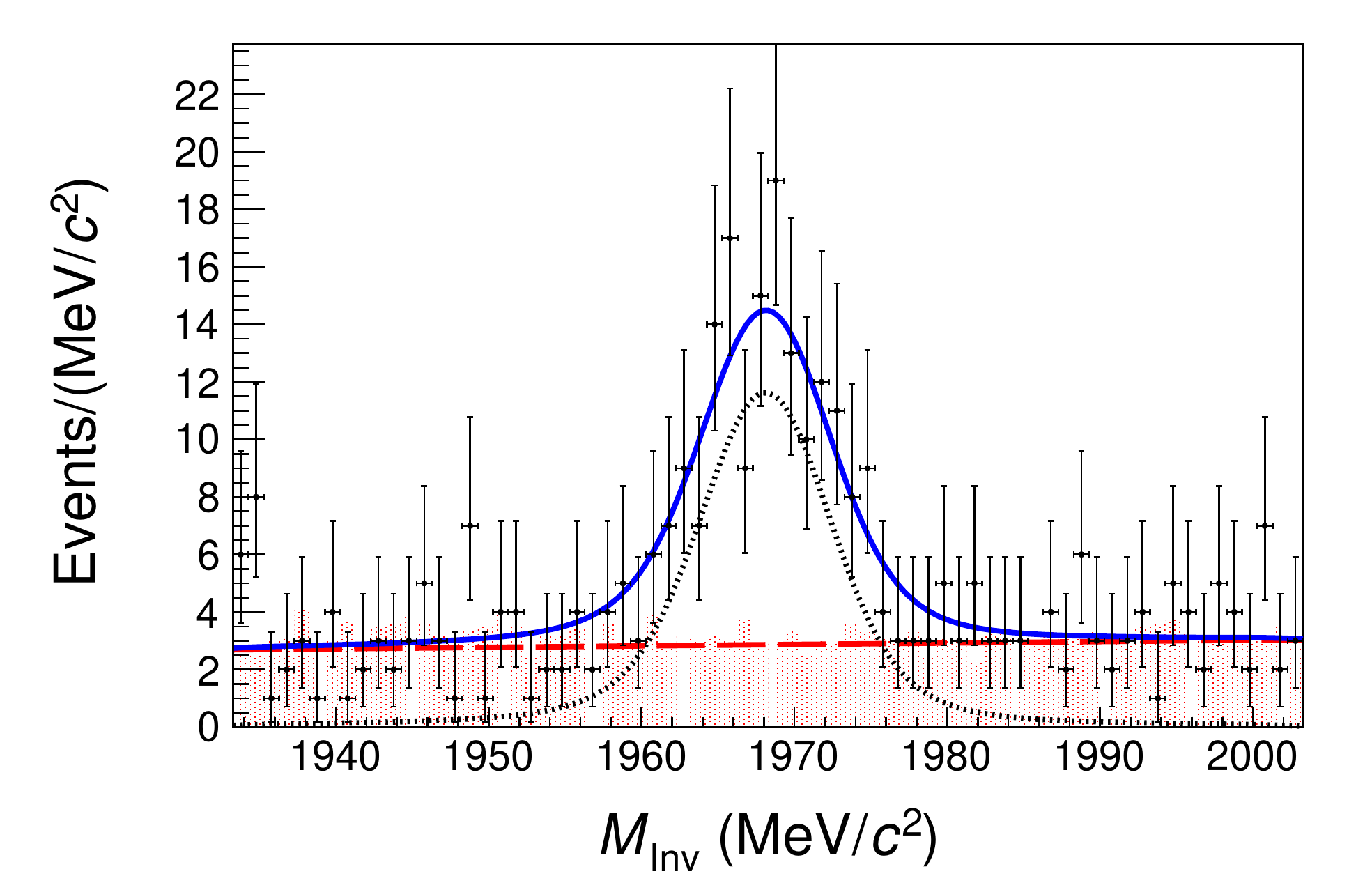} & \includegraphics[width=3in]{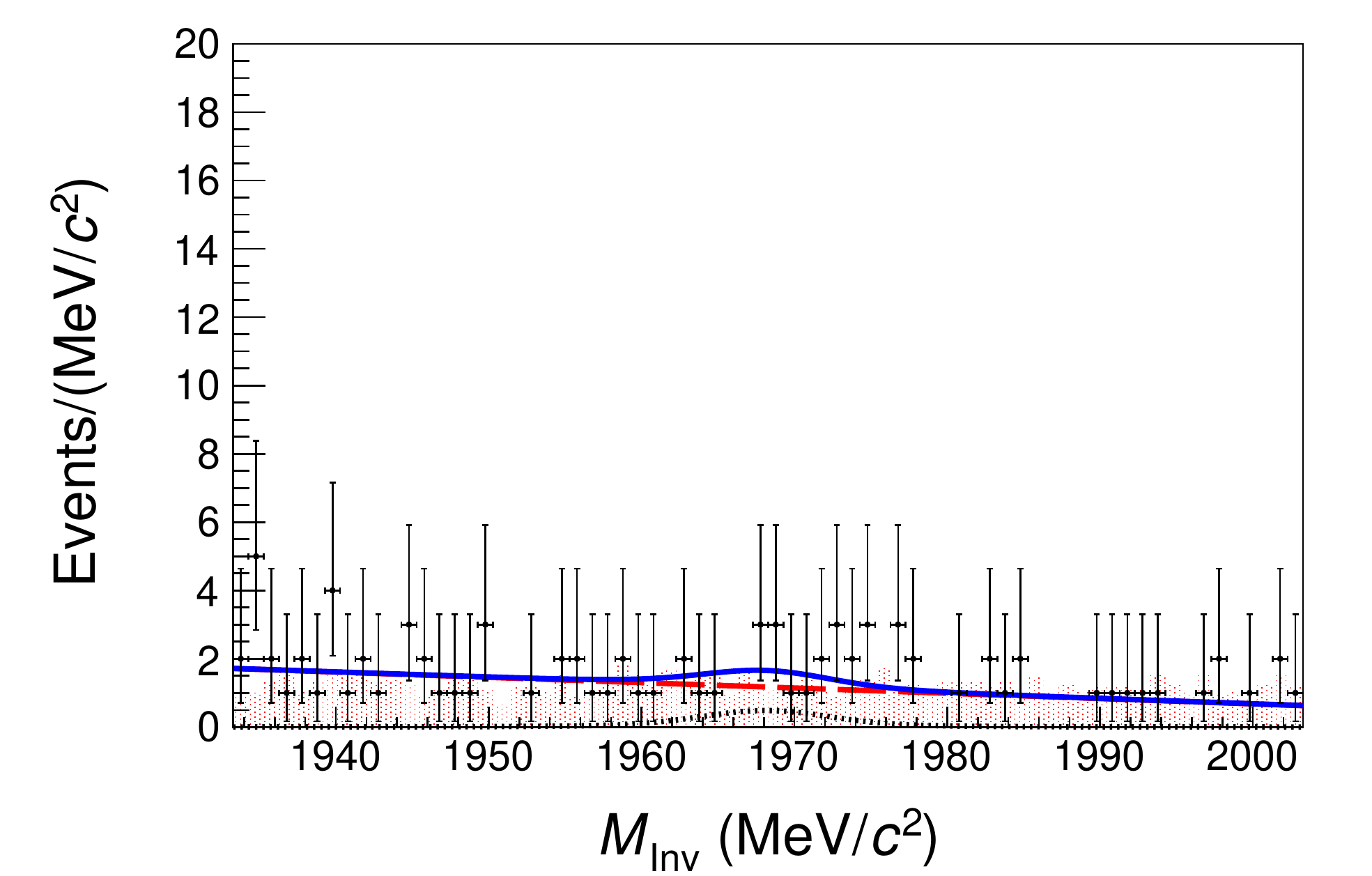}\\
\end{tabular}

\begin{tabular}{cc}
\textbf{RS 600-650 MeV/$c$} & \textbf{WS 600-650 MeV/$c$}\\
\includegraphics[width=3in]{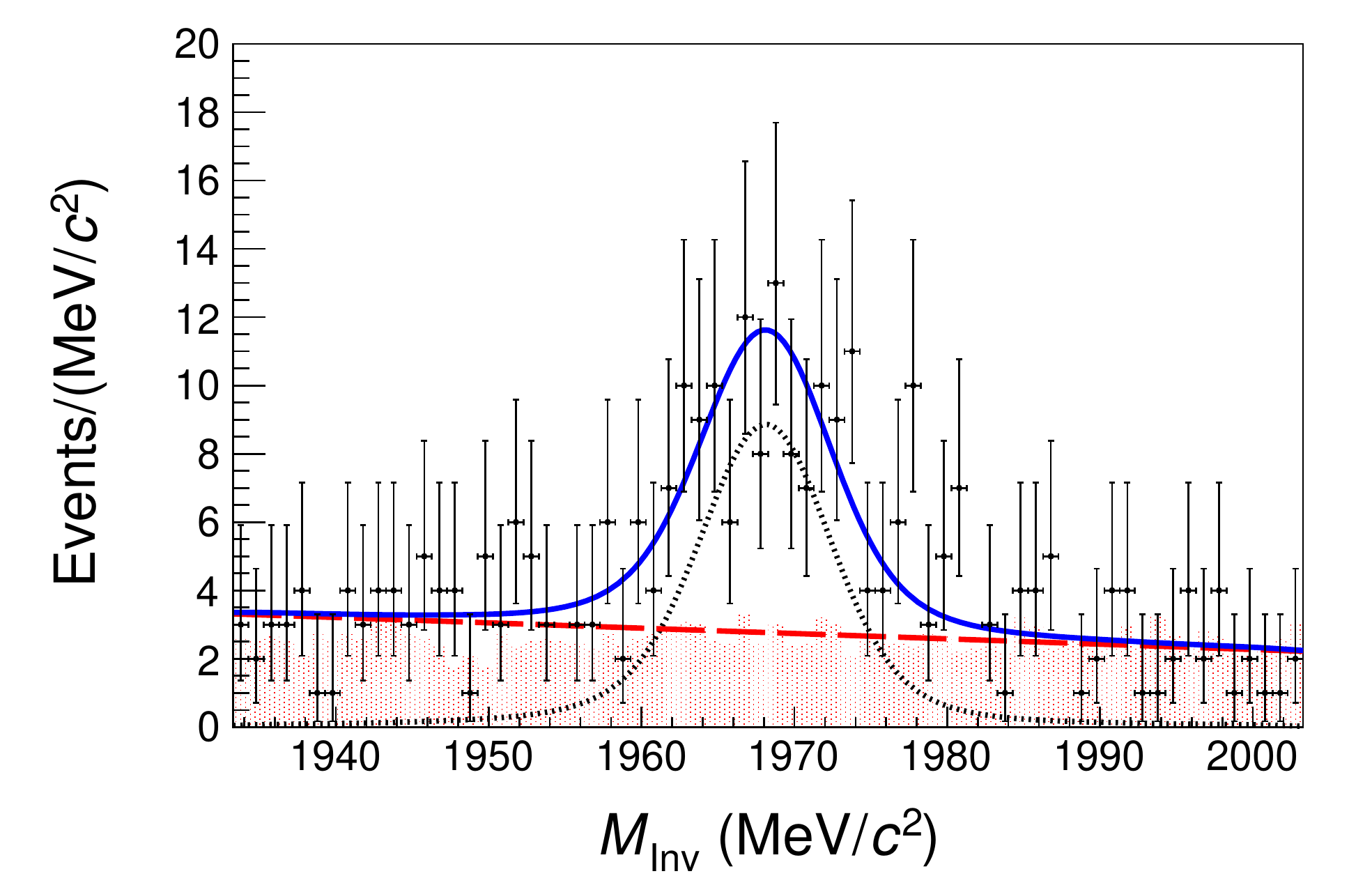} & \includegraphics[width=3in]{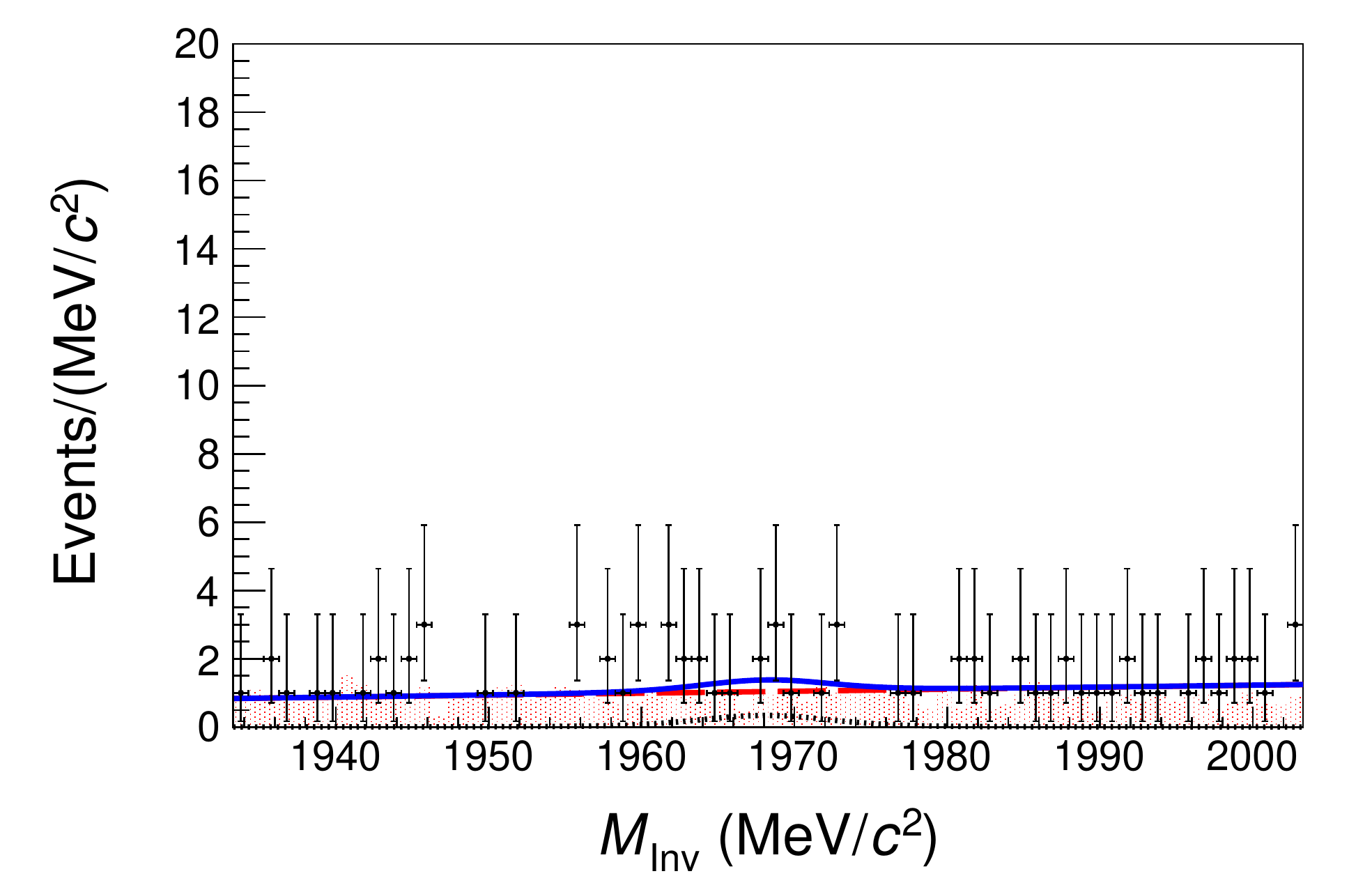}\\
\textbf{RS 650-700 MeV/$c$} & \textbf{WS 650-700 MeV/$c$}\\
\includegraphics[width=3in]{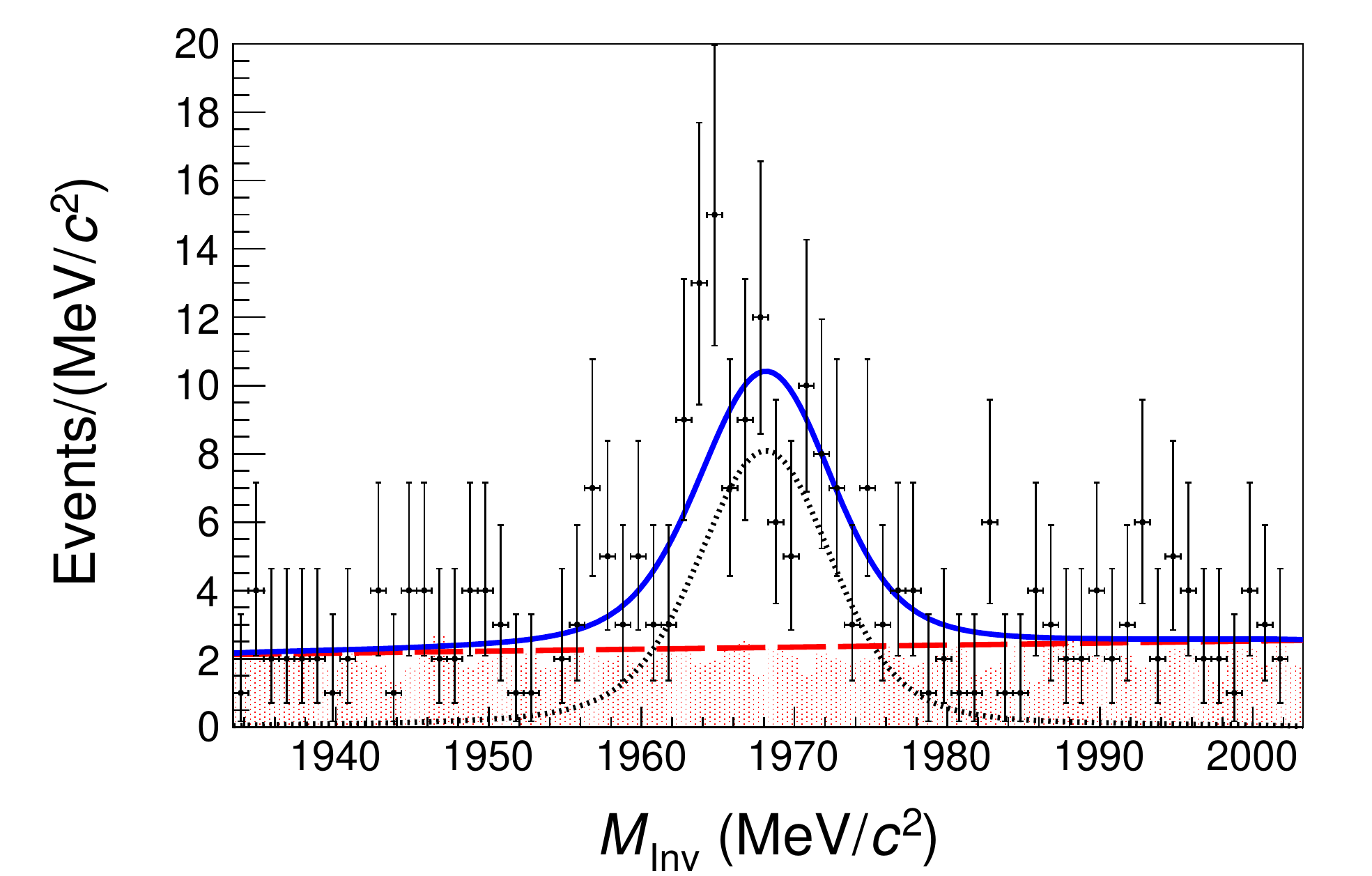} & \includegraphics[width=3in]{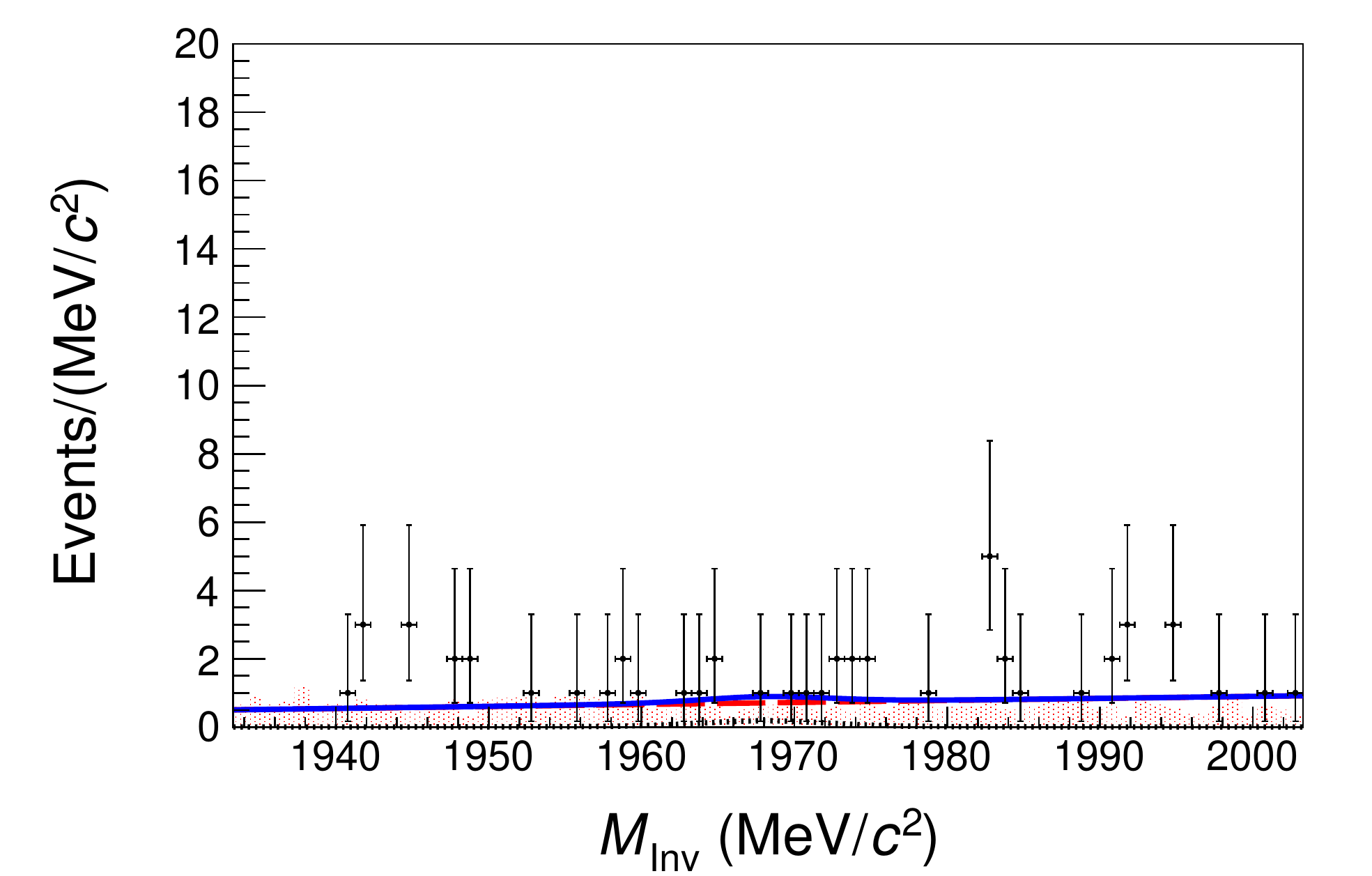}\\
\textbf{RS 700-750 MeV/$c$} & \textbf{WS 700-750 MeV/$c$}\\
\includegraphics[width=3in]{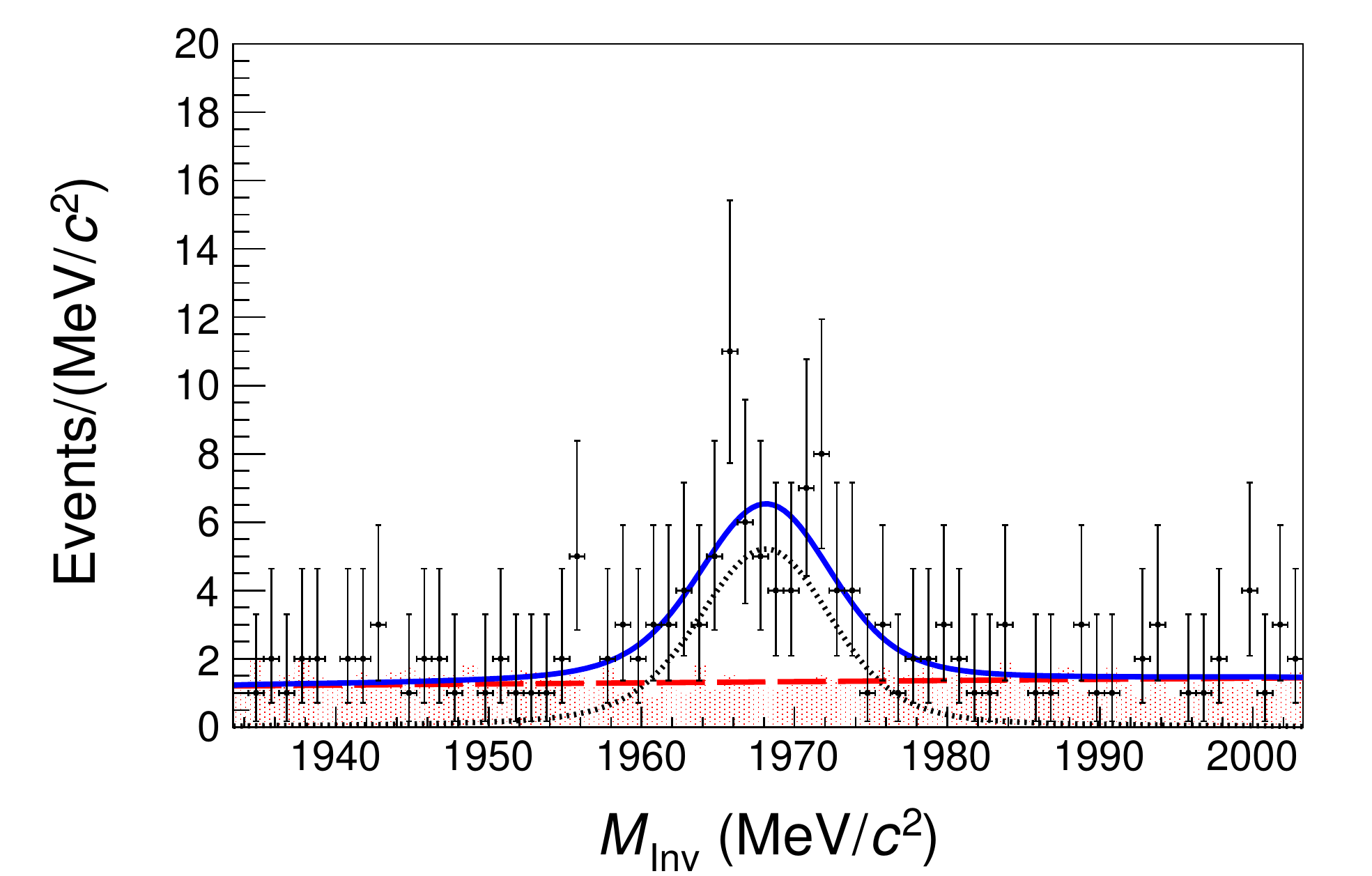} & \includegraphics[width=3in]{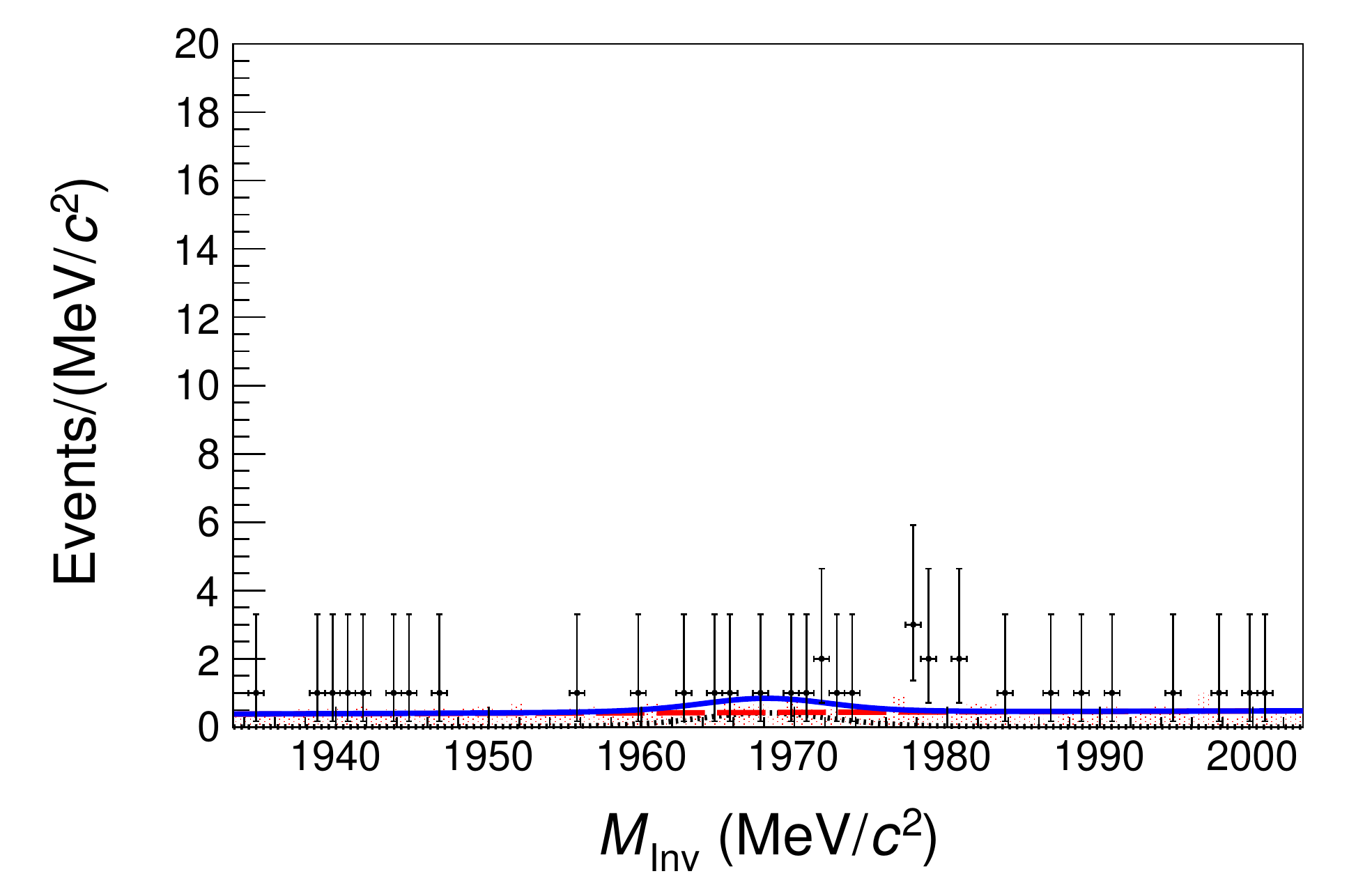}\\
\textbf{RS 750-800 MeV/$c$} & \textbf{WS 750-800 MeV/$c$}\\
\includegraphics[width=3in]{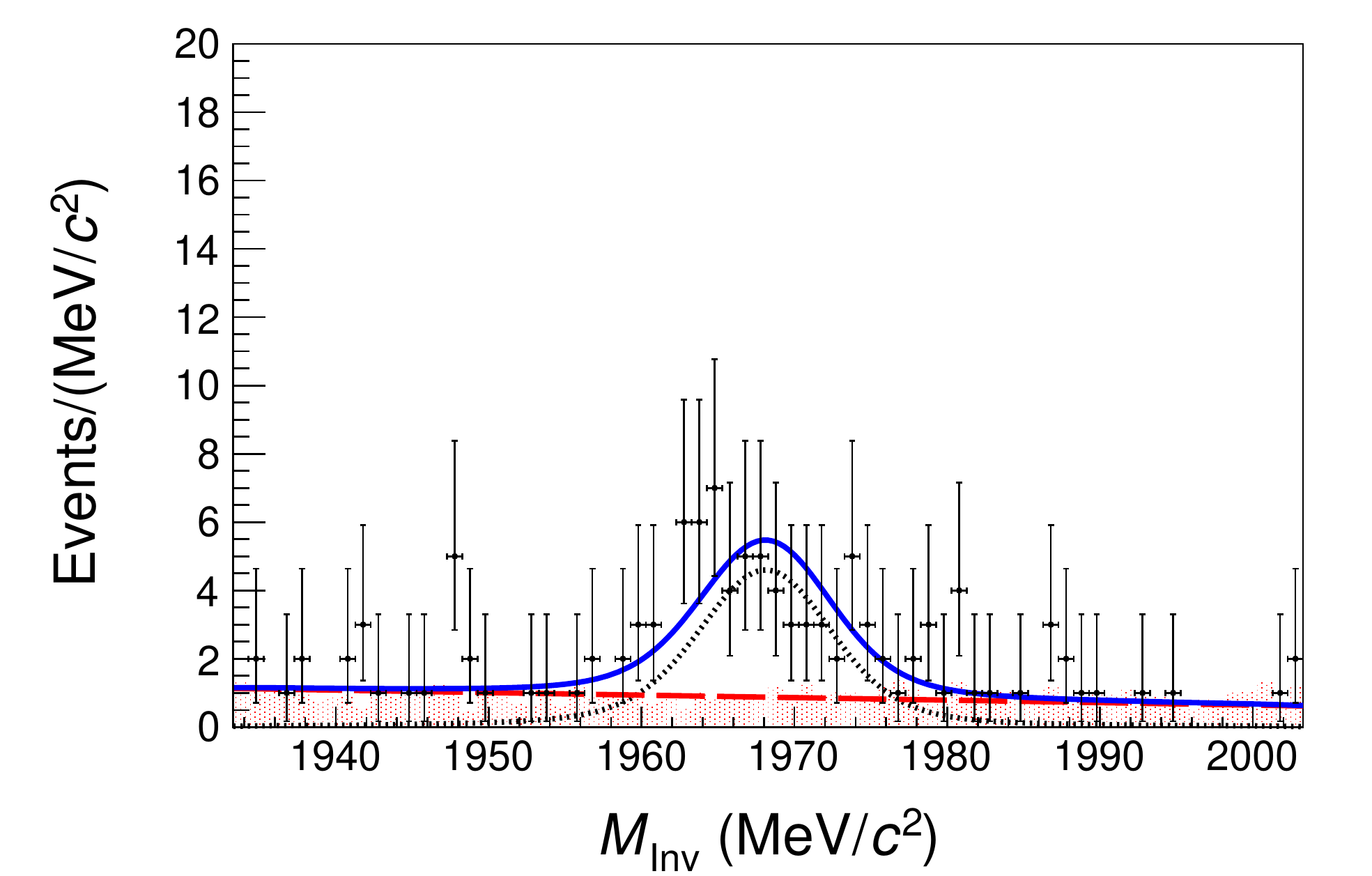} & \includegraphics[width=3in]{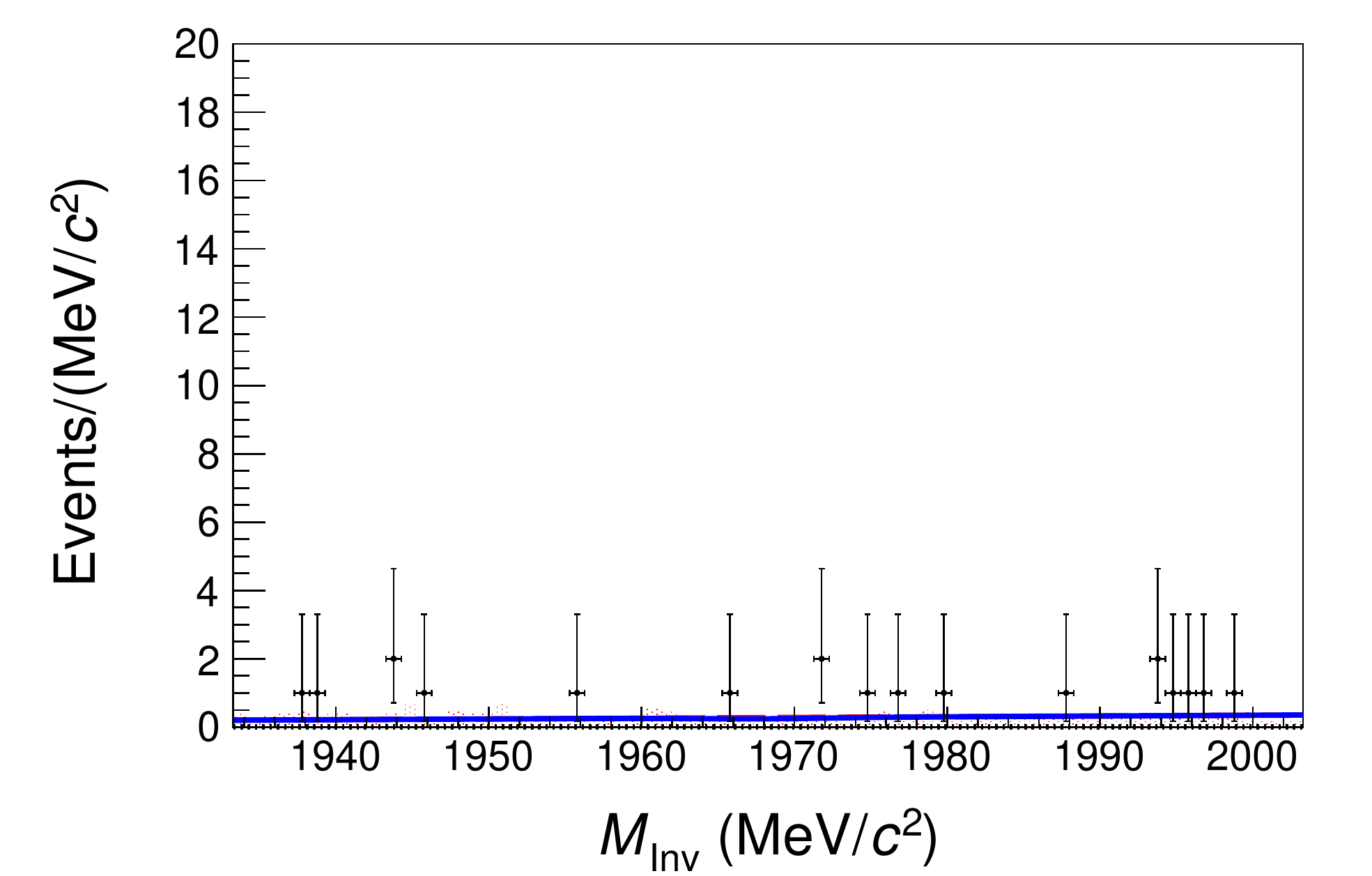}\\
\end{tabular}

\begin{tabular}{cc}
\textbf{RS 800-850 MeV/$c$} & \textbf{WS 800-850 MeV/$c$}\\
\includegraphics[width=3in]{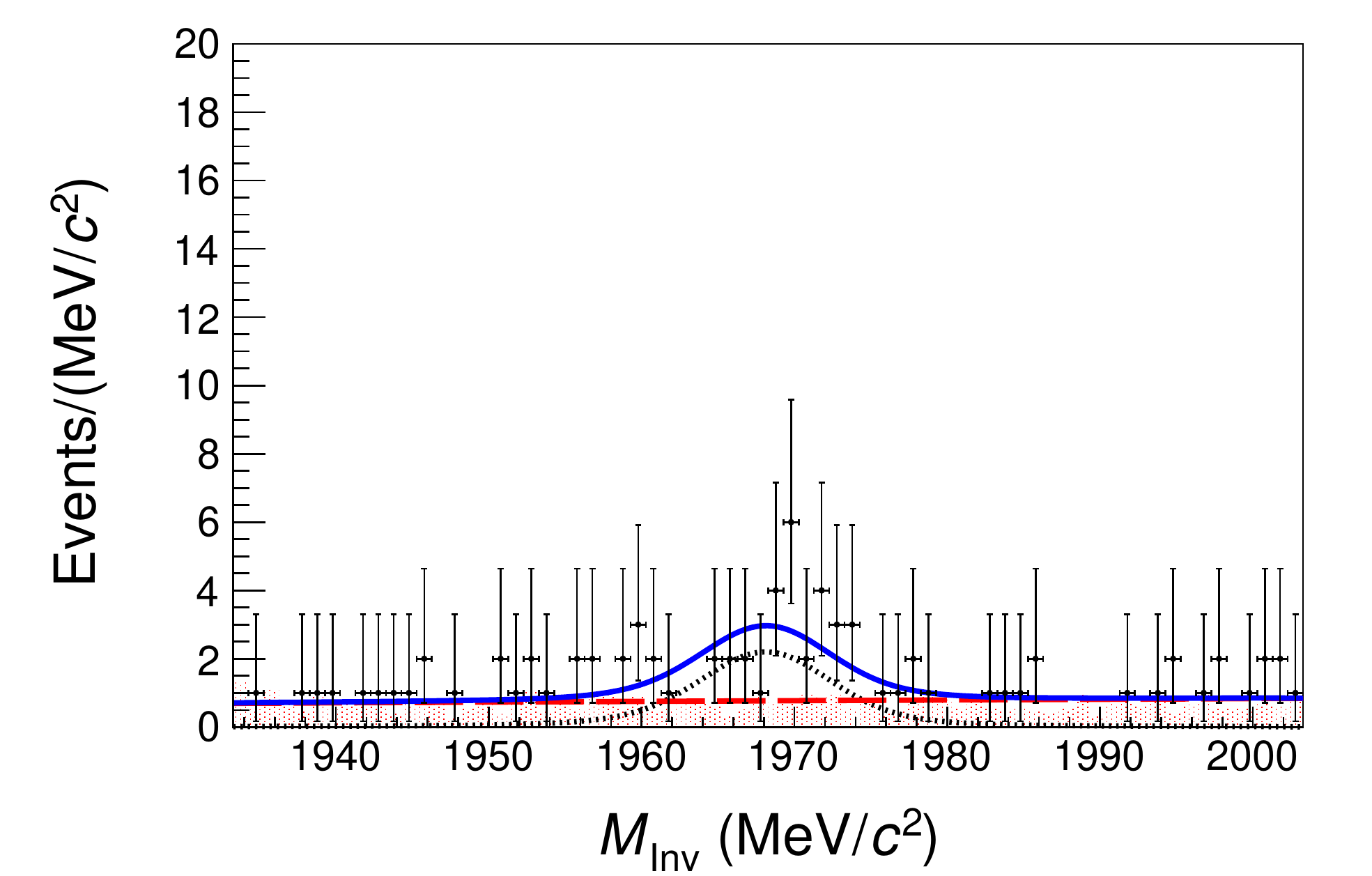} & \includegraphics[width=3in]{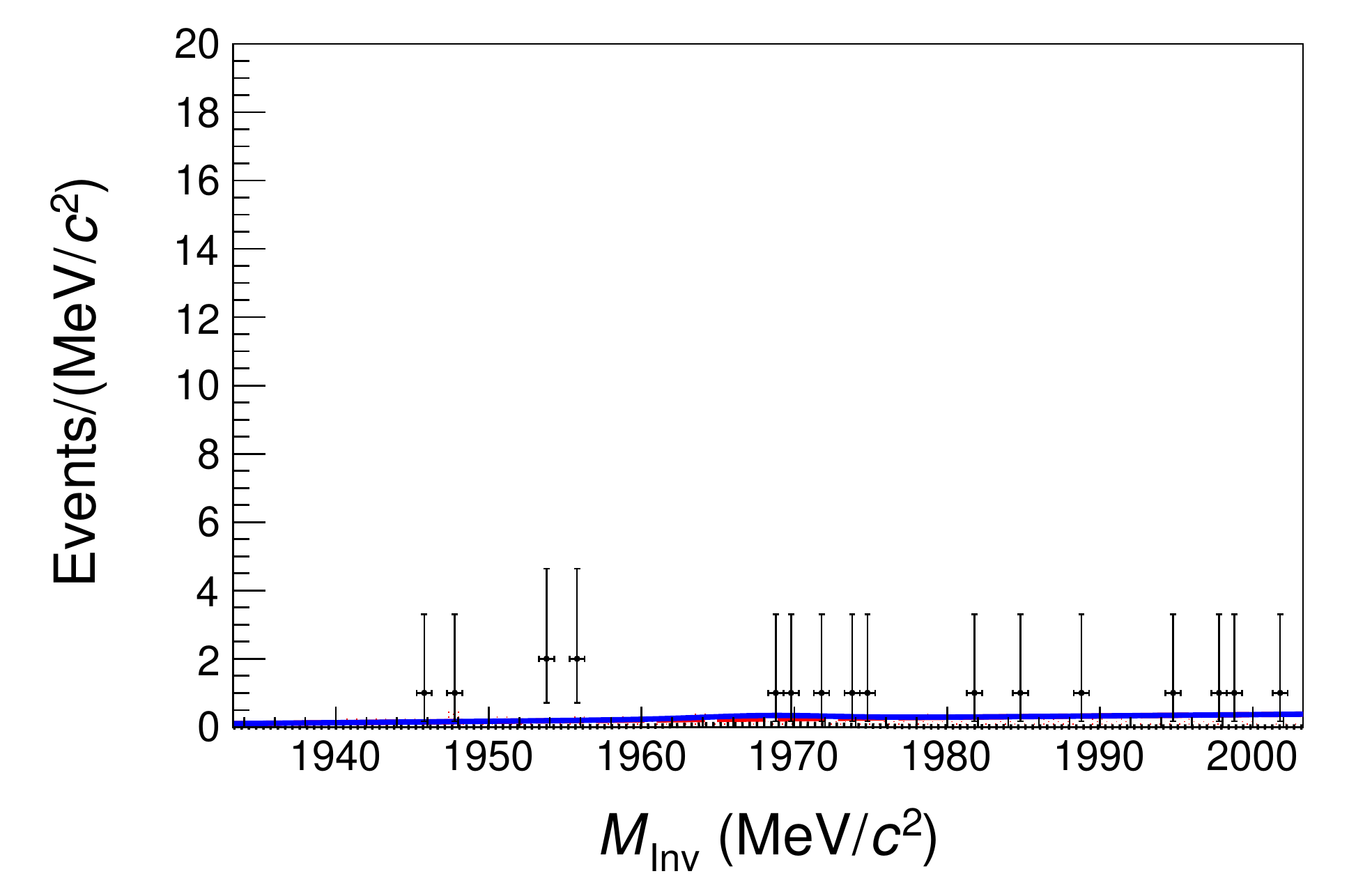}\\
\textbf{RS 850-900 MeV/$c$} & \textbf{WS 850-900 MeV/$c$}\\
\includegraphics[width=3in]{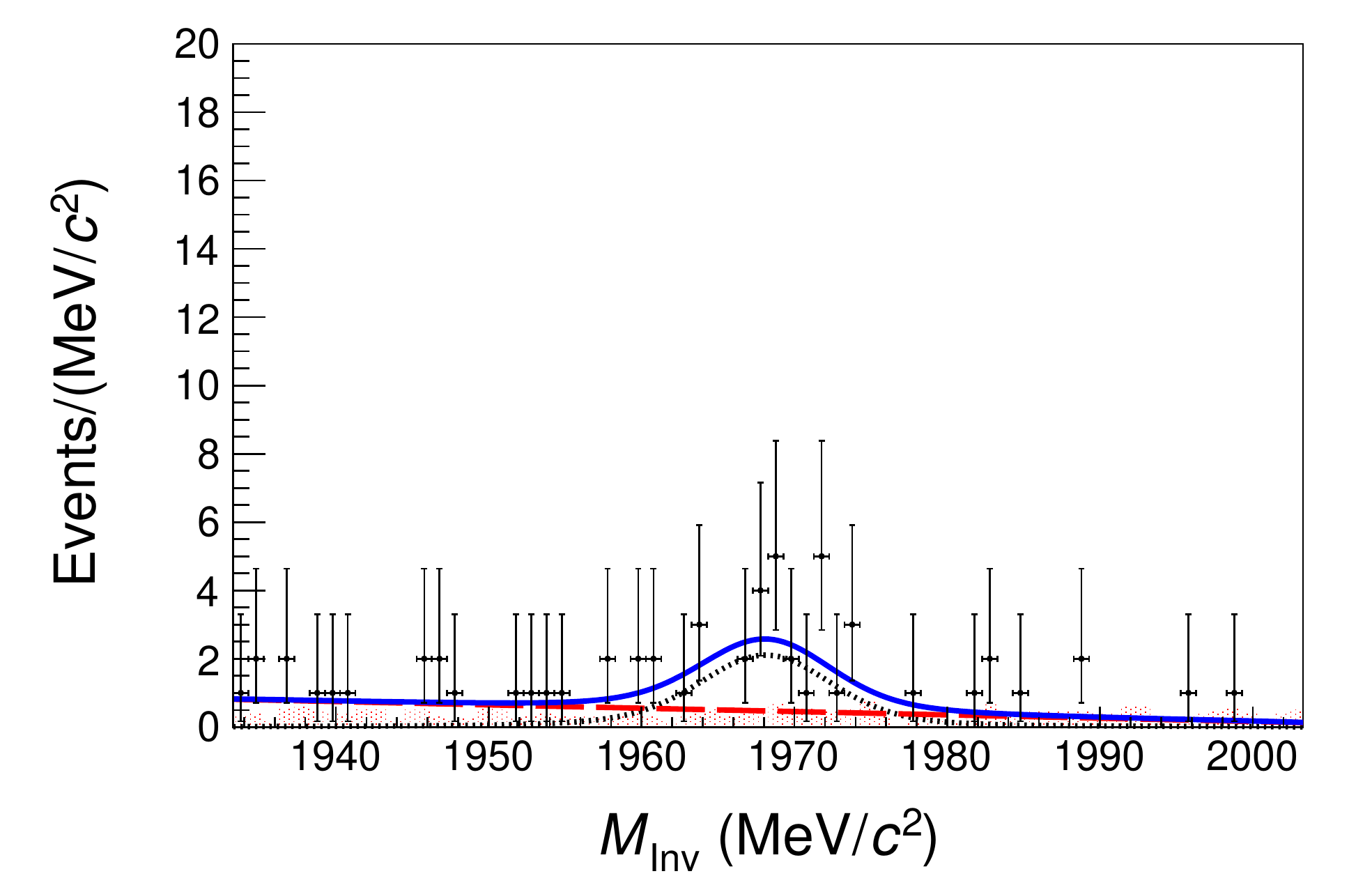} & \includegraphics[width=3in]{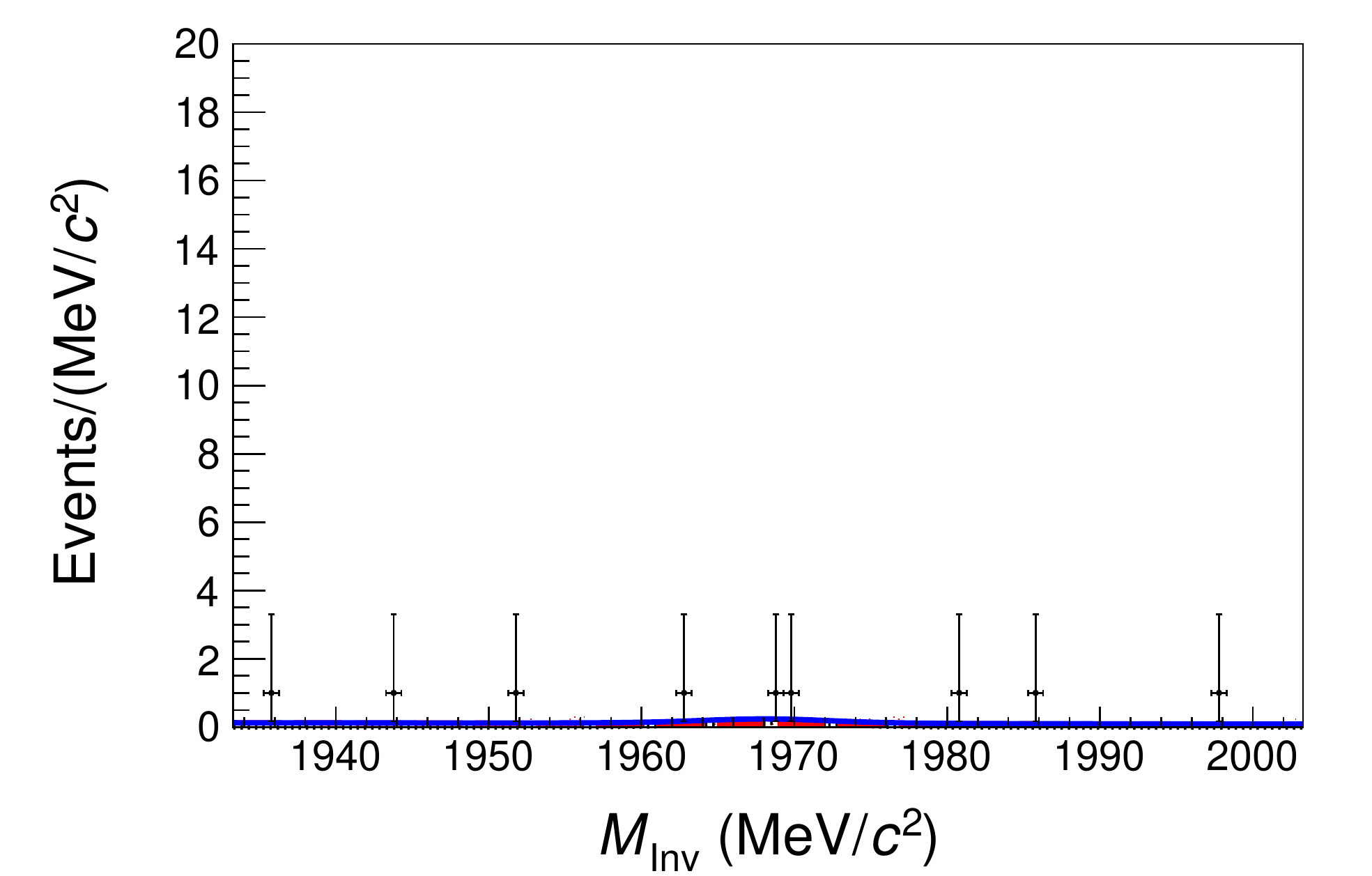}\\
\textbf{RS 900-950 MeV/$c$} & \textbf{WS 900-950 MeV/$c$}\\
\includegraphics[width=3in]{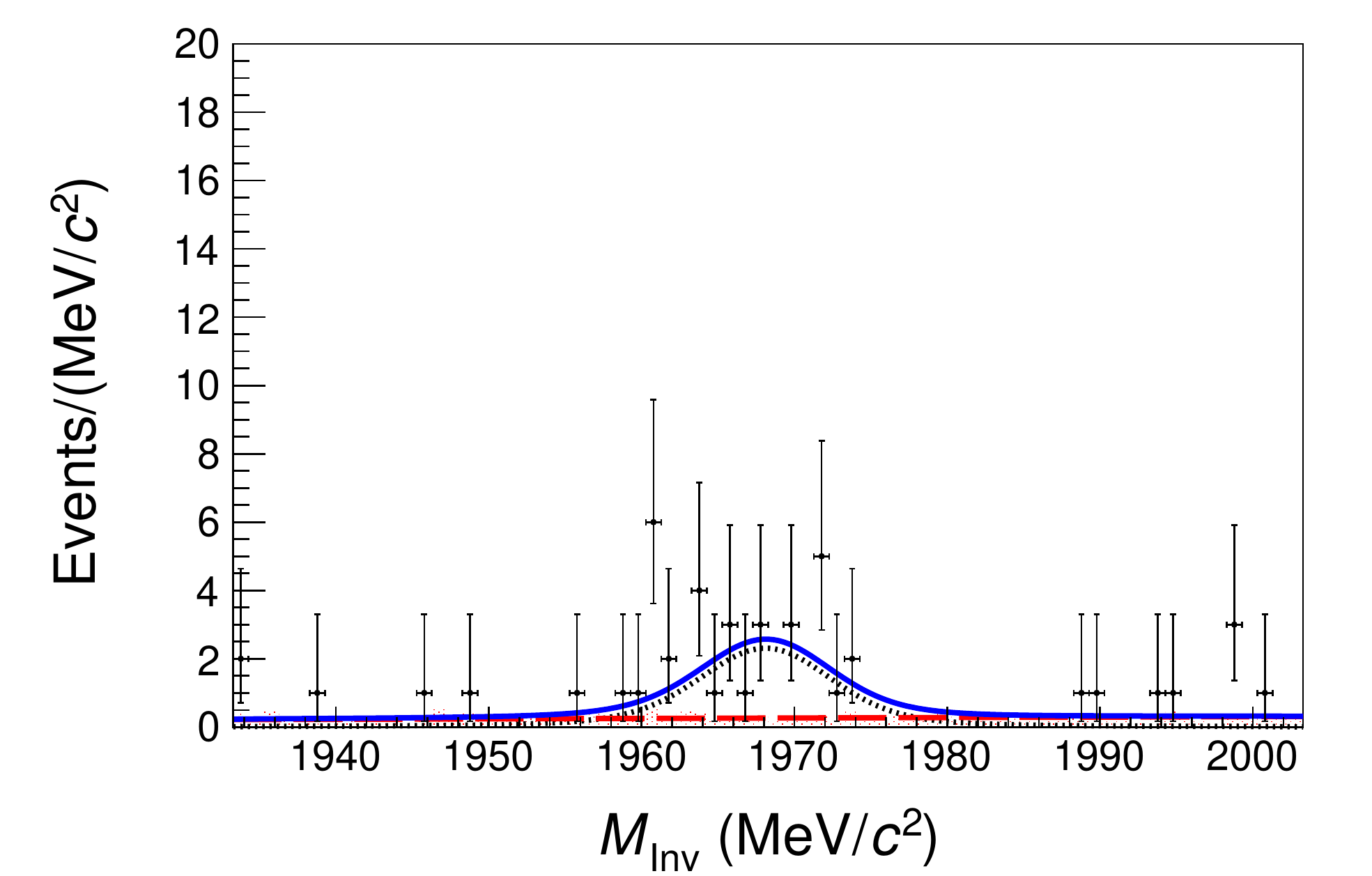} & \includegraphics[width=3in]{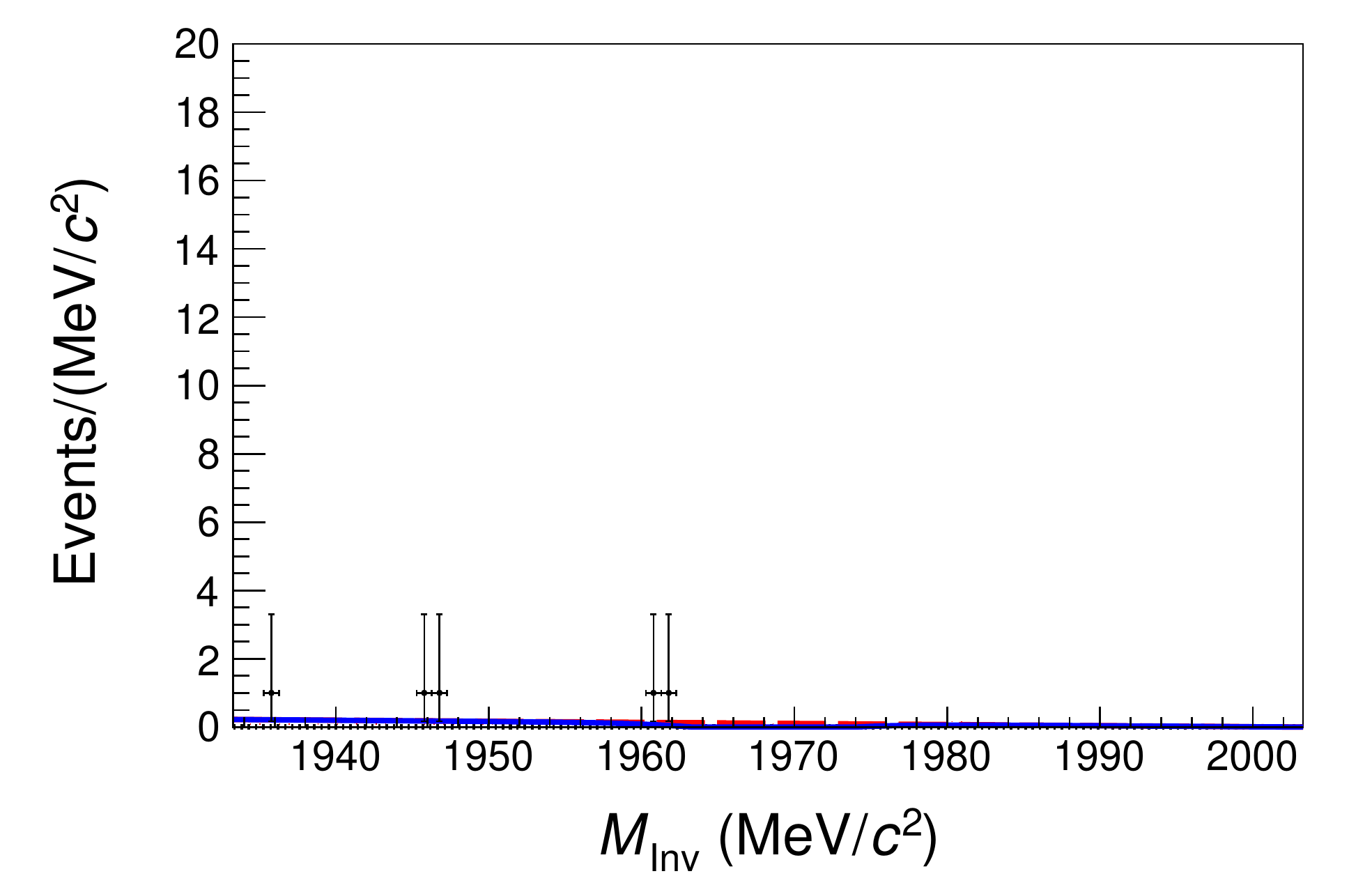}\\
\textbf{RS 950-1000 MeV/$c$} & \textbf{WS 950-1000 MeV/$c$}\\
\includegraphics[width=3in,valign=m]{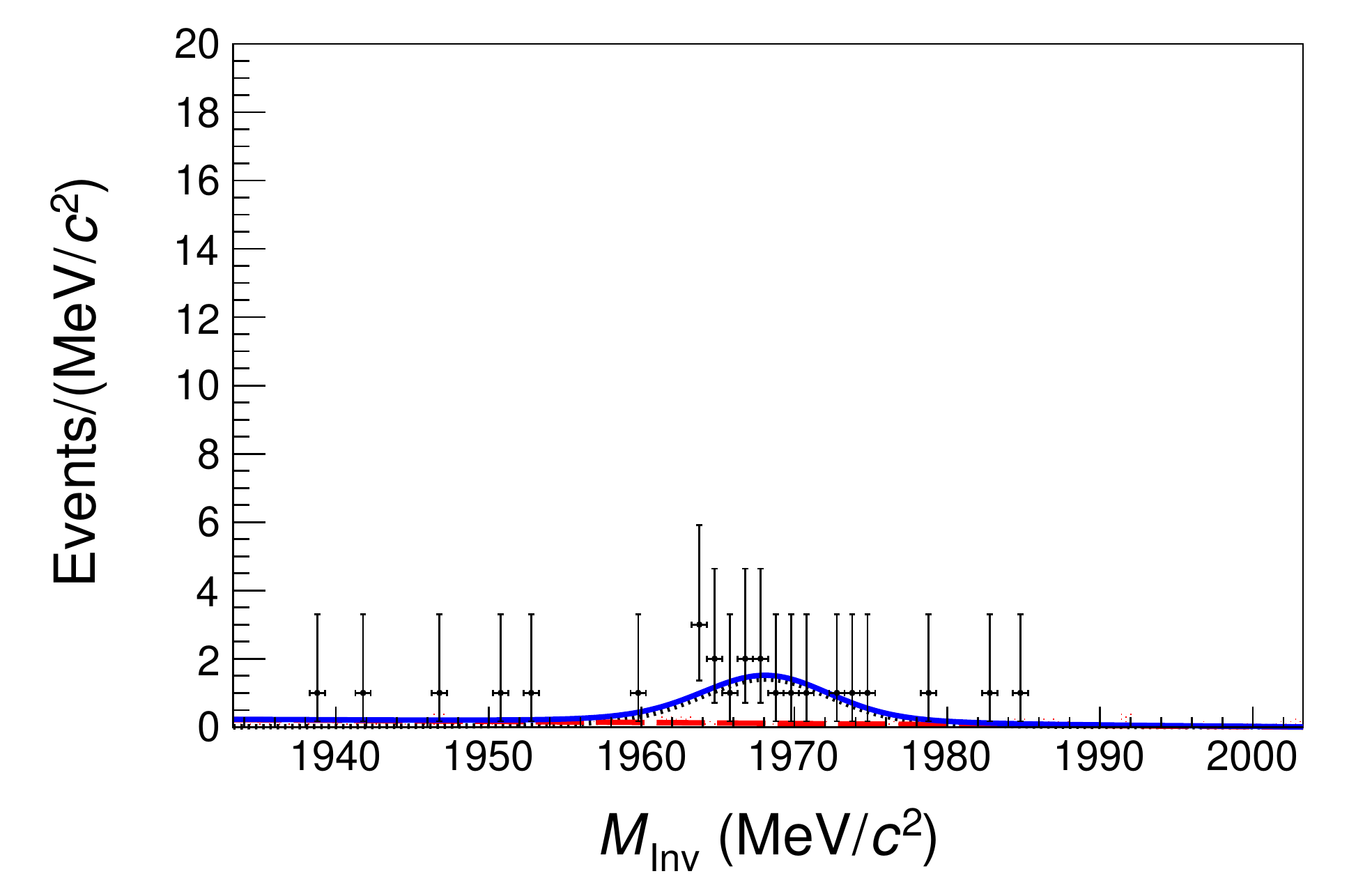} & No entries are seen in the signal region\\
\end{tabular}

\begin{tabular}{cc}
\textbf{RS 1000-1050 MeV/$c$} & \textbf{WS 1000-1050 MeV/$c$}\\
\includegraphics[width=3in,valign=m]{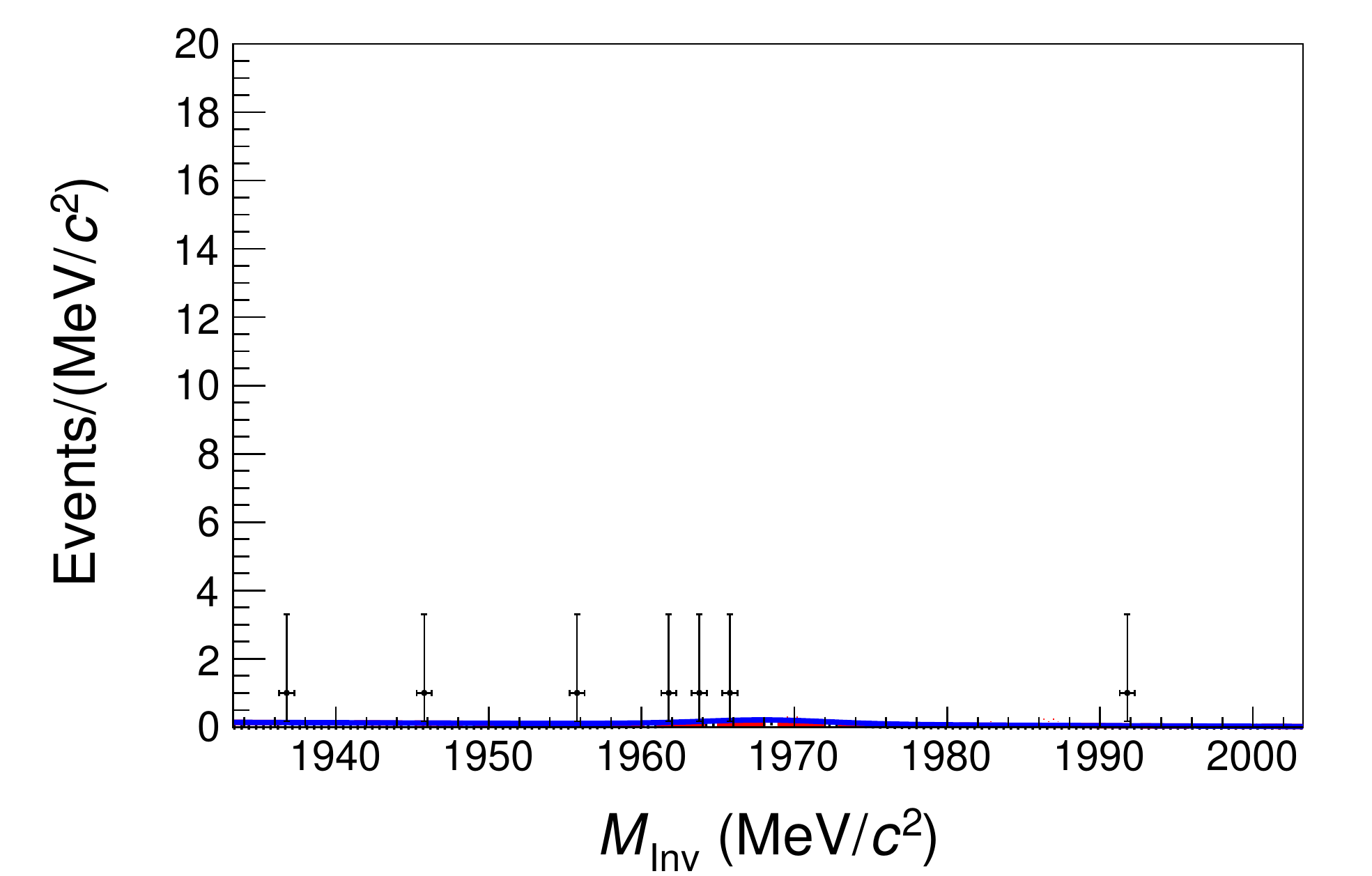} & No entries are seen in the signal region\\
\textbf{RS 1050-1100 MeV/$c$} & \textbf{WS 1050-1100 MeV/$c$}\\
\includegraphics[width=3in,valign=m]{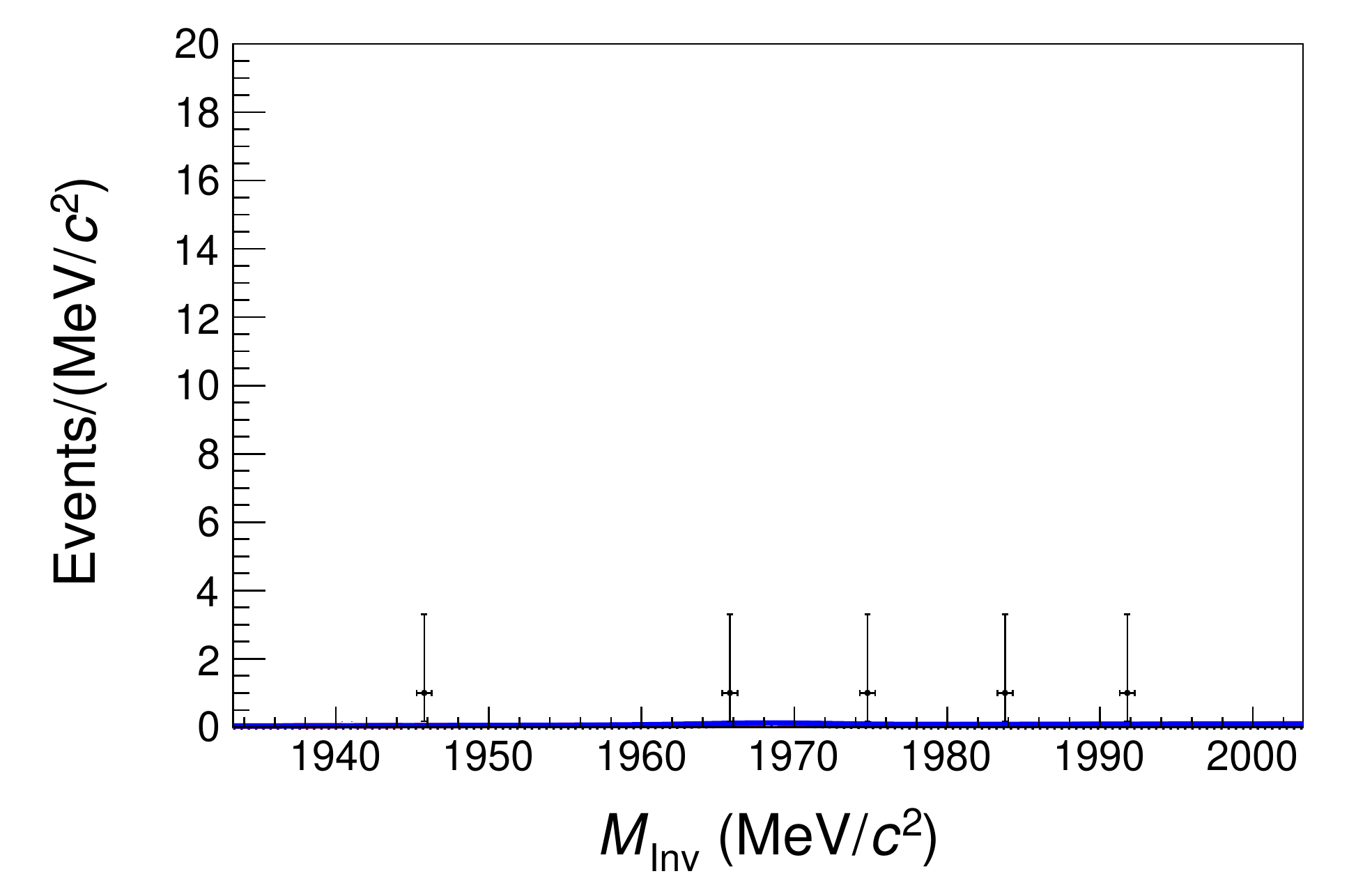} & No entries are seen in the signal region
\end{tabular}

\pagebreak
\raggedright

\subsubsection{$\EcmC$ Data $\pi$ ID Fits}
\label{subsubsec:4230DataPiIDFits}
\centering
\begin{tabular}{cc}
\textbf{RS 200-250 MeV/$c$} & \textbf{WS 200-250 MeV/$c$}\\
\includegraphics[width=3in]{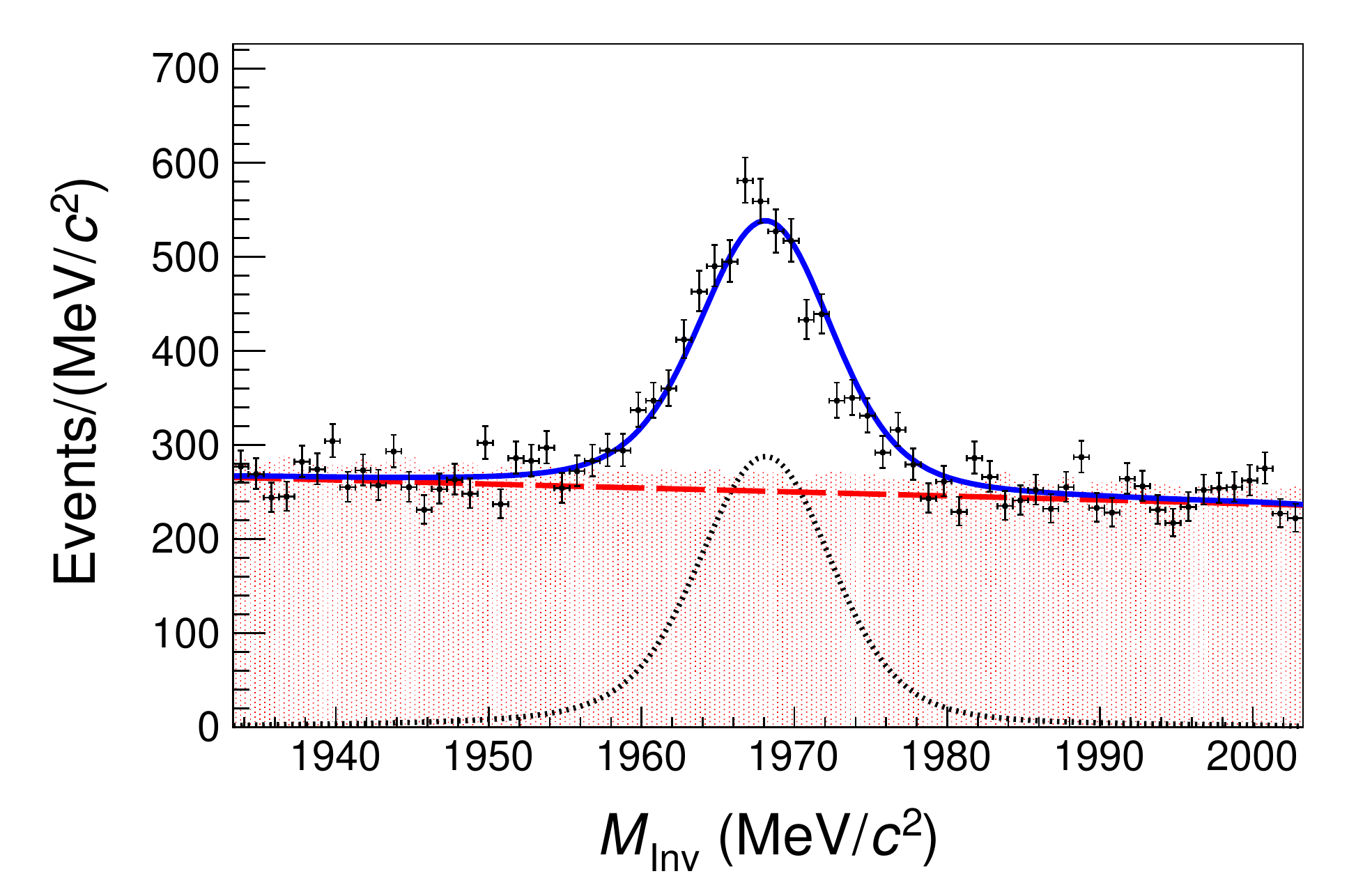} & \includegraphics[width=3in]{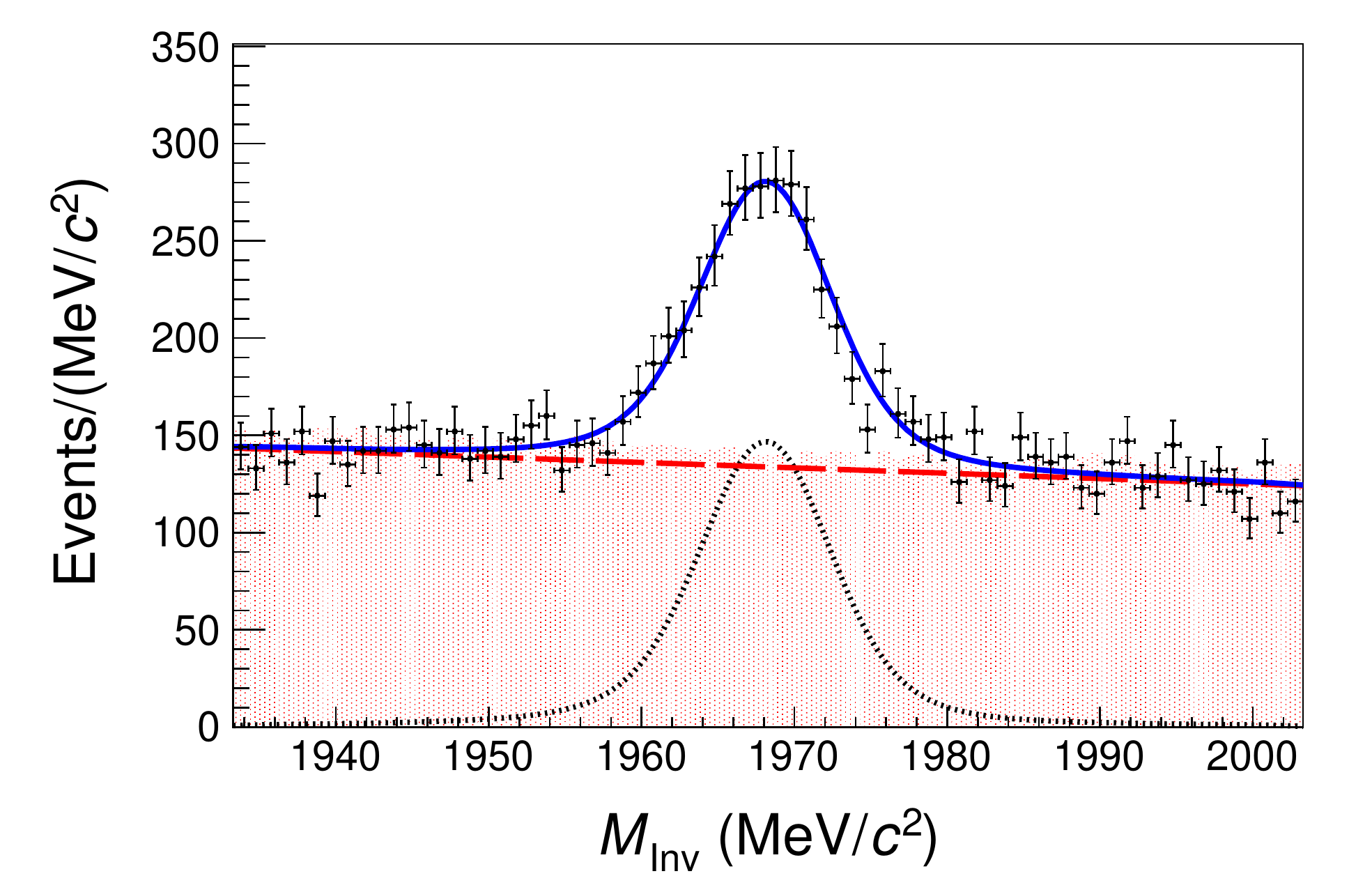}\\
\textbf{RS 250-300 MeV/$c$} & \textbf{WS 250-300 MeV/$c$}\\
\includegraphics[width=3in]{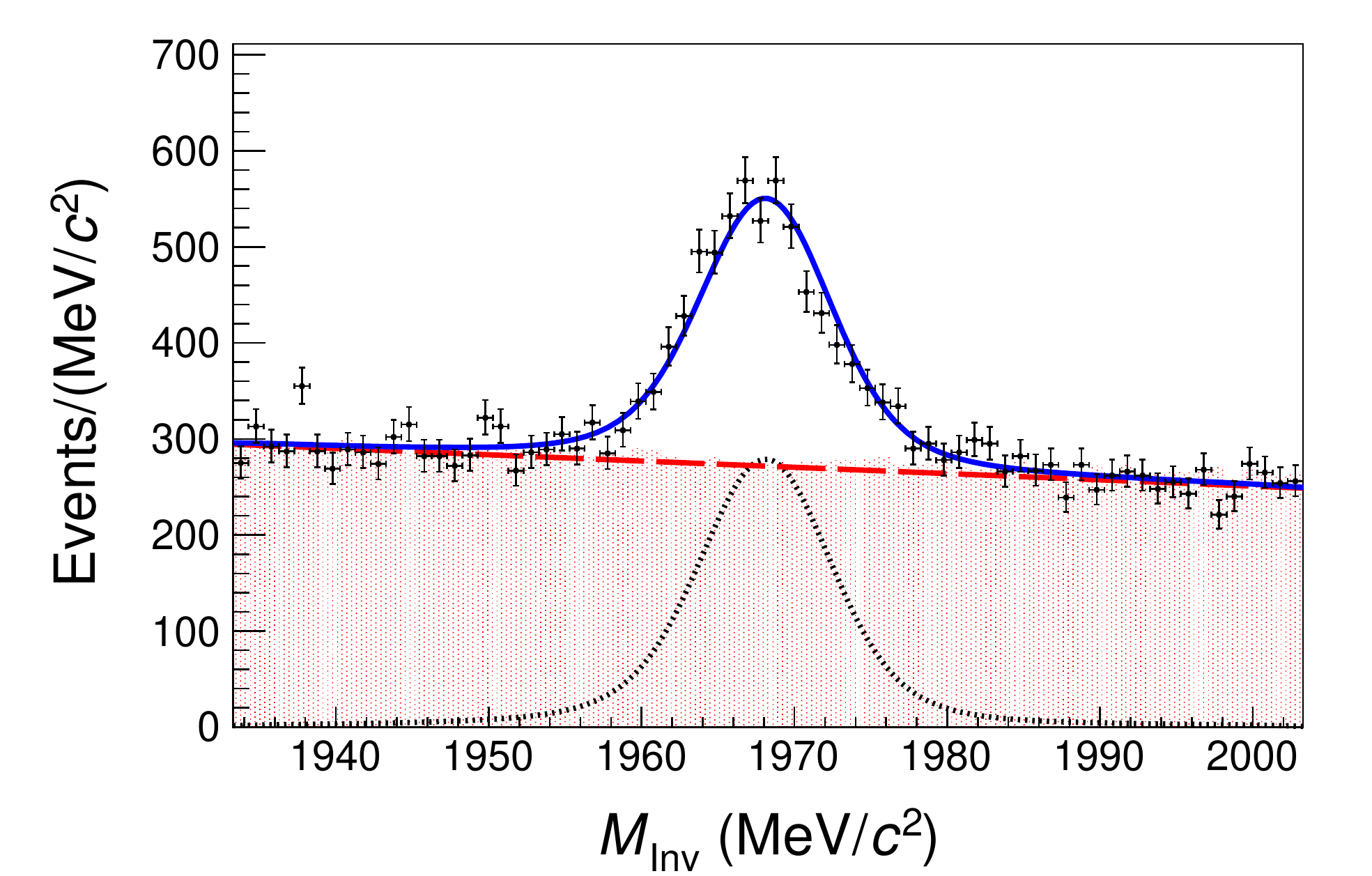} & \includegraphics[width=3in]{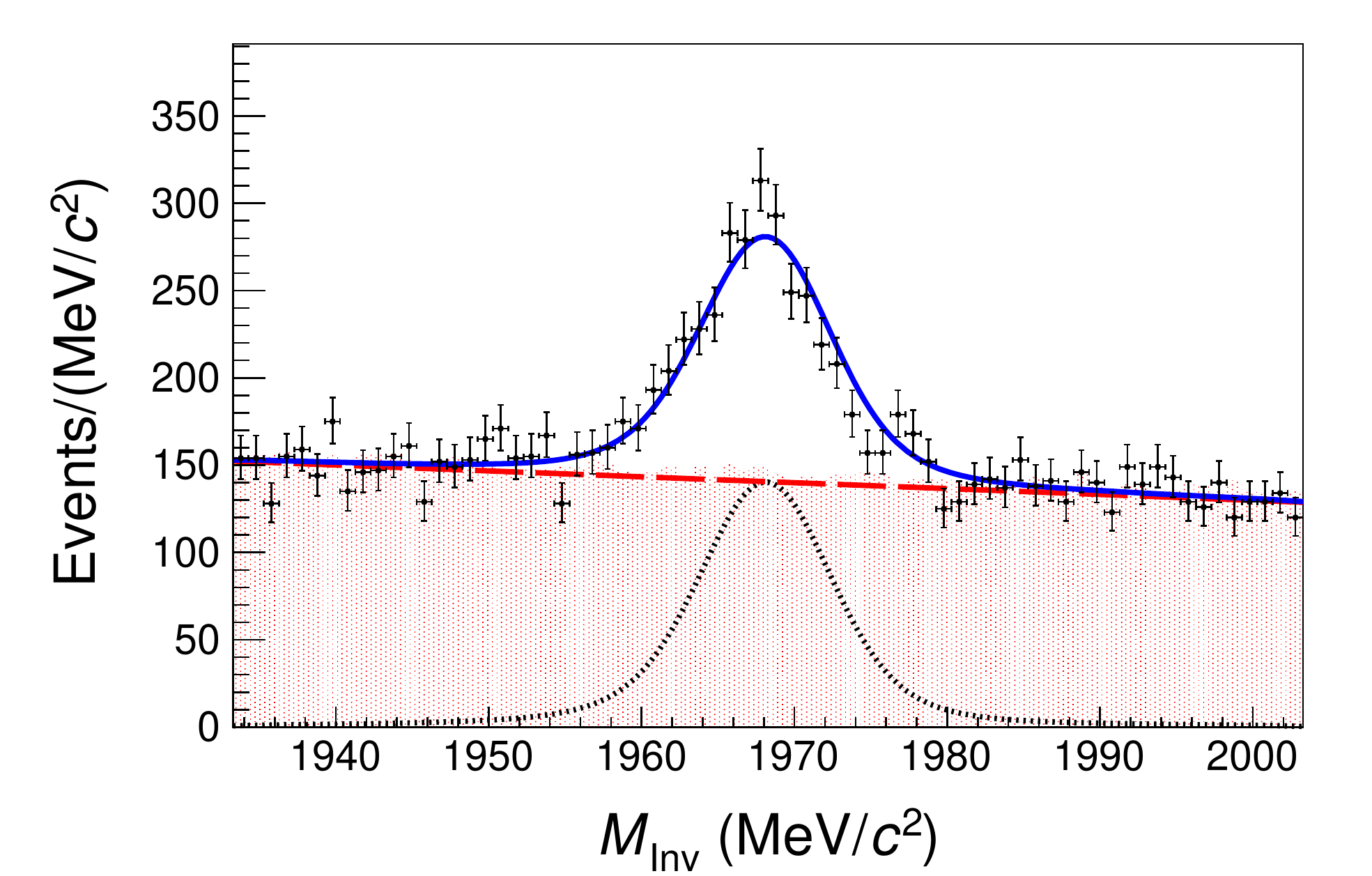}\\
\textbf{RS 300-350 MeV/$c$} & \textbf{WS 300-350 MeV/$c$}\\
\includegraphics[width=3in]{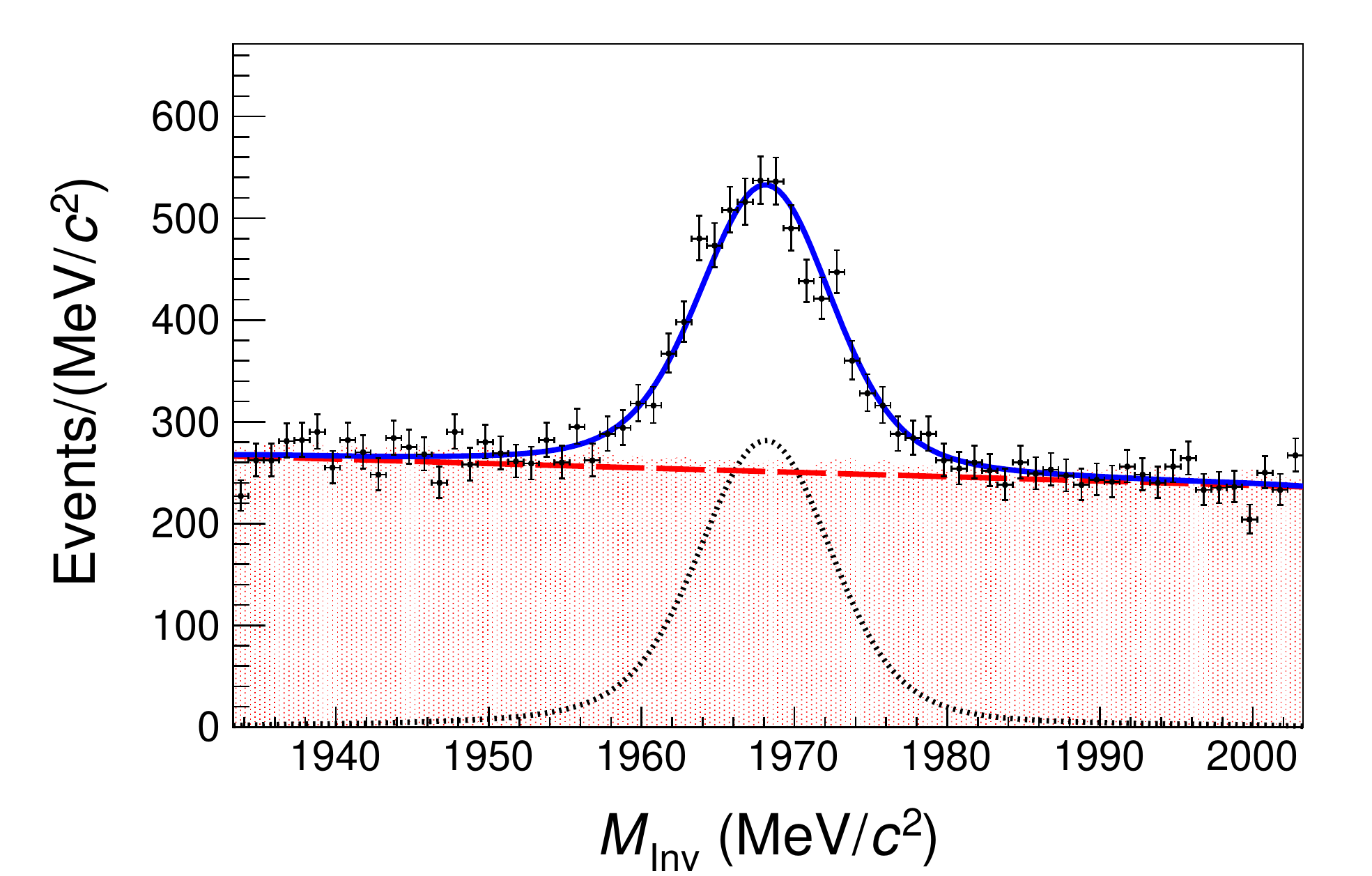} & \includegraphics[width=3in]{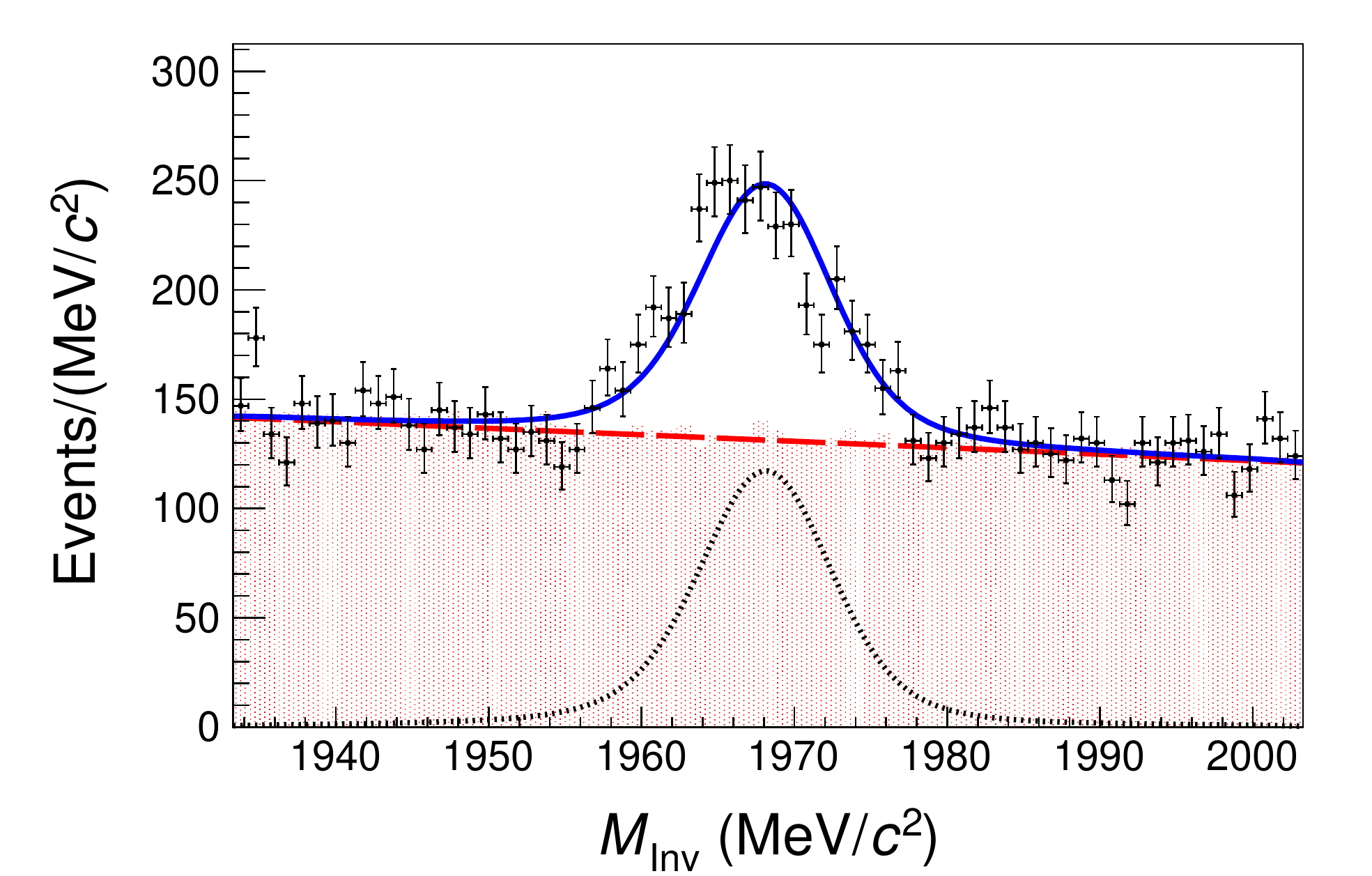}\\
\textbf{RS 350-400 MeV/$c$} & \textbf{WS 350-400 MeV/$c$}\\
\includegraphics[width=3in]{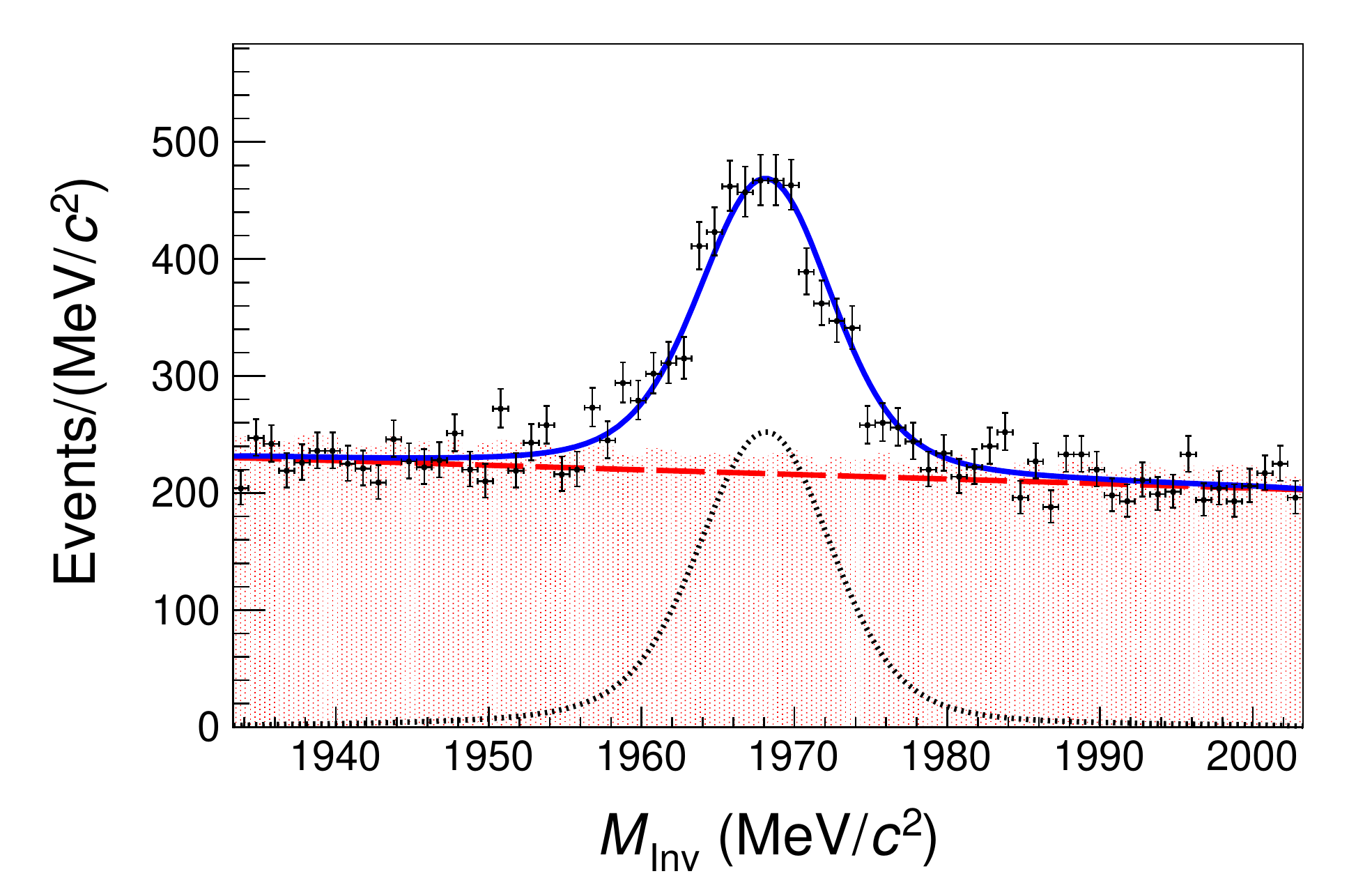} & \includegraphics[width=3in]{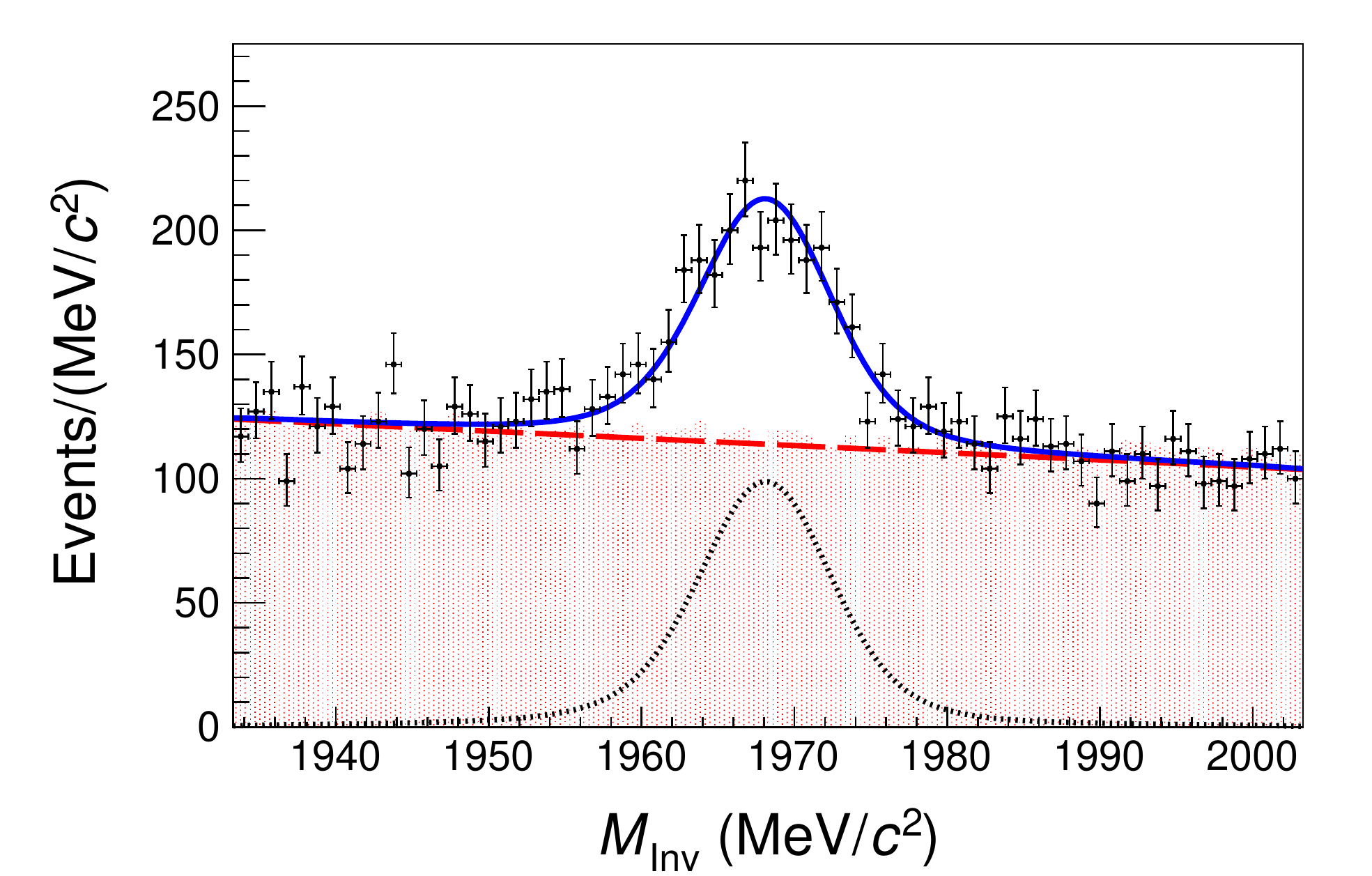}
\end{tabular}
\pagebreak

\begin{tabular}{cc}
\textbf{RS 400-450 MeV/$c$} & \textbf{WS 400-450 MeV/$c$}\\
\includegraphics[width=3in]{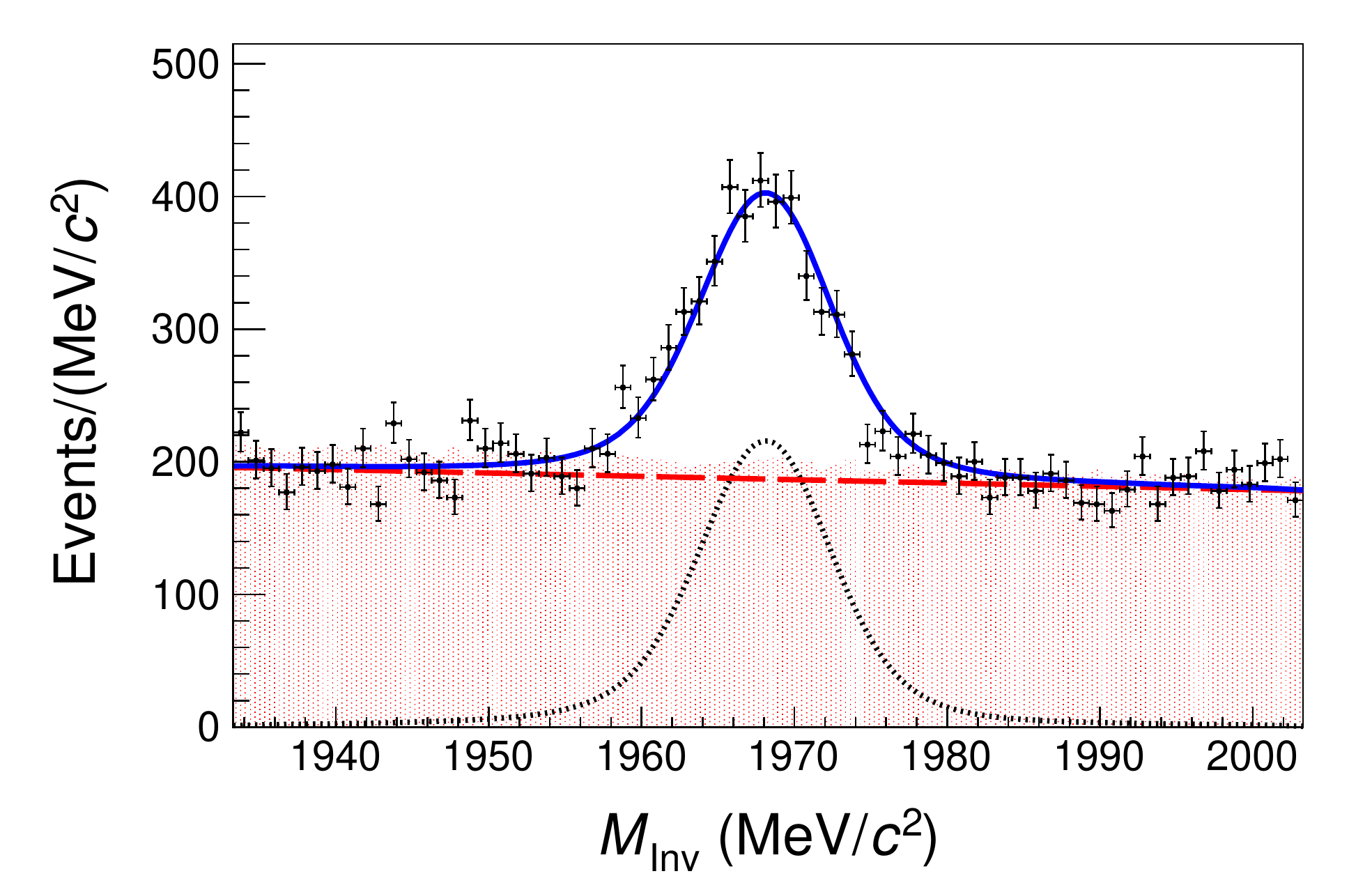} & \includegraphics[width=3in]{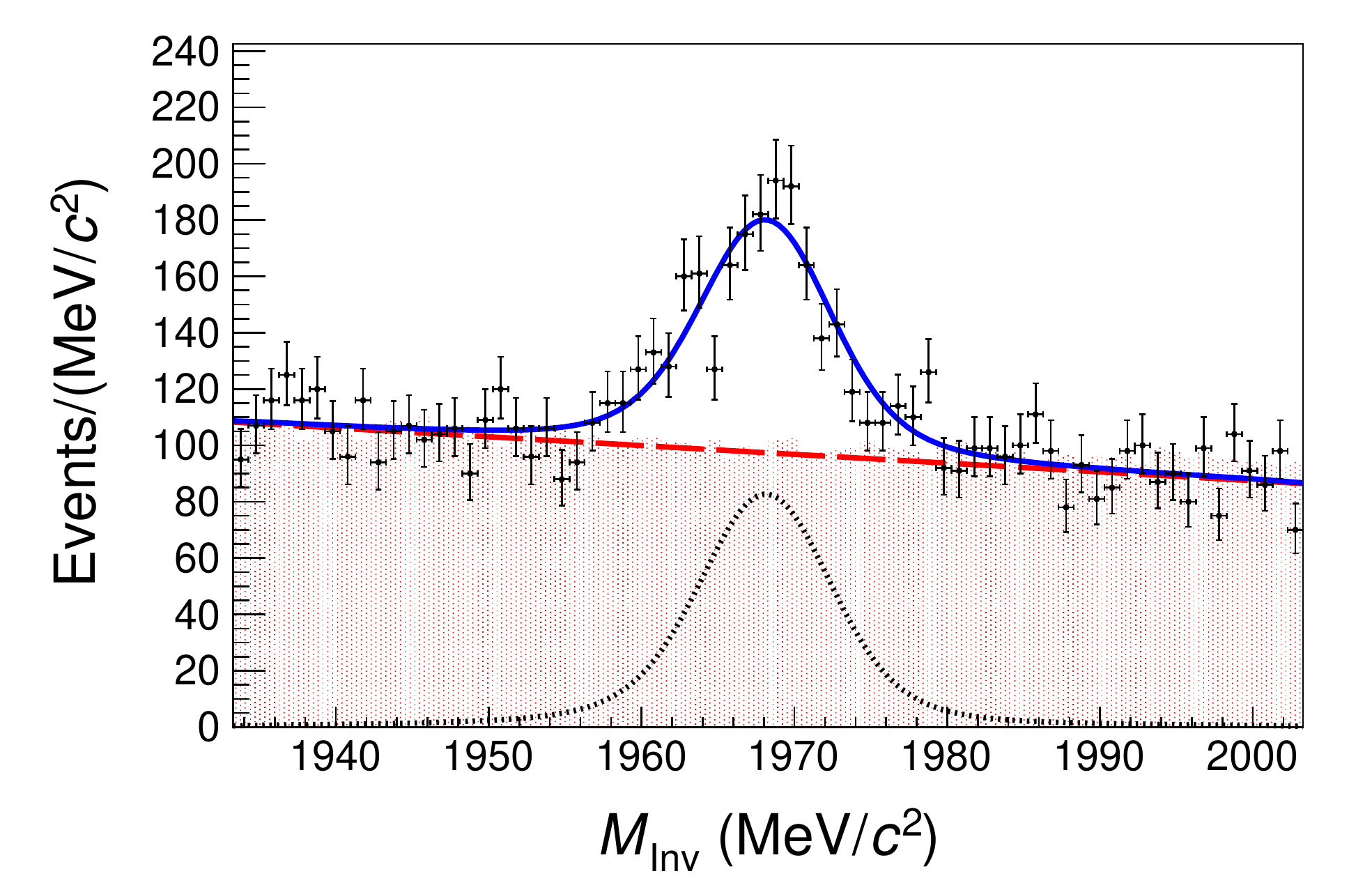}\\
\textbf{RS 450-500 MeV/$c$} & \textbf{WS 450-500 MeV/$c$}\\
\includegraphics[width=3in]{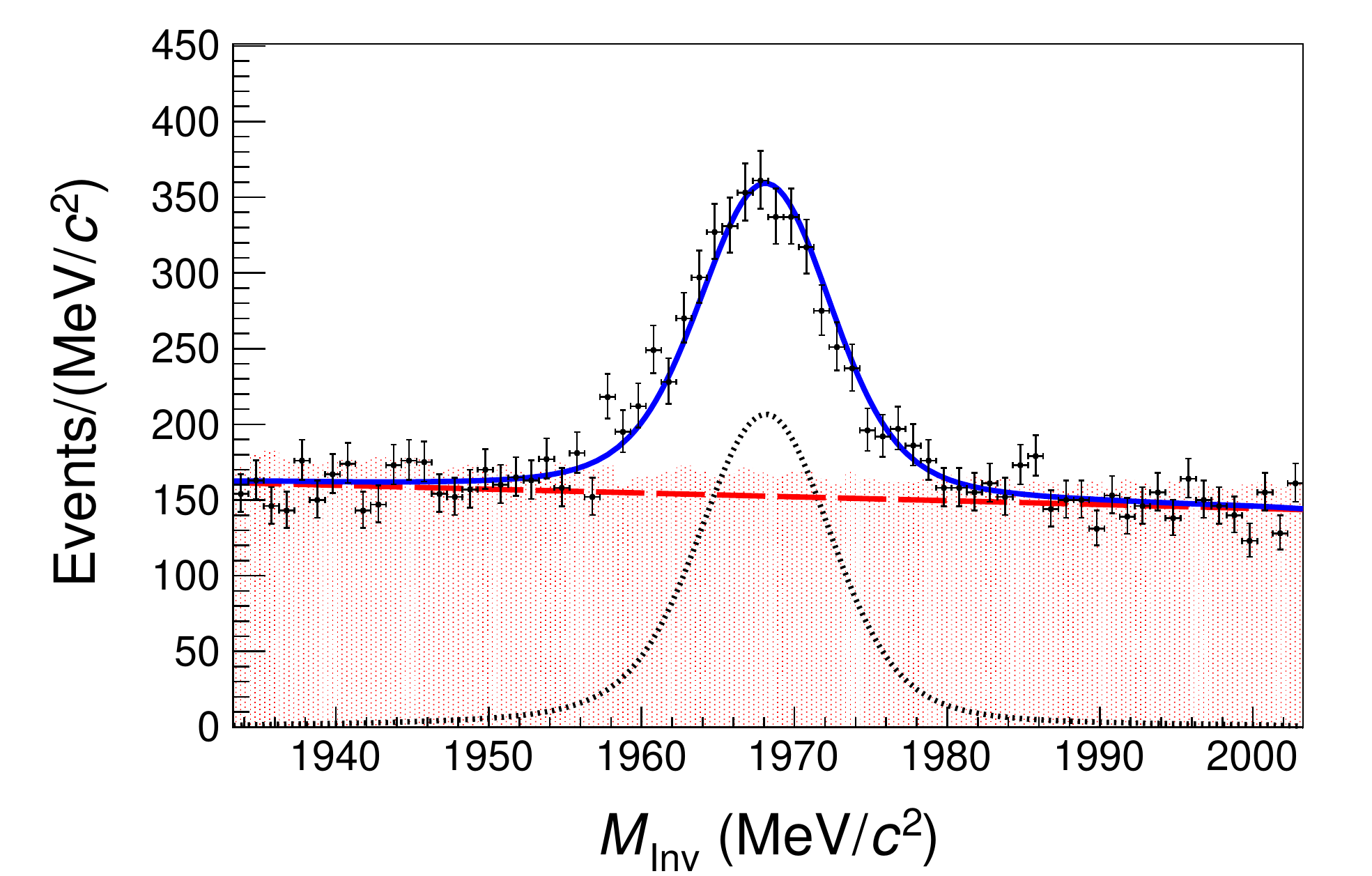} & \includegraphics[width=3in]{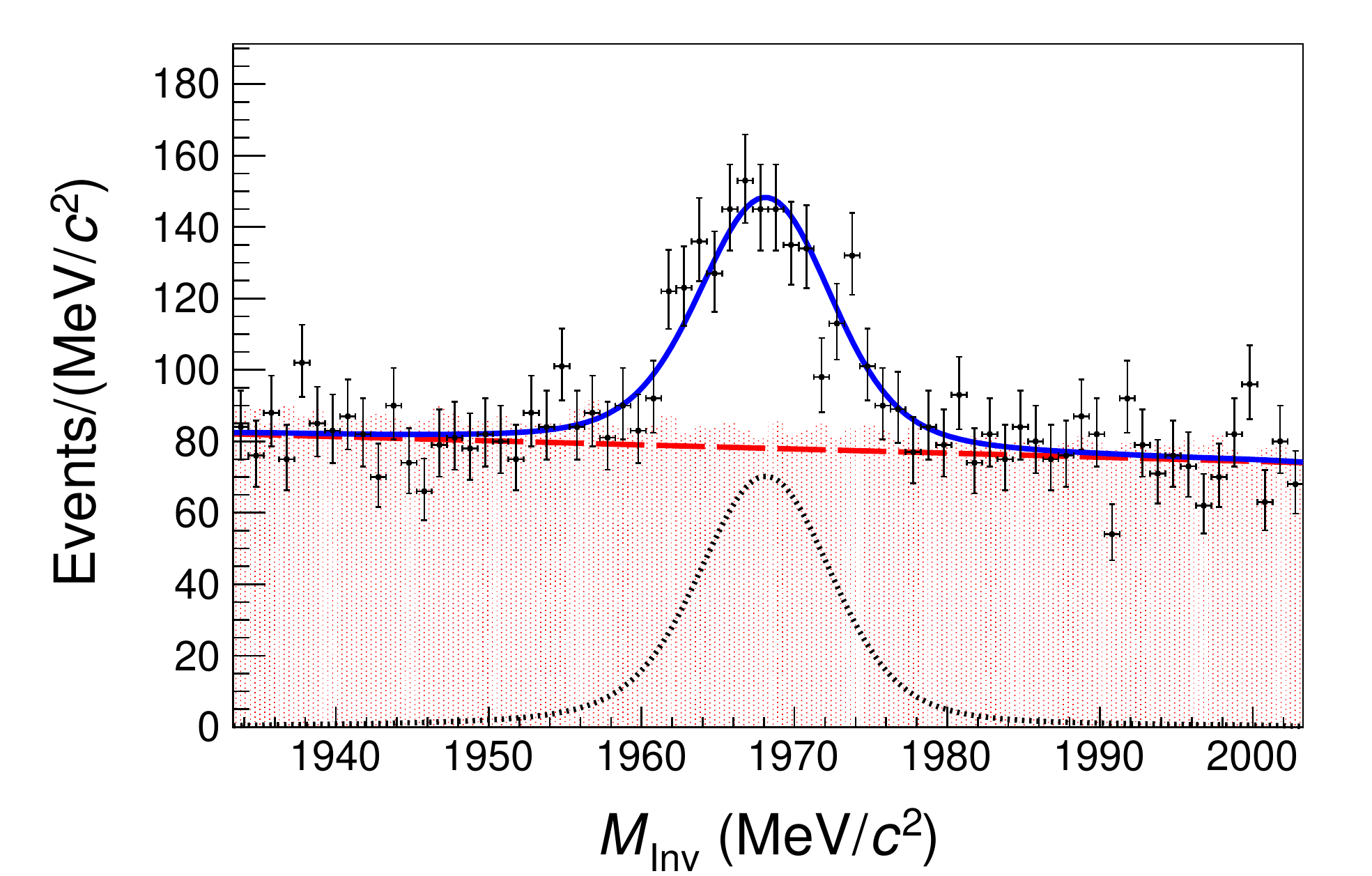}\\
\textbf{RS 500-550 MeV/$c$} & \textbf{WS 500-550 MeV/$c$}\\
\includegraphics[width=3in]{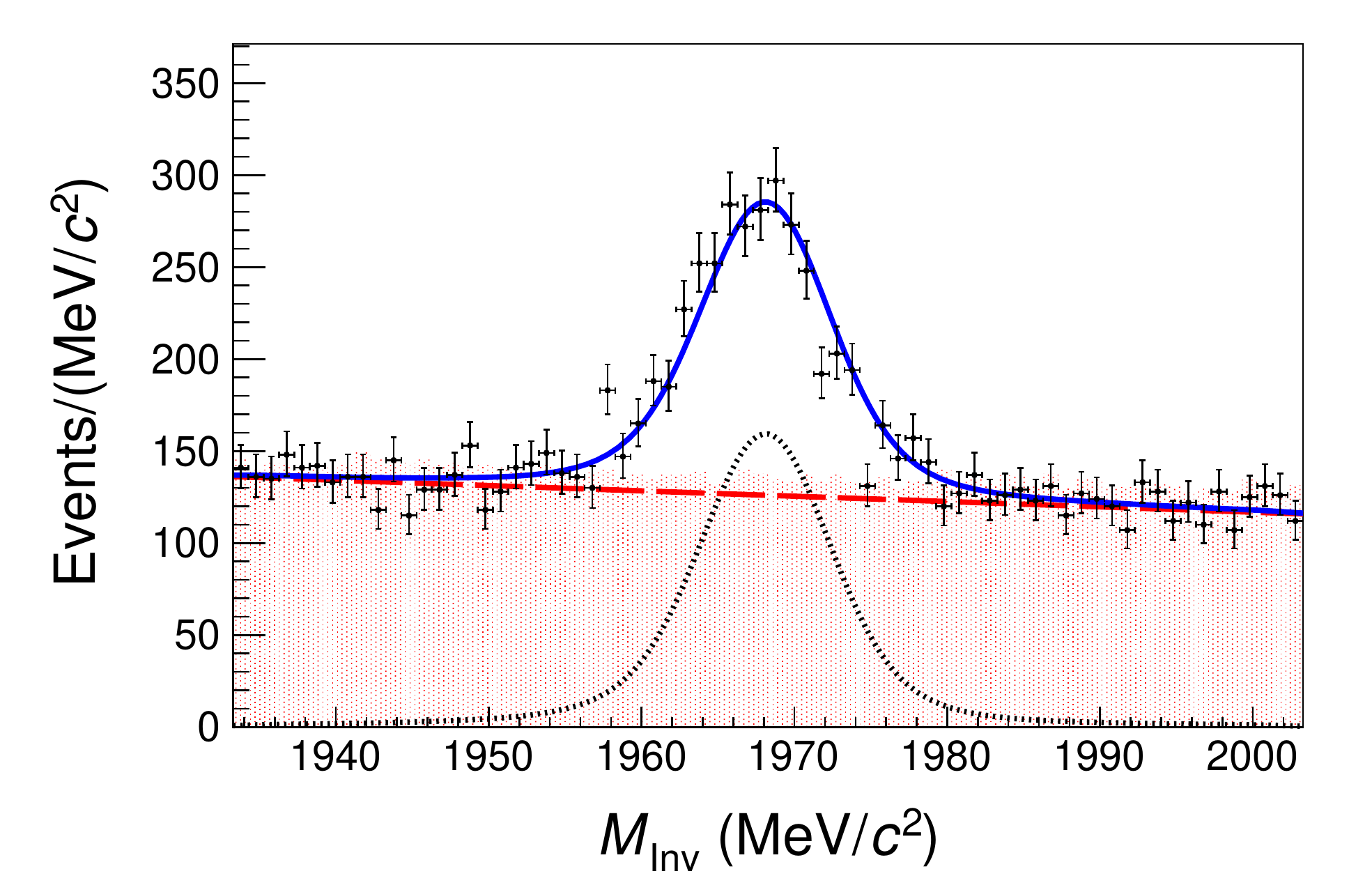} & \includegraphics[width=3in]{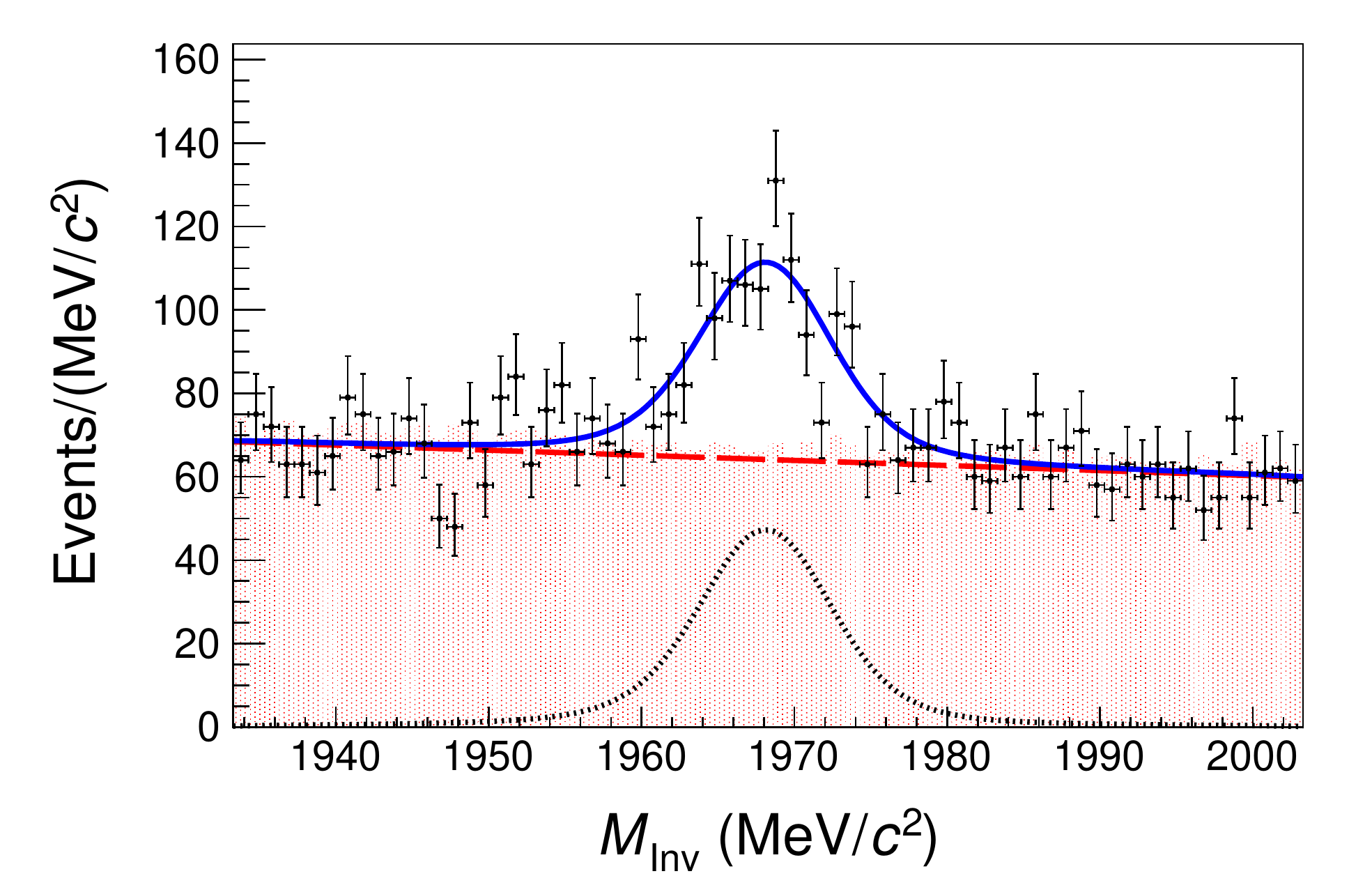}\\
\textbf{RS 550-600 MeV/$c$} & \textbf{WS 550-600 MeV/$c$}\\
\includegraphics[width=3in]{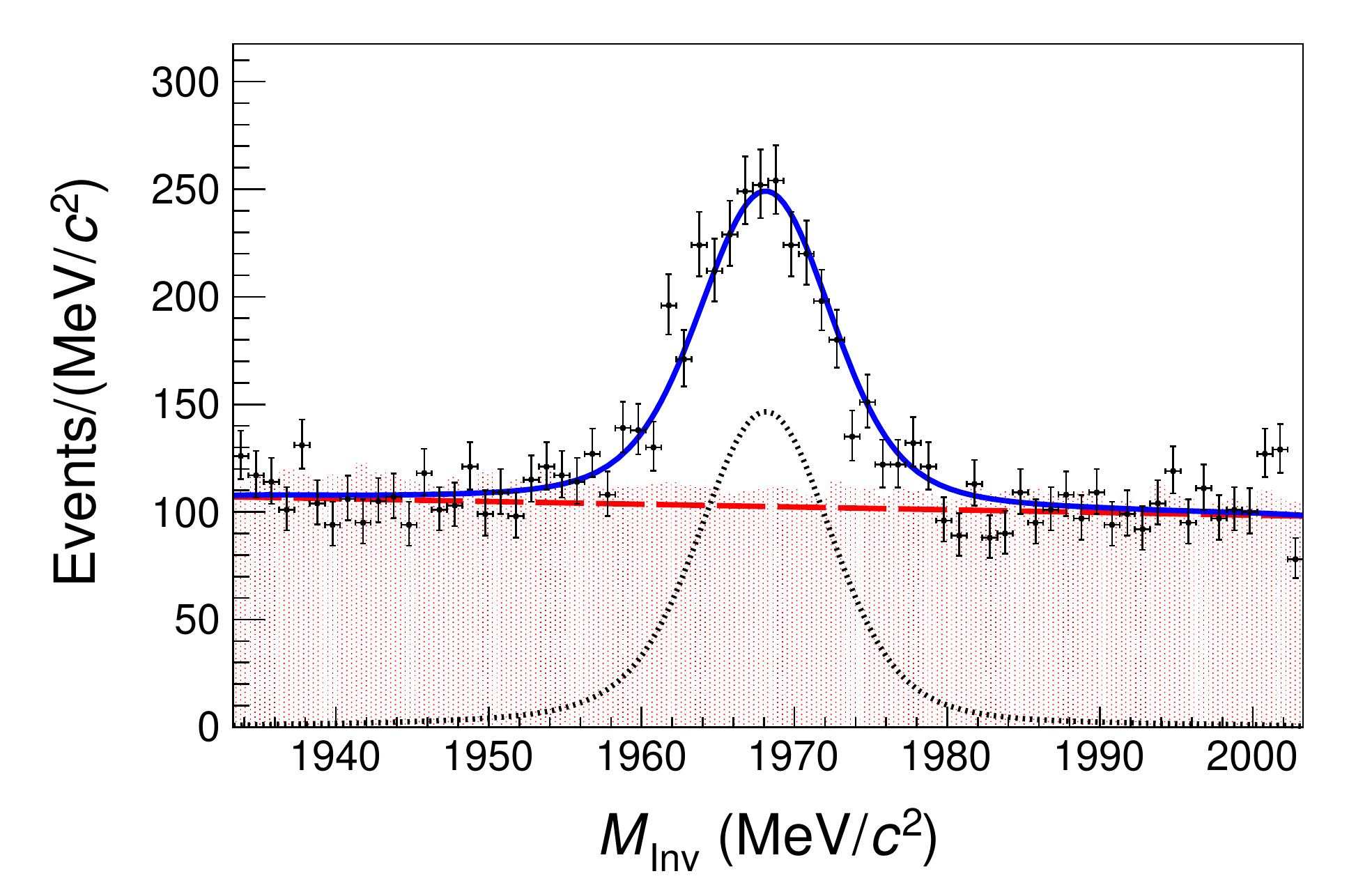} & \includegraphics[width=3in]{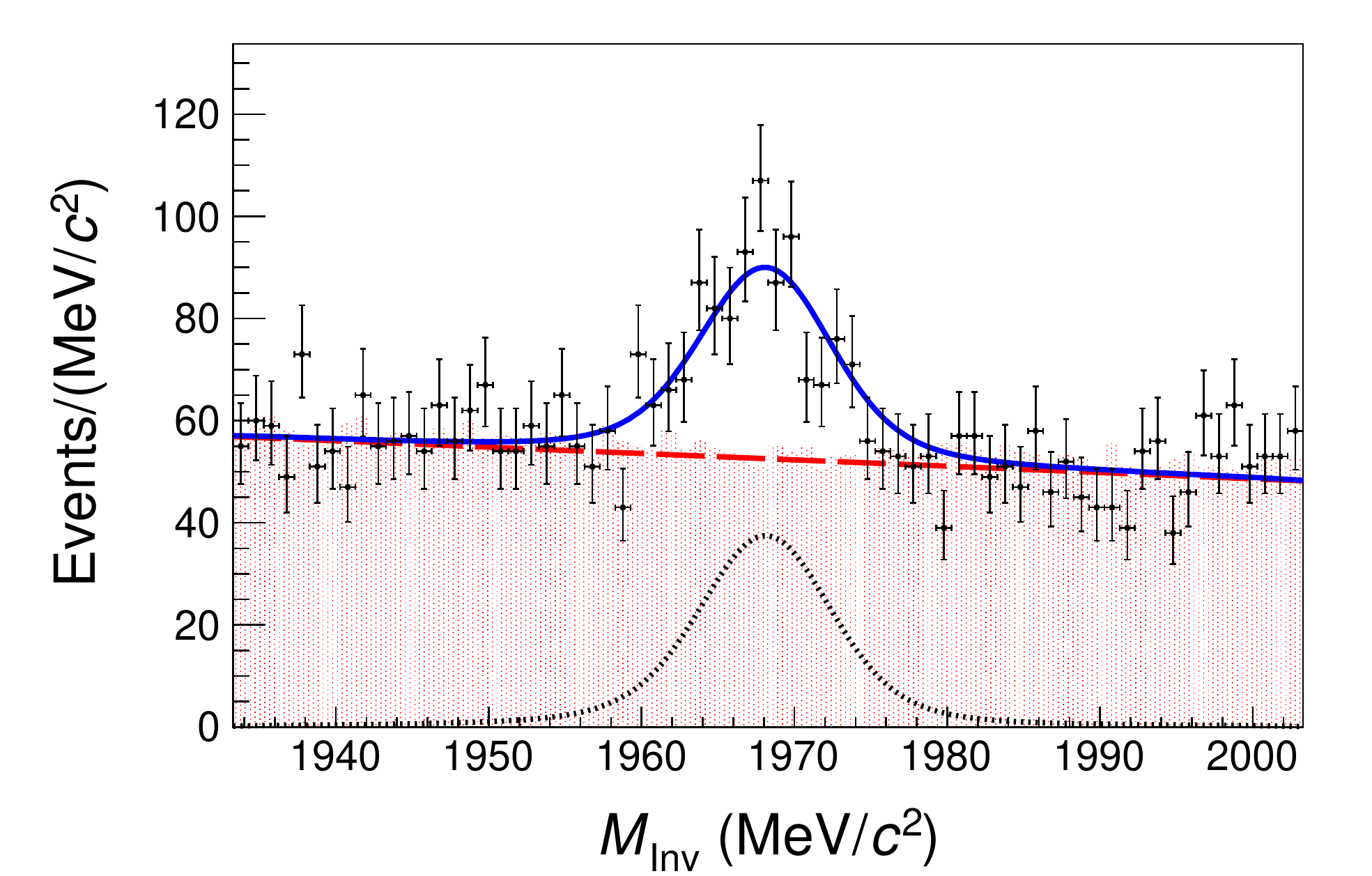}
\end{tabular}

\begin{tabular}{cc}
\textbf{RS 600-650 MeV/$c$} & \textbf{WS 600-650 MeV/$c$}\\
\includegraphics[width=3in]{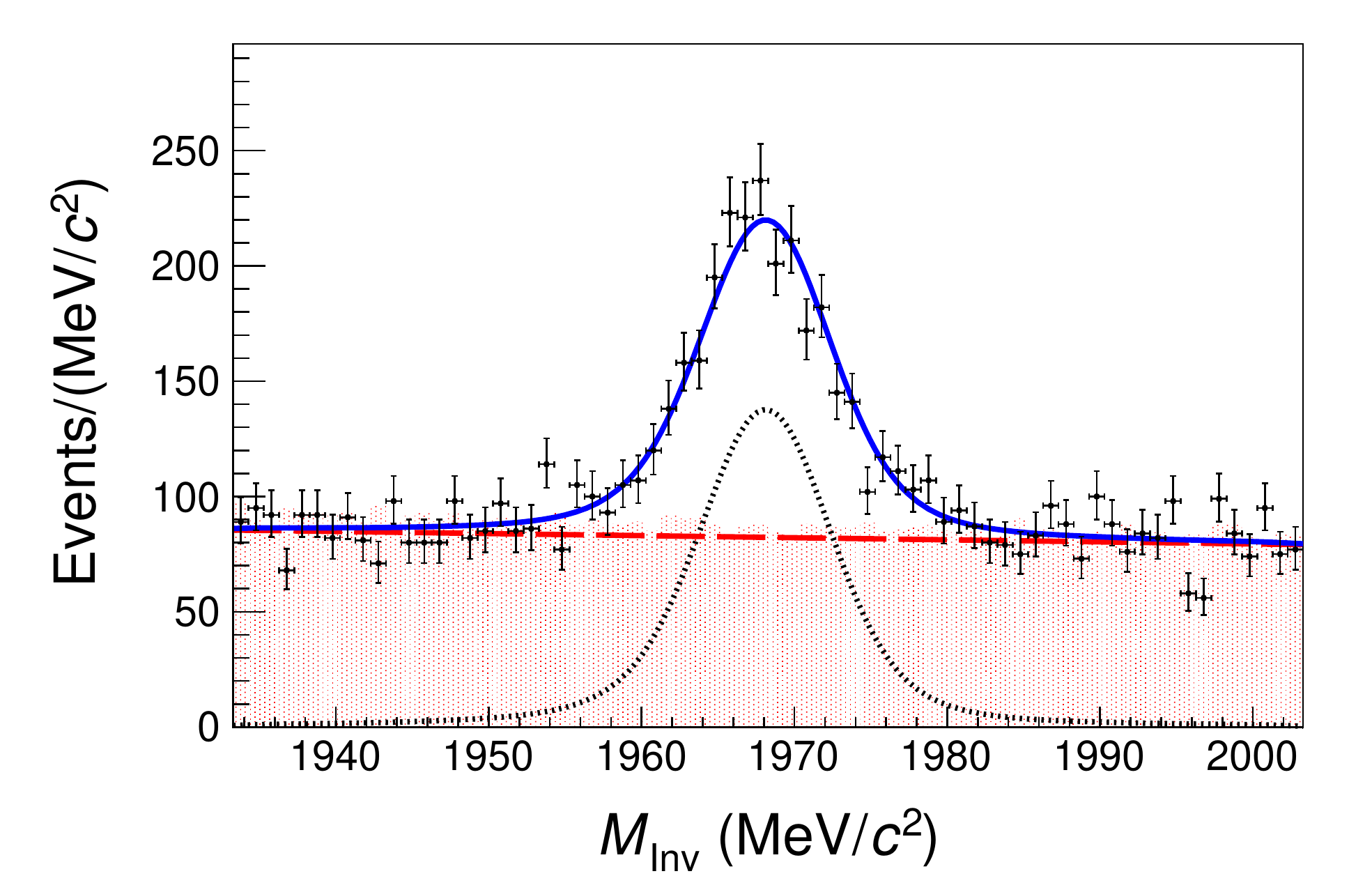} & \includegraphics[width=3in]{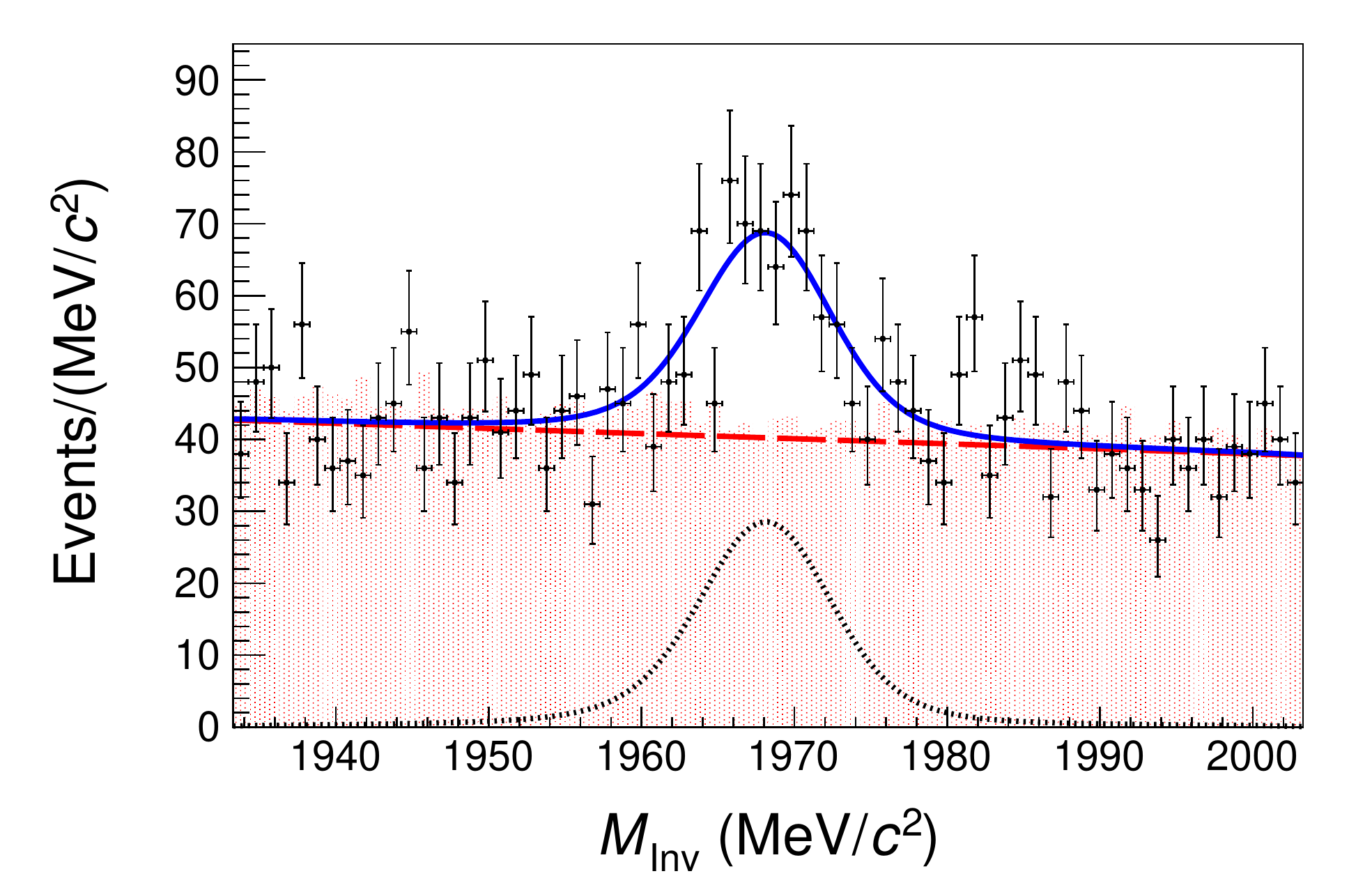}\\
\textbf{RS 650-700 MeV/$c$} & \textbf{WS 650-700 MeV/$c$}\\
\includegraphics[width=3in]{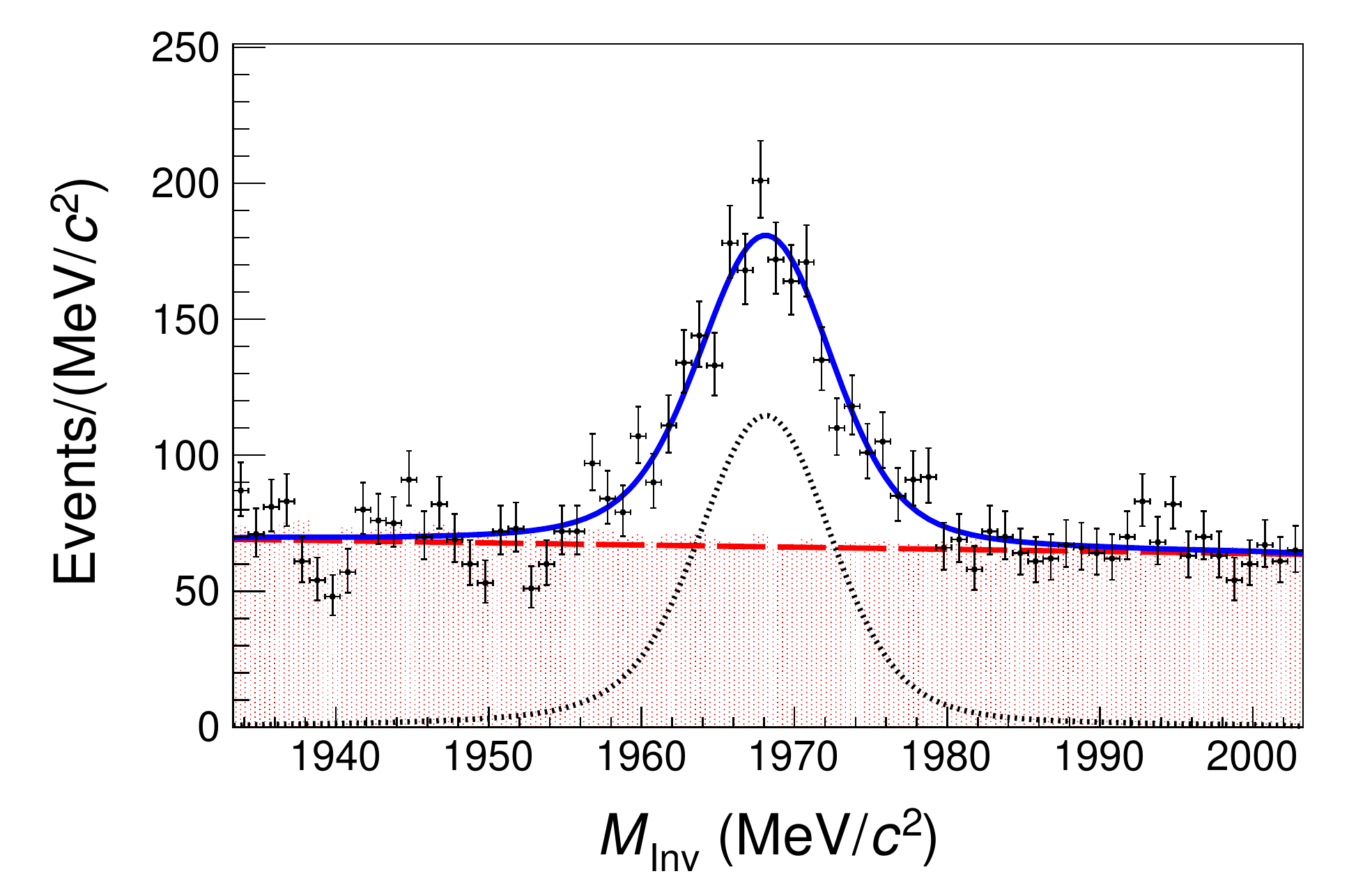} & \includegraphics[width=3in]{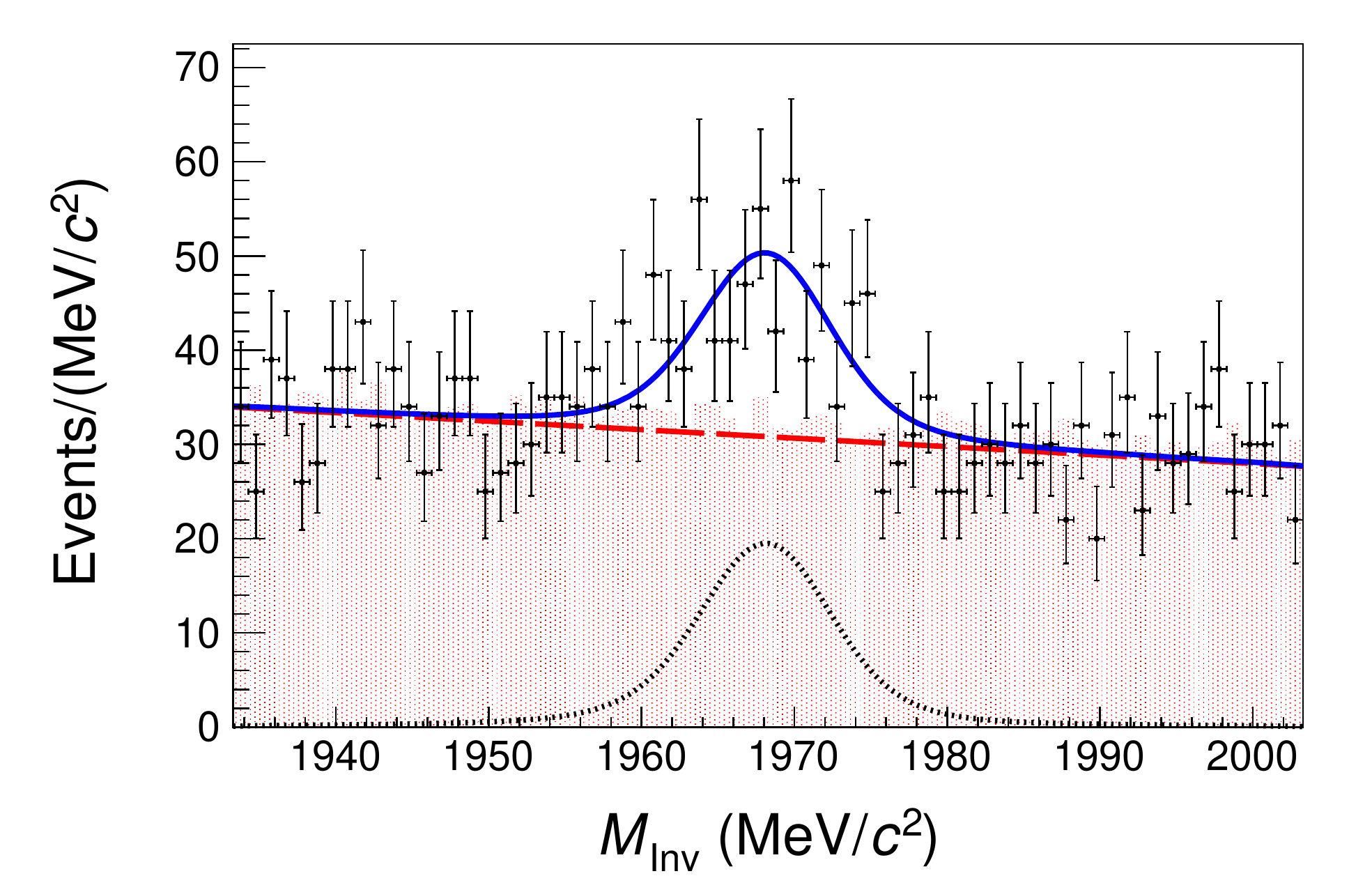}\\
\textbf{RS 700-750 MeV/$c$} & \textbf{WS 700-750 MeV/$c$}\\
\includegraphics[width=3in]{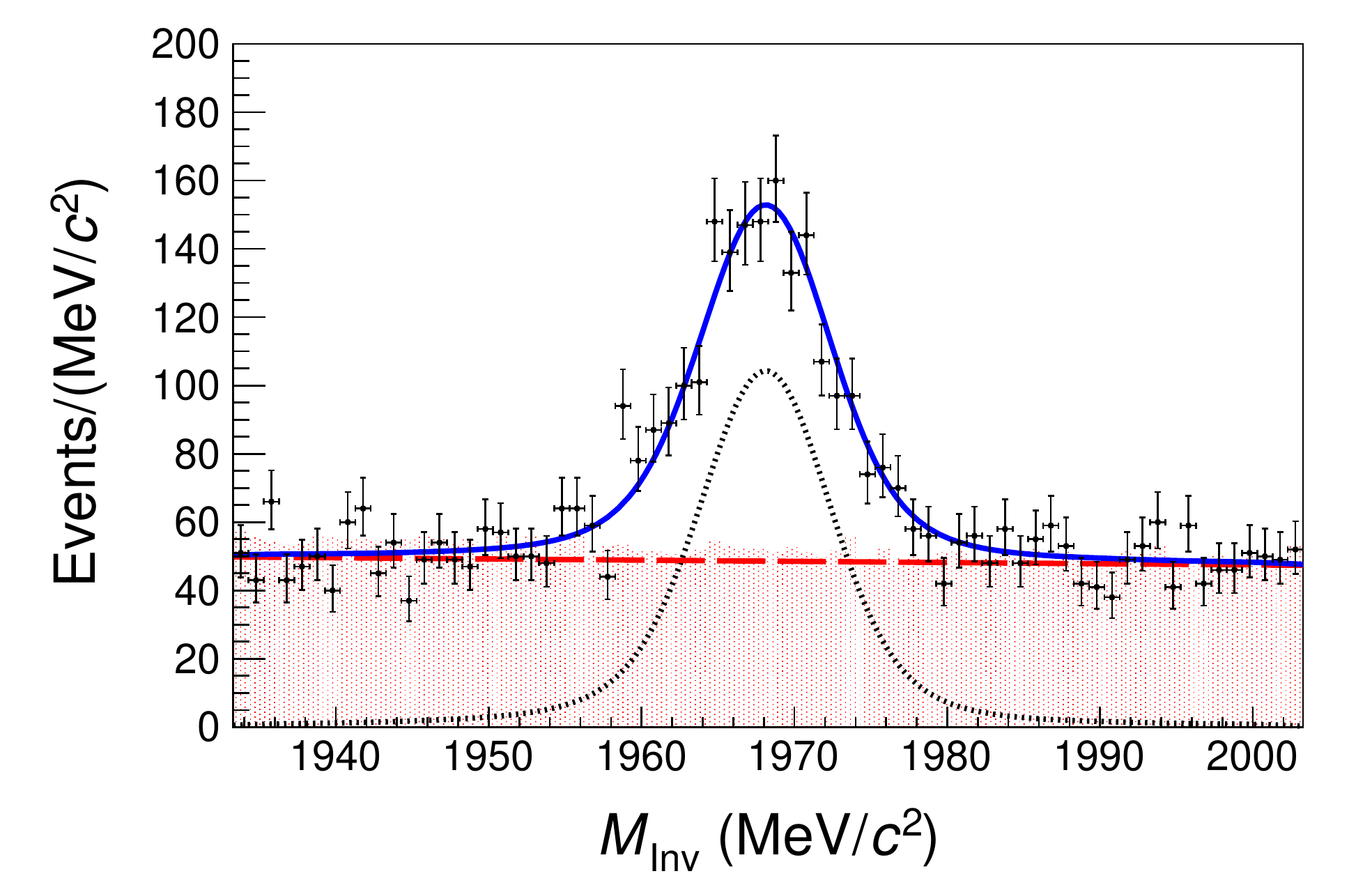} & \includegraphics[width=3in]{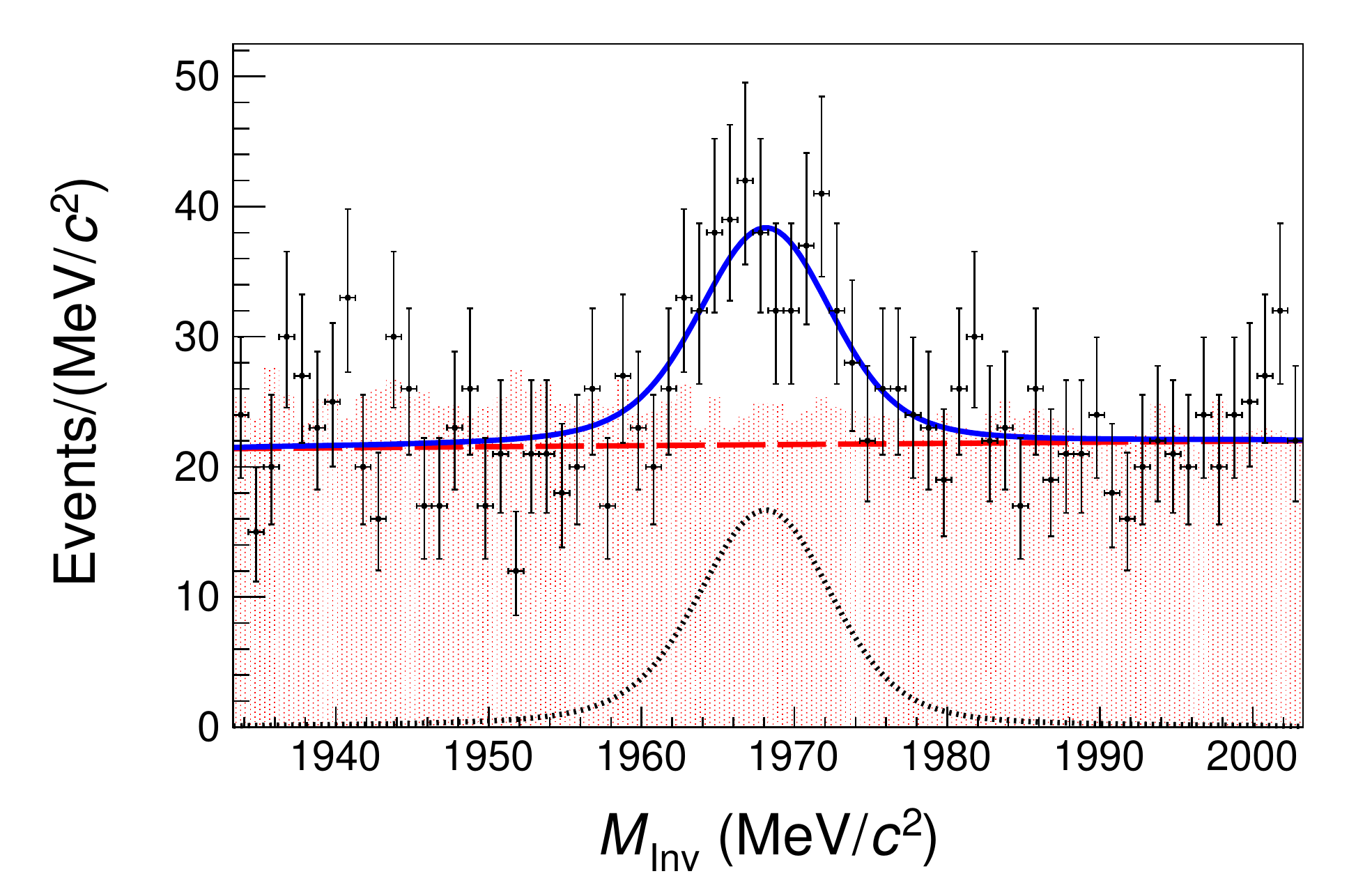}\\
\textbf{RS 750-800 MeV/$c$} & \textbf{WS 750-800 MeV/$c$}\\
\includegraphics[width=3in]{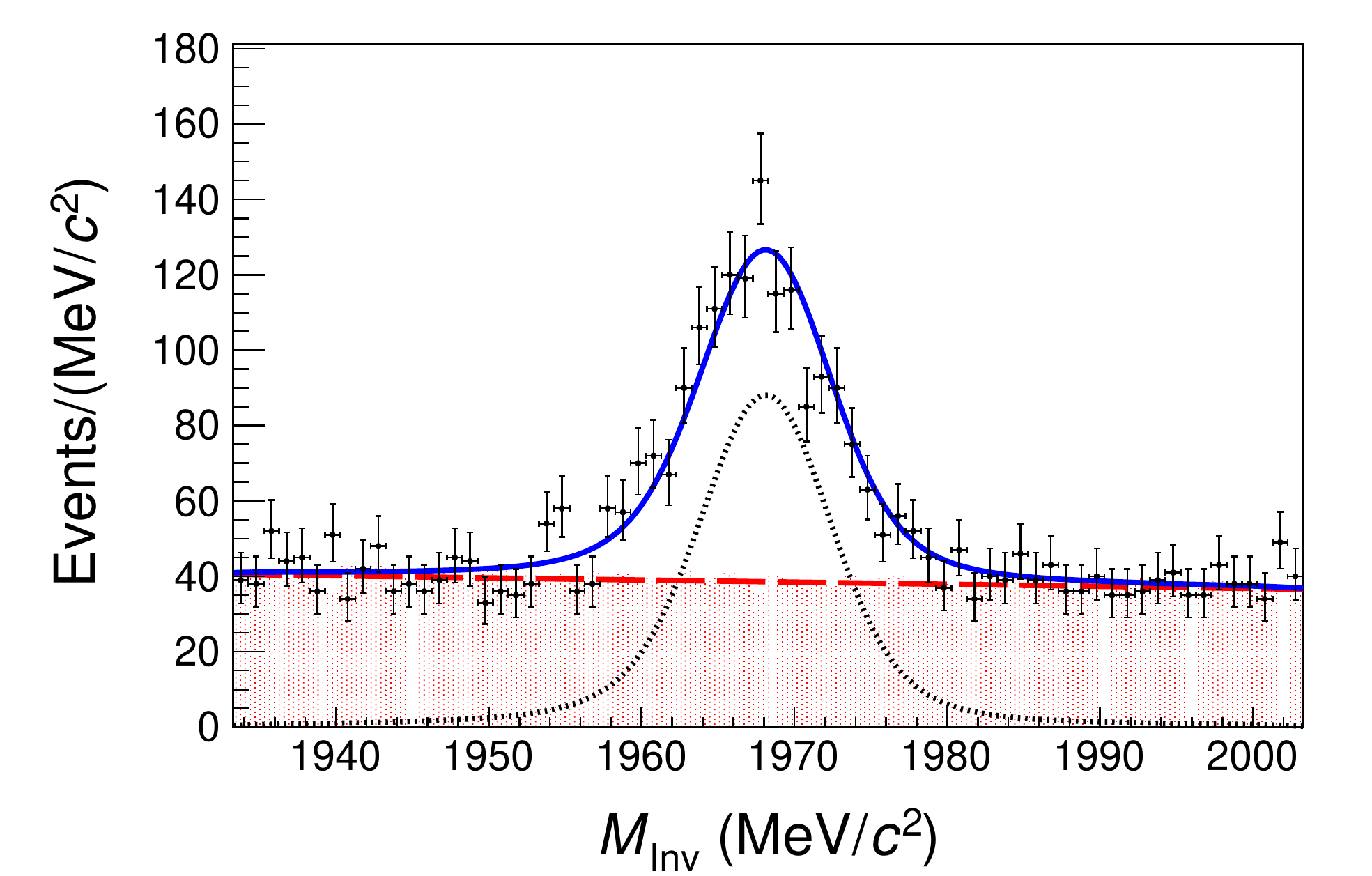} & \includegraphics[width=3in]{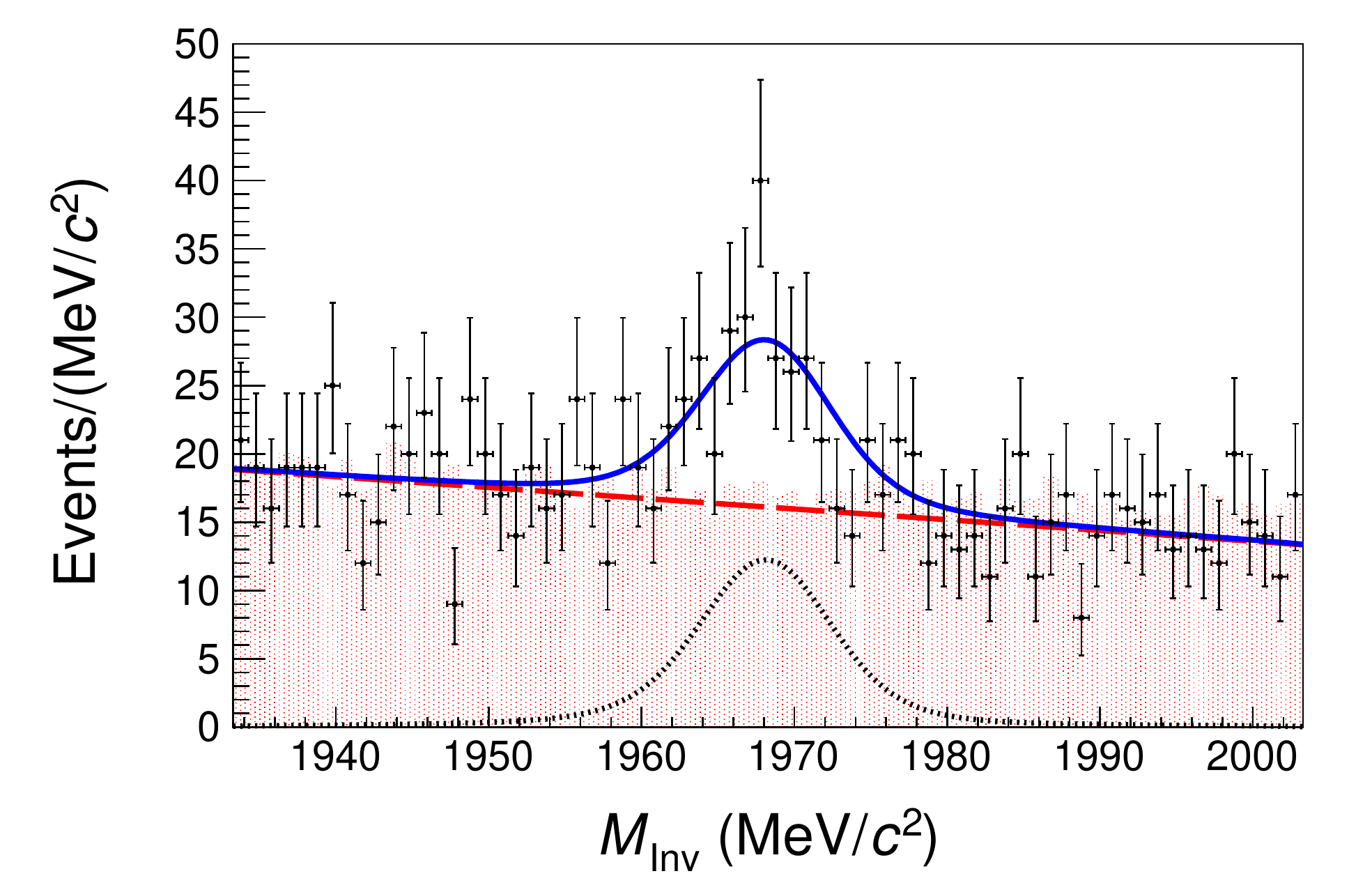}\\
\end{tabular}

\begin{tabular}{cc}
\textbf{RS 800-850 MeV/$c$} & \textbf{WS 800-850 MeV/$c$}\\
\includegraphics[width=3in]{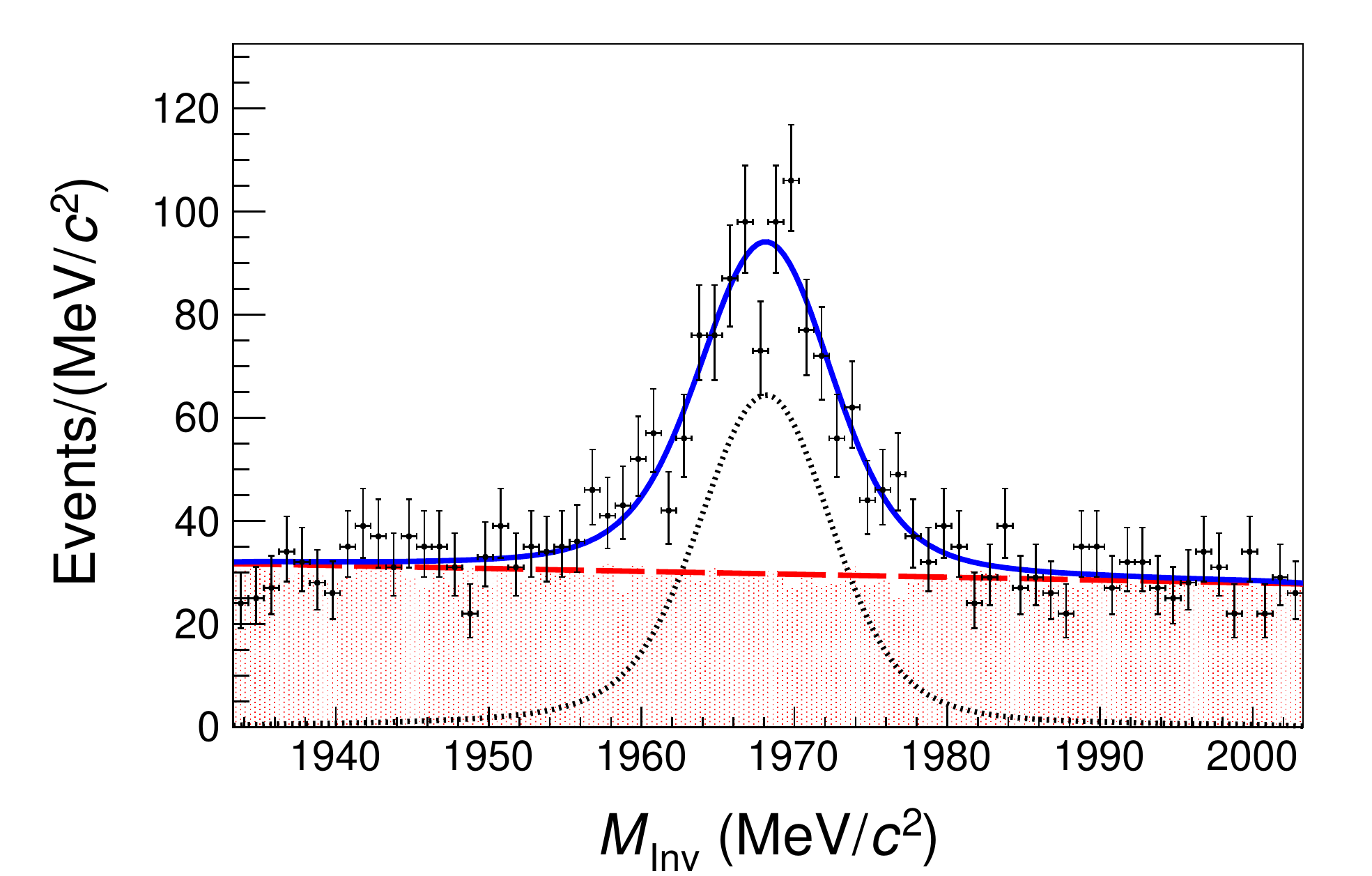} & \includegraphics[width=3in]{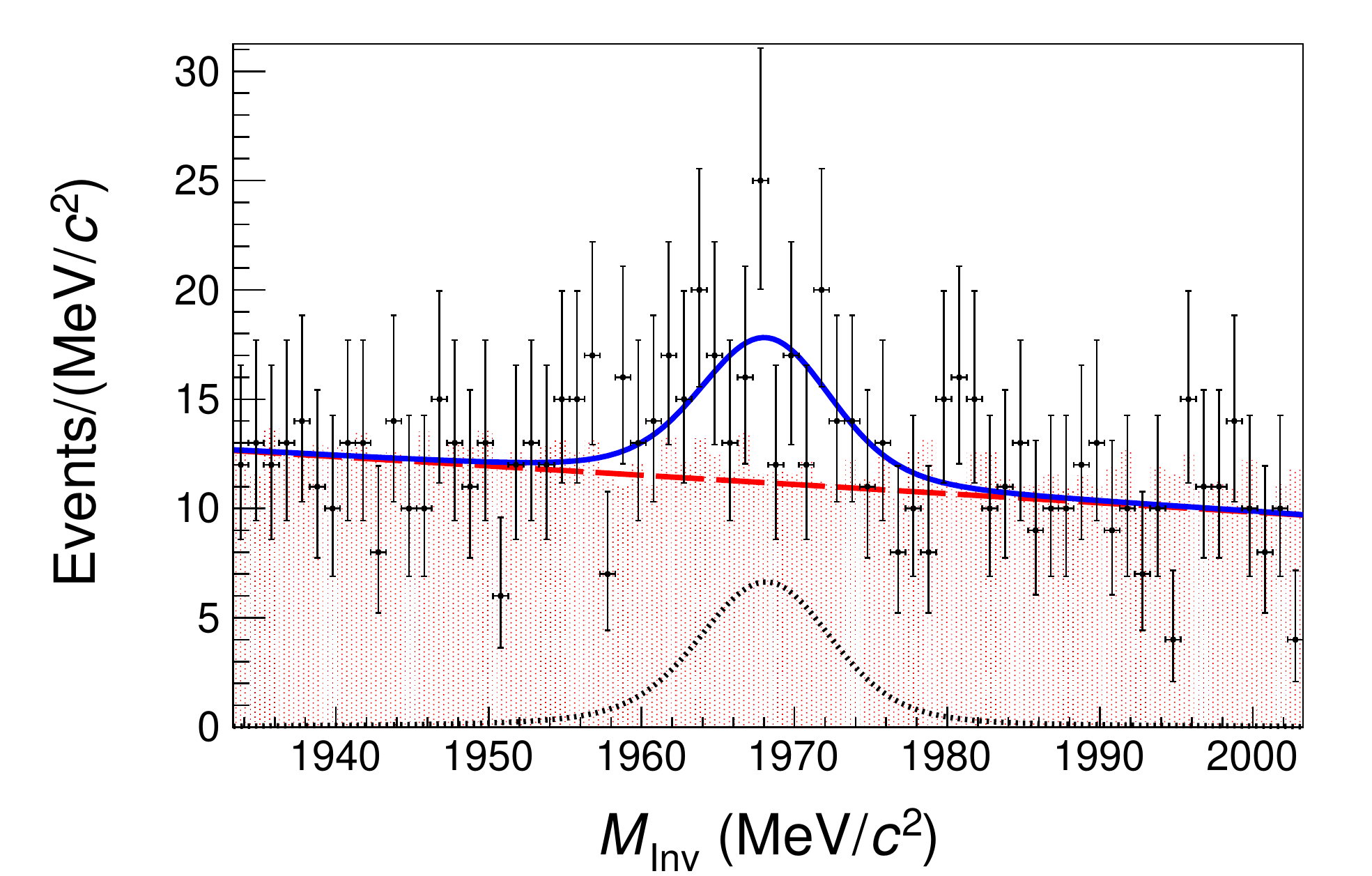}\\
\textbf{RS 850-900 MeV/$c$} & \textbf{WS 850-900 MeV/$c$}\\
\includegraphics[width=3in]{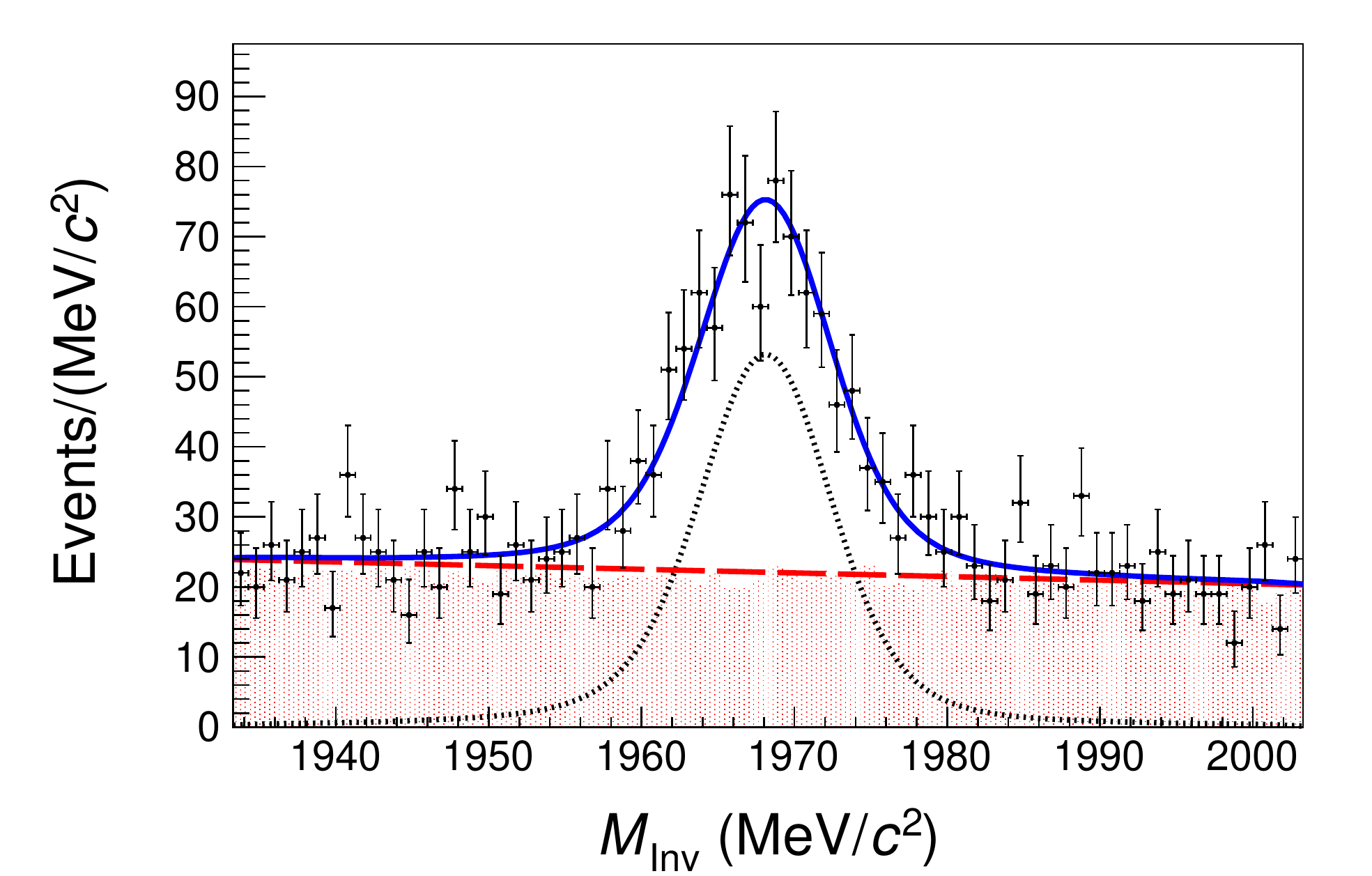} & \includegraphics[width=3in]{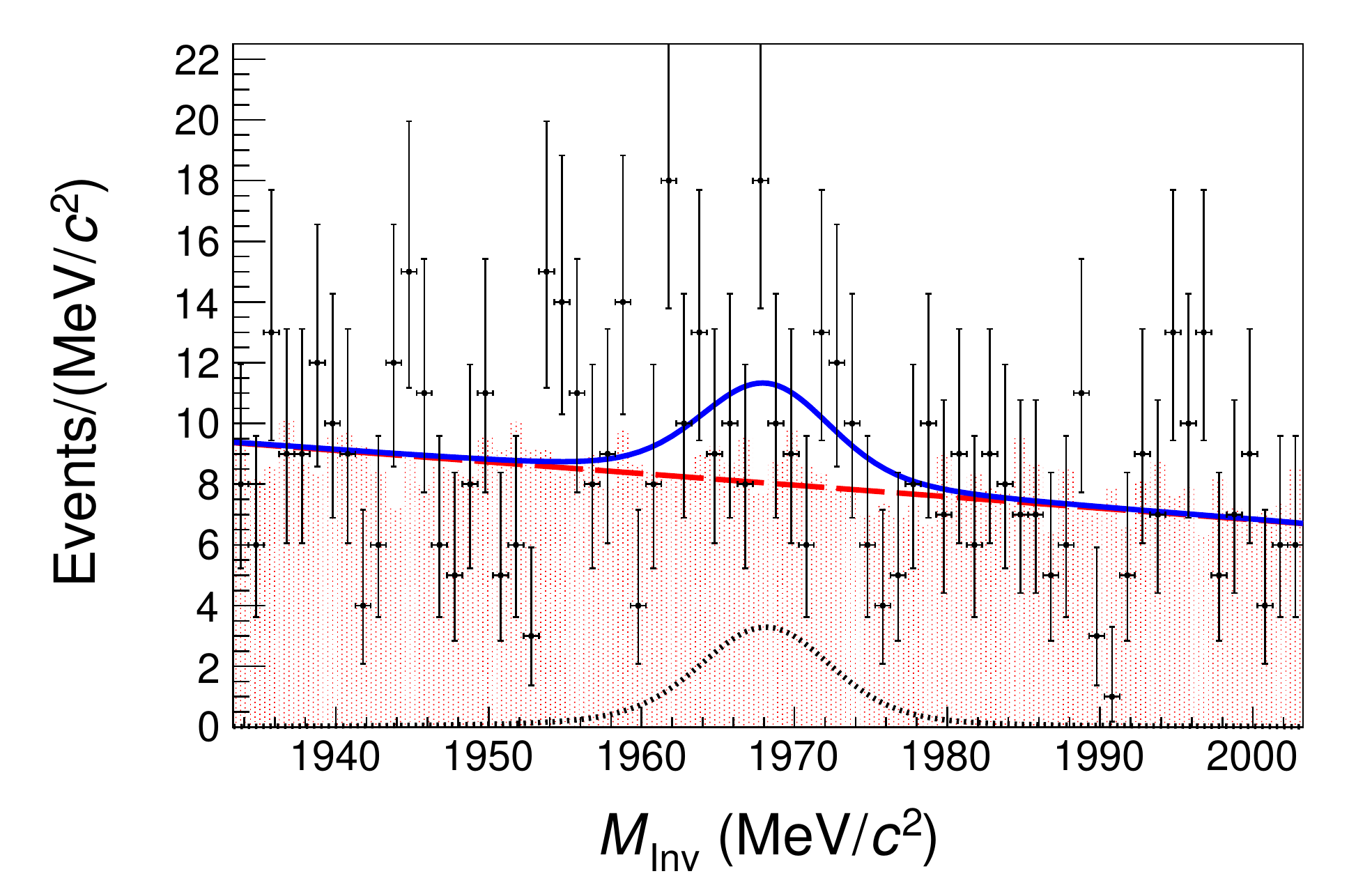}\\
\textbf{RS 900-950 MeV/$c$} & \textbf{WS 900-950 MeV/$c$}\\
\includegraphics[width=3in]{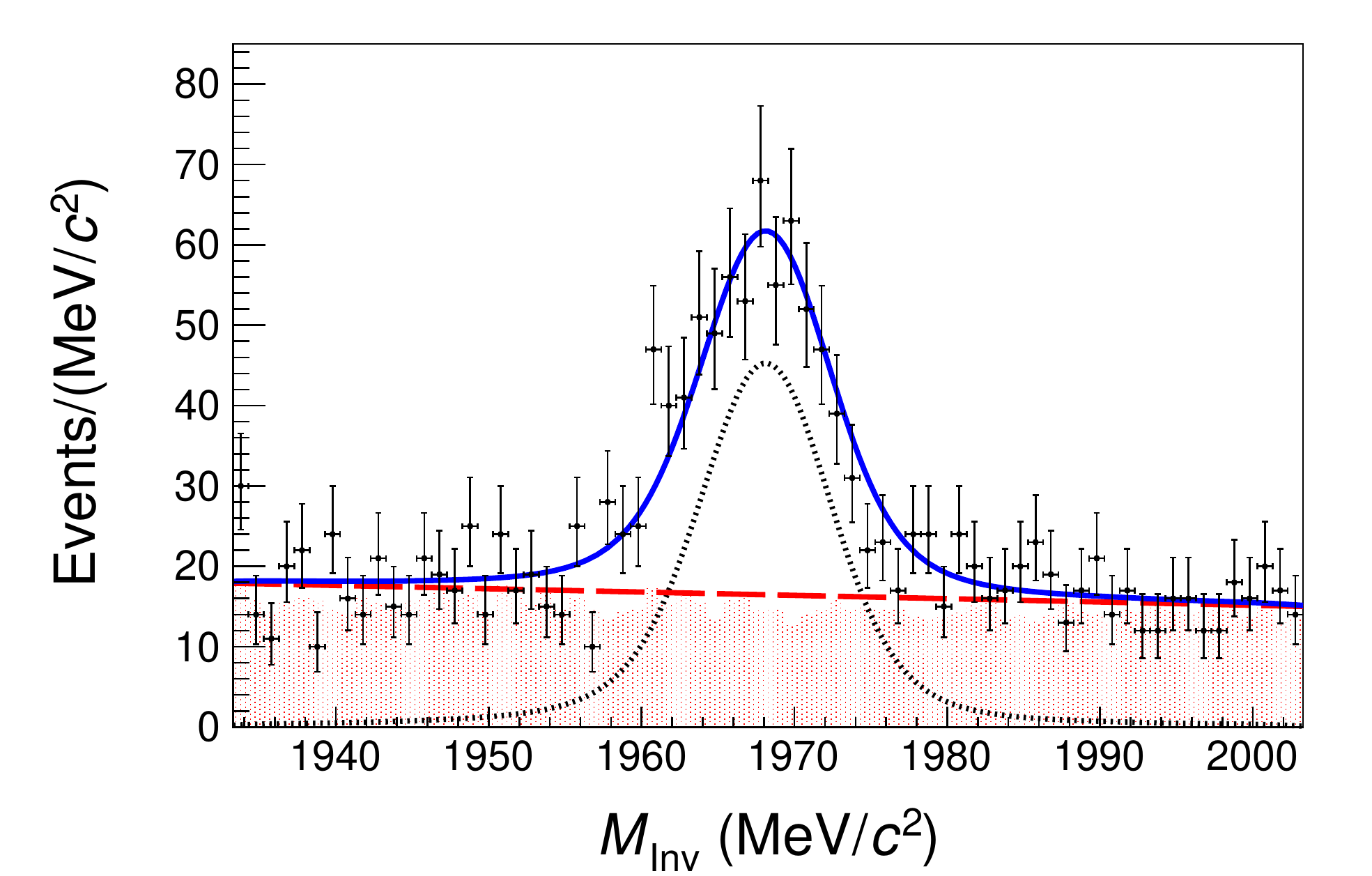} & \includegraphics[width=3in]{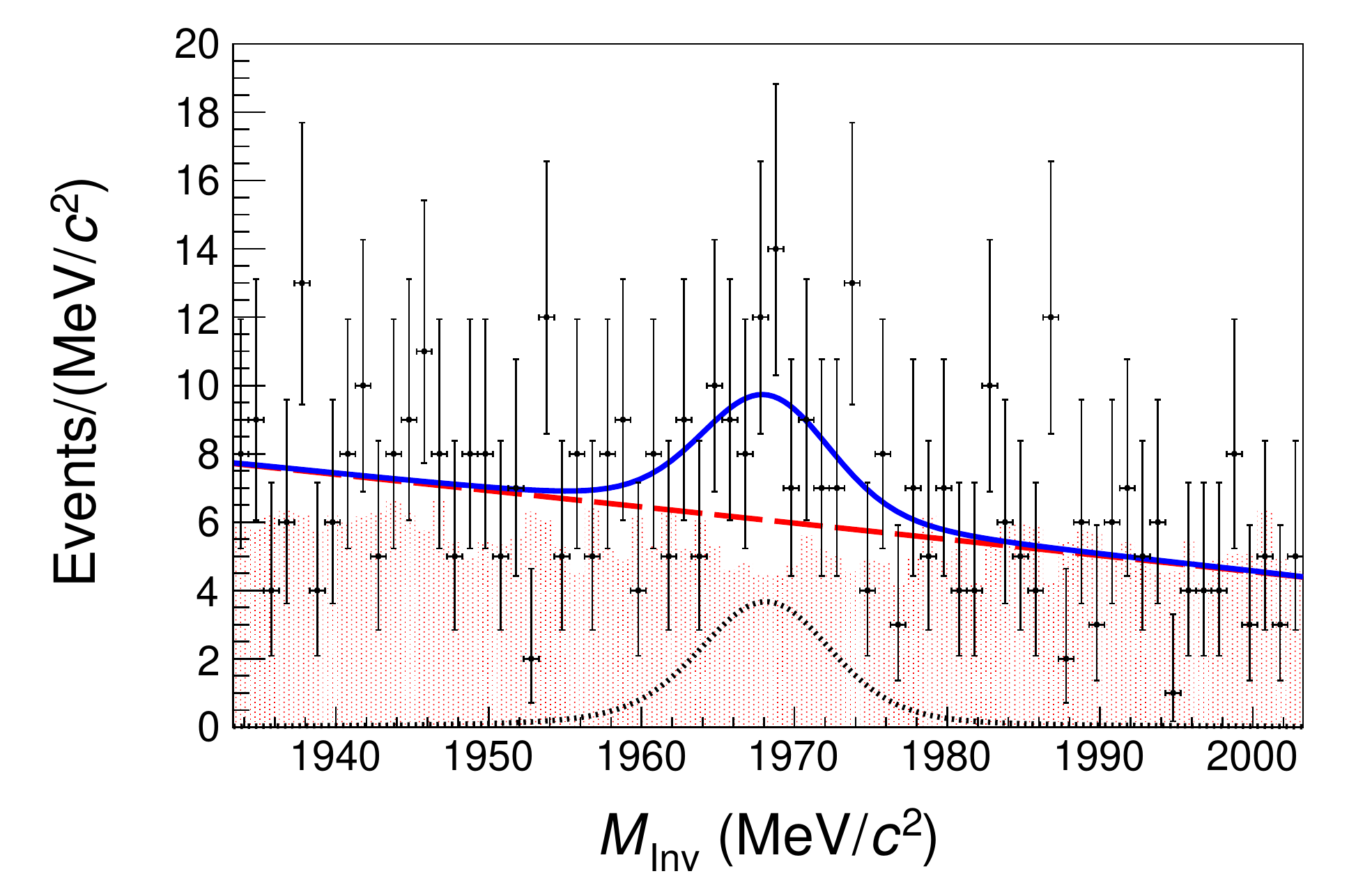}\\
\textbf{RS 950-1000 MeV/$c$} & \textbf{WS 950-1000 MeV/$c$}\\
\includegraphics[width=3in]{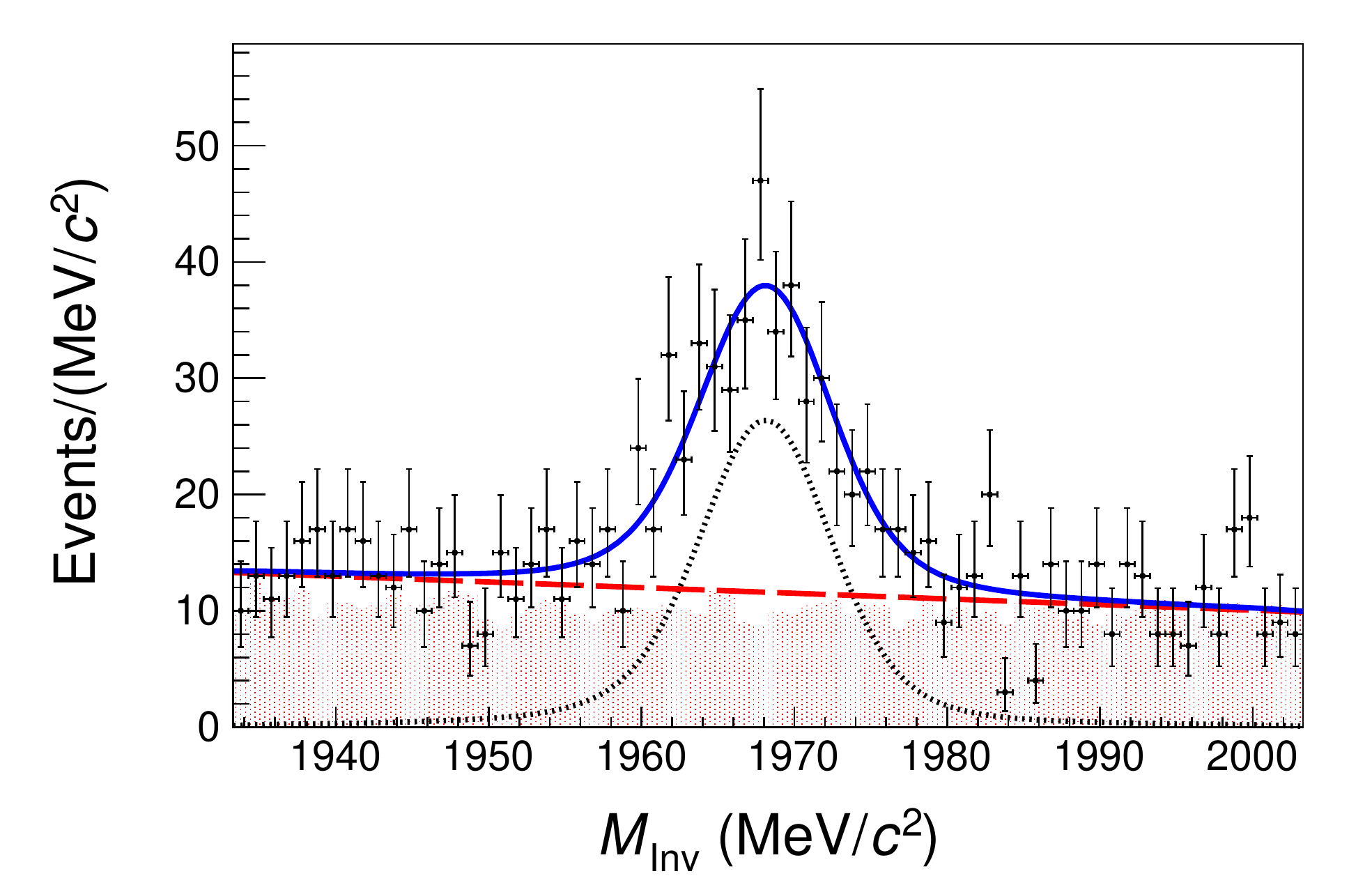} & \includegraphics[width=3in]{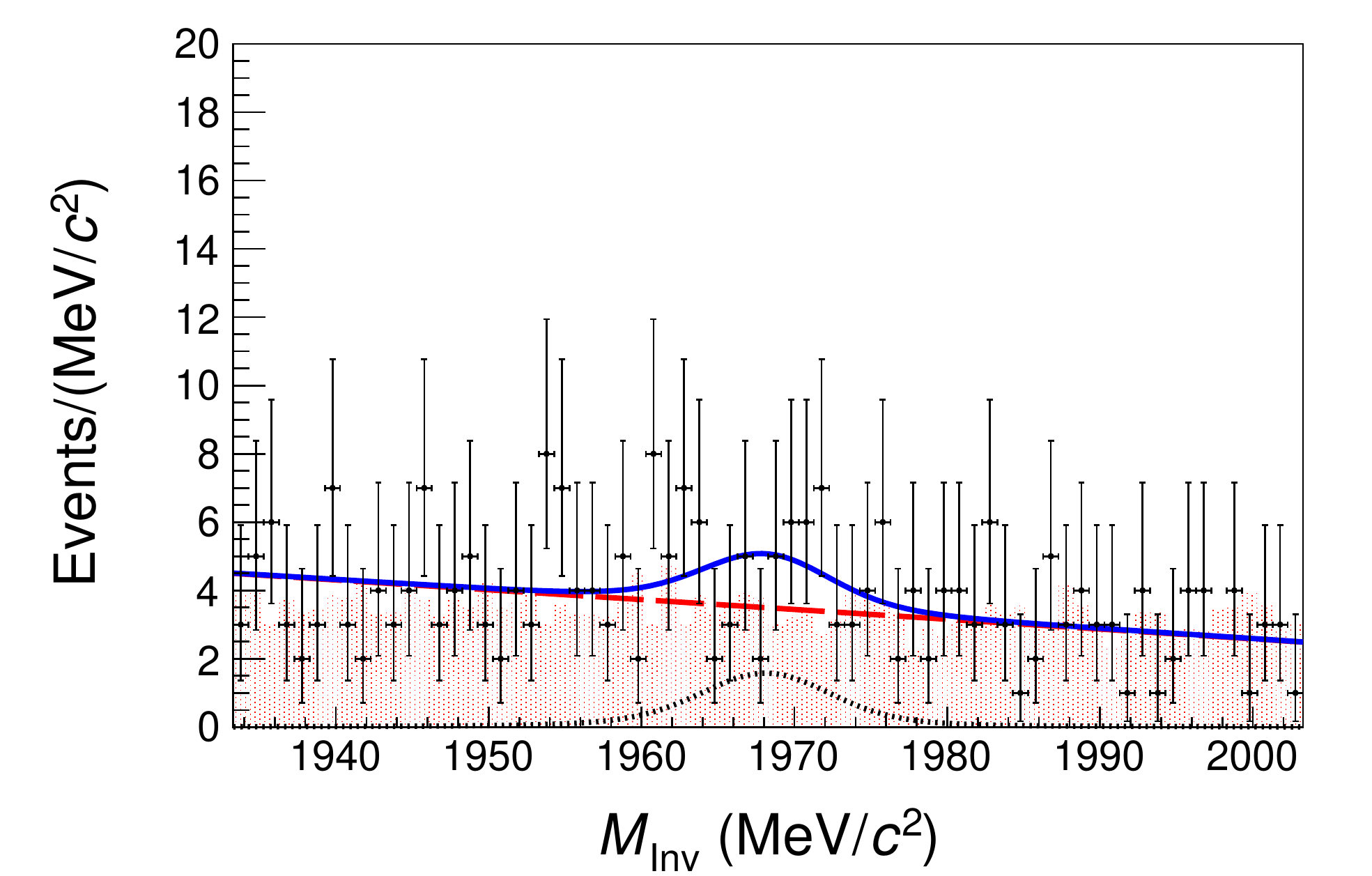}\\

\end{tabular}

\begin{tabular}{cc}
\textbf{RS 1000-1050 MeV/$c$} & \textbf{WS 1000-1050 MeV/$c$}\\
\includegraphics[width=3in]{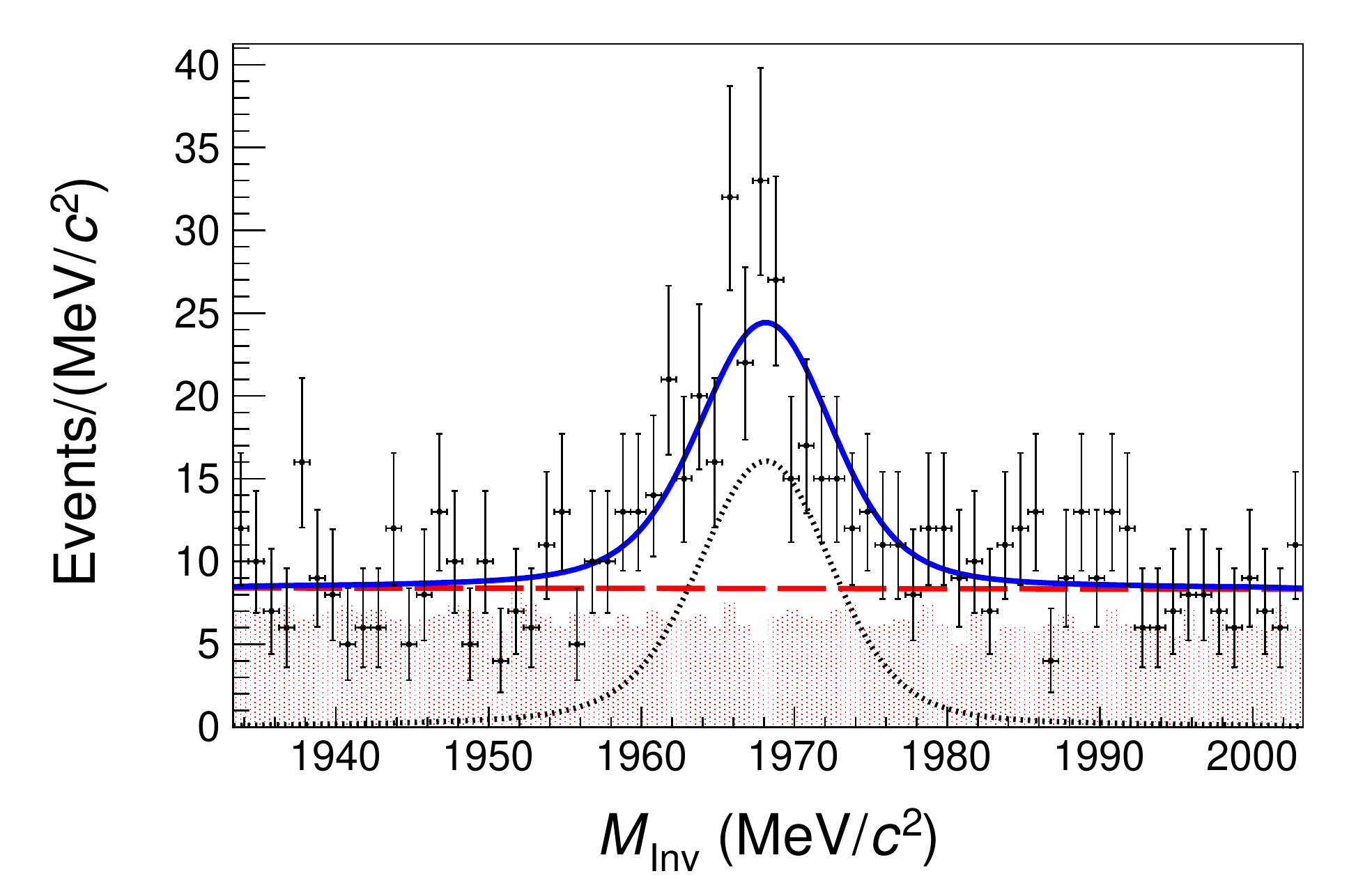} & \includegraphics[width=3in]{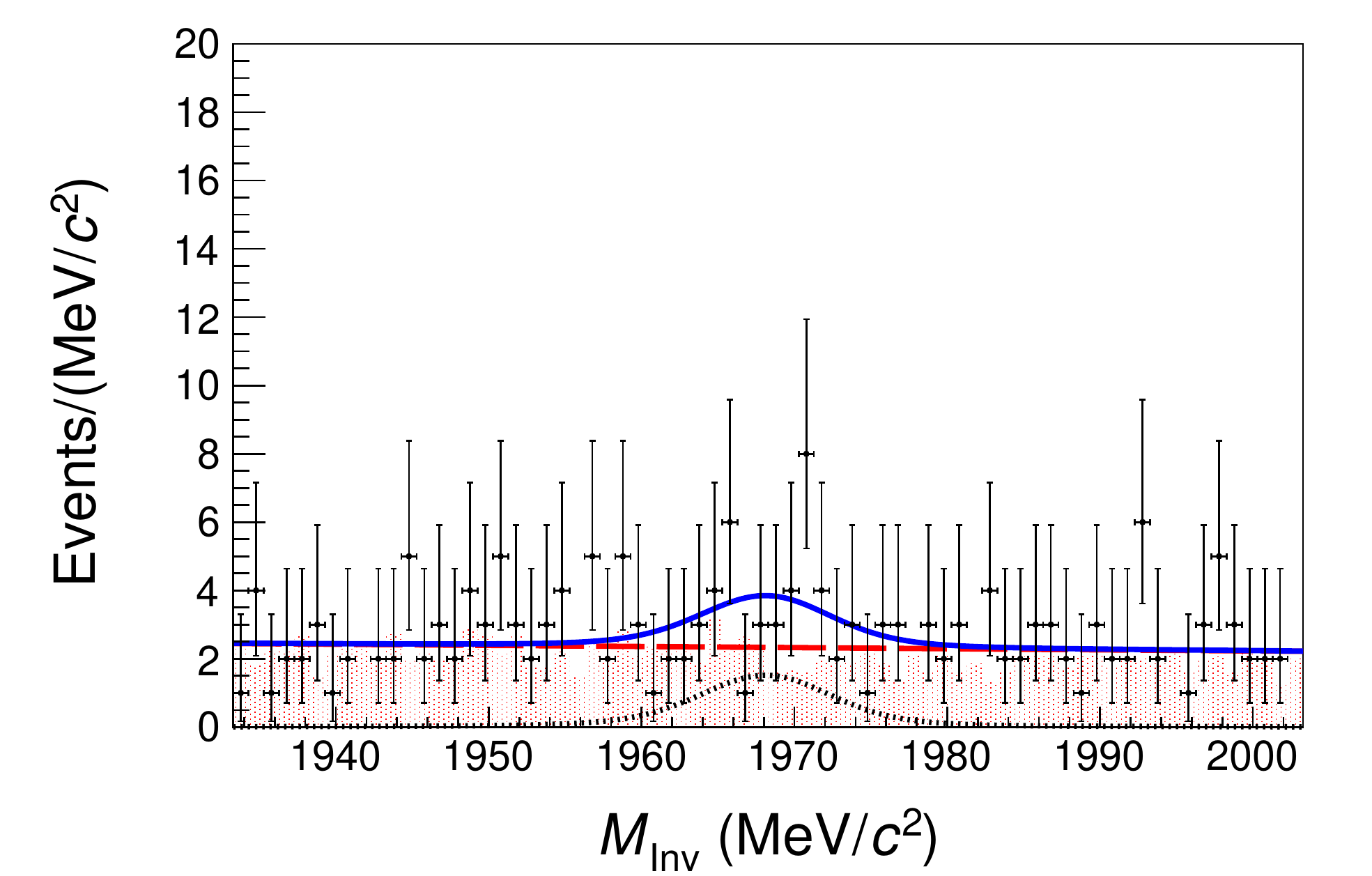}\\
\textbf{RS 1050-1100 MeV/$c$} & \textbf{WS 1050-1100 MeV/$c$}\\
\includegraphics[width=3in]{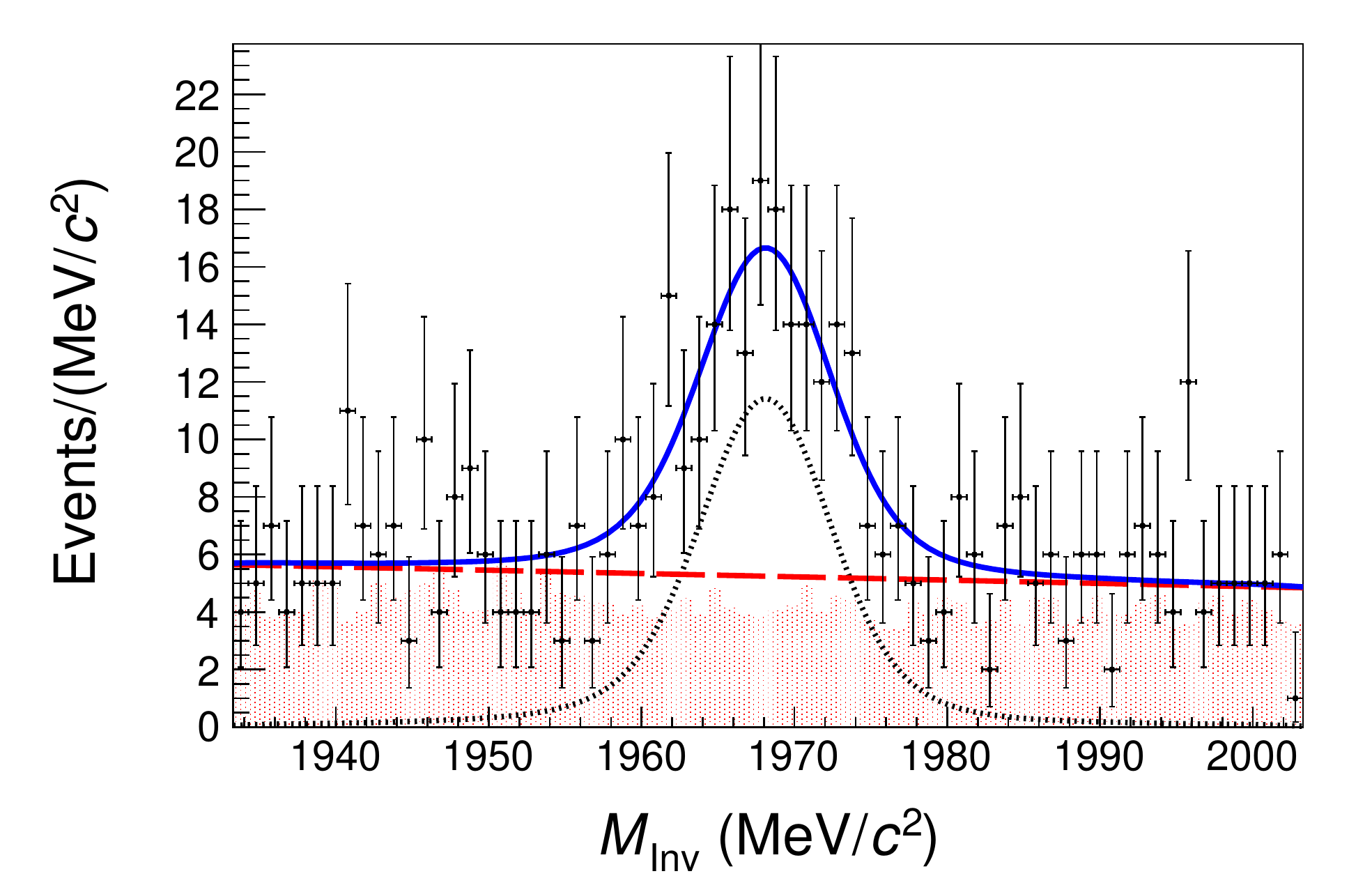} & \includegraphics[width=3in]{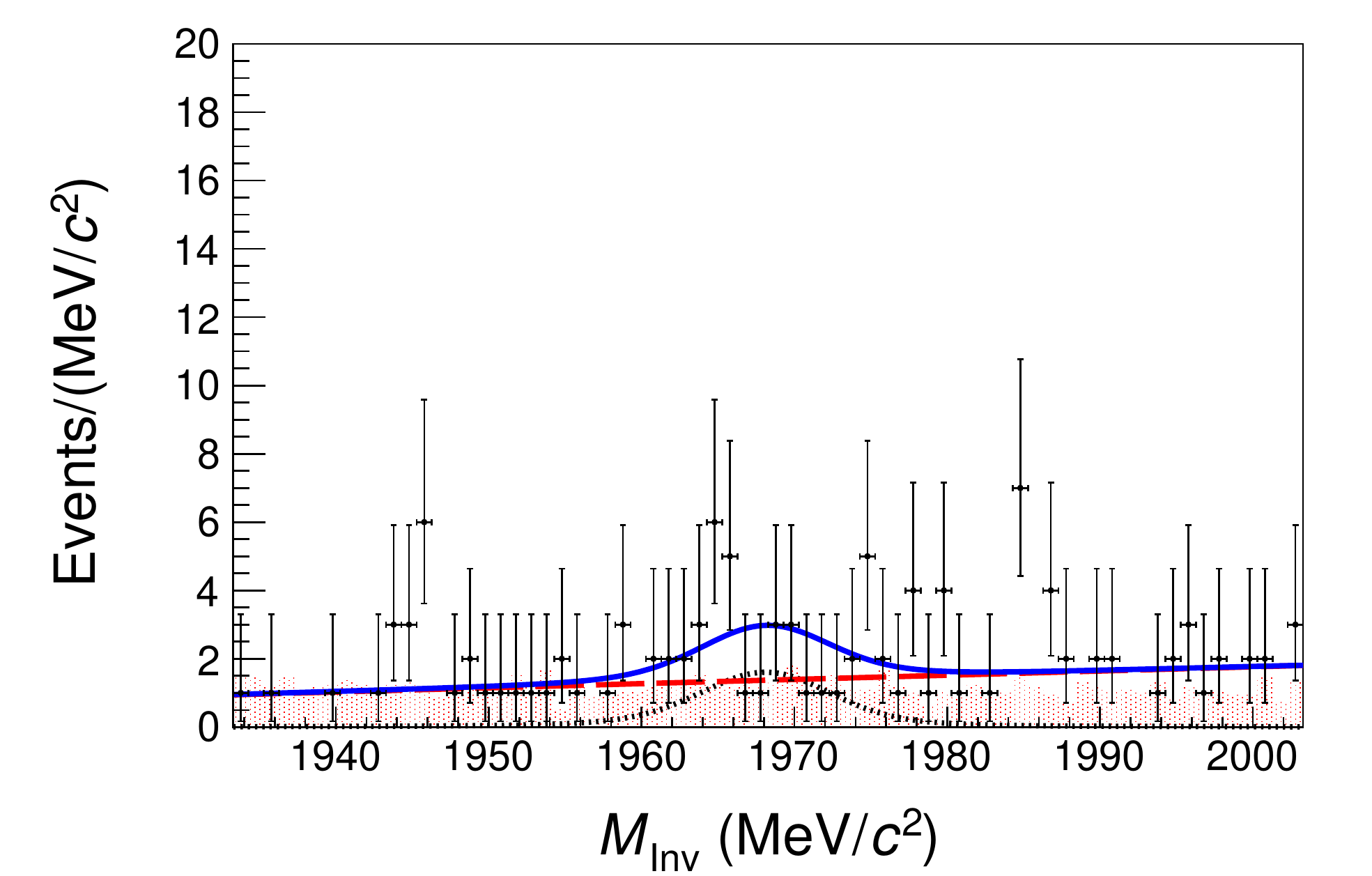}
\end{tabular}

\pagebreak
\raggedright

\subsubsection{$\EcmC$ Data $K$ ID Fits}
\label{subsubsec:4230DataKIDFits}
\centering
\begin{tabular}{cc}
\textbf{RS 200-250 MeV/$c$} & \textbf{WS 200-250 MeV/$c$}\\
\includegraphics[width=3in]{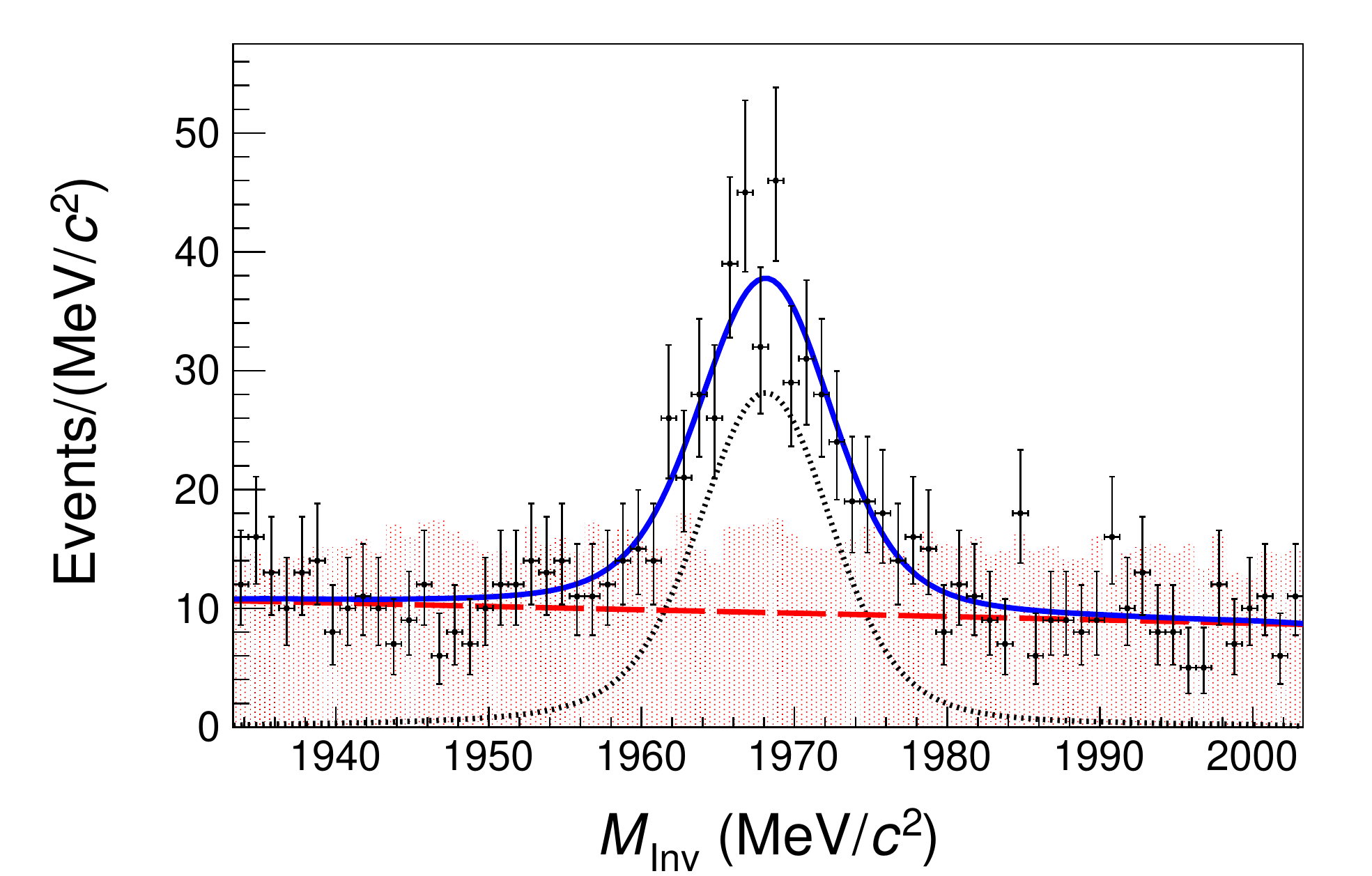} & \includegraphics[width=3in]{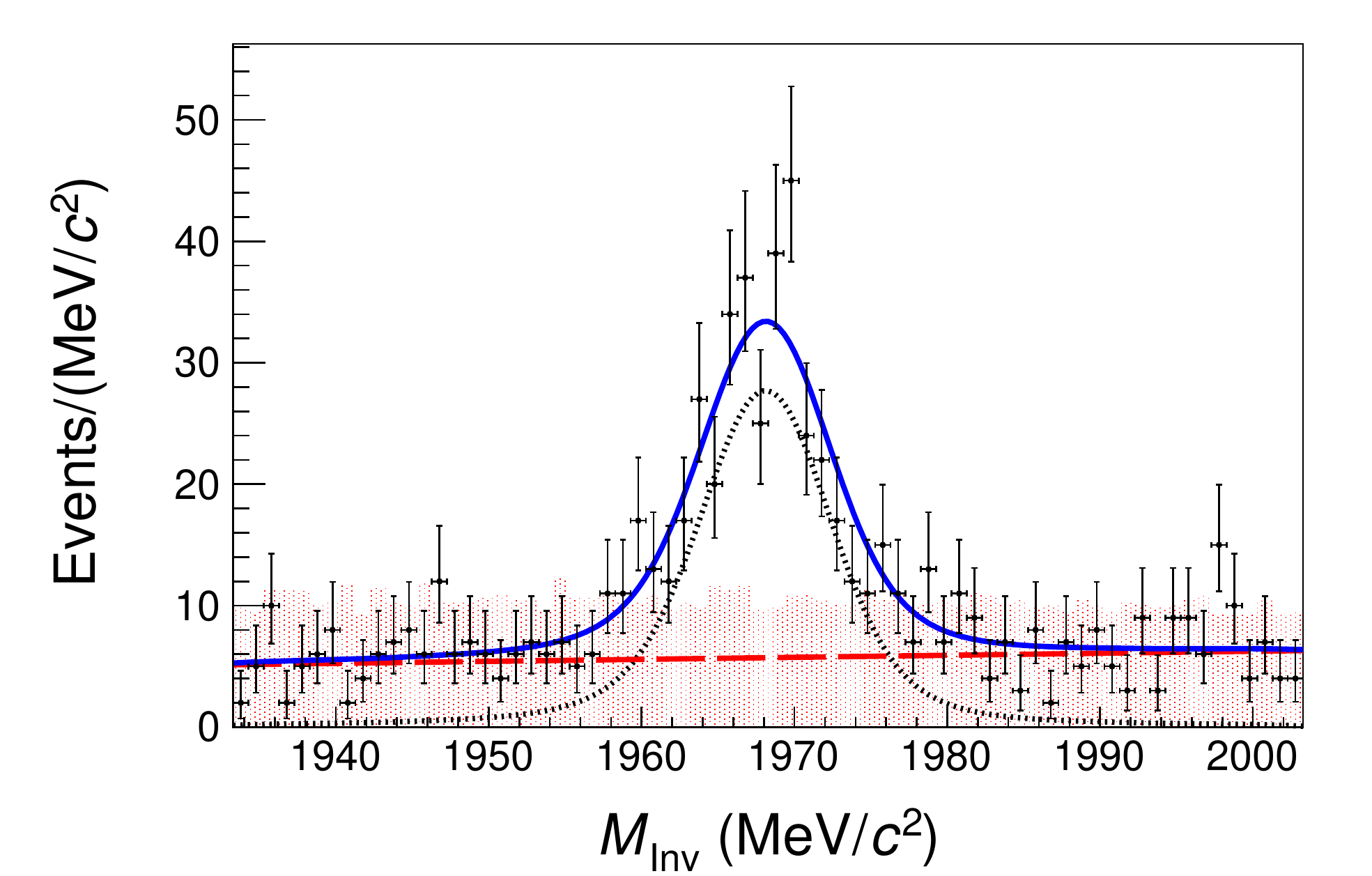}\\
\textbf{RS 250-300 MeV/$c$} & \textbf{WS 250-300 MeV/$c$}\\
\includegraphics[width=3in]{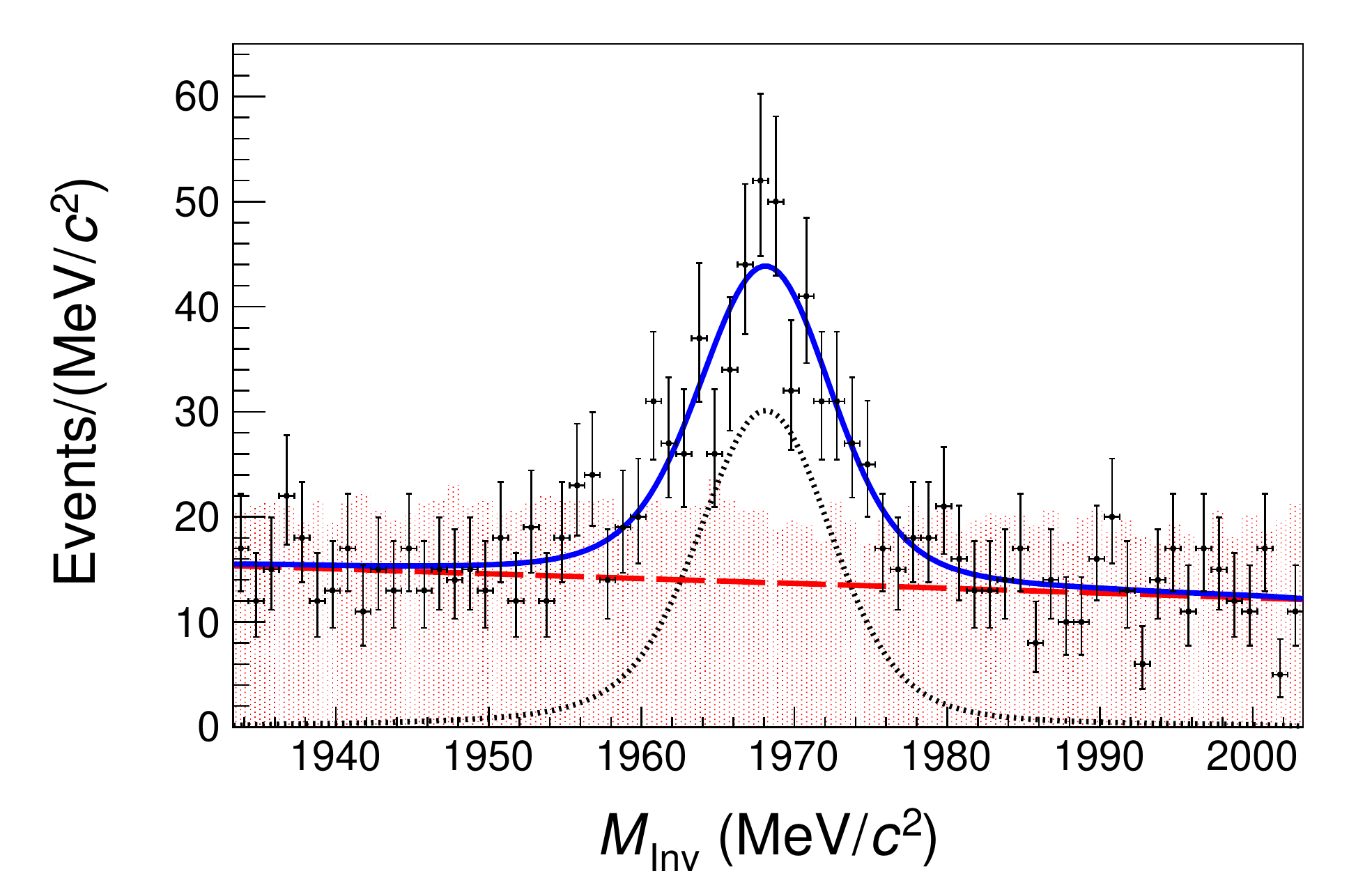} & \includegraphics[width=3in]{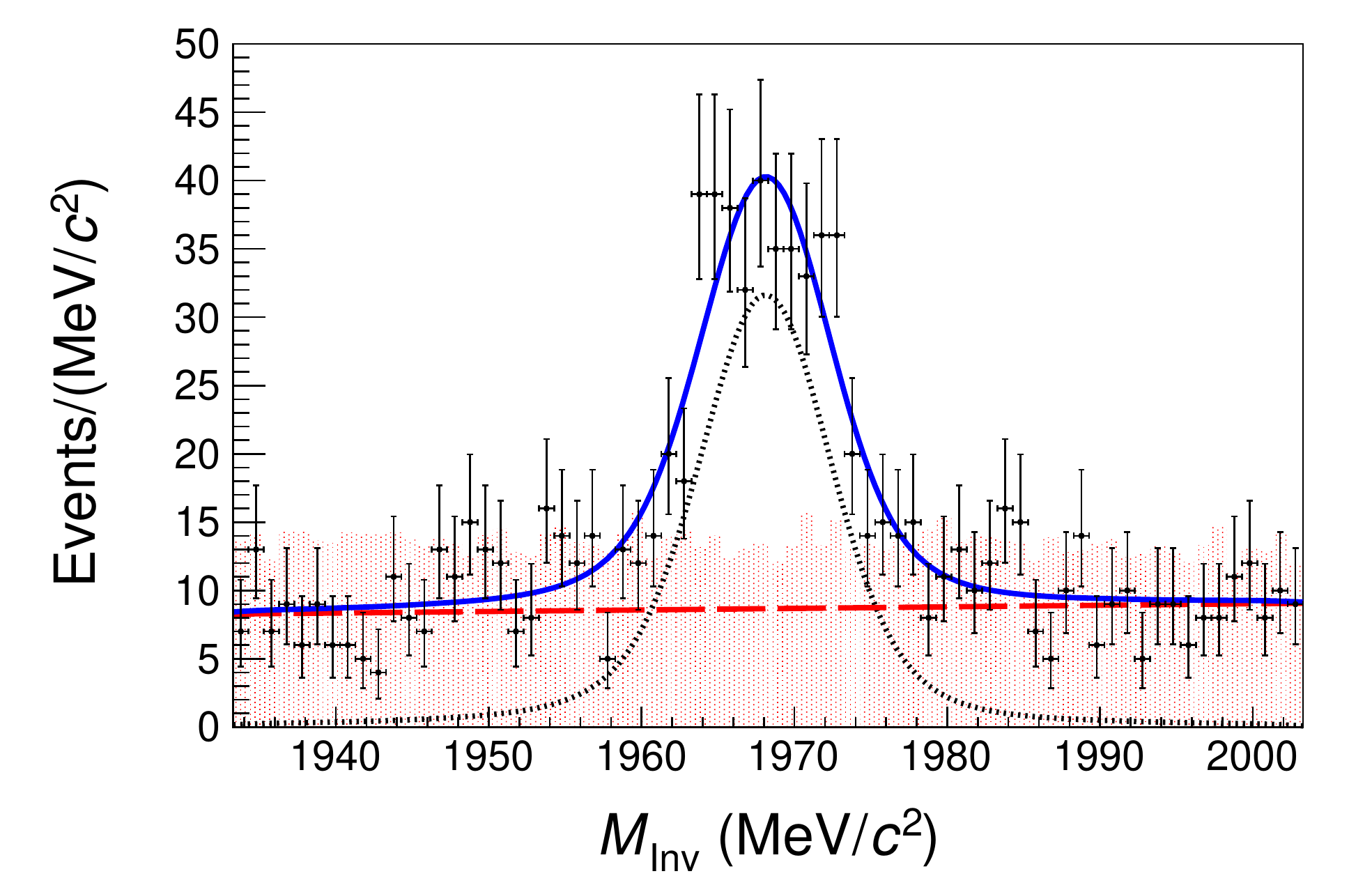}\\
\textbf{RS 300-350 MeV/$c$} & \textbf{WS 300-350 MeV/$c$}\\
\includegraphics[width=3in]{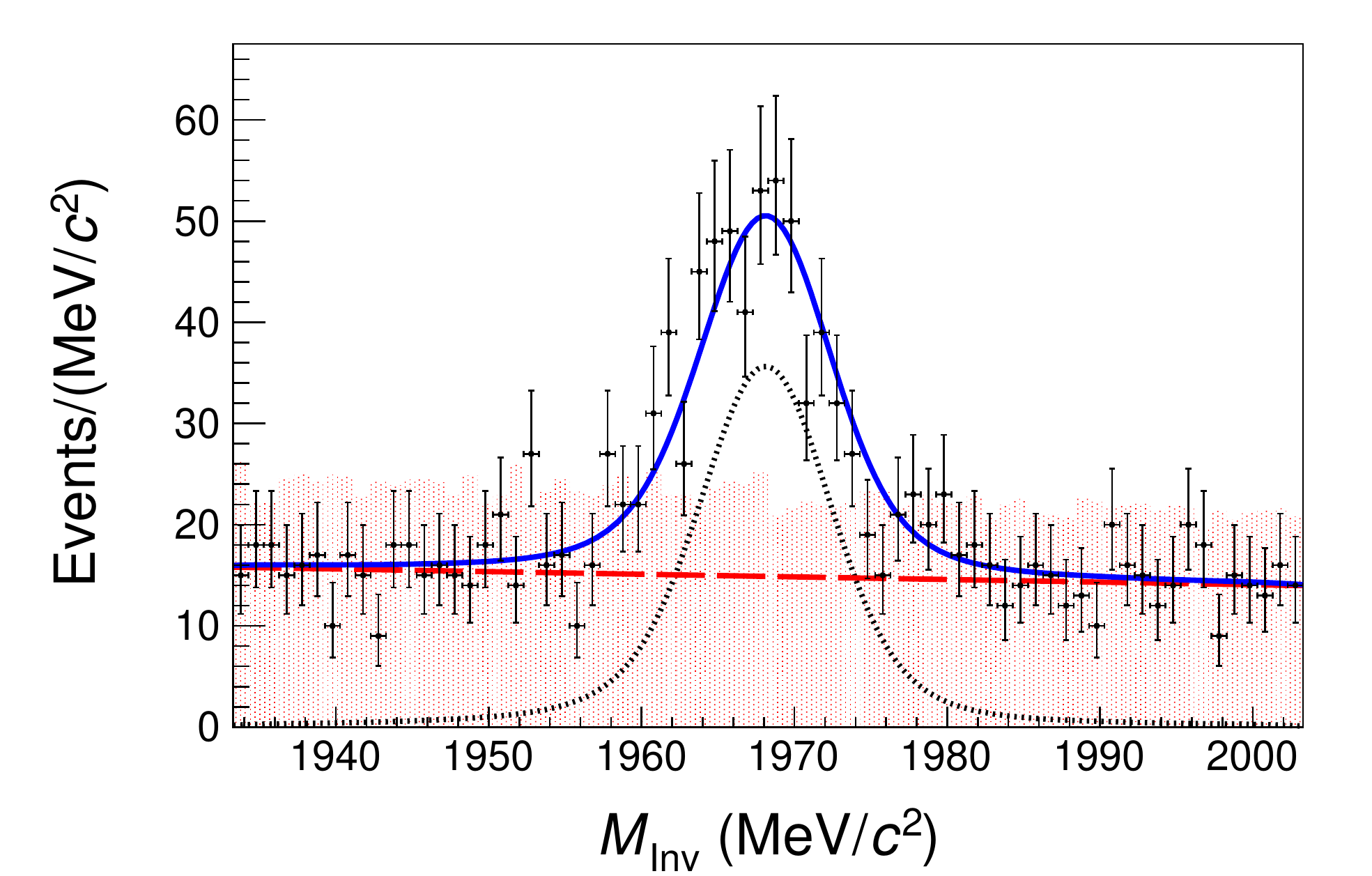} & \includegraphics[width=3in]{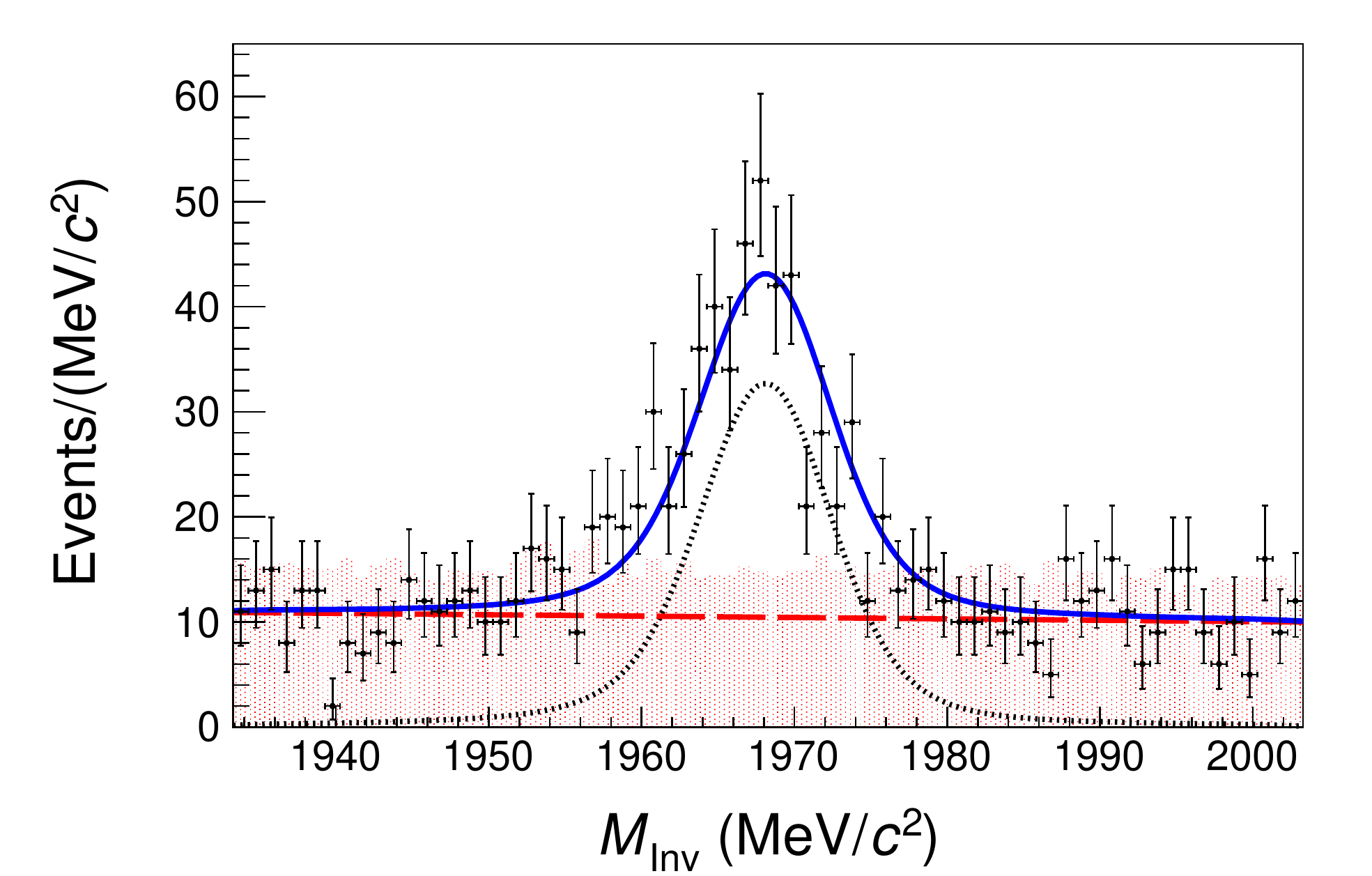}\\
\textbf{RS 350-400 MeV/$c$} & \textbf{WS 350-400 MeV/$c$}\\
\includegraphics[width=3in]{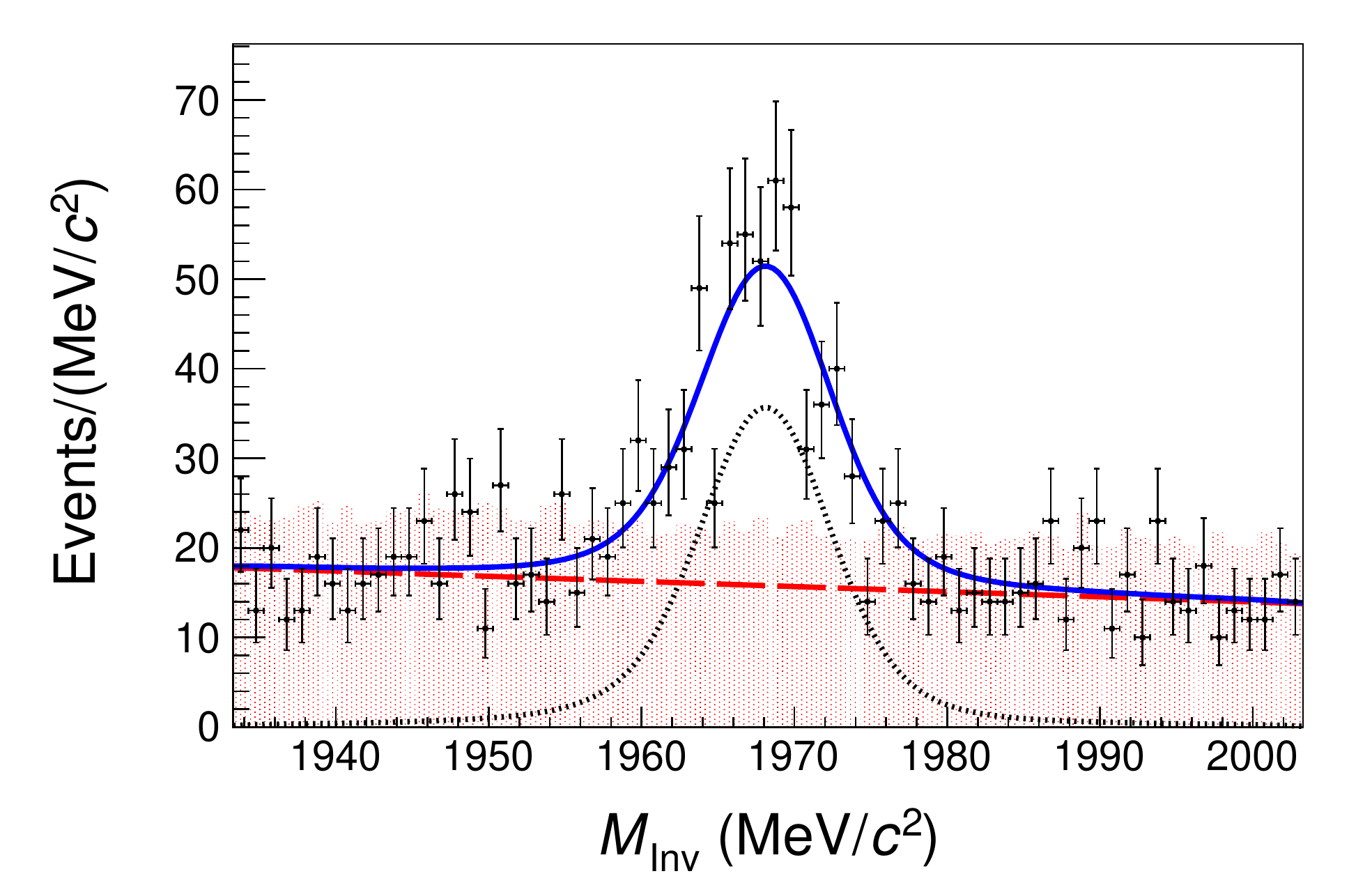} & \includegraphics[width=3in]{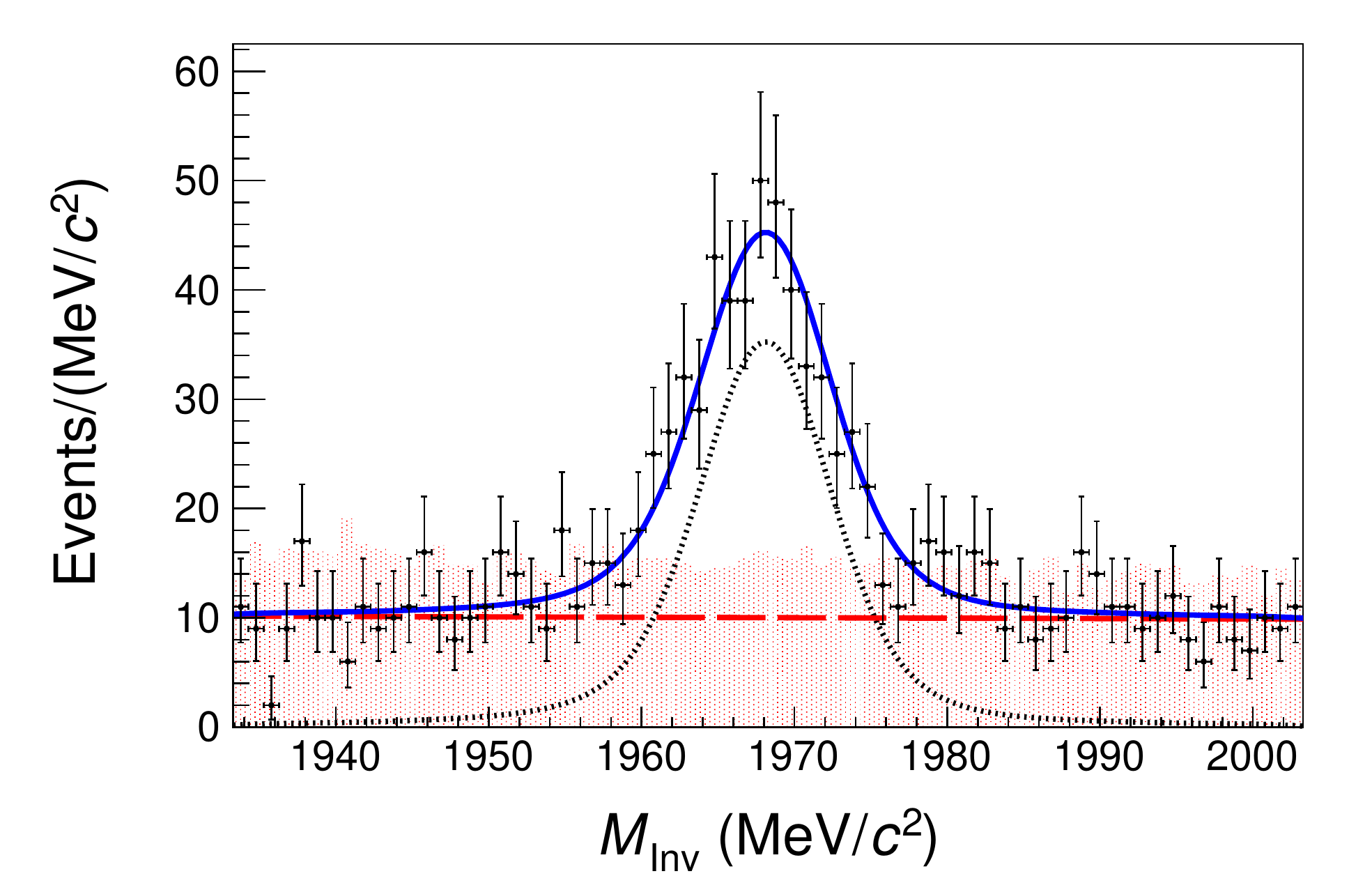}
\end{tabular}
\pagebreak

\begin{tabular}{cc}
\textbf{RS 400-450 MeV/$c$} & \textbf{WS 400-450 MeV/$c$}\\
\includegraphics[width=3in]{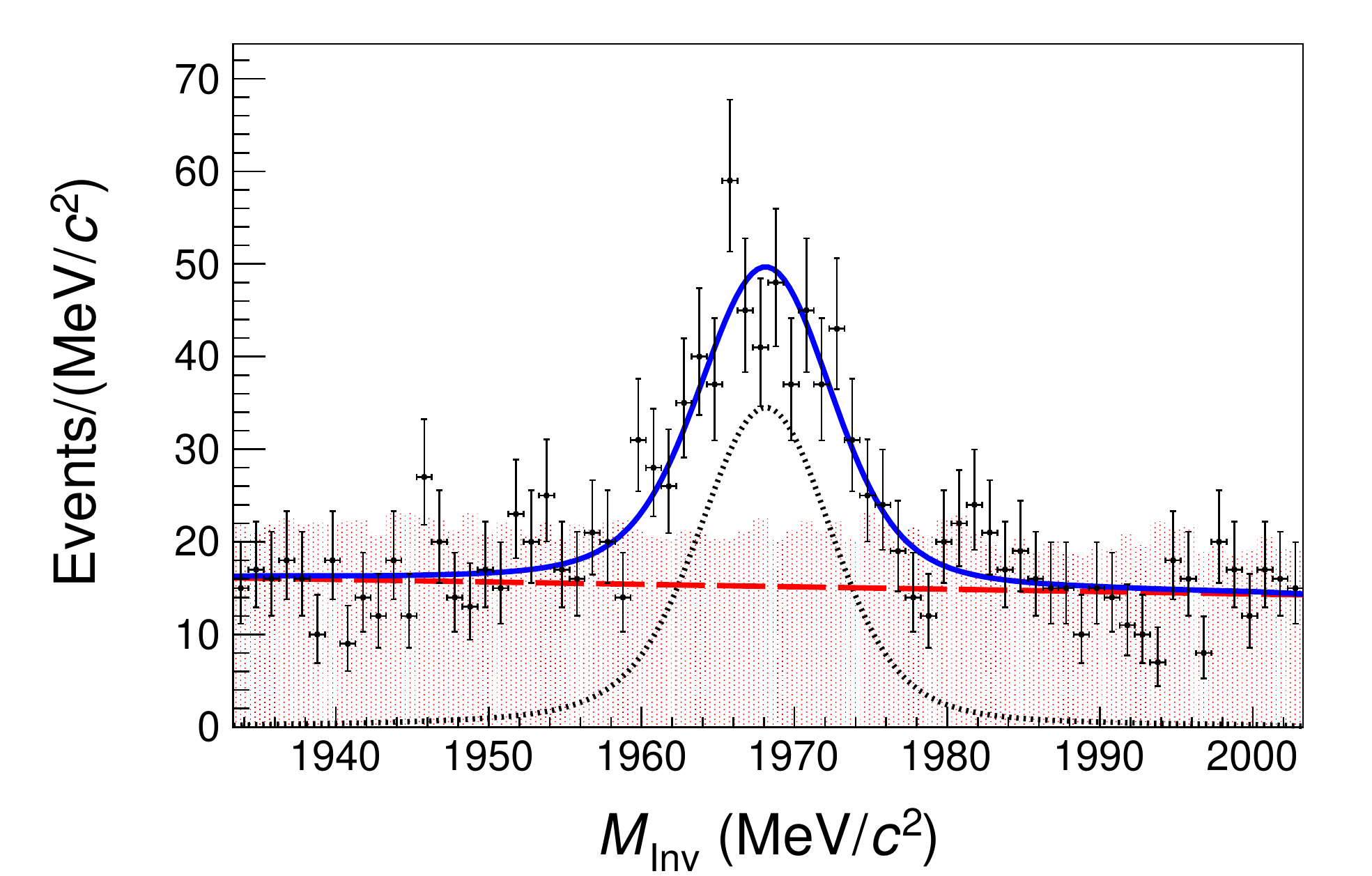} & \includegraphics[width=3in]{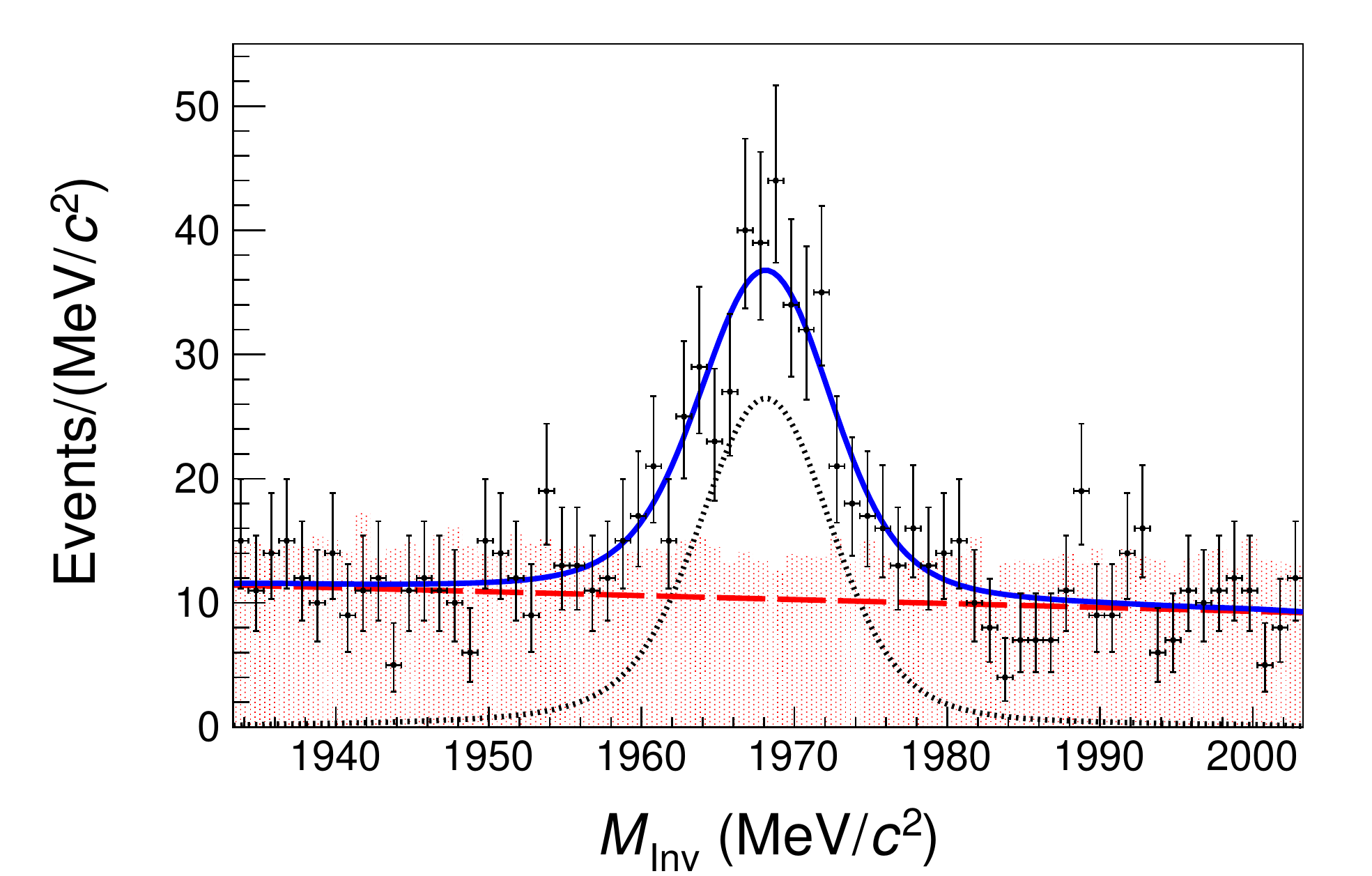}\\
\textbf{RS 450-500 MeV/$c$} & \textbf{WS 450-500 MeV/$c$}\\
\includegraphics[width=3in]{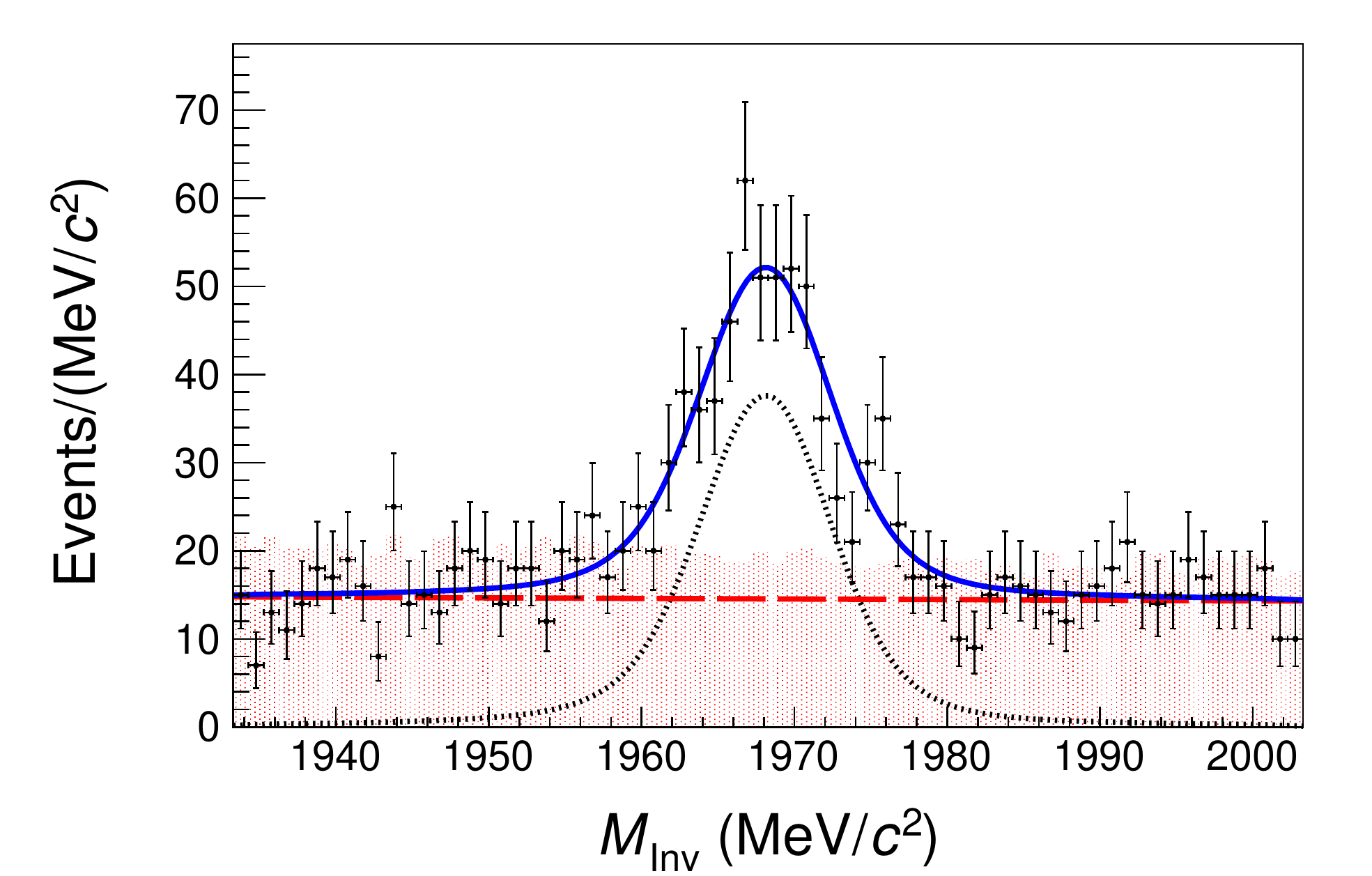} & \includegraphics[width=3in]{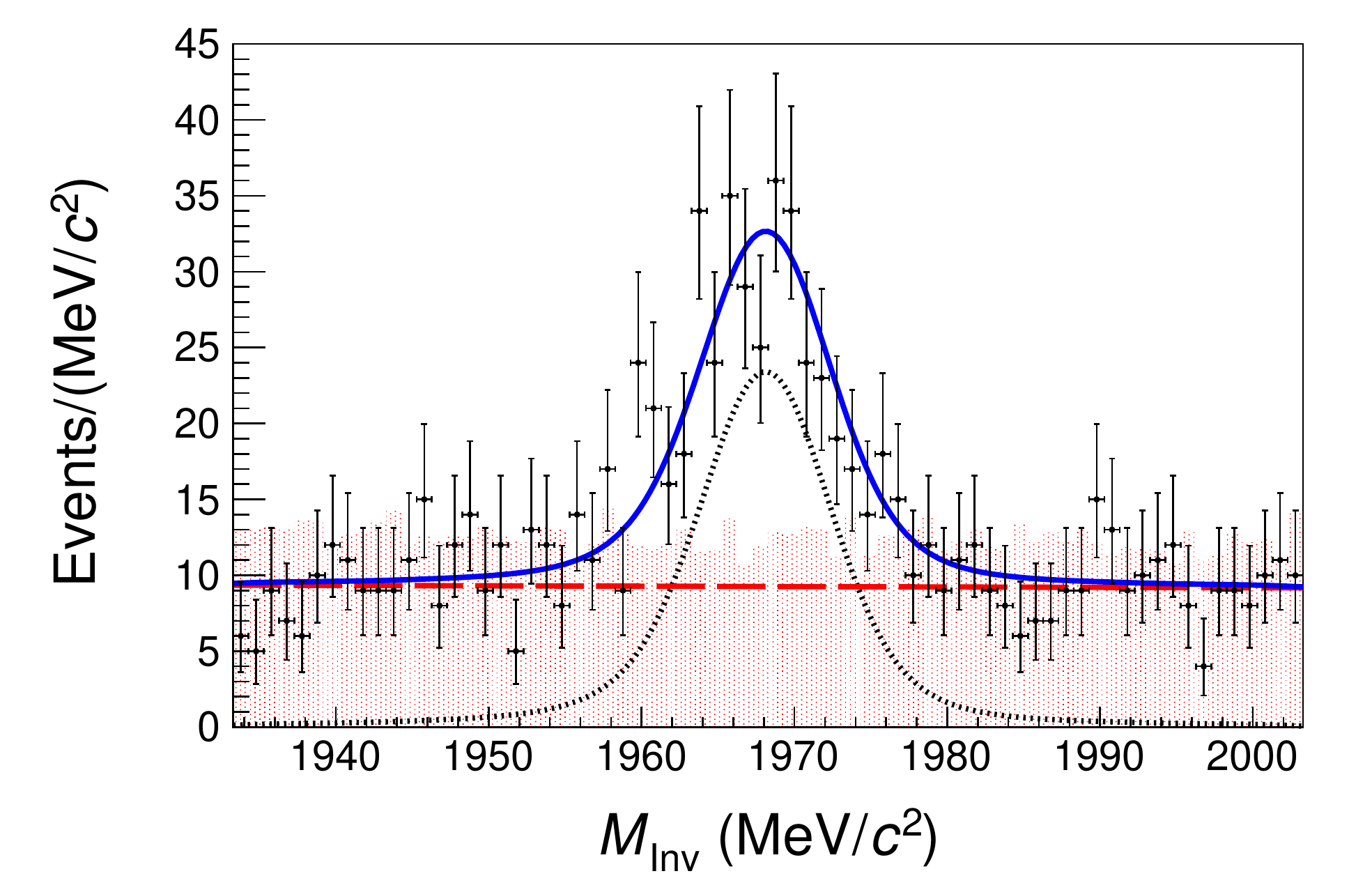}\\
\textbf{RS 500-550 MeV/$c$} & \textbf{WS 500-550 MeV/$c$}\\
\includegraphics[width=3in]{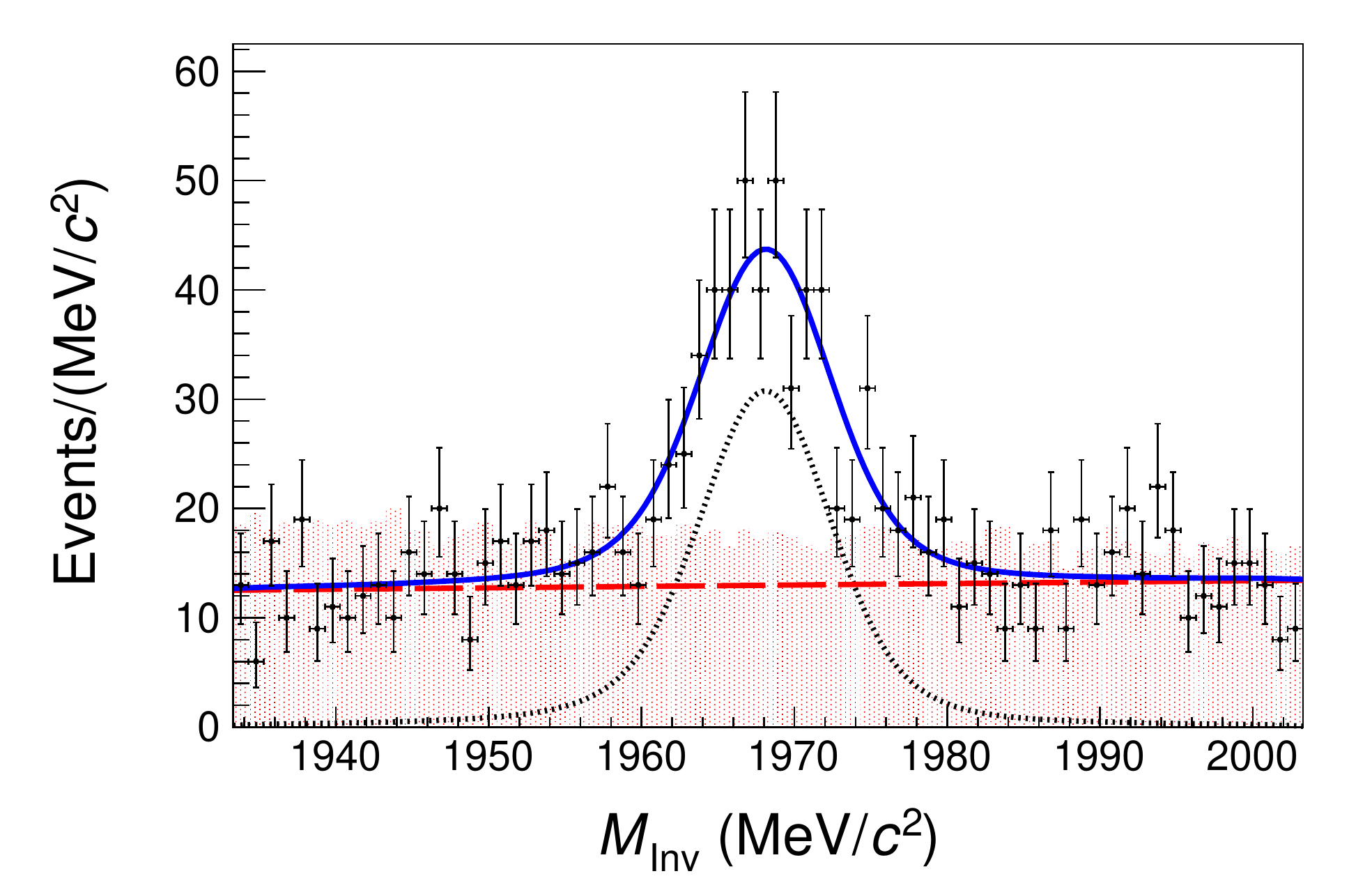} & \includegraphics[width=3in]{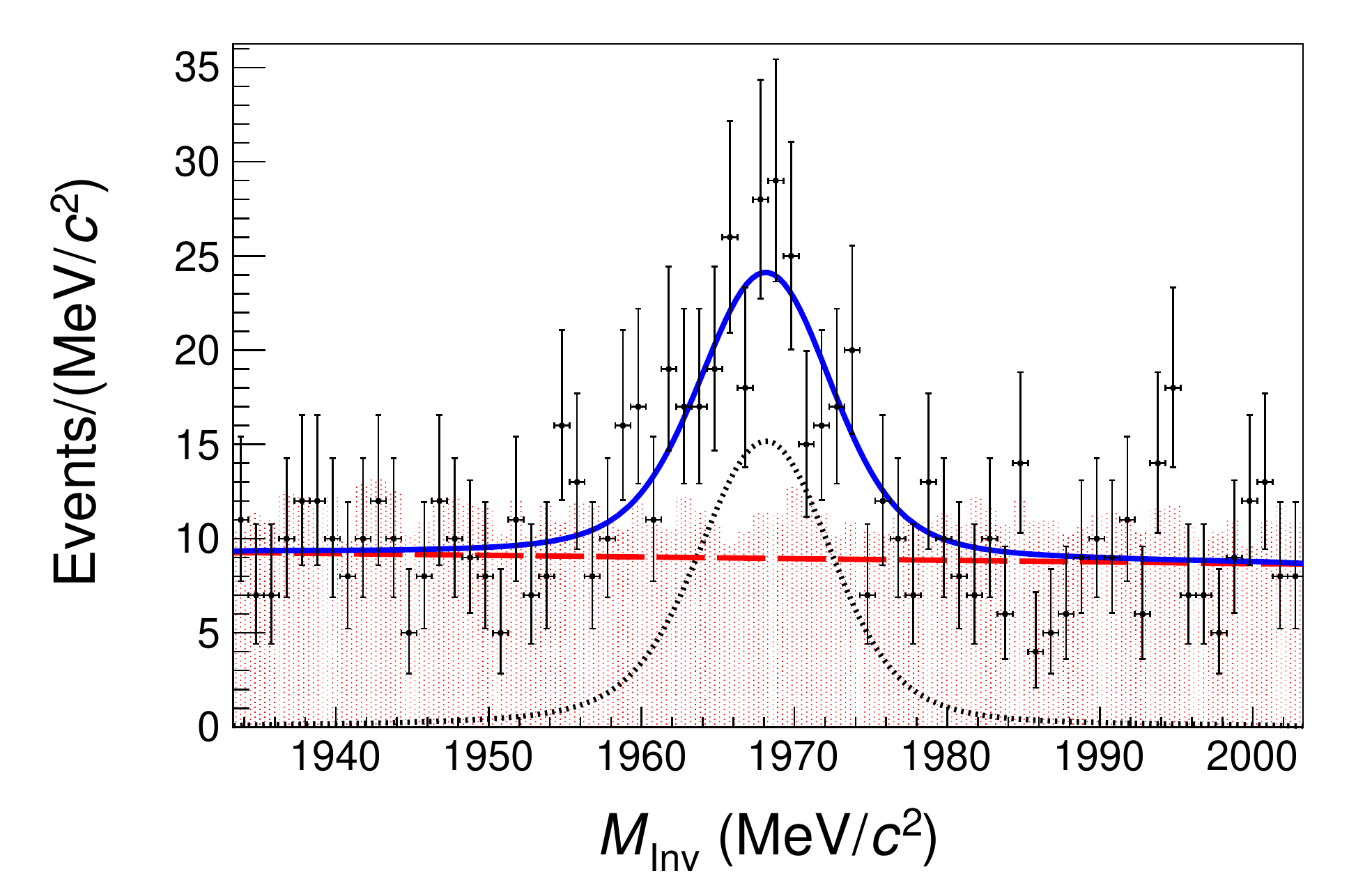}\\
\textbf{RS 550-600 MeV/$c$} & \textbf{WS 550-600 MeV/$c$}\\
\includegraphics[width=3in]{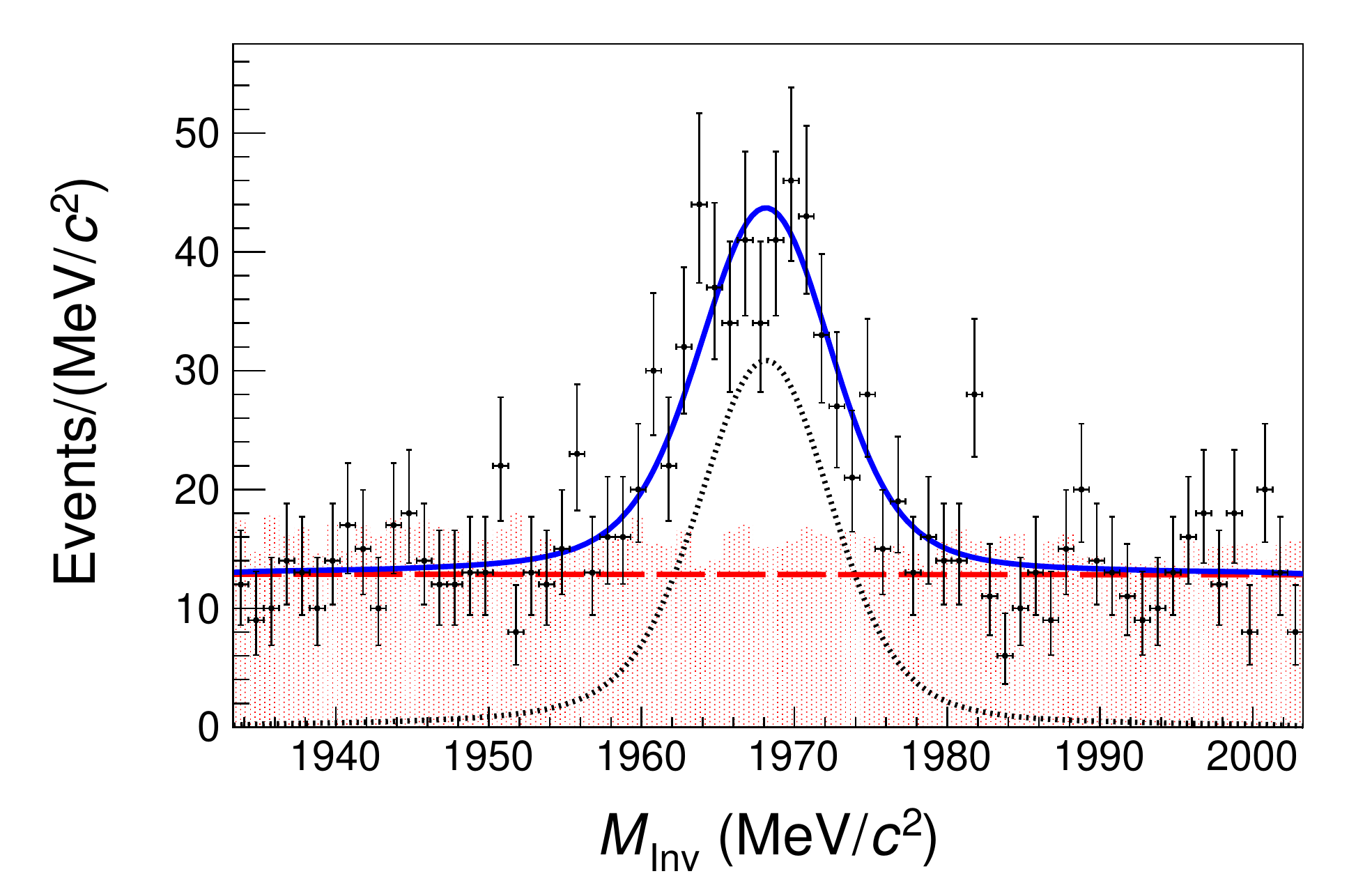} & \includegraphics[width=3in]{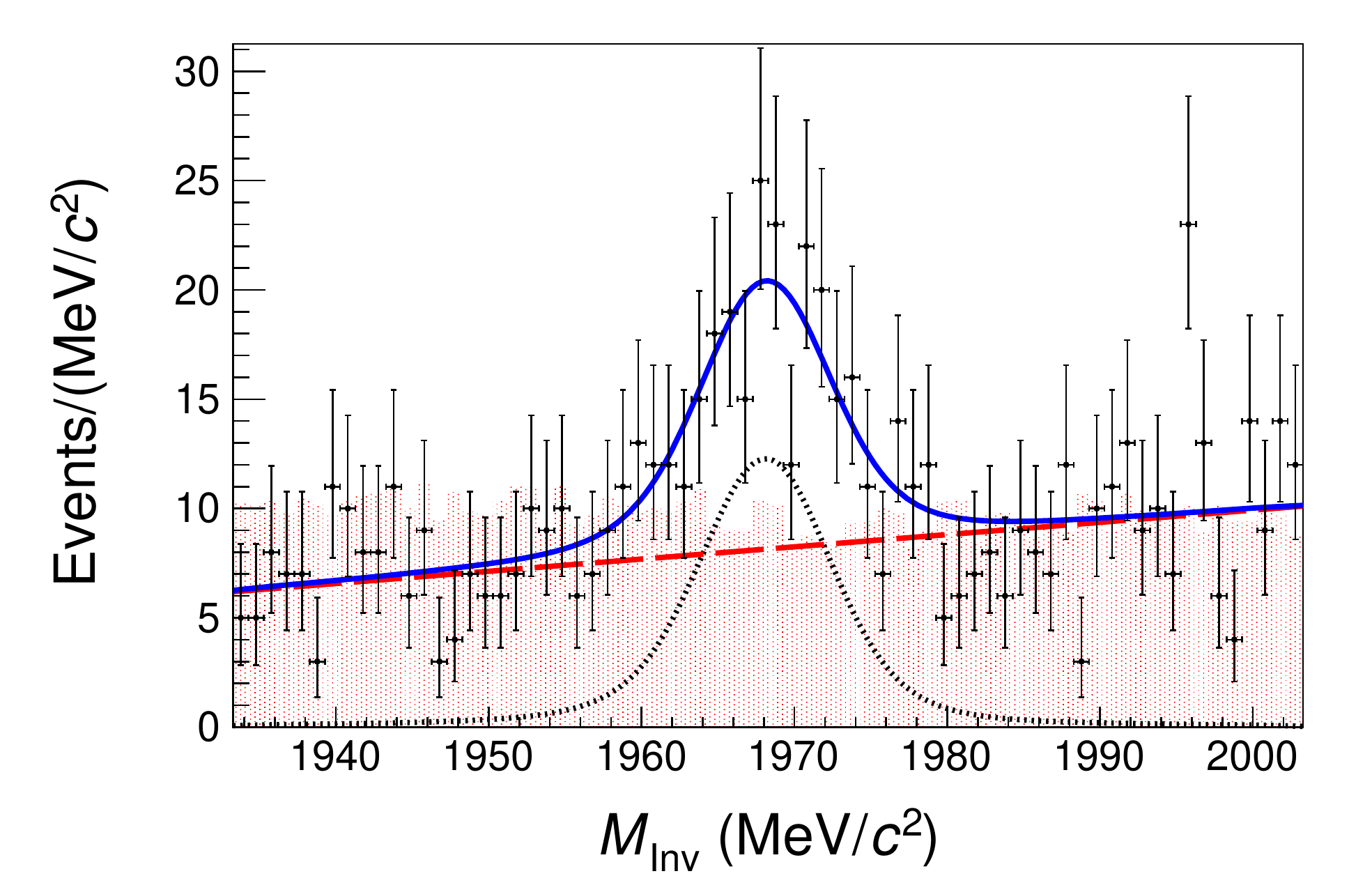}
\end{tabular}

\begin{tabular}{cc}
\textbf{RS 600-650 MeV/$c$} & \textbf{WS 600-650 MeV/$c$}\\
\includegraphics[width=3in]{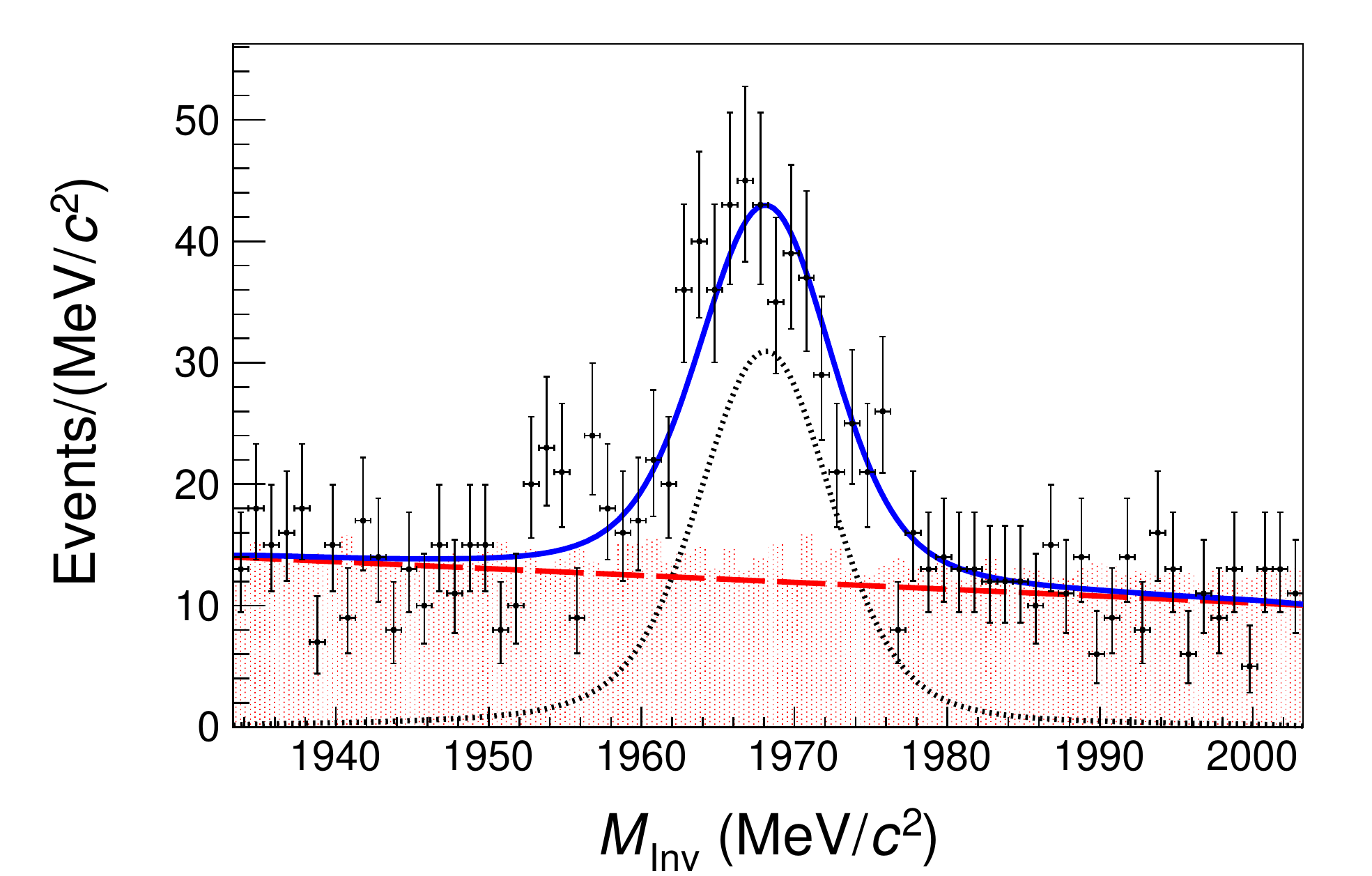} & \includegraphics[width=3in]{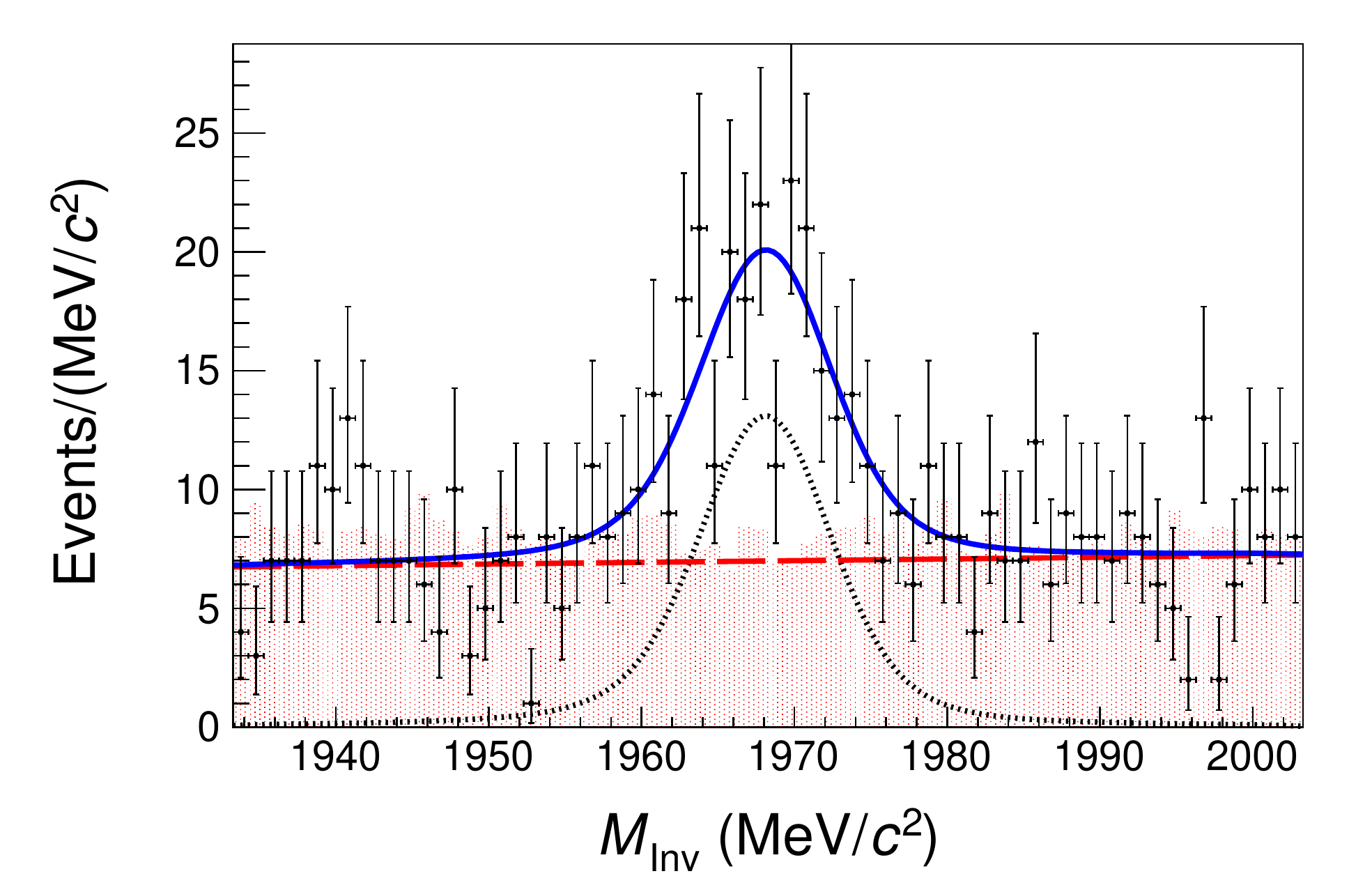}\\
\textbf{RS 650-700 MeV/$c$} & \textbf{WS 650-700 MeV/$c$}\\
\includegraphics[width=3in]{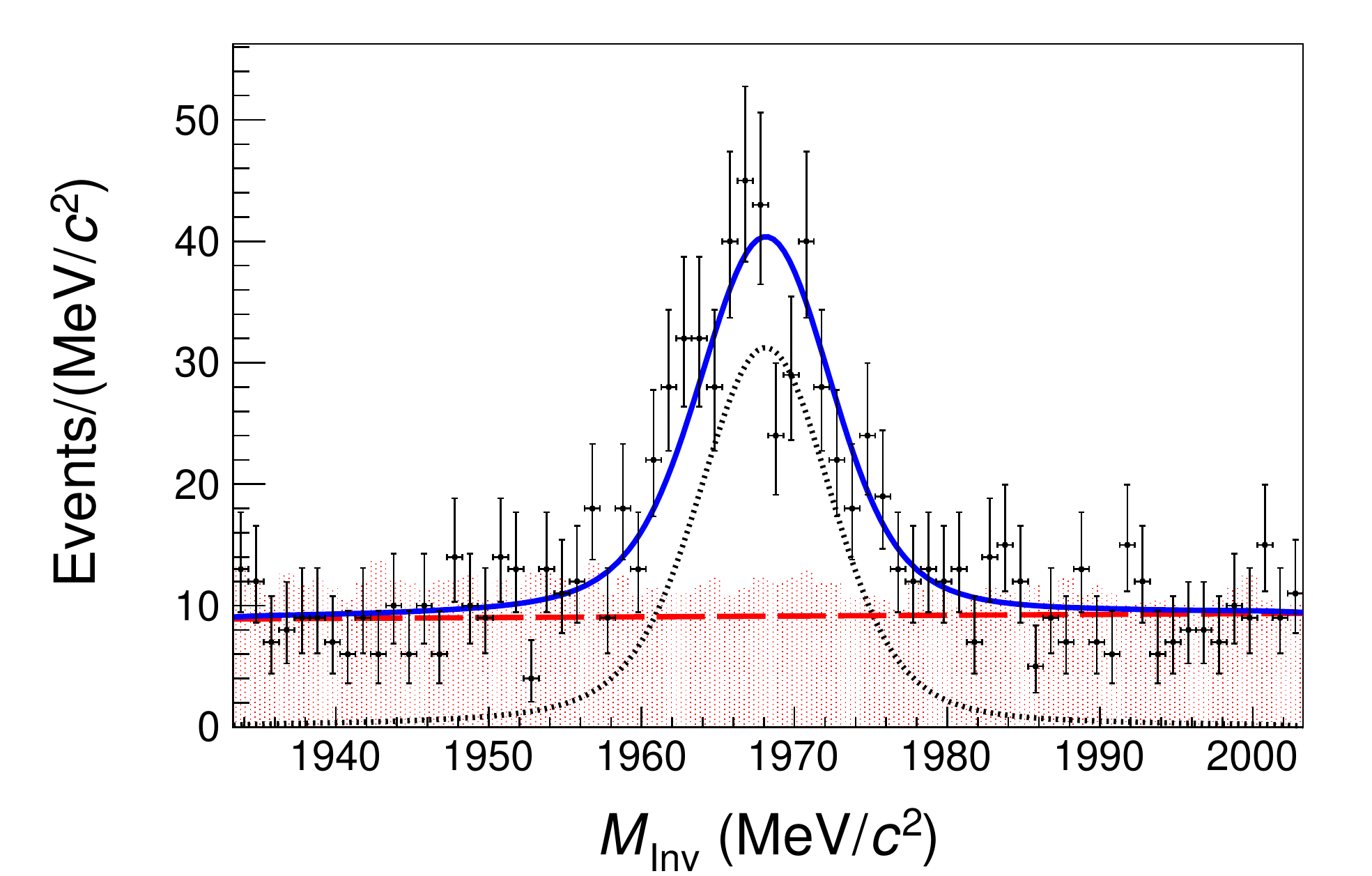} & \includegraphics[width=3in]{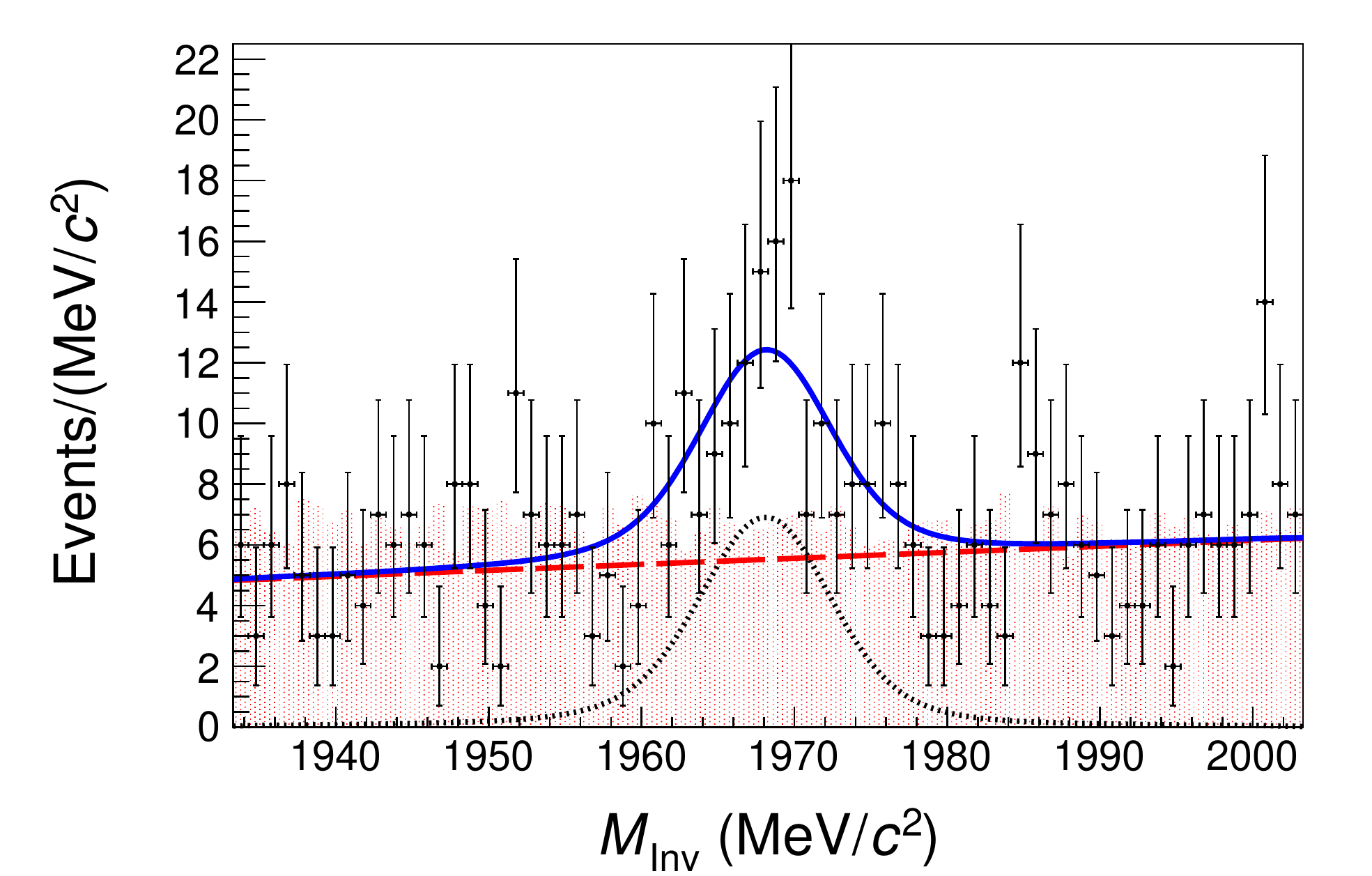}\\
\textbf{RS 700-750 MeV/$c$} & \textbf{WS 700-750 MeV/$c$}\\
\includegraphics[width=3in]{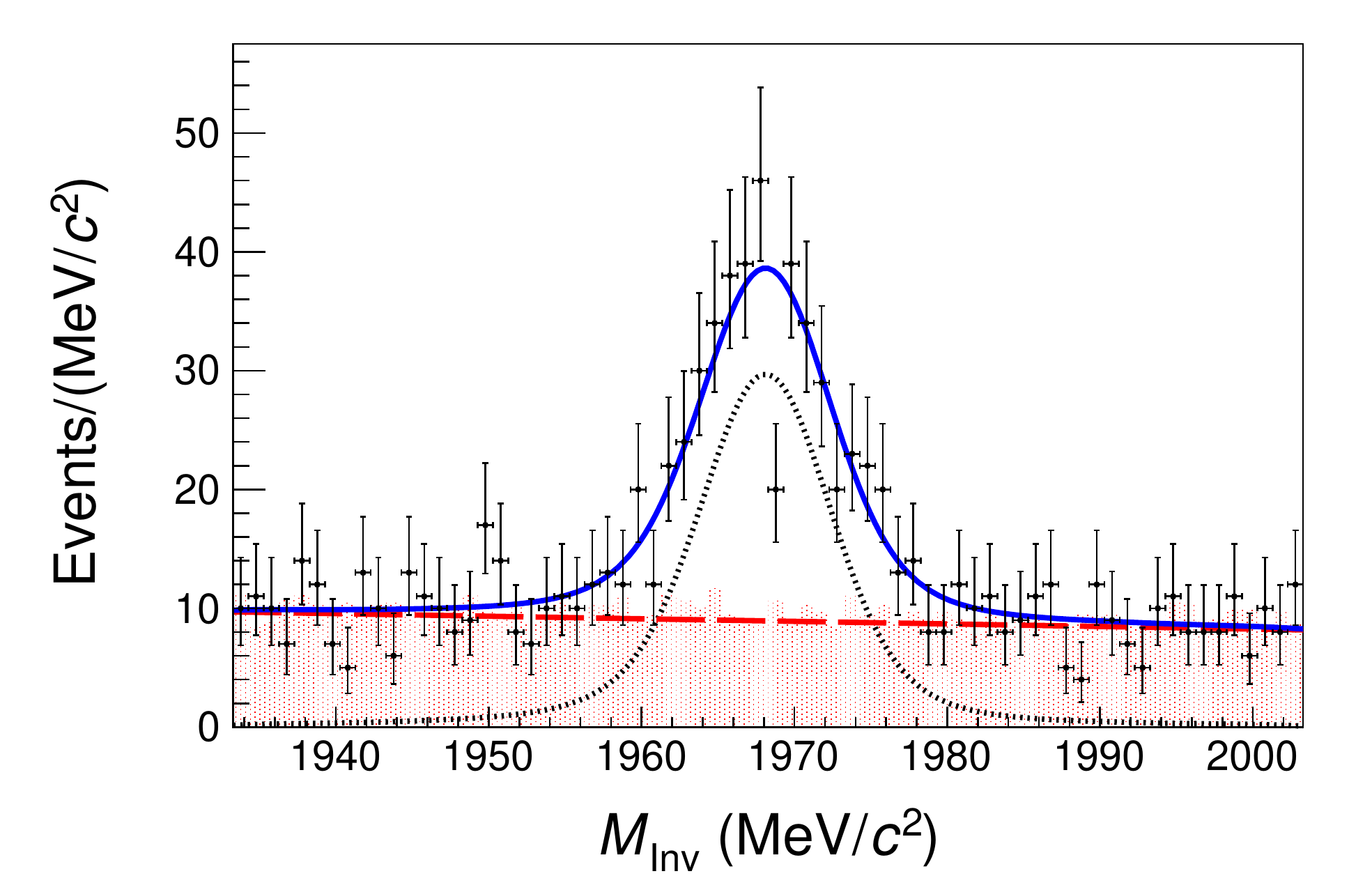} & \includegraphics[width=3in]{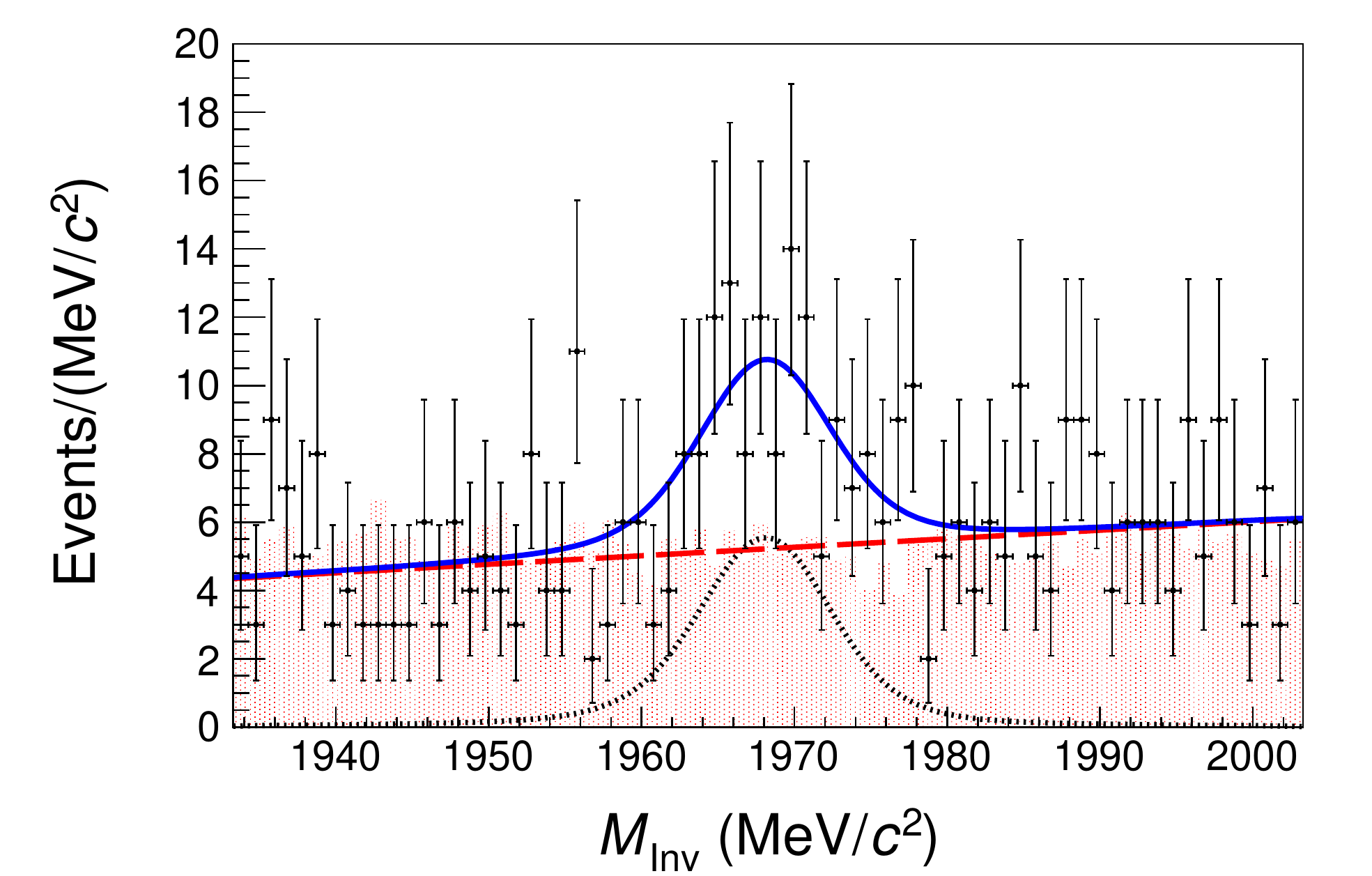}\\
\textbf{RS 750-800 MeV/$c$} & \textbf{WS 750-800 MeV/$c$}\\
\includegraphics[width=3in]{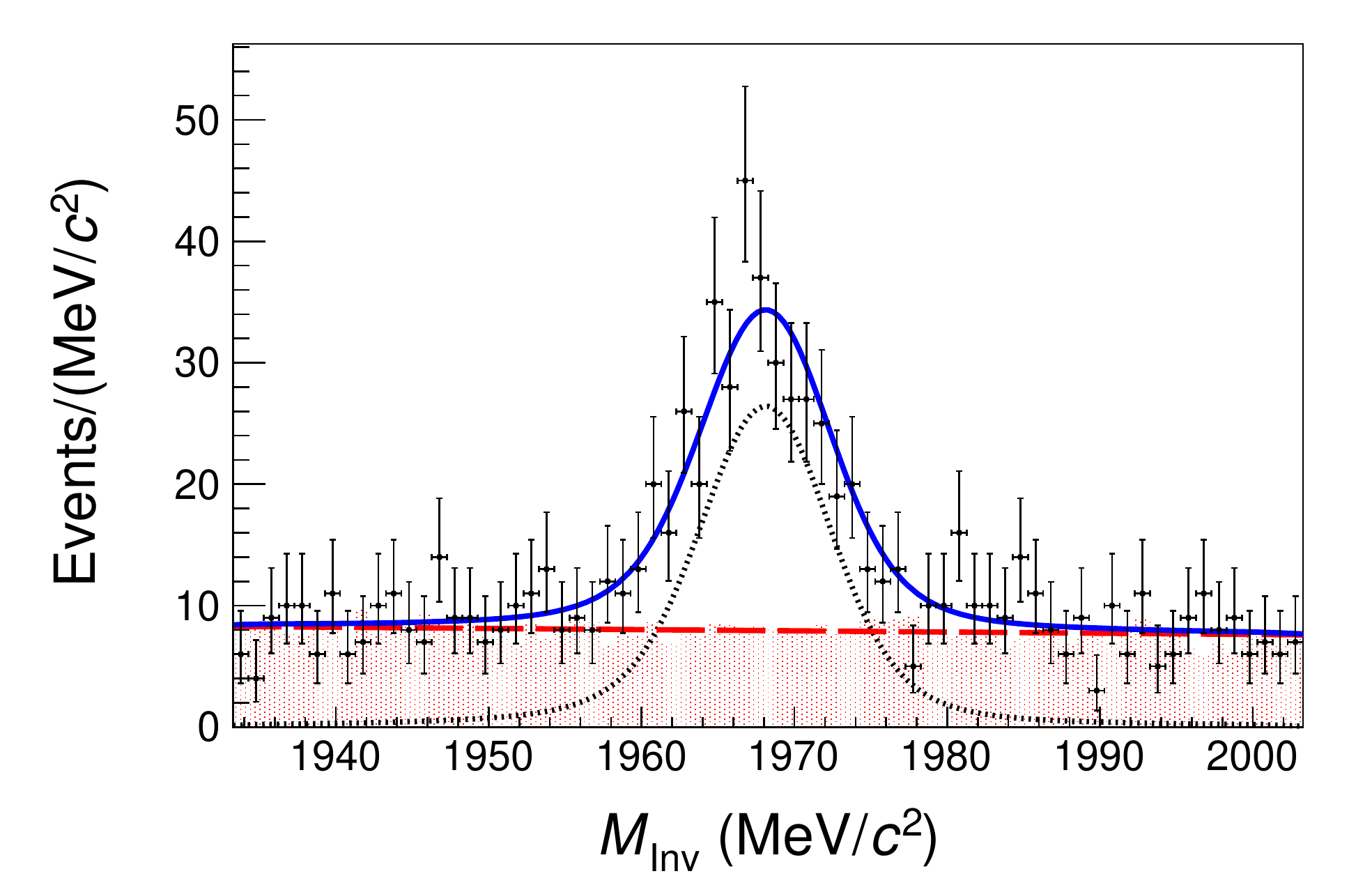} & \includegraphics[width=3in]{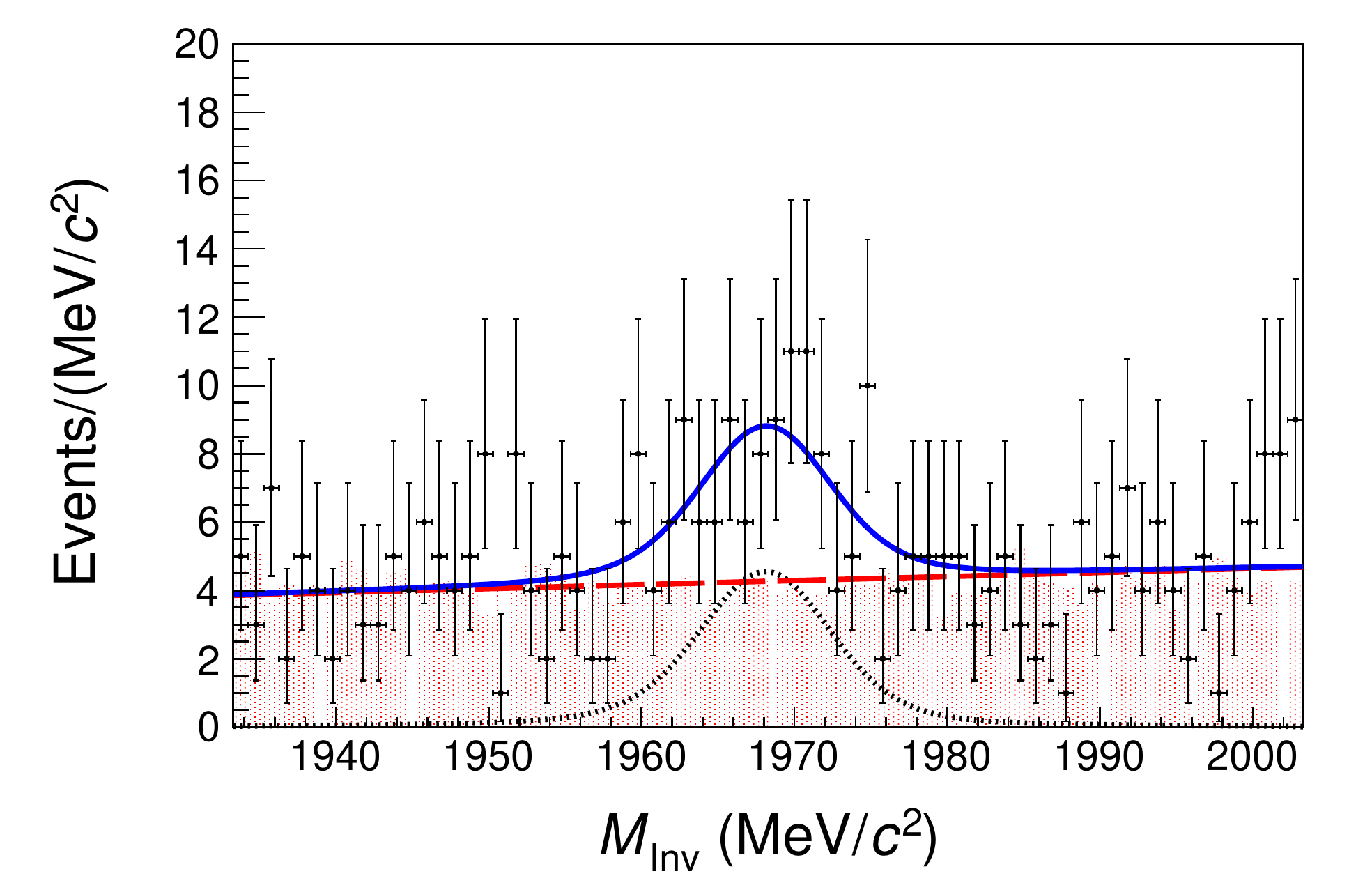}\\
\end{tabular}

\begin{tabular}{cc}
\textbf{RS 800-850 MeV/$c$} & \textbf{WS 800-850 MeV/$c$}\\
\includegraphics[width=3in]{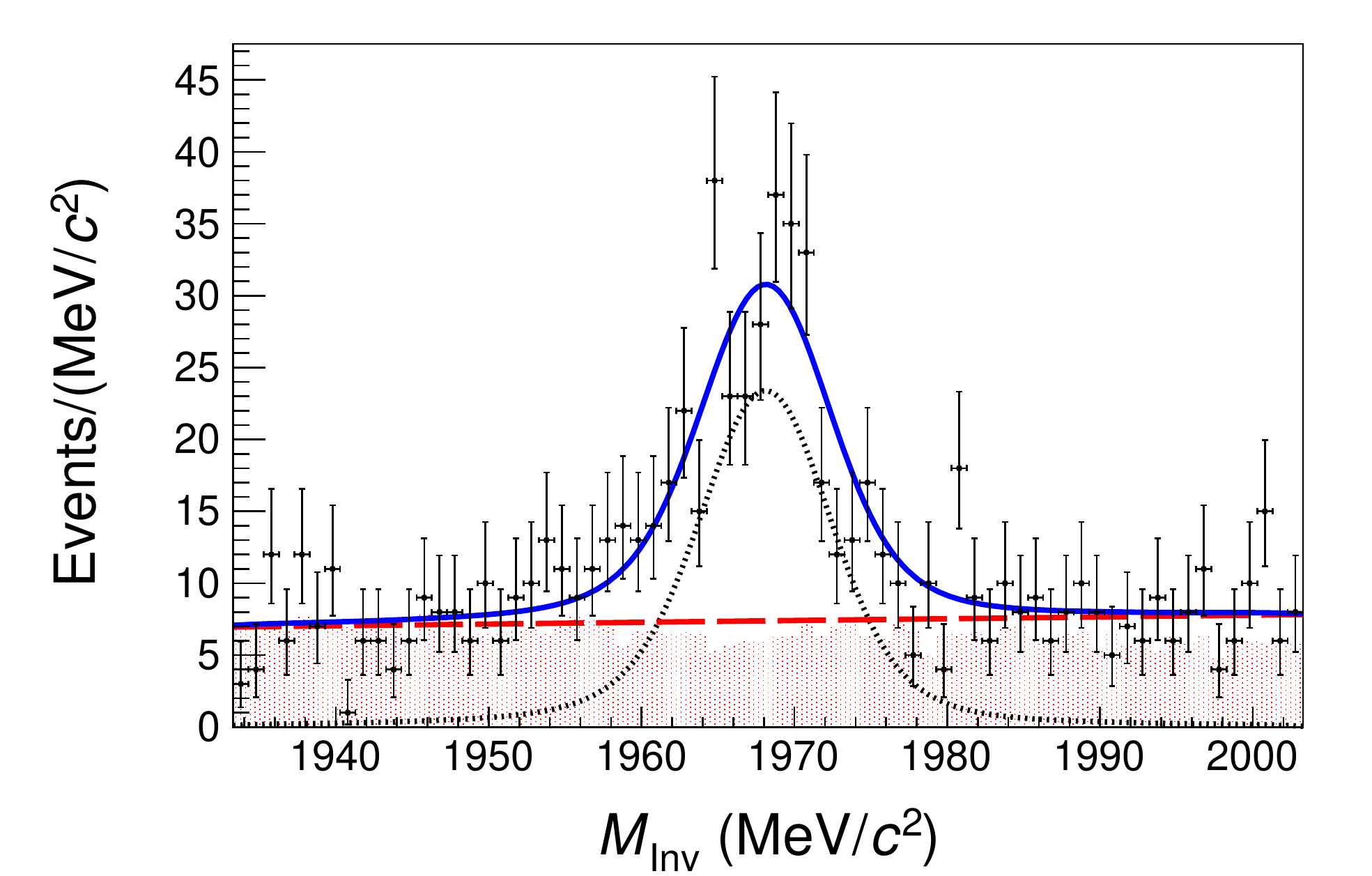} & \includegraphics[width=3in]{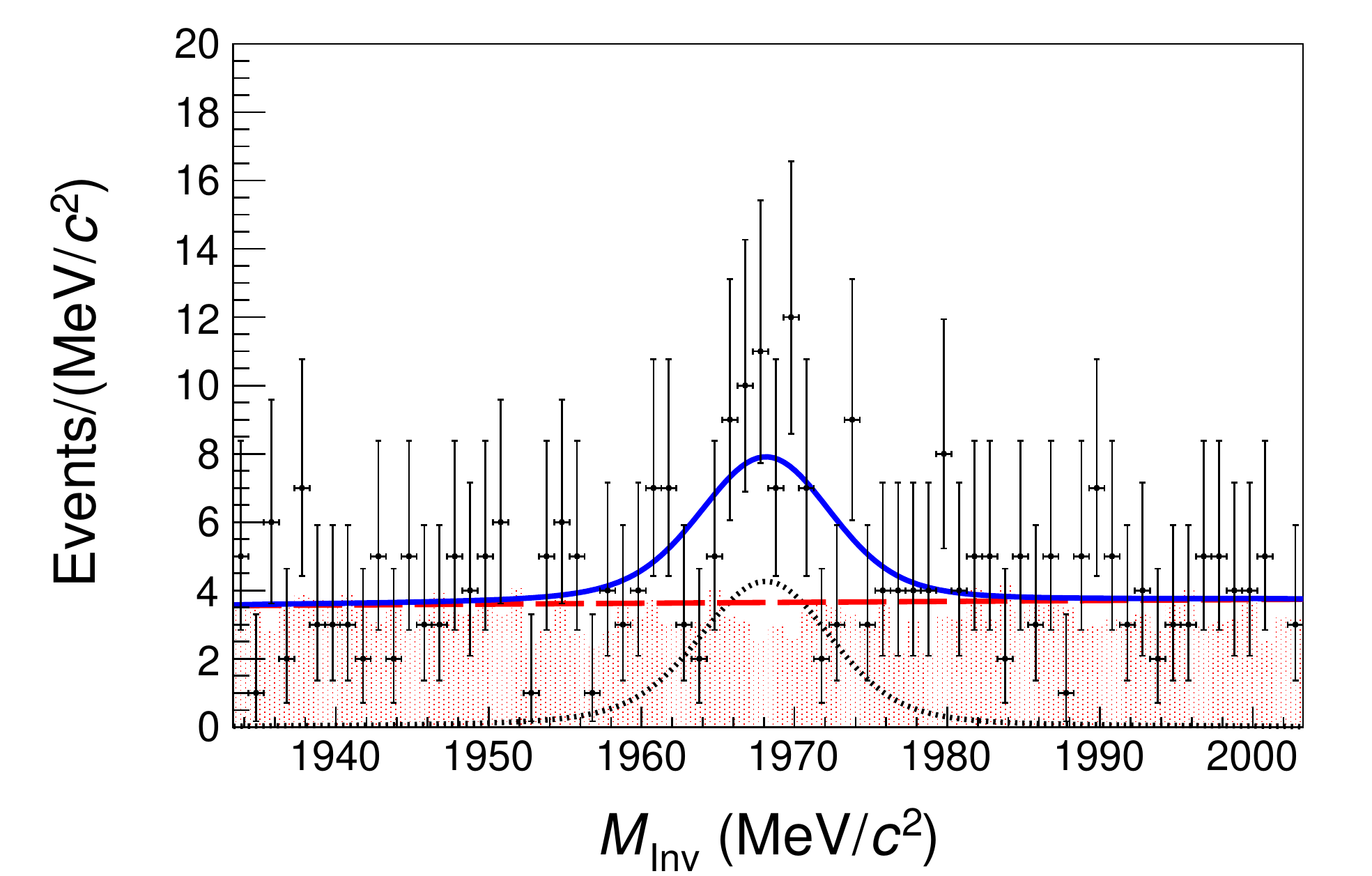}\\
\textbf{RS 850-900 MeV/$c$} & \textbf{WS 850-900 MeV/$c$}\\
\includegraphics[width=3in]{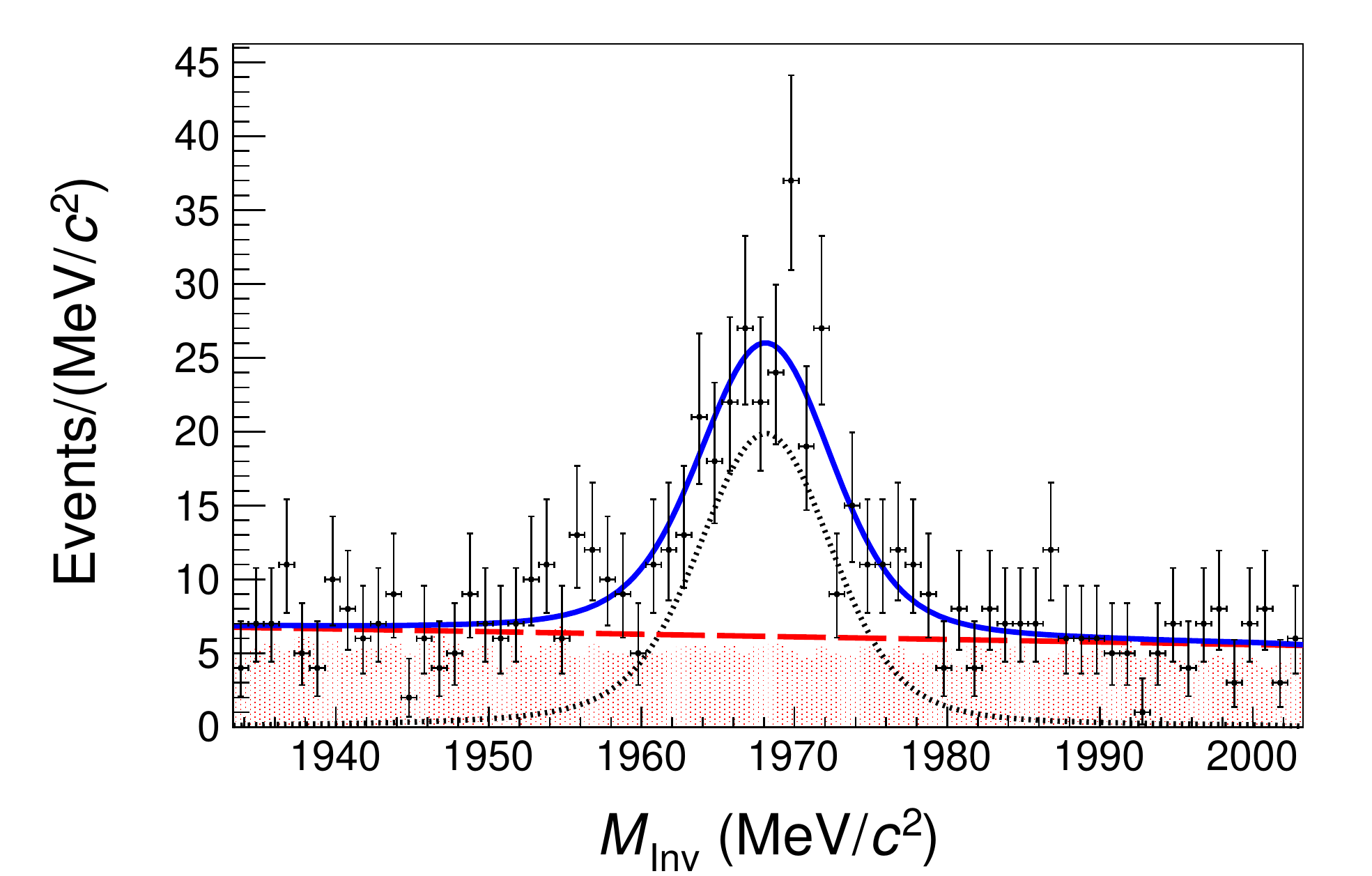} & \includegraphics[width=3in]{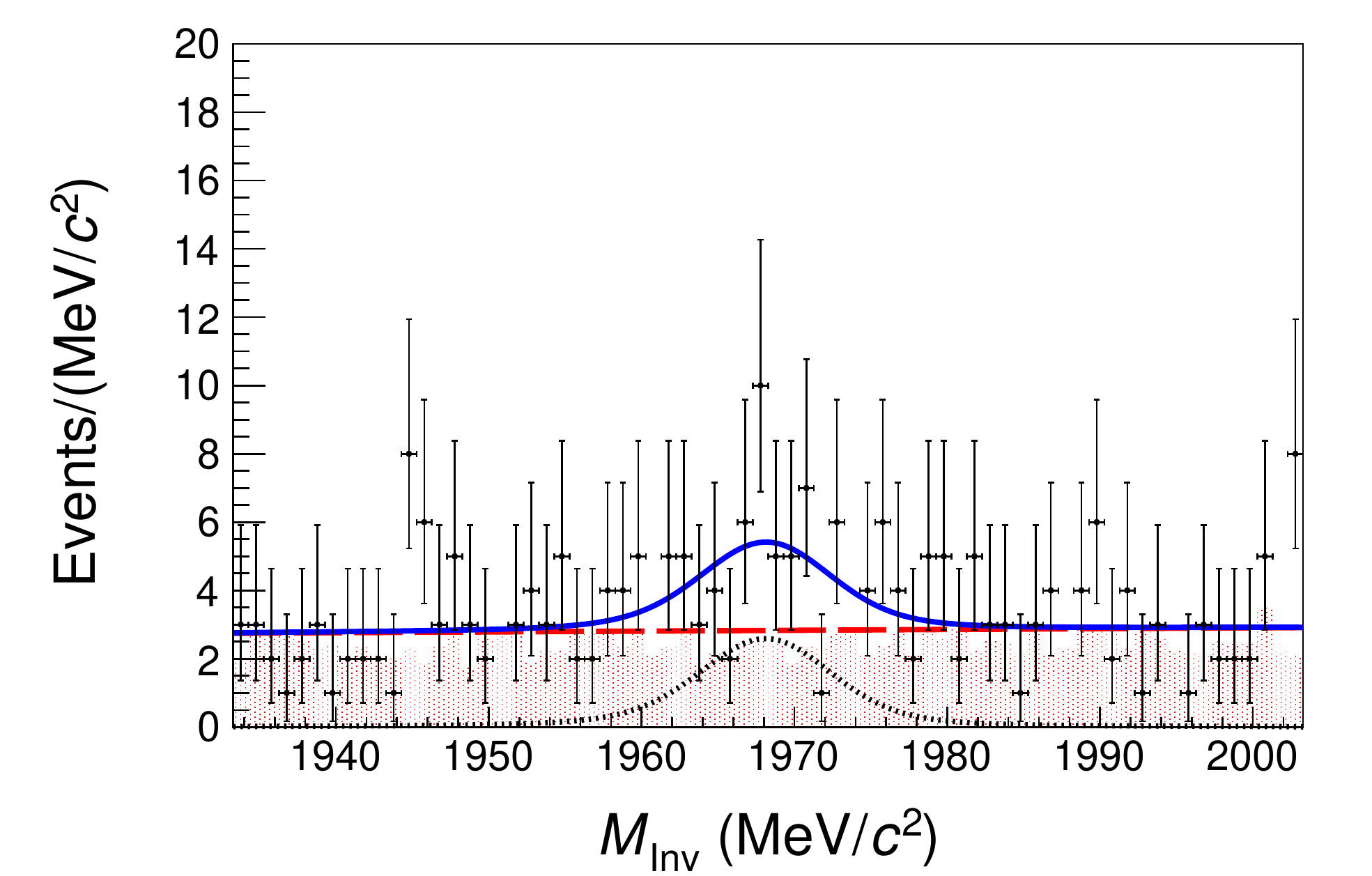}\\
\textbf{RS 900-950 MeV/$c$} & \textbf{WS 900-950 MeV/$c$}\\
\includegraphics[width=3in]{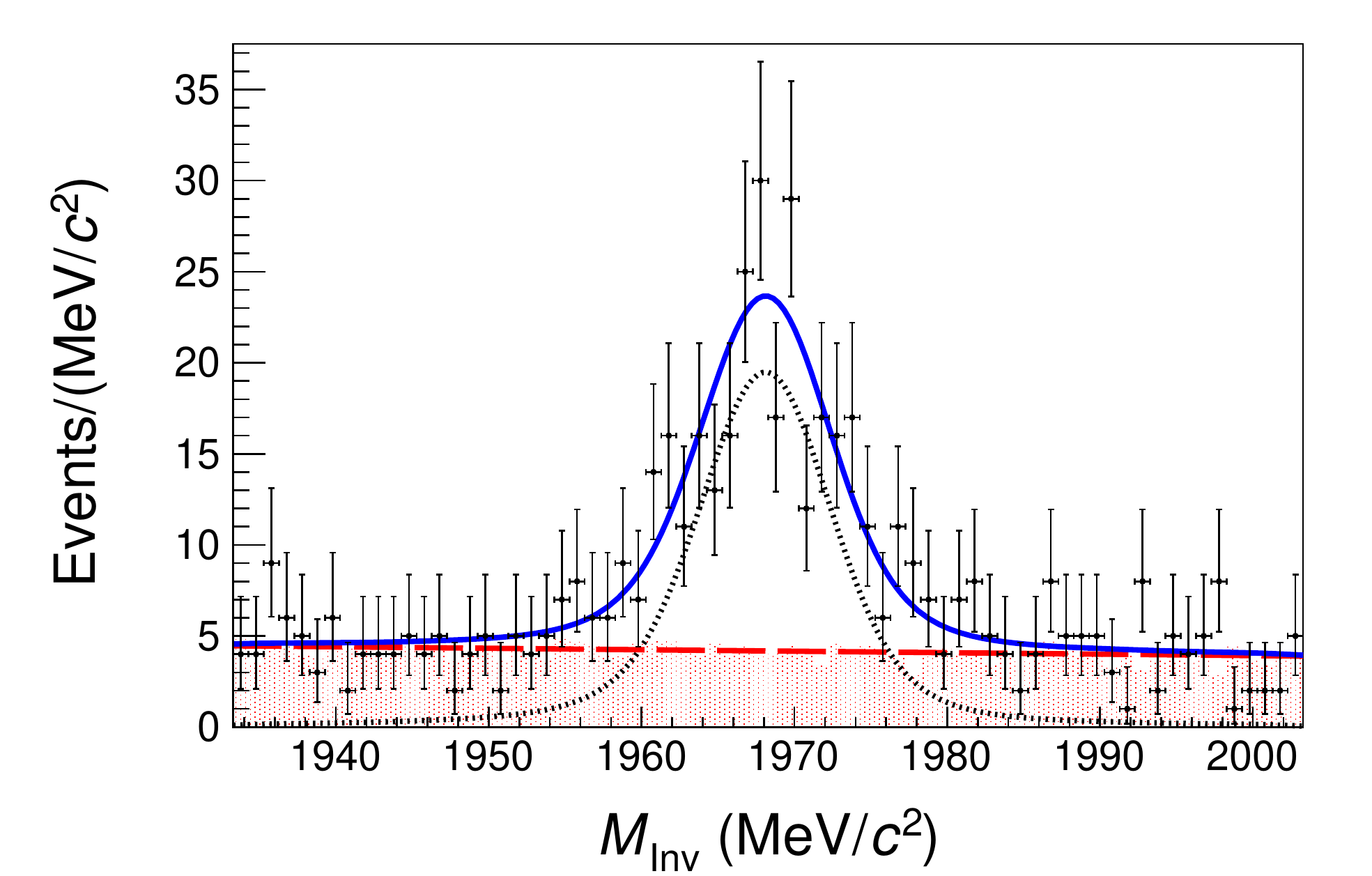} & \includegraphics[width=3in]{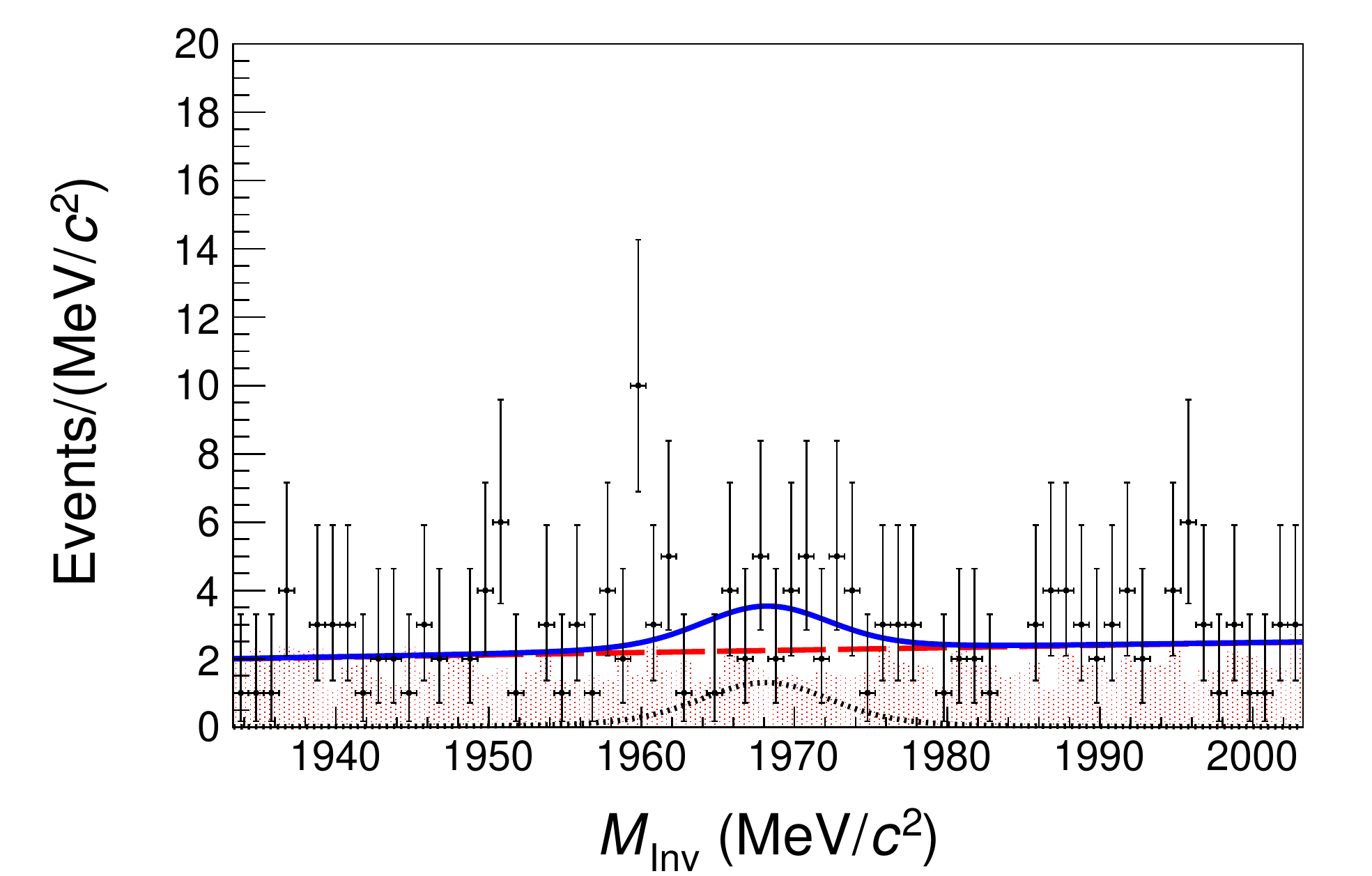}\\
\textbf{RS 950-1000 MeV/$c$} & \textbf{WS 950-1000 MeV/$c$}\\
\includegraphics[width=3in]{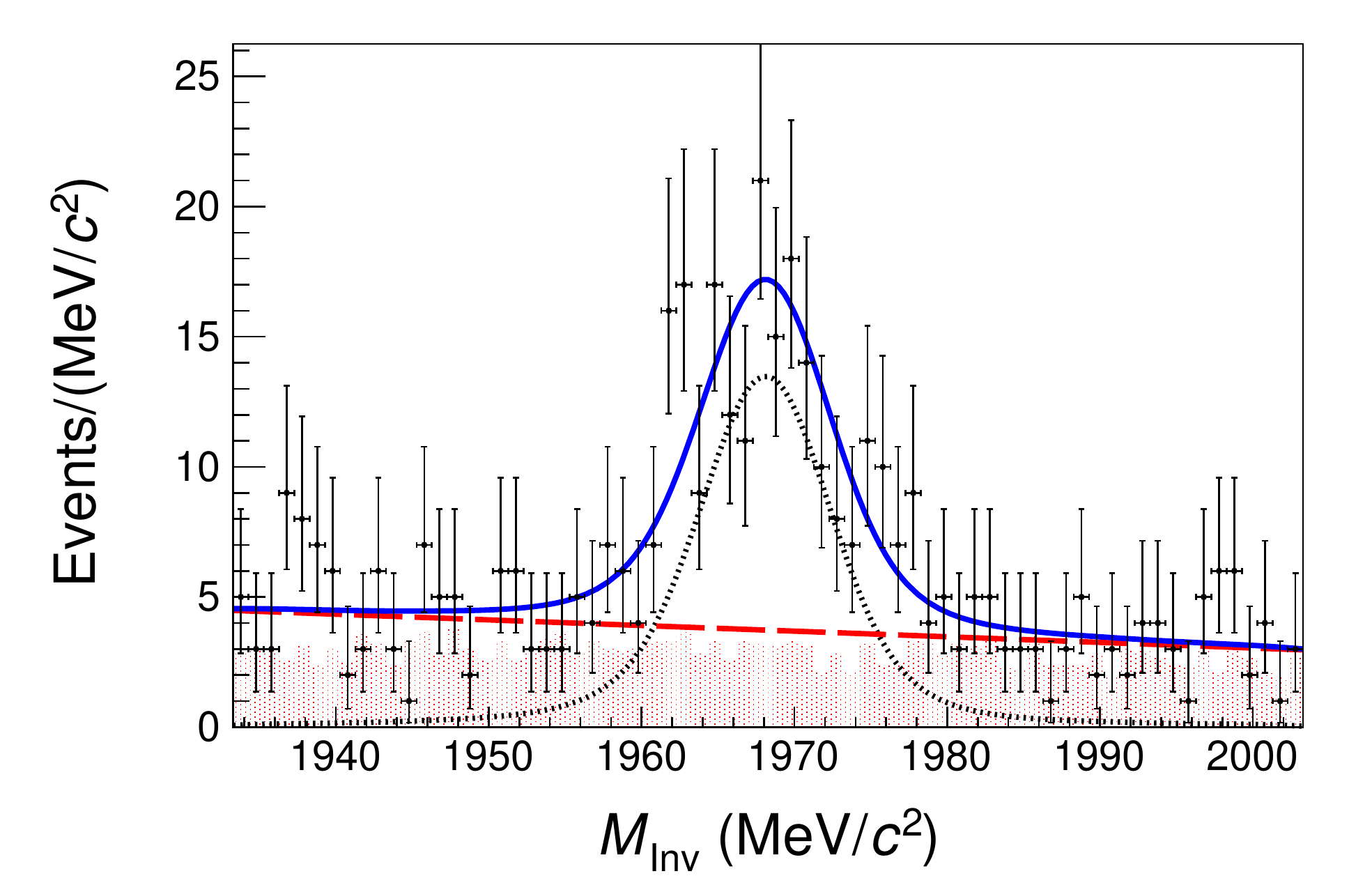} & \includegraphics[width=3in]{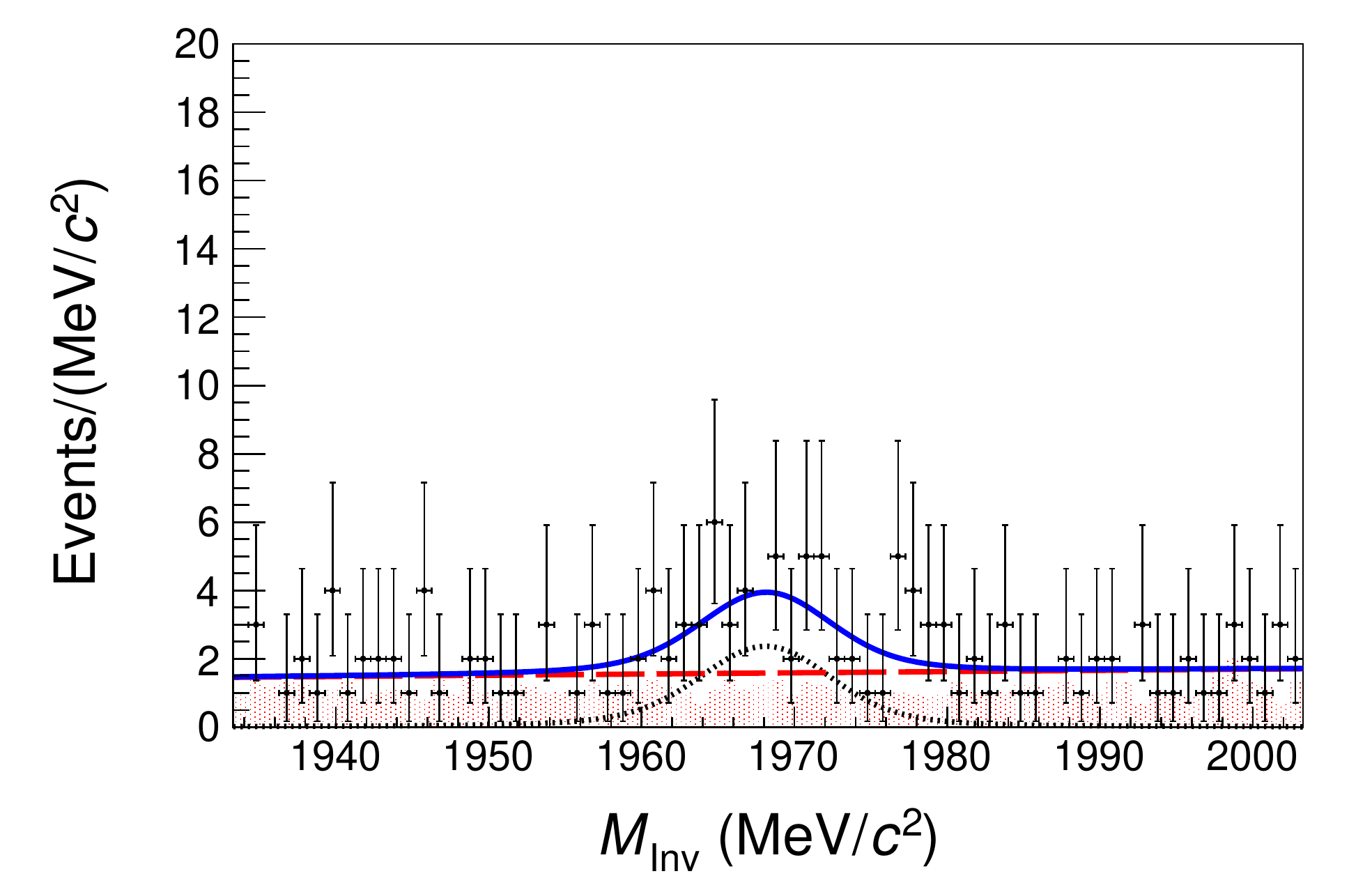}\\
\end{tabular}

\begin{tabular}{cc}
\textbf{RS 1000-1050 MeV/$c$} & \textbf{WS 1000-1050 MeV/$c$}\\
\includegraphics[width=3in]{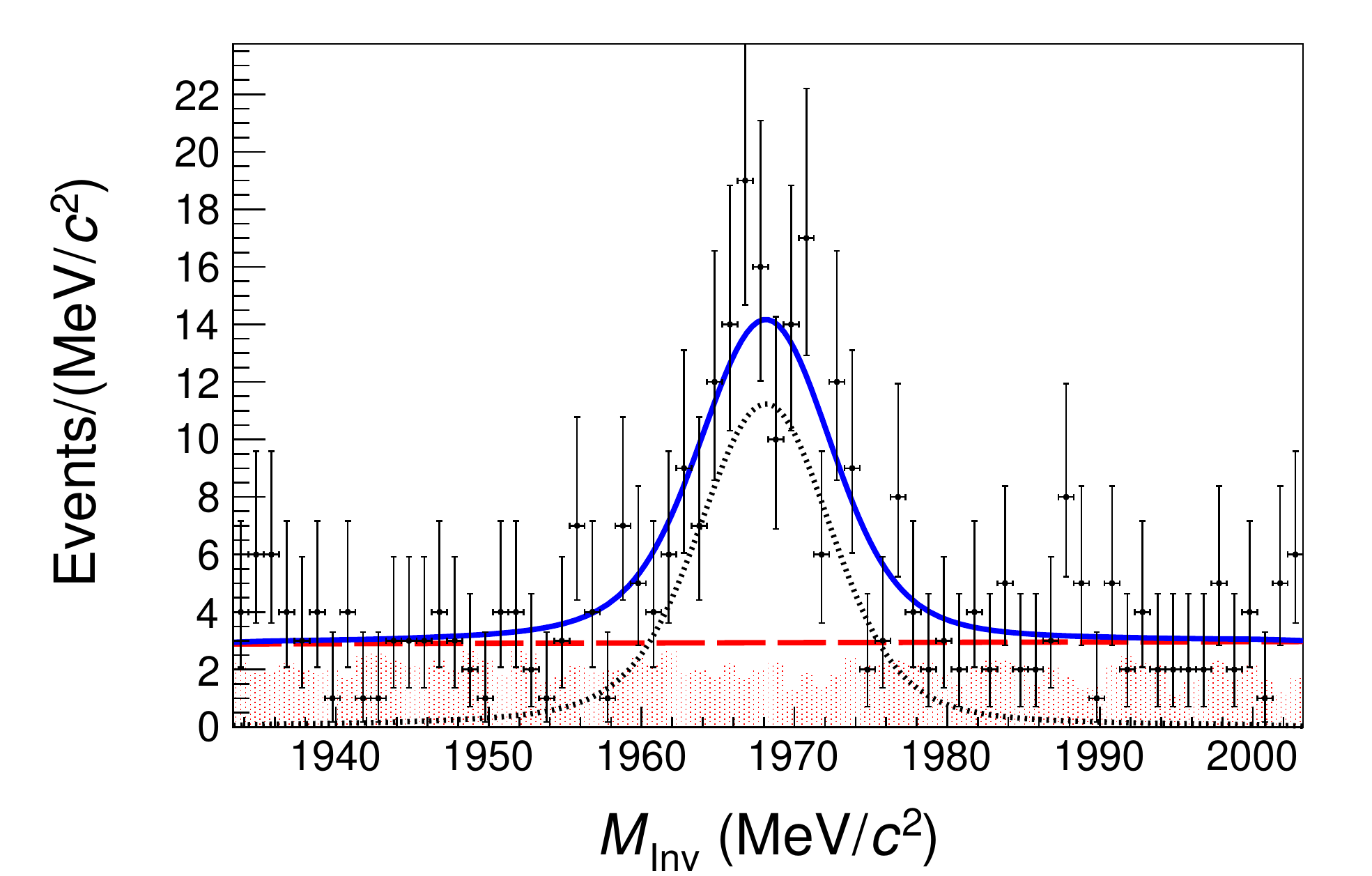} & \includegraphics[width=3in]{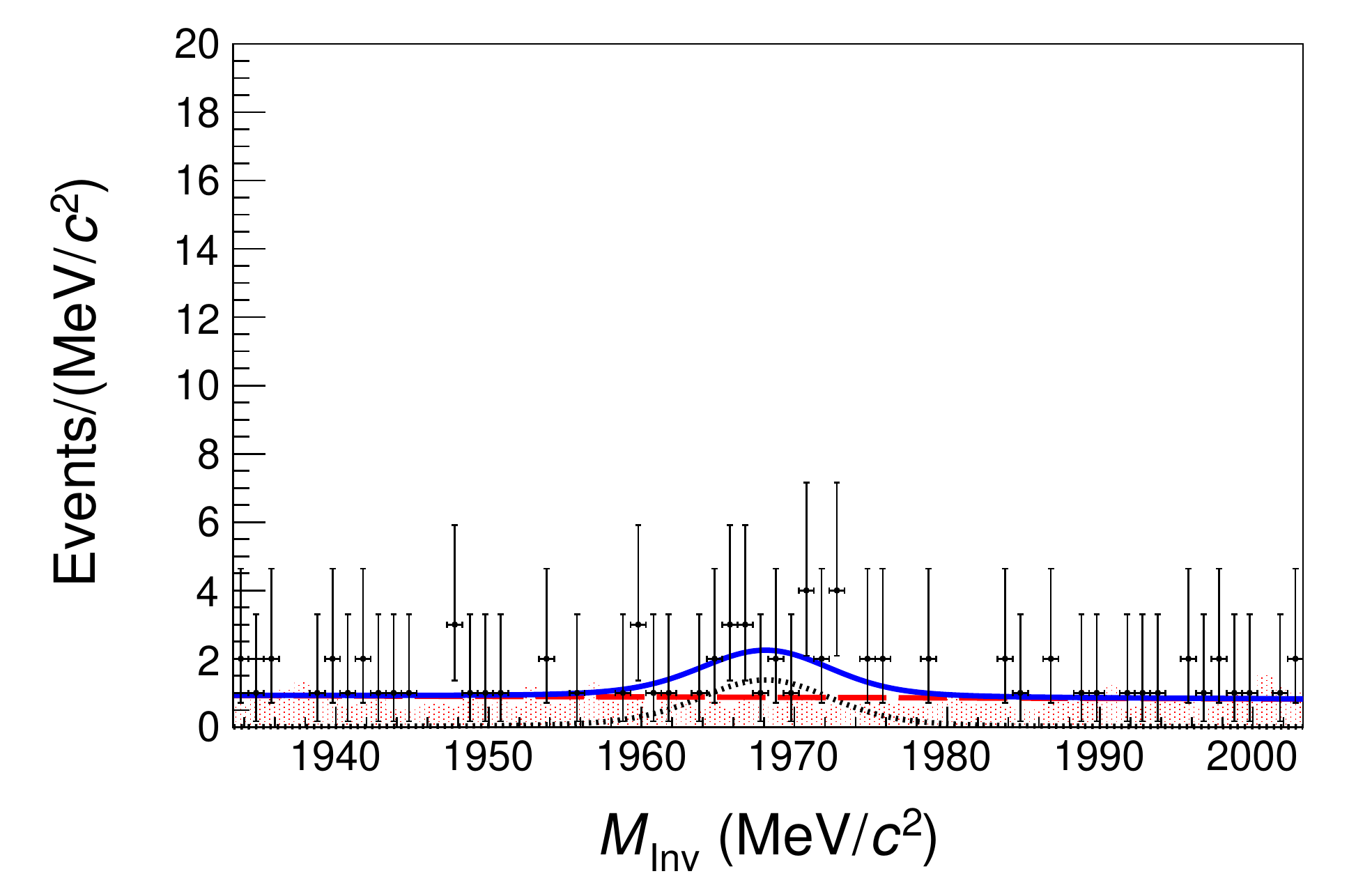}\\
\textbf{RS 1050-1100 MeV/$c$} & \textbf{WS 1050-1100 MeV/$c$}\\
\includegraphics[width=3in]{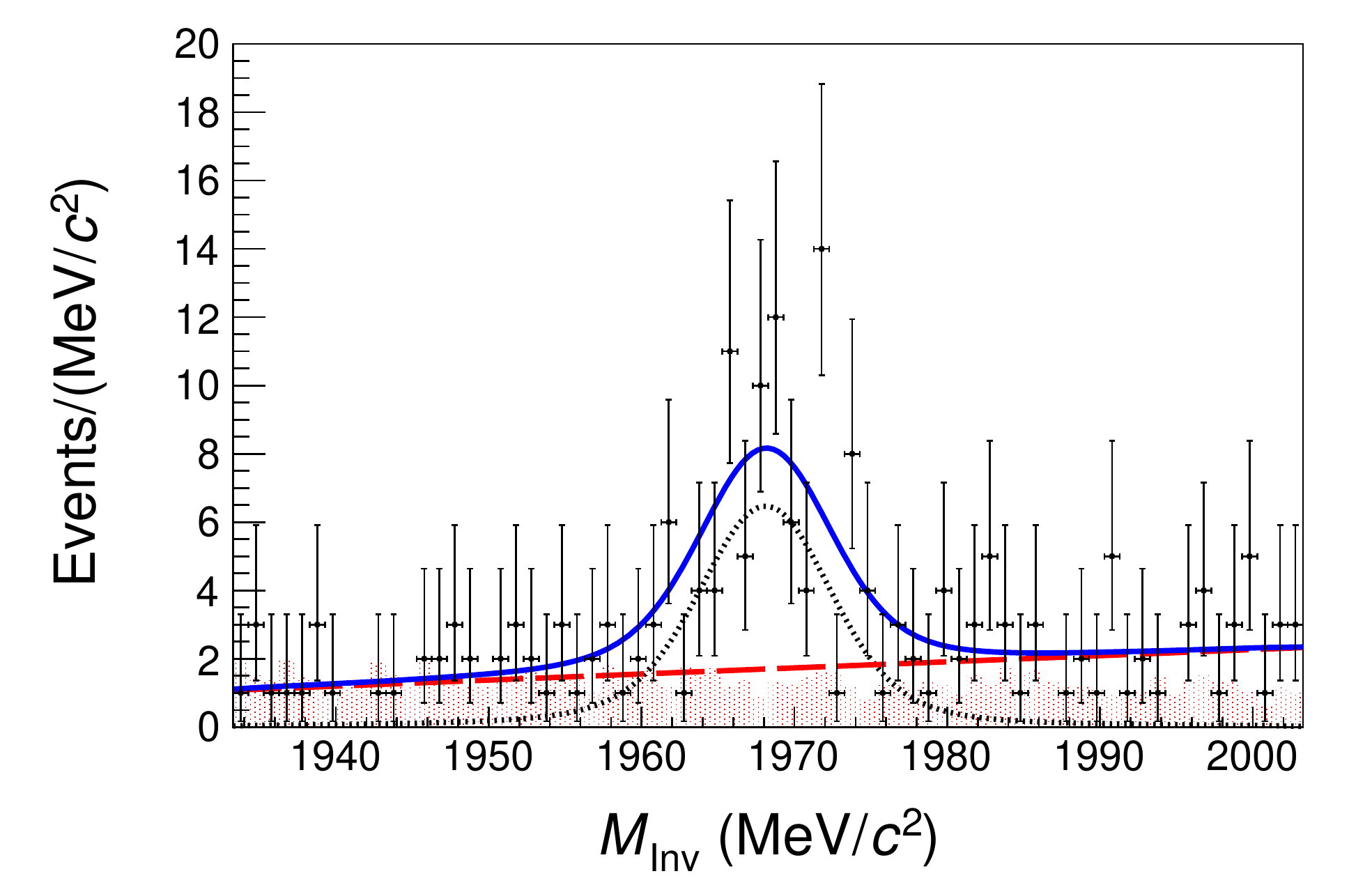} & \includegraphics[width=3in]{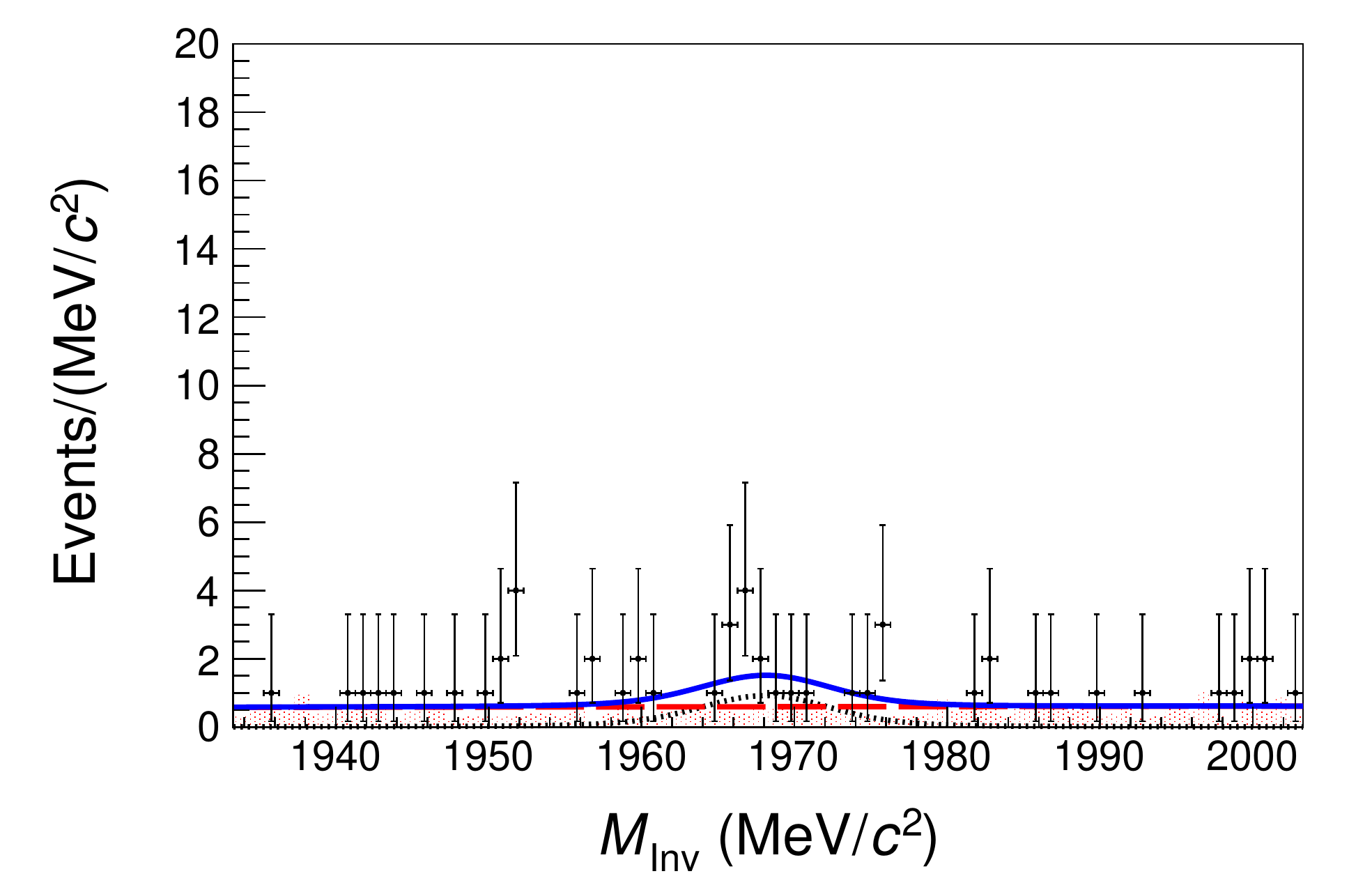}
\end{tabular}

\end{document}